\documentclass[11pt,a4j]{article}

\newif\ifterrapub
 \terrapubfalse  % For Ito's local environment
\newif\ifepsfigure
 \epsfigurefalse % For Ito's local environment  (AI figures)

\long\def\comment#1{}

\usepackage{caption}
 \def\mainword{oscillation}
\def\dualfigwidth{0.990}
\def\dualmedplfigwidth{0.990}
\def\dualmedfigwidth{0.990}
\def\singlefigwidth{0.990}

\def\dualfigwidth{0.990}
\def\dualmedplfigwidth{0.800}
\def\dualmedfigwidth{0.729}
\def\singlefigwidth{0.487}

\long\def\comment#1{}

\newenvironment{mylist}[1]{%
\begin{list}{}{\setlength{\leftmargin}{#1}\setlength{\itemindent}{-#1}}}%
{\end{list}}

\makeatletter
\def\DD{\@ifnextchar [{\D@D}{\D@D[]}}
\def\D@D[#1]#2#3{\frac{d^{#1} #2}{d #3{}^{#1}}}
\def\DP{\@ifnextchar [{\D@P}{\D@P[]}}
\def\D@P[#1]{\@ifnextchar [{\D@Pi[#1]}{\D@PD[#1]}}
\def\D@PD[#1]#2#3{\frac{\partial^{#1} #2}{\partial #3{}^{#1}}}
\def\D@Pi[#1][#2]#3#4{\left( \frac{\partial^{#1} #3}{\partial #4{}^{#1}}\right)_{#2}}
\makeatother
\def\DPsd#1#2#3{\frac{\partial^2 #1}{\partial #2 \partial #3}}

\makeatletter
\def\Hline{%
\noalign{\ifnum0=`}\fi\hrule \@height 0.75pt \futurelet
\reserved@a\@xhline}
\makeatother

\def\ktext{\text{\usefont{T1}{pnc}{m}{it}k}}

\ifterrapub
\else
\fi

\ifterrapub
 \usepackage{url}
\else
\fi

\usepackage{amsmath,amssymb,mathrsfs}

\ifterrapub
\else
\fi

\usepackage{bm}

\usepackage{bbm}

\ifterrapub
\else
 \usepackage[OT2,T1]{fontenc}
 \usepackage[english,russian]{babel}
\fi

\newcommand{\Subr}[1]{}  % for subr2005

\usepackage[pdftex]{graphicx}
\usepackage{times} % <-- This causes the bad proportion of the sffamily fonts!
\usepackage{natbib}
\bibpunct{(}{)}{;}{a}{,}{,}

\usepackage{array,multirow}

\def\supinfo#1{\textrm{Supplementary Information {#1}}}

 \def\mtxtsf#1{{\textsf{#1}}}

 \newcommand{\mysymfigO}{{Fig. }}
 \newcommand{\mysymfigS}{{Figs. }}

 \newcommand{\Oalsqr}{O\left(\alpha^2\right)}
 \newcommand{\Oalcub}{O\left(\alpha^3\right)}
 \newcommand{\Oalqua}{O\left(\alpha^4\right)}
 \newcommand{\Oalhex}{O\left(\alpha^6\right)}
 \newcommand{\Oaloct}{O\left(\alpha^8\right)}

 \newcommand{\Oaldsqr}{O\bigl({\alpha '}^2\bigr)}
 \newcommand{\Oaldcub}{O\bigl({\alpha '}^3\bigr)}
 \newcommand{\Oaldpen}{O\bigl({\alpha '}^5\bigr)}

\newcommand{\tpspcD}{\;\;}
\newcommand{\tpspcE}{\;\;\;\;}

\def\dualfigwidth{0.850}
\def\dualmedplfigwidth{0.850}
\def\dualmedfigwidth{0.850}
\def\singlefigwidth{0.850}

\usepackage[en-US]{datetime2}

\usepackage{color}
\usepackage{hyperref}
\usepackage{xcolor}
\hypersetup{
    colorlinks,
    linkcolor={red!50!black},
    citecolor={blue!50!black},
    urlcolor={blue!80!black}
}
\definecolor{bubbles}{rgb}{0.91, 1.0, 1.0}

\makeatletter
\renewenvironment{quotation}
               {\list{}{\listparindent   0em%
                \itemindent\listparindent \topsep 2.5ex plus 1ex minus .2ex
                \rightmargin\leftmargin  \itemsep 2.5ex plus 1ex minus .2ex
                \parsep        \z@ \@plus\p@}%
                \item\relax}
               {\endlist}
\makeatother

\usepackage{framed}

\usepackage{multicol}

\begin{document}
\selectlanguage{english}

\setcounter{page}{1}
\title{The Lidov--Kozai Oscillation and Hugo von Zeipel${}^\ast$}
\author{Takashi Ito${}^1$ \and Katsuhito Ohtsuka${}^2$}
\date{%
{\small
    $^1$National Astronomical Observatory, Osawa 2--21--1, Mitaka, Tokyo 181--8588, Japan\\%
    $^2$Tokyo Meteor Network at Ohtsuka Dental Clinic, Daisawa 1--27--5, Setagaya, Tokyo 155--0032, Japan
}
}

\selectlanguage{english}

\ifterrapub
\else
 \selectlanguage{english}
\fi

\label{firstpage}
\maketitle

\begin{abstract}
The circular restricted three-body problem,
particularly its doubly averaged version,
has been very well studied in celestial mechanics.
Despite its simplicity, 
circular restricted three-body systems are suited for modeling
the motion of various objects in the solar system,
extrasolar planetary systems, and in many other dynamical systems
that show up in astronomical studies.
In this context,
the so-called \citeauthor{lidov1961}--\citeauthor{kozai1962b} {\mainword}
is well known and applied to various objects.
This makes the orbital inclination and eccentricity of the perturbed body
in the circular restricted three-body system
oscillate with a large amplitude under certain conditions.
It also causes a libration of the perturbed body's argument of pericenter around stationary points.
It is widely accepted that the theoretical framework of this phenomenon
was established independently in the early 1960s 
by a Soviet Union dynamicist (Michail L'vovich Lidov) and
by a Japanese celestial mechanist (Yoshihide Kozai).
Since then, the theory has been extensively studied and developed.
A large variety of studies has stemmed from the original works
by \citeauthor{lidov1961} and \citeauthor{kozai1962b},
now having the prefix of
``\citeauthor{lidov1961}--\citeauthor{kozai1962b}'' or
``\citeauthor{kozai1962b}--\citeauthor{lidov1961}.''
However,
from a survey of past literature published in late nineteenth to early twentieth century,
we have confirmed that there already existed a pioneering work
using a similar analysis of this subject established in that period.
This was accomplished by
a Swedish astronomer,
Edvard Hugo von Zeipel.
In this monograph, we first outline the basic framework of the
circular restricted three-body problem
including typical examples where the
\citeauthor{lidov1961}--\citeauthor{kozai1962b} {\mainword} occurs.
Then, we introduce what was discussed and learned along this line of studies
from the early to mid-twentieth century by summarizing the major works
of \citeauthor{lidov1961}, \citeauthor{kozai1962b}, and relevant authors.
Finally, we make a summary of \citeauthor{vonzeipel1910}'s work, and show
that his achievements in the early twentieth century
already comprehended most of the fundamental and necessary formulations that the 
\citeauthor{lidov1961}--\citeauthor{kozai1962b} {\mainword} requires.
By comparing the works of
\citeauthor{lidov1961}, \citeauthor{kozai1962b}, and
\citeauthor{vonzeipel1910},
we assert that the prefix
``\citeauthor{vonzeipel1910}--\citeauthor{lidov1961}--\citeauthor{kozai1962b}''
should be used for designating this theoretical framework,
and not just
    \citeauthor{lidov1961}--\citeauthor{kozai1962b}  or
    \citeauthor{kozai1962b}--\citeauthor{lidov1961}.
This justifiably shows due respect and appropriately
commemorates these three major pioneers who made significant contributions
to the progress of modern celestial mechanics.
\end{abstract}

\renewcommand{\thefootnote}{\fnsymbol{footnote}}
\footnotetext[1]{%
\noindent
This paper was accepted for publication in
\textit{Monographs on Environment, Earth and Planets\/} on December 4, 2018.
\par
\hspace*{0.5em}
Corresponding author: Takashi Ito (\textsf{tito.geoph.s@95.alumni.u-tokyo.ac.jp})
}
\renewcommand{\thefootnote}{\arabic{footnote}}

\section{Introduction\label{sec:intro}}
Solar system dynamics
has a large diversity of aspects.
We know that it encompasses many complicated and unsolved problems.
But we also know that it is filled with rich and interesting characteristics of
nonlinear dynamical systems.
In spite of the general complexity of solar system dynamics,
it is also true that the orbital motion of many of the solar system objects
can be fairly well approximated by perturbed Keplerian motion, and
the magnitude of perturbation is usually moderate or small.
This is due to the existence of the very strong gravity from a massive central body, the Sun.
The major source of the gravitational perturbation against the two-body
Keplerian motion is the planets.

Having the feature of this kind as a background,
the restricted three-body problem (hereafter referred to as R3BP),
a variant of the general three-body problem,
often becomes a good proxy in solar system dynamics.
In R3BP, the mass of one of the three bodies is assumed to be so small that
it does not affect the motion of the other two bodies at all.
Therefore, R3BP is particularly appropriate when we deal with
the orbital motion of small objects
(such as asteroids, comets, transneptunian objects, natural and
 artificial satellites)
under the perturbation resulting from the major planets.

When the massive two bodies compose a circular binary in R3BP,
the problem is particularly called the circular restricted three-body problem
(hereafter referred to as CR3BP).
In spite of its very simple setting,
we can still use CR3BP as a good proxy in many cases in solar system dynamics.
This is mainly due to the moderate to very small eccentricity of the major planets
in the current solar system.
Thanks to its simple configuration and small degrees of freedom,
CR3BP has played an important role in the development of
analytic perturbation theories, and it is still applied to
many subjects in modern celestial mechanics.
As we will see later in greater detail, we can reduce the degrees of freedom of
CR3BP into unity through the double averaging procedure of
the disturbing function (a function that represents the perturbing force).
This makes the system integrable, and
it enables us to obtain a global picture of the perturbed (third) body's motion,
even when the perturbed body's eccentricity or inclination is
substantially large.

Based on the integrable characteristics of the doubly averaged CR3BP,
the theory of the so-called \citeauthor{lidov1961}--\citeauthor{kozai1962b} {\mainword} has emerged.
This is the major subject of this monograph.
Chronologically speaking,
a Soviet Union dynamicist, Michail L'vovich Lidov,
found in \citeyear{lidov1961} that,
when dealing with the motion of Earth-orbiting satellites under the
perturbation from other objects as CR3BP,
the satellites' argument of pericenter can librate around $\pm \frac{\pi}{2}$
when their initial orbital inclination is larger than a certain value.
The eccentricity and inclination of the perturbed body exhibit a
synchronized periodic oscillation under this circumstance.
Almost at the same time, 
a Japanese celestial mechanist, Yoshihide Kozai,
dealt with the motion of asteroids orbiting inside Jupiter's orbit as CR3BP,
and found in \citeyear{kozai1962b} that,
an asteroid's argument of perihelion can librate around $\pm \frac{\pi}{2}$
when its initial orbital inclination is larger than a certain value.
The two works by \citeauthor{lidov1961} and \citeauthor{kozai1962b}
turned out to be theoretically equivalent, and
the dynamical phenomenon is now collectively referred to as
the \citeauthor{lidov1961}--\citeauthor{kozai1962b} {\mainword}.
Note that
although we basically use the term ``{\mainword}'' in the present monograph,
many other different terms have been used for the same phenomenon
in the literature, such as
``mechanism,'' ``resonance,'' ``cycle,'' ``effect,'' and so on.
See Section \ref{sssec:mainword} for more detail about the choice of terms.

After the series of publications by
\citeauthor{lidov1961} and \citeauthor{kozai1962b} in the 1960s,
this dynamical phenomenon became better known, and found applications
in the fields of astronomy, planetary science, and astronautics.
Their theories have been applied not only to
the long-term motion of Earth-orbiting satellites that \citeauthor{lidov1961} considered or 
the secular asteroidal dynamics that \citeauthor{kozai1962b} pursued in their era,
but also to the motion of other solar system objects
such as irregular satellites, various comets, 
near-Earth asteroids, and transneptunian objects.
In particular,
the discovery of extrasolar planets and their orbital configurations
of a great variety resulted in a recognition that the \citeauthor{lidov1961}--\citeauthor{kozai1962b} {\mainword} has played a significant role in the evolution of these dynamical systems.
The development of the \citeauthor{lidov1961}--\citeauthor{kozai1962b} {\mainword} still goes on,
incorporating higher-order perturbations and more subtle and complicated
physical effects such as general relativity and the combination with mean motion resonances.
Recent application of the \citeauthor{lidov1961}--\citeauthor{kozai1962b} {\mainword} even extends to 
stellar dynamics, and its theory is employed for explaining various problems such as formation of some kind of binaries and triple star systems,
merger mechanism of binary black holes,
modeling the galactic tide,
and so forth.

Part of the purpose of this monograph is to outline
how the \citeauthor{lidov1961}--\citeauthor{kozai1962b} {\mainword} works,
who developed the theory, and in what way.
We will briefly mention what kind of applications have been
considered since the era of \citeauthor{lidov1961} and \citeauthor{kozai1962b}
up to the present.
However, this is not our predominant aim.
In this monograph,
we wish to draw attention to the fact 
that a pioneering work in this line of study had been carried out long before the era of
\citeauthor{lidov1961} and \citeauthor{kozai1962b}.
More specifically, most of the basic ingredients that
\citeauthor{lidov1961} or \citeauthor{kozai1962b} presented for the doubly averaged CR3BP,
including the necessary condition of argument of pericenter's libration,
had been already recognized, quantitatively investigated, and published
on journals by \citeyear{vonzeipel1910}.
This was accomplished by a Swedish astronomer, Hugo von Zeipel.

As far as our investigation shows,
the work by \citeauthor{vonzeipel1910} in \citeyear{vonzeipel1910}
has been ignored and buried in oblivion for a long time,
regardless of its substantial significance and
excellent foresight in solar system dynamics.
The major purpose of this monograph is to validate the correctness of \citeauthor{vonzeipel1910}'s work, and
to redirect the attention of the relevant communities
to this pioneering study that was established and published
at the beginning of the twentieth century.

The complete table of contents for this monograph is in its online version.
\supinfo{1} also gives the same table with specific page number information.
For readers who do not particularly specialize
in the dynamical aspects of astronomy or planetary science,
Section \ref{sec:CR3BP} summarizes what CR3BP is and what kind of phenomena
the \citeauthor{lidov1961}--\citeauthor{kozai1962b} {\mainword} causes,
employing simple numerical demonstrations.
In the following sections,
we will review the achievements of \citeauthor{lidov1961} and \citeauthor{kozai1962b} by summarizing two classic papers:
\citet{kozai1962b} in Section \ref{sec:kozai1962b}, and
\citet{lidov1961}  in Section \ref{sec:lidov1961}.
We will also browse through an earlier work on CR3BP
in the former Soviet Union \citep{moiseev1945a,moiseev1945b}
in Section \ref{sec:lidov1961}.
Finally in Section \ref{sec:vonzeipel},
we summarize \citeauthor{vonzeipel1910}'s work published in \citeyear{vonzeipel1910}.
This is the kernel of this monograph.
Section \ref{sec:discussion} presents discussions, but it also includes
additional matters of even earlier works by \citeauthor{vonzeipel1910}.
Readers that are already familiar with the work by \citeauthor{kozai1962b} or
\citeauthor{lidov1961} may want to skip Sections
\ref{sec:CR3BP}, \ref{sec:kozai1962b}, \ref{sec:lidov1961}, and
proceed straight to Section \ref{sec:vonzeipel}.
Yet, readers should note that
in Sections \ref{sec:vonzeipel} and \ref{sec:discussion}
we often refer to facts, equations, and figures described in
Sections \ref{sec:CR3BP}, \ref{sec:kozai1962b}, \ref{sec:lidov1961}.
Also, note that
Sections \ref{sec:kozai1962b}, \ref{sec:lidov1961}, and \ref{sec:vonzeipel}
are not placed in chronological order.
We place them in the order that these works gained recognition.
Although \citet{lidov1961} was published earlier than \citet{kozai1962b},
\citeauthor{kozai1962b}'s work began gaining attention earlier than \citeauthor{lidov1961}'s work.
Whereas
\citeauthor{vonzeipel1910}'s work was published much earlier than the others,
it has not been recognized to this day.

In this monograph
we basically use the conventional notation for the Keplerian orbital elements:
$a$ for semimajor axis, $e$ for eccentricity, and $I$ or $i$ for inclination.
As for argument of pericenter and longitude of ascending node,
we prefer the notation used for the Delaunay elements ($g$ and $h$, respectively),
rather than the conventional ones ($\omega$ and $\Omega$).
But sometimes we use $\omega$ and $\Omega$,
particularly in Section \ref{sec:lidov1961},
because \citeauthor{lidov1961} uses $\omega$ and $\Omega$, not $g$ and $h$.
We use the standard notations $L, G, H$ for the actions of the Delaunay elements.
We use $l$ for mean anomaly, and $n$ for mean motion.

Note that there are several notations in this monograph that may cause confusion among readers. Examples are:
\begin{itemize}
\item In \citeauthor{vonzeipel1910}'s work,
therefore in our Section \ref{sec:vonzeipel},
he uses the symbol $\Theta$ for denoting one of the Delaunay elements
$(\Theta = G \cos i)$, instead of the conventional notation $H$.
He uses the symbol $H$ for denoting another, different angle in his work.
\item In \citeauthor{kozai1962b}'s work,
therefore in our Section \ref{sec:kozai1962b},
$\Theta$ is used for denoting an important, but totally different quantity.
\item ${\cal H}$, a calligraphic style of $H$, is sometimes used for denoting Hamiltonian in this monograph
(e.g. Eq. \eqref{eqn:def-H-3BP-general}).
\item ${\cal G}$, a calligraphic style of $G$, denotes the gravitational constant in this monograph
(e.g. Eqs. \eqref{eqn:eom-relative-1} and \eqref{eqn:eom-relative-2}).
\end{itemize}
For avoiding potential clutter and confusion,
we try to give definitions of the variables used in this monograph
as clearly as possible
whenever they first show up, or
whenever they are used in different meanings than before.

As for the equation numbering,
we try to follow the ways used in the original literature as much as possible.
More specifically, 
when we cite equations that show up in one of the following literature,
we give them the following designations in this monograph:
``K''  for \citet{kozai1962b},
``L''  for \citet{lidov1961},
``Mb'' for \citet{moiseev1945b}, or
``Z''  for \citet{vonzeipel1910}
$+$
equation number in the original publication
$+$ ``-'' $+$
sequential equation number in this monograph.
Here are some examples of our equation numbering:
\begin{itemize}
\item Eq. (28)   in \citet{kozai1962b}    $\to$ \eqref{eqn:K28}
\item Eq. (2.16) in \citet{moiseev1945b}  $\to$ \eqref{eqn:Mb45-2.16}
\item Eq. (7)    in \citet{lidov1961}     $\to$ \eqref{eqn:L07}
\item Eq. (103)  in \citet{vonzeipel1910} $\to$ \eqref{eqn:Z103}
\end{itemize}
Other equations in this monograph that do not have 
a leading K, L, M, or Z in their equation number are either
those which do not appear in the above literature, or
those which        appear without equation number in the above literature.
Also, sometimes we cite page numbers, section and subsection numbers, and
chapter numbers of the above literature in the same manner such as
``p. K592''    (designating p. 592 of \citet{kozai1962b}) or
``Section Z1'' (designating Section 1 of \citet{vonzeipel1910}).

In this monograph
we cite many publications written in non-English language such as French, German, Swedish, Russian, and Japanese.
We basically reproduced their bibliographic information using their original language in the References section (p. \pageref{pg:beginbibliography}--\pageref{pg:endbibliography}).
However, % unfortunately,
  use of the Cyrillic alphabets and the Japanese characters is prohibited in the main body of the monograph 
  due to a technical limitation about font in the {\LaTeX} typesetting process by the publisher.
Because of this,
for listing the literature that uses the Cyrillic and Japanese characters in the Reference section,
we translated their bibliographic information into English.
But
we believe that the bibliographic information of the non-English literature
written in their original language is quite valuable for the readers of this monograph.
Another point to note is that,
some hyperlinks to the Uniform Resource Locators (URLs)
embedded in the References section do not properly function,
although the URLs themselves are correct.
This is due to another technical limitation in this monograph's {\LaTeX} typesetting process.
For these two reasons
we have created a more complete, alternative bibliography for this monograph using the Cyrillic and Japanese characters with fully functional hyperlinks.
We put it in \supinfo{2} which is free from the technical limitations.
\label{pg:reasonforanotherbib}

This monograph minimizes the use of URLs in the text mainly
due to the technical limitation of embedded hyperlinks mentioned above.
We also wanted to avoid clutter by having many complicated URLs that often become sources of distractions.
Instead,
in \supinfo{3}
we made a list of the URLs of relevant websites that we mention in this monograph,
such as orbital databases of the small solar system bodies.
On the other hand,
most of the literature listed in the References section
of this monograph are accompanied by explicit URLs that are hyperlinked to
each of their online resources.

Now that the \citeauthor{lidov1961}--\citeauthor{kozai1962b} {\mainword} has gained a great popularity,
a number of good review papers and textbooks that deal with the fundamentals
and applications of this phenomenon in substantial depth have been published
\citep[e.g.][]{morbidelli2002,merritt2013,davies2014,naoz2016}.
A doctoral dissertation \citep{antognini2016} that entirely devotes itself
to the study of this phenomenon is also publicly available.
Readers who have a deeper interest in the
\citeauthor{lidov1961}--\citeauthor{kozai1962b} {\mainword}, and
those   who want to seek further applications of the theory,
can consult these works.
In addition, a textbook that totally dedicates itself to the
\citeauthor{lidov1961}--\citeauthor{kozai1962b} {\mainword}
has been recently published \citep{shevchenko2017}.
As expected,
we found that the contents of some part of this monograph
(particularly Sections \ref{sec:kozai1962b} and \ref{sec:lidov1961})
overlap with \citet{shevchenko2017}.
We included these two sections in this monograph to state our own view of
what \citeauthor{kozai1962b} and \citeauthor{lidov1961} achieved,
as well as what they did not,
in the light of \citeauthor{vonzeipel1910}'s publications.
In other words, the major purpose of this monograph is to sketch and highlight the similarities and differences
between the works of \citeauthor{kozai1962b}, \citeauthor{lidov1961},
and \citeauthor{vonzeipel1910}.
Hence we need our Sections \ref{sec:kozai1962b} and \ref{sec:lidov1961}.

\begin{figure*}[t]\centering
\ifepsfigure
  \includegraphics[width=.7\textwidth]{cr3bp_schem.eps} %fig1
\else
 \includegraphics[width=\dualmedplfigwidth\textwidth]{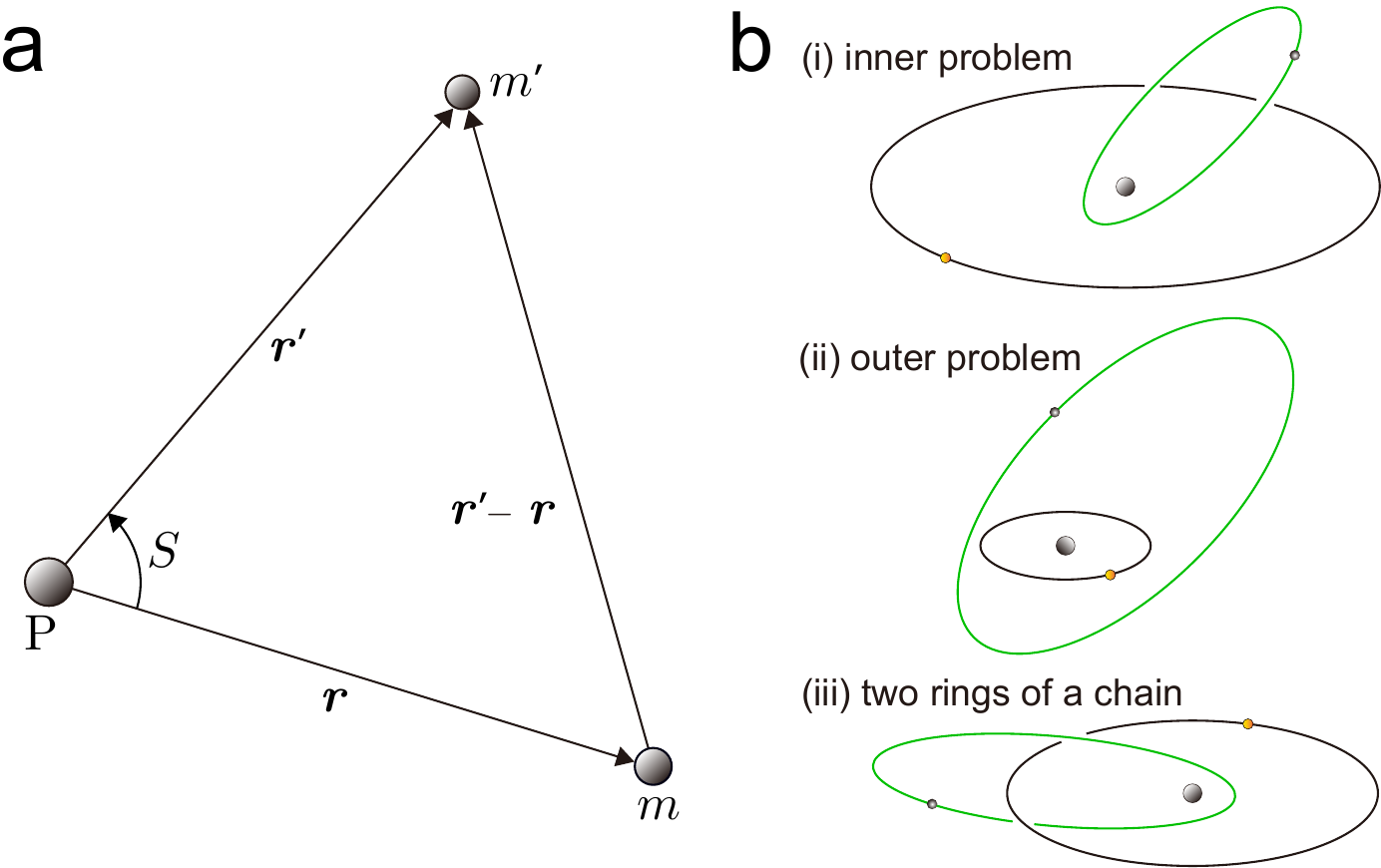} %fig1
\fi
  \caption{%
  Schematic illustrations of the three-body system that we consider in this monograph.
  \mtxtsf{a}: Relative geometric configuration of a three-body system centered at the primary mass (P).
  See Section \protect\ref{ssec:CR3BP-R} for the meanings of the symbols ($\bm{r}$, $\bm{r'}$, $m$, $m'$, $S$).
  \mtxtsf{b}: Three typical patterns of the restricted three-body problem.
\mtxtsf{(i)}   The inner problem where the perturbed body's orbit (the green ellipse) lies inside  the orbit of the perturbing body (the black ellipse).
\mtxtsf{(ii)}  The outer problem where the perturbed body's orbit (the green ellipse) lies outside the orbit of the perturbing body (the black ellipse).
\mtxtsf{(iii)} When two bodies behave like the rings of a chain.
  Note that the phrase ``like the rings of a chain'' is adopted from \citet[][``comme les anneaux d'une cha{\^\i}ne'' on p. 378, and ``comme deux anneaux d'une cha{\^\i}ne'' on p. 413]{vonzeipel1910}.
}
  \label{fig:CR3BP-schematic}
\end{figure*}

\section{Preliminaries: What We Consider\label{sec:CR3BP}}
This section presents a simple illustration of the system that we
deal with in this monograph---the circular restricted three-body problem.
Our intention is to facilitate readers' understanding of what we discuss in later sections.

The two-body problem is integrable, and has an exact analytic
solution---the Keplerian motion described by various conic sections.
However, just by adding one more mass to the system,
the system ceases to have such a general solution.
The three-body problem is not integrable,
and we have no exact analytic solution except in very few special cases.
This fact was already known at the end of the nineteenth century.
\citet{bruns1887} proved the algebraic non-integrability of the general three-body problem.
\citet{poincare1890} soon gave a proof of the non-existence of an integral
in the restricted three-body system: The analytic non-integrability of
the restricted three-body problem was proven.
However, we should recall that
it is this very non-integrability of the three-body problem that
has resulted in a large number of interesting and important aspects of
nonlinear dynamics published in vast amount of the past literature,
such as collisional singularities, periodic orbits, resonances, and chaos.
It is also the reason why the three-body problem has attracted
many scientists from a variety of fields over a long time,
yielding a great deal of achievement.
For the modern progress of the three-body problem in general,
readers can consult many literature
\citep[e.g.][]{valtonen2006,valtonen2016,musielak2014}.
A short summary of the development of studies of the three-body problem
during the late nineteenth and twentieth centuries is
available in \citet[][their Section 3]{ito2007c}.

The dynamics of a three-body system is sometimes highly chaotic.
However, it can also be very regular and nearly integrable,
depending on the mass ratio and the initial orbit configurations
between the three bodies.
Fortunately in the current solar system,
a nearly-integrable hierarchical three-body system often becomes a good proxy
of dynamics.
A hierarchical three-body system comprises
a massive central primary     (the mass $m_0$) accompanied by
a less massive secondary      (the mass $m_1 < m_0$), as well as
an even less massive tertiary (the mass $m_2 < m_1$).
The tertiary mass orbits inside or outside the $(m_0,m_1)$ binary.
Unless
the orbit of the secondary around the primary and
that of      the tertiary  around the primary
get too close or intersect each other,
the two binaries, $(m_0,m_1)$ and $(m_0,m_2)$,
usually behave in the nearly integrable manner
(i.e. close to the Keplerian motion).
In that case,
we can principally obtain their orbital solution through perturbation methods.
The $(m_0,m_1)$ binary would make a pure Keplerian motion if $m_2 \ll m_1$,
not being disturbed by $m_2$ at all,
while the motion of $m_2$ is affected by the $(m_0,m_1)$ binary.
This is the restricted three-body problem (R3BP).
In particular, when the orbit of $m_1$ in the $(m_0,m_1)$ binary is circular,
the system results in the circular restricted three-body problem (CR3BP).
Note that the restricted three-body problem is often dealt with
in a rotating coordinate system where massive bodies ($m_0$ and $m_1$)
always stay on the $x$-axis \citep[e.g.][]{quarles2012}.
However, we do not adopt the rotating coordinate system in this monograph.
Readers can consult \citet{szebehely1967} for more general and detailed
characteristics of the restricted three-body problems,
particularly those considered in a rotating frame.

\subsection{Relative motion\label{ssec:CR3BP-relative}}
Let us briefly summarize how the basic equations of motion that describe
the dynamics of a three-body system are derived in a standard way.
For making our descriptions in this monograph consistent with
conventional literature,
we use the relative coordinates centered on $m_0$,
instead of the Jacobi coordinates \citep[e.g.][]{plummer1960,wisdom1991} or
other canonical coordinates.
The discussion in this subsection follows
\citet[][Chapters X and XII]{brouwer1961},
\citet[][Subsections 9.4 and 11.12]{danby1992},
\citet[][Subsection 6.2]{murray1999}, and
\citet[][Subsection 4.8]{merritt2013} on the whole.

We write the position vectors of the three bodies with masses
$m_0$, $m_1$, $m_2$ with respect to a fixed origin in the inertial
reference frame as
 $\bm{\xi}_0, \bm{\xi}_1, \bm{\xi}_2$.
In addition, we denote
the relative position vector of the secondary mass $m_1$
with respect to the primary mass $m_0$ as
\begin{equation}
  \bm{r}_1 \equiv \bm{\xi}_1 - \bm{\xi}_0
= \left(
    \begin{array}{c}
      x_1 \\
      y_1 \\
      z_1
    \end{array}
  \right) .
  \label{eqn:def-r1-relative}
\end{equation}
Similarly,
the relative position vector of the tertiary mass $m_2$
with respect to the primary mass $m_0$ is denoted as
\begin{equation}
  \bm{r}_2  \equiv \bm{\xi}_2 - \bm{\xi}_0
= \left(
    \begin{array}{c}
      x_2 \\
      y_2 \\
      z_2
    \end{array}
  \right) .
  \label{eqn:def-r2-relative}
\end{equation}
The distance between the secondary and tertiary masses is
\begin{equation}
  \left| \bm{r}_2 - \bm{r}_1 \right| =
  \sqrt{ \left(x_2 - x_1 \right)^2
        +\left(y_2 - y_1 \right)^2
        +\left(z_2 - z_1 \right)^2 } .
\label{eqn:def-r1r2-diff}
\end{equation}

Using
$\bm{\xi}_0, \bm{\xi}_1, \bm{\xi}_2$ and $\bm{r}_1, \bm{r}_2$,
we can determine and reduce the equations of motion of the three bodies in the
inertial reference frame as follows:
\begin{alignat}{1}
  m_0 \DD[2]{\bm{\xi}_0}{t} &=  {\cal G} m_0 m_1 \frac{\bm{r}_1}{r_1^3}
                               +{\cal G} m_0 m_2 \frac{\bm{r}_2}{r_2^3},
\label{eqn:eom-innertial-0} \\
  m_1 \DD[2]{\bm{\xi}_1}{t} &= -{\cal G} m_1 m_0 \frac{\bm{r}_1}{r_1^3}
                               +{\cal G} m_1 m_2 \frac{\bm{r}_2-\bm{r}_1}
                                {\left| \bm{r}_2-\bm{r}_1 \right|^3},
\label{eqn:eom-innertial-1} \\
  m_2 \DD[2]{\bm{\xi}_2}{t} &= -{\cal G} m_2 m_0 \frac{\bm{r}_2}{r_2^3}
                               +{\cal G} m_2 m_1 \frac{\bm{r}_1-\bm{r}_2}
                                {\left| \bm{r}_1-\bm{r}_2 \right|^3} ,
\label{eqn:eom-innertial-2}
\end{alignat}
where ${\cal G}$ is the gravitational constant,
$t$ denotes time,
$r_1 \equiv \left| \bm{r}_1 \right|$, and
$r_2 \equiv \left| \bm{r}_2 \right|$.

Now, let us determine
the equations of motion of the secondary mass
expressed on the relative reference frame centered on the primary mass.
From Eq. \eqref{eqn:def-r1-relative} we have
\begin{equation}
  \DD[2]{\bm{\xi}_1}{t}  = \DD[2]{\bm{r}_1}{t} + \DD[2]{\bm{\xi}_0}{t} .
  \label{eqn:def-r1-accel}
\end{equation}
Similarly,
the equations of motion of the tertiary mass
expressed on the relative reference frame centered on the primary mass are, from Eq. \eqref{eqn:def-r2-relative}
\begin{equation}
  \DD[2]{\bm{\xi}_2}{t}  = \DD[2]{\bm{r}_2}{t} + \DD[2]{\bm{\xi}_0}{t} .
  \label{eqn:def-r2-accel}
\end{equation}
We eliminate $\DD[2]{\bm{\xi}_0}{t}$ from the right-hand sides of
Eqs. \eqref{eqn:def-r1-accel} and \eqref{eqn:def-r2-accel}
using the quantity in the right-hand side of Eq. \eqref{eqn:eom-innertial-0}:
substitution of
$\DD[2]{\bm{\xi}_0}{t} = {\cal G} m_1 \frac{\bm{r}_1}{r_1^3}
                        +{\cal G} m_2 \frac{\bm{r}_2}{r_2^3}$.
Then,
as for the mass $m_1$,
we substitute $\DD[2]{\bm{\xi}_1}{t}$ of Eq. \eqref{eqn:def-r1-accel} into Eq. \eqref{eqn:eom-innertial-1}.
As for the mass $m_2$,
we substitute $\DD[2]{\bm{\xi}_2}{t}$ of Eq. \eqref{eqn:def-r2-accel} into Eq. \eqref{eqn:eom-innertial-2}.
Eventually we obtain their relative equations of motion as:
\begin{alignat}{1}
\DD[2]{\bm{r}_1}{t}
+ {\cal G} \left( m_0 + m_1 \right) \frac{\bm{r}_1}{r_1^3}
  &= {\cal G} m_2 \left(
          \frac{\bm{r}_2-\bm{r}_1}{\left| \bm{r}_2-\bm{r}_1 \right|^3}
         -\frac{\bm{r}_2         }{r_2^3} \right),
  \label{eqn:eom-relative-1} \\
\DD[2]{\bm{r}_2}{t}
+ {\cal G} \left( m_0 + m_2 \right) \frac{\bm{r}_2}{r_2^3}
  &= {\cal G} m_1 \left(
          \frac{\bm{r}_1-\bm{r}_2}{\left| \bm{r}_1-\bm{r}_2 \right|^3}
         -\frac{\bm{r}_1         }{r_1^3} \right) .
  \label{eqn:eom-relative-2}
\end{alignat}

Now it is straightforward to confirm that the terms in the right-hand side of
Eqs. \eqref{eqn:eom-relative-1} and \eqref{eqn:eom-relative-2}
can be rewritten as gradients of certain scalar functions.
Let us write them as $R_1$ and $R_2$.
Their actual forms are as follows:
\begin{alignat}{1}
\DD[2]{\bm{r}_1}{t}
+ {\cal G} \left( m_0 + m_1 \right) \frac{\bm{r}_1}{r_1^3}
  &= \nabla R_1 ,
  \label{eqn:eom-relative-useR-1} \\
\DD[2]{\bm{r}_2}{t}
+ {\cal G} \left( m_0 + m_2 \right) \frac{\bm{r}_2}{r_2^3}
  &= \nabla R_2 ,
  \label{eqn:eom-relative-useR-2}
\end{alignat}
where
\begin{alignat}{1}
  R_1 &\equiv \frac{{\cal G} m_2}{\left| \bm{r}_2 - \bm{r}_1 \right|}
            -{\cal G} m_2 \frac{\bm{r}_1 \cdot \bm{r}_2}{r_2^3},
 \label{eqn:def-R1} \\
  R_2 &\equiv \frac{{\cal G} m_1}{\left| \bm{r}_1 - \bm{r}_2 \right|}
            -{\cal G} m_1 \frac{\bm{r}_2 \cdot \bm{r}_1}{r_1^3} .
 \label{eqn:def-R2}
\end{alignat}

$R_1$ in Eq. \eqref{eqn:def-R1} and
$R_2$ in Eq. \eqref{eqn:def-R2} are called the disturbing function
for the mass $m_1$ and the mass $m_2$, respectively.
They represent the gravitational interaction between $m_1$ and $m_2$.
The major gravitational force exerted from the primary mass $m_0$ is expressed
as the second term in the left-hand side of
Eq. \eqref{eqn:eom-relative-useR-1} or
Eq. \eqref{eqn:eom-relative-useR-2}.
If we ignore the disturbing function $R_1$ from the right-hand side of Eq. \eqref{eqn:eom-relative-useR-1},
the motion of the mass $m_1$ would be the pure Keplerian motion around $m_0$.
Similarly
if we ignore the disturbing function $R_2$ from the right-hand side of Eq. \eqref{eqn:eom-relative-useR-2},
the motion of the mass $m_2$ would be the pure Keplerian motion around $m_0$.

The first terms of the disturbing functions
\eqref{eqn:def-R1} and \eqref{eqn:def-R2} are called the direct part,
representing the major component of the mutual perturbation between $m_1$ and $m_2$.
The second terms are called the indirect part,
which originate from the choice of the coordinate system.
The indirect part would not exist if we took the origin of the coordinate system to be the center of mass \citep[e.g.][]{murray1999,ellis2000}.
Also, the indirect part vanishes or becomes constant
when we carry out an averaging of the system.
Thus they do not contribute to secular dynamics of the system
unless non-negligible mean motion resonances are at work and
the employment of averaging procedure is prohibited.

\subsection{Disturbing function\label{ssec:CR3BP-R}}
Let us restate the equation of motion of the secondary
\eqref{eqn:eom-relative-useR-1}
and that of the tertiary
\eqref{eqn:eom-relative-useR-2} in a more convenient form.
Following the descriptions in conventional textbooks,
we change the notation as follows:
  $m_0 \to M$,
  $m_1 \to m$,
  $m_2 \to m'$,
  $\bm{r}_1 \to \bm{r}$,
  $\bm{r}_2 \to \bm{r'}$,
  $R_1 \to R$, and
  $R_2 \to \widetilde{R}$.
We define the angle between $\bm{r}$ and $\bm{r}'$ as $S$.
We show the geometric configuration of the system under this notation
in a schematic figure (\mysymfigO \ref{fig:CR3BP-schematic}\mtxtsf{a}).

The rewritten equations of motion of the mass $m$ (former $m_1$) become from Eq. \eqref{eqn:eom-relative-useR-1}
\begin{equation}
  \DD[2]{\bm{r}}{t}  + {\cal G} \left( M + m  \right) \frac{\bm{r}}{r^3}   = \nabla R ,
  \label{eqn:eom-relative-useR-conventional-ndash}
\end{equation}
and the rewritten equations of motion of
the mass $m'$ (former $m_2$) become from Eq. \eqref{eqn:eom-relative-useR-2}
\begin{equation}
  \DD[2]{\bm{r}'}{t} + {\cal G} \left( M + m' \right) \frac{\bm{r}'}{{r'}^3} = \nabla \widetilde{R} .
  \label{eqn:eom-relative-useR-conventional-wdash}
\end{equation}
The disturbing function for Eq. \eqref{eqn:eom-relative-useR-conventional-ndash} is, from Eq. \eqref{eqn:def-R1}
\begin{equation}
  R  = \frac{{\cal G} m'}{\Delta}
             -{\cal G} m'  \frac{\bm{r}  \cdot \bm{r}'}{{r'}^3} ,
  \label{eqn:def-R-conventional-ndash}
\end{equation}
and
the disturbing function for Eq. \eqref{eqn:eom-relative-useR-conventional-wdash} is, from Eq. \eqref{eqn:def-R2}
\begin{equation}
  \widetilde{R}
     = \frac{{\cal G} m }{\Delta}
            -{\cal G} m   \frac{\bm{r}' \cdot \bm{r}}{r^3} ,
  \label{eqn:def-R-conventional-wdash}
\end{equation}
with the conventional notation for the mutual distance
\begin{equation}
  \Delta \equiv \left| \bm{r}' - \bm{r}  \right|
         =      \left| \bm{r}  - \bm{r}' \right| .
  \label{eqn:def-Delta-0}
\end{equation}
Note that
$\widetilde{R}$ in
Eqs. \eqref{eqn:eom-relative-useR-conventional-wdash} and
     \eqref{eqn:def-R-conventional-wdash}
is often denoted as $R'$ in the conventional literature.

In this monograph
we will consider only the direct part of the disturbing function
(the first terms in the right-hand sides of
 Eqs. \eqref{eqn:def-R-conventional-ndash} and
      \eqref{eqn:def-R-conventional-wdash}).
The indirect part of the disturbing function
(the second terms in the right-hand sides of
 Eqs. \eqref{eqn:def-R-conventional-ndash} and
      \eqref{eqn:def-R-conventional-wdash})
does not play any significant roles in the doubly averaged system
described below.

We can expand the disturbing function $R$ or $\widetilde{R}$
in an infinite series of orbital elements.
There are several different ways to do this.
Here we consider one of the most straightforward ways:
the expansion using the Legendre polynomials.
Applying the cosine formula to the triangle $m$--P--$m'$ with the angle $S$
in \mysymfigO \ref{fig:CR3BP-schematic}\mtxtsf{a}, we get
\begin{equation}
  \left| \bm{r}' - \bm{r}  \right|^2 = r^2 + {r'}^2 - 2 r r' \cos S .
  \label{eqn:cosineformula}
\end{equation}
Using Eq. \eqref{eqn:cosineformula},
we can expand $\Delta$ in Eq. \eqref{eqn:def-Delta-0} through the Legendre polynomials $P_j$.
When $r < r'$, it becomes
\begin{equation}
\begin{aligned}
  \frac{1}{\Delta} &= \frac{1}{r'}
    \left( 1 - 2 \frac{r}{r'} \cos S + \left( \frac{r}{r'} \right)^2 \right)^{-\frac{1}{2}} \\
  & = \frac{1}{r'}
    \sum_{j=0}^{\infty} \left( \frac{r}{r'} \right)^j P_j (\cos S) .
\end{aligned}
  \label{eqn:Delta-expansion-Legendre-inner}
\end{equation}
On the other hand
when $r > r'$, it becomes
\begin{equation}
\begin{aligned}
  \frac{1}{\Delta} &= \frac{1}{r}
    \left( 1 - 2 \frac{r'}{r} \cos S + \left( \frac{r'}{r} \right)^2 \right)^{-\frac{1}{2}} \\
  & = \frac{1}{r}
    \sum_{j=0}^{\infty} \left( \frac{r'}{r} \right)^j P_j (\cos S) .
\end{aligned}
  \label{eqn:Delta-expansion-Legendre-outer}
\end{equation}

Let us regard
the mass $m$  as the perturbed  body, and
the mass $m'$ as the perturbing body.
The orbital condition $r<r'$ required for the expression of $\Delta$
in Eq. \eqref{eqn:Delta-expansion-Legendre-inner} indicates that
the perturbed body's orbit always stays inside that of the perturbing body
(see \mysymfigO \ref{fig:CR3BP-schematic}\mtxtsf{b(i)}).
We call it the \textit{inner\/} problem.
In this case, the term with $j=0$ in the expansion of $\Delta$ in
Eq. \eqref{eqn:Delta-expansion-Legendre-inner} does not depend on $r$ at all.
As we saw in the equations of relative motion
\eqref{eqn:eom-relative-useR-conventional-ndash},
what matters is not $R$ itself, but its derivative $\nabla R$.
The $j=0$ terms in Eq. \eqref{eqn:Delta-expansion-Legendre-inner}
obviously disappears through this differentiation,
Henceforth we can ignore
the $j=0$ terms in Eq. \eqref{eqn:Delta-expansion-Legendre-inner}
from our discussion.
In addition, we have the relationship
\begin{equation}
  \bm{r} \cdot \bm{r'} = r r' \cos S = r r' P_1 (\cos S),
  \label{eqn:def-rrdcosS}
\end{equation}
and we can apply it to
the indirect part of the disturbing function
(the second term in the right-hand side of Eq. \eqref{eqn:def-R-conventional-ndash}).
Then, we find that the indirect part of the disturbing function
cancels out the term of $j=1$ in Eq. \eqref{eqn:Delta-expansion-Legendre-inner},
and both of them disappear in the expression of $R$.
Therefore for the disturbing function of the inner problem,
we need to consider only the $j \geq 2$ terms in the expansions of
$\Delta$ in Eq. \eqref{eqn:Delta-expansion-Legendre-inner} as
\begin{equation}
\begin{aligned}
  \frac{1}{\Delta} = \frac{1}{r'}
    \sum_{j=2}^{\infty} \left( \frac{r}{r'} \right)^j P_j (\cos S) .
\end{aligned}
  \label{eqn:Delta-expansion-Legendre-inner-nge2}
\end{equation}

When the other orbital condition $(r>r')$ takes place
with the expression of $\Delta$ in Eq. \eqref{eqn:Delta-expansion-Legendre-outer},
the perturbed body's orbit always stays outside that of the perturbing body
(see \mysymfigO \ref{fig:CR3BP-schematic}\mtxtsf{b(ii)}).
We call it the \textit{outer\/} problem.
In this case, unlike the inner case,
the exact cancellation of the indirect part of the disturbing function
does not happen \citep[see][Eq. (6.23) on p. 229]{murray1999}.
Specifically writing down all the relevant terms,
$R$ for the outer problem in Eq. \eqref{eqn:def-R-conventional-ndash} becomes
(omitting the coefficient ${\cal G} m$)
\begin{equation}
  \frac{1}{r}
  \sum_{j=2}^{\infty} \left( \frac{r'}{r} \right)^j P_j (\cos S) 
  + \frac{1}{r}
  + \frac{r'}{r^2} \cos S
  - \frac{r}{{r'}^2} \cos S .
  \label{eqn:MD1999-eq623-rev}
\end{equation}
The second term $\left(\frac{1}{r}\right)$ in Eq. \eqref{eqn:MD1999-eq623-rev} comes
from the $j=0$ term in Eq. \eqref{eqn:Delta-expansion-Legendre-outer},
but it becomes a constant
after we average the disturbing function over the fast-oscillating variables
(consult Section \ref{ssec:CR3BP-averaging} of this monograph for
 the details of the averaging procedure of the disturbing function).
Therefore we do not need to consider this term in the discussion.
The third term comes from the $j=1$ term in Eq. \eqref{eqn:Delta-expansion-Legendre-outer}, and
the fourth term originates from the indirect part in Eq. \eqref{eqn:def-R-conventional-ndash}.
Both of these disappear after the averaging procedure.
As a consequence, it turns out that what we need to consider is only the
$j \geq 2$ terms in the expansion of $\frac{1}{\Delta}$ in Eq. \eqref{eqn:Delta-expansion-Legendre-outer}
for the averaged outer problem.

When the orbit  of the perturbed  body and
           that of the perturbing body cross each other and
behave like the rings of a chain
(i.e. when $r$ can be either smaller or larger than $r'$.
 See \mysymfigO \ref{fig:CR3BP-schematic}\mtxtsf{b(iii)}),
the expansion of the disturbing function using the Legendre polynomials
in Eq. \eqref{eqn:Delta-expansion-Legendre-inner} or
   Eq. \eqref{eqn:Delta-expansion-Legendre-outer} does not work out anymore.
It is because $\frac{r}{r'}$ $\bigl( \mbox{or } \frac{r'}{r} \bigr)$ can exceed unity,
and the infinite series
in Eq. \eqref{eqn:Delta-expansion-Legendre-inner} or
   Eq. \eqref{eqn:Delta-expansion-Legendre-outer} does not converge.
We will briefly mention this case later again 
(Section \ref{ssec:R-general} or Section \ref{ssec:orbitintersection} of this monograph).

Carrying out literal expansions of the disturbing function is a formidable task in general.
But it is relatively simpler in CR3BP,
particularly in its doubly averaged version.
We will see some examples later in this monograph.
In CR3BP, the length of the position vector of the perturber $(r')$
with respect to the primary mass
has a constant value that is equivalent to its semimajor axis, $a'$.
And when $r'=a'$,
we do not need to consider the odd terms $(j=3,5,7,\cdots)$
in the expansion of
Eqs. \eqref{eqn:Delta-expansion-Legendre-inner} and
     \eqref{eqn:Delta-expansion-Legendre-outer} at all,
because they all vanish after the averaging procedure.
Therefore the disturbing function for the inner CR3BP ($r<r'$)
that we consider turns out as, from
Eqs. \eqref{eqn:def-R-conventional-ndash} and
     \eqref{eqn:Delta-expansion-Legendre-inner}:
\begin{equation}
\begin{aligned}
  R = \frac{{\cal G} m'}{a'}
    \sum_{n=1}^{\infty} \left( \frac{r}{a'} \right)^{2n} P_{2n} (\cos S) .
\end{aligned}
  \label{eqn:Delta-expansion-Legendre-iCR3BP}
\end{equation}
On the other hand for the outer CR3BP ($r>r'$), $R$ becomes from
Eqs. \eqref{eqn:def-R-conventional-ndash} and
     \eqref{eqn:Delta-expansion-Legendre-outer} as
\begin{equation}
\begin{aligned}
  R = \frac{{\cal G} m'}{r}
    \sum_{n=1}^{\infty} \left( \frac{a'}{r} \right)^{2n} P_{2n} (\cos S) .
\end{aligned}
  \label{eqn:Delta-expansion-Legendre-oCR3BP}
\end{equation}

Although we do not show it here,
it is clear that we can carry out the expansion of $\widetilde{R}$
in Eq. \eqref{eqn:def-R-conventional-wdash} in a similar manner
whenever necessary.

Let us note in passing that, in the inner problem,
the direct part of the disturbing function $\frac{{\cal G}m'}{\Delta}$ can
be derived in a different, more general way.
Return temporarily to a general three-body system with three masses:
the primary   $m_0$,
    secondary $m_1$, and
    tertiary  $m_2$.
Now let us use the Jacobi coordinates where we
measure $m_1$'s position vector $\tilde{\bm{r}}_1$ from $m_0$, and
measure $m_2$'s position vector $\tilde{\bm{r}}_2$ from
the barycenter of $m_0$ and $m_1$
(here we assume $\tilde{r}_1 < \tilde{r}_2$ for the inner problem).
We define an angle $S_{12}$ as the angle between the vectors $\tilde{\bm{r}}_1$ and $\tilde{\bm{r}}_2$.
In general,
$S_{12}$ is different from $S$ in \mysymfigO \ref{fig:CR3BP-schematic}\mtxtsf{a},
and the origins of $\bm{r}_1$ and $\bm{r}_2$ are different from each other.
Then, the equations of motion of the secondary mass $m_1$ and
the tertiary mass $m_2$ become \citep[e.g.][]{smart1953,brouwer1961,jefferys1966}
\begin{equation}
  \tilde{m}_1 \frac{d^2 \tilde{\bm{r}}_1}{dt^2} = \frac{\partial F}{\partial \tilde{\bm{r}}_1},
\quad
  \tilde{m}_2 \frac{d^2 \tilde{\bm{r}}_2}{dt^2} = \frac{\partial F}{\partial \tilde{\bm{r}}_2},
  \label{eqn:def-EOM=m1-m2}
\end{equation}
where
\begin{equation}
  \tilde{m}_1 = \frac{m_0 m_1}{m_0+m_1}, \quad
  \tilde{m}_2 = \frac{\left(m_0 + m_1\right)m_2}{m_0+m_1+m_2} ,
  \label{eqn:def-Jacobimass}
\end{equation}
are the reduced masses used in the Jacobi coordinates \citep[e.g.][]{wisdom1991,saha1994}.
Here $F$ is the common force function
\begin{equation}
\begin{aligned}
  F &= {\cal G} \left[ \frac{m_0 m_1}{\tilde{r}_1} + \frac{(m_0+m_1) m_2}{\tilde{r}_2}
       \right. \\
    & \quad\quad\quad
     + \left. \frac{1}{\tilde{r}_2} \sum_{j=2}^\infty
        M_j \left( \frac{\tilde{r}_1}{\tilde{r}_2} \right)^j P_j(\cos S_{12}) \right],
\end{aligned}
  \label{eqn:def-F-forcefunction}
\end{equation}
and $M_j$ is the mass factor
\begin{equation}
  M_j = \frac{m_0 m_1 m_2 \left(m_0^{j-1} - \left(-m_1\right)^{j-1} \right)}{\left( m_0 + m_1 \right)^j} .
  \label{eqn:def-Massfunction}
\end{equation}
Using the force function $F$, 
we can construct a Hamiltonian that governs the dynamics of this system.
Assuming $a_1$ and $a_2$ to be the semimajor axes of the orbits of the
secondary and tertiary masses, the Hamiltonian ${\cal H}$ becomes
\citep[e.g.][]{harrington1968,krymolowski1999,beust2006,carvalho2013}:
\begin{equation}
\begin{aligned}
{\cal H}&= \frac{{\cal G} m_0 m_1}{2 a_1}
         + \frac{{\cal G} (m_0+m_1) m_2}{2 a_2} \\
        & \quad\quad\quad
         + \frac{\cal G}{\tilde{r}_2} \sum_{j=2}^\infty
             M_j \left( \frac{\tilde{r}_1}{\tilde{r}_2} \right)^j P_j(\cos S_{12}) .
\end{aligned}
  \label{eqn:def-H-3BP-general}
\end{equation}

The first term of ${\cal H}$ in Eq. \eqref{eqn:def-H-3BP-general} is
responsible for the Keplerian motion of the secondary mass, and
the second term is responsible for that of the tertiary mass.
The third term of ${\cal H}$ represents the mutual interaction
of the secondary and the tertiary, and does not include terms of
$j=0$ or $j=1$.
This is an outcome of the use of the Jacobi coordinates
which subdivides the motions of the three bodies into
two separate binaries and their interactions.

Now, consider a limit where the secondary mass $m_1$ is infinitesimally small.
This corresponds to the restricted inner three-body problem
where $m_2$ serves as the perturbing body.
In this case, we must divide the force function $F$
in Eq. \eqref{eqn:def-F-forcefunction} by $\tilde{m}_1$
in Eq. \eqref{eqn:def-Jacobimass} before taking the mass-less limit.
The normalized third term of $F$ then becomes
\begin{equation}
\begin{aligned}
{} & \frac{F_{\rm 3rd}}{\tilde{m}_1} =  \\
{} & \quad
    \frac{{\cal G} m_2}{\tilde{r}_2} \sum_{j=2}^\infty
    \frac{m_0 ^{j-1} - \left(-m_1\right)^{j-1}}
         {\left( m_0 + m_1 \right)^{j-1}}
    \left( \frac{\tilde{r}_1}{\tilde{r}_2} \right)^j P_j(\cos S_{12}) .
\end{aligned}
  \label{eqn:def-F-3rd-norm}
\end{equation}

Now we can take the limit of $m_1 \to 0$.
This would simultaneously yield the conversions
$\bm{r}_1 \to \bm{r} $,
$\bm{r}_2 \to \bm{r}'$,
$S_{12} \to S$, as well as a replacement of $m_2$ for $m'$
in the previous discussions.
Then we reach an expression equivalent to the direct part of
the disturbing function of the inner case written in the relative coordinates,
such as expressed in 
Eq. \eqref{eqn:Delta-expansion-Legendre-inner-nge2},
or particularly for CR3BP,
Eq. \eqref{eqn:Delta-expansion-Legendre-iCR3BP}.

On the other hand, deriving the disturbing function of the outer case
written in the relative coordinates such as
Eq. \eqref{eqn:Delta-expansion-Legendre-outer} or
Eq. \eqref{eqn:Delta-expansion-Legendre-oCR3BP}
by simply taking a mass-less limit of the Hamiltonian $\cal H$ is difficult,
if not impossible \citep[cf.][]{ito2016}.
This is an example that shows a limitation of the use of
the relative coordinates when developing the disturbing function.
Readers can find newer, more sophisticated methods and techniques for
expanding the disturbing function without using the conventional relative coordinates
\citep[e.g.][]{broucke1981,laskar2010,mardling2013}.

\subsection{Double averaging\label{ssec:CR3BP-averaging}}

Now we calculate the double average of the disturbing function $R$
over mean anomalies of both the perturbed and perturbing bodies.
In general,
averaging of the disturbing function by fast-oscillating variables is carried out
for reducing the degrees of freedom of the system.
In many problems of solar system dynamics,
variation rate of mean anomaly is much larger than that of other elements.
Hence it is justified to eliminate mean anomaly by averaging,
assuming that the other orbital elements do not change over a period of mean anomaly.
The elimination of mean anomaly by averaging procedure can be regarded as a part of canonical transformation
that divides system's Hamiltonian into periodic and secular parts.
Historically speaking,
this procedure was devised by \citet{delaunay1860,delaunay1867}, and
substantially developed by 
\citet{vonzeipel1916a,vonzeipel1916b,vonzeipel1917a,vonzeipel1917b}.
See 
\citet[][Notes and References in Chapter XVII, their pp. 591--593]{brouwer1961} or
\citet[][Subsection 12.4]{goldstein2002} for a more detailed background.

For carrying out the averaging procedure, we have to assume that
there is no major resonant relationship between the mean motions of the perturbed and perturbing bodies.
In other words,
the mean anomalies of the perturbed and perturbing bodies
(referred to as $l$ and $l'$, respectively) must be independent of each other.
Bearing this assumption in mind,
pick the $n$-th term of the disturbing function $R$ for the inner problem
in Eq. \eqref{eqn:Delta-expansion-Legendre-iCR3BP},
and call it $R_{2n}$. We have
\begin{equation}
  R_{2n} = \frac{\mu'}{a'} \left( \frac{r}{a'} \right)^{2n} P_{2n} (\cos S) .
  \label{eqn:Rd-inner-j}
\end{equation}

We first average $R_{2n}$ by mean anomaly of the perturbing body $l'$.
Using the symbols $\left< \right.$ and $\left. \right>$ for averaging, it is
\begin{equation}
  \left< R_{2n} \right>_{l'} = 
  \frac{\mu'}{a'} \left( \frac{r}{a'} \right)^{2n} \left< P_{2n} \right>_{l'},
  \label{eqn:Rd-inner-j-averaged1}
\end{equation}
where
\begin{equation}
  \left< P_{2n} \right>_{l'} = \frac{1}{2\pi} \int^{2\pi}_0 P_{2n} (\cos S) dl'.
  \label{eqn:Pj-avr}
\end{equation}

The angle $S$ is expressed by orbital angles through a relationship \citep[e.g.][Eq. (7) on p. 592]{kozai1962b}
\begin{equation}
\begin{aligned}
  \cos S & = \cos (f+g) \cos (f'+g') \\
         & \quad\quad\quad
         + \cos i \sin (f+g) \sin (f'+g'),
\end{aligned}
  \label{eqn:def-cosS}
\end{equation}
where
$f, f'$ are true anomalies of the perturbed and perturbing bodies,
$g, g'$ are arguments of pericenter of the perturbed and perturbing bodies,
and
$i$ is their mutual inclination measured at the node of the two orbits.
We choose the orbital plane of the perturbing body as a reference plane
for the entire system so that we can measure $g$ and $g'$ from the mutual node.
Note that $g'$ is not actually defined in CR3BP.
Therefore, in Eq. \eqref{eqn:def-cosS} we regard $f'+g'$ as a single,
fast-oscillating variable.
In practice,
we can simply replace $\int dl'$ for $\int df'$ in the discussion here.

To obtain $\left< P_{2n} \right>_{l'}$ of Eq. \eqref{eqn:Pj-avr},
we calculate the time average of $\cos^{2n} S$ by $l'$ as
\begin{equation}
    \left< \cos^{2n} S \right>_{l'}
  = \frac{1}{2\pi} \int^{2\pi}_0 \cos^{2n} S dl' .
  \label{eqn:def-cosS-avr}
\end{equation}
Then we average $\left< R_{2n} \right>_{l'}$ of Eq. \eqref{eqn:Rd-inner-j-averaged1}
by mean anomaly of the perturbed body $l$, as
\begin{equation}
  \left< \left< R_{2n} \right>_{l'} \right>_{l}
  = \frac{\mu'}{a'} \left( \frac{a}{a'} \right)^{2n}
      \frac{1}{2\pi} \int^{2\pi}_0
        \left( \frac{r}{a} \right)^{2n} \left< P_{2n} \right>_{l'} dl .
  \label{eqn:Rd-inner-j-averaged2}
\end{equation}

If we switch the integration variable from mean anomaly $l$ to
eccentric anomaly $u$, Eq. \eqref{eqn:Rd-inner-j-averaged2} becomes
\begin{equation}
\begin{aligned}
{} & \left< \left< R_{2n} \right>_{l'} \right>_{l}
  = \frac{\mu'}{a'} \left( \frac{a}{a'} \right)^{2n} \\
{}& \quad
  \times
      \frac{1}{2\pi} \int^{2\pi}_0
        \left( 1-e \cos u \right)^{2n+1} \left< P_{2n} \right>_{l'} du .
\end{aligned}
  \label{eqn:Rd-inner-j-averaged2-byu}
\end{equation}

Eq. \eqref{eqn:Rd-inner-j-averaged2} or
Eq. \eqref{eqn:Rd-inner-j-averaged2-byu} is the final,
general form of the $n$-th term of the
doubly averaged disturbing function for the inner CR3BP.
If we define the ratio of semimajor axes as $\alpha = \frac{a}{a'} < 1$,
this term has the magnitude of $O\left(\alpha^{2n}\right)$.

We can obtain the doubly averaged disturbing function for the outer CR3BP
in the same way.
In what follows let us denote the disturbing function for the outer CR3BP as $R'$.
From its definition previously expressed in Eq. \eqref{eqn:Delta-expansion-Legendre-oCR3BP},
the direct part of $R'$ becomes as follows:
\begin{equation}
  R' = \frac{\mu'}{r} \sum_{n=1}^\infty
       \left( \frac{a'}{r} \right)^{2n} P_{2n} (\cos S) .
  \label{eqn:Rd-outer}
\end{equation}

Note that our definition of $\frac{1}{\Delta}$ for the outer case
\eqref{eqn:Delta-expansion-Legendre-outer},
and hence also in Eq. \eqref{eqn:Rd-outer}, may be different from what is seen
in conventional textbooks \citep[e.g.][Eq. (6.22) on p. 229]{murray1999}:
The roles of the dashed quantities $(r', \mu')$ may be the opposite.
This difference comes from the fact that
conventional textbooks always assume $\frac{r}{r'}<1$
even in the outer problem,
while we assume $\frac{r}{r'}>1$ for the outer problem.
In other words,
we make it a rule to always use dashed variables
($r', a', l', \mu', \cdots$) for the perturbing body
whether it is located inside or outside the perturbed body.

Similar to the procedures that we went through for the inner CR3BP,
we again assume that there is no major resonant relationship between
mean motions of the perturbed and perturbing bodies.
We then try to get the double average of $R'$ over mean anomalies of both the bodies.
Let us pick the $n$-th term of $R'$ in Eq. \eqref{eqn:Rd-outer},
and call it $R'_{2n}$. We have
\begin{equation}
  R'_{2n} = \frac{\mu'}{r} \left( \frac{a'}{r} \right)^{2n} P_{2n} (\cos S) .
  \label{eqn:Rd-outer-j}
\end{equation}

First we average $R'_{2n}$ by mean anomaly of the perturbing body $l'$.
Similar to Eq. \eqref{eqn:Rd-inner-j-averaged1}, it is
\begin{equation}
  \left< R'_{2n} \right>_{l'} = 
  \frac{\mu'}{r} \left( \frac{a'}{r} \right)^{2n} \left< P_{2n} \right>_{l'} ,
  \label{eqn:Rd-outer-j-averaged1}
\end{equation}
where $\left< P_{2n} \right>_{l'}$ is already defined in Eq. \eqref{eqn:Pj-avr}.

Then we average $\left< R'_{2n} \right>_{l'}$ in Eq. \eqref{eqn:Rd-outer-j-averaged1}
by mean anomaly of the perturbed body $l$, as
\begin{equation}
  \left< \left< R'_{2n} \right>_{l'} \right>_{l}
  = \frac{\mu'}{a'} \left( \frac{a'}{a} \right)^{2n+1}
      \frac{1}{2\pi} \int^{2\pi}_0
        \left( \frac{a}{r} \right)^{2n+1} \left< P_{2n} \right>_{l'} dl .
  \label{eqn:Rd-outer-j-averaged2}
\end{equation}

If we switch the integration variable from $l$ to true anomaly $f$,
Eq. \eqref{eqn:Rd-outer-j-averaged2} becomes
\begin{equation}
\begin{aligned}
{} &  \left< \left< R'_{2n} \right>_{l'} \right>_{l} 
 = \frac{\mu'}{a'}
   \left( \frac{a'}{a} \right)^{2n+1}
   \frac{\left(1-e^2\right)^{-2n+\frac{1}{2}}}{2\pi} \\
{} &  \quad
   \times
     \int^{2\pi}_0
        \left( 1 + e \cos f \right)^{2n-1} \left< P_{2n} \right>_{l'} df .
\end{aligned}
  \label{eqn:Rd-outer-j-averaged2-byf}
\end{equation}

Eq. \eqref{eqn:Rd-outer-j-averaged2} or
Eq. \eqref{eqn:Rd-outer-j-averaged2-byf} is the final,
general form of the $n$-th term of the doubly averaged disturbing function
for the outer CR3BP.
Note that this term has the magnitude of $O\bigl({\alpha '}^{2n+1}\bigr)$, not
                                         $O\bigl({\alpha '}^{2n}  \bigr)$,
when we define $\alpha' = \frac{a'}{a} < 1$.

\label{pg:happycoincidence-intro}
Let us make a couple of additional comments before we move on to the next subsection.
First, we evidently find that argument of pericenter of the perturbing body
$(g')$ is not included in the disturbing function for CR3BP,
because the perturbing body's eccentricity $e'$ is zero.
However, even when the orbit of the perturbing body is not circular $(e'>0)$,
its $g'$ would not show up in the disturbing function
as long as we truncate the doubly averaged disturbing function
at the leading-order, $\Oalsqr$
(note that the truncation of the disturbing function at $\Oalsqr$ is
  often referred to as the quadrupole level (or the quadrupole order) approximation).
This circumstance was named ``a happy coincidence'' by \citet{lidov1976},
and thus the system remains integrable even though $e'>0$.
This ``coincidence'' no longer stands
if we include the terms of $\Oalcub$ or higher in the doubly
averaged disturbing function \citep[e.g.][]{farago2010,lithwick2011}.
The approximation at $\Oalcub$ is called the octupole level (or the octupole order).

\label{pg:limitedmasseffect}
Our second comment is about the fact that the mass of the perturber
$(m')$ does not at all act on the trajectory shape of the perturbed body
that the doubly averaged disturbing function
\eqref{eqn:Rd-inner-j-averaged2} or
\eqref{eqn:Rd-outer-j-averaged2} yields.
As we see from the function form of $\left< \left< R_{2n} \right>_{l'} \right>_{l}$ in Eq. \eqref{eqn:Rd-inner-j-averaged2},
the perturber's mass serves just as a constant factor in the doubly averaged disturbing function, and
its influence is limited to controlling the timescale of the motion of the perturbed body.
This is obvious if we recall the general form of the canonical equations of motion
such as $\DD{G}{t} = \DP{R}{g}$ and $\DD{g}{t} = -\DP{R}{G}$.
This statement is also true even if we consider the doubly averaged general (i.e. not restricted) three-body system,
as long as the central mass is much larger than the perturbing mass.
We can confirm this through the function form of $M_j$ shown in Eq. \eqref{eqn:def-Massfunction}.

As we mentioned, the averaging procedure cannot not be used
when strong mean motion resonances are at work in the considered system.
Also, there may be some conditions that the solution obtained through averaging procedure can deviate from true solution
due to the accumulation of short-term oscillation \citep[e.g.][]{luo2016}.
Nevertheless, the averaging procedure yields a very good perspective
in theoretical studies, as well as a substantially large efficiency in the practical calculation.
Therefore, the averaging procedure is more and more frequently used on
a variety of scenes in modern celestial mechanics \citep[e.g.][]{sanders2007}.

\begin{figure*}[tbhp]\centering
\ifepsfigure
 \includegraphics[width=\dualfigwidth\textwidth]{cr3bp_ex.eps} %fig2
\else
 \includegraphics[width=\dualfigwidth\textwidth]{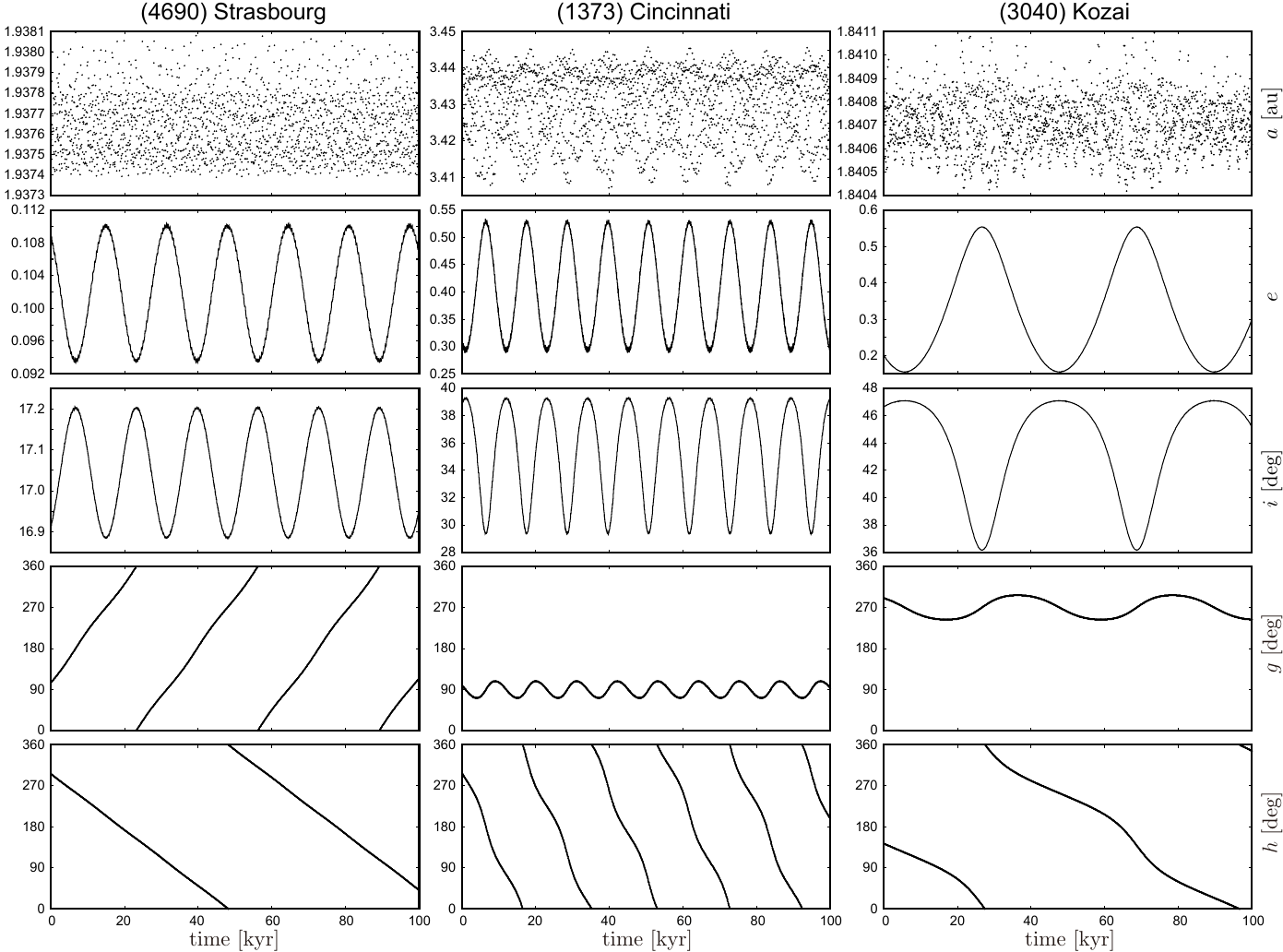} %fig2
\fi
  \caption{%
Numerical solutions of the orbital evolution of three asteroid proxies
under perturbation from a Jupiter-like planet on a circular orbit.
Left:   (4690) Strasbourg.
Middle: (1373) Cincinnati.
Right:  (3040) Kozai.
From the top row, the object's $a$, $e$, $i$, $g$, and $h$ are displayed.
The initial orbital elements of the asteroid proxies are taken from
the JPL Horizons web-interface as of June 7, 2017
(see \supinfo{3} for its specific website as well as other relevant information provided by JPL).
The perturbing planet has the same mass and the same semimajor axis with Jupiter,
but its eccentricity and inclination are both zero.
}
  \label{fig:CR3BP-examples}
\end{figure*}

\subsection{Numerical examples\label{ssec:CR3BP-examples}}
The analytic expression of the disturbing function for CR3BP in
Eqs. \eqref{eqn:Delta-expansion-Legendre-iCR3BP} and
     \eqref{eqn:Delta-expansion-Legendre-oCR3BP},
in particular their lowest-order term $(n=1)$,
plays a central role in the discussions developed in the remainder of this monograph.
Before we move on,
let us show some numerical examples of CR3BP
for giving readers a rough picture of how typical CR3BP systems behave on a long-term, secular timespan.

The first example is a Sun--planet--asteroid system
where the perturbing planet has the same mass and the semimajor axis
as Jupiter, but its orbital eccentricity ($e'$) is zero.
We consider the inner problem, and place three asteroid proxies
as the perturbed body orbiting inside this Sun--planet binary. They are
(4690) Strasbourg,
(1373) Cincinnati, and
(3040) Kozai.
Then we numerically propagate their orbital evolution over 100 kyr
($10^5$ years)
in the future direction by directly integrating the equations of motion
\eqref{eqn:eom-relative-useR-conventional-ndash},
and make a set of plots of their orbital elements 
(\mysymfigO \ref{fig:CR3BP-examples}).
  The nominal stepsize that we use here is 1 day, and
  the data output interval is 100 years.
As for the numerical integrator, we use the Wisdom--Holman symplectic map.
We will explain our numerical method later in more detail
(Section \ref{ssec:Kozai-asteroids}).

Among the three sets of panels in \mysymfigO \ref{fig:CR3BP-examples},
the motion of (4690) Strasbourg shown in the panels at the left exhibits
the most typical behavior in the inner CR3BP.
We find several noticeable characteristics here:
\begin{itemize}
\item Semimajor axis $(a)$ remains almost constant,
      although it shows a short-term oscillation with a small amplitude.
\item Eccentricity $(e)$ and inclination $(i)$ show regular,
      anti-correlated oscillations.
      When $e$ becomes large (or small), $i$ becomes small (or large).
\item Argument of pericenter $(g)$ circulates in the prograde direction.
      Its circulation period has a correlation to the $e$--$i$ oscillation.
\item Longitude of ascending node $(h)$ circulates in the retrograde direction.
      Its circulation period does not seem to have particular correlations to
      $e$, $i$, or $g$.
\end{itemize}
\label{pg:CR3BP-features}

The first characteristic ($a$ being almost constant) originates from the general fact that
the semimajor axis of the perturbed body remains constant in the doubly averaged CR3BP
(i.e. $\left< \left< a \right> \right>$ becomes a constant).
The second and the third characteristics (the regular and correlated oscillations of $e$, $i$, and $g$) typically exemplify
the so-called \citeauthor{lidov1961}--\citeauthor{kozai1962b} {\mainword}
in its circulation regime.
We will explore the further details of these characteristics
in later sections.

On the other hand, the motions of other objects shown in
\mysymfigO \ref{fig:CR3BP-examples} (the middle and the right panels)
look qualitatively different from
that of (4690) Strasbourg, although both of them are regular and
exhibiting the $e$--$i$ anti-correlated oscillation as well.
As for (1373) Cincinnati whose motion is shown in the middle column panels,
the argument of pericenter $g$ librates around $\frac{\pi}{2}$,
instead of circulating from 0 to $2\pi$.
Its oscillation still seems correlated to the $e$--$i$ couple.
As for (3040) Kozai whose motion is shown in the right column panels,
the argument of pericenter $g$ seems to librate around $\frac{3\pi}{2}$
with a similar correlation.

What makes these differences?
The key to understanding things here lies in the difference of their initial orbital inclination.
More specifically, the difference of the vertical component of angular momentum matters.
Looking at a quantity $\left( 1-e^2 \right) \cos^2 i$ which is proportional
to the square of the vertical component of the perturbed body's angular momentum,
it is $\sim 0.91$ for (4690) Strasbourg.
On the other hand it is
$\sim 0.55$ for (1373) Cincinnati, and
$\sim 0.45$ for (3040) Kozai.
Considering the fact that the quantity
$\left( 1-e^2 \right) \cos^2 i$ takes the value between 0 and 1
(as long as the motion is elliptic),
at this point we can deduce that the libration of the argument of pericenter
seen in the motion of (1373) Cincinnati and (3040) Kozai 
takes place
when the vertical component of the asteroids' angular momentum is small.

\begin{figure}[tbhp]\centering
\ifepsfigure
 \includegraphics[width=\singlefigwidth\textwidth]{cr3bp+8p.eps} %fig3
\else
 \includegraphics[width=\singlefigwidth\textwidth]{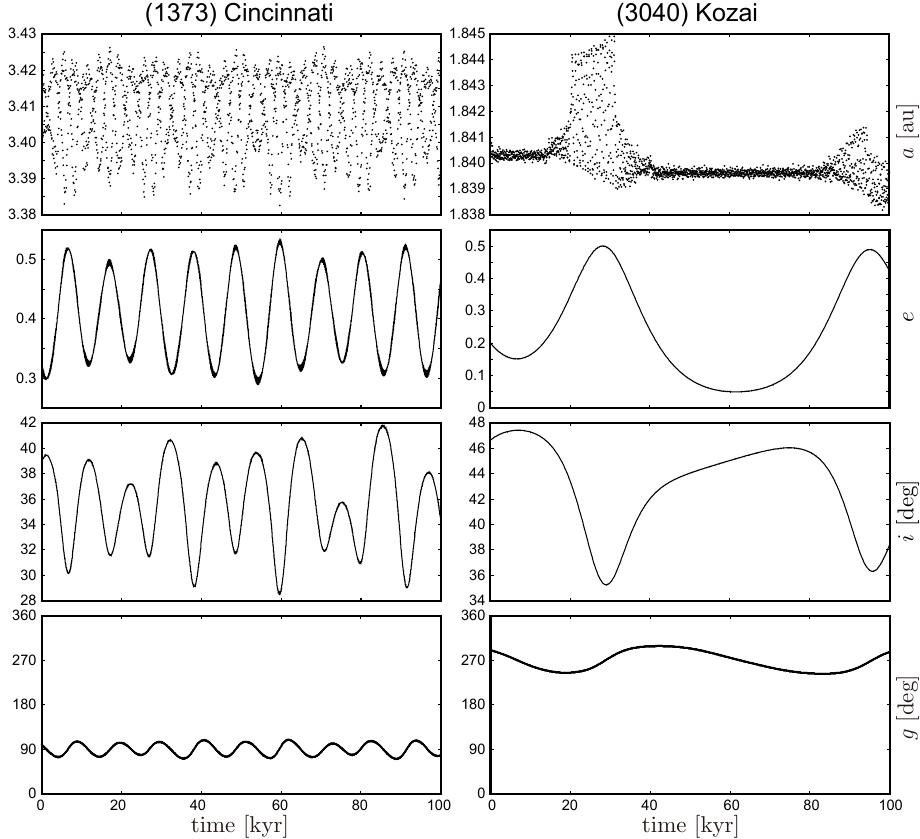} %fig3
\fi
  \caption{%
Numerical solutions of the orbital evolution of two asteroids
under perturbation from all the eight major planets
from Mercury to Neptune.
Left:  (1373) Cincinnati.
Right: (3040) Kozai.
From the top row, the object's $a$, $e$, $i$, and $g$ are displayed.
The initial orbital elements of the asteroids are the same as those used in \mysymfigO \protect{\ref{fig:CR3BP-examples}}.
The planetary masses and initial orbital elements are taken from
the JPL Horizons web-interface as of June 7, 2017.
}
  \label{fig:CR3BP-examples+8P}
\end{figure}

CR3BP is a simple dynamical model.
However, it has the capability to explain many of the fundamental properties
observed in the actual solar system dynamics.
The $e$--$i$--$g$ correlated oscillation seen in
\mysymfigO \ref{fig:CR3BP-examples} is one of them.
For comparison, we carried out another set of numerical propagation of
the orbits of (1373) Cincinnati and (3040) Kozai
starting from the same initial condition as in
\mysymfigO \ref{fig:CR3BP-examples},
but under the perturbation from all the eight major planets
from Mercury to Neptune with their actual orbital elements
(we might want to describe it as a restricted ``8$+$1'' or ``8$+$2''-body system).
We pick the resulting time series of asteroids' $a$, $e$, $i$, and $g$ in
\mysymfigO \ref{fig:CR3BP-examples+8P}.
Comparing the panels that show the motions of
(1373) Cincinnati and
(3040) Kozai
in \mysymfigO \ref{fig:CR3BP-examples} and
in \mysymfigO \ref{fig:CR3BP-examples+8P},
it is obvious that the CR3BP approximation that was employed to draw
\mysymfigO \ref{fig:CR3BP-examples} largely possesses the dynamical characteristics
that the system with the full planetary perturbation
(\mysymfigO \ref{fig:CR3BP-examples+8P}) possesses:
The anti-correlated oscillation between $e$ and $i$,
the coherent oscillation of $g$ with the $e$--$i$ couple,
the libration of (1373) Cincinnati's $g$ at $\frac{ \pi}{2}$, and
the libration of (3040) Kozai's      $g$ at $\frac{3\pi}{2}$.
Their semimajor axes remain almost constant during the integration period
although we see occasional enhancement of the oscillational amplitude of (3040) Kozai's $a$.
This comparison literally tells us that CR3BP is still useful
in solar system dynamics in spite of its structural simplicity.
It helps us understand the dynamical nature of the
motion of various objects that compose hierarchical three-body systems.
This is particularly true for long term dynamics where only the
secular motion of objects matters.

\begin{figure}[tbhp]\centering
\ifepsfigure
 \includegraphics[width=\singlefigwidth\textwidth]{eLK_ex3.eps}%fig4
\else
 \includegraphics[width=\singlefigwidth\textwidth]{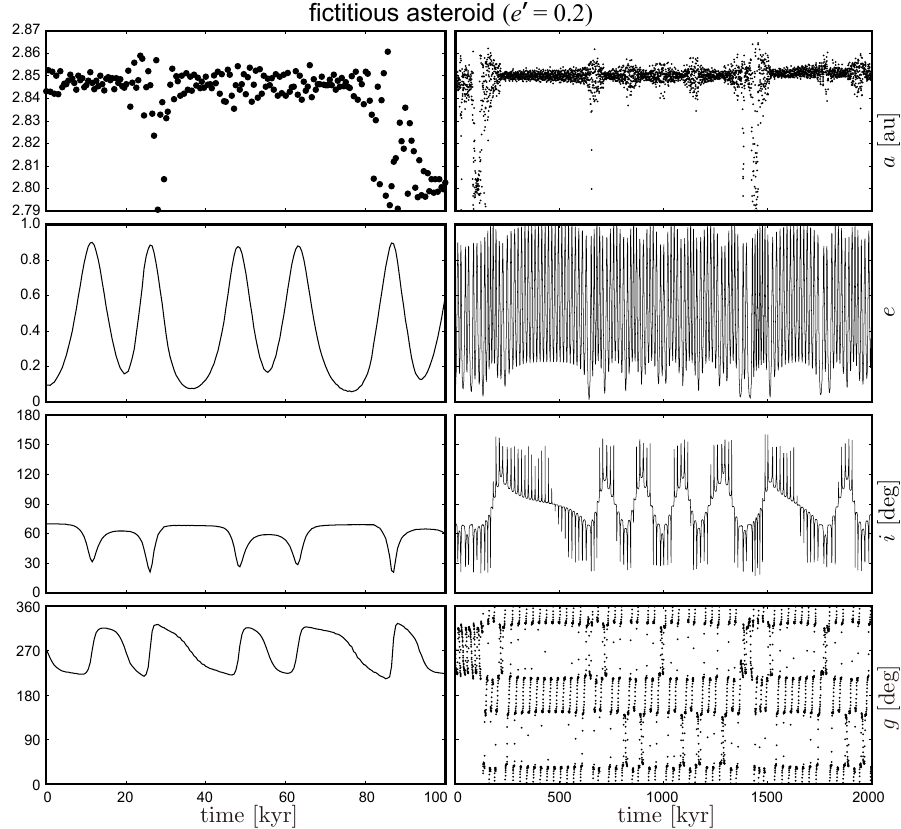}%fig4
\fi
  \caption{%
Numerical solutions of the orbital evolution of a fictitious asteroid
under perturbation from a Jupiter-like planet on an eccentric orbit
$(e' = 0.2)$.
Left:  the evolution during the first  100 kyr.
Right: the evolution during the entire integration period, 2000 kyr.
From the top row, the object's $a$, $e$, $i$, and $g$ are displayed.
The perturbing planet has the same mass and semimajor axis as Jupiter,
and its inclination is zero.
The initial orbital elements of the asteroid are described in the main text.
}
  \label{fig:eLK-example}
\end{figure}

Before closing this section,
we would like to temporarily and intentionally deviate
from the scope of this monograph.
Let us explore just a little the world of non-circular (eccentric) restricted
three-body problem where the perturber's eccentricity $e'$ is not zero.
From a theoretical perspective,
this case is qualitatively different from CR3BP
because the double averaging procedure would not make the system integrable.
This means that, the system's degrees of freedom remain larger than unity
even after the double averaging. As a consequence,
the dynamical behavior of the system can be very different from CR3BP.
As an example, we prepare yet another Sun--planet--asteroid system where
the perturbing planet has the same mass and semimajor axis as Jupiter.
The difference from the examples shown in \mysymfigO \ref{fig:CR3BP-examples} is that
we give the perturbing planet a finite eccentricity: $e' = 0.2$.
The perturbed body is a fictitious asteroid
whose initial orbital elements are
$a=2.80$ au, $e=0.1$, $i=70^\circ$, $h=143^\circ$,
and
$g = 270^\circ = \frac{3\pi}{2}$.
Therefore $\left( 1-e^2 \right) \cos^2 i \sim 0.12$.
We numerically propagate the orbital evolution of this three-body system
by directly integrating the equations of motion
\eqref{eqn:eom-relative-useR-conventional-ndash},
and present the time variation of $a, e, i, g$ of the perturbed asteroid
in \mysymfigO \ref{fig:eLK-example}.
In the first 100 kyr of time evolution
(the four panels in the left column of \mysymfigO \ref{fig:eLK-example}),
the behavior of the perturbed asteroid is somewhat similar to the CR3BP case:
the semimajor axis $a$ remains almost constant except for the occasional and small enhancement of amplitude.
the eccentricity $e$ and inclination $i$ have an anti-correlated oscillation.
The argument of pericenter $g$'s oscillation is also correlated with the $e$--$i$ couple.
$g$ librates around $270^\circ$ during this period.
However, their oscillation is not as regular as what we saw in \mysymfigO \ref{fig:CR3BP-examples}.
Also, we see that the amplitude of the eccentricity variation is very large.
These features become even clearer when we extend the integration period
to 2000 kyr
(the four panels in the right column of \mysymfigO \ref{fig:eLK-example}).
The most intriguing aspect concerns the oscillation of orbital inclination.
Although its oscillation is still anti-correlated to that of the eccentricity $e$,
the inclination $i$ frequently and irregularly exceeds $90^\circ$.
This means that the orbit of the perturbed body flips, and flips back.
The amplitude of the eccentricity variation is remarkably large, and
the argument of pericenter changes its status between libration and circulation.
This kind of behavior is never observed in CR3BP.

The perturbed body's stochastic behavior observed in
\mysymfigO \ref{fig:eLK-example} is typical of the so-called eccentric
\citeauthor{lidov1961}--\citeauthor{kozai1962b} {\mainword}.
This provides us with many clues about the rich dynamical characteristics
that the non-circular (eccentric) restricted three-body problem (ER3BP) has.
It also helps us understand certain dynamical structures of
the solar and other planetary systems that the simple CR3BP cannot explain.
However, ER3BP is clearly out of the scope of this monograph.
Also, we must have a firm and rigorous
understanding of CR3BP before moving on to the world of ER3BP.
Therefore in this monograph we concentrate on the description of CR3BP
where the perturbing body is always on a circular orbit $(e'=0)$.
As a start, we first introduce the classic work of \citeauthor{kozai1962b}
in the following section.

\section{The Work of \protect\citeauthor{kozai1962b}\label{sec:kozai1962b}}
Yoshihide Kozai (1928--2018)%
\footnote{%
Yoshihide Kozai passed away on February 5, 2018, at the age of 89.
It was just two days after
we completed the initial submission of this monograph to the MEEP editorial office.
Obituaries have come from many institutes and organizations such as
American Astronomical Society,
International Astronomical Union, or
the Japan Academy.
See \supinfo{3} for their electronic versions.
} % End of \footnote
was a Japanese celestial mechanist
who is famous for a variety of works on the dynamics of
small solar system bodies, planetary satellites, and
artificial satellites around the Earth.
Several oral history publications are available for \citeauthor{kozai2016}'s research and life,
such as \citet[][an online publication by American Institute of Physics]{devorkin1997} or
\citet[][a series of articles written in Japanese with abstracts in English.]{takahashi2015e-1,takahashi2015e-2,takahashi2015e-3,takahashi2015e-4,takahashi2015e-5}.
More recently, \citeauthor{kozai2016} made a brief summary of
his academic career and personal life \citep{kozai2016}
including his work on the present subject.

One of \citeauthor{kozai2016}'s achievements that made him renowned was
his work on the gravitational potential of the Earth.
Through a detailed analysis of artificial satellite motion,
\citeauthor{kozai1962b} showed that the gravitational potential of the Earth has a non-negligible north-south asymmetry
\citep[e.g.][]{kozai1958,kozai1959a,kozai1959b,kozai1959c,kozai1960,kozai1961}.
Another of his achievements that made him famous in astronomy
worldwide was his work on the very subject of this monograph:
the secular motion of asteroids that have large inclination in the framework of CR3BP.
His celebrated work on this \citep{kozai1962b} was entitled
``Secular perturbation of asteroids with high inclination and eccentricity,''
and was published in \textit{The Astronomical Journal.\/}
The full text of this paper can be accessed through SAO/NASA Astrophysics Data System (hereafter referred to as ADS).
This paper deals with a restricted three-body system including
the central mass (Sun),
a small perturbed body (asteroid), and
a large perturbing body (Jupiter) orbiting on a circular orbit.
The section structure of \citet{kozai1962b} is as follows:
\textit{I.   Introduction,\/}
\textit{II.  Equations of motion,\/}
\textit{III. Stationary point,\/}
\textit{IV.  Disturbing function,\/}
\textit{V.   Case for small $\alpha$,\/}
\textit{VI.  Trajectory,\/} and
\textit{VII. Remarks.\/}
This paper was more widely and quickly recognized than any other literature that dealt with a similar subject,
and consequently,
the influence of this paper on later studies is huge.
As a result, the citation frequency of this paper is very high,
and is still increasing now
(see the descriptions later such as in Section \ref{sssec:lidov-citation}).
\citet{kozai1962b} was also selected as one of the 53 
``Selected Fundamental Papers Published this Century in the Astronomical Journal and the Astrophysical Journal,'' \citep{abt1999}.

\subsection{Purpose, method, findings\label{ssec:K62-purpose}}
\citeauthor{kozai1962b}'s purpose, method, and findings in his work are
concisely summarized in his abstract.
Though it may be unnecessary for some readers, we reproduce it here:
\begin{quotation}
``Secular perturbations of asteroids with high inclination and
eccentricity moving under the attraction of the sun and Jupiter are
studied on the assumption that Jupiter's orbit is circular.
After short-periodic terms in the Hamiltonian are eliminated,
the degree of freedom for the canonical equations of motion
can be reduced to 1.
\par
Since there is an energy integral, the equations can be solved by
quadrature. When the ratio of the semimajor axes of the asteroid and
Jupiter takes a very small value, the solutions are expressed by
elliptic functions.
\par
When the $z$ component of the angular momentum (that is, Delaunay's $H$)
of the asteroid is smaller than a certain limiting value, there are
both a stationary solution and solutions corresponding to libration
cases. The limiting value of $H$ increases as the ratio of the semimajor
axes increases, i.e., the corresponding limiting inclination drops
from $39^{\circ}.2$ to $1^{\circ}.8$ as the ratio of the axes
increases from 0.0 to 0.95.'' (abstract, p. K591)
\end{quotation}

We will see what each of his points means in what follows.
Note that the specific value of ``limiting inclination'' $(39^{\circ}.2)$
that \citeauthor{kozai1962b} mentions in the above third paragraph
corresponds to $\cos^{-1} \sqrt{\frac{3}{5}}$, as we will see soon.

The first section (\textit{``I. Introduction''\/}) seems like
an extended abstract, where \citeauthor{kozai1962b} explains
more about each of the major points mentioned in the abstract. 
In the first paragraph of this section,
\citeauthor{kozai1962b} mentions the fact that conventional perturbation
theories such as those exploiting the Laplace coefficients
basically assume that the eccentricity and inclination of objects are small.
He writes:
\begin{quote}
``The stability of the solar system has been proved in the sense that no
secular change occurs in the semimajor axes of planetary orbits, and
that secular changes of the eccentricities and inclinations are
limited within certain small domains. However, the classical theory of
secular perturbations for the eccentricity and inclination is based on
the assumption that the squares of the eccentricity and inclination
are negligible. Although this assumption may be reasonable for major
planets, it may not be for some asteroids.'' (p. K591)
\end{quote}

As \citeauthor{kozai1962b} wrote in the above,
the major planetary orbits exhibit quasi-periodic oscillations
with eccentricities and inclinations remaining reasonably small for
billions of years \citep[e.g.][]{ito2002a,batygin2008,laskar2009,batygin2015a}.
But this is not always the case for the small solar system bodies.
In the second paragraph of this section
\citeauthor{kozai1962b} points this fact out,
using some symbolic notations ($A$ or $B$) as follows:
\begin{quote}
``The assumption in the classical theory means that a term such as
$B e^2 \sin^2 i \cos 2g$ is negligibly small as compared with the principal
term $A (e^2 - \sin^2 i)$ in the secular part of the disturbing function.
However, as the value of $B$ increases much more rapidly than does
that of $A$ with the ratio of the semimajor axes of the asteroid and
the perturbing planet, the $B$ term cannot be neglected
when the eccentricity and inclination assume large values.
For example, the rate of change of the argument of perihelion, which is
proportional to $A + B \sin^2 i \cos 2g$, may vanish at a certain point when
the inclination of the asteroid takes a reasonably large value.'' (p. K591)
\end{quote}

Readers will later encounter more specific expressions of each of the terms in the disturbing function.
A point to note in the above paragraph is that,
\citeauthor{kozai1962b} mentions a possibility for argument of perihelion
of an asteroid to stay around a fixed value
when its inclination is large enough.

After mentioning a few relevant studies (including \citeauthor{lidov1961}'s work)
in the third paragraph of this section,
\citeauthor{kozai1962b} states his method in the fourth paragraph:
\begin{quote}
``The present paper treats an analytical theory on secular perturbations
of asteroids with high inclination and eccentricity by assuming that
only Jupiter, moving in a circular orbit, is the disturbing body.
This theory may, of course, be applied also to comets or satellites
disturbed by the sun.'' (p. K591)
\end{quote}

The fifth paragraph is about the advantage of
\citeauthor{kozai1962b}'s way to expand the disturbing function into
a power series of the ratio of semimajor axes, $\alpha$. He writes:
\begin{quote}
``The conventional technique for developing the disturbing function
cannot be adopted here, since neither the eccentricity nor the
inclination is considered small. Nor can numerical harmonic analysis
be adopted, since variations of orbital elements may not be regarded
as small quantities. Therefore, the disturbing function has to be
developed into a power series of $\alpha$, the ratio of the semimajor axes of
the asteroid and Jupiter, although convergence of the series may be slow.''
(p. K591)
\end{quote}

The sixth paragraph depicts the possibility to reduce
the degrees of freedom of the system by double averaging.
This procedure makes the system integrable,
and we can then obtain a formal solution by quadrature:
\begin{quote}
``Short-periodic terms depending on the two mean anomalies can be
eliminated from the disturbing function by Delaunay's transformations.
The longitudes of the ascending nodes of Jupiter and
the asteroid disappear by the theorem on elimination of nodes.
Therefore, the equations of motion for the asteroid are reduced
to canonical equations of one degree of freedom with a
time-independent Hamiltonian.
Therefore, the equations can be solved by a quadrature.'' (p. K591)
\end{quote}

In the seventh paragraph,
after mentioning the existence of an analytic solution of the system
expressed by an elliptic function,
\citeauthor{kozai1962b} states the major conclusion of his work:
A stationary solution of $(e,g)$ shows up when a constant parameter
$\left( 1-e^2 \right) \cos^2 i$ is smaller than 0.6:
\begin{quote}
``In fact, the solutions can be expressed by elliptic functions
approximately when $\alpha$ takes a very small value.
For this case there are both one stationary and
some libration solutions when $\left(1-e^2\right) \cos^2 i$,
which is constant, is smaller than 0.6.'' (p. K591)
\end{quote}

Finally in the eighth paragraph,
\citeauthor{kozai1962b} states another major conclusion that he obtained:
Dependence of the limiting value of $\left(1-e^2\right) \cos^2 i$
for the stationary solutions to exist on the semimajor axes ratio, $\alpha$.
Here is what he wrote:
\begin{quote}
``As $\alpha$ increases, the upper limit of $\left(1-e^2\right) \cos^2 i$ for the
existence of a stationary solution increases.
When $\alpha$ is 0.85, the limit is as large as 0.90.'' (p. K591)
\end{quote}

\subsection{Equations of motion\label{ssec:Kozai-eom}}
Unlike his well-organized abstract and introduction (Section I),
we have to say that \citeauthor{kozai1962b}'s following four sections
(Sections II, III, IV, V) are poorly organized.
Not only is it not easy to follow the formulations there,
but some descriptions are too terse (or too conceptual) that we do not follow his intention.
However, now that most readers are familiar with the method and conclusion of \citeauthor{kozai1962b}'s work,
let us briefly summarize these sections.

The section ``\textit{II. Equations of motion\/}'' is devoted to describing
the canonical equations of motion. In this section \citeauthor{kozai1962b}
briefly mentions that the doubly averaged CR3BP has just one degree of freedom,
and the equations of motion can be solved by quadrature.
This section starts from \citeauthor{kozai1962b}'s definition of the variables.
He denotes $m$ as asteroid's mass, and $m'$ as Jupiter's mass.
The solar mass is set to unity.
\citeauthor{kozai1962b} uses the Delaunay elements defined as follows:
\begin{equation}
\begin{aligned}
  L &= \ktext a^{\frac{1}{2}},  & \tpspcD l &= \mbox{ mean anomaly}, \\
  G &= L \left( 1-e^2 \right)^{\frac{1}{2}},
                                & \tpspcD g &= \mbox{ argument of perihelion}, \\
  H &= G \cos i,                & \tpspcD h &= \mbox{ longitude of ascending node}, \\
\end{aligned}
  \tag{K01-\arabic{equation}}
  \stepcounter{equation}
  \label{eqn:K01}
\end{equation}
where
$l$, $g$, $h$ are the canonical coordinates and $L$, $G$, $H$ are the corresponding conjugate momenta.
$\ktext$ is the Gaussian gravitational constant,
which is practically equivalent to $\sqrt{{\cal G}}$ that we used
in our Section \ref{sec:CR3BP} \citep[cf.][their p. 57]{brouwer1961}.
Note that
\citeauthor{kozai1962b}'s definition of $L$ in Eq. \eqref{eqn:K01} seems
slightly different from those in standard textbooks, $L = \sqrt{\mu a}$
\citep[e.g.][p. 161]{boccaletti1996},
although they are fundamentally equivalent.

\citeauthor{kozai1962b} expresses all the quantities of Jupiter with
primes such as $L'$ and $l'$. Also, a variable $\ktext'$ is defined as
\begin{equation}
  {\ktext '}^2 = \frac{\ktext^2}{1+m'} \frac{{m'}^2}{m^2}
            = \ktext^2 {\mu '}^2 \frac{1+m'}{m^2},
  \tag{K02-\arabic{equation}}
  \stepcounter{equation}
  \label{eqn:K02}
\end{equation}
where $\mu'$ is the reduced mass of Jupiter,
\begin{equation}
  \mu' = \frac{m'}{1 + m'} .
  \label{eqn:reduced-Jmass}
\end{equation}

Next,
\citeauthor{kozai1962b} defines
the coordinates of Jupiter with the Sun at the origin, and
that of the asteroid with the barycenter of Jupiter and the Sun at the origin.
It is nothing but the Jacobi coordinates that we mentioned in
Section \ref{ssec:CR3BP-R}, but note that \citeauthor{kozai1962b} assumes
that the asteroid's orbit is located inside Jupiter's orbit (i.e. $r < r'$).
Then,
\citeauthor{kozai1962b} expresses the Hamiltonian $F$ of the system as follows:
\begin{equation}
\begin{aligned}
  F &= \frac{\ktext^4}{2 L^2}
     + \frac{m}{\mu'} \frac{{\ktext '}^4}{2 {L'}^2} \\
    & \quad
     + \ktext^2 \mu' \left\{
      \left[ r^2 - 2 r r' \frac{s}{1+m'} + \left( \frac{r'}{1+m'}\right)^2 \right]^{-\frac{1}{2}} \right. \\
    & \quad\quad
      \left. - \frac{r}{r'} \frac{s}{1+m'} \right\} ,
\end{aligned}
  \tag{K03-\arabic{equation}}
  \stepcounter{equation}
  \label{eqn:K03}
\end{equation}
where
\begin{equation}
  s = \frac{x x' + y y' + z z'}{r r'} .
  \tag{K04-\arabic{equation}}
  \stepcounter{equation}
  \label{eqn:K04}
\end{equation}

Readers should find the equivalence between the Hamiltonian $F$
in Eq. \eqref{eqn:K03} and ${\cal H}$ in Eq. \eqref{eqn:def-H-3BP-general}.
The third term of the right-hand side of Eq. \eqref{eqn:K03} is
not yet expanded into the Legendre polynomials.
Since $s$ in Eq. \eqref{eqn:K04} corresponds to
    $\cos S_{12}$ in Eq. \eqref{eqn:def-H-3BP-general}
(or $\cos S$      in Eq. \eqref{eqn:def-rrdcosS}),
the last  term of the right-hand side of Eq. \eqref{eqn:K03} is
equivalent to the indirect part of the disturbing function expressed
in Eq. \eqref{eqn:def-R-conventional-ndash}.
Also, keep in mind that the indirect part cancels out with the $P_1$ term
in the expansion using the Legendre polynomials,
as we already saw in Eq. \eqref{eqn:def-H-3BP-general}.

After expressing the Hamiltonian $F$ in Eq. \eqref{eqn:K03},
\citeauthor{kozai1962b} tries to reduce the degrees of freedom
of the system through two steps.
First, \citeauthor{kozai1962b} applies ``Jacobi's elimination of the nodes''
to the Hamiltonian $F$.
It is known that the Hamiltonian in the system considered includes
$h$ and $h'$ only in the form of $h-h'$ \citep[e.g.][]{nakai1985}.
By choosing the invariable plane as a reference plane,
the Hamiltonian $F$ acquires the rotation symmetry around
the total angular momentum vector
\citep[e.g.][]{jacobi1843a,jacobi1843b,charlier1902,charlier1907,jefferys1966}.
This circumstance is typically expressed 
by a relationship
\begin{equation}
  h - h' = \pi .
  \label{eqn:eliminationofthenodes}
\end{equation}
This relationship enables us to eliminate both $h$ and $h'$ from the Hamiltonian,
therefore the conjugate momenta $H$ and $H'$ become constants of motion.
As a consequence, the original Hamiltonian
$F(L,G,H,L',G',H', l,g,h,l',g',h')$ in Eq. \eqref{eqn:K03}
with six degrees of freedom is converted into
$F(L,G,  L',G',    l,g,  l',g'   ) $ with four degrees of freedom.

Note that \citeauthor{kozai1962b} actually assumed
\begin{equation}
  h = h' ,
  \tag{K05-\arabic{equation}}
  \stepcounter{equation}
  \label{eqn:K05}
\end{equation}
as a consequence of Jacobi's elimination of the nodes,
not Eq. \eqref{eqn:eliminationofthenodes} that is widely recognized.
Eq. \eqref{eqn:K05} certainly eliminates both $h$ and $h'$ from the Hamiltonian
as long as it contains $h$ and $h'$ just in the form of $h-h'$, and
the conclusion that \citeauthor{kozai1962b} stated would not be affected.
However, we have not found any expressions similar to Eq. \eqref{eqn:K05} in other literature, and
we do not know \citeauthor{kozai1962b}'s intention.

The second step that \citeauthor{kozai1962b} took in order to reduce
the degrees of freedom of the system is double averaging.
As we already summarized its concept in Section \ref{ssec:CR3BP-averaging},
fast-oscillating variables can be eliminated from Hamiltonian by averaging.
In the present case,
the mean anomalies of asteroid $(l)$ and Jupiter $(l')$ can be eliminated.
Then their conjugate momenta $L$ and $L'$ become constants of motion.
As we mentioned in Section \ref{ssec:CR3BP-averaging},
the elimination of fast-oscillating variables by an averaging procedure is a part of canonical transformation.
Therefore \citeauthor{kozai1962b} puts a superscript $\ast$ on the variables
that have gone through averaging as being canonically transformed.
Now, the  Hamiltonian
$F     (L,G,  L',G',    l,g,  l',g'         )$ with four degrees of freedom is
transformed into a new Hamiltonian
$F^\ast(G^\ast, {G'}^\ast, g^\ast, {g'}^\ast)$ with two degrees of freedom.
\citeauthor{kozai1962b} expresses the new Hamiltonian $F^\ast$ as follows:
\begin{equation}
  F^\ast = \frac{\ktext^4}{2 {L^\ast}^2} + m' W^\ast ,
  \tag{K08-\arabic{equation}}
  \stepcounter{equation}
  \label{eqn:K08}
\end{equation}
with
\begin{equation}
  W^\ast = \frac{\ktext^2}{4\pi^2} \int_0^{2\pi} \int_0^{2\pi}
      \frac{1}{\left( {r'}^2 - 2 r r' s + r^2\right)^\frac{1}{2}} dl dl' .
  \tag{K09-\arabic{equation}}
  \stepcounter{equation}
  \label{eqn:K09}
\end{equation}

Note that in the original work by \citeauthor{kozai1962b},
the left-hand side of Eq. \eqref{eqn:K09} is $W$, not $W^\ast$.
However, we believe this is a simple typographic error
because the right-hand side of Eq. \eqref{eqn:K09} is doubly averaged,
as is the new Hamiltonian $F^\ast$ appearing in Eq. \eqref{eqn:K08}.
Therefore we replace $W$ for $W^\ast$ in the following discussion.

$F^\ast$ in Eq. \eqref{eqn:K08} does not include the Hamiltonian that drives the Keplerian motion of Jupiter.
$W^\ast$ in Eq. \eqref{eqn:K09} expresses the perturbation Hamiltonian,
but it contains just the direct part of the disturbing function;
the indirect part
(i.e. the last term of the right-hand side of Eq. \eqref{eqn:K03}) is omitted.
\citeauthor{kozai1962b} then assumes that Jupiter's eccentricity is negligibly small,
and that its argument of perihelion ${g'}^\ast$ and
its conjugate momentum ${G'}^\ast$ disappear from $F^\ast$.
This makes the degrees of freedom unity, and \citeauthor{kozai1962b} gives
the canonical equations of motion of the asteroid as
\begin{equation}
  \DD{G^\ast}{t} =  m' \DP{W^\ast}{g^\ast}, \quad
  \DD{g^\ast}{t} = -m' \DP{W^\ast}{G^\ast},
  \tag{K10-\arabic{equation}}
  \stepcounter{equation}
  \label{eqn:K10}
\end{equation}
with an integral
\begin{equation}
  W^\ast = \mbox{ const.}
  \tag{K11-\arabic{equation}}
  \stepcounter{equation}
  \label{eqn:K11}
\end{equation}
As 
\citeauthor{kozai1962b} writes at the end of this section,
in principle we can solve Eq. \eqref{eqn:K10} by quadrature.

\subsection{Stationary point\label{ssec:Kozai-stationary}}
Next in ``\textit{III. Stationary point,\/}''
\citeauthor{kozai1962b} gives his estimate on the location of
the stationary points that the perturbation part of the doubly averaged
Hamiltonian $W^\ast$ can have. At these stationary points,
$g^\ast$ and $e^\ast$ (therefore $G^\ast$) of the perturbed body are
supposed to be constant.
What \citeauthor{kozai1962b} employs here is a numerical analysis,
not an analytical treatment.
This is perhaps not what many readers would anticipate him to do.

\citeauthor{kozai1962b} first states that,
$W^\ast$ in Eq. \eqref{eqn:K09} takes the following form when $e'=0$:
\begin{equation}
  W^\ast = \sum_{j=0} A_j \left( \alpha, G^\ast, H \right) \cos 2 j g^\ast,
  \tag{K12-\arabic{equation}}
  \stepcounter{equation}
  \label{eqn:K12}
\end{equation}
with
\begin{equation}
  \alpha = \left( \frac{\ktext' L^\ast}{\ktext {L'}^\ast} \right)^2 .
  \tag{K13-\arabic{equation}}
  \stepcounter{equation}
  \label{eqn:K13}
\end{equation}

It is clear that $\alpha$ in Eq. \eqref{eqn:K13} is practically
equivalent to the ratio of semimajor axes between the perturbed and
perturbing body, $\frac{a}{a'}$.
Meanwhile $A_j$ in Eq. \eqref{eqn:K12} is a coefficient depending on $\alpha$, $G^\ast$, and $H$.
\citeauthor{kozai1962b} did not give any proof of Eq. \eqref{eqn:K12}
or specific function form of $A_j$ at all at this point.
Its function form, however, is revealed in later sections
when he presents an analytic expansion of $W^\ast$.

Once admitting that the expansion form of Eq. \eqref{eqn:K12} is valid,
we can accept \citeauthor{kozai1962b}'s statement that
$\DD{g^\ast}{t}$ vanishes when $\sin 2g^\ast = 0$
owing to the canonical equations of motion \eqref{eqn:K10}.
More specifically writing,
from Eq. \eqref{eqn:K12} and the first equation of Eq. \eqref{eqn:K10} we have
\begin{equation}
\begin{aligned}
  \DP{W^\ast}{g^\ast}
  & =  \sum_{j=0}    A_j \left(\alpha, G^\ast, H\right) \DP{}{g^\ast} \cos 2 j g^\ast \\
  & = -\sum_{j=0} 2j A_j \left(\alpha, G^\ast, H\right) \sin 2 j g^\ast ,
\end{aligned}
\label{eqn:Kozai-DPWg}
\end{equation}
which indicates that $\sin 2g^\ast = 0$ is a condition for $W^\ast$
to be stationary somewhere in phase space.
$\sin 2g^\ast = 0$ means $\cos 2g^\ast = +1$ or $\cos 2g^\ast = -1$.
Then, from the second equation of Eq. \eqref{eqn:K10} we get
\begin{equation}
  \DP{W^\ast}{G^\ast}
    = \sum_{j=0} \DP{A_j \left(\alpha, G^\ast, H \right)}{G^\ast} \cos 2 j g^\ast .
  \label{eqn:Kozai-DPWG}
\end{equation}
This result means that the considered system has a stationary point under either of the following conditions:
\stepcounter{equation}
\begin{alignat}{1}
  \sum_{j=0}      \DP{A_j}{G^\ast} &= 0 \;\; (\mbox{when } \cos 2 g^\ast = +1),  \tag{K14-\arabic{equation}}
  \label{eqn:K14} \\
  \sum_{j=0}(-1)^j\DP{A_j}{G^\ast} &= 0 \;\; (\mbox{when } \cos 2 g^\ast = -1).
\stepcounter{equation}
  \tag{K15-\arabic{equation}}
  \label{eqn:K15}
\end{alignat}

Here \citeauthor{kozai1962b} also puts another inequality
\begin{equation}
  H \leq G^\ast \leq L^\ast ,
  \tag{K16-\arabic{equation}}
  \stepcounter{equation}
  \label{eqn:K16}
\end{equation}
which seems obvious for us because 
$\sqrt{1 - e^2} \leq 1$ and
$|\cos i|       \leq 1$ as long as we consider elliptic orbits.

\begin{table}[t]\centering
\vspace{-1.5mm}
\caption[]{%
Reproduction of Table I of \citet[][p. K592]{kozai1962b}.
The values in the column named as ``$i_0$'' are obtained from
\citeauthor{kozai1962b}'s ``numerical harmonic analysis.''
The values in the column named as ``$i_0$ approx'' are from
the analytic expansion of $W^\ast$ that
\citeauthor{kozai1962b} accomplished in the later section
(see our p. \pageref{ssec:Kozai-stationary} for detail).
$\left(\frac{H_0}{L^\ast}\right)^2$ is related to $\cos i_0$
through Eq. \eqref{eqn:K17}.
The values of $i_0$ that are numerically obtained are later plotted
as our \mysymfigO \ref{fig:I02-table} (p. \pageref{fig:I02-table})
for a comparison with \citeauthor{vonzeipel1910}'s achievement.
}
\label{tbl:Kozai-table1}
\begin{tabular}[hbtp]{ccrc}
\hline
$\alpha$ & $\left(\frac{H_0}{L^\ast}\right)^2$ & \multicolumn{1}{c}{$i_0$} & $\substack{i_0 \\ {\mathrm{approx}}}$ \\
\hline
0.00 &  0.60 000 &	$39^\circ.231$ & $39^\circ.231$ \\
0.05 &	0.60 116 &	$39^\circ.164$ & $39^\circ.164$ \\
0.10 &	0.60 464 &	$38^\circ.960$ & $38^\circ.960$ \\
0.15 &	0.61 043 &	$38^\circ.620$ & $38^\circ.620$ \\
0.20 &	0.61 849 &	$38^\circ.146$ & $38^\circ.146$ \\
0.25 &	0.61 880 &	$37^\circ.536$ & $37^\circ.535$ \\
0.30 &	0.64 133 &	$36^\circ.791$ & $36^\circ.790$ \\
0.35 &	0.65 599 &	$35^\circ.911$ & $35^\circ.905$ \\
0.40 &	0.67 274 &	$34^\circ.894$ & $34^\circ.875$ \\
0.45 &	0.69 154 &	$33^\circ.738$ & $33^\circ.694$ \\
0.50 &	0.71 230 &	$32^\circ.437$ & $32^\circ.355$ \\
0.55 &	0.73 495 &	$30^\circ.986$ & $30^\circ.860$ \\
0.60 &	0.75 940 &	$29^\circ.374$ & $29^\circ.239$ \\
0.65 &	0.78 556 &	$27^\circ.586$ & $27^\circ.566$ \\
0.70 &	0.81 330 &	$25^\circ.600$ & $25^\circ.925$ \\
0.75 &	0.84 252 &	$23^\circ.380$ & $24^\circ.410$ \\
0.80 &	0.87 305 &	$20^\circ.874$ & $23^\circ.078$ \\
0.85 &	0.90 488 &	$17^\circ.964$ & $21^\circ.926$ \\
0.90 &	0.94 581 &	$13^\circ.460$ & $20^\circ.963$ \\
0.95 &	0.99 900 &	$ 1^\circ.811$ & $\cdots$       \\
\hline
\end{tabular}
\end{table}

When $\cos 2 g^\ast = +1$,
\citeauthor{kozai1962b} claims that $W^\ast$ does not have any
stationary points according to his numerical analysis.
Literally citing his description:
\begin{quote}
``It has been proved numerically that Eq. (K14) does not have such a solution
except for $H = G^\ast = 0$ and
the equation $d g^\ast/dt=0$ has no meaningful solution other than
$\sin 2g^\ast = 0$, at least when $\alpha$ is less than 0.8.'' (p. K593)
\end{quote}

However, details of \citeauthor{kozai1962b}'s numerical analysis are not presented in his paper at all.
Note also that we changed the original expression ``Eq. (14)'' into ``Eq. (K14)'' in the above citation for clarifying that this equation denotes Eq. \eqref{eqn:K14}.
We will continue to adopt this manner throughout the rest of this monograph.

When $\cos 2 g^\ast = -1$,
\citeauthor{kozai1962b} describes the condition for $W^\ast$ to have
stationary points as follows:
\begin{quote}
``Equation (K15) has a solution when $H$ is equal to or smaller than
  a limiting value $H_0$.
  When $H$ is equal to $H_0$, the stationary solution appears at $G^\ast=1$.
  As $H$ decreases, Eq. (K14) has a smaller value of $G^\ast$ as the root,
  and when $H$ is zero, $G^\ast=0$ corresponds to the stationary value.
  When $H$ is equal to $H_0$, the corresponding inclination is derived by
\begin{equation}
  H_0 = L^\ast \cos i_0 .
  \tag{K17-\arabic{equation}}
  \stepcounter{equation}
  \label{eqn:K17}
\end{equation}
  Both $H_0$ and $i_0$ depend on $\alpha$ and are derived by
  numerical harmonic analysis of $d W^\ast/dG^\ast$.
  The results are given in Table I and as a solid line in Fig. 1.'' (p. K593)
\end{quote}
\label{pg:K593}

For facilitating reader's understanding of the above quoted part,
we have reproduced \citeauthor{kozai1962b}'s Table I
as our Table \ref{tbl:Kozai-table1}.
By mentioning his results obtained through the analytic expansion of $W^\ast$
up to $\Oaloct$ which is not yet presented at this point in his paper,
\citeauthor{kozai1962b} continues as follows:
\begin{quote}
``Besides the numerical harmonic analysis, values of $i_0$ are derived
analytically by developing the disturbing function into power series of
$\alpha$ up to the eighth degree, shown in the last column of Table I
and as a broken line in Fig. 1.
Comparison of the two lines in Fig. 1 shows that
the analytical method can provide rather good values
for $i_0$ up to $\alpha=0.7$.'' (p. K593)
\end{quote}
\label{pg:numericalharmonicanalysis}

We have to say that, 
we do not feel that there will be many readers who correctly understand
\citeauthor{kozai1962b}'s logic and intention at this point,
as his explanations are lame.
Also, the sudden appearance of the result obtained from his own
analytic expansion of the disturbing function at this point seems odd.
We believe that readers of this monograph
who return to this section after going through
\citeauthor{kozai1962b}'s paper will find that their understanding is much deeper.

\citeauthor{kozai1962b} concludes this section with the following paragraph.
It mentions an important, quantitative conclusion on the largest value of
$i_0$ and its dependence on $\alpha$. However, at this point there is
no explanation as to how \citeauthor{kozai1962b} reached this result,
or what kind of value ``$39^\circ.2$'' means:
\begin{quote}
``In the first approximation, $i_0$ and $H_0$ do not depend on
Jupiter's mass $m'$. The value of $i_0$ drops from $39^\circ.2$ to
$1^\circ.8$ as $\alpha$ increases from 0.0 to 0.95.
However, there are few asteroids that have $H$ smaller than $H_0$.
When $\alpha$ is larger than 0.95,
there may be a stationary solution for any value of $H$.'' (p. K593)
\end{quote}
\label{pg:first39appearance}

\subsection{Disturbing function\label{ssec:Kozai-R}}
In the next section ``\textit{IV. Disturbing function,\/}''
\citeauthor{kozai1962b} presents his detailed calculation on the
analytic expansion of the disturbing function up to $\Oalsqr$.
He uses the notation $R$ for the direct part of the disturbing function.
Its definition is the same as in our Section \ref{sec:CR3BP}:
\begin{equation}
  R = \frac{\ktext^2 m'}{\left(r^2 - 2 r r' s + {r'}^2\right)^{\frac{1}{2}}} % \\
    = \frac{\ktext^2 m'}{r'} \sum_{j=0}^\infty P_j (s)
         \left( \frac{r}{r'} \right)^j ,
  \tag{K18-\arabic{equation}}
  \stepcounter{equation}
  \label{eqn:K18}
\end{equation}
where $s$, which is seen in Eq. \eqref{eqn:K03} and is equivalent to $\cos S$ in our Section \ref{ssec:CR3BP-R}, is expressed as
\begin{equation}
\begin{aligned}
  s &=        \cos (f+g) \cos (f'+g') \\
    &\quad\quad\quad
     + \cos i \sin (f+g) \sin (f'+g') .
\end{aligned}
  \tag{K07-\arabic{equation}}
  \stepcounter{equation}
  \label{eqn:K07}
\end{equation}
Note that the expression of $s$ in Eq. \eqref{eqn:K07} is an outcome
of the assumption that the reference plane of the system coincides
with the perturber's orbit.

As we mentioned in our Section \ref{sec:CR3BP},
the $P_1$ term can be dropped from Eq. \eqref{eqn:K18}.
Also, after the averaging procedure using the mean anomaly $l'$ of
the perturbing body, all the odd-order terms
$(j=3,5,7,\ldots)$
disappear if the perturbing body is on a circular orbit $(e'=0)$.
Hence \citeauthor{kozai1962b} describes the major part of
the disturbing function $R_1$ that is averaged by $l'$ as follows:
\begin{equation}
\begin{aligned}
  R_1 &= \frac{1}{2\pi} \int_0^{2\pi} (R)_{e'=0} dl' \\
      &= \frac{\ktext^2 m'}{a'} \sum_{j=0}^\infty P_{2j} (s_1)
         \left( \frac{r}{a'} \right)^{2j} .
\end{aligned}
  \tag{K19-\arabic{equation}}
  \stepcounter{equation}
  \label{eqn:K19}
\end{equation}
Recall that $r'$ is now equal to $a'$, the perturbing body's semimajor axis.
Note also that \citeauthor{kozai1962b} did not give any definitions of $s_1$
in Eq. \eqref{eqn:K19}.
We can say it is a symbolic expression for the averaged value of $s$ by the mean anomaly $l'$ of the perturbing body such as
\begin{equation}
  s_1 \equiv \frac{1}{2\pi} \int_0^{2\pi} s dl'
           = \left< s \right>_{l'} ,
  \label{eqn:def-s1}
\end{equation}
which is practically equivalent to $\left< \cos S \right>_{l'}$
seen in our Section \ref{ssec:CR3BP-R}
(see Eq. \eqref{eqn:def-cosS-avr} for comparison).
A confusing point in \citeauthor{kozai1962b}'s notation here is that,
the averaged values of $s$ never actually show up in the form of
Eq. \eqref{eqn:def-s1}: They show up in the form of even powers such as
$\left< s^2 \right>_{l'},
 \left< s^4 \right>_{l'},
 \left< s^6 \right>_{l'},
 \left< s^8 \right>_{l'}$,
but \citeauthor{kozai1962b} denotes them as
$s_1^2, s_1^4, s_1^6, s_1^8$.

After introducing Eq. \eqref{eqn:K19},
\citeauthor{kozai1962b} presents the specific function forms of 
$s_1^2, s_1^4, s_1^6, s_1^8$ in Eq. (K20) together with
the averaged values of the Legendre polynomials of
the corresponding order,
$P_2(s_1),
 P_4(s_1),
 P_6(s_1),
 P_8(s_1)$ in Eq. (K21).
\citeauthor{kozai1962b} did not show the specific definition of $P_{2j} (s_1)$, but it is as follows:
\begin{equation}
  P_{2j} (s_1) \equiv \left< P_{2j} (s) \right>_{l'}
               =      \frac{1}{2\pi} \int_0^{2\pi} P_{2j} (s) dl' .
  \label{eqn:def-Ps-Kozai}
\end{equation}
We do not reproduce the specific forms of
$s_1^2,    s_1^4,    s_1^6,    s_1^8$ and
$P_2(s_1), P_4(s_1), P_6(s_1), P_8(s_1)$ in this monograph
because of their complexity.
See Eqs. (K20) and (K21) for the detail.

The next step is to average the disturbing function by the mean anomaly $l$ of the perturbed body.
\citeauthor{kozai1962b} carried this task out using one of the formulas
devised by \citet[][see Eq. (K22) which we do not reproduce here]{tisserand1889}.
The resulting doubly averaged disturbing function $W^\ast$ is very complicated,
but we venture to reproduce it here.
First, remark \citeauthor{kozai1962b}'s abbreviated notations
\begin{equation}
  \theta = \frac{H}{G^\ast}, \quad
  \eta   = \frac{G^\ast}{L^\ast} .
  \tag{K24-\arabic{equation}}
  \stepcounter{equation}
  \label{eqn:K24}
\end{equation}
It is obvious that
$\theta$ is practically equivalent to $\cos i$, and
$\eta$   is practically equivalent to $\sqrt{1-e^2}$,
if we ignore their difference denoted by $\ast$.
Using the notations defined by Eq. \eqref{eqn:K24},
$W^\ast$ becomes up to $\Oaloct$ as
\begin{align*}
&W^\ast
 = \frac{\ktext^2}{a'} \alpha^2 \left\{
  \frac{1}{16}
  \left[ -\left( 1 -3\theta^2 \right) \left( 5 -3\eta^2 \right) \right.
                                \right. \\
&\; \left .
 +15 \left( 1-\theta^2 \right) \left( 1-\eta^2 \right) \cos 2g^\ast
  \right] \\
&\; +\frac{9}{2^{12}} \alpha^2 \left[ 
       \left( 3 -30\theta^2 +35\theta^4 \right)
       \left( 63 -70\eta^2 + 15\eta^4 \right) \right. \\
&\; -140 \left( 1-\theta^2 \right) \left( 1 -7\theta^2 \right)
       \left( 1-\eta^2 \right) \left( 3 -\eta^2 \right) \cos 2 g^\ast \\
&\;                            \left.
  +735 \left( 1-\theta^2\right)^2 \left(1-\eta^2 \right)^2 \cos 4g^\ast
                             \right] \\
&\; +\frac{5}{2^{17}} \alpha^4 \left[
       -10 \left( 5 - 105 \theta^2 + 315 \theta^4 - 231 \theta^6 \right)
                             \right. \\
&\quad \times \left( 429 -693 \eta^2 + 315\eta^4 - 35\eta^6 \right) \\
&\; + 315 \left( 1-\theta^2 \right)
          \left( 1-18\theta^2 +33\theta^4 \right) \left( 1-\eta^2 \right) \\
&\quad \times 
           \left( 143 -110\eta^2 + 15\eta^4 \right) \cos 2 g^\ast \\
&\; -4158\left(1-\theta^2\right)^2 \left( 1-11\theta^2\right) \left( 1-\eta^2\right)^2 \\
&\quad \times 
  \left( 13 - 3\eta^2 \right) \cos 4 g^\ast \\
&\;                             \left.
+99099 \left(1-\theta^2\right)^3 \left(1-\eta^2 \right)^3 \cos 6 g^\ast
                                \right] \\
&\; +\frac{175}{2^{28}} \alpha^6 \\
&\quad \times \left[
  7 \left(35 -1260\theta^2 +6930\theta^4 - 12012\theta^6 +6435\theta^8\right)
              \right. \\
&\quad \times
    \left(12155 - 25740\eta^2 + 18018\eta^4 -4620\eta^6 + 315\eta^8 \right) \\
&\; -27720\left(1-\theta^2\right) \left(1 -33\theta^2 +143\theta^4 -143\theta^6\right) \\
&\quad \times
    \left(1-\eta^2\right) \left( 221 -273\eta^2 +91\eta^4 -7\eta^6 \right)
    \cos 2 g^\ast \\
&\; +396396 \left(1-\theta^2\right)^2 \left( 1 -26\theta^2 +65\theta^4 \right) \left( 1-\eta^2\right)^2 \\
&\quad \times
  \left( 17 - 10\eta^2 +\eta^4 \right) \cos 4 g^\ast \\
&\; -490776 \left( 1-\theta^2\right)^3 \left( 1-15\theta^2\right) \left( 1-\eta^2\right)^3 \\
&\quad \times
  \left( 17 - 3\eta^2\right) \cos 6 g^\ast \\
&\; \left.     \left.
+15643485 \left( 1-\theta^2 \right)^4 \left( 1-\eta^2 \right)^4 \cos 8 g^\ast
               \right]
    \right\} .
  \tag{K23-\arabic{equation}}
  \stepcounter{equation}
  \label{eqn:K23}
\end{align*}
Through our own algebraic manipulation \citep{ito2016},
we have confirmed that there is no miscalculation or typographic error
in the expansion of $W^\ast$ in Eq. \eqref{eqn:K23}.

For reference, 
let us take just the terms in the leading-order 
$\Oalsqr$ out of the expression of $W^\ast$ in Eq. \eqref{eqn:K23}.
Translating $\theta$ and $\eta$ into the standard notations of
orbital elements using $e$ and $i$, and
using ${\cal G}$ instead of $\ktext$,
and ignoring all $\ast$ from the symbols for simplicity,
this quantity becomes
\begin{equation}
\begin{aligned}
{} & W^\ast_{\Oalsqr}
=
\frac{{\cal G}}{a'} \left( \frac{a}{a'} \right)^2 \\
{} & \quad \times
\frac{1}{16}
\left[
  15 e^2 \sin^2 i \cos 2 g - \left(3e^2+2 \right) \left( 3\sin^2 i -2 \right)
\right] ,
\end{aligned}
  \label{eqn:R2-final}
\end{equation}
which we often find in modern literature
\citep[e.g.][Eq. (15) on p. 448]{naoz2016}.

Using the leading-order terms of the disturbing function
described in Eq. \eqref{eqn:R2-final},
we can write down the canonical equations of motion for $G$ and $g$ as follows:
\begin{alignat}{1}
  \DD{G}{t}
&= m' \DP{W^\ast_{\Oalsqr}}{g} \nonumber \\
&= - \frac{{\cal G} m'}{a'} \left( \frac{a}{a'}\right)^2
     \cdot \frac{15}{8} e^2 \sin^2 i \sin 2g, 
\label{eqn:dGdt-quadrupole} \\
  \DD{g}{t} &= -m' \DP{W^\ast_{\Oalsqr}}{G} \nonumber \\
            &= \frac{{\cal G} m'}{a'} \left( \frac{a}{a'}\right)^2
               \cdot \frac{ 3}{8G}
    \left[  \left( 5\cos^2 i - \left(1-e^2\right) \right) \right. \nonumber \\
  & \quad\quad
    \left.
          -5\left(  \cos^2 i - \left(1-e^2\right) \right) \cos 2g \right] .
\label{eqn:dgdt-quadrupole}
\end{alignat}

The set of equations
\eqref{eqn:dGdt-quadrupole} and
\eqref{eqn:dgdt-quadrupole}
is a simplified version of the canonical equations of motion
whose general form is Eq. \eqref{eqn:K10}.
They are also seen in conventional literature
\citep[e.g.][Eqs. (5) and (6) on their p. 127]{kinoshita1999}.
Note that we ignored all $\ast$ from the symbols in
Eqs. \eqref{eqn:dGdt-quadrupole} and
     \eqref{eqn:dgdt-quadrupole}
except for $W^\ast_{\Oalsqr}$.
\label{pg:dG-dt-2n=2}

In Section \ref{ssec:Kozai-stationary}
(p. \pageref{ssec:Kozai-stationary} of this monograph)
we introduced \citeauthor{kozai1962b}'s estimate that
$W^\ast$ can have stationary points when $\cos 2 g^\ast = -1$.
Let us see where they are located at the $\Oalsqr$ level approximation
using Eq. \eqref{eqn:dgdt-quadrupole}.

At the stationary points, we have $\DD{g^\ast}{t} = 0$.
From Eq. \eqref{eqn:dgdt-quadrupole}, this means
\begin{equation}
  10 \cos^2 i - 6 \left( 1-e^2 \right)   = 0 .
  \label{eqn:K-pre26}
\end{equation}

Here let us notify readers that \citeauthor{kozai1962b} defined an
important parameter in his discussion at the end of his Section \textit{IV\/}.
It is denoted as $\Theta$, and expressed as follows:
\begin{equation}
  \Theta = \left( \frac{H}{L^\ast} \right)^2 .
  \tag{K26-\arabic{equation}}
  \stepcounter{equation}
  \label{eqn:K26}
\end{equation}
This variable is roughly equivalent to $\left(1-e^2 \right) \cos^2 i$
and a constant,
because both $H$ and $L^\ast$ are constant as we saw
in the previous discussion.
Using $\Theta$ in Eq. \eqref{eqn:K26},
we can rewrite the condition \eqref{eqn:K-pre26} as follows:
\begin{equation}
  \left( 1-e^2 \right)^2 = \frac{5}{3} \Theta .
  \label{eqn:Theta-53}
\end{equation}
Since we have $0 \leq \left( 1-e^2 \right)^2 \leq 1$
as long as we consider elliptic orbits,
the condition \eqref{eqn:Theta-53} holds true only when
\begin{equation}
  \Theta \leq \frac{3}{5} .
  \label{eqn:Theta-max}
\end{equation}
In other words, the doubly averaged disturbing function $W^\ast_{\Oalsqr}$
in Eq. \eqref{eqn:R2-final} cannot have stationary points
unless the condition \eqref{eqn:Theta-max} is satisfied.
Recalling the definition of $\Theta$ in Eq. \eqref{eqn:K26},
Eq. \eqref{eqn:Theta-max} means that there is a threshold value of $H$
only below which the system can have stationary points.
\citeauthor{kozai1962b} designated it as $H_0$
(see his p. K593 and p. \pageref{pg:K593} of this monograph), and
its actual expression is
\begin{equation}
  H_0 = \sqrt{\frac{3}{5}} L^\ast,
  \label{eqn:def-H0-Kozai}
\end{equation}
from Eqs. \eqref{eqn:K26} and \eqref{eqn:Theta-max}.

The threshold value $H_0$ can be translated into
a threshold value of orbital inclination of the perturbed body, $i_0$.
As we showed before,
\citeauthor{kozai1962b} defined $i_0$ in Eq. \eqref{eqn:K17}
as the value that realizes $G^\ast = 1$.
This obviously happens when $e=0$.
\citeauthor{kozai1962b} also defined the corresponding threshold
$\Theta_0$ just after Eq. \eqref{eqn:K26}.
They yield the relationship
\begin{equation}
  \Theta_{e=0} = \Theta_0 = \left( \frac{H_0}{L^\ast} \right)^2 = \cos^2 i_0 .
  \label{eqn:def-Theta0-Kozai}
\end{equation}
Substituting
Eq. \eqref{eqn:def-H0-Kozai} into Eq. \eqref{eqn:def-Theta0-Kozai},
the actual value of $i_0$ is calculated as
\begin{equation}
  i_0 = \cos^{-1} \sqrt{\frac{3}{5}} = 39.^\circ 2315205 \cdots ,
  \label{eqn:imin-R2}
\end{equation}
at the $\Oalsqr$ approximation.
\citeauthor{kozai1962b} develops the same discussion at the 
$\Oaloct$ approximation based on his calculation result,
Eq. \eqref{eqn:K23}. He writes:
\begin{quote}
``The limiting value of $H$ is derived from the equation
\begin{equation*}
  \left( \DP{W^\ast}{G^\ast} \right)_{\cos 2g^\ast = -1, \eta=1} = 0,
\end{equation*}
that is
\begin{align}
& -5\Theta + 3
+ \frac{15}{32} \left( -49 \Theta^2 + 46 \Theta - 5 \right) \alpha^2 
\nonumber \\
& \quad
+ \frac{175}{512} \left( -297\Theta^3 + 417\Theta^2 -143\Theta + 7\right) \alpha^4
\nonumber \\
& \quad\quad
+ \frac{18375}{65536} \left( -1573\Theta^4 +2974\Theta^3 -1738\Theta^2 \right.
\nonumber \\
& \quad\quad\quad \left.
+320\Theta - 9 \right) \alpha^6 = 0 ,
  \tag{K25-\arabic{equation}}
  \stepcounter{equation}
  \label{eqn:K25}
\end{align}
'' (p. K594)
\end{quote}
Note that in Eq. \eqref{eqn:K25}, all $\Theta$ should be replaced for
$\Theta_0$ due to the condition $\eta=1$ (or $e=0$).

\citeauthor{kozai1962b} then continues:
\begin{quote}
``Equation \eqref{eqn:K25} gives the limiting value $\Theta_0$ corresponding
to $H_0$ as a function of $\alpha$.
When $\alpha$ is zero, $\Theta_0$ is equal to 0.6.'' (p. K594)
\end{quote}

Equation \eqref{eqn:K25} is still complicated, and
it is not easy to derive an inverse form such as
$\Theta_0(i_0) = f (\alpha)$ where $f$ is some function.
It may be possible to solve Eq. \eqref{eqn:K25} as a cubic equation
of $\alpha^2$, but we do not know how \citeauthor{kozai1962b} obtained
the list of the values tabulated in his Table I
(in the ``$i_0$ approx'' column in Table \ref{tbl:Kozai-table1}).

\begin{figure}[t]\centering
\ifepsfigure
 \includegraphics[width=\singlefigwidth\textwidth]{fig_K25.eps} %fig5
\else
 \includegraphics[width=\singlefigwidth\textwidth]{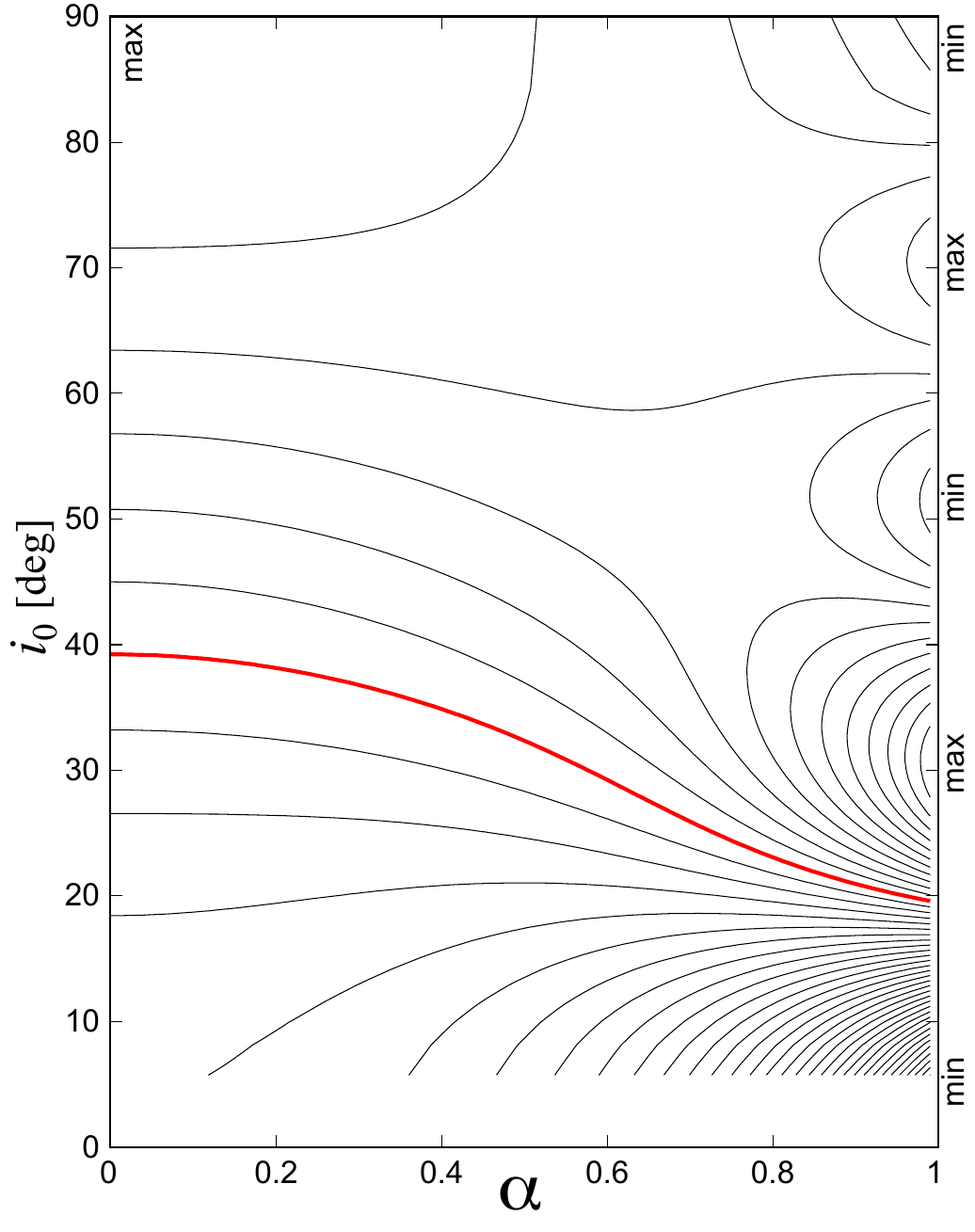} %fig5
\fi
  \caption{%
Equi-value contours plotted on the $(\alpha, i_0)$ plane
produced from the left-hand side of Eq. \eqref{eqn:K25}.
The values of $i_0$ are translated from those of $\Theta_0$ through Eq. \eqref{eqn:def-Theta0-Kozai}.
The thick red line indicates the location of the solution of
Eq. \eqref{eqn:K25}, $f_1 (\alpha,\Theta_0) = 0$.
Note that we have added the labels ``\mtxtsf{min}'' and ``\mtxtsf{max}''
near the locations of local minima and
local maxima of the function $f_1 (\alpha,\Theta_0)$,
because it is hard to recognize them just from the grayscale contours.
}
  \label{fig:K25}
\end{figure}

Let us symbolically write Eq. \eqref{eqn:K25} as $f_1 (\alpha,\Theta_0) = 0$.
Instead of directly solving Eq. \eqref{eqn:K25},
we have made a plot of equi-value contours that $f_1 (\alpha,\Theta_0)$
creates on the $(\alpha, \Theta_0)$ plane (\mysymfigO \ref{fig:K25}).
Among the contours seen in \mysymfigO \ref{fig:K25},
we marked a particular contour that denotes the region
$f_1 (\alpha,\Theta_0) = 0$ with red.
We have confirmed that the values plotted on the red curve
in \mysymfigO \ref{fig:K25} is consistent with those tabulated in the
``$i_0$ approx'' column in \citeauthor{kozai1962b}'s Table I.

The thick red line in \mysymfigO \ref{fig:K25},
which denotes the approximate analytic solution of Eq. \eqref{eqn:K25},
tells us that the threshold inclination $i_0$ monotonically decreases from
$\cos^{-1} \sqrt{\frac{3}{5}} \sim 39.^\circ 2$ to much lower values
as $\alpha$ increases.
This is consistent with \citeauthor{kozai1962b}'s statement that we cited on
p. \pageref{pg:first39appearance} of this monograph.
This fact means that in the doubly averaged inner CR3BP,
the larger $\alpha$ gets,
the more easily the stationary points of the motion of
the perturbed body can take place even with smaller $i$.
We will see this feature later again in \citeauthor{vonzeipel1910}'s work
(Section \ref{sec:vonzeipel} of this monograph).

\subsection{Solution at quadrupole level\label{ssec:Kozai-quadrupole}}

In \citeauthor{kozai1962b}'s next section ``\textit{V. Case for small $\alpha$,\/}''
he just picks the lowest-order $\Oalsqr$ terms of
the doubly averaged disturbing function \eqref{eqn:R2-final},
and discusses its characteristics.
This is the quadrupole level approximation, and
it is valid only when $\alpha \ll 1$.
\citeauthor{kozai1962b}'s main interest in this section is to derive
an analytic, time-dependent solution of orbital elements governed by
the doubly averaged disturbing function at the quadrupole level approximation.
On the way
he gives considerations on possible solutions in several special cases,
and in particular, on their trajectory shapes.
Frankly speaking, 
we feel that this part of the section is poorly organized,
partly because of too many conversions of variables and too terse literal descriptions.
Therefore we just make a brief summary of \citeauthor{kozai1962b}'s
categorization of trajectories, and return to the same subject again
when we introduce \citeauthor{lidov1961}'s work
(Section \ref{sec:lidov1961} of this monograph).

\clearpage

\citeauthor{kozai1962b} begins this section with a simple statement:
\begin{quote}
``When $a$ is small enough so that we can neglect $\alpha^2$
in the braces $\{ \}$ in $W^\ast$ \eqref{eqn:K23},
Eqs. \eqref{eqn:K10} can be integrated by using
an elliptic function of Weierstrauss.'' (p. K594)
\end{quote}

Then \citeauthor{kozai1962b} transforms the canonical equation of motion
for $\DD{G^\ast}{t}$ in Eq. \eqref{eqn:K10} into a form
that uses different variables. He shows that the energy integral $W^\ast$
in Eq. \eqref{eqn:K11} is expressed in the following form
\begin{align}
 &   -\left( 1 - \frac{3\Theta}{x}\right)\left(5-3 x\right) \nonumber \\
 & \quad\quad
   +15\left( 1 -  \frac{\Theta}{x}\right)\left(1-x  \right) \cos 2g^\ast = C,
  \tag{K27-\arabic{equation}}
  \stepcounter{equation}
  \label{eqn:K27}
\end{align}
with a new variable $x$ defined as
\begin{equation}
  x = \eta^2 .
  \tag{K28-\arabic{equation}}
  \stepcounter{equation}
  \label{eqn:K28}
\end{equation}
In other words, $x = 1-e^2$.
He also defines its initial value $x_0$ as follows:
\begin{equation}
  x_0 = \left . x \right|_{g^\ast = 0}.
  \label{eqn:def-x0-kozai}
\end{equation}
Then the constant $C$ is expressed by $x_0$ and $\Theta$ as
\begin{equation}
  C = 10 - 12 x_0 + 6 \Theta .
  \tag{K29-\arabic{equation}}
  \stepcounter{equation}
  \label{eqn:K29}
\end{equation}

\citeauthor{kozai1962b} also introduces another variable $y$ as
\begin{equation}
  y = 3 x^2 - x \left( 5 + 5\Theta - 2x_0 \right) + 5 \Theta .
  \tag{K31-\arabic{equation}}
  \stepcounter{equation}
  \label{eqn:K31}
\end{equation}
Applying Eqs. \eqref{eqn:K27}, \eqref{eqn:K29} and \eqref{eqn:K31}
to the equation of motion for $\DD{G^\ast}{t}$ in Eq. \eqref{eqn:K10},
\citeauthor{kozai1962b} obtains the following ordinary differential
equation for $x$:
\begin{equation}
  \DD{x}{t} = \mp \frac{3}{2} n \alpha^3 m'
              \sqrt{2 \left( x - x_0 \right) y},
  \tag{K30-\arabic{equation}}
  \stepcounter{equation}
  \label{eqn:K30}
\end{equation}
where $n$ is the mean motion of the perturbed body.
In the right-hand side of Eq. \eqref{eqn:K30},
the negative sign corresponds to the positive values of $\sin 2g^\ast$, and
the positive sign corresponds to the negative values of $\sin 2g^\ast$.

\citeauthor{kozai1962b} claims that the solution of the differential equation
\eqref{eqn:K30} is categorized into four types,
depending on the value of $x_0$
(i.e. the initial eccentricity value at $g^\ast = 0$).
\citeauthor{kozai1962b} carries this procedure out by solving an equation
\begin{equation}
  y = 0 ,
  \label{eqn:kozai-y=0}
\end{equation}
which leads to $\DD{x}{t}=0$ from Eq. \eqref{eqn:K30}.
Let us put brief descriptions of what he wrote for the four cases:

\label{pg:Kozais4cases}
\begin{mylist}{1.0em}
\item {\textit{Case 1.\/} (when $x_0 = \Theta$)} \\
In this case, the solution of Eq. \eqref{eqn:kozai-y=0} is
either $x=\Theta$ or $x=\frac{5}{3}$.
As $x$ cannot exceed 1, this means $x$ is always equal to $\Theta$.
Since $x = 1-e^2$ and $\Theta = \left( 1-e^2 \right) \cos^2 i$
by their definitions, $\cos^2 i$ is always unity.
This means that the inclination $i$ of the perturbed body is always zero.
\item {\textit{Case 2.\/} (when $x_0 = 1$)} \\
In this case, the solution of Eq. \eqref{eqn:kozai-y=0} is
either $x=1$ or $x=\frac{5}{3}\Theta$.
If $x=1$, perturbed body is always on a circular orbit $(e=0)$.
On the other hand if $x=\frac{5}{3}\Theta$,
there is a stationary point at $\cos 2g^\ast = -1$
only when $\Theta < \Theta_0 = \frac{3}{5}$.
\item {\textit{Case 3.\/} (when $\Theta < x_0 < 1$)} \\
In this case, one of the solutions of Eq. \eqref{eqn:kozai-y=0} lies
in the range of $\Theta \leq x \leq 1$, and the other is $x > 1$
(which is not valid).
This case embraces the most ordinary trajectories
where $g^\ast$ makes a circulation from 0 to $2 \pi$.
\item {\textit{Case 4.\/} (when $x_0 > 1$)} \\
In this case, both the solutions of Eq. \eqref{eqn:kozai-y=0} lie
in the range of $\frac{5}{3}\Theta \leq x \leq 1$.
So a dynamically meaningful solution can exist only when
  $\Theta < \Theta_0 = \frac{3}{5}$.
However, $x$ cannot exceed 1 by its definition \eqref{eqn:K28},
neither can $x_0$ by its definition \eqref{eqn:def-x0-kozai}.
Therefore we do not exactly understand
\citeauthor{kozai1962b}'s assumption $(x_0 > 1)$ in this case.
\end{mylist}
\label{pg:trajectorycase-kozai}

\label{pg:typosinKozai1962b}
After these categorization of characteristic solutions,
\citeauthor{kozai1962b} slightly changes the course of discussion.
He begins introducing time-dependent analytic solution of
the equations of motion expressed by an elliptic function.
First, \citeauthor{kozai1962b} makes the following statement:
\begin{quote}
``In each case Eq. \eqref{eqn:K30} can be solved by an elliptic function of
  Weierstrauss $\wp$,'' (p. K595).
\end{quote}
More specifically, \citeauthor{kozai1962b} introduces another set of variable conversions
$(x,t) \to (z,t^\ast)$ as
\begin{equation}
\begin{aligned}
  z      &= x - \frac{5}{9} \left( 1+\Theta \right) - \frac{1}{9} x_0 , \\
  t^\ast &= -\frac{3\sqrt{6}}{4} n m' \alpha^3 t ,
\end{aligned}
  \tag{K38-\arabic{equation}}
  \stepcounter{equation}
  \label{eqn:K38}
\end{equation}
and he converts the differential equation \eqref{eqn:K30} into the following one:
\begin{equation}
  \DD{z}{t^\ast} = \pm \sqrt{4\left(z-z_0\right)
                              \left(z-z_1\right)
                              \left(z-z_2\right)} ,
  \tag{K39-\arabic{equation}}
  \stepcounter{equation}
  \label{eqn:K39}
\end{equation}
where
\begin{equation}
\begin{aligned}
-z_0    &=  z_1 + z_2 \\
        &=  \frac{5}{9}   \left( 1+\Theta \right) -\frac{8}{9} x_0 , \\
z_1 z_2 &= -\frac{50}{81} \left( 1+\Theta \right)^2 \\
        &\quad\quad
         + \frac{25}{81} x_0 \left( 1+\Theta \right)
         + \frac{ 7}{81} x_0^2 + \frac{5}{3} \Theta .
\end{aligned}
  \tag{K41-\arabic{equation}}
  \stepcounter{equation}
  \label{eqn:K41}
\end{equation}
Then \citeauthor{kozai1962b} says that solution of Eq. \eqref{eqn:K39}
can be expressed using Weierstrass's elliptic function $\wp$ as
\begin{equation}
  z = \wp(t^\ast) .
  \tag{K40-\arabic{equation}}
  \stepcounter{equation}
  \label{eqn:K40}
\end{equation}
Consult \citet{southard1965} or \citet{weisstein2017} 
for more detailed information on the function $\wp$.

Note that the original
Eqs. \eqref{eqn:K38} and
     \eqref{eqn:K41} contain
typographic errors in \citet{kozai1962b},
and we have already rectified them in the above:
The original Eq. \eqref{eqn:K38} has 
$+ \frac{1}{9} x_0$ instead of the correct
$- \frac{1}{9} x_0$ in its right-hand side.
Also,
the second equation of the original Eq. \eqref{eqn:K41} has
$+ \frac{ 5}{81} x_0 \left( 1+\Theta \right)$ instead of the correct
$+ \frac{25}{81} x_0 \left( 1+\Theta \right)$ in its right-hand side.
Hiroshi Kinoshita kindly notified us of the typographic errors,
and we have confirmed the correctness of this information through
our own algebra.

The solution $z$ of the differential equation \eqref{eqn:K40}
must be translated into the solution $x$ of Eq. \eqref{eqn:K30} as $x (t^\ast)$.
$x(t^\ast)$ is supposed to express the time-dependent solution of $e$ as $e(t^\ast)$.
Then, the solution for $g^\ast$ is subsequently obtained from Eqs. \eqref{eqn:K27} and \eqref{eqn:K29} as
\begin{equation}
  \cos 2 g^\ast = \frac{Q(x)}{5\left(x-\Theta\right)\left(1-x\right)},
  \tag{K42-\arabic{equation}}
  \stepcounter{equation}
  \label{eqn:K42}
\end{equation}
where
\begin{equation}
  Q(x) = -x^2 + \left[ 5 \left(1+\Theta\right) - 4x_0\right] x - 5\Theta .
  \tag{K43-\arabic{equation}}
  \stepcounter{equation}
  \label{eqn:K43}
\end{equation}
We can obtain the time-dependent solution for $i$ from the conservation of
$\Theta$ such as
\begin{equation}
  i = \cos^{-1} \frac{H}{G^\ast} .
  \tag{K44-\arabic{equation}}
  \stepcounter{equation}
  \label{eqn:K44}
\end{equation}

Note again that the original Eq. \eqref{eqn:K43}
in \citet{kozai1962b} contains a typographic error:
The first term of the right-hand side of the original Eq. \eqref{eqn:K43} is
expressed as $x^2$, as opposed to the correct $-x^2$.
This information was also communicated by Hiroshi Kinoshita.

\citeauthor{kozai1962b} also gives a set of differential equations
for obtaining time-dependent solutions for the mean anomaly $l^\ast$ and
the longitude of ascending node $h^\ast$ of the perturbed body.
They are as follows:
\stepcounter{equation}
\begin{alignat}{1}
  \DD{l^\ast}{t} &= n + \frac{3}{8} \frac{n m' \alpha^3}{\eta}
                   \left( x - 3\Theta - \frac{Q(x)}{1-x} \right) ,
  \tag{K45-\arabic{equation}}
  \label{eqn:K45} \\
  \DD{h^\ast}{t} &=   - \frac{3}{8} \frac{n m' \alpha^3 \theta}{\eta}
                   \left[ 5 - 3x - 5\left(1-x\right) \cos 2g^\ast \right]
  \nonumber \\
                 &=   - \frac{3}{8} \frac{n m' \alpha^3 \theta}{\eta}
                   \left[ 5 - 3x -  \frac{Q(x)}{x-\Theta} \right] .
\stepcounter{equation}
  \tag{K46-\arabic{equation}}
  \label{eqn:K46}
\end{alignat}

Note that \citeauthor{kozai1962b} uses the variable $t$ in the left-hand side of
Eqs. \eqref{eqn:K45} and \eqref{eqn:K46} instead of $t^\ast$.
Also,
although we guess that he reached these solutions through analytic quadrature,
there is no mention of the detail as to
how \citeauthor{kozai1962b} derived Eqs. \eqref{eqn:K45} and \eqref{eqn:K46}.
Readers can consult later studies along this subject for more detail
\citep[e.g.][]{vashkovyak1999,kinoshita2007a}.
There is also a short discussion about this subject later in this monograph
(Section \ref{sssec:later-kozai} on p. \pageref{sssec:later-kozai}).

\subsection{Equi-potential trajectories\label{ssec:Kozai-equi-R}}

We presume that one of the features that makes \citeauthor{kozai1962b}'s work
a historic standard is his use of the so-called equi-potential contours
described in his section ``\textit{VI. Trajectory.\/}''
Drawing equi-potential contours (trajectories) of the averaged disturbing function
helps us grasp a global picture of the perturbed body's motion in phase space
without actually seeking specific time-dependent solutions.
On diagrams with equi-potential contours,
we can see the existence or non-existence of stationary points,
the oscillating amplitude of orbital elements, and
the location of separatrix if any, at a glance.
\citeauthor{kozai1962b} describes what he did as follows:
\begin{quote}
``Trajectories of Eqs. \eqref{eqn:K10} can be plotted on the
$\left(2g^\ast, 1-e^2\right)$ plane by using the energy integral $W^\ast$ \eqref{eqn:K23}.
In Figs. 2 through 5 the trajectories are shown for $\alpha=0$.'' (p. K596)
\end{quote}

\begin{figure*}[t]\centering
\ifepsfigure
 \includegraphics[width=.94\textwidth]{replot_kozai1962b.eps} %fig6
\else
 \includegraphics[width=.94\textwidth]{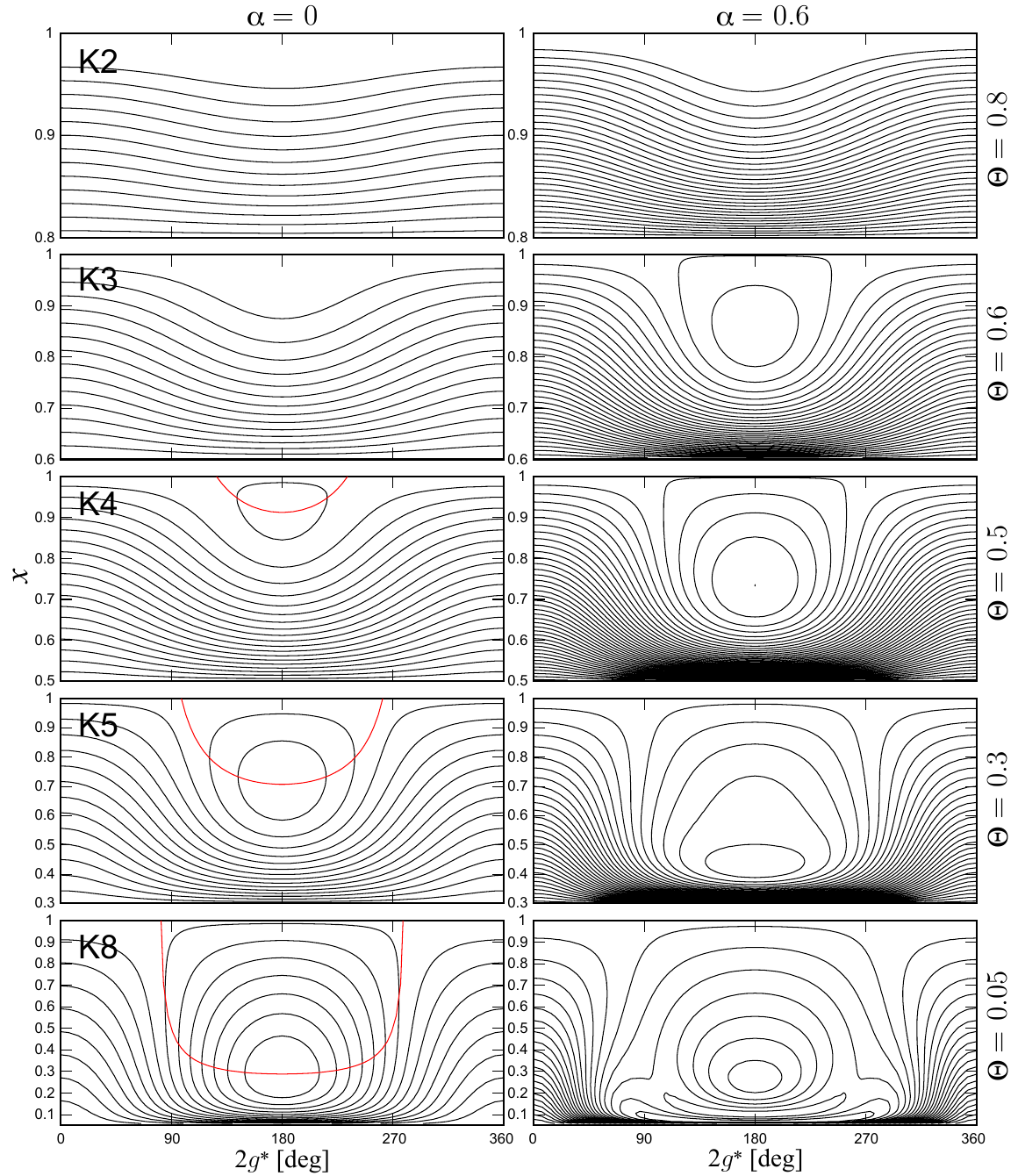} %fig6
\fi
  \caption{%
Equi-potential contours of the doubly averaged disturbing function for
the inner CR3BP calculated through the analytic expansion that
\protect\citet{kozai1962b} gave.
The horizontal axis is $2 g^\ast$, and the vertical axis is $x = 1-e^2$.
The five panels in the left column are the results assuming $\alpha = 0$
with $\Theta = 0.8, 0.6, 0.5, 0.3, 0.05$.
They are reproductions of \protect\citeauthor{kozai1962b}'s original figures
(\mysymfigS K2, K3, K4, K5, K8) where only the lowest-order terms of
\protect\citeauthor{kozai1962b}'s expansion \protect\eqref{eqn:K23} is included.
The red lines in the bottom three panels indicate the location where $\protect\DD{g^\ast}{t} = 0$ is realized.
On the other hand,
the five panels in the right column are those when $\alpha = 0.6$
for the same values of $\Theta$.
We placed them for comparison with the panels in the left column.
In this case all the terms appearing in
\protect\citeauthor{kozai1962b}'s expansion \protect\eqref{eqn:K23} up to $\Oaloct$ are used.
Note that the smallest value of $x$ is always equal to $\Theta$.
This is evident from their definitions ($x = 1-e^2$, and $\Theta = x \cos^2 i$) and the fact that the minimum of $x$ is realized when $\cos^2 i = 1$, i.e. when $x = \Theta$.
}
  \label{fig:replot_kozai1962}
\end{figure*}

Needless to say, this procedure is possible because
the doubly averaged CR3BP has just one degree of freedom, and
because the energy integral $W^\ast$ \eqref{eqn:K23} itself is constant.
We reproduced \citeauthor{kozai1962b}'s
\mysymfigS K2, K3, K4, K5, and K8 as the left column panels of
our \mysymfigO \ref{fig:replot_kozai1962}.
Note that \citeauthor{kozai1962b} did not mention anything
on his \mysymfigO K8 in his paper.
We do not know why, and
we have no idea either as to why only \mysymfigO K8 is isolated,
placed beyond \mysymfigS K6 and K7.

\citeauthor{kozai1962b}'s \mysymfigO K2
(corresponding to our reproduction shown in the       top panel in the left column of \mysymfigO \ref{fig:replot_kozai1962})
is for a system with $\Theta = 0.8$ where stationary points would not show up
at the quadrupole level approximation.
All the trajectories exhibit circulation of $2g^\ast$ from 0 to $2\pi$.
Note that \citeauthor{kozai1962b} added arrows on each of the contours
in his original figures so that readers can understand the direction of motion.
Our reproduction does not include these arrows.

\citeauthor{kozai1962b}'s \mysymfigO K3
(corresponding to our reproduction shown in the second panel from the top in the left column of \mysymfigO \ref{fig:replot_kozai1962})
is for a system with $\Theta = 0.6$,
the limiting value that changes the dynamical characteristics of the system
at the quadrupole level approximation.
Past the limiting value,
we come to visually recognize stationary points on the diagram.
An example result with $\Theta = 0.5$ is presented in \citeauthor{kozai1962b}'s \mysymfigO K4
(corresponding to our reproduction shown in the third panel from the top in the left column of \mysymfigO \ref{fig:replot_kozai1962}).
In \citeauthor{kozai1962b}'s original \mysymfigO K4,
it is noticeable that one of the contours reaches the upper boundary of the panel where $x = 1-e^2 = 1$.
This means that this contour denotes a separatrix of the motion.
Note that our reproduction in \mysymfigO \ref{fig:replot_kozai1962} does not
have an explicit separatrix on it because of the automatic choice of contour
interval by the plotting application that we used.
The existence of stationary points and the separatrix get more obvious as $\Theta$ gets smaller.
An example result when $\Theta = 0.3$ is presented in \citeauthor{kozai1962b}'s \mysymfigO K5
(corresponding to our reproduction shown in the fourth panel from the top in the left column of \mysymfigO \ref{fig:replot_kozai1962}).

In \citeauthor{kozai1962b}'s \mysymfigS K4, K5, and K8,
he added the boundaries where $\DD{g^\ast}{t} = 0$ as broken lines.
We reproduced the boundaries in red in the corresponding panels
of \mysymfigO \ref{fig:replot_kozai1962}.
Here is how \citeauthor{kozai1962b} calculated the location of the boundaries.
From the canonical equation of motion for $g^\ast$ 
\eqref{eqn:dgdt-quadrupole}
at the quadrupole level approximation,
$\DD{g^\ast}{t}$ vanishes when the following relationship is satisfied:
\begin{equation}
  \cos 2g^\ast = \frac{5\Theta-   x^2}{5\left(\Theta-   x^2\right)} .
  \tag{K36-\arabic{equation}}
  \stepcounter{equation}
  \label{eqn:K36}
\end{equation}
Eq. \eqref{eqn:K36} yields the condition that realizes $\DD{g^\ast}{t} = 0$ as
\begin{equation}
  x % = \eta^2
    = \sqrt{ \frac{5\Theta\left(1 -\cos 2g^\ast \right)}{1 -5\cos 2g^\ast} } .
  \label{eqn:boader-dgdt=0}
\end{equation}
Equation \eqref{eqn:boader-dgdt=0} denotes the location of the boundaries
on the $(2 g^\ast, x)$ plane shown in \mysymfigO \ref{fig:replot_kozai1962}.

\citeauthor{kozai1962b} also gave the maximum oscillation amplitude of
$2 g^\ast$ utilizing Eq. \eqref{eqn:K36}.
When $\Theta < \frac{3}{5}$ and the doubly averaged disturbing function has stationary points,
$g^\ast$ librates between the two points that realize $\DD{g^\ast}{t} = 0$ centered at $2g^\ast = \pi$.
At these two points, $2g^\ast$ is bound by the condition \eqref{eqn:K36}.
And from the function form of Eq. \eqref{eqn:K36},
the smallest value of $2 g^\ast$ (or the largest value of $\pi - 2 g^\ast$) takes place when $x^2 = 1$.
Substitution of $x^2 = 1$ into Eq. \eqref{eqn:K36}
yields the largest oscillation amplitude of $2 g^\ast$ centered at $2g^\ast = \pi$ as
\begin{equation}
  2 \left(
    \pi       - \cos^{-1} \frac{5\Theta - 1}{5\left(\Theta - 1\right)}
    \right) .
  \tag{K37-\arabic{equation}}
  \stepcounter{equation}
  \label{eqn:K37}
\end{equation}

\clearpage

In \mysymfigO \ref{fig:replot_kozai1962},
we can visually confirm some of the characteristic trajectories
that \citeauthor{kozai1962b} categorized in his previous section
(see p. \pageref{pg:trajectorycase-kozai} of this monograph).
Case 1 $(x_0 = \Theta)$ which corresponds to the status of $i=0$ is
realized at the lower boundaries of the panels in
\citeauthor{kozai1962b}'s \mysymfigS K2, K3, K4, K5, K8.
As $\Theta = \left(1-e^2\right) \cos^2 i$ is a constant,
$i=0$ automatically means that $e$ takes its maximum and also is a constant.
Since $x = 1-e^2$, this trajectory occupies the lower boundary of each panel.
\label{pg:kozai_upperlowerboundaries}

As for Case 2 $(x_0 = 1)$, particularly when $x=1$,
the perturbed body is always on a circular orbit $(e=0)$.
Therefore the inclination $i$ is also a constant.
This trajectory corresponds to the upper boundary of each of the panels,
as $x = 1$ is its maximum value.
Note that when $\Theta < \frac{3}{5}$,
the upper boundary of $x=1$ is divided into two parts:
where $2 g^\ast$ circulates from 0 to $2\pi$, and
where $2 g^\ast$ librates with the amplitude given by Eq. \eqref{eqn:K37}.
Hence it can be misunderstood that the upper boundary of the
$(2g^\ast, x)$ diagram composes a singularity, but it is not.
Citing \citeauthor{kozai1962b}'s words:
\begin{quote}
``Two bifurcation points on a line $x=1$ are not actual singularities
since they disappear when the coordinates are transformed into polar ones 
$(e,g^\ast)$. In fact, for $x=1$, the circular orbit,
$g^\ast$ cannot be defined.
At the bifurcation points $\DD{g^\ast}{t}$ vanishes.'' (p. K596)
\end{quote}
Note that what \citeauthor{kozai1962b} refers to as ``the bifurcation points''
above are the contact points of the separatrix and the upper boundary of the panels.

In relation to his \mysymfigS K4, K5, and K8,
\citeauthor{kozai1962b} left a description on the circumstance when $\Theta$ is small:
\begin{quote}
``As the value of $\Theta$ decreases, the libration region becomes wide.
Even when $\Theta$ is zero, there are trajectories of both libration and
complete revolution. For an extreme case the orbit oscillates between
a circular perpendicular one and a parabolic one of zero inclination.
Although the assumption $\alpha=0$ may not be valid for this case,
the results may confirm Lidov's numerical work (1962).'' (p. K597)
\end{quote}

It is interesting that
\citeauthor{kozai1962b} cites \citeauthor{lidov1961}'s numerical work here.
We will mention this issue later in this monograph
(Section \ref{sssec:i11n2kozai} on p. \pageref{sssec:i11n2kozai}).
As for the $\Theta = 0$ case, see
\citeauthor{lidov1961}'s discussion that we summarize
in Section \ref{ssec:Lidovdiagram} of this monograph
(p. \pageref{pg:line-B-O-E}).

When making his \mysymfigS K2, K3, K4, K5, K8, 
\citeauthor{kozai1962b} assumed $\alpha \ll 1$.
This is equivalent to the quadrupole level approximation that ignores
all the higher-order terms than $\Oalsqr$ in the disturbing function.
When the higher-order terms in the disturbing function are included,
the limiting value of $\Theta$ as well as the shape of
the equi-potential trajectories change.
This is what \citeauthor{kozai1962b} previously mentioned
in ``\textit{III. Stationary point\/}''
(see p. \pageref{pg:numericalharmonicanalysis} of this monograph).
He describes it again as follows:
\begin{quote}
``As the value of $\alpha$ increases, for a fixed value of $\Theta$,
amplitudes of $x$ become large and the libration region expands
as is expected.'' (p. K597)
\end{quote}
By comparing the panels in the left  column (for $\alpha = 0$)
              and those in the right column (for $\alpha = 0.6$)
in \mysymfigO \ref{fig:replot_kozai1962},
we can easily see that \citeauthor{kozai1962b}'s above statement is true.

There is a point to notice in the bottom right panel of
\mysymfigO \ref{fig:replot_kozai1962}
which is for $(\alpha, \Theta) = (0.6, 0.05)$.
In this panel
we recognize several local extremums in the region of $x \lesssim 0.1$
along $2 g^\ast = 180^\circ$,
in addition to the major local minimum located around
$(2 g^\ast, x) = (180^\circ, 0.25)$.
We presume they are artificial features, not real.
$x \sim 0.1$ means $e \sim 0.95$.
It is known that the analytically expanded doubly averaged disturbing function
for CR3BP
can sometimes give spurious local extremums
when the eccentricity of the perturbed body is very large,
even if its truncation order is as high as $\alpha^8$ \citep[e.g.][]{ito2016}.
The apparent local extremums seen in this panel are probably one of these.

\subsection{Actual asteroids\label{ssec:Kozai-asteroids}}

\begin{figure}[htbp]\centering
\ifepsfigure
 \includegraphics[width=\singlefigwidth\textwidth]{replot_kozai1962b_2ast.eps} %fig7
\else
 \includegraphics[width=\singlefigwidth\textwidth]{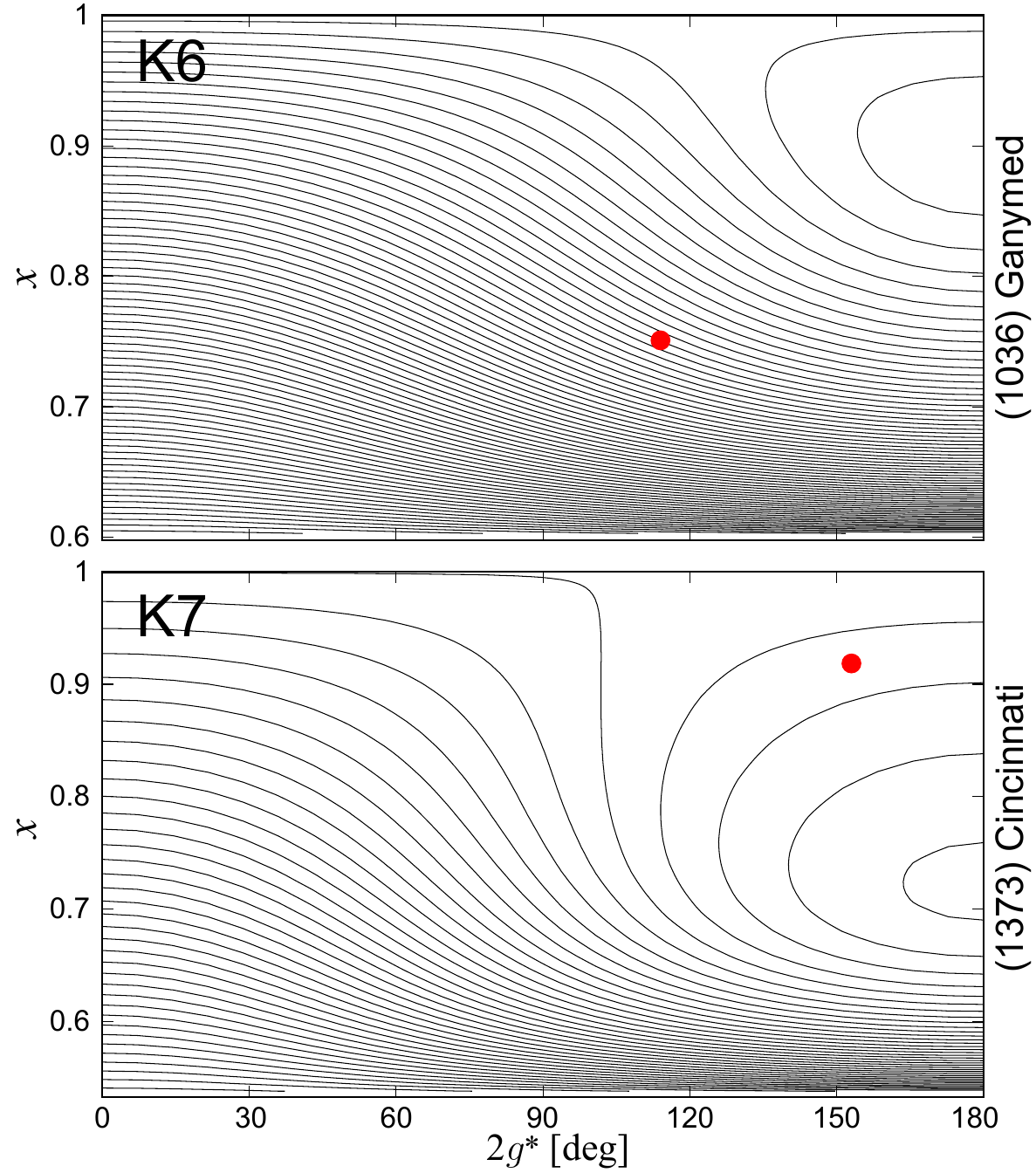} %fig7
\fi
  \caption{%
Reproduction of \protect\citeauthor{kozai1962b}'s
\mysymfigS K6 (upper) and K7 (lower).
The upper panel shows the equi-potential contours of the doubly averaged
disturbing function        for an asteroid      (1036) Ganymed, and
the lower panel shows that for another asteroid (1373) Cincinnati.
They are calculated through the analytic expansion of the doubly averaged
disturbing function up to $\Oaloct$ presented in Eq. \eqref{eqn:K23}.
The axes are common to \mysymfigO \protect\ref{fig:replot_kozai1962} except that
the range of $2 g^\ast$ is just between 0 and $180^\circ$
(following \protect\citeauthor{kozai1962b}'s way).
The parameters $\Theta$ and $\alpha$ are adopted from
\protect\citeauthor{kozai1962b}'s description:
$(\Theta,\alpha) = (0.5979, 0.5123)$ for Ganymed, and
$(\Theta,\alpha) = (0.5325, 0.6569)$ for Cincinnati.
The red points in each of the panels indicate the ``present'' location of
the asteroids that \protect\citeauthor{kozai1962b} mentioned:
$(2g^\ast,x) = (246^\circ, 0.7510)$ for Ganymed, and
$(2g^\ast,x) = (207^\circ, 0.9184)$ for Cincinnati.
However, note that the actual location of the asteroids' $2g^\ast$
in these panels are not the written values, but
$114^\circ (= 360^\circ - 246^\circ)$ for Ganymed, and
$153^\circ (= 360^\circ - 207^\circ)$ for Cincinnati.
This is also what \protect\citeauthor{kozai1962b} did in his paper.
}
  \label{fig:replot_kozai1962_fig67}
\end{figure}
\clearpage

\citeauthor{kozai1962b} then moves on to drawing equi-potential contours
for actual objects in the solar system.
His objects are two asteroids, (1036) Ganymed and (1373) Cincinnati,
whose $\alpha$ values are not negligibly small $(\alpha > 0.5)$.

\citeauthor{kozai1962b} presented the resulting equi-potential plots
in his \mysymfigS K6 and K7 (p. K597), and
we reproduce them in our \mysymfigO \ref{fig:replot_kozai1962_fig67}.
His description on the motion of these asteroids are short.
As for (1036) Ganymed, he writes:
\begin{quote}
``It is not in the libration region. The eccentricity and inclination
oscillate, respectively, between 0.3 and 0.55 and between $23^\circ$
and $48^\circ$, whereas the present values are 0.5 and $27^\circ$.''
(p. K597)
\end{quote}
This means that, while the doubly averaged disturbing function for
Ganymed has a stationary point, its argument of perihelion does not actually
librate, but circulates.
On the other hand as for (1373) Cincinnati, \citeauthor{kozai1962b} writes:
\begin{quote}
``The present values of the eccentricity and inclination are, respectively,
0.29 and $42^\circ$. and they oscillate between 0.25 and 0.6 and
between $25^\circ$ and $42^\circ$.
The motion of the argument of perihelion is limited between $60^\circ$
and $120^\circ$.''
(p. K597).
\end{quote}
Note that the period placed in
``0.29 and $42^\circ$. and they $\cdots$'
must be \citeauthor{kozai1962b}'s typographic error.
It should be replaced for a comma as
``0.29 and $42^\circ$, and they $\cdots$ .''

\citeauthor{kozai1962b}'s above description on (1373) Cincinnati is
now considered to be the first ``discovery'' of a small solar system body
whose argument of pericenter librates \citep{marsden1999}.
The behavior difference between the two asteroids,
although their disturbing potentials both have stationary points (local minima)
as seen in the right edge of \mysymfigO \ref{fig:replot_kozai1962_fig67},
comes from the difference of the asteroids' initial orbital energy.
\citeauthor{kozai1962b} did not explicitly mention this point.
The libration of Cincinnati's $g$ was
later confirmed by a direct numerical integration of equations of motion
including perturbation from four giant planets and Pluto
\citep[][p. 210, although no figure or table was given]{marsden1970}.
\label{pg:kozai-cincinnati}

At this point let us somewhat deviate from what \citeauthor{kozai1962b}
achieved, and let us reproduce his result through a different method in different coordinates.
This is to confirm the correctness and accuracy of \citeauthor{kozai1962b}'s
analytic theory.

As for the method,
we resort to numerical quadrature defined in Eq. \eqref{eqn:K09}.
Numerical quadrature is expected to yield a more accurate result than a truncated analytic expansion of the disturbing function,
particularly when $\alpha$ is large.
Technically speaking,
by using the fact that perturber's orbit is circular in CR3BP,
we can turn the double integral \eqref{eqn:K09} into a single integral
by employing the complete elliptic integral of the first kind
\citep[e.g.][]{bailey1983,quinn1990b,bailey1992}.
This conversion makes the quadrature \eqref{eqn:K09} faster and more accurate.
Our numerical quadrature including the calculation of the elliptic integral was
achieved using the functions implemented in GNU Scientific Library
\citep[][GSL 1.16]{galassi2009}.
Note that as we already mentioned on p. \pageref{pg:numericalharmonicanalysis},
\citeauthor{kozai1962b} himself seemed to carry out the numerical quadrature
under the name of ``numerical harmonic analysis.''
\citeauthor{kozai1962b} wrote that he used it for calculating the values listed in his Table K1.

As for the coordinate system, we use polar coordinates $\left(e\cos g^\ast, e\sin g^\ast\right)$
instead of \citeauthor{kozai1962b}'s rectangular coordinates $\left(2g^\ast, 1-e^2\right)$.
The polar coordinates of this type have several advantages over rectangular coordinates.
First, periodic trajectories appear as closed curves,
not being intersect by diagram borders which happens with rectangular coordinates.
Second, the $e=0$ area is represented by a single point at the origin $(0,0)$,
and not extended into a line as in rectangular coordinates.
Since $g^\ast$ is an angle that intrinsically rotates,
we believe that the polar coordinates of this type are more appropriate in this problem than rectangular coordinates.
The polar coordinates of this type have often been used in prior literature along these lines
\citep[e.g.][]{froeschle1991,michel1996,michel1997a,michel1997c,michel1998,hamilton1997,wan2007}.
We can also regard the coordinates $\left(e\cos g^\ast, e\sin g^\ast\right)$
as a variant of Poincar\'e coordinates \citep[e.g.][]{subr2005,chenciner2015}.
A disadvantage of using this type of coordinates is that,
equi-potential contours sometimes get too crowded and difficult to see
in the large $e$ region near the outer boundary.
Rectangular coordinate diagrams such as $\left(2g^\ast, 1-e^2\right)$ or 
$\bigl(g^\ast,\sqrt{1-e^2}\bigr)$ have an advantage in this case.
Later in this monograph we will show some of our calculation results in the
$\bigl(g^\ast,\sqrt{1-e^2}\bigr)$ rectangular coordinates (p. \pageref{fig:Rmap-deVico}).
Note that in what follows in this monograph,
we write $g^\ast$ just as $g$ for simplicity.
Therefore, the coordinate system we most often employ in this monograph is
$\left(e\cos g, e\sin g\right)$.

We chose three asteroids as examples of our demonstration:
(1036) Ganymed, (1373) Cincinnati, and (3040) Kozai.
As we have seen,
Ganymed and Cincinnati are what \citeauthor{kozai1962b} dealt with.
(3040) Kozai is what we dealt with in our \mysymfigO \ref{fig:CR3BP-examples}
(p. \pageref{fig:CR3BP-examples}).
Incidentally, (3040) Kozai was named after Yoshihide Kozai.
The name was proposed by James G. Williams who found that the argument of perihelion of this asteroid is in libration.
See 
MPC (Minor Planet Circulars) 9770 (1985 July 2)
and \citet{milani1989b} for more detail.
Note also that \citet[][in his abstract]{milani1989b}
introduced the term ``Kozai class,'' and
used it for the objects whose orbital behavior seems to
``be protected from node crossing by secular resonances and $e$--$\omega$ coupling (Kozai class).''

\begin{figure*}[htbp]\centering
\ifepsfigure
 \includegraphics[width=\dualfigwidth\textwidth]{eqR_kozai_inner.eps} %fig8
\else
 \includegraphics[width=\dualfigwidth\textwidth]{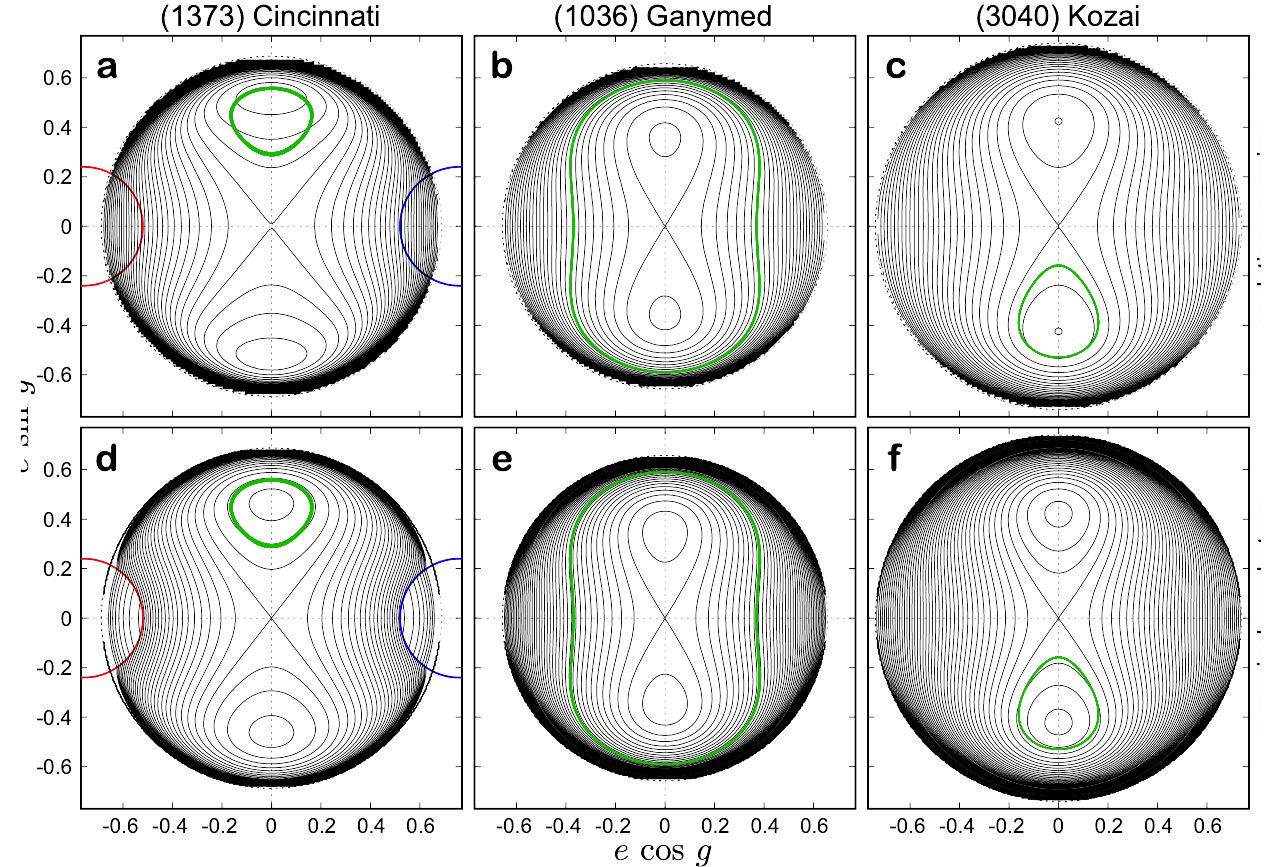} %fig8
\fi
  \caption{%
The equi-potential trajectories of three asteroids
plotted on the $(e\cos g, e\sin g)$ plane in the CR3BP framework.
Top    (\mtxtsf{a}, \mtxtsf{b}, \mtxtsf{c}):
Equi-potential contours calculated through the analytically
expanded doubly averaged disturbing function
up to $\Oaloct$ presented in Eq. \eqref{eqn:K23}.
Bottom (\mtxtsf{d}, \mtxtsf{e}, \mtxtsf{f}):
Equi-potential contours calculated through numerical quadrature defined
in Eq. \protect\eqref{eqn:K09}.
\mtxtsf{a} and \mtxtsf{d} are for (1373) Cincinnati,
\mtxtsf{b} and \mtxtsf{e} are for (1036) Ganymed, and
\mtxtsf{c} and \mtxtsf{f} are for (3040) Kozai.
Consult the caption of \mysymfigO \protect\ref{fig:replot_kozai1962_fig67}
as for the $(\Theta,\alpha)$ values of (1373) Cincinnati and (1036) Ganymed.
(3040) Kozai's values used here are $(\Theta,\alpha) = (0.452, 0.354)$.
The green dots are the actual trajectories of the asteroids
obtained through direct numerical integration of the equations of motion.
Note that the green numerical trajectories
shown in the panel set (\mtxtsf{a}, \mtxtsf{b}, \mtxtsf{c}) and
those in the panel set (\mtxtsf{d}, \mtxtsf{e}, \mtxtsf{f}) are identical.
In the numerical integration,
the initial values of the asteroids' orbital elements are taken from
the JPL Horizons web-interface as of February 16, 2016.
The initial locations of each of the asteroids are as follows:
$(e,g) = (0.3151321,  99^{\circ}.948105)$ for Cincinnati,
$(e,g) = (0.5338748, 129^{\circ}.934759)$ for Ganymed, and
$(e,g) = (0.2005303, 288^{\circ}.967682)$ for Kozai.
Note also that the numerical data for (1373) Cincinnati and (3040) Kozai is 
equivalent to what was used in \mysymfigO \protect\ref{fig:CR3BP-examples},
except that the integration period is longer here (5 million years).
The red and the blue partial circles in \mtxtsf{a} and \mtxtsf{d} represent
the conditions where the orbits of the perturbed and perturbing
bodies intersect each other
at the ascending  node (red) and
at the descending node (blue) of the perturbed body.
The black dashed circles represent the theoretically largest eccentricity of
the perturbed body in each system.
See Section \protect\ref{ssec:R-general} for more rigorous definitions of
the red and blue partial circles as well as the black dashed circles.
}
  \label{fig:xy-inner}
\end{figure*}

We show our calculation results in \mysymfigO \ref{fig:xy-inner}.
The trajectories seen in
the upper three panels \mtxtsf{a}, \mtxtsf{b}, \mtxtsf{c} are those
obtained through \citeauthor{kozai1962b}'s analytic expansion of
the doubly averaged disturbing function up to $\Oaloct$, Eq. \eqref{eqn:K23}.
The trajectories seen in
the lower three panels \mtxtsf{d}, \mtxtsf{e}, \mtxtsf{f} are those
obtained from the numerical quadrature.
Overall, these two sets of equi-potential contours look very similar,
indicating the high accuracy of \citeauthor{kozai1962b}'s analytic theory.
We may see a small discrepancy in the locations of the equilibrium points
for (1373) Cincinnati along the direction of $g = \pm\frac{\pi}{2}$
(the panels \mtxtsf{a} and \mtxtsf{d}). This kind of discrepancy would
possibly be eliminated if we use even higher-order analytic expansions of
the disturbing function than $\Oaloct$.

We see red and blue partial circles in the panels for (1373) Cincinnati
(\mtxtsf{a} and \mtxtsf{d}). As we will see in more detail later
(Sections \ref{ssec:R-general} and \ref{ssec:orbitintersection} of this monograph),
they correspond to the condition of orbit intersection
where the orbits of the perturbing and perturbed bodies intersect each other.
The disturbing function becomes non-holomorphic on these lines, and
they form a set of borders on the $\left(e \cos g, e \sin g\right)$ plane.
The numerical quadrature (the panel \mtxtsf{d}) clearly yields these borders,
while the analytic expansion of the disturbing function (the panel \mtxtsf{a}) does not.
This difference typically shows the limitation of the simple analytic expansion of the disturbing function in this type.
The line of orbit intersection in the panels \mtxtsf{e} and \mtxtsf{f} are
out of the panel ranges, and
we do not have any visual confirmations.

For comparing the equi-potential trajectories that the doubly averaged disturbing function creates with what the actual, unaveraged ``raw'' CR3BP yields,
we carried out a set of direct numerical integration of
the equations of motion of these three asteroids.
The equations of motion we integrated correspond to
      Eq. \eqref{eqn:eom-relative-useR-conventional-ndash}
in Section \ref{sec:CR3BP} 
(see p. \pageref{eqn:eom-relative-useR-conventional-ndash} of this monograph).
The system we considered includes asteroids (as mass-less particles),
the Sun with its current mass, and Jupiter with its current mass
on a circular orbit with semimajor axis $a' = 5.2042$ au.
As for the numerical integration scheme, we employed the so-called
Wisdom--Holman symplectic map \citep{wisdom1991,wisdom1992a}
implemented as the \mtxtsf{SWIFT} package \citep{levison1994}.
We have modified this code and used it in our previous works
\citep[e.g.][]{strom2005,ito2010},
so we can be assured of the correctness of our numerical integration.
  The nominal stepsize of the numerical integration here is 1 day, and
  the total integration time is 5 million years with
  a data output interval of 500 years.
\label{pg:swift_integrator}

The resulting numerical trajectories are shown as green dots
in each of the panels of \mysymfigO \ref{fig:xy-inner}.
Overall they seem well suited on the equi-potential contours
obtained from the doubly averaged disturbing function.
Here, let us recall several characteristic features exhibited
in \mysymfigO \ref{fig:CR3BP-examples} (p. \pageref{fig:CR3BP-examples}).
Among the features, the coupled oscillation of $e$ and $g$ is
reasonably explained by the trajectories that the green dots make
on \mysymfigO \ref{fig:xy-inner}.
The coupling of $e$ and $i$ is now obvious from the fact that
$\Theta = \left( 1-e^2 \right) \cos^2 i$ remains constant
in the doubly averaged CR3BP
as we saw in Section \ref{ssec:Kozai-R}.

\begin{figure}[htbp]\centering
\ifepsfigure
 \includegraphics[width=\singlefigwidth\textwidth]{Strasbourg.eps} %fig9
\else
 \includegraphics[width=\singlefigwidth\textwidth]{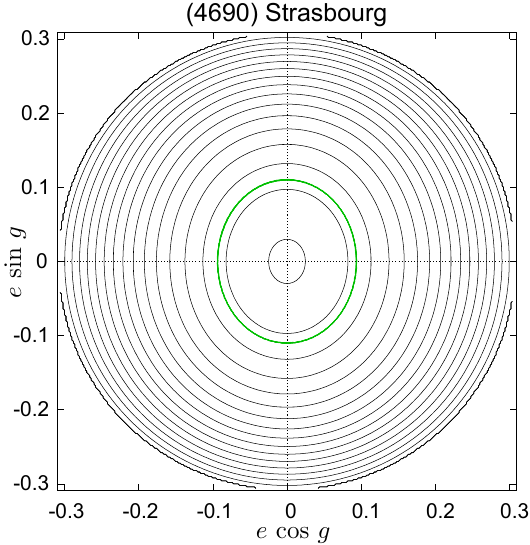} %fig9
\fi
  \caption{%
The motion of the asteroid (4690) Strasbourg on the $(e\cos g, e\sin g)$ plane
in the CR3BP framework.
The black curves are the equi-potential contours calculated
from the numerical quadrature defined by Eq. \eqref{eqn:K09}.
The parameter values employed here are $(\Theta, \alpha) = (0.9045, 0.3726)$.
The green dots are the actual trajectories of the asteroid
obtained through direct numerical integration of the equations of motion.
The numerical data is identical to what was shown in the left panels of
\mysymfigO \ref{fig:CR3BP-examples}.
The initial values of the asteroids' orbital elements are adopted from
the JPL Horizons web-interface as of June 7, 2017, and
the initial location is $(e,g) = (0.1089756, 105^{\circ}.515364)$.
}
  \label{fig:Strasbourg}
\end{figure}

All the asteroids' equi-potential diagrams shown in \mysymfigO \ref{fig:xy-inner}
possess stationary points because their $\Theta$ values are smaller
than the critical value $\left(\Theta = \frac{3}{5}\right)$
of the quadrupole level approximation.
Let us bring up another example where $\Theta$ is larger.
The object is the asteroid (4690) Strasbourg which we already dealt with
in the left panels of \mysymfigO \ref{fig:CR3BP-examples}.
In \mysymfigO \ref{fig:Strasbourg}
we plotted the equi-potential contours for this asteroid
through the numerical quadrature.
Then we superposed the trajectory obtained from
direct numerical integration over the equi-potential contours in green.
The numerical integration data are identical to what was shown
in the left panels of \mysymfigO \ref{fig:CR3BP-examples}.
As is indicated in the caption of \mysymfigO \ref{fig:Strasbourg},
$\Theta$ of this asteroid is as large as 0.9.
With this parameter value,
the asteroid's equi-potential diagram on the
$(e \cos g, e \sin g)$ plane does not possess any local extremums
except for a minimum at the origin $(0,0)$.
Naturally, the asteroid's argument of perihelion $g$ just circulates
quite regularly from 0 to $2\pi$.

\subsection{Remarks on future prospects\label{ssec:kozai-remarks}}
In his last section ``\textit{VII. Remarks,\/}''
after briefly mentioning the similarity and difference between his work and
a previous work \citep{brouwer1947},
\citeauthor{kozai1962b} makes the following comment in its second paragraph:
\begin{quote}
``The theory discussed in the present paper can be applied to the actual
  asteroid motion with some restrictions, as Jupiter's eccentricity and
  other perturbing planets have been ignored.'' (p. K598)
\end{quote}

From the next paragraph \citeauthor{kozai1962b} begins stating several issues
that he did not take care of in his theoretical model:
(i)   potential influence of Jupiter's eccentricity $e'$,
(ii)  indirect perturbation from other planets,
(iii) motion of satellites around an oblate planet, and
(iv)  motion of satellites for which the orbital period of the Sun may not be regarded as short.
We think some of his comments are suggestive,
and they foresee future prospects.
Let us make a brief summary of \citeauthor{kozai1962b}'s statements.

As for the issue (i) of non-zero $e'$,
the major part of \citeauthor{kozai1962b}'s statement is as follows:
\begin{quote}
``When Jupiter's eccentricity is included,
the canonical equations with two degrees of freedom must be solved,
whereas $H$ is still constant and there is an energy integral.
However, it may be very difficult to find any meaningful stationary
solution because of an apparent rapid motion of ${g'}^\ast$
due to a very small inclination.
And $l+g+h$, $g+h$, and $h$ should be adopted instead of $l$, $g$, $h$.
Then Jupiter's orbit can be regarded as known,
although there is no integral corresponding to $H = {\rm const}$
for this case.'' (p. K598)
\end{quote}
Recalling that $H$ is one of the Delaunay elements
$\bigl(H = L \sqrt{1-e^2} \cos i\bigr)$, the way \citeauthor{kozai1962b} wrote
``When Jupiter's eccentricity is included, $\ldots$
  whereas $H$ is still constant,''
makes us guess that he is talking about the quadrupole level approximation
with ``the happy coincidence'' which reduces $H$ to a constant even when $e'>0$
(see p. \pageref{pg:happycoincidence-intro} of this monograph).
However at the quadrupole level approximation,
the doubly averaged disturbing function does not contain $g'$.
Hence we do not know what he meant by
``an apparent rapid motion of ${g'}^\ast$.''
In addition, this description seems inconsistent with the later sentence
``although there is no integral corresponding to $H = {\rm const}$ for this case.''
We do not exactly understand \citeauthor{kozai1962b}'s intention here.

As for the issue (ii) of indirect perturbation from other planets,
\citeauthor{kozai1962b} makes a statement as follows:
\begin{quote}
``When indirect perturbations due to other planets are considered,
the integral of the elimination of nodes does not hold in the form $h=h'$.
However, since Jupiter's orbital plane deviates very little from
the invariable plane, $H$ may be regarded as a stable constant,
especially when the inclination of the asteroid is high.'' (p. K598)
\end{quote}
\citeauthor{kozai1962b} later solved the problem, and
found a way to incorporate more than one perturber
in this line of calculations, although their orbits must be circular.

As for the issue (iii) of the inclusion of planetary oblateness,
\citeauthor{kozai1962b} writes as follows:
\begin{quote}
``When the motion of a satellite around an oblate planet is considered,
the perturbations due to the sun and the oblateness should be
taken into consideration. 
When the equator of the planet coincides with the ecliptic,
the present theory can be applied with little modification,
since both $H$ and $W^\ast$ are constant.
However, since a term of $\cos 2g$ does not appear in the first-order disturbing function due to the oblateness, the limiting value of $H$
for the existence of a stationary solution becomes smaller or even
disappears according to the ratio of the disturbing forces of the sun
and the oblateness.'' (p. K598)
\end{quote}
The possibility that \citeauthor{kozai1962b} brought up above
(the diminishing or disappearance of the limiting value of $H$
 due to the planetary oblateness) is investigated in
\citeauthor{lidov1963a}'s work as well as in some later studies.
See p. \pageref{pg:planetaryoblateness} for more detail.
\label{pg:planetaryoblateness-kozai}

As for the issue (iv), \citeauthor{kozai1962b} did not give much description.
He just stated:
\begin{quote}
``Sometimes,
in the case of a satellite for which the period of one revolution of the sun may not be regarded as short,
the solar mean anomaly $l'$ may not be dropped.
It is impossible, therefore, to make an exact study of a general case.''
(p. K598)
\end{quote}
Although \citeauthor{kozai1962b} did not bring up any actual examples,
we presume that this kind of circumstance would happen
when a satellite's secular orbital variation occurs so quickly
that we cannot regard the orbital period of the perturbing body
(for instance, the Sun) to be negligibly short
compared with the secular timescale.
In this case, the double averaging procedure would be certainly inapplicable,
as \citeauthor{kozai1962b} writes.
He then continues as follows to conclude the paper:
\begin{quote}
``If, however, only the principal terms are taken in the disturbing function,
stationary solutions can be derived.
Lunar orbits provide us with especially interesting problem.'' (p. K598)
\end{quote}
Though \citeauthor{kozai1962b}'s statement may contain material
directing future prospects, we do not go into its detail any further.

Later, \citeauthor{kozai1962b} extended his own work on this subject into various directions
\citep[e.g.][]{kozai1963c,kozai1969a,kozai1979,kozai1985}.
Also, it is worth noting that
the number of asteroids whose argument of perihelion librates
around $g = \pm\frac{\pi}{2}$ has significantly increased
in \citeauthor{kozai1962b}'s publications.
In \citet{kozai1962b} it was just one: (1373) Cincinnati.
In \citet{kozai1979}  it went up to four: (944) Hidalgo, (1373) Cincinnati, (1866) Sisyphus, and (1981) Midas.
In \citet{kozai1980}  it further increased to eight (see his Table I on p. 91).
Much later \citeauthor{kozai1962b} published a concise summary of his work
on this subject \citep{kozai2004} where he reviewed the theoretical framework,
not only of the doubly averaged inner CR3BP but also of the outer problem
where $\alpha = \frac{a}{a'} > 1$.
See Section \ref{sssec:later-kozai} of this monograph
(p. \pageref{sssec:later-kozai}) for more descriptions about it.

\section{The Work of \citeauthor{lidov1961}\label{sec:lidov1961}}
The core achievement of \citeauthor{kozai1962b}'s work that
we browsed through in the previous section,
as is summarized in \citet{marsden1999},
is that
\citeauthor{kozai1962b} discovered the
existence of stationary points of argument of pericenter $g$ of perturbed body in the framework of the doubly averaged inner CR3BP.
\citeauthor{kozai1962b} also quantified the condition for $g$ to be
in libration around the stationary points as $\Theta < \frac{3}{5}$.
\citeauthor{marsden1999} begins his article with the following statement:
\begin{quote}
``The significance of \citeauthor{kozai1962b}'s \citeyearpar{kozai1962b} paper
lies in his showing, for the first time,
that there are circumstances where the argument of pericenter
in a perturbed orbit librates rather than circulates.'' (p. 934)
\end{quote}

However, nowadays it is well known that
``for the first time'' in \citeauthor{marsden1999}'s above statement is
not entirely correct---more and more people have come to know that
there is a study published slightly earlier than
\citeauthor{kozai1962b}'s work,
and it deals with fundamentally the same problem
(the doubly averaged inner CR3BP) and
obtained fundamentally the same result as \citeauthor{kozai1962b}.
The author of this work is Michail L'vovich Lidov (1926--1993),
a scientist who worked on celestial mechanics and astronautics
in the former Soviet Union.
Details of his academic achievement are concisely summarized in his obituary \citep{lidovobituary1994}
and in many other articles \citep[e.g.][]{egorov2001}.
\label{pg:lidovobituary}

Before going into the details of \citeauthor{lidov1961}'s work,
we would like readers to be aware that there has been a rich flow of studies
of the three-body problem in the Russian academic community
since long before \citeauthor{lidov1961}.
Among these, a pair of publications \citep{moiseev1945a,moiseev1945b} occupies the position of a landmark.
We begin this section with a brief introduction of these landmark papers.
Note that there is a publication \citep{vashkovyak2008}
which serves as an excellent review of this line of studies
in the former Soviet Union including
\citeauthor{moiseev1945a}'s and \citeauthor{lidov1961}'s work.
\label{pg:vashkovyak2008}

\subsection{\citeauthor{moiseev1945a}'s work on CR3BP\label{ssec:moiseev1945}}
Nikolay Dmitriyevich Moiseev (1902--1955)
was a celestial mechanist in the former Soviet Union
who produced a large number of publications on the three-body problem
from the 1930s to the 1950s.
Among his work, a pair of papers
written in 1940 and published in 1945
about the general framework of CR3BP is still regarded as a classical standard,
particularly in the field of averaged CR3BP.
There is no English translation of these papers as far as we know,
but the titles may be translated as
``About some primary simplified schemes of celestial mechanics, obtained by means of averaging of restricted circular problem of three bodies,
1. About averaged variants of restricted circular planar problem of three bodies'' \citep{moiseev1945a},
and
``2. About averaged variants of spatial restricted circular problem of three bodies'' \citep{moiseev1945b},
respectively.
The section structure of the two papers is exactly the same as each other.
Their only difference is that
\citet{moiseev1945a} deals with a planar  (two-dimensional)   CR3BP, while
\citet{moiseev1945b} deals with a spatial (three-dimensional) CR3BP.
Let us quickly browse through the contents of these papers.

\citeauthor{moiseev1945a} first describes the basic equations of motion of CR3BP in Sections 1
 (``\textit{Primary equations of restricted planar  circular problem of three bodies\/}'' of the \citeyear{moiseev1945a} paper
and
  ``\textit{Primary equations of restricted spatial circular problem of three bodies\/}'' of the \citeyear{moiseev1945b} paper).
Then in Sections 2 (``\textit{Doubly averaged Gaussian problem\/}'')
he moves on to the doubly averaged CR3BP, and
illustrates the equations of motion and conserved quantities.
\citeauthor{moiseev1945a} names this problem as the doubly averaged Gaussian problem.
The differential equations for the orbital elements of the perturbed body
in the spatial (three-dimensional) CR3BP that \citeauthor{moiseev1945b}
described in Section 1 of his \citeyear{moiseev1945b} paper are
equivalent to what we see in modern textbooks, and they are as follows:
\begin{equation}
  \DD{a}{t} = \frac{2\sqrt{a}}{k\sqrt{m_{\rm s}}} \DP{W_{\rm j}}{M} ,
    \tag{Mb1.28-\arabic{equation}}
    \stepcounter{equation}
    \label{eqn:Mb45-1.28}
\end{equation}
\begin{equation}
  \DD{p}{t} = \frac{2\sqrt{p}}{k\sqrt{m_{\rm s}}} \DP{W_{\rm j}}{\omega} ,
    \tag{Mb1.29-\arabic{equation}}
    \stepcounter{equation}
    \label{eqn:Mb45-1.29}
\end{equation}
\begin{equation}
  \DD{i}{t} = \frac{\cot i}{k\sqrt{m_{\rm s}}\sqrt{p}} \DP{W_{\rm j}}{\omega}
            - \frac{1}{k\sqrt{m_{\rm s}}\sqrt{p}\sin i}\DP{W_{\rm j}}{\Omega},
    \tag{Mb1.30-\arabic{equation}}
    \stepcounter{equation}
    \label{eqn:Mb45-1.30}
\end{equation}
\begin{equation}
  \DD{M}{t} = \frac{k \sqrt{m_{\rm s}}}{a\sqrt{a}}
             -\frac{2\sqrt{a}}{k\sqrt{m_{\rm s}}} \DP{W_{\rm j}}{a},
    \tag{Mb1.31-\arabic{equation}}
    \stepcounter{equation}
    \label{eqn:Mb45-1.31}
\end{equation}
\begin{equation}
  \DD{\omega}{t} =-\frac{2 \sqrt{p}}{k\sqrt{m_{\rm s}}}    \DP{W_{\rm j}}{p}
                  -\frac{\cot i}{k\sqrt{m_{\rm s}}\sqrt{p}}\DP{W_{\rm j}}{i},
    \tag{Mb1.32-\arabic{equation}}
    \stepcounter{equation}
    \label{eqn:Mb45-1.32}
\end{equation}
\begin{equation}
  \DD{\Omega}{t} = \frac{1}{k\sqrt{m_{\rm s}}\sqrt{p \sin i}}
                   \DP{W_{\rm j}}{i} ,
    \tag{Mb1.33-\arabic{equation}}
    \stepcounter{equation}
    \label{eqn:Mb45-1.33}
\end{equation}
where
$m_{\rm s}$ is the central mass of the system,
$k$ is used for describing the gravitational constant $\left(k^2 = {\cal G}\right)$,
$\omega$ is argument of pericenter
(designated as $g$ in \citeauthor{kozai1962b}'s work),
$\Omega$ is longitude of ascending node
(designated as $h$ in \citeauthor{kozai1962b}'s work),
and
$M$ is mean anomaly
(designated as $l$ in \citeauthor{kozai1962b}'s work).
Note that \citeauthor{moiseev1945a} uses semilatus rectum defined as
\begin{equation}
  p = a \left(1-e^2\right) ,
  \label{eqn:def-semilatusrectum}
\end{equation}
instead of eccentricity $e$.
We conjecture that \citeauthor{moiseev1945a} probably followed a custom
to use this element in the Soviet Union academic community at that time.

$W_{\rm j}$ in Eq. \eqref{eqn:Mb45-1.33} is the disturbing function whose general form is
\begin{equation}
  W_{\rm j} = k^2 m_{\rm j} \left[
    \frac{1}{\sqrt{a_{\rm j}^2 + {\rm R}^2 - 2 a_{\rm j} {\rm R} \cos \theta}}
  - \frac{{\rm R} \cos \theta}{a^2_{\rm j}} \right] ,
    \tag{Mb1.11-\arabic{equation}}
    \stepcounter{equation}
  \label{eqn:Mb45-1.11}
\end{equation}
where
${\rm R}$   is the radial distance of the perturbed body from the central mass
(designated as $r$  in \citeauthor{kozai1962b}'s work),
$a_{\rm j}$ is the radius of the circular motion of the perturbing body
(designated as $a'$ in \citeauthor{kozai1962b}'s work), and
$\theta$    is the angle between the positional vector of the perturbed body and that of the perturbing body
(equivalent to the angle $S$ in our \mysymfigO \ref{fig:CR3BP-schematic}).
The subscript $\mathrm{j}$ in Eq. \eqref{eqn:Mb45-1.11} implicitly means that
the perturbing body that \citeauthor{moiseev1945a} considered is Jupiter,
while the perturbed body is regarded to be an asteroid.

\citeauthor{moiseev1945b} states on p. 104 of his \citeyear{moiseev1945b} paper that,
unless the distance between the perturbed and perturbing bodies becomes zero and
    if the distance ${\rm R}$ remains finite,
$W_{\rm j}$ can be expanded into a triple Fourier series as
\begin{equation}
  W_{\rm j} = \sum_{q=0}^\infty \sum_{r=-\infty}^\infty \sum_{s=-\infty}^\infty
    C_{qrs} (a, p, i) \cos \left(q M + r \overline{\Omega} + s \omega \right),
    \tag{Mb1.45-\arabic{equation}}
    \stepcounter{equation}
  \label{eqn:Mb45-1.45}
\end{equation}
where
\begin{equation}
  \overline{\Omega} = \Omega - l_\mathrm{j} ,
    \tag{Mb1.39-\arabic{equation}}
    \stepcounter{equation}
  \label{eqn:Mb45-1.39}
\end{equation}
is the longitude of ascending node of the perturbed body relative to the
(true) longitude of the perturbing body, $l_{\rm j}$.
The conversion \eqref{eqn:Mb45-1.39} and its use in $W_{\rm j}$
practically turns the coordinate system into a rotating frame
with the perturbing body (Jupiter), and eliminates
the explicit time-dependence of the disturbing function $W_{\rm j}$
that is included in $l_\mathrm{j}$.
Note that \citeauthor{kozai1962b} did not explicitly take care of this kind of
coordinate conversion in his work.

Now in \citeauthor{moiseev1945b}'s Section 2,
the double averaging procedure is carried out
against the disturbing function $W_{\rm j}$ of Eq. \eqref{eqn:Mb45-1.11}.
This yields the result as follows:
\begin{equation}
  [W_{\rm j}] = \frac{1}{4\pi^2} \int_{M=0}^{2\pi} \int_{l_\mathrm{j}=0}^{2\pi}
                W_\mathrm{j} d M d l_\mathrm{j} .
  \tag{Mb2.1-\arabic{equation}}
  \stepcounter{equation}
  \label{eqn:Mb45-2.1}
\end{equation}

From Eq. \eqref{eqn:Mb45-1.45} we know that the doubly averaged disturbing function has the general form
\begin{equation}
  [W_{\rm j}] = \sum_{s=-\infty}^\infty C_{00s} (a,p,i) \cos s \omega,
  \tag{Mb2.5-\arabic{equation}}
  \stepcounter{equation}
  \label{eqn:Mb45-2.5}
\end{equation}
with the coefficients $C_{00s}$ as functions of $a$, $p$, and $i$.
Then, the differential equations for the doubly averaged orbital elements
with the doubly averaged disturbing function become:
\begin{equation}
  \DD{a}{t} = 0,
  \tag{Mb2.6-\arabic{equation}}
  \stepcounter{equation}
  \label{eqn:Mb45-2.6}
\end{equation}
\begin{equation}
  \DD{p}{t} = \frac{2\sqrt{p}}{k\sqrt{m_{\rm s}}} \DP{[W_{\rm j}]}{\omega},
  \tag{Mb2.7-\arabic{equation}}
  \stepcounter{equation}
  \label{eqn:Mb45-2.7}
\end{equation}
\begin{equation}
  \DD{i}{t} = \frac{\cot i}{k\sqrt{m_{\rm s}}\sqrt{p}} \DP{[W_{\rm j}]}{\omega},
  \tag{Mb2.8-\arabic{equation}}
  \stepcounter{equation}
  \label{eqn:Mb45-2.8}
\end{equation}
\begin{equation}
  \DD{\omega}{t} =-\frac{2 \sqrt{p}}{k\sqrt{m_{\rm s}}} \DP{[W_{\rm j}]}{p}
                  -\frac{\cot i}{k\sqrt{m_{\rm s}}\sqrt{p}} \DP{[W_{\rm j}]}{i},
  \tag{Mb2.10-\arabic{equation}}
  \stepcounter{equation}
  \label{eqn:Mb45-2.10}
\end{equation}
\begin{equation}
  \DD{\Omega}{t} = \frac{1}{k\sqrt{m_{\rm s}}\sqrt{p \sin i}}
                   \DP{[W_{\rm j}]}{i} .
  \tag{Mb2.11-\arabic{equation}}
  \stepcounter{equation}
  \label{eqn:Mb45-2.11}
\end{equation}

Note that in the right-hand side of the
equations for $\DD{\omega}{t}$ that was originally presented in
\citet[][Eq. (2.10) on p. 106]{moiseev1945b},
the negative sign at the $\DP{[W_{\rm j}]}{p}$ term is somehow missing.
We believe it is just a typographic error,
and added the negative sign in the above Eq. \eqref{eqn:Mb45-2.10}.

Note also that on p. 106 of his \citeyear{moiseev1945b} paper,
we find yet another differential equation in the same section:
\begin{equation}
  \DD{M}{t} = \frac{k \sqrt{m_{\rm s}}}{a\sqrt{a}}
            - \frac{2\sqrt{a}}{k\sqrt{m_{\rm s}}} \DP{[W_{\rm j}]}{a} .
    \tag{Mb2.9-\arabic{equation}}
    \label{eqn:Mb45-2.9}
\end{equation}
However, we do not think this equation is practically meaningful after
the averaging procedure defined as Eq. \eqref{eqn:Mb45-2.1}.
The averaging procedure would eliminate the mean anomaly of the perturbed body $M$
out of description of the system.
So we think
Eq. \eqref{eqn:Mb45-2.9} just has an implication
to express the formal dependence of mean motion $(n = \DD{M}{t})$ of the perturbed body
on the doubly averaged disturbing function and its partial derivative,
$\DP{[W_{\rm j}]}{a}$.

The first constant of integration in this system is semimajor axis $a$,
which is obtained from Eq. \eqref{eqn:Mb45-2.6}.
This is equivalent to the (canonically transformed) constant Delaunay element
$L^\ast$ in \citeauthor{kozai1962b}'s work.
Then, by dividing Eq. \eqref{eqn:Mb45-2.7} by Eq. \eqref{eqn:Mb45-2.8}
\citeauthor{moiseev1945b} obtains
\begin{equation}
  \DD{p}{i} = 2 p \tan i ,
  \tag{Mb2.13-\arabic{equation}}
  \stepcounter{equation}
  \label{eqn:Mb45-2.13}
\end{equation}
which can be simply integrated as
\begin{equation}
  \int \frac{1}{2p}dp = \int \tan i di ,
  \label{eqn:int-p-i-moiseev1945b-1}
\end{equation}
which yields
\begin{equation}
  \ln |p|^\frac{1}{2} = \ln C |\cos i|^{-1}, 
  \label{eqn:int-p-i-moiseev1945b-2}
\end{equation}
with an arbitrary constant $C$.
Thus we find the second constant of integration from Eq. \eqref{eqn:int-p-i-moiseev1945b-2} as
\begin{equation}
  \sqrt{p} \cos i = C = \mbox{\rm constant} .
  \tag{Mb2.14-\arabic{equation}}
  \stepcounter{equation}
  \label{eqn:Mb45-2.14}
\end{equation}
We presume $p \geq 0$ and $0 \leq i \leq \frac{\pi}{2}$.
From the definition of semilatus rectum in Eq. \eqref{eqn:def-semilatusrectum},
the quantity in Eq. \eqref{eqn:Mb45-2.14} turns out to be
the vertical component of the angular momentum of the perturbed body.
This is obviously equivalent to \citeauthor{kozai1962b}'s $\sqrt{\Theta}$.

The third constant of integration is the doubly averaged disturbing function
$[W_{\rm j}]$ itself.
We can confirm the fact that $[W_{\rm j}]$
in Eq. \eqref{eqn:Mb45-2.1} or \eqref{eqn:Mb45-2.5}
is a constant by constructing the following total derivative
\begin{equation}
  \DD{[W_{\rm j}]}{t} = %\DP{[W_{\rm j}]}{a}\DD{a}{t}
                         \DP{[W_{\rm j}]}{p}\DD{p}{t}
                       + \DP{[W_{\rm j}]}{i}\DD{i}{t}
                       + \DP{[W_{\rm j}]}{\omega}\DD{\omega}{t},
  \label{eqn:dWj-dt}
\end{equation}
using the equations from Eq. \eqref{eqn:Mb45-2.6} to Eq. \eqref{eqn:Mb45-2.11}.
Then, let us pick
the expression of $\DD{p}{t}$      from Eq. \eqref{eqn:Mb45-2.7},
          that of $\DD{i}{t}$      from Eq. \eqref{eqn:Mb45-2.8}, and
          that of $\DD{\omega}{t}$ from Eq. \eqref{eqn:Mb45-2.10}.
Substituting these expressions     into Eq. \eqref{eqn:dWj-dt},
we reach the conclusion
\begin{equation}
  \DD{[W_{\rm j}]}{t} = 0 ,
\end{equation}
which automatically means
\begin{equation}
  [W_{\rm j}] = {\mbox{\rm constant}} .
  \tag{Mb2.15-\arabic{equation}}
  \stepcounter{equation}
  \label{eqn:Mb45-2.15}
\end{equation}

Note that in this system $\DD{[W_{\rm j}]}{t}$ does not depend on $\Omega$.
This is due to the averaging procedure
\eqref{eqn:Mb45-2.1} with the relation \eqref{eqn:Mb45-1.39}.
As a result,
the total derivative \eqref{eqn:dWj-dt} does not include
the term $\DP{[W_{\rm j}]}{\Omega}\DD{\Omega}{t}$.

Now we have the three constants of integration
in a system with three degrees of freedom.
This brings us the conclusion that the doubly averaged CR3BP is integrable by quadrature.
Note that in \citeauthor{moiseev1945a}'s discussion
it does not matter whether the system composes the inner problem
or the outer one, as he did not place any specific conditions
on $W_{\rm j}$ (and therefore on $[W_{\rm j}]$)
except that the orbits of the perturbed and perturbing bodies do not
intersect each other.

Although it is not quite relevant to the main discussion,
let us incidentally mention that \citeauthor{moiseev1945b} also derived an expression of
the Jacobi integral of the doubly averaged CR3BP
in a form using orbital elements as
\begin{equation}
  \frac{k^2 m_{\rm s}}{2a} + k\sqrt{m_{\rm s}} n_j \sqrt{p} \cos i + [W_{\rm j}] = {\mbox{\rm constant}} .
  \tag{Mb2.16-\arabic{equation}}
  \stepcounter{equation}
  \label{eqn:Mb45-2.16}
\end{equation}
A similar expression for the doubly averaged planar CR3BP is seen
in his \citeyear{moiseev1945a} paper (Eq. (2.11) on p. 82).

Needless to say,
the fact that the system is integrable does not mean
that the actual procedure of quadrature is simple or easy.
\citeauthor{moiseev1945a} just describes
the formal solutions that would be calculated by quadrature as follows.
The relationship \eqref{eqn:Mb45-2.14} tells us that
there is a formal dependency $p = p(i)$. Also,
the relationship \eqref{eqn:Mb45-2.15} tells us that
there is a formal dependency $\omega = \omega(i)$.
Then, by formally integrating Eq. \eqref{eqn:Mb45-2.8} we get
\begin{equation}
  t - t_0 = \int_{i=i_0}^i \frac{k\sqrt{m_{\rm s}} \sqrt{p} \tan i }{\DP{[W_{\rm j}]}{\omega}} d i ,
  \tag{Mb2.17-\arabic{equation}}
  \stepcounter{equation}
  \label{eqn:Mb45-2.17}
\end{equation}
where $i_0$ is the value of $i$ at $t=t_0$.
Equation \eqref{eqn:Mb45-2.17} can be formally processed by quadrature,
ending up with a function form of $i = i(t)$.
We can obtain formal solutions $p(t)$, $\omega(t)$, and $\Omega(t)$ in a similar way.

The remaining part of \citeauthor{moiseev1945a}'s papers
(his Sections 3--6) is about another class of averaging schemes,
called the singly averaged problems.
In the singly averaged CR3BP, the disturbing function is averaged 
only once either by longitude of the perturbed body or
that of the perturbing body, depending on which varies faster.
Sections 3 (``\textit{External variant of the singly averaged Fatou problem\/}'')
are about a variant of the singly averaging scheme
called the external ``Fatou'' problem \citep{fatou1931}.
In this scheme, the disturbing function is averaged just over the longitude
of the perturbing body whose orbital motion is assumed to be faster than that
of the perturbed body. Therefore we deduce that the orbit of the perturbed body
is located outside that of the perturbing body,
hence the problem is called external.
On the other hand,
Sections 4 (``\textit{Internal variant of the singly averaged problem\/}'')
are about the other variant of the singly averaged CR3BP,
called the internal problem.
In this case the disturbing function is averaged just by longitude of
the perturbed body whose orbital motion is faster than that of the perturbing body.
This implies that the orbit of the perturbed body is located inside that of
the perturbing body, hence the problem is called internal.
The last two sections
(Section 5 ``\textit{Singly averaged Delaunay--Hill problem\/}'' and
 Section 6 ``\textit{Generalized singly averaged Delaunay--Hill problem\/}'')
are about the singly averaged schemes extended to systems with 
a mean motion commensurability between the mean motions of the two objects
using the so-called Delaunay--Hill method
\citep[cf.][]{grebenikov1970,singh1977}.

Although the singly averaged CR3BP may be theoretically interesting,
it seems that they are not very often considered in modern celestial mechanics anymore.
A possible reason is that, since
the singly averaged CR3BP in three dimensions is not generally integrable,
people would choose direct numerical integration rather than analytic method.
This is a difference from the doubly averaged CR3BP which is principally integrable.
Thus we put the singly averaged CR3BP out of the scope of this monograph.
Interested readers may consult a recent study that deals with the eccentric R3BP in the singly averaged method \citep{domingos2013}.
\citet[][Subsection 3.1, his p. 28]{shevchenko2017} also has a description of the singly averaged R3BP.

\subsection{Publications by \citeauthor{lidov1961}\label{ssec:publicationsbylidov}}
Equipped with the knowledge of the doubly averaged CR3BP in its general form
that \citeauthor{moiseev1945a} summarized,
let us move on to the introduction of \citeauthor{lidov1961}'s work.

Michail L'vovich Lidov's landmark paper \citep{lidov1961} was published
in the Russian language,
sixteen years after the work by \citeauthor{moiseev1945a}.
It was soon translated into English and published in two different journals.
The first one, which seems much better known than the other, is entitled
``The evolution of orbits of artificial satellites of planets
  under the action of gravitational perturbations of external bodies,''
published in \textit{Planetary and Space Science\/} \citep{lidov1962}.
Compared with this, the existence of the second translation is much less known.
It is entitled
``Evolution of orbits of artificial satellites of planets as affected by gravitational perturbations of external bodies,''
published in \textit{AIAA Journal Russian Supplement\/} \citep{lidov1963}.
See the References section of
this monograph for more detailed bibliographic records of these translations.
From our own personal point of view, the second translation
\citep{lidov1963} seems more accurate and closer to the original version
\citep{lidov1961} than the first translation \citep{lidov1962},
in terms of its English presentation as well as its
choice of technical terms used in celestial mechanics.
For example, the very first sentence of \citet{lidov1961} is translated
as follows in \citet{lidov1962}:
\begin{quote}
``Until recently, in works devoted to the evolution of the orbits of artificial
satellites, investigations have been made in detail of the influence, on the
orbit of the satellite, of the difference of the gravitational field of the
Earth and the central and the influence of the braking of the satellite in
the Earth's atmosphere.''
\end{quote}
while the same sentence is translated in \citet{lidov1963} as:
\begin{quote}
``Until very recently, the writers of papers on the evolution of the orbits
of artificial satellites studied in detail the influence exerted by the
departure of the earth's gravitational field from a central field and
by atmospheric drag.''
\end{quote}
We prefer the latter, and our summary of
\citeauthor{lidov1961}'s \citeyearpar{lidov1961} work
is based on the descriptions given in \citet{lidov1963}.
The original section titles that we sometimes refer to are
also transcribed from \citet{lidov1963}.

Incidentally, it is interesting to note that the AIAA
version of the English translation \citep{lidov1963}
has five additional paragraphs
between the front matter (title and the author list) and
the first paragraph of the main text (see his p. 1985).
The original paper \citep{lidov1961} does not contain these,
neither does the other English translation \citep{lidov1962}.
Judging from the contents of the additional five paragraphs,
they seem to be a kind of general summary of the paper
added by the translators or the reviewer of the translation.
Later we briefly mention this point 
(p. \pageref{sssec:i11n2kozai} of this monograph).

Let us also note that there is a pair of subsequent publications
that has a very close relationship to \citeauthor{lidov1961}'s original paper.
One of them \citep[][written in Russian]{lidov1963a} is      a chapter of a       proceedings volume \citep{subbotin1963}.
The other   \citep[][written in English]{lidov1963b} is also a chapter of another proceedings volume \citep{roy1963}.
The contents of the two articles are almost identical to each other, and
they practically serve as a supplement of \citet{lidov1961}:
Both of them contain several subjects that \citet{lidov1961} did not discuss,
such as the equi-potential diagram for perturbed body
     or the effect of planetary oblateness on the motion of satellites.
Hence in this monograph we essentially bunch the achievements by
\citet{lidov1961} and \citet{lidov1963a,lidov1963b} together,
regarding them as a series of work.
Section \ref{ssec:lidov1963} of this monograph is a brief summary of
what \citeauthor{lidov1961} has further achieved in these two publications.

\subsection{\citeauthor{lidov1961}'s motivation and background\label{ssec:lidovmotivation}}
In the introductory part that appears before Section 1 of his paper,
\citeauthor{lidov1961} emphasizes the importance of including the gravitational
influence of other celestial bodies than the Earth on the change of
artificial satellites' orbit. This was of less serious concern at that time,
compared with the effect of the Earth's atmospheric drag or
higher-order harmonics of the Earth's gravitational field.
\citeauthor{lidov1961} mentions the fact that the perigee height of an
American artificial satellite (Vanguard I) exhibited a non-negligible change \citep{musen1960}.
\citeauthor{lidov1961} then mentions that a Soviet Union spacecraft Luna-3 \citep{sedov1960}
bound for the Moon significantly reduced its perigee height
compared with its initial value after orbiting the Earth eleven times, and
eventually hit the Earth's atmosphere.
\citeauthor{lidov1961} also conjectures that 
another American orbiter, Explorer VI \citep{sonett1960},
may have been under a similar influence of the Moon and the Sun.

Among the above-mentioned three spacecrafts,
the incident concerning Luna-3 is probably the best known \citep[e.g.][]{batygin2018}.
\citeauthor{lidov1961} describes the incident as follows:
\begin{quote}
``In practice,
  the first space flight in which orbital variation was seen to be influenced
  to any notable degree by the gravitational attraction of the moon and the sun
  was that of the Soviet automatic interplanetary station (Lunik III)
  launched on October 4, 1959.
  The earth orbit into which the station transferred after approaching
  the moon evolved,
  such that in spite of the fact that the initial perigee height had been
  of the order of $47 \times 10^3$ km, by the time 11 revolutions had been
  completed the perigee height predicted by computation
  was less than the radius of the earth and the station had re-entered
  the earth's atmosphere.''
  \citep[][the first paragraph in the left column on p. 1986]{lidov1963}
\end{quote}

Note that ``Lunik III'' in the above is equivalent to Luna-3.
This spacecraft was also known as the ``automatic interplanetary station'' at that time \citep[e.g.][p. 34]{harvey2007}.
There is a more detailed literal description about its orbital evolution in
\citet[][pp. 9--10]{sedov1960}%
\footnote{%
Let us cite two paragraphs from \citet{sedov1960} that are about
the orbital evolution of Luna-3 (``third space rocket'')
from its launch on October 4, 1959,
  to the end state at the end of March, 1960.
The following English translation is adopted from \citet{sedov1961c}:
\begin{quote}
``During the flight of the automatic interplanetary station
from Earth to Moon the orbital inclination to the equatorial plane was
$65^\circ$. After perturbation by the Moon the further path
under influence of the Earth's gravity followed on orbit
near to elliptical with an inclination to the equator approximately
$80^\circ$. Calculation of the further path shows that
the Sun and Moon influence the orbit of the
automatic interplanetary station in such a way
that the orbital inclination varies irregularly
and gradually diminishes. At the tenth loop
the inclination is $48^\circ$. At the eleventh loop,
under the influence of the Moon the orbital
inclination increases again to $57^\circ$.
It is a striking fact that the minimum distance from
the Earth as a result of the influence of the Sun
and Moon falls off from loop to loop.
Calculation shows that after completing the eleventh
revolution at the end of March 1960 the
automatic interplanetary station entered the
Earth's atmosphere in the northern hemisphere
and terminated its existence.

  This fact is connected with the shape of the
orbit and the nature of its situation relative to
the Earth and the Sun. This effect, unexpected
at first sight, is subject only to Newtonian
forces. It is obvious that similar effects must be
taken into account in the theoretical analysis
of problems of the structure of planetary
systems and of orbital characteristics of different
planets and their satellites in the solar system.
As a result of perturbations set up by
the Sun there is an evolution of the orbit which
may lead to collision of the satellite with the
parent planet; hence only satellites with certain
specific types of orbit may persist over a
prolonged period.'' \citep[][his pp. 112--113]{sedov1961c}
\end{quote}

A point to note here is that
the long-term orbital evolution of Luna-3 including
its final fate to hit the Earth's atmosphere
at the end of 1960 March was not actually observed,
despite that some later literature states it was.
This is due to the fact that the communication to Luna-3 was lost
just a few weeks after its launch \citep[e.g.][p. 19]{johnson1979}.
As \citeauthor{sedov1961c} mentioned above, and
as \citeauthor{lidov1961} wrote ``predicted by computation'' above,
Lunar-3's dynamical evolution was predicted by later calculations
except for the radical change of its inclination
observed in the early phase of the mission.
Due to the lack of observational evidence,
there is even a claim that the spacecraft kept drifting around the
Earth until late 1961 or later \citep[][p. 181]{caidin1963}.
\citet[][p. 9]{shevchenko2017} explicitly ascribed Luna-3's incident to
the \citeauthor{lidov1961}--\citeauthor{kozai1962b} {\mainword}.
}% End of \footnote
.
\label{pg:sedovswork}

After giving the historic background,
\citeauthor{lidov1961} states his basic assumption throughout his paper
as follows:
\begin{quote}
``The basic assumption made in this paper is that there is a
small enough ratio between the height of the apocenter
of the satellite and the distance from the perturbing body to
the central body around which the satellite is revolving.
This assumption naturally restricts the class of orbits to
which our formulas can be applied;''
\citep[][the sixth paragraph in the left column on p. 1986]{lidov1963}
\end{quote}

We can rephrase that \citeauthor{lidov1961} assumes
the ratio between the orbital distance of perturbed body $r$ from
the central mass and that of perturbing body $r_k$ is small.
Although \citeauthor{lidov1961} did not use any specific equations
when stating this assumption, it would be expressed as follows:
\begin{equation}
  \frac{r}{r_k} \ll 1 .
  \label{eqn:assumptioninL61-p1987}
\end{equation}

As long as the eccentricity of perturbing body can be ignored,
the assumption expressed by Eq. \eqref{eqn:assumptioninL61-p1987} is
practically equivalent to $\alpha = \frac{a}{a'} \ll 1$
that \citeauthor{kozai1962b} presumed.
In this regard, \citeauthor{lidov1961}'s discussion is limited to
the quadrupole level approximation at $\Oalsqr$.

\subsection{Basic equations, force components\label{ssec:forceSTW}}
\citeauthor{lidov1961}'s Sections
1 (``\textit{Formulation of the problem and system of notation,\/}'') and
2 (``\textit{Basic equations and perturbing forces\/}'')
are devoted to the
formulation of the basic equations of motion used in his work.
Unlike \citeauthor{kozai1962b},
\citeauthor{lidov1961} did not use the Hamiltonian formalism in this study.
Instead, \citeauthor{lidov1961}'s work starts from the so-called
Gauss's form of equations \citep[e.g.][]{brouwer1961,battin1987} where
the time derivative of the orbital elements are expressed
through three components of the perturbing force, $S$, $T$, and $W$.
$S$ is the force component projected on the radius vector,
$T$ is the component projected on a perpendicular direction to $S$ in the plane of the osculating ellipse, and
$W$ is the component projected on a perpendicular direction to the plane of the osculating ellipse.
In what follows let us transcribe the equations of motion 
from \citeauthor{lidov1961}'s Section 2.
Note that 
the independent variable is not the time $t$ in the following equations,
but the true anomaly $\vartheta$ of the perturbed body:
\begin{equation}
\begin{aligned}
  \DD{p}{\vartheta} &= \frac{2 r^3 \gamma}{\mu} T, \\
  \DD{e}{\vartheta} &= \frac{  r^2 \gamma}{\mu}
   \left[ S \sin \vartheta +
\left( 1 + \frac{r}{p} \right) T \cos \vartheta + e \frac{r}{p} T \right], \\
  \DD{\omega}{\vartheta} &= \frac{r^2 \gamma}{\mu e}
    \left[-S \cos \vartheta + \left( 1 + \frac{r}{p} \right) T \sin \vartheta
    \right. \\
& \left.
  \quad\quad\quad\quad\quad\quad\quad\quad\quad\quad
  - e \frac{r}{p} W \cot i \sin u \right], \\
  \DD{\Omega}{\vartheta} &= \frac{r^3 \gamma}{\mu p} W \frac{\sin u}{\sin i}, \\
  \DD{i}{\vartheta} &= \frac{r^3 \gamma}{\mu p} W \cos u,
\end{aligned}
  \tag{L01-\arabic{equation}}
  \stepcounter{equation}
  \label{eqn:L01}
\end{equation}
where
\begin{equation}
  \gamma = \frac{1}{1 + \frac{r^2}{\mu e}S \cos \vartheta - \frac{r^2}{\mu e} \left( 1 + \frac{r}{p} \right) T \sin \vartheta },
  \tag{L02-\arabic{equation}}
  \stepcounter{equation}
  \label{eqn:L02}
\end{equation}
is a parameter close to 1, $u$ is argument of latitude
(sum of argument of pericenter and true anomaly: $u = \omega + \vartheta$),
and $\mu$ is equal to ${\cal G} \times\!$ central mass.
\citeauthor{lidov1961} then develops the perturbing force components 
$(S, T, W)$ into powers of $\frac{r}{r_k}$, and picks terms
in the order of $\frac{r}{r_k}$ and $\left(\frac{r}{r_k}\right)^2$.
In principle, the perturbing force $\bm{F}^{(k)}$
acting on a satellite staying at a position $\bm{r}$ from the central body
under the influence of a perturbing body with a potential factor
$\mu_k$ ($= {\cal G} \times\!$ mass of the perturbing body) and
a position $\bm{r}_k$ is expressed as
\begin{equation}
  \bm{F}^{(k)} = \mu_k \left( \frac{\bm{r}_k - \bm{r}}{\left| \bm{r}_k - \bm{r} \right|^3} - \frac{\bm{r}_k}{\left|\bm{r}_k\right|^3} \right) .
  \tag{L03-\arabic{equation}}
  \stepcounter{equation}
  \label{eqn:L03}
\end{equation}

Now, assume in Eq. \eqref{eqn:L03}
that $\bm{F}^{(k)}$ can be expanded in terms of $\frac{r}{r_k}$ as
\begin{equation}
  \bm{F}^{(k)} = \sum_{i=1}^\infty \bm{F}_i^{(k)} .
  \label{eqn:L1961-p1987nnm1}
\end{equation}
Then the first-order $(i=1)$ term of $\bm{F}^{(k)}$ becomes
\begin{equation}
  \bm{F}_1^{(k)} = \frac{\mu_k}{r_k^2} \left[
         3 \frac{\bm{r}_k}{r_k} \frac{\left(\bm{r}\cdot\bm{r}_k\right)}{r_k^2}
         - \frac{\bm{r}}{r_k} \right],
  \tag{L04-\arabic{equation}}
  \stepcounter{equation}
  \label{eqn:L04}
\end{equation}
and the second-order term becomes
\begin{equation}
\begin{aligned}
{} & \bm{F}_2^{(k)} = \\
{} & \;
  \frac{\mu_k}{r_k^2} \left[
    \left(-\frac{3}{2}\frac{r^2}{r_k^2} + \frac{15}{2} \frac{\left(\bm{r}\cdot\bm{r}_k\right)^2}{r_k^4} \right) \frac{\bm{r}_k}{r_k}
-3 \frac{\left( \bm{r}\cdot\bm{r}_k \right)}{r_k^2} \frac{\bm{r}}{r_k} \right].
\end{aligned}
  \tag{L05-\arabic{equation}}
  \stepcounter{equation}
  \label{eqn:L05}
\end{equation}

\citeauthor{lidov1961} then introduces a new coordinate system.
The axis 1 of this system starts from the central mass to the pericenter of the satellite ($=$ the perturbed body),
the axis 2 is orthogonal to the axis 1 and directed along the motion of the satellite and lies in the plane of the orbit, and
the axis 3 is oriented normal to the plane of the satellite's orbit.
$\xi_1$, $\xi_2$, $\xi_3$ are the direction cosines of
the position vector $\bm{r}_k$ in this coordinate system.
Then, the force components $(S, T, W)$ in the first-order of $\frac{r}{r_k}$ are expressed as:
\begin{equation}
\begin{aligned}
  S_1^{(k)} &= \frac{\mu_k}{r_k^2} \frac{r}{r_k} \left[
      3 \xi^2 \cos^2 \left( \vartheta - \vartheta_\xi \right) - 1 \right] , \\
  T_1^{(k)} &= -3\frac{\mu_k}{r_k^2} \frac{r}{r_k} \xi^2
      \cos \left( \vartheta - \vartheta_\xi \right)
      \sin \left( \vartheta - \vartheta_\xi \right), \\
  W_1^{(k)} &=  3 \frac{\mu_k}{r_k^2} \frac{r}{r_k} \xi_3 \xi
      \cos \left( \vartheta - \vartheta_\xi \right) .
\end{aligned}
  \tag{L06-\arabic{equation}}
  \stepcounter{equation}
  \label{eqn:L06}
\end{equation}

In the second-order they become
\begin{equation}
\begin{aligned}
  S_2^{(k)} &= \frac{15}{2} \frac{\mu_k}{r_k^2} \frac{r^2}{r_k^2}
    \left[   \xi^3 \cos^3 \left( \vartheta - \vartheta_\xi \right)
-\frac{3}{5} \xi   \cos   \left( \vartheta - \vartheta_\xi \right) \right] , \\
  T_2^{(k)} &=-\frac{15}{2} \frac{\mu_k}{r_k^2} \frac{r^2}{r_k^2}
    \left[ \xi^3 \cos^2 \left( \vartheta - \vartheta_\xi \right)
                 \sin^2 \left( \vartheta - \vartheta_\xi \right) \right. \\
& \left.
\quad\quad\quad\quad\quad\quad\quad\quad\quad\quad
-\frac{1}{5} \xi \sin   \left( \vartheta - \vartheta_\xi \right) \right] , \\
  W_2^{(k)} &= \frac{15}{2} \frac{\mu_k}{r_k^2} \frac{r^2}{r_k^2} \xi_3
    \left[ \xi^2 \cos^2 \left( \vartheta - \vartheta_\xi \right) -\frac{1}{5} \right] , \\
\end{aligned}
  \tag{L07-\arabic{equation}}
  \stepcounter{equation}
  \label{eqn:L07}
\end{equation}
where $\xi$ is defined as $\xi = \sqrt{\xi_1^2 + \xi_2^2}$, and
$\vartheta_\xi$ is the projected true anomaly of the vector
$\bm{r}_k$ on the plane of the satellite orbit.
We have the pair of relationships
\begin{equation}
  \sin \vartheta_\xi = \frac{\xi_2}{\xi} , \quad
  \cos \vartheta_\xi = \frac{\xi_1}{\xi} .
  \label{eqn:L1961-p1987nnm2}
\end{equation}

\citeauthor{lidov1961} then introduces two variables as%
\footnote{%
\citeauthor{lidov1961}'s use of the large triangle symbol
in this section is very confusing,
because he uses the same symbol $\Delta$ for several different purposes
such as in
Eqs. \eqref{eqn:def-LidovDelta},
\eqref{eqn:redef-Lidov-r},
\eqref{eqn:L1961-p1988intext1},
\eqref{eqn:L11}, and
\eqref{eqn:L15}.
For minimizing the possible confusion, in this monograph
we use a slanted $\varDelta$ for the symbols defined in
\eqref{eqn:def-LidovDelta} and \eqref{eqn:redef-Lidov-r},
   and a roman   $\Delta$    for others.
} % End of \footnote
\begin{equation}
  \varDelta_k = 1 + e_k \cos \vartheta_k, \quad
  \varDelta   = 1 + e   \cos \vartheta  ,
  \label{eqn:def-LidovDelta}
\end{equation}
where $\vartheta_k$ is true anomaly of the perturbing body.
Note that at this point \citeauthor{lidov1961} does not assume
anything about the perturbing body's eccentricity, $e_k$.

Using $\varDelta_k$ and $\varDelta$ in Eq. \eqref{eqn:def-LidovDelta},
$r_k$ and $r$ are expressed as
\begin{equation}
  r_k = \frac{p_k}{\varDelta_k}, \quad
  r   = \frac{p}{\varDelta},
  \label{eqn:redef-Lidov-r}
\end{equation}
with semilatus rectums $p_k$ and $p$.
Applying Eqs.
\eqref{eqn:L1961-p1987nnm2},
\eqref{eqn:def-LidovDelta},
\eqref{eqn:redef-Lidov-r}
to the expressions of
$S_1^{(k)}$,
$T_1^{(k)}$,
$W_1^{(k)}$,
$S_2^{(k)}$,
$T_2^{(k)}$,
$W_2^{(k)}$
in Eqs. \eqref{eqn:L06} and \eqref{eqn:L07},
\citeauthor{lidov1961} obtained another form of the force components as
\begin{equation}
\begin{aligned}
  S_1^{(k)} &=  3\frac{\mu_k}{p_k^2}\frac{p}{p_k}
   \left[ \beta_1 \cos^2 \vartheta + 2 \beta_3 \cos \vartheta \sin \vartheta
   \right. \\
  & \left.
\quad\quad\quad\quad\quad\quad\quad\quad
         -\frac{\beta_6}{3} + \beta_2 \sin^2 \vartheta \right]
             \frac{1}{\varDelta} , \\
  T_1^{(k)} &= -3\frac{\mu_k}{p_k^2}\frac{p}{p_k}
   \left[ \left(\beta_1 - \beta_2 \right) \cos \vartheta \sin \vartheta
   \right. \\
  & \left.
\quad\quad\quad\quad\quad\quad\quad\quad
           + \beta_3 \left( \sin^2 \vartheta - \cos^2 \vartheta \right) \right]
             \frac{1}{\varDelta} , \\
  W_1^{(k)} &=  3\frac{\mu_k}{p_k^2}\frac{p}{p_k}
          \left[ \beta_5 \cos \vartheta + \beta_4 \sin \vartheta \right]
             \frac{1}{\varDelta} ,
\end{aligned}
  \tag{L08-\arabic{equation}}
  \stepcounter{equation}
  \label{eqn:L08}
\end{equation}
and
\begin{equation}
\begin{aligned}
  S_2^{(k)} &=  \frac{15}{2}\frac{\mu_k}{p_k^2}\frac{p^2}{p^2_k}
   \left[   \gamma_1 \cos^3 \vartheta
        + 3 \gamma_3 \cos^2 \vartheta \sin \vartheta
   \right. \\
  & \left.
\quad\quad\quad\quad\quad
        + 3 \gamma_6 \cos \vartheta \sin^2 \vartheta
        +   \gamma_2 \sin^3 \vartheta
   \right. \\
  & \left.
\quad\quad\quad\quad\quad\quad\quad
        - \frac{3}{5} \alpha_1 \cos \vartheta
        - \frac{3}{5} \alpha_2 \cos \vartheta \right]
             \frac{1}{\varDelta^2} , \\
  T_2^{(k)} &= -\frac{15}{2}\frac{\mu_k}{p_k^2}\frac{p^2}{p^2_k}
   \left[ - \gamma_3 \cos^3 \vartheta
   \right. \\
  & \left.
          + \left( \gamma_1 -2\gamma_6 \right) \cos^2 \vartheta \sin \vartheta
          + \left(2\gamma_3 - \gamma_2 \right) \cos \vartheta \sin^2 \vartheta
   \right. \\
  & \left.
\quad\quad
          + \gamma_6 \sin^3 \vartheta
        - \frac{1}{5} \alpha_1 \sin \vartheta
        + \frac{1}{5} \alpha_2 \sin \vartheta \right]
             \frac{1}{\varDelta^2} , \\
  W_2^{(k)} &=  \frac{15}{2}\frac{\mu_k}{p_k^2}\frac{p^2}{p^2_k}
   \left[ \gamma_4 \cos^2 \vartheta
        +2\gamma_7 \cos   \vartheta \sin \vartheta
   \right. \\
  & \left.
\quad\quad\quad\quad\quad\quad\quad\quad
        + \gamma_5 \sin^2 \vartheta
        - \frac{\alpha_3}{5} \right]
             \frac{1}{\varDelta^2} ,
\end{aligned}
  \tag{L09-\arabic{equation}}
  \stepcounter{equation}
  \label{eqn:L09}
\end{equation}
with a new set of coefficients $\alpha_i, \beta_i, \gamma_i$
which are functions of $\xi_i$ $(i=1 \ldots)$ and $\varDelta_k$:
\begin{equation}
\begin{aligned}
          \alpha_1 &= \xi_1             \varDelta_k^4,
 &\tpspcD \alpha_2 &= \xi_2             \varDelta_k^4,
 &\tpspcD \alpha_3 &= \xi_3             \varDelta_k^4, \\
          \beta_1  &= \xi_1^2           \varDelta_k^2,
 &\tpspcD \beta_2  &= \xi_2^2           \varDelta_k^3,
 &\tpspcD \beta_3  &= \xi_1   \xi_2     \varDelta_k^3, \\
          \beta_4  &= \xi_2   \xi_3     \varDelta_k^3,
 &\tpspcD \beta_5  &= \xi_1   \xi_3     \varDelta_k^3,
 &\tpspcD \beta_6  &=                   \varDelta_k^3, \\
          \gamma_1 &= \xi_1^3           \varDelta_k^4,
 &\tpspcD \gamma_2 &= \xi_2^3           \varDelta_k^4,
 &\tpspcD \gamma_3 &= \xi_1^2 \xi_2     \varDelta_k^4, \\
          \gamma_4 &= \xi_1^2 \xi_3     \varDelta_k^4,
 &\tpspcD \gamma_5 &= \xi_2^2 \xi_3     \varDelta_k^4,
 &\tpspcD \gamma_6 &= \xi_2^2 \xi_1     \varDelta_k^4, \\
          \gamma_7 &= \xi_1 \xi_2 \xi_3 \varDelta_k^4 .
\end{aligned}
  \tag{L10-\arabic{equation}}
  \stepcounter{equation}
  \label{eqn:L10}
\end{equation}

\subsection{Three assumptions\label{ssec:lidov-asummptions}}
In his Section 3
(``\textit{Basic assumptions for obtaining approximate formulas\/}''),
\citeauthor{lidov1961} describes three basic assumptions that are required
for deriving the averaged equations of motion in his later sections.

\paragraph{Assumption 1.} %
The first assumption is that $\gamma$ in Eq. \eqref{eqn:L02} is always unity.
We think it is rather easy for readers to understand his intention
by literally citing his words:
\begin{quote}
''It follows from Eqs. (L02) and (L06) that the maximum deviation of
the value of $\gamma$ from unity is defined to the first-order
by a parameter proportional to
  $\frac{\mu_k}{\mu} \left( \frac{a}{r_k}\right)^3 \frac{1}{e}$
in the case of small eccentricities and by a parameter proportional to
  $\frac{\mu_k}{\mu} \left( \frac{a}{r_k}\right)^3 \frac{1}{1-e}$
in the case of eccentricities close to unity
(here $a$ is the semimajor axis of the orbit).
For example, consider the case of perturbation by the moon of an earth
satellite whose semimajor axis is of the order of 30--$40 \times 10^3$ km.
In this case
$$
  \frac{\mu_k}{\mu} \left( \frac{a}{r_k}\right)^3 \sim 10^{-5}
$$

\hspace*{1em}
This estimate shows that for a broad class of satellite orbits,
with practically all values of eccentricity, we can set with a
certain degree of approximation $\gamma = 1$. The derivations that
follow will be based on equations in which $\gamma = 1$.''
\citep[][p. 1988]{lidov1963}
\end{quote}

\paragraph{Assumption 2.} %
\citeauthor{lidov1961}'s second assumption is quite linguistically described.
Our understanding about it is as follows.
The variation component of orbital elements of the satellites
(such as $\Delta p$, $\Delta e$, $\cdots$)
over its one revolution can be expressed in series such as
\begin{equation}
  \Delta p = \sum_{j} \Delta_j p, \quad
  \Delta e = \sum_{j} \Delta_j e, \quad \cdots ,
  \label{eqn:L1961-p1988intext1}
\end{equation}
and their $j$-th order terms
(such as $\Delta_j p$ or $\Delta_j e$ in Eq. \eqref{eqn:L1961-p1988intext1})
are derived only from the $j$-th order component of the
expanded perturbing force $\bm{F}_j$
on the right-hand side of Eq. \eqref{eqn:L1961-p1987nnm1}.
For example,
$\Delta_1 p$ is derived just from $\bm{F}_1$,
$\Delta_2 e$ is derived just from $\bm{F}_2$, and so on.
Let us cite \citeauthor{lidov1961}'s words concerning the second assumption:
\begin{quote}
``In the approximate treatment we have adopted, the
equations describing the process become linear with a linear
dependence on the perturbing forces. Thus in determining the secular
change of the elements for one revolution of the satellite under
the influence of a sum of accelerations $\sum_{i=1}^n \bm{F}_i$ from
a wide variety of perturbing factors, we can independently
determine the deviations under the influence of each of such factors,
and obtain the general result by the simple summation
$\Delta p = \sum \Delta p_i$, $\Delta e = \sum \Delta e_i$, $\ldots$ .
In particular, we can also consider as independent the deviations of
the elements under the influence of each term of series (L03) of the
expansion of the perturbing acceleration in powers of $\frac{r}{r_k}$.''
\citep[][p. 1988]{lidov1963}
\end{quote}

\paragraph{Assumption 3.} %
\citeauthor{lidov1961}'s third assumption is that
the coefficients $\alpha_i, \beta_i, \gamma_i$ that show up
in Eq. \eqref{eqn:L10} can be Taylor-expanded
by a small time interval $\Delta t$,
and that the first two terms of the series yield approximations
accurate enough for the purpose of his discussion.
Let us cite \citeauthor{lidov1961}'s words concerning the third assumption.
In the following,
$t^\ast$ is the time when satellite's true anomaly satisfies $\vartheta = \pi$,
i.e. the time corresponding to satellite's position at its apocenter:
\begin{quote}
``The variability in the course of one revolution of the quantities
$\alpha_i, \beta_i, \gamma_i$,
determined by formulas \eqref{eqn:L10}, with fixed orbital elements,
is related to the motion of the perturbing body in absolute space.

\hspace*{1em}
The third important assumption we shall make in developing
our approximate formulas is that the values
$\alpha_i, \beta_i, \gamma_i$,
can be represented in the form of the series
\begin{equation}
\begin{aligned}
  \alpha_i &= \alpha_i^\ast + \left( \DD{\alpha_i}{t}\right)^\ast \Delta t
            + \frac{1}{2} \left( \DD[2]{\alpha_i}{t}\right)^\ast
              \left( \Delta t \right)^2 \\
        {} & \quad\quad + \cdots  \\
  \beta_i  &= \beta_i^\ast  + \left( \DD{\beta_i}{t}\right)^\ast \Delta t
            + \frac{1}{2} \left( \DD[2]{\beta_i}{t}\right)^\ast
              \left( \Delta t \right)^2 \\
        {} & \quad\quad + \cdots  \\
  \gamma_i &= \gamma_i^\ast + \left( \DD{\gamma_i}{t}\right)^\ast \Delta t
            + \frac{1}{2} \left( \DD[2]{\gamma_i}{t}\right)^\ast
              \left( \Delta t \right)^2 \\
        {} & \quad\quad + \cdots  \\
\end{aligned}
  \tag{L11-\arabic{equation}}
  \stepcounter{equation}
  \label{eqn:L11}
\end{equation}

We also assume that in an interval of time equal to half the
period of revolution of the satellite,
with a degree of approximation adequate for study purposes, we can approximate
$\alpha_i, \beta_i, \gamma_i$,
by the small number of terms in the series (by two, in the present paper).
Here $\Delta t = t - t^\ast$, where $t^\ast$ is a fixed instant of
time interval corresponding to the given orbit.''
\citep[][p. 1988. A more specific definition of $t^\ast$ shows up in p. 1989]{lidov1963}
\end{quote}

\subsection{Averaged equations of motion\label{ssec:Lidov-averaging}}
In his Section 4
(``\textit{Formulas for the variation in orbital elements for one revolution of the satellite\/}''),
\citeauthor{lidov1961} carries out an averaging of the satellite's orbital elements
over its one revolution.
This procedure is practically equivalent to the single averaging procedure
described in \citeauthor{moiseev1945a}'s work.
The procedure begins with substituting the expanded force components
$S$, $T$, $W$ expressed as Eqs. \eqref{eqn:L08} and \eqref{eqn:L09}
into the equations of motion \eqref{eqn:L01}, and then integrating them
from 0 to $2\pi$ with respect to the true anomaly of the satellite.
It may sound odd for readers to hear that 
\citeauthor{lidov1961} employed true anomaly (actually, argument of latitude)
as the integration variable.
However in \citeauthor{lidov1961}'s averaging procedure,
the actual integration procedures using the argument of latitude are
ingeniously encapsulated through a Taylor-expanded conversion formula
involving the true anomaly $\vartheta$ and the time increment $\Delta t$
(Eqs. (L13) and (L14), although we do not reproduce them here).
We presume this conversion turns the integrations equivalent to
those using mean anomaly as the integration variable.
As a result, we can obtain integral formulas for the
variation of the satellite's orbital elements for its one revolution such as
$\Delta_1^{(k)} p$
in the first-order approximation, and
$\Delta_2^{(k)} p$
in the second-order approximation.
Example expressions are shown in \citeauthor{lidov1961}'s Eq. (L12)
although we do not reproduce them in this monograph due to their complexity.

Next, \citeauthor{lidov1961} assumes that
the variations of orbital elements in each order
 such as $\Delta_1^{(k)}p$ or $\Delta_2^{(k)}p$
can be Taylor-expanded into a power series of the time increment, $\Delta t$.
This is related to the aforementioned assumption 3 that the coefficients
$\alpha_i$, $\beta_i$, $\gamma_i$ can be Taylor-expanded
in the form of Eq. \eqref{eqn:L11}.
As a result, 
the variation of the satellite's orbital element, for example $p$, is expressed as
\begin{equation}
\begin{aligned}
  \Delta_1^{(k)} p &= \Delta_{11}^{(k)} p + \Delta_{12}^{(k)}p + \Delta_{13}^{(k)}p + \cdots, \\
  \Delta_2^{(k)} p &= \Delta_{21}^{(k)} p + \Delta_{22}^{(k)}p + \Delta_{23}^{(k)}p + \cdots,
\end{aligned}
  \tag{L15-\arabic{equation}}
  \stepcounter{equation}
  \label{eqn:L15}
\end{equation}
where the second terms in the right-hand sides
($\Delta_{12}^{(k)}p$ and $\Delta_{22}^{(k)}p$)
correspond to the terms of $O\left(\Delta t\right)$ in Eq. \eqref{eqn:L11}, and
the third terms
($\Delta_{13}^{(k)}p$ and $\Delta_{23}^{(k)}p$)
correspond to the terms of $O\left(\Delta t^2\right)$.
\citeauthor{lidov1961} shows in his Section 4 the specific forms of
all the relevant terms
($\Delta_{11}a$,
 $\Delta_{11}e$,
 $\Delta_{11}i$,
 $\Delta_{11}\Omega$,
 $\Delta_{11}\omega$,
 $\Delta_{12}a$,
 $\Delta_{12}e$,
 $\Delta_{12}i$,
 $\Delta_{12}\Omega$,
 $\Delta_{12}\omega$,
 $\Delta_{21}a$,
 $\Delta_{21}e$,
 $\Delta_{21}i$,
 $\Delta_{21}\Omega$,
 $\Delta_{21}\omega$)
as Eqs. (L17), (L18) and (L19),
which we do not reproduce here due to their complexity.
Note that the superscript $(k)$ on
$\Delta_1^{(k)}, \Delta_{12}^{(k)}, \Delta_{13}^{(k)}, \cdots$, denotes
the influence from the $k$-th perturber.
But now that \citeauthor{lidov1961} considers just one perturber,
the superscript $(k)$ is omitted, and
$\Delta^{(k)}_{11}a$ is just written as $\Delta_{11}a$, and so forth.

At the end of his Section 4, \citeauthor{lidov1961} gave several
considerations on the results that he obtained in this section.
The most important one we think is that
both $\Delta_{11} a$ and $\Delta_{21} a$ become zero.
\citeauthor{lidov1961} depicts the reason as follows:
\begin{quote}
``1. The increments $\Delta_{11} a$ and $\Delta_{21} a$ are equal to zero
because the increments $\Delta_{11} x$ and $\Delta_{21} x$ are
for a perturbing point [that are] motionless in absolute space.
For a motionless perturbing body
the motion of the satellite occurs in a conservative field of force.
Since it is assumed in addition that the orbital elements do not vary
in the course of one revolution, then on completion of the revolution
the satellite is at the same point in space and the increment of the total
energy will therefore equal to zero.''
\citep[][p. 1990. We added the content in \mbox{[}\mbox{]} for better clarity]{lidov1963}
\end{quote}

Although there is no mention on what $\Delta_{11} x$ and $\Delta_{21} x$
in the above quotation are in \citeauthor{lidov1961}'s original description,
we suspect that they are generalized symbols
that denote the satellite's orbital elements.
If we generalize Eq. \eqref{eqn:L15} using $x$, it would become
(removing all superscripts $k$):
\begin{equation}
\begin{aligned}
  \Delta_1 x &= \Delta_{11} x + \Delta_{12} x + \Delta_{13} x + \Delta_{14} x + \cdots , \\
  \Delta_2 x &= \Delta_{21} x + \Delta_{22} x + \Delta_{23} x + \Delta_{24} x + \cdots .
\end{aligned}
  \label{eqn:L15-general-x}
\end{equation}

We presume that \citeauthor{lidov1961}'s above statement,
particularly the ``motionless'' part,
comes from the fact that the first terms in the right-hand side of
Eq. \eqref{eqn:L15-general-x} ($\Delta_{11} x$ and $\Delta_{21} x$)
are in the order of $\left(\Delta t\right)^0 = O(1)$.
In other words, these quantities (belonging to the perturbing body) are $\Delta t$-independent:
no matter how long or short $\Delta t$ is, these quantities would not change.
Therefore, the circumstance is equivalent to having a perturbing body
that is fixed in absolute space without any motion,
which leads to \citeauthor{lidov1961}'s above conclusion.
Note that the variation components of other orbital elements such as
$\Delta_{j1} p,
 \Delta_{j1} i,
 \Delta_{j1} \omega,
 \Delta_{j1} \Omega,
 \ldots$
$(j = 1, 2, 3, \ldots)$ are not zero because of the integrated force
(or moment) exerted on the satellite over its one revolution.

After some mathematical preparations in his Section 5
(``\textit{Computation of the parameters $\alpha_i$, $\beta_i$, and $\gamma_i$ and their derivatives for the case in which the perturbing body moves in an ellipse\/}'') based on the assumption that the orbital motion of the perturbing body is Keplerian,
\citeauthor{lidov1961} extends the formulation of orbital change of the satellite over its several $(N)$ revolutions in his Section 6
(``\textit{Formulas for variation in orbital elements during several revolutions of the satellites\/}'').
The mathematical expositions in these two sections are quite detailed and lengthy.
We will not go into them because they are out of the scope of this monograph.
Major results exhibited in \citeauthor{lidov1961}'s Section 6 are summarized in the form of
finite integrals using the perturbing body's $u$ (argument of latitude)
from $u_0$ to $u_N$, corresponding to the time from $t_0$ to $t_N$,
such as Eqs. (L45), (L46) and (L47).

The preparations carried out in his Sections 5 and 6 allow
\citeauthor{lidov1961} to move on to the next step of his approximation:
deriving the secular change of the satellite's orbital elements
during one revolution of the perturbing body.
This procedure is carried out in his Section 7
(``\textit{Treatment of the equations for secular variation in elements during the period of revolution of the perturbing body\/}'').
This section starts with deriving the doubly averaged equations of
motion of the satellite based on the achievement obtained in the previous two sections (Sections 6 and 7).
More specifically, substitution of $u_N = u_0 + 2\pi$ into the finite
integrals calculated in Section 6,
from $\int_{u_0}^{u_N}$ to $\int_{u_0}^{u_0+2\pi}$, is applied.
After deriving the doubly averaged differential equations for the satellite's orbital elements,
\citeauthor{lidov1961} moves on to an analysis of the characteristics of the satellite's orbital motion.
He particularly focuses on the conserved quantities of the three-body system
that he considers.
At this point, we finally find ourselves able to compare
\citeauthor{lidov1961}'s and \citeauthor{kozai1962b}'s results in detail.

Let us show
the resulting doubly averaged differential equations of the satellite's orbital
elements in \citeauthor{lidov1961}'s Section 7.
He depicted them as
``the equations for the variation in the orbital elements [of satellite]
  for one revolution, averaged for the period of revolution of the
  perturbing body'' \citep[][p. 1994]{lidov1963}.
Their actual forms are:
\begin{equation}
\begin{aligned}
  \frac{\delta a}{\delta N} &= 0, \\
  \frac{\delta e}{\delta N} &= \frac{Ae \sqrt{\varepsilon}}{2} \sin ^2 i \sin 2 \omega, \\
  \frac{\delta i}{\delta N} &= -\frac{A}{2} \frac{1-\varepsilon}{\sqrt{\varepsilon}} \sin i \cos i \sin 2 \omega, \\
  \frac{\delta \Omega}{\delta N} &= -\frac{A \cos i}{\sqrt{\varepsilon}}
    \left[ (1-\varepsilon) \sin^2 \omega + \frac{\varepsilon}{5} \right], \\
  \frac{\delta \omega}{\delta N} &=  \frac{A}{\sqrt{\varepsilon}}
    \left[ \left(\cos^2 i - \varepsilon\right) \sin^2 \omega + \frac{2\varepsilon}{5} \right],
\end{aligned}
  \tag{L54-\arabic{equation}}
  \stepcounter{equation}
  \label{eqn:L54}
\end{equation}
where
\begin{equation}
  \varepsilon   = 1 - e^2   , \quad
  \varepsilon_k = 1 - e_k^2 , \quad
  p_k = a_k \varepsilon_k   ,
  \label{eqn:L1961-def-varepsilon}
\end{equation}
\begin{equation}
  A = \frac{15 \pi}{2} \frac{\mu_k}{\mu} \left( \frac{a}{p_k} \right)^3 \varepsilon_k^{\frac{3}{2}} .
  \tag{L55-\arabic{equation}}
  \stepcounter{equation}
  \label{eqn:L55}
\end{equation}

Recall that $N$ denotes the number of revolutions of the satellite
during one revolution of the perturbing body.
The increment of the satellite's longitude over its $N$ revolutions is $2\pi N$.
It advances the satellite's time by $2\pi N /n$
(where $n$ is the satellite's mean motion).
Therefore we see that the differentiation $\frac{\delta}{\delta N}$
in Eq. \eqref{eqn:L54} is directly associated with the
ordinary time differentiation $\DD{}{t}$ as
\begin{equation}
  \frac{\delta}{\delta N} = \frac{2\pi}{n} \frac{d}{dt} .
  \label{eqn:dN-lidov1961}
\end{equation}

Using the relationship \eqref{eqn:dN-lidov1961},
it is straightforward to show that the differential equation for
the satellite's argument of pericenter $\omega$
(the last one in Eq. \eqref{eqn:L54}) is equivalent to the
canonical equation of motion for $g$ at the quadrupole level
(Eq. \eqref{eqn:dgdt-quadrupole}
 on p. \pageref{pg:dG-dt-2n=2} of this monograph).
It is also easy to construct a differential equation for 
$G = \sqrt{\mu a\left(1-e^2\right)}$ from Eq. \eqref{eqn:L54}, and
to confirm its equivalence to the canonical equation of motion,
Eq. \eqref{eqn:dGdt-quadrupole}.

\subsection{Constants of motion\label{ssec:Lidov-constants}}
Of all ten sections of \citeauthor{lidov1961}'s paper
in \citeyear{lidov1961},
we regard Section 7 to contain the most important contents.
In this section, \citeauthor{lidov1961} first presents three constants of
motion that the doubly averaged inner CR3BP possesses.
Then he gives considerations on the solutions of the doubly averaged
equations of motion \eqref{eqn:L54} for several special cases,
based on the three constants of motion.

As \citeauthor{lidov1961} writes, and
as is clear from the first equation of Eq. \eqref{eqn:L54},
the satellite's semimajor axis $a$ is constant in his approximation.
This is the first constant of motion.
The second constant is what \citeauthor{lidov1961} calls $c_1$ defined as
\begin{equation}
  c_1 = \left(1-e^2 \right) \cos^2 i ,
  \label{eqn:def-Lidov-c1}
\end{equation}
or in \citeauthor{lidov1961}'s original expression, it is defined as
\begin{equation}
  \varepsilon = \frac{c_1}{\cos^2 i} .
  \tag{L58-\arabic{equation}}
  \stepcounter{equation}
  \label{eqn:L58}
\end{equation}
This quantity is equivalent to \citeauthor{kozai1962b}'s $\Theta$
(see Eq. \eqref{eqn:K26} on p. \pageref{eqn:K26} of this monograph)
as well as to the square of \citeauthor{moiseev1945b}'s $C$
(see Eq. \eqref{eqn:Mb45-2.14} on p. \pageref{eqn:Mb45-2.14} of this monograph).
It is easy to confirm that $c_1$ is actually a constant
in the framework of \citeauthor{lidov1961}'s approximation as follows:
\begin{enumerate}
\item Differentiate $c_1$ of Eq. \eqref{eqn:L58} by the time $t$ to obtain $\DD{c_1}{t}$ .
\item Convert $\frac{\delta e}{\delta t}$ and $\frac{\delta i}{\delta t}$ into
$\DD{e}{t}$ and $\DD{i}{t}$ using the relationship \eqref{eqn:dN-lidov1961} .
\item Substitute $\DD{e}{t}$ and $\DD{i}{t}$ into $\DD{c_1}{t}$ .
\end{enumerate}
And we will find $\DD{c_1}{t} = 0$.

The third constant of motion, which \citeauthor{lidov1961} calls $c_2$,
makes his work unique and distinguished from other classic studies.
In the work of \citeauthor{kozai1962b} and that of \citeauthor{moiseev1945a},
the third constant of motion is the doubly averaged disturbing function itself.
In \citeauthor{kozai1962b}'s work, it is $W^\ast$ in Eq. \eqref{eqn:K23}
or Eq. \eqref{eqn:R2-final} in the quadrupole level approximation.
In \citeauthor{moiseev1945a}'s work,
it is $[W_{\rm j}]$ in Eq. \eqref{eqn:Mb45-2.15}.
In contrast,
\citeauthor{lidov1961} defines his third constant $c_2$ as follows:
\begin{equation}
  c_2 = \left( 1-\varepsilon \right) \left( \frac{2}{5} - \sin^2 i \sin^2 \omega \right) ,
  \label{eqn:def-Lidov-c2}
\end{equation}
or in \citeauthor{lidov1961}'s original expression, it is defined as
\begin{equation}
  1-\varepsilon = \frac{c_2}{\frac{2}{5} - \sin^2 i \sin^2 \omega} .
  \tag{L59-\arabic{equation}}
  \stepcounter{equation}
  \label{eqn:L59}
\end{equation}
Actually, we have not found any information or descriptions as to
how \citeauthor{lidov1961} devised, derived, or reached the function form
of Eq. \eqref{eqn:L59} in any of the relevant publications.
An interesting point about $c_2$ is that, $c_2$ is not independent from $c_1$.
$c_1$ and $c_2$ are connected to each other through the following relationship:
\begin{equation}
  c_2 = \left( 1 - \frac{c_1}{\cos^2 i} \right)
        \left( \frac{2}{5} - \left( 1- \frac{c_1}{\varepsilon} \right) \sin^2 \omega \right) .
  \label{eqn:c2-depend-lidov1961}
\end{equation}
Similar to the discussion about $c_1$,
it is again straightforward to confirm that $c_2$ is a constant of motion
in the framework of \citeauthor{lidov1961}'s approximation as follows:
\begin{enumerate}
\item Differentiate $c_2$ of Eq. \eqref{eqn:L59} by the time $t$ to obtain $\DD{c_2}{t}$ .
\item Convert $\frac{\delta e}{\delta t}$, $\frac{\delta i}{\delta t}$,
     $\frac{\delta \omega}{\delta t}$ into
     $\DD{e}{t}$, $\DD{i}{t}$, $\DD{\omega}{t}$ using the relationship
     \eqref{eqn:dN-lidov1961}.
\item Substitute $\DD{e}{t}$, $\DD{i}{t}$, $\DD{\omega}{t}$
      into $\DD{c_2}{t}$ .
\end{enumerate}
And we will find $\DD{c_2}{t} = 0$.
The actual values of $c_1$ and $c_2$ are determined by
the initial values of $\varepsilon$, $i$, and $\omega$ as
\begin{equation}
\begin{aligned}
  c_1 & = \varepsilon_0 \cos^2 i_0 , \\
  c_2 & = \left( 1-\varepsilon_0 \right) \left( \frac{2}{5} - \sin^2 i_0 \sin^2 \omega_0 \right) ,
  \end{aligned}
  \tag{L60-\arabic{equation}}
  \stepcounter{equation}
  \label{eqn:L60}
\end{equation}
where
$\varepsilon_0$, $i_0$, $\omega_0$ are the initial values of
$\varepsilon$,   $i$,   $\omega$,  respectively.

After introducing $c_1$ and $c_2$, 
\citeauthor{lidov1961} briefly mentions how to obtain
a time-dependent solution for the orbital elements.
\citeauthor{lidov1961} states that
the equations \eqref{eqn:L58} and \eqref{eqn:L59}
determine the dependencies of two functions as
\begin{equation}
  \varepsilon = f_1 (\omega), \quad
  \cos i      = f_2 (\omega) .
  \label{eqn:L61-e-omega-i-dependence}
\end{equation}
Note that $f_1$ is mistyped as $f_i$ in \citet[][p. 1994, below Eq. (L60)]{lidov1963}.
Then \citeauthor{lidov1961} expresses the solution by definite integrals,
in other words, by quadrature. Literally citing his words,
``the solution of the entire problem is obtained by using the two definite integrals,'' \citep[][the right column on p. 1994]{lidov1963}.
Their actual forms are as follows:
\begin{equation}
  N - N_0 = \frac{1}{A} \int^\omega_{\omega_0}
    \frac{\sqrt{\varepsilon}}{\left(\cos^2 i - \varepsilon\right) \sin^2 \omega + \frac{2\varepsilon}{5}} d \omega ,
  \tag{L61-\arabic{equation}}
  \stepcounter{equation}
  \label{eqn:L61}
\end{equation}
\begin{equation}
  \Omega - \Omega_0 = -{A} \int^N_{N_0}
    \frac{\cos i}{\sqrt{\varepsilon}}
      \left[ \left(1 - \varepsilon\right) \sin^2 \omega + \frac{\varepsilon}{5} \right] dN .
  \tag{L62-\arabic{equation}}
  \stepcounter{equation}
  \label{eqn:L62}
\end{equation}

The formal operation of the quadrature depicted by
Eqs. \eqref{eqn:L61} and \eqref{eqn:L62} are principally equivalent to
what \citeauthor{moiseev1945a} showed by his Eq. \eqref{eqn:Mb45-2.17} and
the discussion that follows it
(see p. \pageref{eqn:Mb45-2.17} of this monograph).

\subsection{The \citeauthor{lidov1961} diagram\label{ssec:Lidovdiagram}}
As is better known today,
there is a major advantage to exploit $c_2$ as the third constant of motion
of the doubly averaged inner CR3BP
compared with the use of the averaged disturbing function itself.
At the quadrupole level approximation,
we can immediately predict if the perturbed body's argument of pericenter
librates around $\pm \frac{\pi}{2}$ or circulates from 0 to $2\pi$
just from its $c_2$ value
without producing an equi-potential diagram.
This prediction is not straightforward
if we use the averaged disturbing function itself as the third constant of motion.
In this subsection let us consider this issue in more detail
along with \citeauthor{lidov1961}'s discussion.
\citet[][his Section 3.2.3 on p. 36]{shevchenko2017} also has a dedicated description for this subject.

\begin{figure}[htbp]\centering
\ifepsfigure
 \includegraphics[width=\singlefigwidth\textwidth]{fig_lidovdiagram3.eps} %fig10
\else
 \includegraphics[width=\singlefigwidth\textwidth]{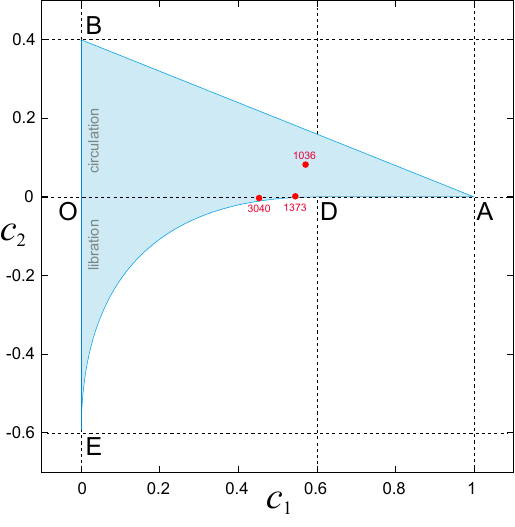} %fig10
\fi
  \caption{%
The \citeauthor{lidov1961} diagram that indicates the possible range of $(c_1, c_2)$
in the doubly averaged inner CR3BP at the quadrupole level approximation.
This figure is practically a reproduction of
what \citet[][\mysymfigO 1 on p. L29]{lidov1961} showed.
The line $c_2 = 0$ (the line A--D--O) makes an important boundary of the perturbed body's secular motion.
When the parameter set $(c_1, c_2)$ of the perturbed body is
above the boundary and located in the triangle $\bigtriangleup$AOB,
its argument of pericenter exhibits a circulation from 0 to $2 \pi$.
Meanwhile
it makes a libration around $+\frac{\pi}{2}$ or $-\frac{\pi}{2}$
when the parameter set $(c_1, c_2)$ of the perturbed body is
below the boundary and located in the concave triangle $\bigtriangleup$DOE
(except when it is just on the line O--D).
The red dots show the locations of the parameter sets $(c_1, c_2)$ of
the three asteroids presented in \mysymfigO \protect{\ref{fig:xy-inner}}:
(1036) Ganymed,
(1373) Cincinnati, and
(3040) Kozai.
Their orbital elements are as of February 16, 2014, 00:00:00 CT,
adopted from the JPL Horizons web-interface.
}
  \label{fig:lidovdiagram}
\end{figure}

From its definition \eqref{eqn:L58}, $c_1$ takes the value from 0 to 1.
From its definition \eqref{eqn:L59},
$c_2$ takes the value from $-\frac{3}{5}$ to $\frac{2}{5}$.
However since $c_2$ depends on $c_1$ as seen in Eq. \eqref{eqn:c2-depend-lidov1961},
the dynamically possible area on the $(c_1,c_2)$ plane is not a simple rectangle, and
it shapes the cyan-hatched region A--B--O--E--D--A shown in \mysymfigO \ref{fig:lidovdiagram}.
The coordinates of each of the points are
A$\bigl(1, 0\bigr)  $,
B$\bigl(0, \frac{2}{5}\bigr)$,
O$\bigl(0, 0\bigr)  $,
E$\bigl(0, -\frac{3}{5}\bigr)$, and
D$\bigl(\frac{3}{5}, 0\bigr)$.
Let us call this type of plot on the $(c_1,c_2)$ plane as the
\citeauthor{lidov1961} diagram,
as it first appeared in \citeauthor{lidov1961}'s work.

The borders of the closed shape A--B--O--E--D--A in
\mysymfigO \ref{fig:lidovdiagram} are defined by the solutions of
Eqs. \eqref{eqn:L54} in several special cases.
The descriptions on these special cases in \citet[][p. L26--L30]{lidov1961}
are quite elaborate and well organized.
These solutions also help us understand the dynamical characteristics of
the system described by Eq. \eqref{eqn:L54} in non-special, ordinary cases.
We first follow \citeauthor{lidov1961}'s description concerning the function form of the borders.
\label{pg:lidov-specialsolutions}

\paragraph{The straight line A--B} %
This line represents the maximum of $c_2$ as a function of $c_1$.
From Eqs. \eqref{eqn:L58} and \eqref{eqn:L59},
we know that $c_2$ takes the maximum when $\sin^2 i = 0$.
This yields the maximum value of $e$ through Eq. \eqref{eqn:L58},
and it is a constant because $c_1$ is constant.
As a result, the line A--B indicates a planar motion ($i = 0$ or $\pi$)
of the perturbed body with a fixed eccentricity.
From Eq. \eqref{eqn:c2-depend-lidov1961},
we can see that $c_2$ and $c_1$ are connected to each other along this line
via the following relationship:
\begin{equation}
  c_2 = \frac{2}{5}\left(1-c_1\right) .
  \label{eqn:c2-AB-lidov1961}
\end{equation}
Note that the motion of the perturbed body along this line corresponds to
what takes place at the outer boundary of the $(e \cos g, e \sin g)$
diagram that we showed in the previous section
such as \mysymfigO \ref{fig:xy-inner} on p. \pageref{fig:xy-inner}.
It also corresponds to the lower boundaries in \citeauthor{kozai1962b}'s
\mysymfigS K2--K8.
See \mysymfigS \ref{fig:replot_kozai1962},
               \ref{fig:replot_kozai1962_fig67},
and the discussion on p. \pageref{pg:kozai_upperlowerboundaries} of this monograph.

\paragraph{The vertical straight line B--O--E} %
This line is realized when $c_1 = 0$.
As $c_1$ is proportional to the perpendicular component of
the angular momentum vector of the perturbed body,
$c_1 = 0$ means that the motion of the perturbed body takes place on a plane
orthogonal to the orbital plane of the perturbing body
$(\cos i = 0)$, except in the case of the very special configuration of $e=1$
which we do not consider in this monograph.
\citeauthor{lidov1961} gave a detailed consideration on this class of
satellite orbits with $\cos i = 0$ (p. L26--L28), and
concluded that in most cases these orbits will turn into extremely
eccentric ones until the satellite falls onto the central body
within a finite length of time.
\citet{lidov1963a,lidov1963b} gave a supplemental illustration for
elucidating the evolution of these type of orbits
(see Section \ref{ssec:lidov1963} of this monograph).
\label{pg:line-B-O-E}

\paragraph{The curved line E--D} %
This curve represents the minimum of $c_2$ as a function of $c_1$
between $0 \leq c_1 \leq \frac{3}{5}$.
From the dependency of $c_2$ on $c_1$ seen in Eq. \eqref{eqn:c2-depend-lidov1961},
we can derive the function form of $c_2$ on this line
through the following two steps.
\citeauthor{lidov1961} considers $c_2$ as a function of two variables,
$\varepsilon$ $(= 1-e^2)$ and $\omega$.
He first fixes $\varepsilon$ and searches for the value of $\omega$ that gives a local minimum of $c_2$.
Then \citeauthor{lidov1961} varies $\varepsilon$ and searches for its value that gives the minimum of $c_2$.

When $\varepsilon$ is fixed as a constant,
we know that the lowest value of $c_2$ occurs when $\sin^2 \omega = 1$
from  Eq. \eqref{eqn:c2-depend-lidov1961}.
In this case, Eq. \eqref{eqn:c2-depend-lidov1961} becomes as follows:
\begin{equation}
  c_2 = - \left( 1 - \varepsilon \right) \left( \frac{3}{5} - \frac{c_1}{\varepsilon} \right) .
  \label{eqn:c2-min1-lidov1961}
\end{equation}
Then, consider $c_2$ in Eq. \eqref{eqn:c2-min1-lidov1961} as a function of
$\varepsilon$, and search for its local extremums
by expressing the partial derivative of $c_2$ with respect to $\varepsilon$ as
\begin{equation}
  \DP{c_2}{\varepsilon} = \frac{3}{5} - \frac{c_1}{\varepsilon^2} .
  \label{eqn:c2-min2-lidov1961}
\end{equation}
Putting $\DP{c_2}{\varepsilon} = 0$ tells us that $c_2$ takes a local extremum at
\begin{equation}
  \varepsilon = \sqrt{\frac{5}{3}c_1} .
  \label{eqn:c2-min4-lidov1961}
\end{equation}
This turns out to be a minimum, because
from Eq. \eqref{eqn:c2-min2-lidov1961} we always have
\begin{equation}
  \DP[2]{c_2}{\varepsilon} = \frac{2 c_1}{\varepsilon^3} > 0 .
  \label{eqn:c2-min3-lidov1961}
\end{equation}
For $\varepsilon$ to satisfy Eq. \eqref{eqn:c2-min4-lidov1961},
$c_1$ cannot be larger than $\frac{3}{5}$
because $\varepsilon$ cannot exceed 1.
Substituting $\varepsilon$ of
Eq. \eqref{eqn:c2-min4-lidov1961} into
Eq. \eqref{eqn:c2-min1-lidov1961},
we have the function form of $c_2$ on the curve E--D as
\begin{equation}
  c_2 = -\left(\sqrt{c_1} - \sqrt{\frac{3}{5}}\right)^2 .
  \label{eqn:c2-ED-lidov1961}
\end{equation}

Having $\varepsilon$ in the form of Eq. \eqref{eqn:c2-min4-lidov1961} in hand,
it is straightforward to confirm that $\frac{\delta \omega}{\delta N}$
in Eq. \eqref{eqn:L54} vanishes when initially $\sin^2 \omega = 1$.
This fact means that the perturbed body's argument of pericenter $\omega$ stays
stationary at $\omega = \pm\frac{\pi}{2}$
when the system's parameter set $(c_1,c_2)$ lies somewhere on the line E--D.
$\frac{\delta e}{\delta N}$ also becomes zero at these points,
indicating that the satellite's eccentricity $e$ stays stationary too.
From Eq. \eqref{eqn:c2-min4-lidov1961} or Eq. \eqref{eqn:L54}
we have $\cos^2 i = \frac{3}{5} \varepsilon$ in this case.
As we mentioned above,
the range of $c_1$ should be between 0 and $\frac{3}{5}$
for this case to happen.
This is equivalent to the condition that \citeauthor{kozai1962b}
gave as Eq. \eqref{eqn:Theta-max} on p. \pageref{eqn:Theta-max},
$\Theta \leq \frac{3}{5}$.

In short, motion of the perturbed body on the line E--D in the
\citeauthor{lidov1961} diagram is equivalent to
the motion at the stationary points with $g = \pm \frac{\pi}{2}$
on the $(e \cos g, e \sin g)$ plane such as \mysymfigO \ref{fig:xy-inner}
(p. \pageref{fig:xy-inner} of this monograph).
Since the perturbed body's eccentricity and argument of pericenter are both constant in this case,
the secular motion of its orbit would be a simple precession around the
normal vector of the reference place (i.e. the perturbing body's orbital plane).
\citeauthor{lidov1961} literally describes the circumstance as follows:
\begin{quote}
``4. When $\omega_0$ and $i_0$ satisfy the conditions $\cos \omega_0 = 0$ and
$\cos^2 i_0 = \frac{3}{5} \varepsilon_0$ and $\varepsilon_0$ is arbitrary,
system (L54) has a solution representing, on the average,
an orbit with constant elements
$\omega = \omega_0 = \pm (\pi/2)$,
$\varepsilon = \varepsilon_0$,
$i = i_0 = \varphi = \varphi_0$
where $\sin^2 \varphi_0 = 1 - \frac{3}{5} \varepsilon_0$.
[$\cdots$]
In this case the entire evolution, on the average, consists in the
rotation of the orbit around a normal to the plane of the motion
of the perturbing body.''
\citep[][p. 1995, the point 4 in the right column]{lidov1963}
\end{quote}

\paragraph{The horizontal line O--D--A} %
This line is realized when $c_2 = 0$.
From the definition of $c_2$ in Eq. \eqref{eqn:L59},
we know that there can be two subcases here.
It is either when
\begin{equation}
  \sin^2 i \sin^2 \omega = \frac{2}{5} ,
  \label{eqn:c2=0-subcase1}
\end{equation}
or when
\begin{equation}
  \varepsilon = 1 .
  \label{eqn:c2=0-subcase2}
\end{equation}

In the first subcase when Eq. \eqref{eqn:c2=0-subcase1} holds true, we have
\begin{equation}
  \sin^2 \omega = \frac{2}{5\sin^2 i}      \geq \frac{2}{5},
  \label{eqn:c2=0-subcase1-2over5-omega}
\end{equation}
and
\begin{equation}
  \sin^2 i      = \frac{2}{5\sin^2 \omega} \geq \frac{2}{5} .
  \label{eqn:c2=0-subcase1-2over5-inc}
\end{equation}
Also, by Eq. \eqref{eqn:c2=0-subcase1}
$\frac{\delta \omega}{\delta N}$ in Eq. \eqref{eqn:L54} can be written as
\begin{equation}
  \frac{\delta \omega}{\delta N} = A \frac{1-\varepsilon}{\sqrt{\varepsilon}}
       \left( \sin^2 \omega - \frac{2}{5} \right) .
  \tag{L73-\arabic{equation}}
  \stepcounter{equation}
  \label{eqn:L73}
\end{equation}
From Eq. \eqref{eqn:L73} and Eq. \eqref{eqn:c2=0-subcase1-2over5-omega}, we know that
$\frac{\delta \omega}{\delta N} \geq 0$ is always true in this subcase.

Meanwhile from the equation of $\frac{\delta e}{\delta N}$ in Eq. \eqref{eqn:L54},
the satellite's $e$ would increase while $\omega < \frac{\pi}{2}$,
and it would take its maximum at $\omega = \frac{\pi}{2}$.
At this point, the satellite's inclination takes its minimum value that satisfies
\begin{equation}
  \sin^2 i = \frac{2}{5} ,
  \label{eqn:c2=0-subcase1-sin2i=2over5}
\end{equation}
from Eq. \eqref{eqn:c2=0-subcase1-2over5-inc},
and $\varepsilon$ also takes its minimum value as
\begin{equation}
  \varepsilon = \frac{5}{3} \varepsilon_0 \cos^2 i_0 ,
  \label{eqn:c2=0-subcase1-varepsilon-min}
\end{equation}
from \eqref{eqn:L58},
where $\varepsilon_0$ and $i_0$ denote the initial values of
$\varepsilon$ and $i$, respectively.
The maximum value of eccentricity $e$ in this case is calculated from
Eq. \eqref{eqn:c2=0-subcase1-varepsilon-min}.
Past this point, $e$ decreases while $\omega > \frac{\pi}{2}$,
and it would gradually approach the point of $e=0$.
However, since the variation rates
$\frac{\delta e}{\delta N}$      in Eq. \eqref{eqn:L54} and
$\frac{\delta \omega}{\delta N}$ in Eq. \eqref{eqn:L73}
both become small when $e \to 0$ (or $\varepsilon \to 1$),
the secular motion of the satellite in phase space
near the point $e=0$ would be very slow.
Formally, it takes an infinite amount of time to reach the point $e=0$.

From today's perspective,
this subcase corresponds to the motion on a separatrix that separates
the motion of the satellite's argument of pericenter
into libration and circulation.
Note that this subcase is realized only on the line O--D
while $c_1 \leq \frac{3}{5}$.
Actually when Eq. \eqref{eqn:c2=0-subcase1} holds true, $c_1$ is written as
\begin{equation}
  c_1 = \varepsilon - \frac{2 \varepsilon}{5 \sin^2 \omega} .
  \label{eqn:c2=0-subcase1-c1}
\end{equation}
Partial derivatives of $c_1$ by $\varepsilon$ and by $\sin^2 \omega$ become
\begin{equation}
\begin{aligned}
 \DP{c_1}{\varepsilon} &= 1 - \frac{2}{5 \sin^2 \omega}, \\
 \DP{c_1}{\left(\sin^2 \omega \right)} &= \frac{2\varepsilon}{5 \sin^4 \omega}.
\end{aligned}
 \label{eqn:c2=0-subcase1-c1-derivative}
\end{equation}
Both quantities are always greater than or equal to zero
because of Eq. \eqref{eqn:c2=0-subcase1-2over5-omega}.
This means that $c_1$ becomes its maximum when both $\varepsilon$ and
$\sin^2 \omega$ take their largest values:
$\varepsilon=1$ and $\sin^2 \omega = 1$.
Substituting these values into Eq. \eqref{eqn:c2=0-subcase1-c1},
we see that the maximum value of $c_1$ in this subcase is $\frac{3}{5}$.

On the other hand,
the second subcase expressed by Eq. \eqref{eqn:c2=0-subcase2}
takes place when initially $\varepsilon=1$ (or $e=0$).
In this case
both $\frac{\delta e}{\delta N}$ and $\frac{\delta i}{\delta N}$
in Eq. \eqref{eqn:L54} become zero, and the orbit of the satellite
remains circular with a fixed inclination.
This status can be realized anywhere on the line O--D--A,
whatever value $c_1$ takes between 0 to 1.
The motion of the satellite along this line corresponds to the origin $(0,0)$
in the $(e \cos g, e \sin g)$ diagrams such as
\mysymfigO \ref{fig:xy-inner} on p. \pageref{fig:xy-inner}.
It also corresponds to the upper boundaries in \citeauthor{kozai1962b}'s
\mysymfigS K2--K8
 (see our \mysymfigS \ref{fig:replot_kozai1962} and
                     \ref{fig:replot_kozai1962_fig67}).
Also, note that \citeauthor{kozai1962b} left brief descriptions
on perturbed body's motion in several cases
(see \citeauthor{kozai1962b}'s p. K594--K595 or p. \pageref{pg:Kozais4cases} of this monograph)
which include the two subcases discussed here.

\subsection{$c_2$ as a flag of $\omega$-libration\label{ssec:Lidov-c2}}

In the later part of his Section 7, \citeauthor{lidov1961} presents
a description of the satellite's secular motion on the $(c_1,c_2)$ plane
in more general cases, not only on the border of the hatched shape
A--B--O--E--D--A of \mysymfigO \ref{fig:lidovdiagram}.
\citeauthor{lidov1961} explains that $c_2$ can be used as a flag
as to whether the satellite's argument of pericenter librates or circulates.

\paragraph{When $c_2 < 0$} %
This case is easy to understand.
From the definition of $c_2$ in Eq. \eqref{eqn:L59}, we have
\begin{equation}
  \sin^2 i \sin^2 \omega > \frac{2}{5} .
  \label{eqn:c2negative-subcase1}
\end{equation}
Similar to the discussion when we derived
Eqs. \eqref{eqn:c2=0-subcase1-2over5-omega} and
     \eqref{eqn:c2=0-subcase1-2over5-inc},
the inequality \eqref{eqn:c2negative-subcase1} naturally yields
\begin{equation}
  \sin^2 \omega > \frac{2}{5\sin^2 i}      \geq \frac{2}{5} .  % , \quad
  \label{eqn:c2negative-subcase1-2over5}
\end{equation}
The inequality \eqref{eqn:c2negative-subcase1-2over5} means that
$\omega$ is confined in a range
centered at the position that satisfies $\sin \omega = \pm 1$
(i.e. $\omega = \pm \frac{\pi}{2}$), and
it cannot complete a circulation from 0 to $2\pi$.
Therefore we conclude that the motion of the satellite's argument of
pericenter $\omega$ librates when $c_2 < 0$.
On the \citeauthor{lidov1961} diagram (\mysymfigO \ref{fig:lidovdiagram})
this happens in the lower concave triangle $\bigtriangleup$DOE
(recall that this triangle does not include the line O--D).
As we mentioned before, the range of $c_1$ inside the triangle
$\bigtriangleup$DOE is between 0 and $\frac{3}{5}$.
Therefore we can state that $\omega$ exhibits libration
only when both the conditions $c_1 < \frac{3}{5}$ and $c_2 < 0$ are satisfied
at the same time,
as long as the quadrupole level approximation is applied.

\paragraph{When $c_2 > 0$} %
In this case we focus on the sign of
$\frac{\delta \omega}{\delta N}$ in Eq. \eqref{eqn:L54}.
From the last equation of Eq. \eqref{eqn:L54}, it is clear that the quantity
\begin{equation}
  \left(\cos^2 i - \varepsilon\right) \sin^2 \omega + \frac{2\varepsilon}{5},
  \tag{L75-\arabic{equation}}
  \stepcounter{equation}
  \label{eqn:L75}
\end{equation}
is equal to $\frac{\sqrt{\varepsilon}}{A} \frac{\delta \omega}{\delta N}$,
and $\frac{\sqrt{\varepsilon}}{A}$ is always positive.
Hence the quantity \eqref{eqn:L75} determines the sign of $\frac{\delta \omega}{\delta N}$.
Here \citeauthor{lidov1961} introduces two subcases:
when $\frac{c_1}{\varepsilon^2} \geq 1$, and
when $\frac{c_1}{\varepsilon^2} <     1$.
The first subcase (when $\frac{c_1}{\varepsilon^2} \geq 1$) is easier to understand.
From the definition of $c_1$ in Eq. \eqref{eqn:L58},
the first term of Eq. \eqref{eqn:L75} can be rewritten as
\begin{equation}
    \left(\cos^2 i - \varepsilon\right) \sin^2 \omega
  = \varepsilon \left( \frac{c_1}{\varepsilon^2} - 1\right) \sin^2 \omega .
  \label{eqn:lidov1961-deltawGT0-1}
\end{equation}
The right-hand side of Eq. \eqref{eqn:lidov1961-deltawGT0-1} is
larger than, or equal to, zero
if $\frac{c_1}{\varepsilon^2} \geq 1$.
This means that
$\frac{\delta \omega}{\delta N} \geq 0$ is always true in this subcase, and
$\omega$ monotonically circulates without a libration.
There is no restriction on the value of $c_1$ here.

In the second subcase (when $\frac{c_1}{\varepsilon^2} < 1$),
it is wiser to once return to the definitions of
$c_1$ in Eq. \eqref{eqn:L58} and
$c_2$ in Eq. \eqref{eqn:L59}.
Then, $\sin^2 i$ and $\sin^2 \omega$ can be expressed
by using $c_1$, $c_2$, $\omega$ as follows:
\begin{equation}
\begin{aligned}
  \sin^2 i &= 1 - \frac{c_1}{\varepsilon}, \\
  \left( 1-\cos^2 i \right) \sin^2 \omega
           &= \frac{2}{5} - \frac{c_2}{1 - \varepsilon} .
\end{aligned}
  \label{eqn:c2negative-subcase2}
\end{equation}
Substituting the expressions of $\sin^2 i$ and $\sin^2 \omega$
in Eq. \eqref{eqn:c2negative-subcase2} into
Eq. \eqref{eqn:L75}, \citeauthor{lidov1961} found that
the following inequality must hold
for $\frac{\delta \omega}{\delta N} \geq 0$ to be true:
\begin{equation}
  \frac{\frac{2}{5} - \frac{c_2}{1-\varepsilon}}{1-\frac{c_1}{\varepsilon}}
  \leq
  \frac{\frac{2}{5}}{1-\frac{c_1}{\varepsilon^2}} .
  \tag{L76-\arabic{equation}}
  \stepcounter{equation}
  \label{eqn:L76}
\end{equation}
Now that we assume $\frac{c_1}{\varepsilon^2} < 1$, and
$0 \leq \varepsilon \leq 1$ by the definition of $\varepsilon$, we have
the following relationship:
\begin{equation}
  1 - \frac{c_1}{\varepsilon} \geq 1 - \frac{c_1}{\varepsilon^2} > 0 .
  \label{eqn:c2negative-subcase2-c1inq1}
\end{equation}
From Eq. \eqref{eqn:c2negative-subcase2-c1inq1},
we see that the inequality \eqref{eqn:L76} holds always true
as long as $c_2 > 0$.
This means that, in this subcase too,
$\frac{\delta \omega}{\delta N} \geq 0$ is true
regardless of the value of $c_1$.
Thus $\omega$ monotonically circulates without libration when $c_2 > 0$.

By now, we see that it is proven that the satellite's argument of pericenter
$\omega$ librates   when $c_2 < 0$ and $c_1 < \frac{3}{5}$,
while it circulates when $c_2 > 0$.
$c_2=0$ is a special case that happens exactly on the line O--D--A in
the \citeauthor{lidov1961} diagram, as we have already seen.
We might want to say that $\omega$ still librates on the separatrix
that overlaps with the line O--D.

The following explains how to utilize the \citeauthor{lidov1961} diagram.
In \mysymfigO \ref{fig:lidovdiagram},
we plotted the actual parameter values $(c_1,c_2)$ of the three asteroids
as a set of examples
whose trajectories we integrated in \mysymfigO \ref{fig:xy-inner}:
(1373) Cincinnati,
(1036) Ganymed, and
(3040) Kozai.
The locations of their $(c_1,c_2)$ are designated by the red points
in \mysymfigO \ref{fig:lidovdiagram}.
We see that Ganymed is clearly located inside the triangle $\bigtriangleup$AOB,
indicating that this asteroid's argument of pericenter circulates.
It is consistent with the numerical results presented in \mysymfigO \ref{fig:xy-inner}.
On the other hand,
Cincinnati and Kozai are located inside the concave triangle $\bigtriangleup$DOE (although they are very close to the boundary O--D--A).
This indicates that arguments of pericenter of these asteroids
librate around $\omega = \frac{\pi}{2}$ or $-\frac{\pi}{2}$.
As we saw, this is exemplified by the numerical results presented
in \mysymfigO \ref{fig:xy-inner} where we see $\omega$ of
  Cincinnati librates around $\omega = \frac{ \pi}{2}$, and that of
       Kozai librates around $\omega = \frac{3\pi}{2}$.

Here let us note two things.
First, the parameter values of $(c_1, c_2)$ of the three asteroids
plotted in \mysymfigO \ref{fig:lidovdiagram} may not be perfectly precise.
The hatched shape in \mysymfigO \ref{fig:lidovdiagram} was drawn
from the doubly averaged disturbing function.
Meanwhile, the location of the red points of the actual asteroids
were calculated using their osculating, non-averaged orbital elements
taken from the JPL Horizons web-interface.
The short-term oscillation of the asteroids' osculating elements blurs
the location of their $(c_1, c_2)$ in general.
Second, even if we use averaged orbital elements of the asteroids for
locating their $(c_1,c_2)$ combinations, we should be aware that
the borders $c_2 \lessgtr 0$ may not rigidly work out for
distinguishing their motion between libration and circulation.
This is because the \citeauthor{lidov1961} diagram of
\mysymfigO \ref{fig:lidovdiagram} is produced
based on the quadrupole level approximation.
It uses the truncated doubly averaged disturbing function at $\Oalsqr$.
Actually, $\alpha^2$ of the three asteroids with respect to
Jupiter is not negligibly small. Specifically,
$\alpha^2 \sim 0.13$ for (3040) Kozai,
$\alpha^2 \sim 0.26$ for (1036) Ganymed, and
$\alpha^2 \sim 0.43$ for (1373) Cincinnati.
Therefore their secular motion may not be well represented by the quadrupole level approximation.

In the remaining part of his Section 7,
\citeauthor{lidov1961} calculates the maximum and minimum values of
$\varepsilon$ and their dependence on $c_1$ and $c_2$ when $c_2 > 0$
(Eqs. (L77)--(L80), although we do not reproduce them here). In this case
the maximum of $\varepsilon$ (and the minimum of $e$) occurs when $\omega = 0$ or $\pi$, and
the minimum of $\varepsilon$ (and the maximum of $e$) occurs when $\omega = \pm \frac{\pi}{2}$.
\citeauthor{lidov1961} also presented similar equations for the case of $c_2 < 0$
(Eqs. (L82)--(L85), not reproduced here).
The end part of Section 7 is devoted to presenting
several plots of extreme values of the satellite's eccentricities and
arguments of pericenter as $c_1$ and $c_2$ being parameters
(\mysymfigS L2, L3, L4 on pp. L32--L33, not reproduced here).

\citeauthor{lidov1961}'s last three sections are as follows:
Section  8 ``\textit{Estimates of oscillations in the pericenter height of a satellite orbit during the period of revolution of the perturbed body,\/}
 Section  9 ``\textit{Method of computing the revolution of the orbit of an artificial Earth satellite using approximate formulas,\/}
and
 Section 10 ``\textit{Comparison of the results of computation by approximate formulas with the precise solution of the problem by numerical integration of the differential equations.\/}
They are largely devoted to descriptions of the practical computation method of
the satellite's orbital evolution, as well as a comparison of
the results from \citeauthor{lidov1961}'s analytic approximation and
that of the direct numerical integration of the equations of motion.
Although these sections probably had practical importance in
satellite dynamics at that time,
nowadays we think that most of these sections can be replaced
for more efficient numerical procedures.
Therefore we do not deal with these sections in this monograph.

\subsection{Newly added issues in \citet{lidov1963a,lidov1963b}\label{ssec:lidov1963}}
At the end of this section,
let us mention a few issues that show up only in \citeauthor{lidov1961}'s
supplementary publications \citep{lidov1963a,lidov1963b},
and not in his main paper in \citeyear{lidov1961}.

One of them is the treatment of a special case when the orbital inclination
$i$ of the perturbed body (satellite) is just $90^\circ$
\citep[][p. 174]{lidov1963b}.
This corresponds to the motion that happens on the line B--O--E of
the \citeauthor{lidov1961} diagram (\mysymfigO \ref{fig:lidovdiagram}).
For this case,
\citeauthor{lidov1963b} gave a detailed illustration as to the dependence of
the motion of argument of pericenter $\omega$ of the perturbed body on its
initial value, $\omega_0$. He even gave an analytic solution that represents
the time evolution of $\omega$ using an elliptic integral of the first kind
\citep[][Eq. (15) and a table on p. 175]{lidov1963b}.
As an actual example, 
\citeauthor{lidov1963b} succeeded in obtaining a quantitative
estimate that an Earth-orbiting satellite whose semimajor axis and
eccentricity are the same as those of the Moon and whose inclination
with respect to the ecliptic is $90^\circ$,
can fall back to the surface of the Earth
within $\sim$52 revolutions ($\sim$4 years).
\citeauthor{lidov1963b} also carried out the numerical integration of the
equations of motion of the satellite.
He did a similar calculation using the singly averaged approximation.
Both calculations demonstrate how quickly the minimum distance
from the satellite to the Earth's surface becomes reduced
\citep[][\mysymfigO 7 on p. 178]{lidov1963b}.
His numerical result (54 or 55 revolutions until the fall of the satellite)
agrees well with his analytic estimate ($\sim$52 revolutions).
This anecdote is cited by \citet[][see p. 95--100 of its English translation,
   \citet{beletsky2001}]{beletsky1972},
by \citet[][his p. 9--10]{vashkovyak2008}, and
by \citet[][his p. 7--8]{shevchenko2017} with a famous cartoon
where a man (a caricature of \citeauthor{lidov1963b}) is about to release
the Moon on the $i=90^\circ$ orbit around the Earth.

Another subject that \citet{lidov1963a,lidov1963b} newly added is
the treatment of oblateness of the central mass
\citep[][Section ``\textit{3. The influence of the noncentricity of the gravitational field,\/}'' p. 176--177]{lidov1963b}.
The set of equations of motion that \citeauthor{lidov1963b} introduced
includes the effect of oblateness of the central body
through parameters $\delta$ and $\beta$
\citep[][Eqs. (1) and (2) on p. 169]{lidov1963b}.
As a consequence, although no detail is given,
\citeauthor{lidov1963b} claimed that he found that
the existence or non-existence of stationary points of
perturbed body's argument of pericenter $\omega$ depends on $\beta$,
i.e. on the oblateness of the central body.
He concluded that the stationary points of $\omega$ happen only when the effect of oblateness is weak.
Readers must remember that \citeauthor{kozai1962b} also mentioned this point
(see p. \pageref{pg:planetaryoblateness-kozai} of this monograph)
but without showing anything specific or quantitative.
However, we do not know how \citeauthor{lidov1963b} derived
the terms that represent the oblateness of the central body
appearing in his equations of motion,
because \citeauthor{lidov1963b} did not give any details about the derivation.
\label{pg:planetaryoblateness}

 \section{The Work of \citeauthor{vonzeipel1910}\label{sec:vonzeipel}}
Edvard Hugo von Zeipel (1873--1959)
was a Swedish scientist in astronomy and mathematics.
As is mentioned in his biographical sketches such as \citet{solc2014},
\citeauthor{vonzeipel1908} is famous for his pioneering achievements in many fields of theoretical astronomy,
in particular,
for his early work on the singularities of the $N$-body problem \citep{vonzeipel1908}.
\citet{diacu1996} made a brief but excellent review of his achievement on this subject.
\citeauthor{vonzeipel1908}'s name is also embedded in a theorem
called the \citeauthor{vonzeipel1910} theorem related to his later work
on the calculation of radiative equilibrium of a rotating star
\citep[e.g.][]{vonzeipel1924a,vonzeipel1924b,vonzeipel1924c}.
Nowadays, we are probably aware of his name most often in celestial mechanics,
particularly related to perturbation theories:
\citeauthor{vonzeipel1910} constructed a canonical perturbation method
\citep{vonzeipel1916a,vonzeipel1916b,vonzeipel1917a,vonzeipel1917b}
which is now called the \citeauthor{vonzeipel1910} method.
In this series of work, he devised a method for introducing
generating functions to carry out canonical transformations
proposed by \citet{delaunay1860,delaunay1867} in order to systematically
reduce the degrees of freedom of systems by the averaging procedure
\citep[for more detail, see][]{brouwer1961,meffroy1966,boccaletti1998}.
These works are fundamental milestones in modern celestial mechanics,
followed by numerous confirmations and extensions
\citep[e.g.][]{brown1933,hori1966,hori1967,musen1968,deprit1969,yuasa1971}.

\citeauthor{vonzeipel1910}'s work on perturbation theory in celestial mechanics is very popular, and its reputation is well established.
However it is almost not known at all that,
while making the epoch-making achievements mentioned above,
\citeauthor{vonzeipel1910} also worked on a theoretical subject
that is equivalent to what we have already discussed in this monograph---the
doubly averaged CR3BP,
in particular when inclination of the perturbed body is large.
In this section we will summarize
the major contents of one of \citeauthor{vonzeipel1910}'s publications about this subject.
Our aim is to show that the basic theoretical structure of the
\citeauthor{lidov1961}--\citeauthor{kozai1962b} {\mainword}
was already established
in the early twentieth century
almost at the same depth or deeper.
The publication that we are going to focus on from now is
\citet{vonzeipel1910}, entitled
``Sur l'application des s\'eries de M. Lindstedt \`a l'\'etude du mouvement des com\`etes p\'eriodiques,``
written in French and published in
\textit{Astronomische Nachrichten.\/}
The full text of this paper is available on ADS.
As far as we know from the records on ADS and
on Web of Science (hereafter referred to as WoS),
this paper has been cited only once in a modern science context,
only by \citet[][through which we learned of the existence of \citeauthor{vonzeipel1910}'s contribution]{bailey1996}.
\label{pg:firstcitation-vZ10}

Note that as far as we can ascertain,
\citeauthor{vonzeipel1910} did not mention his first name (Edvard) at all
in any of his own publications.
We follow his way and refer to him as Hugo von Zeipel in the title of this monograph.
Note also that his name is sometimes mentioned and cited just as Zeipel \citep[e.g.][]{merman1982,orlov1965b,bailey1996}.
But we call him \citeauthor{vonzeipel1910} throughout the monograph.
See \supinfo{4} for a more detailed discussion and consideration on this subject.

\subsection{Purpose, method, findings\label{ssec:Z10-purpose}}
As is obvious from its title (which can be translated into English as
``On the application of the Lindstedt series to study the motion of periodic comets''),
\citeauthor{vonzeipel1910}'s major objective in this publication
is to establish a mathematical representation
of the motion of periodic comets as precisely as possible
using a method called the Lindstedt series.
\citeauthor{vonzeipel1910}'s interests particularly lies
in the orbital variation of comets that have a large orbital inclination.
The beginning part of the paper entitled ``\textit{Introduction\/}''
(pp. Z345--Z347, before his Chapter I begins) well summarizes
\citeauthor{vonzeipel1910}'s purpose, method, and important findings.
For facilitating an understanding of \citeauthor{vonzeipel1910}'s intentions,
we believe that literally citing several paragraphs from the ``\textit{Introduction\/}'' is appropriate.
Note that the English translation from French appearing in what follows
are all due to the present authors of this monograph.
We have added several expressions in $[\; ]$ for supplementing an understanding of the translation.

At the beginning, \citeauthor{vonzeipel1910} states the standard method for
constructing analytic solutions of planetary motion at his time and its limitation:
\begin{quote}
``It is well known that in calculating the motion of planets, some series, called by Mr. Poincar\'e, the Lindstedt series, can be used.
Elements of the planets are thus developed in powers of a small quantity $\mu$ of the order of planetary masses.
The coefficients of the various powers of $\mu$ are Fourier series of a number of linearly dependent arguments of time.
In the series, it is essentially assumed that the masses of the planets are relatively small compared with that of the Sun.
But in the applications made so far, it was also recognized that
eccentricities and inclinations of the orbits are small.'' (p. Z345)
\end{quote}

Let us just say at this point that calculating the Lindstedt series is
a way of constructing approximate periodic solutions.
We will see more detail about this in Section \ref{ssec:Z10Lindstedt}.

There, \citeauthor{vonzeipel1910} mentions the difficulty of
constructing general theories of planetary orbital motion
when their eccentricity $e$ and inclination $I$ are arbitrary:
\begin{quote}
``Trying to study, by means of series that analogues the motion of a planetary system, when eccentricities and inclinations are arbitrary, we face the difficulty in solving the equations of secular variations in a general way.'' (p. Z345)
\end{quote}

Next,
\citeauthor{vonzeipel1910} states a possible direction for solving this problem.
In what follows $R$ denotes the secular part of the disturbing function.
He claims that he shows a possibility of constructing the Lindstedt series around $R$'s local extremums,
not just around $R$'s minimum at $e=0$ and $I=0$ when an object's $e$ and $I$ are as small as the major planets, but
at other places in phase space even when object's $e$ or $I$ is not small.
Here is \citeauthor{vonzeipel1910}'s statement:
\begin{quote}
``We know that the function $R$ is minimum when the eccentricities and inclinations becomes zero.
Because of this property of the function $R$, it is possible to form the series of Lindstedt representing the motion of the planets, whose eccentricities and inclinations are small.
But, the function $R$ often owns other maxima and minima.
When the value of $R$ is close to such a maximum or minimum value, it is still possible to calculate the Lindstedt series representing the motion.
In the orbits thus obtained, the eccentricities and inclinations can be considerable. This is what I propose to show in this memoire.'' (p. Z345)
\end{quote}

Now, \citeauthor{vonzeipel1910} describes the dynamical model that
he employs in this study---CR3BP.
He also mentions the integrability of the secular version of this problem:
\begin{quote}
``In order not to complicate the exposition, I will confine myself first to a special case assuming that an infinitely small mass (asteroid, comet, meteorite or satellite) is attracted to the Sun and a perturbing planet moving around the Sun in a circle.
In the study of the motion of such a body, it shall be reduced to a system of canonical differential equations with three degrees of freedom, and falling within the general type of equations, which has been studied by Mr. Poincar\'e in his work
{\scriptsize $\gg$}Les M\'ethodes Nouvelles de la M\'ecanique C\'eleste{\scriptsize $\ll$}, vol. II, Chapter XI.
The equations of secular variations form a canonical system, whose degree of freedom is unity.
They can therefore be integrated by a quadrature and by trigonometric series of a single argument.'' (pp. Z345--Z346)
\end{quote}

However, as we are already aware, formal integrability of the doubly averaged
CR3BP does not mean that we can immediately obtain actual time-dependent solutions with explicit function forms.
\citeauthor{vonzeipel1910} mentions this point, and
claims that in this study he limits himself to only finding approximate
solutions around the local extremums of $R$:
\begin{quote}
``But, although the existence of these series is demonstrated, we do not know in general how to form their coefficients analytically.
Thus, trying to calculate the Lindstedt series, we are forced, even in this relatively simple case, to limit the problem and assume that we are in the vicinity of a maximum or minimum value of the secular part of the disturbing function.'' (p. Z346)
\end{quote}

After briefly mentioning the contents of his Chapters II and III,
\citeauthor{vonzeipel1910} describes the contents of his Chapter IV
that contains the most important aspects of his study.
It begins with the following sentence:
\begin{quote}
``Research on the maxima and minima of the secular part of the disturbing function, described in Chapter IV, shows that orbits with large inclinations and always having small eccentricities can exist only outside the disturbing planet.'' (p. Z346)
\end{quote}

We find this statement true if we remember the studies
by \citeauthor{lidov1961} and \citeauthor{kozai1962b}.
When the inclination of the perturbed body is larger than a certain value
($i_0$ in \citeauthor{kozai1962b}'s work),
stationary points of the argument of pericenter appear with separatrix.
In this case,
the eccentricity of the perturbed bodies cannot always stay small
(see \mysymfigO \ref{fig:xy-inner}\mtxtsf{a} on p. \pageref{fig:xy-inner} of this monograph).
Therefore it is concluded that, for the perturbed body to always have
both a large inclination and a small eccentricity,
it cannot be in an inner CR3BP system.
In other words, \citeauthor{vonzeipel1910}'s
statement above depicts the circumstance that the outer CR3BP realizes.
We will later return to this point in Section \ref{ssec:ocr3bp} of this monograph.

Immediately after the above sentence,
\citeauthor{vonzeipel1910} continues his statement as follows.
We are certain that readers of this monograph will notice the equivalence of this statement to what was obtained by
\citeauthor{lidov1961} and \citeauthor{kozai1962b}:
\begin{quote}
``At inside the orbit of this planet, there can be asteroids with inclinations exceeding a certain limit, depending on the ratio $\alpha$ of the two major axes.
This limit is about $39^\circ$ from the Sun, and it decreases when $\alpha$ increases, and it vanishes when $\alpha = 1$.
For orbits with large inclinations and
inside the disturbing planet,
the upper limit of the eccentricity is considerable---even if the lower limit is small---and it [the upper limit] approaches unity when the inclination is close to $90^\circ$.
Such an orbit will be, under certain conditions, disrupted by the disturbing planet.'' (p. Z346)
\end{quote}

The above phrase ``about $39^\circ$'' must ring a bell with readers,
because it is equivalent to the limiting inclination that
\citeauthor{lidov1961} and \citeauthor{kozai1962b} discussed:
$\cos^{-1} \sqrt{\frac{3}{5}} \sim 39^\circ.23$.

Then, \citeauthor{vonzeipel1910} moves on to stating his findings
about a categorization of stable cometary orbits
along with the motion of their argument of perihelion $g$.
Note that his following statement depicts three patterns of $g$'s motion:
Libration around $\pm\frac{\pi}{2}$,
libration around $0$ or $\pi$, and
circulation from $0$ to $2\pi$.
This means that his statement is not only about the inner CR3BP
but also about the outer CR3BP, and
also about the systems with orbit intersection:
\begin{quote}
``The studies mentioned in Chapter IV also helped me establish that there can be orbits of comets in stable motion, dependent on 6 arbitrary constants of integration. $[\ldots]$
As for the distance $g$ from the perihelion to the node, there is a first type of comets in stable motion, for which $g$ is always close to $\pm 90^\circ$, a second type, where $g$ is always close to $0^\circ$ or $180^\circ$, and finally a third type, where this angle $g$ is driven by a mean motion.'' (pp. Z346--Z347)
\end{quote}

\citeauthor{vonzeipel1910} also mentions the possible extension of
his theory to more general systems that contain more than one perturbers
whose orbits are nearly circular and planar:
\begin{quote}
``The results, which we have arrived [at] in the simple case, where there is only one disturbing planet, are also valid in the more general case, where an infinitely small mass is attracted to the Sun and a number of planets, whose masses, eccentricities and inclinations are small.'' (p. Z347)
\end{quote}

\citeauthor{vonzeipel1910} states the following theorem at the end of his ``\textit{Introduction.\/}''
Here he mentions the dependence of the limiting inclination
on the ratio of the semimajor axes, $\alpha$:
\begin{quote}
``The study of secular perturbations in this general case leads us to the following theorem:
\par
\hspace*{1em}
\textit{For the eccentricity of the orbit of the infinitely small mass, being small at some point, is still small, it is necessary and sufficient that the inclination is situated within certain limits, which are functions of the planetary masses and the ratio of the major axes.}
\par
\hspace*{1em}
\textit{
A slight resistance against the motion, which has the effect of reducing the major axis and eccentricity, also tends to establish limits for the inclination.
}'' (p. Z347)
\end{quote}

\subsection{Equations of motion of perturbed body\label{ssec:Z10eom}}
Following the \textit{Introduction,\/}
\citeauthor{vonzeipel1910} presents four major chapters
(Chapters I, II, III, IV) and a short, additional chapter
at its end (Chapter V). From the viewpoint of comparison between 
\citeauthor{vonzeipel1910}'s work and
\citeauthor{lidov1961}'s or \citeauthor{kozai1962b}'s work,
we would say that the most important aspects of \citeauthor{vonzeipel1910}'s work
are in its Chapter IV where he searches local extremums of
the doubly averaged disturbing function of CR3BP with specific forms.
Other chapters do not particularly deal with the disturbing function with specific forms.
\citeauthor{vonzeipel1910}'s Chapter I is placed for describing the general equations of motion for CR3BP.
In Chapter II \citeauthor{vonzeipel1910} develops general framework
to apply the Lindstedt series to the disturbing function.
Chapter III serves as a general study of the secular part of
the CR3BP disturbing function including the case of orbit intersection.
In what follows we summarize \citeauthor{vonzeipel1910}'s
Chapter   I in our Section  \ref{ssec:Z10eom},
Chapter  II in     Sections \ref{ssec:Z10Lindstedt} and \ref{ssec:Z10eom-secular},
Chapter III in     Section  \ref{ssec:R-general},
Chapter  IV in     Sections \ref{ssec:icr3bp}, \ref{ssec:ocr3bp}, \ref{ssec:orbitintersection}, and
Chapter   V in     Section  \ref{ssec:Z10-extention}.

\citeauthor{vonzeipel1910}'s first chapter
``\textit{Chapitre I. Equations diff\'erentielles du mouvement,\/}''
contains four sections (Z1--Z4).
This chapter is devoted to describing the equations of motion of
the perturbed body in CR3BP using the Hamiltonian formalism.
\citeauthor{vonzeipel1910} first describes the canonical equations of motion
using the ordinary Delaunay elements.
Then he introduces another set of equations of motion
using a set of Poincar\'e-like canonical variables.
At this point, no averaging is applied to any variables or equations.

\label{pg:vonzeipelsunitdef}
Throughout his paper,
the unit of length that \citeauthor{vonzeipel1910} uses is
the constant distance between Jupiter and the Sun.
The unit of mass is the total mass of Jupiter and the Sun.
The unit of time is defined so that the Gauss constant becomes 1.
In what follows,
$\mu$ denotes the mass of Jupiter, and
$H$ appearing in the disturbing function designates the angle between the
radius vectors of the perturbed body and that of the perturbing body at the Sun
(equivalent to the angle $S$ in our \mysymfigO \ref{fig:CR3BP-schematic}).
The resulting equations of motion are:
\begin{equation}
\begin{aligned}
  \DD{x}{t}  &= -\DP{\Phi}{x'}, &\tpspcE \DD{x'}{t} &=  \DP{\Phi}{x}, \\
  \DD{y}{t}  &= -\DP{\Phi}{y'}, &\tpspcE \DD{y'}{t} &=  \DP{\Phi}{y}, \\
  \DD{z}{t}  &= -\DP{\Phi}{z'}, &\tpspcE \DD{z'}{t} &=  \DP{\Phi}{z},
\end{aligned}
  \tag{Z01-\arabic{equation}}
  \stepcounter{equation}
  \label{eqn:Z01}
\end{equation}
with
\begin{equation}
\begin{aligned}
  \Phi & = -\frac{1}{2} \left( {x'}^2+{y'}^2+{z'}^2 \right) + \frac{1}{r} \\
       & \quad
         + \mu\left( \frac{1}{\sqrt{1-2r\cos H + r^2}}
                    -r \cos H - \frac{1}{r} \right) ,
  \label{eqn:vZ10-Psi-hamiltonian}
\end{aligned}
\end{equation}
where $x,y,z$ are the coordinates of the perturbed body,
   $x',y',z'$ are their time derivatives (i.e. velocity), and
$r$ is the radial distance of the perturbed body from the Sun.

Readers may find it slightly odd here to find a second term
$\left(+\frac{1}{r}\right)$ and the last one in the parentheses of the third term
$\left(-\frac{1}{r} \times \mu\right)$ in Eq. \eqref{eqn:vZ10-Psi-hamiltonian}.
Indeed from these two terms,
we can make up a new term $\frac{1-\mu}{r}$.
By the definition of the unit of mass, $1-\mu$ is the Sun's mass.
This new term can be considered as a potential that drives
the Keplerian motion of the perturbed body around the Sun.
However, the time average of $\frac{1}{r}$ becomes 
$\left< \frac{1}{r} \right> = \frac{1}{a}$,
which is a constant in the doubly averaged CR3BP.
Therefore, the terms proportional to $\frac{1}{r}$
in Eq. \eqref{eqn:vZ10-Psi-hamiltonian} merely serve as
additional constants, and they have no influence on the following discussions.

Next, \citeauthor{vonzeipel1910} introduces a set of canonical equations of
motion of the perturbed body using the Delaunay elements
$L, G, \Theta, l, g, \theta$.
Note that \citeauthor{vonzeipel1910} uses a variable $\theta$ for
longitude of ascending node, instead of the modern standard notation, $h$.
He also uses a variable $\Theta$ instead of the conventional notation, $H$.
The equations are:
\begin{equation}
\begin{aligned}
 \DD{L}{t} &= +\DP{\Phi}{l},           &\tpspcE \DD{l}{t} &= -\DP{\Phi}{L}, \\
 \DD{G}{t} &= +\DP{\Phi}{g},           &\tpspcE \DD{g}{t} &= -\DP{\Phi}{G}, \\
 \DD{\Theta}{t} &= +\DP{\Phi}{\theta}, &\tpspcE \DD{\theta}{t} &= -\DP{\Phi}{\Theta} ,
\end{aligned}
  \tag{Z02-\arabic{equation}}
  \stepcounter{equation}
  \label{eqn:Z02}
\end{equation}
where
\begin{alignat}{1} 
  \Phi   &= \Phi_0 + \mu \Phi_1,
  \label{eqn:vZ10-PsiAll} \\
  \Phi_0 &= \frac{1}{2L^2} ,
  \label{eqn:vZ10-Psi0} \\
  \Phi_1 &= \frac{1}{\Delta} -r \cos H -\frac{1}{r} .
  \label{eqn:vZ10-Psi1}
\end{alignat}

Definition of $\Delta$ in Eq. \eqref{eqn:vZ10-Psi1} somehow does not
show up in this section
(it shows up much later in Section Z13 in Chapter III, p. Z368).
But we see $\Delta$'s specific form in the third term in Eq. \eqref{eqn:vZ10-Psi-hamiltonian}:
it is the distance between the perturbing body and the perturbed body as follows:
\begin{equation}
  \Delta = \sqrt{1 - 2r\cos H + r^2} .
  \label{eqn:vZ10-defDelta}
\end{equation}

Since the length of the positional vector of the perturbing body $(r')$ is fixed to unity in \citeauthor{vonzeipel1910}'s work,
$\Delta$ in Eq. \eqref{eqn:vZ10-defDelta} looks slightly different
from what we usually see in the modern literature.
\citeauthor{vonzeipel1910}'s unit definition $(a'=1)$ thus
sometimes causes confusion.
We may rather want to understand that quantities concerning length are
implicitly normalized by Jupiter's semimajor axis
in \citeauthor{vonzeipel1910}'s descriptions,
such as $a = \frac{a}{a'}$ or $r = \frac{r}{a'}$.
\label{pg:vZ10-lengthsarenormalized}

Then, \citeauthor{vonzeipel1910} points out that
the longitude of the perturbing body is expressed as $\pm t$ in this system,
as is obvious from his definition of units.
Here
$+t$ corresponds to the prograde   orbital motion of the perturbing planet, and
$-t$ corresponds to the retrograde orbital motion.
As is well known, the disturbing function of this kind of system contains
$\pm t$  (for the perturbing body) and
$\theta$ (for the perturbed body)
only as the combination of $\theta - (\pm t) = \theta \mp t$
\citep[e.g.][]{brouwer1961,nakai1985}.
Therefore \citeauthor{vonzeipel1910} defines a new variable
$\theta' = \theta \mp t$
instead of $\theta$, and brings it into the equations of motion.
Let us cite \citeauthor{vonzeipel1910}'s original description:
\label{pg:def-thetadash}
\begin{quote}
``The angle $\theta$ and the longitude of the disturbing planet are reckoned
on the plane of the orbit of this planet in a sense that, at the nodes,
[they] make the acute angle $I$ with the direction of angles $l$ and $g$.
The longitude of the disturbing planet is represented by $\pm t$,
following whether this planet orbits in the direct way or in the indirect way.
\par
\hspace*{1em}
That being so, it is obvious that the disturbing function contains two
longitudes $\theta$ and $t$ in combination of $\theta \mp t$.
We consequently assume
$$
  \theta' = \theta \mp t .
$$
\par
\hspace*{1em}
Equations (Z02) still survives if we write $\theta'$ and $\Phi \pm \Theta$
instead of $\theta$ and $\Theta$.'' (pp. Z348--Z349)
\end{quote}

From the modern viewpoint,
the conversion from $\theta$ to $\theta'$ is equivalent to a
coordinate conversion of the inertial frame into the rotating frame of
the perturber's orbital motion \citep[e.g.][]{szebehely1967}.
This conversion generates an additional term in
the Hamiltonian of the system (such as $\pm x_2$ in $F_0$ in
Eq. \eqref{eqn:vZ10-F0}).
However, terms of this kind
would not affect the secular motion of the perturbed body at all.

Next in Section Z2,
\citeauthor{vonzeipel1910} introduces another series of canonical equations of
motion using a new set of canonical variables
$(x_1, x_2, x_3, y_1, y_2, y_3)$
that are a combination of the conventional Delaunay elements as:
\begin{equation}
\begin{aligned}
  x_1 &= L,       &\tpspcE  x_2 &= \Theta,  &\tpspcE   x_3 &= L - G, \\
  y_1 &= l + g,   &\tpspcE  y_2 &= \theta', &\tpspcE   y_3 &= -g, 
\end{aligned}
  \label{eqn:vZ10-xyzvars}
\end{equation}
together with
\begin{equation}
  \xi = \sqrt{2x_3} \cos y_3, \quad \eta = \sqrt{2x_3} \sin y_3 .
  \label{eqn:vZ10-xi+eta}
\end{equation}

The variables in Eq. \eqref{eqn:vZ10-xyzvars} and Eq. \eqref{eqn:vZ10-xi+eta}
satisfy the following canonical equations of motion:
\begin{equation}
\begin{aligned}
  \DD{x_\nu}{t} &=  \DP{F}{y_\nu}, &\tpspcE \DD{y_\nu}{t} &= -\DP{F}{x_\nu}, \\
  \DD{\xi }{t}  &=  \DP{F}{\eta},  &\tpspcE \DD{\eta}{t}  &= -\DP{F}{\xi} ,
\end{aligned}
  \tag{Z03-\arabic{equation}}
  \stepcounter{equation}
  \label{eqn:Z03}
\end{equation}
where $\nu = 1, 2$ with
\begin{alignat}{1}
  F   &= F_0 + \mu F_1,
 \label{eqn:vZ10-F-all} \\
  F_0 &= \frac{1}{2x_1^2} \pm x_2,
  \label{eqn:vZ10-F0} \\
  F_1 &= \frac{1}{\Delta} - r \cos H - \frac{1}{r} .
  \label{eqn:vZ10-F1}
\end{alignat}

He then states that
the perturbation Hamiltonian $F_1$ in Eq. \eqref{eqn:vZ10-F1}
can be formally expanded in a Fourier series as
\begin{equation}
\begin{aligned}
  F_1 &= \sum C_{m_1,m_2} \left(\xi,\eta\right) \cos \left(m_1 y_1 + m_2 y_2\right) \\
      &+ \sum S_{m_1,m_2} \left(\xi,\eta\right) \sin \left(m_1 y_1 + m_2 y_2\right) .
\end{aligned}
  \tag{Z04-\arabic{equation}}
  \stepcounter{equation}
  \label{eqn:Z04}
\end{equation}
Note that the coefficients $C_{m_1,m_2}$ and $S_{m_1,m_2}$ depend
also on the variables $x_1$ and $x_2$.

In Section Z3, \citeauthor{vonzeipel1910} introduces yet another set of
canonical variables $\left(x'_1, x'_2, x'_3, y'_1, y'_2, y'_3\right)$ as
\begin{equation}
\begin{aligned}
  x'_1 &= L,              &\tpspcD x'_2 &= L - \Theta,  &\tpspcD x'_3 &= G - \Theta, \\
  y'_1 &= l + g + \theta',&\tpspcD y'_2 &=-g - \theta', &\tpspcD y'_3 &= g,
\end{aligned}
  \label{eqn:vZ10-xyzvars-dash}
\end{equation}
together with
\begin{equation}
  \xi' = \sqrt{2x'_3} \cos y'_3, \quad
 \eta' = \sqrt{2x'_3} \sin y'_3 ,
  \label{eqn:vZ10-xi+eta-dash}
\end{equation}
and a relationship that shows their canonical equivalence
\begin{equation}
  x'_1 y'_1 + x'_2 y'_2 + x'_3 y'_3 = Ll + Gg + \Theta \theta' .
\end{equation}
The variables in Eqs. \eqref{eqn:vZ10-xyzvars-dash} and \eqref{eqn:vZ10-xi+eta-dash}
may come in handy when the orbital inclination of the perturbed body is small.
\citeauthor{vonzeipel1910} later uses them in his Section Z25, Chapter IV, p. Z413
(see also p. \pageref{sssec:motionneare=kd} of this monograph).
The variables $x'_1$, $x'_2$, $y'_1$, $y'_2$, $\xi'$, $\eta'$ meet the following equations of motion:
\begin{equation}
\begin{aligned}
  \DD{x'_\nu}{t} &=  \DP{F}{y'_\nu}, &\tpspcE \DD{y'_\nu}{t} &= -\DP{F}{x'_\nu}, \\
  \DD{\xi' }{t}  &=  \DP{F}{\eta'},  &\tpspcE \DD{\eta'}{t}  &= -\DP{F}{\xi' },
\end{aligned}
  \tag{Z05-\arabic{equation}}
  \stepcounter{equation}
\label{eqn:Z05}
\end{equation}
where $\nu = 1, 2$ with
\begin{alignat}{1}
  F   &= F_0 + \mu F_1,
\label{eqn:vZ10-F-dash} \\
  F_0 &= \frac{1}{2{x'}^2_1} \pm x'_1 \mp x'_2,
\label{eqn:vZ10-F0-dash} \\
  F_1 &= \frac{1}{\Delta} - r \cos H - \frac{1}{r} .
\label{eqn:vZ10-F1-dash}
\end{alignat}

Again, \citeauthor{vonzeipel1910} shows a Fourier-expanded
formal form of $F_1$ of Eq. \eqref{eqn:vZ10-F1-dash} as follows
\begin{equation}
\begin{aligned}
  F_1 &= \sum C'_{m'_1,m'_2} \left(\xi',\eta'\right) \cos \left(m'_1 y'_1 + m'_2 y'_2\right) \\
      &+ \sum S'_{m'_1,m'_2} \left(\xi',\eta'\right) \sin \left(m'_1 y'_1 + m'_2 y'_2\right) .
\end{aligned}
  \tag{Z06-\arabic{equation}}
  \label{eqn:Z06}
\end{equation}
\label{pg:def-xidashetadash}

Section Z4 is devoted to describing some symmetric characteristics
of the coefficients
$C_{m_1,m_2}$, $S_{m_1,m_2}$, $C'_{m'_1,m'_2}$, $S'_{m'_1,m'_2}$
seen in Eqs. \eqref{eqn:Z04} and \eqref{eqn:Z06},
although we do not reproduce them here.
Section Z4 also contains several conversion formulas between the
variables defined in Sections Z2--Z3 and the conventional Kepler orbital
elements $a$, $e$, $I$, $g$
(we do not reproduce them here. See Eqs. (Z07) on p. Z351).
Incidentally, note that \citeauthor{vonzeipel1910} uses the notation $I$
for orbital inclination, not $i$.

\subsection{The Lindstedt series\label{ssec:Z10Lindstedt}}
In his next chapter entitled
``\textit{Chapitre II. Calcul des s\'eries de M. Lindstedt,\/}''
\citeauthor{vonzeipel1910} brings the Lindstedt series into
the equations of motion described in his Chapter I.
Eventually, a set of secular equations of motion with just one degree
of freedom is derived.
This chapter contains eight sections (Z5--Z12), and its latter part is
filled with detailed mathematical expositions of the Lindstedt series
without specifying the function form of the disturbing function.
The structure of this chapter is as follows.
\citeauthor{vonzeipel1910} first assumes
that the secular disturbing function $R$ has a local minimum or maximum
somewhere in phase space. Then, he gives a general proof that
it is possible to formulate a periodic solution around
the local minimum or maximum using the Lindstedt series.
\citeauthor{vonzeipel1910} does not specify the function form of
the disturbing function throughout the procedure,
so his result applies to the general CR3BP.
The validity of the assumption---whether or not $R$ has a local minimum or maximum other than at the origin $(\xi,\eta)=(0,0)$, and if it does,
where the local extremums are---is discussed later in his Chapters III and IV.
Readers who want to quickly see \citeauthor{vonzeipel1910}'s major conclusion
on the existence of periodic solutions in CR3BP that can be
directly compared with \citeauthor{lidov1961}'s and
\citeauthor{kozai1962b}'s work might want to skip
Sections \ref{ssec:Z10Lindstedt}, 
         \ref{ssec:Z10eom-secular},
         \ref{ssec:R-general}, and jump to
Section  \ref{ssec:icr3bp} or \ref{ssec:ocr3bp}
where we summarize the contents of \citeauthor{vonzeipel1910}'s Chapter IV.

Before we go into \citeauthor{vonzeipel1910}'s Chapter II,
let us briefly review what the Lindstedt series is.
The study of this series began with a publication by a Swedish scientist,
Anders Lindstedt (1854--1939).
It is a short paper of less than five pages, entitled
``{\"U}eber die Integration einer f{\"u}r die St{\"o}rungstheorie wichtigen Differentialgleichung,''
published in \textit{Astronomische Nachrichten\/} \citep{lindstedt1882a}.
This work was immediately followed by a series of
publications by \citeauthor{lindstedt1882a} himself
\citep[e.g.][]{lindstedt1882b,lindstedt1883a,lindstedt1884} and others, such as
\citet{poincare1886},
\citet{bohlin1888}, and
\citet{gylden1891,gylden1893}.
See also \citet[][p. 102]{garding1998} for the interrelationships of these authors.
Eventually \citeauthor{poincare1892} organized and published the theoretical method \citep{poincare1892},
and the method is now also known as
the Lindstedt--Poincar{\'e} method \citep[e.g.][]{amore2005,yu2017} or
the Poincar{\'e}--Lindstedt method \citep[e.g.][]{khrustalev2001}.
Readers can consult
\citet[][their Section 6.6 on p. 33]{boccaletti1998} or
\citet[][their Chapter 2 on p. 9]{marinca2011}
for brief but excellent reviews about what the Lindstedt series is,
as well as how it was developed in the history of perturbation theory studies.
As for the interrelationship of the Lindstedt series to Poincar{\'e}'s work,
a long review \citep{chenciner2015} in a book \citep{duplantier2015} is
very thorough and worth a read.

The general objective to employ the Lindstedt series in this line of study is
to obtain a time-dependent solution of an autonomous conservative
dynamical system
without generating the so-called artificial mixed secular terms.
For example, consider a one-dimensional system controlled by
a variable $\varrho$ described by the differential equation
\begin{equation}
  \DD[2]{\varrho}{t} + \varrho = \epsilon f(\varrho,\epsilon) ,
  \label{eqn:Lindstedt-example}
\end{equation}
with a small constant parameter $\epsilon$.
$f$ is a function of $\varrho$ including $\epsilon$.
In the original publication by \citeauthor{lindstedt1882a},
$f$ in Eq. \eqref{eqn:Lindstedt-example} shows up
as an expanded form in an infinite series from the beginning.
In recent times,
the so-called simplified Duffing equation is often employed
for explaining how the Lindstedt series works, putting
$f = - \epsilon \varrho^3 $ \citep[e.g.][]{nayfeh1973,grimshaw1991,bush1992}.
In any case,
we want to obtain the time-dependent solution $\varrho(t)$ of
Eq. \eqref{eqn:Lindstedt-example} that is free of mixed secular terms.
For this purpose, the system's frequency $\omega$ is expanded
using the small parameter $\epsilon$ as
\begin{equation}
  \omega = 1 + \epsilon \omega_1 + \epsilon^2 \omega_2 + \epsilon^3 \omega_3 + \cdots .
  \label{eqn:DuffingEq-omega}
\end{equation}
Using the expansion \eqref{eqn:DuffingEq-omega}
the time variable $t$ can be ``stretched'' to a new variable $\tau$ as
\begin{equation}
   \tau = \omega t .
  \label{eqn:DuffingEq-tau}
\end{equation}
The usual expansion of the variable $\varrho$ by $\epsilon$ becomes
\begin{equation}
  \varrho = \varrho_0 + \epsilon \varrho_1 + \epsilon^2 \varrho_2 + \epsilon^3 \varrho_3 + \cdots ,
  \label{eqn:DuffingEq-q}
\end{equation}
and a Taylor-expansion of $f$ by $\epsilon$ is also necessary.
Substituting these expansions
into the original equation \eqref{eqn:Lindstedt-example},
a series of differential equations for each order of
$O(\epsilon^0)$, $O(\epsilon^1)$, $O(\epsilon^2)$, $\ldots$, are obtained.

The solution of the zero-th order $(\varrho_0)$ is often trivially obtained
without difficulty.
When solving the differential equation of the next order $O(\epsilon^1)$,
we carefully determine $\omega_1$ so that no mixed secular term shows up
in the first-order solution $(\varrho_1)$.
Repeating similar procedures, we can principally obtain asymptotic solutions
$\varrho_2, \varrho_3, \ldots ,$ together with the frequency components
$\omega_1, \omega_2, \ldots ,$
without any mixed secular terms appearing in the final form.

Nowadays, the use of the Lindstedt series is so common in modern
celestial mechanics that people may be unaware of using it
even when the series is embedded in the perturbation method that is employed.
As an example to illustrate this situation, 
let us quote from \citet{mardling2013} that mentioned
\citeauthor{hori1966}'s \citeyearpar{hori1966}
famous canonical perturbation theory:
\begin{quote}
``In fact, Hori's averaging process is simply a version of the better-known
 \textit{Lindstedt--Poincar\'e\/} method for correcting the frequencies of
  a forced non-linear oscillator.'' (the third paragraph of p. 2189)
\end{quote}
In passing, let us mention that the Lindstedt series is often discussed intensively in connection with the KAM (Kolmogorov--Arnold--Moser) theorem
\citep[e.g.][]{lichtenberg1992,gentile1996,jorba1999,merritt1999}.

\subsection{Secular equations of motion\label{ssec:Z10eom-secular}}
In his Chapter II,
\citeauthor{vonzeipel1910} formally exploits the Lindstedt series
in order to derive the secular equations of motion of the perturbed body.
On the path where he follows the method of Lindstedt,
the small parameter that he uses for the expansions is Jupiter's mass, $\mu$.
In his Section Z5 (pp. Z351--Z353),
\citeauthor{vonzeipel1910} expands the orbital variables defined in
Eqs. \eqref{eqn:vZ10-xyzvars} and \eqref{eqn:vZ10-xi+eta} by $\mu$ as follows:
\begin{equation}
\begin{aligned}
  x_\nu &= \;\;\;\;\;\;\;\; x^0_\nu + \mu x^1_\nu + \mu^2 x^2_\nu + \cdots, \\
  y_\nu &= w_\nu + y^0_\nu + \mu y^1_\nu + \mu^2 y^2_\nu + \cdots, \\
  \xi   &= \;\;\;\;\;\;\;\; \xi^0  + \mu \xi^1   + \mu^2 \xi^2   + \cdots, \\
  \eta  &= \;\;\;\;\;\;\;\; \eta^0 + \mu \eta^1  + \mu^2 \eta^2  + \cdots,
\end{aligned}
  \tag{Z08-\arabic{equation}}
  \stepcounter{equation}
  \label{eqn:Z08}
\end{equation}
where $\nu=1,2$.
Note that not all the superscripts $(1, 2, \cdots)$ in Eq. \eqref{eqn:Z08}
denote powers of the variable.
For example, $x_\nu^2 \neq (x_\nu)^2$, $\eta^2 \neq (\eta)^2$.
The superscripted variables such as $x_0^1, y_1^2, \xi^1$, $\eta^2$, are
new, independent variables except for $\mu$ (e.g. $\mu^2 \equiv (\mu)^2$).
This is a confusing notation which is used throughout \citet{vonzeipel1910}.

Now, \citeauthor{vonzeipel1910} states that the variables
$x_\nu^i$, $y_\nu^i$, $\xi^i$, $\eta^i$  $(i=1, 2, \cdots)$
expressed as Eq. \eqref{eqn:Z08} are Fourier series
with the following three arguments
\begin{alignat}{1}
  w_1 &= nt + c_1,
  \label{eqn:vZ10-def-w-w1-w2=w1} \\
  w_2 &= (q \mp 1) t + c_2,
  \label{eqn:vZ10-def-w-w1-w2=w2} \\
  w   &= \sigma t + c .
  \label{eqn:vZ10-def-w-w1-w2=w}
\end{alignat}

As we will see shortly, each of $w_1$, $w_2$, $w$ implies different timescales
that the system possesses.
The quantities $c_1$, $c_2$, $c$
are arbitrary constants, and $n$ is a constant independent of $\mu$.
He expands the coefficients $q$ and $\sigma$ as follows:
\begin{equation}
  \begin{aligned}
    q       &= \mu q^1      + \mu^2 q^2      + \cdots , \\
    \sigma  &= \mu \sigma^1 + \mu^2 \sigma^2 + \cdots .
  \end{aligned}
  \tag{Z09-\arabic{equation}}
  \stepcounter{equation}
  \label{eqn:Z09}
\end{equation}
Note again that $q^2 \neq (q)^2$ and $\sigma^2 \neq (\sigma)^2$.

\citeauthor{vonzeipel1910} then determines the value of unknown variables
by substituting the above expansions into the equations of motion
\eqref{eqn:Z03}, and equating the members of the same order with
respect to $\mu$---the standard procedure when employing the Lindstedt series.
As a result, the zero-th order terms give the following equations:
\begin{equation}
\begin{aligned}
    n \DP{x_1^0}{w_1} \mp \DP{x_1^0}{w_2}    &= 0, \\
    n \DP{x_2^0}{w_1} \mp \DP{x_2^0}{w_2}    &= 0, \\
n + n \DP{y_1^0}{w_1} \mp \DP{y_1^0}{w_2}    &= \frac{1}{\left(x_1^{0}\right)^3}, \\
\mp 1+ n \DP{y_2^0}{w_1} \mp \DP{y_2^0}{w_2} &= \mp 1, \\
    n \DP{\xi^0}{w_1}  \mp \DP{\xi^0}{w_2}   &= 0, \\
    n \DP{\eta^0}{w_1} \mp \DP{\eta^0}{w_2}  &= 0 .
  \label{eqn:vZ10-lindstedt-zeroth}
\end{aligned}
\end{equation}

At this point, \citeauthor{vonzeipel1910} brings up two assumptions (p. Z352).
Assumption (i): $x_1^0$ is a constant so that
\begin{equation}
  n = \left(x_1^0\right)^{-3} .
  \label{eqn:vZ1910-p352-noname02}
\end{equation}
By its definition in Eq. \eqref{eqn:vZ10-xyzvars},
we know that $x_1$ is just a function of semimajor axis $a$ of the perturbed body.
And, as we are already well aware of,
$a$ of the perturbed body is one of the conserved quantities
in the doubly averaged CR3BP.
Hence we believe this assumption is valid.
Assumption (ii):
The zero-th order part of each of the variables
$x_2^0$, $y_1^0$, $y_2^0$, $\xi^0$, $\eta^0$ are functions just of $w$,
not of $w_1$ or $w_2$.
To check whether this assumption is valid or not,
we need to give consideration to the timescales that each of
$w$, $w_1$, $w_2$ has.
As for $w_1$,
from Eq. \eqref{eqn:vZ10-def-w-w1-w2=w1},
     Eq. \eqref{eqn:vZ1910-p352-noname02}, and the assumption (i)
we know its variation timescale is as follows:
\begin{equation}
  \frac{1}{n} = \left(x_1^0\right)^3 \sim L^3 .
  \label{eqn:vZ1910-timescale-w1}
\end{equation}
We can regard that the dimension of this quantity is $O(1)$, since
there is no small parameter $\mu$ included in
$x^0_\nu$ in Eq. \eqref{eqn:Z08}.
Now it is clear that $n$ in Eq. \eqref{eqn:vZ10-def-w-w1-w2=w1} has
a meaning of the mean motion of the perturbed body, and
the timescale $n^{-1}$ is related to its orbital period.

As for $w_2$,
the timescale of $w_2$ is $\frac{1}{q \mp 1}$ from the definition
\eqref{eqn:vZ10-def-w-w1-w2=w2}.
And, since $q$ is a small quantity of $O\left(\mu\right)$ by Eq. \eqref{eqn:Z09},
we find the timescale of $w_2$ close to 1.
We can consider that the quantity $\frac{1}{q \mp 1}$ has a relation to the mean motion of Jupiter,
because Jupiter's mean motion $n_{\rm J}$ is $1$ as follows.
First, from Kepler's third law we have
\begin{equation}
  n_{\rm J}^2 a_{\rm J}^3 = {\cal G} \left( m_{\rm Sun} + \mu\right) ,
  \label{eqn:kepler3rd-zeipel}
\end{equation}
where $m_{\rm Sun}$ is the Sun's mass.
However, both ${\cal G}$ and Jupiter's semimajor axis $a_{\rm J}$
are set to 1 by \citeauthor{vonzeipel1910}'s definition
(see p. \pageref{pg:vonzeipelsunitdef} of this monograph).
Also, the sum of the solar mass and Jupiter's mass $(m_{\rm Sun} + \mu)$ is
set to 1 by his definition.
These facts yield $n_{\rm J} = 1$ through Eq. \eqref{eqn:kepler3rd-zeipel}.

On the other hand, the timescale of $w$ is $\sigma^{-1}$
by Eq. \eqref{eqn:vZ10-def-w-w1-w2=w}.
This is a quantity of $O\left(\mu^{-1}\right)$ according to Eq. \eqref{eqn:Z09}.
This means that the variable $w$ changes much more slowly than $w_1$ or $w_2$.
The variation timescale of $w$ is related to the perturbation from Jupiter
on the perturbed body, and it would be infinitely long when $\mu \to 0$.
Therefore we can say that the variable $w$ describes 
the secular orbital change of the perturbed body.
Thus, \citeauthor{vonzeipel1910}'s assumption (ii) indicates that
the zero-th order quantities $x_2^0$, $y_1^0$, $y_2^0$, $\xi^0$, $\eta^0$ are
slow-oscillating variables having a timescale of $O\left(\mu^{-1}\right)$.
In the doubly averaged CR3BP where the motion of the perturbed body is
just under weak to moderate perturbation as \citeauthor{vonzeipel1910} considers, this assumption is justified.

\citeauthor{vonzeipel1910} now moves on to the first-order solution.
By equating the terms of $O\left(\mu^1\right)$ appearing in the expanded Eq. \eqref{eqn:Z03},
he obtains the following set of equations:
\begin{equation}
\begin{aligned}
    n \DP{x_1^1}{w_1} \mp \DP{x_1^1}{w_2}                         &=  \DP{F_1^0}{y_1^0}, \\
    n \DP{x_2^1}{w_1} \mp \DP{x_2^1}{w_2} +\sigma^1 \DP{x_2^0}{w} &=  \DP{F_1^0}{y_2^0}, \\
    n \DP{y_1^1}{w_1} \mp \DP{y_1^1}{w_2} +\sigma^1 \DP{y_1^0}{w} &= -\frac{3}{\left(x_1^0\right)^4} x_1^1 - \DP{F_1^0}{x_1^0}, \\
q^1+n \DP{y_2^1}{w_1} \mp \DP{y_2^1}{w_2} +\sigma^1 \DP{y_2^0}{w} &= -\DP{F_1^0}{x_2^0}, \\
    n \DP{\xi^1}{w_1}  \mp \DP{\xi^1}{w_2}  + \sigma_1 \DD{\xi^0}{w}  &=  \DP{F_1^0}{\eta^0}, \\
    n \DP{\eta^1}{w_1} \mp \DP{\eta^1}{w_2} + \sigma_1 \DD{\eta^0}{w} &= -\DP{F_1^0}{\xi^0},
\end{aligned}
  \tag{Z10-\arabic{equation}}
  \stepcounter{equation}
  \label{eqn:Z10}
\end{equation}
where $F_1^0$ is the zero-th order part of $F_1$ in Eq. \eqref{eqn:Z04}:
\begin{equation}
\begin{aligned}
F_1^0
  & = \sum C_{m_1,m_2}\left(x_1^0,x_2^0,\xi^0,\eta^0\right)  \\
  & \qquad \times \cos \left[m_1\left(w_1+y_1^0\right)+m_2\left(w_2+y_2^0\right)\right] \\
  & + \sum S_{m_1,m_2}\left(x_1^0,x_2^0,\xi^0,\eta^0\right) \\
  & \qquad \times \sin \left[m_1\left(w_1+y_1^0\right)+m_2\left(w_2+y_2^0\right)\right] .
\end{aligned}
\label{eqn:vZ10-F1-zeroth}
\end{equation}

From the function form of $F_1^0$ in Eq. \eqref{eqn:vZ10-F1-zeroth}, and
from the above mentioned two assumptions (i) and (ii),
we can say that $F_1^0$ is a periodic function of
the fast-oscillating variables $w_1$ and $w_2$.
In addition, \citeauthor{vonzeipel1910} states as follows:
\begin{quote}
``Since $x_\nu^1$, $y_\nu^1$, $\xi^1$, $\eta^1$, $F_1^0$ are periodic
  functions of $w_1$ and $w_2$ of period $2\pi$, the derivatives of
  these functions with respect to $w_1$ or $w_2$ contain no term
  independent of $w_1$ and of $w_2$.'' (p. Z352)
\end{quote}

We understand that the above statement is a common assumption to make
when exploiting the Lindstedt series:
The zero-th order part of the variables are functions only of $w$ (with a slow variation),
while their higher-order parts are functions only of $w_1$ and $w_2$ (with a fast variation).
Then, according to the standard procedure when using the Lindstedt series,
\citeauthor{vonzeipel1910} equates the terms that do not contain
$w_1$ or $w_2$ in Eq. \eqref{eqn:Z10}.
As a consequence, he obtains the following set of equations:
\begin{equation}
\begin{aligned}
    \sigma^1 \DD{x_2^0}{w}  &= 0, \\
    \sigma^1 \DD{y_1^0}{w}  &= -\frac{3}{\left(x_1^0\right)^4}[x_1^1] - \DP{R}{x_1^0}, \\
q^1+\sigma^1 \DD{y_2^0}{w}  &= -\DP{R}{x_2^0},  \\
    \sigma^1 \DD{\xi^0}{w}  &=  \DP{R}{\eta^0}, \\
    \sigma^1 \DD{\eta^0}{w} &= -\DP{R}{\xi^0},
\end{aligned}
  \tag{Z11-\arabic{equation}}
  \stepcounter{equation}
  \label{eqn:Z11}
\end{equation}
where $[x_1^1]$ denotes the secular part of $x_1^1$ that is independent of $w_1$ or $w_2$.
$R$ is defined as
\begin{equation}
  R = C_{0,0} \left( x_1^0, x_2^0, \xi^0, \eta^0 \right) ,
 \label{eqn:vZ10-def-R}
\end{equation}
and it is nothing but the secular disturbing function.
The substitution of $m_1 = m_2 = 0$ into Eq. \eqref{eqn:vZ10-F1-zeroth} yields this.

It is evident that $x_2^0$ is a constant from the first equation in \eqref{eqn:Z11},
$\sigma^1 \DD{x_2^0}{w} = 0$.
It is also important to recall that $x_2^0$ is defined as the secular part
of one of the Delaunay elements, $\Theta$,
which is proportional to $\sqrt{1-e^2} \cos I$
(see Eqs. \eqref{eqn:vZ10-xyzvars} and \eqref{eqn:Z08} for the definitions).

So far, \citeauthor{vonzeipel1910} has found two constants of integration
($x_1^0$ and $x_2^0$) in the considered system.
In addition, we should recall that the secular disturbing function $R$
in Eq. \eqref{eqn:vZ10-def-R} is also a constant due to the conservative
characteristics of mutual potentials between the three bodies.
As a result, the degrees of the system's freedom can be reduced to one, and
the system becomes integrable.
The equations of motion for the considered system with one degree of freedom
are the last two in Eq. \eqref{eqn:Z11}:
\begin{equation}
  \sigma^1 \DD{\xi^0}{w}  =  \DP{R}{\eta^0}, \quad
  \sigma^1 \DD{\eta^0}{w} = -\DP{R}{\xi^0},
  \tag{Z12-\arabic{equation}}
  \stepcounter{equation}
  \label{eqn:Z12}
\end{equation}
with the secular disturbing function $R$.

In the short Section Z6 (pp. Z353--Z354),
\citeauthor{vonzeipel1910} repeats what he has already said. He writes:
\begin{quote}
``The integration of this system is theoretically very simple.
It may, indeed, be carried out by a quadrature under the integral
$$
  R = \mbox{const.}
$$
\hspace{1em}
However in practice,
it is difficult to integrate equations (Z12) in any generality,
since the function $R$ is very complicated.
But fortunately there are extended cases where the integration is even
practically simple enough.
This happens if the original values of $\xi^0$ and $\eta^0$ are
in the vicinity of a maximum or minimum value of $R$
regarded as [a] function of $\xi^0$ and of $\eta^0$.

\hspace{1em}
It is well known that $R \left(x_1^0, x_2^0, \xi, \eta \right)$, regarded as a function
of $\xi$ and $\eta$, possesses a minimum value for $\xi=\eta=0$ at least
for small inclinations, that is to say, if the ratio $x_1^0 : x_2^0$ is
close to unity.
This minimum corresponds to series of the form (Z08),
which represents the motion of asteroids with small eccentricities
and small inclinations.'' (p. Z353)
\end{quote}

\citeauthor{vonzeipel1910} then makes a quick summary of
what will be discussed in the remaining part of Chapter II:
Calculation of possible periodic orbits of comets using the Lindstedt series
around new local extremums of $R$ that were not known at his time.
Let us cite his words again:
\begin{quote}
``A thorough study of the function $R$, to which we will dedicate Chapter IV,
first shows that the function $R$ has, for all the values of the
parameters $x_1^0$ and $x_2^0$ in the domain $0 < x_2^0 < x_1^0$,
a minimum value at the point $\xi = \eta = 0$.
However, this study shows on the other hand that the function $R$ possesses
other minima that are yet unknown.
The positions of these new minima depend on the values of the parameters
$x_1^0$ and $x_2^0$.
They are located either on the axis of $\xi$ or on the axis of $\eta$,
so that the line of [cometary] apsides coincides with, or is perpendicular to,
the line of nodes, in the corresponding orbits.
These new minimum values correspond, as we will show,
to the series of the form (Z08) that is easy to calculate, representing
the motion of comets, whose perihelia are approximately fixed relative
to the nodes.'' (pp. Z353--Z354)
\end{quote}

At the end of Section Z6
\citeauthor{vonzeipel1910} mentions a possible periodic cometary orbit
with zero inclination as follows:
\begin{quote}
``Finally,
we also show that the function $R$ sometimes reaches a maximum value,
if the inclination is zero.
In the vicinity of this maximum there exist series similar to the series (Z08) of a very simple form, which can be applied to the study of the motion of comets, whose orbits are always only slightly inclined and whose perihelia turn under secular perturbations.'' (p. Z354)
\end{quote}

\citeauthor{vonzeipel1910} spends the rest of Chapter II
(Sections Z7--Z12) on extensive calculations that aim at applying
the Lindstedt series to the equations of motion that have been derived so far.
As we previously mentioned,
the function form of the disturbing function is not specified here.
Therefore
\citeauthor{vonzeipel1910}'s result here is applicable to general, arbitrary cases of doubly averaged CR3BP,
irrespective whether it is the inner problem or the outer problem.

In Section Z7,
\citeauthor{vonzeipel1910} considers a general case
when $R$ has a local maximum or minimum at $(\xi,\eta) = (\xi^{0.0}, 0)$,
where $\xi^{0.0}$ is the main term of $\xi^0$
when $\sigma^1$, $\xi^0$, $\eta^0$ are expanded using the Lindstedt series as
\begin{equation}
\begin{aligned}
  \sigma^1 &= \sigma^{1.0} +\varepsilon^2 \sigma^{1.2} +\varepsilon^4 \sigma^{1.4} + \cdots , \\
  \xi^0    &=    \xi^{0.0} +\varepsilon   \xi^{0.1}    +\varepsilon^2 \xi^{0.2} + \cdots , \\
  \eta^0   &= \quad\quad\quad
                            \varepsilon   \eta^{0.1}   +\varepsilon^2 \eta^{0.2} + \cdots .
\end{aligned}
  \tag{Z13-\arabic{equation}}
  \stepcounter{equation}
  \label{eqn:Z13}
\end{equation}
He shows that in this case it is possible to
construct approximate periodic solutions for $\xi^0$ and $\eta^0$.

In Section Z8,
\citeauthor{vonzeipel1910} further moves on to a general procedure to obtain
$q^1$ and $y_2^0$ that are related to the longitude of ascending node 
$(\theta)$ of the perturbed body (comet).
In Section Z9,
he shows a general demonstration in order to prove
the theoretical possibility of successively determining all the coefficients
that appear in the Lindstedt series \eqref{eqn:Z08} and \eqref{eqn:Z09}
up to any orders under the assumption stated in Sections Z7 and Z8.
The resulting solution indicates that argument of perihelion of
the perturbed body librates with a small amplitude around $g=0$ or $\pi$ 
if $R$ has a local extremum at the point $(\xi,\eta) = (\xi^{0.0}, 0)$.

Section Z10 goes to a similar exposition as in Section Z7.
\citeauthor{vonzeipel1910} shows the existence of the Lindstedt series
when $R$ has a local extremum at $(\xi,\eta) = (0, \eta^{0.0})$.
Here, $\eta^{0.0}$ is the main term of $\eta$ when it is expanded as
\begin{equation}
\begin{aligned}
  \xi^0    &= \quad\quad\quad
                           \varepsilon \xi^{0.1} +\varepsilon^2 \xi^{0.2} + \cdots , \\
  \eta^0   &=  \eta^{0.0} +\varepsilon \eta^{0.1} +\varepsilon^2 \eta^{0.2} + \cdots .
\end{aligned}
  \label{eqn:Z13rev-eta}
\end{equation}
The resulting solution indicates that the comet's argument of perihelion
librates with a small amplitude around $g=\pm \frac{\pi}{2}$ in this case.
Note that \citeauthor{vonzeipel1910} did not explicitly show the expansion forms of Eq. \eqref{eqn:Z13rev-eta}.

\label{pg:SectionZ11}
In Section Z11,
\citeauthor{vonzeipel1910} describes the case when $R$ has a maximum or minimum
at $\xi = \eta = 0$.
In this case, the comet's $g$ has a mean motion, i.e. $g$ circulates from 0 to $2\pi$.
In Section Z12, \citeauthor{vonzeipel1910} mentions a special case when the cometary inclination becomes zero.
In this case the comet's $g$ exhibits circulation from 0 to $2 \pi$.

Considering \citeauthor{vonzeipel1910}'s viewpoint and purpose,
we can say that the existence of the Lindstedt series
in the problem that he dealt with is practically
equivalent to the theoretical possibility of constructing approximate,
periodic solutions of the cometary motion.
His aim in his Chapter II is to quantitatively
demonstrate that the Lindstedt series
exists around local extremums of the secular disturbing function $R$ of the
doubly averaged CR3BP,
even when the perturbed body's eccentricity or inclination is substantially large.
Nowadays, we are well aware of this result through the works of
\citeauthor{lidov1961} and \citeauthor{kozai1962b}
who rigorously showed the existence of periodic solutions around
the local extremums of $R$.
But this fact was not obvious in \citeauthor{vonzeipel1910}'s era.

In modern celestial mechanics,
the Lindstedt series is categorized as a method employed in
classical perturbation theory, rather than in canonical perturbation theory
\citep[e.g.][]{boccaletti1998}.
Since \citeauthor{vonzeipel1910}'s discussion began with
the canonical equations of motion
(see Eq. \eqref{eqn:Z01} or Eq. \eqref{eqn:Z02}),
introducing the Lindstedt series might seem odd in the modern viewpoint.
However, several years later \citeauthor{vonzeipel1910} devised
the canonical version of his perturbation method,
which is now called the \citeauthor{vonzeipel1910} method.
As is well recognized, his accomplishment was published
as a series of publications over more than 300 pages in total
\citep{vonzeipel1916a,vonzeipel1916b,vonzeipel1917a,vonzeipel1917b}.
In this regard, the detailed mathematical expositions described in
\citeauthor{vonzeipel1910}'s Chapter II can be regarded as a prototype study
that was eventually developed into the canonical \citeauthor{vonzeipel1910} method.
Note, however, that our strongest interest in this monograph does not lie
in the detailed mathematical expositions presented in
\citeauthor{vonzeipel1910}'s Chapter II.
Our interest is focused on what he achieved in his Chapter IV
which has a direct relevance to the later studies by
\citeauthor{lidov1961} or \citeauthor{kozai1962b}.
Therefore we do not go into \citeauthor{vonzeipel1910}'s Chapter II any further.
\label{pg:fromprototypetocomplete}

\subsection{Secular disturbing function: General case\label{ssec:R-general}}
\citeauthor{vonzeipel1910}'s next chapter is entitled
``\textit{Chapitre III. \'Etude de la partie s\'eculaire de la fonction perturbatrice,\/}'' containing three sections (Z13--Z15).
Here he describes the general characteristics of
the secular disturbing function for CR3BP.
His particular emphasis lies on the relationship between the
function form of the disturbing function and its dependence on the
relative orbital configuration of the perturbing and perturbed bodies.
\citeauthor{vonzeipel1910}'s treatment of the disturbing function still
remains rather general in this chapter, and does not deal with any kind of
specific expansion of the disturbing function by a series.
This means that his result in this chapter is applicable to general cases
of doubly averaged CR3BP,
irrespective whether it is the inner or the outer problem.

The first section in this chapter (Section Z13) seems a preparation for the rest of the chapter.
First, he introduces an important parameter $k$ as follows:
\begin{equation}
  \frac{x_2^0}{x_1^0} = \sqrt{1-e^2} \cos I = k .
  \tag{Z42-\arabic{equation}}
  \stepcounter{equation}
  \label{eqn:Z42}
\end{equation}

The parameter $k$ is constant because both $x_1^0$ and $x_2^0$ are constant
in \citeauthor{vonzeipel1910}'s approximation
(see Eqs. \eqref{eqn:vZ1910-p352-noname02} and \eqref{eqn:Z11}).
Later in this monograph,
$k$ in Eq. \eqref{eqn:Z42} turns out to be a key parameter which determines
the fundamental characteristics of the doubly averaged disturbing function.
We can express $k$ using \citeauthor{moiseev1945a}'s $C$ in Eq. \eqref{eqn:Mb45-2.14}
as $k = \frac{C}{\sqrt{a}}$.
Since the perturbed body's semimajor axis $a$ is constant in the considered system,
it means that $k$ is only proportional to $C$.
And, $k^2 = \left(1-e^2\right) \cos^2 I$ is equivalent to
\citeauthor{kozai1962b}'s $\Theta$ in Eq. \eqref{eqn:K26} as well as to
\citeauthor{lidov1961}'s   $c_1$   in Eq. \eqref{eqn:L58}.
Since \citeauthor{vonzeipel1910} assumes that the perturbed body's inclination $I$ is not dull
(i.e. $I \leq \frac{\pi}{2}$), both $k$ and $k^2$ range between 0 and 1.

From its definition in Eq. \eqref{eqn:vZ10-def-R},
we know that the secular disturbing function $R$ is a function of
$x_1^0$, $x_2^0$, $\xi^0$, $\eta^0$ only.
And, from the definitions of $x_1$, $x_2$, $\xi$, $\eta$ in
Eqs. \eqref{eqn:vZ10-xyzvars} and \eqref{eqn:vZ10-xi+eta},
$R$ turns out to be a function of $a^0$, $e^0$, $I^0$, and $g^0$.
They are the zero-th order quantities of $a$, $e$, $I$, $g$
when they are expanded in the Lindstedt series with $\mu$.
But we already know that semimajor axis $a$ becomes a constant in the secular system that we consider.
Also, $e$ and $I$ depend on each other through the constant parameter $k$ defined in Eq. \eqref{eqn:Z42}.
We then know that $R$ is a function just of $e$ and $g$ with two constant parameters, $a$ and $k$.
\citeauthor{vonzeipel1910} points out that the range of $e$ is as follows:
\begin{equation}
  0 \leq g \leq 2\pi, \quad
  0 \leq e \leq k' \equiv \sqrt{1-k^2} ,
  \tag{Z43-\arabic{equation}}
  \stepcounter{equation}
  \label{eqn:Z43}
\end{equation}
where $k'$ is the largest value of $e$ when $\cos I = 1$.
Since $R = R(e,g)$ itself is a constant of motion
(see the discussion following Eq. \eqref{eqn:vZ10-def-R}),
$R$ becomes independent from $g$ when $e = k'$;
in other words, when $e$ is a constant.
\label{pg:border-kdash}

The secular disturbing function $R$ is formally obtained by averaging
the perturbation Hamiltonian $F_1$ in Eq. \eqref{eqn:vZ10-F1}.
\citeauthor{vonzeipel1910} points out that
$F_1$'s second term $\left(-r \cos H\right)$ does not leave any secular component
after the averaging procedure.
Averaging $F_1$'s third term $\left(-\frac{1}{r}\right)$ just leaves a constant, $-\frac{1}{a}$.
Hence he expresses the secular disturbing function $R$
just by using the first term in Eq. \eqref{eqn:vZ10-F1} as
\begin{equation}
\begin{aligned}
  R  & = \frac{1}{4\pi^2} \int_0^{2\pi} \int_0^{2\pi} \frac{dl d\theta'}{\Delta}
 \nonumber \\
     & = \frac{1}{4\pi^2} \int_0^{2\pi} \int_0^{2\pi} \frac{1-e\cos u}{\Delta} du d\theta' ,
\end{aligned}
  \tag{Z44-\arabic{equation}}
  \stepcounter{equation}
  \label{eqn:Z44}
\end{equation}
where $\Delta$ is defined in Eq. \eqref{eqn:vZ10-defDelta} and
$\theta'$ is the longitude of the perturbed body defined in his Chapter I
(see p. \pageref{pg:def-thetadash} of this monograph).
$R$ in Eq. \eqref{eqn:Z44} is equivalent to the doubly averaged 
direct part of the disturbing function that we dealt with in
Sections \ref{ssec:CR3BP-R} and \ref{ssec:CR3BP-averaging} of this monograph.

At the end of Section Z13, \citeauthor{vonzeipel1910} makes the following
statement about the regularity of $R$ in the region where
orbit intersection between the perturbed body and the perturbing body does not happen:
\begin{quote}
``The function $R$ is, without ambiguity, given by the formula (Z44)
whenever the orbits do not intersect.
Obviously $R$ is holomorphic for all values of $e$, $g$, $a$, $k$
in the domain (Z43),
which corresponds to an orbit that does not encounter that of Jupiter.'' (p. Z368)
\end{quote}
\label{pg:entirelyholomorphic1}

\begin{figure*}[bhtp]\centering
\ifepsfigure
 \includegraphics[width=\dualfigwidth\textwidth]{fig_Z1-5.eps} %fig11
\else
 \includegraphics[width=\dualfigwidth\textwidth]{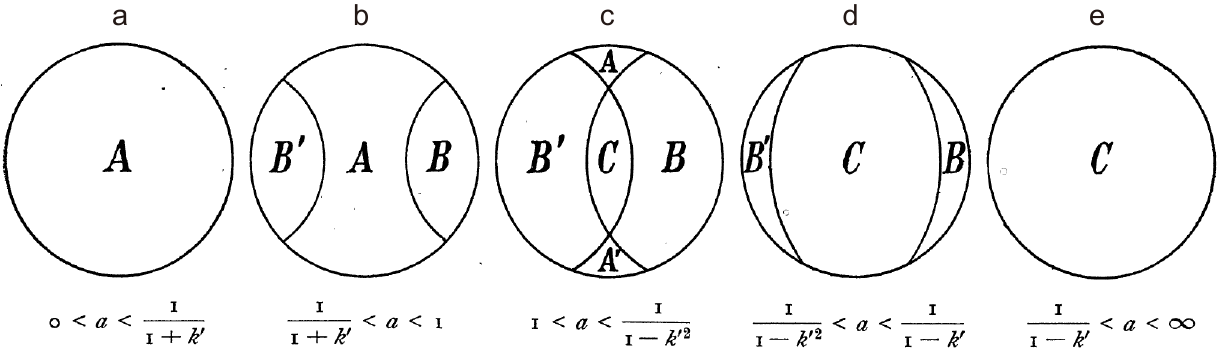} %fig11
\fi
  \caption{%
  A transcription of \citeauthor{vonzeipel1910}'s \mysymfigS Z1--Z5 on his pp. Z369--Z370
(Credit: John Wiley and Sons. Reproduced with permission).
  \mtxtsf{a} is \citeauthor{vonzeipel1910}'s \mysymfigO Z1,
  \mtxtsf{b} is Z2,
  \mtxtsf{c} is Z3,
  \mtxtsf{d} is Z4, and
  \mtxtsf{e} is Z5.
  Each of the panels represents the trajectories
  that the circles \eqref{eqn:Z45} and \eqref{eqn:Z46} make
  on the $(x,y)=(e \cos g, e \sin g)$ plane.
  Note that the symbol ``I'' in the numerators of the figure labels designates
  perturbing body's semimajor axis $\left(a'\right)$
  which is defined as 1 in \citet{vonzeipel1910}.
  On the other hand, the same symbol ``I'' in the denominators is
  a non-dimensional number, just equivalent to 1.
  For example,
  $\frac{{\rm I}}{{\rm I}+k'}$ in \mtxtsf{a} actually means
  $\frac{a'}{1+k'}$.
  Note also that in this figure \citeauthor{vonzeipel1910} assumes
  $k' = \frac{1}{2}$ for an expository purpose.
  }
  \label{fig:vZ10-f1-5}
\end{figure*}

\label{pg:orbitintersect}
Section Z14 is devoted to describing the characteristics of $R$
when the orbits of the perturbed and perturbing bodies intersect.
Since \citeauthor{vonzeipel1910} mentions this subject 
later again in Section Z24 in Chapter IV,
let us just make a brief summary.
The orbit intersection between the perturbing body on a circular orbit
and the perturbed body can happen
at the heliocentric distance of $a'$ (the semimajor axis of the perturbing body).
Let us cite \citeauthor{vonzeipel1910}'s original words on this subject:
\begin{quote}
``It is important to study the character of the function $R$ if the orbits intersect. Obviously, for the orbits that intersect, it is necessary and sufficient that one or other of the following conditions is met:
\begin{align}
 {\mbox{1) } }  1 &=  \frac{a\left(1-e^2\right)}{1 \pm e \cos g} \nonumber \\
 {\mbox{2) } }  e &=  k'                              \nonumber
\end{align}
and that, moreover, $a$ satisfies the inequality
$$
 \frac{1}{1+k'} < a < \frac{1}{1-k'} .
$$

The first of these conditions expresses that $r=1$ when $w=-g$ or when $w=\pi-g$.'' (p. Z369)
\end{quote}
Recall that $w$ seen in \citeauthor{vonzeipel1910}'s above statement is true anomaly.
Note also that by the definition of the unit of mass,
his above equation uses ``1'' on the left-hand side of the condition ``1)'' instead of $a'$.
This distance condition is seen in modern literature
\citep[e.g.][]{babadzhanov1992,farinella2001,jopek2017}
in a more general form such as
\begin{equation}
  a' = \frac{a \left(1-e^2\right)}{1 \pm e \cos g} .
  \label{eqn:vZ10-crosscond}
\end{equation}
In the denominator of the right-hand term of Eq. \eqref{eqn:vZ10-crosscond},
the positive sign applies when the orbit intersection occurs
at the ascending  node of the perturbed body (when $w = -g$).
The negative sign applies when the orbit intersection occurs
at the descending node of the perturbed body (when $w = \pi - g$).

As for the condition ``2) $e=k'$'' in
\citeauthor{vonzeipel1910}'s description, we do not follow him well. He wrote
``for the orbits that intersect, it is necessary and sufficient that one or other of the following conditions is met.''
However, the condition $e=k'$ just describes the maximum eccentricity of
the perturbed body, not making any constraints on the orbit (in other words, semimajor axis) of the perturbing body.
Therefore we have no idea why \citeauthor{vonzeipel1910} described $e=k'$
as one of the necessary and sufficient conditions for an orbit intersection
to happen between the perturbed and perturbing bodies.

The inequality that \citeauthor{vonzeipel1910} leaves in the
above statement can also be more generally rewritten
using the semimajor axis $a'$ of the perturbing body as
\begin{equation}
  \frac{a'}{1+k'} < a < \frac{a'}{1-k'} .
  \label{eqn:vZ10-a-crossing}
\end{equation}
We can interpret the condition \eqref{eqn:vZ10-a-crossing} as follows.
For an orbit intersect to occur,
the smallest value of the perihelion distance of the perturbed body $a\left(1-k'\right)$
must be smaller than the semimajor axis of the perturber $a'$
(i.e. $a < \frac{a'}{1-k'}$), and
the largest value of the aphelion distance $a\left(1+k'\right)$ of the perturbed body
must be larger  than $a'$
(i.e. $\frac{a'}{1+k'} < a$).

Here \citeauthor{vonzeipel1910} introduces a pair of new variables
$x = e \cos g$ and $y = e \sin g$ for rewriting the above condition ``1)''
(equivalent to Eq. \eqref{eqn:vZ10-a-crossing}).
Let us literally cite his words:
\begin{quote}
``By introducing the variables
$$
  x = e\cos g, \quad
  y = e\sin g
$$
the relation 1) can be written
\begin{equation}
   \left( x \pm \frac{1}{2a} \right)^2 + y^2
 = \left(   1 - \frac{1}{2a} \right)^2 .
  \tag{Z45-\arabic{equation}}
  \stepcounter{equation}
  \label{eqn:Z45}
\end{equation}

\hspace*{1em}
This is the equation of two circles whose centers are at the points $x=\mp\frac{1}{2a}$, $y=0$ and which pass, one by the point $x=-1$, $y=0$, the other by the point $x=1$, $y=0$.

\hspace*{1em}
The position of the circle (Z45) relative to the circle
\begin{equation}
  x^2 + y^2 = {k'}^2
  \tag{Z46-\arabic{equation}}
  \stepcounter{equation}
  \label{eqn:Z46}
\end{equation}
that limits the domain (Z43), depends on the values of $a$ and $k$.''
(pp. Z369--Z370)
\end{quote}

As before, Eq. \eqref{eqn:Z45} can be rewritten in a more general form
using $a'$ as
\begin{equation}
   \left( x \pm \frac{a'}{2a} \right)^2 + y^2
 = \left(   1 - \frac{a'}{2a} \right)^2 .
  \label{eqn:Z45-general}
\end{equation}

As \citeauthor{vonzeipel1910} writes,
Eq. \eqref{eqn:Z45-general} produces a pair of circles on the $(x,y)$ plane
whose center and radii depend just on $\frac{a}{a'}$.
The geometric circumstance of the orbit intersection is expressed on the $(x,y)$
plane through the locations of the circles \eqref{eqn:Z45} and
another circle \eqref{eqn:Z46}.
The circle \eqref{eqn:Z46} represents the outermost boundary of
the possible motion range of the perturbed body on the $(x,y)$ plane.

\citeauthor{vonzeipel1910} graphically illustrates the circumstance
that the circles \eqref{eqn:Z45} and \eqref{eqn:Z46} depict
in his \mysymfigS Z1--Z5 (pp. Z369--Z370), using $k' = \frac{1}{2}$ as an example
(note that the value of $k'$ just affects the circle \eqref{eqn:Z46},
 not the circles \eqref{eqn:Z45}).
We transcribed these figures as \mysymfigO \ref{fig:vZ10-f1-5}.
Let us explain, in a more general way,
the geometric condition that each panel of \mysymfigO \ref{fig:vZ10-f1-5} implies.
Recalling the fact that $k'$ denotes the maximum of the perturbed body's
eccentricity in the considered system,
we use the following notation:
\begin{alignat}{1}
  Q_{\rm max}    &= a \left(1+k'\right),
  \label{eqn:vZ10-def-Qmax} \\
  q_{\rm min}    &= a \left(1-k'\right),
  \label{eqn:vZ10-def-qmin} \\
  \ell_{\rm min} &= a \left( 1-{k'}^2 \right),
  \label{eqn:vZ10-def-ellmin}
\end{alignat}
where
$Q_{\rm max}$     is the largest  value of the perturbed body's apocenter distance,
$q_{\rm min}$     is the smallest value of the perturbed body's pericenter distance, and
$\ell_{\rm max}$  is the smallest value of the perturbed body's semilatus rectum.
\begin{itemize}
\item \mysymfigO \ref{fig:vZ10-f1-5}\mtxtsf{a}: $0 < a < \frac{a'}{1+k'}$.
We can interpret this as $0 < a$ and $Q_{\rm max} < a'$.
$0<a$ is obvious, and
$Q_{\rm max} < a'$ is satisfied if
the perturbed body's orbit always lies inside that of the perturbing body.
Hence in this case, the two orbits never intersect.
This orbit configuration is similar to that between the Atira asteroids
(also known as interior-Earth objects) and the Earth's orbit
\citep[e.g.][]{greenstreet2012,ribeiro2016,delafuentemarcos2018}.
\item \mysymfigO \ref{fig:vZ10-f1-5}\mtxtsf{b}: $\frac{a'}{1+k'} < a < a'$.
We can interpret this as $a' < Q_{\rm max}$ and $a < a'$.
In this case, the two orbits can intersect.
This orbit configuration is similar to that between the Aten near-Earth asteroids and the Earth's orbit \citep[e.g.][]{shoemaker1979,bottke2002}.
\item \mysymfigO \ref{fig:vZ10-f1-5}\mtxtsf{c}: $a' < a < \frac{a'}{1-{k'}^2}$.
We can interpret this as $a' < a$ and $\ell_{\rm min} < a'$.
In this case, the two orbits can intersect each other.
\item \mysymfigO \ref{fig:vZ10-f1-5}\mtxtsf{d}: $\frac{a'}{1-{k'}^2} <a< \frac{a'}{1-k'}$.
We can interpreted this as $a' < \ell_{\rm min}$ and $a' > q_{\rm min}$.
If we combine the two conditions \mtxtsf{c} and \mtxtsf{d},
and write them as $a' < a$ and $a' > q_{\rm min}$,
this orbit configuration is now close to that between the Apollo near-Earth asteroids and the Earth's orbit \citep[e.g.][]{shoemaker1979,bottke2002}.
\item \mysymfigO \ref{fig:vZ10-f1-5}\mtxtsf{e}: $\frac{a'}{1-k'} < a < \infty$ .
We can interpret this as $a' < q_{\rm min}$ and $a < \infty$.
$a < \infty$ is obvious, and
the perturbed body's orbit always lies outside that of the perturbing body
if $a' < q_{\rm min}$.
Hence in this case, the two orbits never intersect each other.
This orbit configuration is close to that between the Amor near-Earth asteroids and the Earth's orbit
\citep[e.g.][]{shoemaker1979,bottke2002}.
\end{itemize}
\label{pg:vZ10-orbitconfigurations}

\begin{figure}[htbp]\centering
\ifepsfigure
 \includegraphics[width=\singlefigwidth\textwidth]{fig_mp-circles.eps} %fig12
\else
 \includegraphics[width=\singlefigwidth\textwidth]{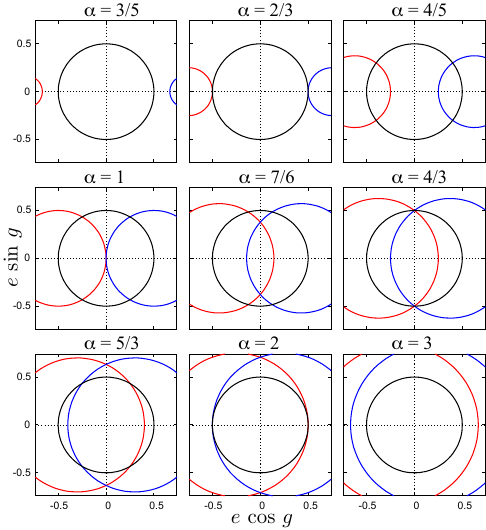} %fig12
\fi
  \caption{%
  Similar to \mysymfigO \protect\ref{fig:vZ10-f1-5},
  we depict circles described by Eqs. \eqref{eqn:Z45} and \eqref{eqn:Z46}
  with several more values of $\alpha$
  on the $(x,y)=(e \cos g, e \sin g)$ plane.
  The circles drawn in red  correspond to the negative sign
  in the first term of the left-hand side of Eq. \eqref{eqn:Z45}.
  The circles drawn in blue correspond to the positive sign
  in the first term of the left-hand side of Eq. \eqref{eqn:Z45}.
  The black solid circles at the center are a representation of Eq. \eqref{eqn:Z46}.
  The parameter $k' = \frac{1}{2}$ is common to all the panels.
}
  \label{fig:vZ10-f1-5-replot}
\end{figure}

To facilitate the reader's understanding,
based on Eqs. \eqref{eqn:Z45} and \eqref{eqn:Z46} we present,
as \mysymfigO \ref{fig:vZ10-f1-5-replot},
diagrams of this kind with several more values of $\alpha$.
We adopted $k'=\frac{1}{2}$ following
\citeauthor{vonzeipel1910}'s figures.
Now we can better see the relation between the $\alpha$ values and the topological
configurations of circles.
Among the nine values of $\alpha = \frac{a}{a'}$ in \mysymfigO \ref{fig:vZ10-f1-5-replot},
the following four depict configurations when
the relative topology of the three circles changes:
$\alpha = \frac{2}{3} = \frac{1}{1+k'}$,
$\alpha = 1$,
$\alpha = \frac{4}{3} = \frac{1}{1-{k'}^2}$, and
$\alpha = 2 = \frac{1}{1-k'}$.

As seen in \mysymfigO \ref{fig:vZ10-f1-5},
\citeauthor{vonzeipel1910} named each of the domains as $A$, $A'$, $B$, $B'$, $C$.
In the rest of Section Z14, he tries to prove the following theorem:
\begin{quote}
``\textit{%
The function $R$ is represented in various domains $A$, $B$, $C$, $A'$, $B'$ by different analytic functions.
Each of these functions is holomorphic in the corresponding domain and on its borders (except the circle $e=k'$).}'' (p. Z369)
\end{quote}
\label{pg:entirelyholomorphic2}

The doubly averaged disturbing function $R$ is originally defined
as the double integral expressed in Eq. \eqref{eqn:Z44}.
As a preparation to prove the above theorem,
\citeauthor{vonzeipel1910} expresses $R$
in a single integral by performing the integration with respect to $\theta'$.
Although we do not go into his proof of the theorem in this monograph,
let us quickly introduce his preparation part because the result is used
in a later section (Section Z20; discussed on p. \pageref{pg:Z20} of this monograph).

\citeauthor{vonzeipel1910} first introduces a pair of parameters $\tau$ and $V$ as
\begin{alignat}{1}
  \left(1+r^2\right) \tau \cos V & = 2 r \cos (w+g),
  \label{eqn:vZ10-p370-nn1-cos} \\
  \left(1+r^2\right) \tau \sin V & = 2 r \sin (w+g) \cos I ,
  \label{eqn:vZ10-p370-nn1-sin}
\end{alignat}
and expresses $\Delta$ in the denominator of Eq. \eqref{eqn:Z44} as follows:
\begin{equation}
  \Delta^2 = \left( 1+r^2 \right) \left[ 1-\tau \cos \left(\theta'+V\right) \right] .
  \tag{Z47-\arabic{equation}}
  \stepcounter{equation}
  \label{eqn:Z47}
\end{equation}
Then the following equation for $\tau^2$
\begin{equation}
  \tau^2 = \frac{4r^2}{\left(1+r^2\right)^2} \left[1-\sin^2 I \sin^2 (w+g)\right] ,
  \tag{Z48-\arabic{equation}}
  \stepcounter{equation}
  \label{eqn:Z48}
\end{equation}
or that for ${\tau '}^2 \equiv 1 - \tau^2$
\begin{equation}
  {\tau '}^2 = 1- \tau^2 = \frac{\left(1-r^2\right)^2 + 4r^2 \sin^2 I \sin^2 (w+g)}{\left(1+r^2\right)^2} ,
  \tag{Z49-\arabic{equation}}
  \stepcounter{equation}
  \label{eqn:Z49}
\end{equation}
are derived.
Now \citeauthor{vonzeipel1910} defines a function $F(\tau)$ as
\begin{equation}
\begin{aligned}
{} & \frac{1}{2\pi} \int_0^{2\pi} \frac{\sqrt{1+r^2}}{\Delta} \theta' \\
{} & \quad
 = \frac{1}{2\pi} \int_0^{2\pi} \frac{d\theta'}{\sqrt{1-\tau\cos\left(\theta'+V\right)}}
 = F(\tau) .
\end{aligned}
  \tag{Z50-\arabic{equation}}
  \stepcounter{equation}
  \label{eqn:Z50}
\end{equation}

Note that $F(\tau)$ is newly defined in Eq. \eqref{eqn:Z50},
but the notation may be confusing.
This $F$ is different from the Hamiltonian in Eq. \eqref{eqn:vZ10-F-all}.
Also, note that $F$ is a function of $\tau$ only, not that of $V$,
due to the relation between $V$ and $\tau$
shown in Eqs. \eqref{eqn:vZ10-p370-nn1-cos} and \eqref{eqn:vZ10-p370-nn1-sin}.

Now \citeauthor{vonzeipel1910} adopts $F(\tau)$ defined in Eq. \eqref{eqn:Z50}
for the literal definition of $R$.
Changing the integration variable used in the averaging procedure
from mean anomaly $l$ to true anomaly $w$ through the commonly used relationship
\begin{equation}
  dl = \frac{r^2}{a^2} \frac{dw}{\sqrt{1-e^2}},
  \label{eqn:dl2dw}
\end{equation}
he obtains $R$ in the form of a single integral as
\begin{equation}
  R = \frac{1}{2\pi} \int_0^{2\pi} \frac{r^2}{a^2} \frac{F(\tau)}{\sqrt{1-e^2} \sqrt{1+r^2}} dw .
  \tag{Z51-\arabic{equation}}
  \stepcounter{equation}
  \label{eqn:Z51}
\end{equation}

Note that in Eqs. \eqref{eqn:dl2dw} and \eqref{eqn:Z51}
we use $a$, not $\alpha = \frac{a}{a'}$.
This is for explicitly displaying the quantity $\frac{r}{a}$ as
being non-dimensional.
However this is just for a formal consistency, because
we can practically regard $a$ and $r$ are both non-dimensional
from the beginning in \citeauthor{vonzeipel1910}'s theory.
We already mentioned this point in the discussion
following Eq. \eqref{eqn:vZ10-defDelta}
on p. \pageref{pg:vZ10-lengthsarenormalized} of this monograph.
\label{pg:vZ10-lengthsarenormalized-2}

\citeauthor{vonzeipel1910} continues as follows:
\begin{quote}
``It is well known and easy to verify that $F$ is represented,
if $0 \leq \tau < 1$, by the hypergeometric series,
\begin{equation}
  F (\tau) = F \left( \frac{1}{4}, \frac{3}{4}, 1, \tau^2 \right),
  \tag{Z52-\arabic{equation}}
  \stepcounter{equation}
  \label{eqn:Z52}
\end{equation}
or by the [complete] elliptic integral of the first kind,
\begin{equation}
  F(\tau) = \frac{2}{\pi}\frac{1}{\sqrt{1+\tau}} \int_0^{\frac{\pi}{2}} \frac{d\varphi}{\sqrt{1-\frac{2\tau}{1+\tau} \sin^2 \varphi}} .
\mbox{\rm ''}
  \tag{Z53-\arabic{equation}}
  \stepcounter{equation}
  \label{eqn:Z53}
\end{equation}
(p. Z371)
\end{quote}
\citeauthor{vonzeipel1910} still continues:
\begin{quote}
``It is also known (see Picard: Trait\'e d'Analysis vol.~III, page 273) that $F(\tau)$ is of the form
\begin{equation}
  F (\tau) = A \left({\tau '}^2\right) \log {\tau '}^2 + B \left({\tau '}^2\right) ,
  \tag{Z54-\arabic{equation}}
  \stepcounter{equation}
  \label{eqn:Z54}
\end{equation}
in the neighborhood of $\tau=1$, and both $A$ and $B$ are developed in powers of $\tau^2$ with
\begin{equation}
  A(0) \neq 0 .
  \tag{Z55-\arabic{equation}}
  \stepcounter{equation}
  \label{eqn:Z55}
\end{equation}

Then $A$, $B$ and $\log {\tau '}^2$ are real, if $\tau^2$ is real and $>0$.
Finally we evidently have
\begin{equation}
  A(0) < 0
  \tag{Z56-\arabic{equation}}
  \stepcounter{equation}
  \label{eqn:Z56}
\end{equation}
because, if $A(0)>0$, we would have $F(\tau) =-\infty$ for $\tau=1$, which is impossible since $F(\tau)>0$.''
(p. Z371)
\end{quote}
\label{pg:def-A0}

Nowadays, approximating complete elliptic integrals by a logarithmic function
like this is a common idea \citep[e.g.][]{hastings1955,cody1965a,cody1965b}.
Note that although we found an expression similar to Eq. \eqref{eqn:Z54} in
\citet[][Chapter XI, sub-chapter II, Section 13, p. 273, as suggested by \citeauthor{vonzeipel1910}]{picard1896},
\citeauthor{picard1896}'s original equation is in a more general form.
Also, the function 
that \citeauthor{vonzeipel1910} depicted as ``$\log$'' in Eq. \eqref{eqn:Z54} 
turned out to be a natural logarithm $(\ln)$ in \citeauthor{picard1896}'s original expression.

Let us wrap up our introduction of \citeauthor{vonzeipel1910}'s Section Z14
by citing one of his paragraphs on pp. Z374--Z375.
It is about the general characteristics of the secular disturbing function
that he deals with:
\begin{quote}
``According to the formula (Z66),
the function $R$ is thus given by two different analytic functions $R'$ and $R''$ in the two domains $A$ and $B$ in Fig.~Z2.
Of these functions, $R''$ is holomorphic in the domain $A$, and $R''$ [is holomorphic] in the domain $B$.
Subsequently, the functions $R'$ and $R''$ are holomorphic also on the arcs of the circles (Z45), which separate the domains $A$ and $B$ from each other, except at the points of intersection of the curves (Z45) and (Z46).
Finally, we have $R'= R''$ on the boundary that separates the domains $A$ and $B$.

\hspace{1em}
In the previous demonstration we have assumed
$$
  \frac{1}{1+k'} < a < 1 .
$$

\hspace{1em}
But it is obviously possible to use the same method almost without change for dealing with
$$
  1 < a < \frac{1}{1+{k'}^2}
$$
and
$$
  \frac{1}{1-{k'}^2} < a < \frac{1}{1-k'}
$$
represented by Figs. Z3 and Z4 on page Z369.
To calculate $R$, it is necessary to use different analytic functions in the domains $A$, $B$, $C$, $A'$, $B'$.
These functions are holomorphic in each of the domains $A$, $B$, $C$, $A'$, $B'$ which they belong to, and also on the borders of this area with an exception of the circle (Z46).''
(pp. Z374--Z375)
\end{quote}

Note that we have not transcribed \citeauthor{vonzeipel1910}'s Eq. (Z66) in this monograph.
Note also that in all the three inequalities in the above quotation,
``1'' in the numerators (not in the denominators) designates
the perturbing body's semimajor axis, $a'$.
Therefore, the three inequalities can be respectively rewritten as
either of the following combinations:
\begin{equation}
\begin{array}{cccl}
   \frac{a'}{1+k'}   < a < a',
&               a'   < a < \frac{a'}{1+{k'}^2},
&  \frac{a'}{1-{k'}^2} < a < \frac{a'}{1-k'} ,
\nonumber
\end{array}
\end{equation}
or
\begin{equation}
\begin{array}{cccl}
   \frac{1}{1+k'}   < \alpha < 1,
&               1   < \alpha < \frac{1}{1+{k'}^2},
&  \frac{1}{1-{k'}^2} < \alpha < \frac{1}{1-k'} .
\nonumber
\end{array}
\end{equation}
See also \mysymfigO \ref{fig:vZ10-f1-5}'s caption and the ensuing discussion.

In the final section (Z15) of Chapter III,
\citeauthor{vonzeipel1910} mentions the behavior of $R$ close to the
outer boundary circle represented by $e = x^2 + y^2 = k'$ \eqref{eqn:Z46}
in a very general way.
Although in this monograph we do not go into the contents of this section,
we will return to this subject later
(Section \ref{sssec:motionneare=kd} on p. \pageref{sssec:motionneare=kd})
with a more specific function form of $R$ and numerical examples.
\label{pg:motionneare=kd-general}

Readers should recall that in his Chapter III,
\citeauthor{vonzeipel1910}'s description is still general.
No function form is specified for the secular disturbing function $R$.
\citeauthor{vonzeipel1910} mentions the specific function form of $R$
for the first time in the following chapter (Chapter IV).

\subsection{Secular disturbing function: Inner case\label{ssec:icr3bp}}
Among \citeauthor{vonzeipel1910}'s work in \citeyear{vonzeipel1910}, the chapter
``\textit{Chapitre IV. Maxima et minima de la partie s\'eculaire de la fonction perturbatrice\/}''
comprises the most important component for us.
In this chapter he shows an analysis of the doubly averaged disturbing function $R$
with specific function forms.
This is what we can directly compare with the later work of
\citeauthor{lidov1961} and \citeauthor{kozai1962b} in detail.

This chapter is very long, occupying half of the entire article
(37 pages out of 74 pages) including ten sections (from Z16 to Z25).
We would like to categorize the sections into three parts.
For this purpose, we think it is best to cite the beginning part of this chapter in Section Z16:
\begin{quote}
``In Chapter II we have shown that the integration of the equations (Z12) of the secular variations is much simplified in the vicinity of a maximum or minimum value of the function $R$.
It is therefore important to find the points $(e,g)$ of the domain (Z43) for which the function $R$ reaches its maxima and minima.

\hspace{1em}
In the study, that we have in view, it is advantageous to distinguish three cases, characterized by the position of the nodes of the orbit of the infinitesimal mass:
\begin{enumerate}
\item Both nodes are located inside  the orbit of Jupiter, so that the orbit of the comet remaining in its plane can be reduced to the Sun without intersecting Jupiter's orbit.
\item Both nodes are located outside the orbit of Jupiter, so that the orbit of the comet remaining in its plane can be reduced to an infinitely large circle around the Sun as the center without touching the orbit Jupiter.
\item Of the two nodes, one is located inside, [and] the other [is located] outside, the orbit of Jupiter.
The two orbits behave like the rings of a chain.''
\end{enumerate}
(p. Z378)
\end{quote}
\label{pg:orbit-threecases}

In the above,
the case 1 is about the inner CR3BP which is discussed in Sections Z16--Z21.
The case 2 is about the outer CR3BP which is discussed in Sections Z22--Z23.
The case 3
corresponds to a system where orbit intersection between the perturbed and perturbing bodies can happen.
This is discussed in Sections Z24--Z25.
In the present subsection (Section \ref{ssec:icr3bp}),
we will summarize \citeauthor{vonzeipel1910}'s Sections Z16--Z21 that deal with the case 1.

\subsubsection{Expansion of $R$ by $\alpha$\label{sssec:Z10-expansion}}
\citeauthor{vonzeipel1910} first assumes in Section Z16
that $\alpha = \frac{a}{a'}$ is very small.
This is equivalent to adopting the quadrupole level approximation, and
the expansion of $R$ in the series of $\alpha$ can be truncated at lower-order without losing accuracy.
The method that he took for the expansion of $R$
follows \citet{tisserand1889} that exploits
the Hansen coefficients $X_0^{2m,2i}$.
Omitting the detail, let us transcribe \citeauthor{vonzeipel1910}'s
expansion result as follows:
\begin{equation}
\begin{aligned}
  R &= \left[ \frac{1}{\Delta} \right]_{l,\theta'} % \\
     = 1 +      \sum_{m=1}^\infty \Bigl( A_{0.0}^{(2m)} X_0^{2m,0} \\
    & \qquad\qquad
             + 2\sum_{i=1}^m A_{i.j}^{(2m)}X_0^{2m,2i} \cos 2ig \Bigr) \alpha^{2m},
\end{aligned}
  \tag{Z72-\arabic{equation}}
  \stepcounter{equation}
  \label{eqn:Z72}
\end{equation}
where $\left[ \frac{1}{\Delta} \right]_{l,\theta'}$ means that the quantity $\frac{1}{\Delta}$ is averaged both by $l$ and by $\theta'$.
This operation is equivalent to $\left<\left< \frac{1}{\Delta} \right>_{l'}\right>_{l}$ that we employed on p. \pageref{eqn:Rd-inner-j-averaged2} of this monograph.

The Hansen coefficients are expressed as
\begin{equation}
\begin{aligned}
  X_0^{2m,2i} &= \frac{(2m+2i+1)!}{(2m+1)!(2i)!}\left(\frac{e^2}{4}\right)^i \\
  & % \qquad
               \times F\left(i-m,i-m-\frac{1}{2},2i+1,e^2\right),
\end{aligned}
  \tag{Z71-\arabic{equation}}
  \stepcounter{equation}
  \label{eqn:Z71}
\end{equation}
with
\begin{equation}
\begin{aligned}
  A_{i,j}^{(2m)} & = k_{i,j}^{(2m)}\left(\frac{\sin^2 I}{4}\right)^i \\
  & % \qquad
           \times F\left(i-m,i+m+\frac{1}{2},2i+1,\sin^2 I\right),
\end{aligned}
  \tag{Z74-\arabic{equation}}
  \stepcounter{equation}
  \label{eqn:Z74}
\end{equation}
and
\begin{equation}
  k_{i,j}^{(2m)} = \frac{(2m+2i)!(2m)!}{2^{4m}(2i)!(m+i)!(m-i)!(m!)^2}.
  \tag{Z73-\arabic{equation}}
  \stepcounter{equation}
  \label{eqn:Z73}
\end{equation}
Note that $F$ in Eq. \eqref{eqn:Z71} and Eq. \eqref{eqn:Z74} is the hypergeometric series of Gauss.

\citeauthor{vonzeipel1910} regards the expansion \eqref{eqn:Z72} as a series of $\alpha^2$, and writes down $R$ as
\begin{equation}
  R = 1 + R_2 \alpha^2 + R_4 \alpha^4 + \cdots ,
  \tag{Z75-\arabic{equation}}
  \stepcounter{equation}
  \label{eqn:Z75}
\end{equation}
where $R_2$, $R_4$, $\cdots$ are functions of $e$ and $g$.
The lowest-order term of $R$ (except for the constant 1) becomes
\begin{equation}
\begin{aligned}
  R_2 &= A_{0.0}^{(2)} X_0^{2.0} + 2A_{1.1}^{(2)}X_0^{2.2} \cos 2g \\
      &= \frac{1}{4}\left(1-\frac{3}{2}\sin^2 I \right)
                   \left(1+\frac{3}{2}e^2 \right) \\
      & \qquad
        + \frac{15}{16}\sin^2 I \cdot e^2 \cos 2g \\
      &=  \frac{1}{8} \left( -1 + \frac{3k^2}{1-e^2} \right)
                    \left(1+\frac{3}{2}e^2 \right) \\
      & \qquad
        + \frac{15}{16} \left(1-\frac{k^2}{1-e^2} \right) e^2 \cos 2g .
\end{aligned}
  \tag{Z76-\arabic{equation}}
  \stepcounter{equation}
  \label{eqn:Z76}
\end{equation}

It is clear that $R_2$ in Eq. \eqref{eqn:Z76} is equivalent to
what we saw in \citeauthor{kozai1962b}'s work
(Eq. \eqref{eqn:R2-final} of this monograph).

Note that
\citeauthor{vonzeipel1910} actually used $a$
instead of $\alpha = \frac{a}{a'}$
throughout his discussion of the inner problem.
For example, the original form of Eq. \eqref{eqn:Z75} is written as follows:
\begin{equation}
  R = 1 + R_2 a^2 + R_4 a^4 + \cdots .
  \label{eqn:Z75-original}
\end{equation}
The apparent difference between Eq. \eqref{eqn:Z75} and Eq. \eqref{eqn:Z75-original}
comes from the fact that \citeauthor{vonzeipel1910} fixes
the perturber's semimajor axis $\left(a'\right)$ as 1 (the unit of length in his theory).
However,
for keeping the generality and compatibility with the later studies by
\citeauthor{lidov1961} and \citeauthor{kozai1962b},
we use the variable of $\alpha$ instead of just $a$.

Next, using the polar coordinate variables $x = e\cos g$ and $y = e\sin g$,
\citeauthor{vonzeipel1910} expands $R$ of Eq. \eqref{eqn:Z75}
into a two-variable Taylor series of $x^2$ and $y^2$ as
\begin{equation}
  R = R_{0,0} + R_{2.0} x^2 + R_{0.2} y^2 + \cdots .
  \tag{Z77-\arabic{equation}}
  \stepcounter{equation}
  \label{eqn:Z77}
\end{equation}

The coefficients $R_{2.0}$ and $R_{0.2}$ are expanded into
the power series of $\alpha^2$ as
\begin{equation}
\begin{aligned}
  R_{2.0} &= \frac{3}{4} \alpha^2 + P_4 \alpha^4 + P_6 \alpha^6 + \cdots , \\
  R_{0.2} &= \frac{15}{8}\left(k^2-\frac{3}{5}\right)\alpha^2 + Q_4 \alpha^4 + Q_6 \alpha^6 + \cdots ,
\end{aligned}
  \tag{Z78-\arabic{equation}}
  \stepcounter{equation}
  \label{eqn:Z78}
\end{equation}
where $P_{2i}$ and $Q_{2i}$ $(i=2,3,\cdots)$ are polynomials of $k^2$, although
\citeauthor{vonzeipel1910} did not show their actual forms.

\citeauthor{vonzeipel1910} did not show the specific form of the first term $R_{0.0}$ in Eq. \eqref{eqn:Z77}.
But we can calculate it by actually carrying out
the two-variable Taylor expansion. We get:
\begin{equation}
  R_{0.0} = 1 + \frac{3k^2 -1}{8} \alpha^2 + \cdots .
  \label{eqn:R00-inner-maple}
\end{equation}
However,
constant terms such as $R_{0.0}$ do not matter in the following discussions.
This is because \citeauthor{vonzeipel1910}'s
major goal was to search for local extremums of $R$
by calculating its partial derivatives such as
$\DP{R}{x}$ or $\DP{R}{y}$.

Now \citeauthor{vonzeipel1910} defines a new quantity $k^2_{0.2}$
as the root of the following equation
\begin{equation}
  R_{0.2} = 0 .
  \label{eqn:vZ10-R02=0}
\end{equation}
When $\alpha \ll 1$ (i.e. the quadrupole level approximation),
we know $k^2_{0.2} = \frac{3}{5}$ from Eq. \eqref{eqn:Z78}.
This number must already be familiar to the readers.
\citeauthor{vonzeipel1910} further expands $k^2_{0.2}$ 
into the power series of $\alpha^2$ as
\begin{equation}
  k^2_{0.2} = \frac{3}{5} + K_2 \alpha^2 + K_4 \alpha^4 + \cdots ,
  \tag{Z79-\arabic{equation}}
  \stepcounter{equation}
  \label{eqn:Z79}
\end{equation}
where $K_2$, $K_4$, $\cdots$ are some numerical coefficients
whose actual form \citeauthor{vonzeipel1910} did not show.
$k^2_{0.2}$ later plays a fundamental role in locating the local extremums of $R$.

\subsubsection{Search for local extremums\label{sssec:iCR3BP-extremum}}
In the rest of Section Z16, \citeauthor{vonzeipel1910} tries to find
local extremums of $R$ on the $(x,y)=(e\cos g, e\sin g)$ plane
by assuming $\alpha \ll 1$.
This means that he assumes $k_{0.2}^2 = \frac{3}{5}$.
His general approach here is divided into two stages:
\begin{enumerate}
\item He first locates local extremums of $R$ by calculating its
      first derivatives, such as $\DP{R}{x}$.
\item He then determines whether the discovered local extremum is
      a local minimum, a local maximum, or a saddle point
      by inspecting the sign of the second derivatives of $R$
      such as $\DP[2]{R}{x}$.
\end{enumerate}
As we see in what follows,
\citeauthor{vonzeipel1910}'s search areas for the local extremums are
the origin $(x,y)=(0,0)$,
the outer boundary $x^2+y^2={k'}^2$, and
the $y$-axis.

\paragraph{At the origin\label{pg:iCR3BP-origin}}
\citeauthor{vonzeipel1910} first considers the behavior of $R$ at the origin, $(x,y)=(0,0)$.
His conclusion is as follows:
\begin{itemize}
  \item The origin is a local minimum of $R$ when $k^2 > k^2_{0.2}$.
  \item The origin is a saddle point  of $R$ when $k^2 < k^2_{0.2}$.
\end{itemize}
\citeauthor{vonzeipel1910} confirms this as follows.
At the origin, from Eqs. \eqref{eqn:Z77} and \eqref{eqn:Z78} we have
\begin{equation}
  \DP{R}{x} = \DP{R}{y} = 0 .
  \label{eqn:vZ10-inner-Rxy0-1}
\end{equation}
Also, we have
\begin{alignat}{1}
  \DP[2]{R}{x} &= 2 R_{2.0} \sim \frac{3}{2}\alpha^2,
  \label{eqn:vZ10-inner-Rxy0-2-R20} \\
  \DP[2]{R}{y} &= 2 R_{0.2} \sim \frac{15}{4} \left( k^2 - \frac{3}{5} \right) \alpha^2 ,
  \label{eqn:vZ10-inner-Rxy0-2-R02}
\end{alignat}
at $(x,y) = (0,0)$ when $\alpha \ll 1$.
The relationships
\eqref{eqn:vZ10-inner-Rxy0-2-R20} and
\eqref{eqn:vZ10-inner-Rxy0-2-R02} hold true at the origin
because the higher-order expansion terms appearing in the Taylor series
\eqref{eqn:Z77} all vanish in the second derivatives,
just leaving the terms of $2 R_{2.0}$ and $2 R_{0.2}$ as in
Eqs. \eqref{eqn:vZ10-inner-Rxy0-2-R20} and
     \eqref{eqn:vZ10-inner-Rxy0-2-R02}.

Now from
Eqs. \eqref{eqn:vZ10-inner-Rxy0-2-R20} and
     \eqref{eqn:vZ10-inner-Rxy0-2-R02},
when $k^2 > k^2_{0.2}$ we have
\begin{equation}
  \DP[2]{R}{x} > 0 , \quad
  \DP[2]{R}{y} > 0 .
  \label{eqn:vZ10-inner-Rxy0-2a}
\end{equation}
This indicates that $R$ has a local minimum at the origin.
It is obvious from Eq. \eqref{eqn:vZ10-inner-Rxy0-2-R20}
that $\DP[2]{R}{x} > 0$ is true
regardless of the value of $k^2$ in this approximation.

On the other hand when $k^2 < k^2_{0.2}$, from
Eqs. \eqref{eqn:vZ10-inner-Rxy0-2-R20} and
     \eqref{eqn:vZ10-inner-Rxy0-2-R02} we have at $(x,y) = (0,0)$
\begin{equation}
  \DP[2]{R}{x} > 0, \quad
  \DP[2]{R}{y} < 0 .
  \label{eqn:vZ10-inner-Rxy0-3}
\end{equation}
From Eqs. \eqref{eqn:vZ10-inner-Rxy0-2a} and \eqref{eqn:vZ10-inner-Rxy0-3}
we know that
$R$ has a saddle point (``minimax'' in \citeauthor{vonzeipel1910}'s terminology)
at the origin.

\paragraph{At the outer boundary\label{pg:iCR3BP-kd}}
\citeauthor{vonzeipel1910} then moves on to a consideration of $R$
on its outer boundary. As we saw before
(Section \ref{ssec:R-general} on p. \pageref{pg:border-kdash}),
$R$ has a constant value on the outer boundary $x^2 + y^2 = {k'}^2$,
and it is independent of $g$ on this circle.
By calculating the partial derivative $\frac{\partial R}{\partial (e^2)}$
and substituting $e^2 = {k'}^2= 1-k^2$ into it,
on the boundary circle \citeauthor{vonzeipel1910} obtains (p. Z381)
\begin{equation}
\begin{aligned}
{} & \left. \frac{\partial R}{\partial \left(e^2\right)} \right|_{e^2={k'}^2}
 = \left. \alpha^2 \frac{\partial R_2}{\partial \left(e^2\right)} \right|_{e^2={k'}^2} + \cdots \\
{} & \quad
 = \alpha^2 \frac{3}{16k^2}\left[5-k^2-5\left(1-k^2\right)\cos 2g\right] + \cdots .
\end{aligned}
  \label{eqn:vZ10-nn381}
\end{equation}

As long as $\alpha$ is small and the higher-order terms than $\alpha^2$ can be ignored,
it is clear that the right-hand side of Eq. \eqref{eqn:vZ10-nn381} is always positive (with the smallest value of $4 k^2$).
Hence, at the quadrupole level approximation we have
\begin{equation}
  \left. \frac{\partial R}{\partial \left(e^2\right)} \right|_{e^2={k'}^2} > 0 ,
  \label{eqn:vZ10-nn381-positive}
\end{equation}
on the outer boundary.
Equation \eqref{eqn:vZ10-nn381-positive} indicates that,
just inside the outer boundary,
$R$ continues to increase along the radial direction 
on the $(x,y)=(e \cos g, e \sin g)$ plane until it reaches $e = k'$.
In this sense,
the outer boundary circle expressed by $x^2 + y^2 = {k'}^2$ can be regarded
as a kind of local maximum of $R$.
\label{pg:R_outerboundary-i}

Note that
\citeauthor{vonzeipel1910} frequently uses $\DP{R}{(e^2)}$,
instead of $\DP{R}{e}$,
for determining whether or not a particular point (or a curve) of $R$ makes a local extremum.
It may seem strange because the considered phase space is
described by the variables $(x,y) = (e\cos g, e\sin g)$.
But we suspect his use of $\DP{R}{\left(e^2\right)}$ is justified for two reasons.
First, the function form of $R$ in Eq. \eqref{eqn:Z76} contains the perturbed body's
eccentricity only in the form of $e^2$.
Second, whether we use $\DP{R}{\left(e^2\right)}$ or $\DP{R}{e}$ would not matter
for the purpose of locating local extremums of $R$.
This is due to the identity
\begin{equation}
  \DP{R}{e} = 2e \DP{R}{\left(e^2\right)} .
  \label{eqn:DPRe-DPRe2}
\end{equation}
Therefore, we see that
$\DP{R}{\left(e^2\right)} \lesseqgtr 0$ always leads to $\DP{R}{e} \lesseqgtr 0$
as long as $e > 0$.
Thus we find that
searching for the locations where $\DP{R}{e}                = 0$ is realized is equivalent to
searching for the locations where $\DP{R}{\left(e^2\right)} = 0$ is realized.

\paragraph{Along the $y$-axis\label{pg:iCR3BP-xy=0e02}}
Now \citeauthor{vonzeipel1910} moves on to the
search for local maxima or minima which may be located somewhere
between the origin $(x,y)=(0,0)$ and the outer boundary circle
$\bigl( x^2+y^2={k'}^2 \bigr)$.
For this purpose, the following equations must be solved in general:
\begin{equation}
  \DP{R}{\left(e^2\right)} = 0 , \quad
  \DP{R}{g} = 0 .
  \tag{Z80-\arabic{equation}}
  \stepcounter{equation}
  \label{eqn:Z80}
\end{equation}
We should bear in mind that $R \sim R_2$ as long as $\alpha \ll 1$.

It is clear that the differential operation $\DP{R_2}{g}$ eliminates the
first term of $R_2$ in Eq. \eqref{eqn:Z76}, and just leaves its second term.
The second term of $R_2$ in Eq. \eqref{eqn:Z76} yields a factor of $\sin 2g$
through the operation $\DP{R_2}{g}$, so the roots of the equation
$\DP{R}{g} = 0$ in Eq. \eqref{eqn:Z80} occur when $\sin 2g = 0$.
This means $\cos 2g = +1$ or $\cos 2g = -1$.

When $\cos 2g = +1$,
$\DP{R}{\left(e^2\right)}$ becomes (in the second equation from the top on p. Z382)
\begin{equation}
\begin{aligned}
  \DP{R}{\left(e^2\right)}
    &= \alpha^2 \DP{R_2}{\left(e^2\right)} + \cdots \\
    &= \frac{3}{4} \alpha^2 + \Oalqua .
\end{aligned}
  \label{eqn:vZ10-nn382-1}
\end{equation}

The quantity in the right-hand side of Eq. \eqref{eqn:vZ10-nn382-1} is always greater than zero when $\alpha \ll 1$.
This indicates that the first equation of Eq. \eqref{eqn:Z80} does not have any solution.
It means that, along the axis that satisfies $\cos 2g = +1$,
$R$ (or $R_2$) monotonically increases from the origin
toward the outer boundary without any local extremums.

On the other hand when $\cos 2g = -1$,
$\DP{R}{\left(e^2\right)}$ becomes (in the third equation on p. Z382)
\begin{equation}
\begin{aligned}
  \DP{R}{\left(e^2\right)}
  &= \alpha^2 \DP{R_2}{\left(e^2\right)} + \cdots \\
  &= \alpha^2 \left( -\frac{9}{8} + \frac{15}{8} k^2 \frac{1}{\left(1-e^2\right)^2} \right) + \Oalqua .
\end{aligned}
  \label{eqn:vZ10-nn382-2}
\end{equation}
The quantity in the right-hand side of Eq. \eqref{eqn:vZ10-nn382-2} can be either positive or negative
depending on the value of $k^2$.
It leaves a possibility for local extremums to occur.
Then, \citeauthor{vonzeipel1910} expresses the solution of the equation
  $\DP{R}{\left(e^2\right)} = 0$
as:
\begin{equation}
  e^2 = e^2_{0.2} = 1-\sqrt{\frac{5}{3}k^2} + c_2 \alpha^2 + c_4 \alpha^4 + \cdots ,
  \tag{Z81-\arabic{equation}}
  \stepcounter{equation}
  \label{eqn:Z81}
\end{equation}
where $c_2$, $c_4$, $\cdots$ are polynomials of $k$
whose function form he did not show.
Equation \eqref{eqn:Z81} means that $R$ can have a pair of local extremums
at the points $(x,y) = (0, \pm e_{0.2})$, or in other words,
at            $(e,g) = (e_{0.2}, \pm \frac{\pi}{2})$.
\label{pg:iCR3BP-k2l35}
In p. Z382 \citeauthor{vonzeipel1910} also introduces the equation
\begin{equation}
  e^2_{0.2} = 0 ,
  \label{eqn:vZ10-e02=0}
\end{equation}
together with a formal solution that satisfies Eq. \eqref{eqn:vZ10-e02=0} as
\begin{equation}
  k^2 = \overline{k_{0.2}}^2
      = \frac{3}{5} + K'_2 \alpha^2 + K'_4 \alpha^4 + \cdots ,
  \label{eqn:vZ10-nn382-3}
\end{equation}
where $K'_2$, $K'_4$ are some numerical coefficients.

Note that in \citeauthor{vonzeipel1910}'s original notation (p. Z382),
$\overline{k_{0.2}}^2$ in Eq. \eqref{eqn:vZ10-nn382-3} is denoted as ${k'}^2_{0.2}$.
However, ${k'}^2_{0.2}$ is used later for a different purpose
(Eq. \eqref{eqn:Z100} on p. \pageref{sssec:oCR3BP-extremum} of this monograph).
For avoiding the probable confusion that readers might have,
in this monograph we have altered the notation
from ${k'}^2_{0.2}$ to $\overline{k_{0.2}}^2$ in Eq. \eqref{eqn:vZ10-nn382-3}.
Note also that \citeauthor{vonzeipel1910} did not mention at all
what $K'_2$, $K'_4$, $\cdots$ in Eq. \eqref{eqn:vZ10-nn382-3} are.
Their function forms are not shown either.
We interpret they are some numerical coefficients
similar to those $(K_2, K_4, \cdots)$ in Eq. \eqref{eqn:Z79}.

From Eq. \eqref{eqn:Z81} it is straightforward to understand that
$k^2 < \frac{3}{5}$ is necessary for $e^2_{0.2}$ to be positive\footnote{%
Actually,
\citeauthor{vonzeipel1910}'s original logic behind these equations
described on p. Z382 does not seem quite straightforward.
First, his equation \eqref{eqn:vZ10-nn382-3} means that,
when the value of $k^2$ approaches $\overline{k_{0.2}}^2$,
the two solutions $(x,y)=(0, \pm e_{0.2})$ of
the equation $\DP{R}{\left(e^2\right)} = 0$
would both approach the origin $(x,y)=(0,0)$.
We already learned the behavior of $R$ at the origin
(p. \pageref{pg:iCR3BP-origin}).
Then on p. Z382,
\citeauthor{vonzeipel1910} mentions the equivalence of
$k^2_{0.2}$            in Eq. \eqref{eqn:Z79} and
$\overline{k_{0.2}}^2$ in Eq. \eqref{eqn:vZ10-nn382-3}.
$k^2_{0.2}$ is the threshold value of $k^2$ that determines the sign of $R_{0.2}$ in Eq. \eqref{eqn:Z78},
hence it determines the type of local extremum at the origin $(0,0)$.
The origin would become a local minimum of $R$ when $k^2 > k^2_{0.2}$.
On the other hand,
$\overline{k_{0.2}}^2$ is the threshold value of $k^2$ that determines the value of $e_{0.2}$ in Eq. \eqref{eqn:Z81}.
$e_{0.2}$ cannot exist as a real number if $k^2 > \overline{k_{0.2}}^2$, and
$\DP{R}{\left(e^2\right)}$ would become positive all along the direction of
$\cos 2g = -1$.
Therefore
the origin would become a local minimum of $R$ when $k^2 > \overline{k_{0.2}}^2$.
From this functional compatibility between $k^2_{0.2}$ and $\overline{k_{0.2}}^2$,
\citeauthor{vonzeipel1910} concludes the equivalence
\begin{equation}
  \overline{k_{0.2}}^2 \equiv k^2_{0.2} ,
  \label{eqn:vZ10-nn382-4b}
\end{equation}
although his description on p. Z382 is very terse.%
}, % End of \footnote
as long as $\alpha \ll 1$.
When $k^2 =    \frac{3}{5}$, we have $e^2_{0.2} = 0$, and
the positions of the local extremums converge into the origin $(0,0)$.
When $k^2 > \frac{3}{5}$, we have $e^2_{0.2} < 0$, and
$e_{0.2}$ cannot be a real number.
Therefore, we find $k^2 < \frac{3}{5}$ as a condition for local extremums
to show up on the $y$-axis (but except at the origin).
It is also obvious that the value $\frac{3}{5}$ is equivalent to
$k^2_{0.2}$ described in Eq. \eqref{eqn:Z79} as long as $\alpha \ll 1$.

Now \citeauthor{vonzeipel1910} starts his examination of
the type of the local extremums at $(x,y) = (0, \pm e_{0.2})$;
in other words, at $(e,g) = (e_{0.2}, \pm\frac{\pi}{2})$.
Which type of local extremum would $R$ take at these points, local minima or maxima?

\paragraph{Type of the local extremum on the $y$-axis\label{par:iCR3BP-type-ext}}
For examining the characteristics of the points
$(e,g) = (e_{0.2}, \pm\frac{\pi}{2})$ as local extremums of $R$,
\citeauthor{vonzeipel1910} focuses on the sign of
the second derivatives of $R$ at $\cos 2g = -1$.
They are $\DP[2]{R}{g}$, $\DPsd{R}{e}{g}$, and $\DP[2]{R}{e}$.

As for $\DP[2]{R}{g}$, from Eq. \eqref{eqn:Z76} we have
\begin{equation}
  \left. \DP[2]{R}{g} \right|_{\cos 2g = -1} = \frac{15}{4}e^2 \sin^2 I,
  \label{eqn:vZ10-DP2Rig2}
\end{equation}
which is clearly larger than, or equal to, zero.

As for $\DPsd{R}{e}{g}$, it is also clear that
\begin{equation}
  \left. \DPsd{R}{e}{g} \right|_{\cos 2g = -1} = 0 .
  \label{eqn:vZ10-DP2Rieg}
\end{equation}

\label{pg:iCR3BP-dp2Rde2}
$\DP[2]{R}{e}$ is slightly complicated, and
\citeauthor{vonzeipel1910}'s original description seems too terse.
Let us try to add a complementary description as to
how we should deal with $\DP[2]{R}{e}$.

\citeauthor{vonzeipel1910} shows an identity on p. Z383 as
\begin{equation}
         \frac{\partial^2 R}{\partial^2 e}
 \equiv 2\frac{\partial   R}{\partial \left(e^2\right)}
   + 4e^2\frac{\partial^2 R}{\partial \left(e^2\right)^2} .
  \label{eqn:vZ10-nn383-1}
\end{equation}
Adopting Eq. \eqref{eqn:vZ10-nn383-1} for $\DP[2]{R}{e}$, we get
\begin{equation}
\begin{aligned}
  \left. \DP[2]{R}{e} \right|_{\cos 2g = -1} 
                & = \frac{3}{4\left( 1 - e^2\right)^3} \\
                \times
                &
    \left( 3 e^6 - 9 e^4 + 9 e^2 + 5 k^2 \left(3e^2 +1\right) -3 \right) .
\end{aligned}
  \label{eqn:vZ10-DP2Rie2}
\end{equation}

By substituting the eccentricity value ($e = e_{0.2}$ of Eq. \eqref{eqn:Z81})
at the considered points
into Eq. \eqref{eqn:vZ10-DP2Rie2}, we get
\begin{equation}
  \left. \DP[2]{R}{e} \right|_{\cos 2g = -1, e = e_{0.2}}
    = \frac{9}{5} \sqrt{\frac{15}{k^2}} - 9 .
  \label{eqn:vZ10-DP2Rie2-kk}
\end{equation}

Equation \eqref{eqn:vZ10-DP2Rie2-kk} means that, at $\cos 2g = -1$,
$\DP[2]{R}{e}$ monotonically decreases as $k^2$ increases.
But it remains positive while $k^2 < \frac{3}{5}$.
Here we must remember that, as mentioned before,
$k^2 < \frac{3}{5}$ is the necessary condition
for $e^2_{0.2}$ to exist as a positive number (see Eq. \eqref{eqn:Z81}).
Therefore in this range of $k^2$, we have the following:
\begin{equation}
  \left. \DP[2]{R}{e} \right|_{\cos 2g = -1, e = e_{0.2}} > 0 .
  \label{eqn:vZ10-DP2Rie2-kk-gt0}
\end{equation}

In summary, the three second derivatives
\eqref{eqn:vZ10-DP2Rig2},
\eqref{eqn:vZ10-DP2Rieg}, and
\eqref{eqn:vZ10-DP2Rie2-kk-gt0}
indicate that $R$ takes local minima on the $(x,y)$ plane when $\cos 2g = -1$.
Their coordinates are $(x,y) = (0, \pm e_{0.2})$;
in other words, $(e,g) = (e_{0.2}, \pm\frac{\pi}{2})$.

Taking into account all the results that have been obtained,
\citeauthor{vonzeipel1910} states the following proposition:
\begin{quote}
``That being so, we can state the following proposition as demonstrated:
\par
\hspace*{1em}
\textit{If $\alpha$ is small and $k^2_{0.2}<k^2<1$, the function $R$ possesses one minimum value in the domain (Z43), located at the point $e=0$.
--- Instead, if $\alpha$ being small, we have $0<k^2<k^2_{0,2}$, then $R$ has a minimax at the origin and its only minimum value in the domain (Z43) at the two symmetric points $g=\pm\frac{\pi}{2}$, $e=e_{0.2}$.
--- Finally, the constant value that $R$ takes on the circle $e=k'$ is the largest value in the domain (Z43); and there is no other maximum value for $R$ in this domain.}'' (p. Z383)
\end{quote}

\begin{figure}[htbp]\centering
\ifepsfigure
 \includegraphics[width=\singlefigwidth\textwidth]{fig_Z6-7.eps} %fig13
\else
 \includegraphics[width=\singlefigwidth\textwidth]{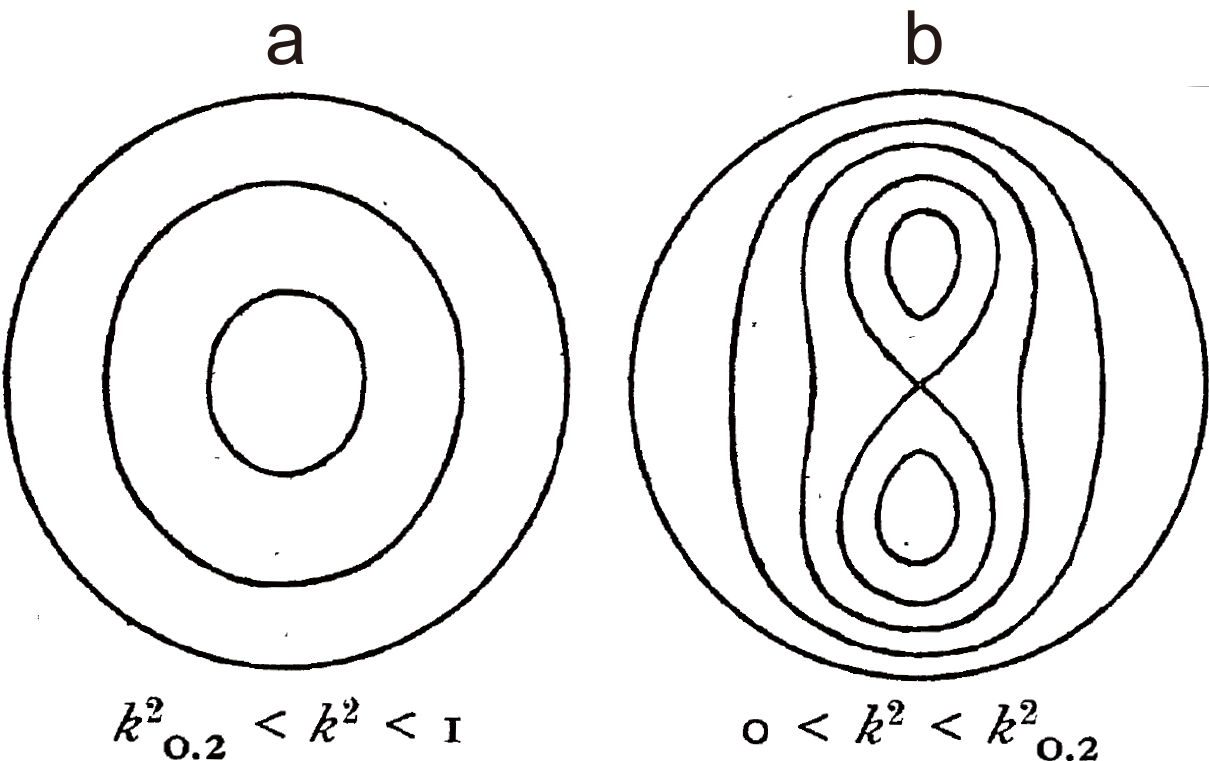} %fig13
\fi
  \caption{%
Transcription of \citeauthor{vonzeipel1910}'s \mysymfigS Z6 and Z7 on his p. Z384
(Credit: John Wiley and Sons. Reproduced with permission).
  They represent schematic trajectories of equi-$R$ contours of
  the doubly averaged inner CR3BP on the $(e \cos g, e \sin g)$ plane.
  \mtxtsf{a}: \citeauthor{vonzeipel1910}'s \mysymfigO Z6,
  showing the status when $k^2_{0.2} < k^2 < 1$.
  \mtxtsf{b}: \citeauthor{vonzeipel1910}'s \mysymfigO Z7,
  showing the status when $0 < k^2 < k^2_{0.2}$.
  }
  \label{fig:vZ10-f6-7}
\end{figure}

After these considerations, 
\citeauthor{vonzeipel1910} presents two schematic plots of equi-$R$ curves
on the $(e \cos g, e \sin g)$ plane as his \mysymfigS Z6 and Z7
for illustrating the circumstances.
We transcribed them as our \mysymfigO \ref{fig:vZ10-f6-7}.
Through his figures, it is obvious that
the topology of the equi-$R$ contours qualitatively changes across $k^2 = k^2_{0.2}$.
Recalling the fact that $k^2_{0.2} = \frac{3}{5}$ when $\alpha \ll 1$,
it is evident that \citeauthor{vonzeipel1910}'s
schematic plots in \mysymfigO \ref{fig:vZ10-f6-7} are exactly about
what \citeauthor{lidov1961} and \citeauthor{kozai1962b} reached
at the quadrupole level approximation.
See \mysymfigO \ref{fig:xy-inner} of this monograph for comparison.

\subsubsection{When $\alpha$ is not so small\label{sssec:iCR3BP-nosmall-alpha}}
In the subsequent sections from Z17 to Z21,
\citeauthor{vonzeipel1910} deals with the doubly averaged inner CR3BP
when 
the semimajor axis ratio
$\alpha = \frac{a}{a'}$ is not negligibly small.
He takes care of this problem through a kind of numerical method.
His discussions in what follows begin with an examination of the behavior of $R$ at the origin $(x,y)=(0,0)$.
Then he moves on to a consideration of the dynamical characteristics of
several asteroids that actually exist in the solar system.
After that, he returns to considering the behavior of $R$ along the $y$-axis.

\paragraph{At the origin\label{pg:iCR3BP-origin-largealpha}}
\citeauthor{vonzeipel1910} first goes on to consider
whether or not $R$ takes a local minimum at the origin $(0,0)$,
even when $\alpha$ is not so small.
This is the objective of his Section Z17.
For this purpose \citeauthor{vonzeipel1910} expands
$R_{2.0}$ and $R_{0.2}$ using the Laplace coefficients,
in a different way from what he demonstrated
in Eqs. \eqref{eqn:Z72}, \eqref{eqn:Z77}, and \eqref{eqn:Z78}.
This is because \citeauthor{vonzeipel1910} thought the series expansion
Eq. \eqref{eqn:Z72} converges too slowly when $\alpha$ is not small.
Nowadays, this is rather a famous fact in the context of comparison between
expansions of the disturbing function using the Legendre polynomials and
that using the Laplace coefficient \citep[e.g.][]{farago2010}.
The advantage of the former expansion is that there is no restriction in
the value range of eccentricity in the secular form of the disturbing function,
whereas the latter requires truncation at some order of eccentricity.
As a result,
\citeauthor{vonzeipel1910}'s expansions of $R_{2.0}$ and $R_{0.2}$ in this section seem limited to $O\left(e^2\right)$.
We do not get into the detail of the elaborate formulations
that \citeauthor{vonzeipel1910} develops in Section Z17, and
we confine ourselves to introducing only the major formulas that he showed.
\citeauthor{vonzeipel1910}'s formulations in this section
rely largely on the legacy of celestial mechanics developed
in the nineteenth century, in particular, by Tisserand and Jacobi.

\citeauthor{vonzeipel1910} first introduces a set of new coefficients
$c^{i,k}$ and $e^{i,k}$, and shows how $R_{2.0}$ and $R_{0.2}$ are
expressed by these. $c^{i,k}$ and $e^{i,k}$ are defined as follows
(pp. Z383--Z384):

\begin{alignat}{1}
& \left[ 1+\alpha^2-2\alpha\left(\mu\cos M+\nu\cos N\right) \right]^{-\frac{3}{2}} \nonumber \\
  & \quad= c^{0.0} + 2c^{1.0} \cos M + 2c^{0.1} \cos N \nonumber \\
  & \quad\quad
      + 4c^{1.1} \cos M \cos N + \cdots ,
  \label{eqn:vZ10-def-cik} \\
& \left[ 1+\alpha^2-2\alpha\left(\mu\cos M+\nu\cos N\right) \right]^{-\frac{5}{2}} \nonumber \\
  & \quad= e^{0.0} + 2e^{1.0} \cos M + 2e^{0.1} \cos N \nonumber \\
  & \quad\quad
    + 4e^{1.1} \cos M \cos N + \cdots .
  \label{eqn:vZ10-def-eik}
\end{alignat}
where
\begin{alignat}{1} 
   M = u + \theta', \quad N = u - \theta' , 
  \label{eqn:vZ10-nn386-01} \\
  \mu + \nu = 1, \quad \mu - \nu = k ,
  \label{eqn:vZ10-nn384-01}
\end{alignat}
and $u$ is eccentric anomaly.
Note that
$e^{i,k}$ in Eq. \eqref{eqn:vZ10-def-eik} is not directly related to eccentricity $e$.
Note also that $\mu$ in Eq. \eqref{eqn:vZ10-nn384-01} is not Jupiter's mass.
$\nu$ in Eq. \eqref{eqn:vZ10-nn384-01} is also a newly defined parameter,
and it does not have anything to do with what showed up in earlier chapters.

Using eccentric anomaly $u$,
\citeauthor{vonzeipel1910} obtains an expression of
the doubly averaged disturbing function $R$ as
\begin{align}
  R &= \frac{1}{4\pi^2}\int_0^{2\pi}\int_0^{2\pi}\frac{1}{\Delta}dld\theta' \nonumber \\
    &= \frac{1}{4\pi^2}\int_0^{2\pi}\int_0^{2\pi}\frac{1-e\cos u}{\Delta} du d\theta' ,
  \tag{Z83-\arabic{equation}}
  \stepcounter{equation}
  \label{eqn:Z83}
\end{align}
with
\begin{align}
  \Delta^2 & = 1 + r^2 \nonumber \\
           & - 2r \left[ \cos(w+g)\cos \theta' - \sin(w+g) \sin\theta' \cos I \right] .
  \label{eqn:vZ10-nn384-03}
\end{align}
Note that $w$ is the true anomaly of the perturbed body.

Now \citeauthor{vonzeipel1910} puts $g=0$ in order to obtain the expression of
$R_{2.0}$ that showed up in Eq. \eqref{eqn:Z77}.
We presume that the substitution of $g=0$ into
Eq. \eqref{eqn:Z83} is justified,
because $R_{2.0}$ serves as the amplitude of the $x^2$-component of $R$
along the $x$-axis by its definition \eqref{eqn:Z77}. Therefore
only $R_{2.0}$ matters along the $x$-axis $\left(g=0, \pi\right)$, and
only $R_{0.2}$ matters along the $y$-axis $\left(g=\pm\frac{\pi}{2}\right)$
as long as $e^2$ (hence $x^2$ and $y^2$) is small.
By putting $g=0$, \citeauthor{vonzeipel1910} obtains an expression for
$\Delta$ in Eq. \eqref{eqn:vZ10-nn384-03} as
\begin{equation}
\begin{aligned}
  \Delta^2 & = 1 + \alpha^2(1-e\cos u)^2 \\
           & - 2\alpha \left[ \mu\cos\left(u+\theta'\right)+\nu\cos\left(u-\theta'\right)-e\cos\theta' \right] .
\end{aligned}
  \label{eqn:vZ10-nn384-04}
\end{equation}

Then he defines a new variable
\begin{equation}
  \Delta^2_0 = 1 + \alpha^2
             - 2\alpha \left[ \mu\cos\left(u+\theta'\right)+\nu\cos\left(u-\theta'\right) \right] ,
  \label{eqn:vZ10-nn385-01}
\end{equation}
and obtains the following expansion
\begin{equation}
\begin{aligned}
 \frac{1-e\cos u}{\Delta}
   = \frac{1}{\Delta_0}
    + e \left(
     - \frac{\cos u}{\Delta_0}
     + \frac{\alpha^2 \cos u}{\Delta_0^3}
     - \frac{\alpha \cos \theta'}{\Delta_0^3} \right) \\
    + e^2 \left(
      -\frac{3}{2}\frac{\alpha^2 \cos^2 u}{\Delta_0^3}
      +           \frac{\alpha \cos u \cos \theta'}{\Delta_0^3}
      +\frac{3}{2}\frac{\alpha^4 \cos^2 u}{\Delta_0^5} \right. \\
       \left.
      - 3         \frac{\alpha^3 \cos u \cos \theta'}{\Delta_0^5}
      +\frac{3}{2}\frac{\alpha^2 \cos^2 \theta'}{\Delta_0^5}
         \right)
      + \cdots .
\end{aligned}
  \label{eqn:vZ10-nn385-02}
\end{equation}

Now, we find that $\Delta_0^{-3}$ and $\Delta_0^{-5}$ appearing in
Eq. \eqref{eqn:vZ10-nn385-02} can be expressed
by the coefficients $c^{i,k}$ and $e^{i,k}$ defined in
Eqs. \eqref{eqn:vZ10-def-cik} and \eqref{eqn:vZ10-def-eik}. 
\citeauthor{vonzeipel1910} eventually finds the expanded form of $R_{2.0}$ as follows:
\begin{equation}
\begin{aligned}
  R_{2.0} =& -\frac{3}{4}\alpha^2 \left( c^{0.0} + c^{1.1} \right)
             +\frac{1}{2}\alpha   \left( c^{1.0} + c^{0.1} \right) \\
           & +\frac{3}{4}\alpha^2 \left( 1 + \alpha^2 \right)
                             \left( e^{0.0} + e^{1.1} \right) \\
           & -\frac{3}{2}\alpha^3 \left( e^{1.0} + e^{0.1} \right) .
\end{aligned}
  \tag{Z85-\arabic{equation}}
  \stepcounter{equation}
  \label{eqn:Z85}
\end{equation}

A similar procedure is applied to $R_{0.2}$.
Since $R_{0.2}$ serves as the amplitude of the $y^2$-component of $R$
along the $y$-axis by its definition \eqref{eqn:Z72},
we should assume $g = \frac{\pi}{2}$ or $g = -\frac{\pi}{2}$
for obtaining the expression of $R_{0.2}$.
By placing $g = \frac{\pi}{2}$ and going through the same procedure,
\citeauthor{vonzeipel1910} reaches the following expression for $R_{0.2}$ as
\begin{equation}
\begin{aligned}
  R_{0.2} =& -\frac{3}{4}\alpha^2     \left(c^{0.0}-c^{1.1}\right)
             +\frac{3}{4}\alpha k     \left(c^{1.0}-c^{0.1}\right) \\
           & -\frac{1}{4}\alpha       \left(c^{1.0}+c^{0.1}\right)
             +\frac{3}{4}\alpha^4     \left(e^{0.0}-e^{1.1}\right) \\
           & -\frac{3}{2}\alpha^3 k   \left(c^{1.0}-c^{0.1}\right)
             +\frac{3}{4}\alpha^2 k^2 \left(c^{0.0}-c^{1.1}\right) .
\end{aligned}
  \tag{Z87-\arabic{equation}}
  \stepcounter{equation}
  \label{eqn:Z87}
\end{equation}

Now, he brings up the following recursion formulas
between $c^{i.k}$ and $e^{i.k}$ by citing Jacobi's
\textit{Gesammelte Werke,\/} vol. 6, p. 140 \citep{weierstrass1891}:
\begin{alignat}{1}
& 3 c^{0.0}+c^{1.1} \nonumber \\
  & \quad = 3\left(1+\alpha^2\right)  \left(e^{0.0}-e^{1.1}\right)
    -6\alpha\left(\mu-\nu\right)\left(e^{1.0}-e^{0.1}\right),
  \label{eqn:vZ10-p386-nn01} \\
& 3 c^{0.0}-c^{1.1} \nonumber \\
  & \quad = 3\left(1+\alpha^2\right)  \left(e^{0.0}+e^{1.1}\right)
    -6\alpha\left(\mu+\nu\right)\left(e^{1.0}+e^{0.1}\right).
  \label{eqn:vZ10-p386-nn02}
\end{alignat}
Using Eqs. \eqref{eqn:vZ10-p386-nn01} and \eqref{eqn:vZ10-p386-nn02},
\citeauthor{vonzeipel1910} simplifies Eqs. \eqref{eqn:Z85} and \eqref{eqn:Z87} as follows:
\begin{equation}
\begin{aligned}
  R_{2.0} &= -\alpha^2 c^{1.1} +\frac{1}{2}\alpha  \left(c^{1.0}+c^{0.1}\right), \\
  R_{0.2} &=  \alpha^2 c^{1.1} -\frac{1}{4}\alpha  \left(c^{1.0}+c^{0.1}\right)  \\
          &               +\frac{3}{4}\alpha k\left(c^{1.0}-c^{0.1}\right)
                          -\frac{3}{4}\alpha^2\left(1-k^2\right)
                                         \left(e^{0.0}-e^{1.1}\right) .
\end{aligned}
  \tag{Z88-\arabic{equation}}
  \stepcounter{equation}
  \label{eqn:Z88}
\end{equation}

Then \citeauthor{vonzeipel1910} expands $c^{i.k}$ and $e^{i.k}$
into a power series of $\alpha$ assuming $|\alpha| < 1$
(here he cites \citet[][p. 111]{poincare1907}),
but not in the manner of Eq. \eqref{eqn:Z72} because of its slow convergence
when $\alpha$ is not small.
Instead, he introduces new coefficients $b^{i.k}$ defined as
\begin{equation}
\begin{aligned}
  \left[ 1 + \alpha^2 - 2\alpha\left(\mu\cos M + \nu\cos N\right)\right]^{-\frac{1}{2}} \\
 = b^{0.0} + 2b^{1.0} \cos M + 2b^{0.1} \cos N \\
 + 4b^{1.1} \cos M \cos N + \cdots ,
\end{aligned}
  \label{eqn:vZ10-def-bik}
\end{equation}
and calculates $c^{i.k}$ and $e^{i.k}$ using the recursion formulas provided by Jacobi
(\textit{Gesammelte Werke,\/} vol.~6, p. 142) as
\begin{equation}
\begin{aligned}
  \varepsilon & = \alpha + \frac{1}{\alpha}, \\
  c^{0.0} + c^{1.1} &= \frac{ \varepsilon\left(b^{0.0}-3b^{1.1}\right)-2 \left(b^{1.0}+b^{0.1}\right)}
                            {\alpha\left(\varepsilon^2-4\right)}, \\
  c^{0.0} - c^{1.1} &= \frac{ \varepsilon\left(b^{0.0}+3b^{1.1}\right)-2k\left(b^{1.0}-b^{0.1}\right)}
                            {\alpha\left(\varepsilon^2-4k^2\right)}, \\
  c^{1.0} + c^{0.1} &= \frac{2\left(b^{0.0}-3b^{1.1}\right)-\varepsilon\left(b^{1.0}+b^{0.1}\right)}
                            {\alpha\left(\varepsilon^2-4\right)}, \\
  c^{1.0} - c^{0.1} &= \frac{2k\left(b^{0.0}+3b^{1.1}\right)-\varepsilon\left(b^{1.0}-b^{0.1}\right)}
                            {\alpha\left(\varepsilon^2-4k^2\right)}, \\
  e^{0.0} - e^{1.1} &= \frac{\varepsilon\left(3c^{0.0}+c^{1.1}\right)+2k\left(c^{1.0}-c^{0.1}\right)}
                            {3\alpha\left(\varepsilon^2-4k^2\right)} .
\end{aligned}
  \tag{Z89-\arabic{equation}}
  \stepcounter{equation}
  \label{eqn:Z89}
\end{equation}

Here \citeauthor{vonzeipel1910} obtains the expressions of
$b^{0.0}$, $b^{1.0}$, $b^{0.1}$, $b^{1.1}$ 
with the aid of \citet[][note that \citeauthor{vonzeipel1910} cites pp. 444--447 (Sections 191 and 192 in Chapter XXVIII) of this book, but no equivalent expression to Eq. \eqref{eqn:Z90} is found there]{tisserand1889} as follows:
\begin{equation}
\begin{aligned}
  b^{0.0} &= \frac{1}{2}b^{(0)}
             +\sum_{m=1}^\infty b^{(2m)}   Q_{0.0}^{(2m  )}, \\
  b^{0.1} &=  \sum_{m=1}^\infty b^{(2m-1)} Q_{0.1}^{(2m-1)}, \\
  b^{1.0} &=  \sum_{m=1}^\infty b^{(2m-1)} Q_{1.0}^{(2m-1)}, \\
  b^{1.1} &=  \sum_{m=1}^\infty b^{(2m)}   Q_{1.1}^{(2m)},
\end{aligned}
  \tag{Z90-\arabic{equation}}
  \stepcounter{equation}
  \label{eqn:Z90}
\end{equation}
where $b^{(i)}$ are (according to \citeauthor{vonzeipel1910}) the Laplace coefficients.
The definition of $Q_{i.j}$ are shown later.

Note that what \citeauthor{vonzeipel1910} calls the Laplace coefficients
$b^{(i)}$ in Eq. \eqref{eqn:Z90} seems slightly different from what we now see
in modern textbooks \citep[e.g.][p. 495]{brouwer1961}
because there is no subscript such as $b^{(i)}_s$.
From the definition of $b^{i.k}$ in Eq. \eqref{eqn:vZ10-def-bik},
we suspect that \citeauthor{vonzeipel1910}'s $b^{(i)}$ is equivalent to
$b^{(i)}_{\frac{1}{2}}$ in modern textbooks: The Laplace coefficients with the smallest subscript.

Next \citeauthor{vonzeipel1910} calculates the function $Q_{i.j}^{(n)}$ in Eq. \eqref{eqn:Z90}
which depends only on $\mu$ and $\nu$.
For this purpose he employed the following relationships
\eqref{eqn:Z(c)}, \eqref{eqn:Z(c)}, \eqref{eqn:Z(c)}
originally seen in \citet[][Chapter XXVIII, p. 447, p. 452, and p. 456]{tisserand1889}:
\begin{equation}
  2Q_{i.j}^{(n)} = R_{i.j}^{(n)} - R_{i.j}^{(n-2)} ,
  \tag{Z(c)-\arabic{equation}}
  \stepcounter{equation}
  \label{eqn:Z(c)}
\end{equation}
\begin{equation}
\begin{aligned}
 R_{i,j}^{(n)}
  & = c_{i.j}^{(n)} \mu^i \nu^j & \\
  & \times F^2
    \left( \frac{i+j-n}{2},\frac{i+j+n+2}{2},j+1,\nu \right) ,
\end{aligned}
  \tag{Z(d)-\arabic{equation}}
  \stepcounter{equation}
  \label{eqn:Z(d)}
\end{equation}
\begin{equation}
  c_{i,j}^{(n)} = \frac{\Pi \left( \frac{n+i+j}{2} \right)
                        \Pi \left( \frac{n-i+j}{2} \right) }
                     { \left[\Pi(j)\right]^2 
                        \Pi \left( \frac{n+i-j}{2} \right)
                        \Pi \left( \frac{n-i-j}{2} \right) } ,
  \tag{Z(e)-\arabic{equation}}
  \stepcounter{equation}
  \label{eqn:Z(e)}
\end{equation}
where
\begin{equation}
  \Pi(s) = 1 \cdot 2 \cdot 3 \cdots s = s! .
\end{equation}

\begin{figure}[t]\centering
\ifepsfigure
 \includegraphics[width=\singlefigwidth\textwidth]{R202.eps} %fig14
\else
 \includegraphics[width=\singlefigwidth\textwidth]{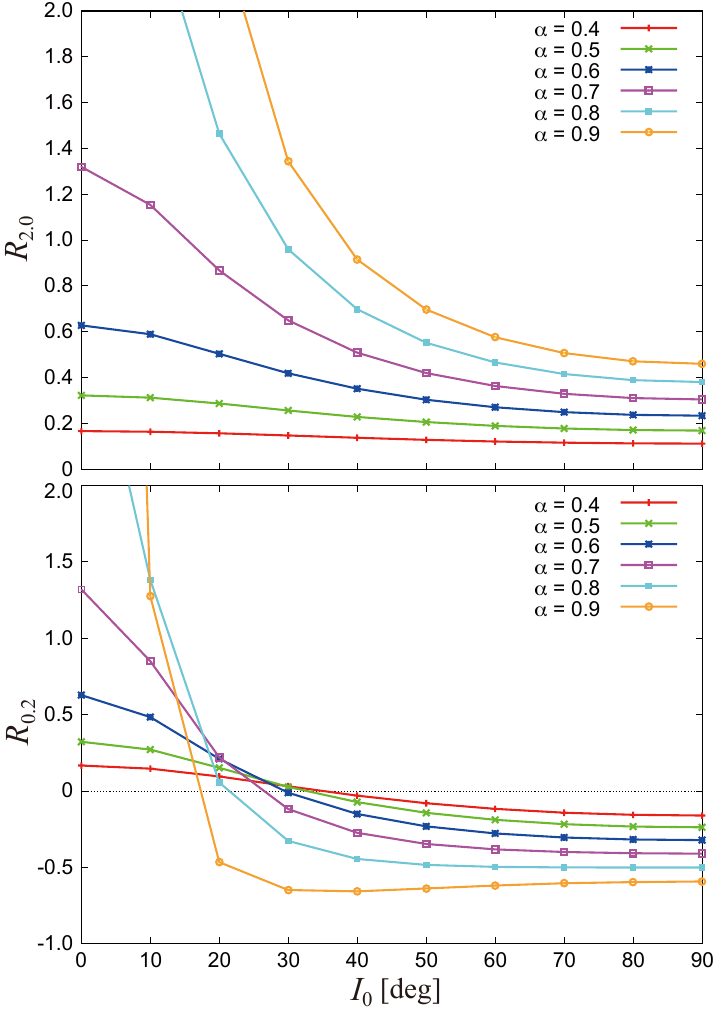} %fig14
\fi
  \caption{%
  Dependence of the numerical values of $R_{2.0}$ (upper) and $R_{0.2}$ (lower)
  on $I_0 = \cos^{-1} k$ and on $\alpha$.
  The plots are based on the values tabulated on an unnumbered table
  in \citet[][pp. Z389--Z390]{vonzeipel1910}.
  }
  \label{fig:R020-table}
\end{figure}
\clearpage

Using the formulas \eqref{eqn:Z88}, \eqref{eqn:Z89}, \eqref{eqn:Z90},
\eqref{eqn:Z(c)}, \eqref{eqn:Z(d)}, \eqref{eqn:Z(e)},
\citeauthor{vonzeipel1910} calculates the actual numerical values of
$R_{2.0}$ and $R_{0.2}$.
He tabulated them on an unnumbered table on pp. Z389--Z390.
The table shows the values of $R_{2.0}$ and $R_{0.2}$
having $I_0 \left(= \cos^{-1} k\right)$ and $\alpha$ as parameters.
The range of $I_0$ is from 0 to $90^\circ$ with an interval of $10^\circ$, and
the range of $\alpha$ (or $a$ in \citeauthor{vonzeipel1910}'s notation) is
from 0.4 and 0.9 with an interval of 0.1.
Instead of just transcribing the table in this monograph,
we make visual plots of their values in our \mysymfigO \ref{fig:R020-table}.

Looking at the upper panel of \mysymfigO \ref{fig:R020-table},
it is clear that $R_{2.0} > 0$ is true everywhere in the given parameter range
($0.4 \leq \alpha \leq 0.9$ and $0 \leq I_0 \leq \frac{\pi}{2}$).
\citeauthor{vonzeipel1910} also gave a further guess about $R_{2.0}$:
\begin{quote}
``It is even likely that $R_{2.0}$ remains $>0$, whenever $0 < \alpha \leq 1$.''
(p. Z389)
\end{quote}

Although \citeauthor{vonzeipel1910} placed the above statement here just
as a conjecture without a rigorous proof,
it seems that this conjecture is probably true.
Indeed at each of the $I_0$ values,
$R_{2.0}$ monotonically increases as $\alpha$ increases
over the entire range of $I_0$.
It is unlikely that $R_{2.0}$ becomes negative when $\alpha$ gets larger than 0.9.
On the other hand when $\alpha$ becomes smaller than 0.4,
the quadrupole level approximation $\left(\alpha^2 \ll 1\right)$ would be applicable,
and $R_{2.0}$ would remain positive according to Eq. \eqref{eqn:Z78}.

In contrast to $R_{2.0}$,
$R_{0.2}$ takes both positive and negative values
(see the lower panel of \mysymfigO \ref{fig:R020-table}).
Let us cite \citeauthor{vonzeipel1910}'s description
about the behavior of $R_{0.2}$:
\begin{quote}
``It also appears that the equation $R_{0.2}=0$ defines a function $k^2_{0.2}$ of $a$, increases with $a$ (at least as $a \leq 0.9$) and takes the value $k^2_{0.2}=+\frac{3}{5}$ for $a=0$, and probably the value $k^2_{0.2} = +1$ for $a=1$.
We finally have
$R_{0.2}>0$ when $k^2_{0.2} < k^2 < 1$, and
$R_{0.2}<0$ when $0         < k^2 < k^2_{0.2}$.''
(p. Z389)
\end{quote}

By interpolating the numerical values of $R_{0.2}$,
\citeauthor{vonzeipel1910} calculated the values of
$I_{0.2} = \cos ^{-1} k_{0.2}$ that realize $R_{0.2} = 0$ at each $\alpha$.
He tabulated them in an unnumbered table on p. Z389.
Here, we should recall the fact that
the origin $(x,y) = (\xi, \eta) = (0,0)$ changes its characteristics
as a local extremum of $R$ according to the sign of $R_{0.2}$.
This is obvious from the definition of $R_{2.0}$ and $R_{0.2}$ in Eqs. \eqref{eqn:Z77} and \eqref{eqn:Z78}.
From the development form in Eq. \eqref{eqn:Z77}
this fact should remain true even when $\alpha$ is not so small.
He then states a theorem as follows:
\begin{quote}
``The function $k_{0.2}$ being thus defined, we can state the following theorem:
\par
\hspace*{1em}
  \textit{At the origin $\xi = \eta = 0$,
  the function $R$ has a minimum if $k^2_{0.2} < k^2 < 1$,
              and a saddle point if $0 < k^2 < k^2_{0.2}$.}'' (p. Z390)
\end{quote}
\label{pg:theorem-Z390}

\begin{figure}[t]\centering
\ifepsfigure
 \includegraphics[width=\singlefigwidth\textwidth]{I02color.eps} %fig15
\else
 \includegraphics[width=\singlefigwidth\textwidth]{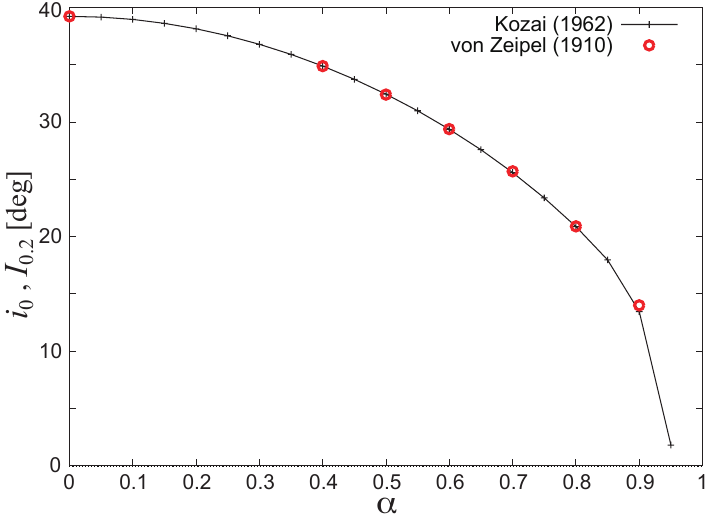} %fig15
\fi
  \caption{%
   Dependence of \citeauthor{vonzeipel1910}'s $I_{0.2}$ and 
                 \citeauthor{kozai1962b}'s $i_0$ on $\alpha$.
   Red open circles denote the values of $I_{0.2} = \cos^{-1} k_{0.2}$ tabulated in \protect{\citet[][p. Z399]{vonzeipel1910}}.
   Black lines with $+$ denote the numerically calculated values of $i_0$ in
   \citet[][his Table I, the third column. It is transcribed in our Table \ref{tbl:Kozai-table1} on p. \pageref{tbl:Kozai-table1} of this monograph]{kozai1962b}.
  }
  \label{fig:I02-table}
\end{figure}

In order to visually confirm
\citeauthor{vonzeipel1910}'s result in comparison with modern knowledge,
in \mysymfigO \ref{fig:I02-table} we plotted the values of $I_{0.2}$ that
he tabulated as a function of $\alpha$.
For comparison, in \mysymfigO \ref{fig:I02-table} we also plotted
\citeauthor{kozai1962b}'s limiting inclination $i_0$
calculated through his numerical harmonic analysis
(see the third column of \citeauthor{kozai1962b}'s Table I in p. K592,
 as well as \citeauthor{kozai1962b}'s \mysymfigO 1 in p. K593.
 See also our Table \ref{tbl:Kozai-table1} on p. \pageref{tbl:Kozai-table1} of this monograph).
As we have learned, both
\citeauthor{vonzeipel1910}'s $I_{0.2}$ and
\citeauthor{kozai1962b}'s $i_0$ represent the threshold value of
the perturbed body's inclination whether or not the origin $(0,0)$ makes a saddle point.
The comparison seen in \mysymfigO \ref{fig:I02-table} evidently tells us that
\citeauthor{vonzeipel1910}'s $I_{0.2}$ almost exactly matches
\citeauthor{kozai1962b}'s    $i_0$ up to $\alpha = 0.9$.
This fact proves very well the correctness of
\citeauthor{vonzeipel1910}'s theory and the calculation
for the existence of local extremums in $R$ even when $\alpha$ is not small.
Hence, the above theorem that \citeauthor{vonzeipel1910} states turns out
not only qualitatively true but also quantitatively accurate.

Right after stating the above theorem,
by recalling the result presented in his Section Z11
where he described the characteristics of
the origin $(x,y)=(0,0)$ as a local extremum of $R$
(see our Section \ref{ssec:Z10eom-secular}, in particular p. \pageref{pg:SectionZ11}),
\citeauthor{vonzeipel1910} states yet another theorem:
\begin{quote}
``By comparing this result to that of no. Z11, we can obviously express the theorem in question in the following way:
\par
\hspace*{1em}
  \textit{%
  Suppose that the eccentricity of the orbit of an infinitely small mass
  located at the interior of the perturbing planet is small at a given moment;
  for this eccentricity to be always be small,
  it is necessary and sufficient that $I < I_{0.2}$.}
\par
\hspace*{1em}
  In this statement we have neglected the perturbing mass and the square of the  eccentricity.'' (p. Z390)
\end{quote}

Note that ``no. Z11'' in the above citation denotes \citeauthor{vonzeipel1910}'s Section Z11. We added the prefix Z for clarity.
Similar expressions (such as ``no. Z7'' or ``no. Z16'') appear several times in the remaining part of this monograph when we cite \citeauthor{vonzeipel1910}'s original descriptions.

Having \citeauthor{vonzeipel1910}'s previous theorem (p. \pageref{pg:theorem-Z390}) in mind,
it is now obvious that his above theorem is true too.
When $I > I_{0.2}$, the origin $(x,y)=(0,0)$ becomes a saddle point, and
the perturbed body cannot always stay near the origin with a small eccentricity.
The origin $(0,0)$ becomes a local minimum only when $I < I_{0.2}$, and
the perturbed body can always stay near the origin with a small eccentricity
if its initial eccentricity is small.
The two circumstances were already depicted in \citeauthor{vonzeipel1910}'s
schematic illustration that we transcribed before (\mysymfigO \ref{fig:vZ10-f6-7} of this monograph).

\label{pg:Z1910-actualasteroids}
\paragraph{Actual asteroids in the solar system}
In Section Z18
\citeauthor{vonzeipel1910} tries to evaluate how large an asteroid's
eccentricity can increase when its inclination $I$ exceeds $I_{0.2}$.
Although we do not go into the contents of Section Z18 in this monograph,
readers with an interest might want to compare his discussion
with modern analytic studies on a similar topic such as 
\citet[][their Eq. (31) on p. 71]{kinoshita2007a} and
\citet[][his Eqs. (28)--(30) on p. 3613]{antognini2015}.
After that, in Section Z19 \citeauthor{vonzeipel1910} picks several asteroids
that were actually recognized by his time from an ephemeris.
He considers them as candidate objects that may have a large secular oscillation
of eccentricity due to their large inclination.
In what follows, we present a brief summary of the contents of his Section Z19.

As \citeauthor{vonzeipel1910} wrote (p. Z392),
orbital elements of 665 asteroids are listed in
Berliner Astronomisches Jahrbuch f\"ur 1911 (pp. (2)--(35)).
\nocite{berlinerastronjahrbuch1911}
Among them, he found that only six asteroids have $I_0$ that is larger than their $I_{0.2}$. They are
  (2) Pallas,
(183) Istria,
(473) Nolli,
(531) Zerlina,
(582) Olympia, and
(594) Mireille.
He points out a possibility that these asteroids
have a secular variation of eccentricity with a large amplitude.
Citing his words:
\begin{quote}
''Among the asteroids, in the number of 665, whose elements are in the Berliner Jahrbuch for the year 1911, there are only six whose quantity
$$
  I_0 = \arccos \left( \sqrt {1-e^2} \cos I \right)
$$
exceeds the limit $I_{0.2}$.
Here are the planets in question
\begin{center}
(2), (183), (473), (531), (582), (594).
\end{center}

\hspace*{1em}
We thus have
\begin{align}
  I_0 & = 38^\circ.0 & I_{0.2} & = 32^\circ.3 & \mbox{ for (594)} \nonumber \\
  I_0 & = 36.7       & I_{0.2} & = 31.5       & \mbox{ for   (2)} \nonumber
\end{align}
while for the other four planets, the difference $I_0 - I_{0.2}$ is less considerable.

\hspace*{1em}
For the two planets (594) and (2), the amplitude of the secular perturbations of the eccentricity must be considerable.'' (p. Z392)
\end{quote}

\begin{figure}[t]\centering
\ifepsfigure
 \includegraphics[width=\singlefigwidth\textwidth]{fig_pallas.eps}%fig16
\else
 \includegraphics[width=\singlefigwidth\textwidth]{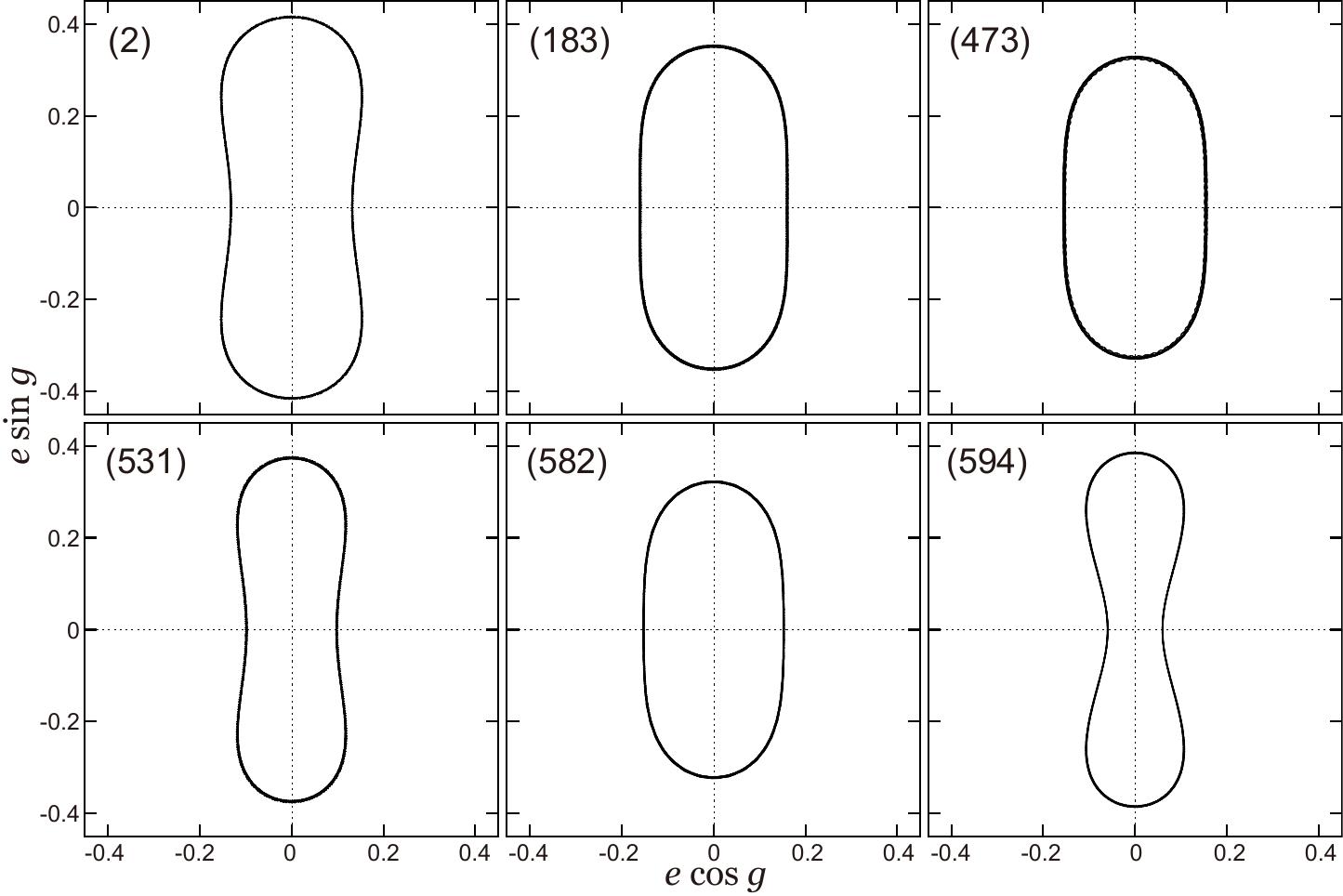}%fig16
\fi
  \caption{%
  Numerically obtained trajectories of the six actual asteroids on the
  $(e\cos g, e\sin g)$ plane
  having Jupiter on a circular orbit as the perturber.
  Upper, from the left: (2) Pallas, (183) Istria, (473) Nolli.
  Lower, from the left: (531) Zerlina, (582) Olympia, (594) Mireille.
}
  \label{fig:aasteroids-ecosg}
\end{figure}

For demonstrating and confirming \citeauthor{vonzeipel1910}'s prediction,
we carried out a set of direct numerical integration of the orbit propagation of
these six asteroids in the framework of CR3BP.
Our numerical integration started from their orbital elements that had been 
published in Berliner Astronomisches Jahrbuch f{\"u}r 1911.
We placed Jupiter on a circular orbit as the perturber.
The numerical integration scheme, stepsize, and data output interval are
all common to what was used to draw \mysymfigO \ref{fig:xy-inner}
(see p. \pageref{fig:xy-inner} of this monograph for details).
The total integration period is one million years, and
the result is shown as \mysymfigO \ref{fig:aasteroids-ecosg}.
We summarized the parameters $k^2$, $\alpha$, and $c_2$ for the six asteroids
in Table \ref{tbl:Zeipel-6asteroids}.
In this table, $\alpha = \frac{a}{a'}$ is the semimajor axes ratio between the asteroid and Jupiter.
$c_2$ is what \citeauthor{lidov1961} devised in Eq. \eqref{eqn:def-Lidov-c2}.
\label{pg:vonzeipel-actualasteroids}

As seen in \mysymfigO \ref{fig:aasteroids-ecosg},
we can say that \citeauthor{vonzeipel1910}'s prediction
``the amplitude of the secular perturbations of the eccentricity must be considerable'' turns out to be largely correct, not only for (2) Pallas and (594) Mireille
but also for the other four asteroids.
Although the $k^2$ values are larger than
the critical value
$\left( k^2_{0.2} = \frac{3}{5} \right)$ of the quadrupole level approximation
(see Table \ref{tbl:Zeipel-6asteroids})
for all the six asteroids,
it seems that most of the asteroids probably possess stationary points
on their disturbing potential along the axis of $g = \pm \frac{\pi}{2}$.
This is because these asteroids have non-negligible $\alpha$ values,
and we cannot simply apply the thresholds that are valid only
at the quadrupole level approximation.

Incidentally,
let us note that among the six asteroids that \citeauthor{vonzeipel1910} picked,
two of them (Pallas and Zerlina) are now categorized in the Pallas family
\citep[e.g.][]{nesvorny2015}.
Mireille may be one of the Pallas family members too \citep[e.g.][his \mysymfigO 1]{kozai1979}.
We believe that
this is the reason why their orbital elements are close to each other.

\citeauthor{vonzeipel1910} also briefly mentions the status of ``the comet Tempel''
in this section as follows:
\begin{quote}
``Regarding the periodic comets, it must be mentioned that the quantity $I_0-I_{0.2}$ is generally $>0$.
It is only the comet Tempel that is [an] exception.
We have, in fact, for this comet
$$
  I_0 = 25^\circ.9 \quad I_{0.2} = 26^\circ.9 .
$$
'' (p. Z392)
\end{quote}
However currently, neither of the comets named ``Tempel'' has that large inclination:
 9P/Tempel 1 has $I \sim 10^\circ.5$, and
10P/Tempel 2 has $I \sim 12^\circ.0$
(from
the JPL Small-Body Database Search Engine).
Therefore we do not go into the above statement any further.

\begin{table}\centering
\footnotesize
\caption[]{%
Parameters $k^2$, $\alpha$, $c_2$ of the six asteroids that
\citeauthor{vonzeipel1910} chose as examples in Section Z19.
We calculated these values from the published orbital elements on \textit{Berliner Astronomisches Jahrbuch f{\"u}r 1911.\/}
}
\label{tbl:Zeipel-6asteroids}
\resizebox{240pt}{!}{%
\begin{tabular}[hbtp]{lccc}
\hline
\multicolumn{1}{c}{name}
               & $k^2$     & $\alpha$ & $c_2$     \\ % $e$      &   $i$  &    $g$ \\
\hline
  (2) Pallas   & 0.6371525 & 0.532757 & 0.0116807 \\ % 0.239089 & 34.708 & 309.027 \\
(183) Istria   & 0.7039341 & 0.536969 & 0.0250717 \\ % 0.349427 & 26.433 & 262.362 \\ 
(473) Nolli    & 0.7316754 & 0.572902 & 0.0161331 \\ % 0.255639 & 27.776 &  57.111 \\
(531) Zerlina  & 0.6540550 & 0.538855 & 0.0068172 \\ % 0.189308 & 34.550 &  53.862 \\
(582) Olympia  & 0.7123067 & 0.503593 & 0.0126522 \\ % 0.226086 & 29.955 & 308.554 \\
(594) Mireille & 0.6207999 & 0.505205 & 0.0151757 \\ % 0.349443 & 32.762 &  76.005 \\
\hline
\end{tabular}
}% End of \resizebox
\end{table}

\paragraph{Along the $y$-axis\label{pg:iCR3BP-xy=0e02-largealpha}}
Back to the search for local extremums of $R$, in Section Z20
\citeauthor{vonzeipel1910} focuses on demonstrating that $R$ possesses
local minima at the two symmetric points $(x,y) = (0, \pm e_{0.2})$,
even when $\alpha$ is not small.
In this section, \citeauthor{vonzeipel1910} assumes $k^2$ is small.
He does not give general demonstrations for arbitrary values of $k^2$,
writing as follows at the beginning of this section:
\begin{quote}
``We have demonstrated at no.~Z16 that the function $R$ possesses minimum values at the two symmetric points
$$
  x=0, \quad
  y=\pm e_{0.2}
$$
if $a$ is small and if
\begin{equation}
  R_{0.2} < 0 .
  \tag{Z92-\arabic{equation}}
  \stepcounter{equation}
  \label{eqn:Z92}
\end{equation}

These minima, do they exist for any values of $a$ when the inequality (Z92) is satisfied?

I did not succeed in demonstrating it generally.

But I will now show that, [for] $a$ having any values, these minima on the axis of $y$ exist, if $k^2$ is small enough.'' (pp. Z392--Z393)
\end{quote}

\citeauthor{vonzeipel1910}'s
mathematical demonstration presented in Section Z20 is rather brief
and concise compared with those in other sections with extreme details.
Formulas presented in Section Z20 show up mostly in their final form,
and a large part of their derivations is omitted.
In some part of what follows
we try to complement what \citeauthor{vonzeipel1910} presented in this section.
This is for facilitating the reader's understanding of his original intention.

\label{pg:Z20}
Section Z20 starts from the expression of
the doubly averaged disturbing function $R$ using a single integral,
rather than a double integral such as Eq. \eqref{eqn:Z44}.
\citeauthor{vonzeipel1910} already carried out a discussion on the
characteristics of $R$ expressed as a single integral as Eq. \eqref{eqn:Z51}
in Section Z14, Chapter III (p. \pageref{eqn:Z47} of this monograph).
Now, changing the integration variable in Eq. \eqref{eqn:Z51}
from true anomaly $w$ to eccentric anomaly $u$,
he formally expresses $R$ as
\begin{equation}
  R = \frac{1}{2\pi a}\int^{2\pi}_0 \frac{r}{\sqrt{1+r^2}} F(\tau) du .
  \tag{Z93-\arabic{equation}}
  \stepcounter{equation}
  \label{eqn:Z93}
\end{equation}

Let us bear in mind that we are now searching local minima along the axis
where $\cos 2g = -1$ is satisfied.
Substituting $g = \pm\frac{\pi}{2}$ into Eq. \eqref{eqn:Z48},
\citeauthor{vonzeipel1910} obtains an expression of $\tau$
(at $g = \pm\frac{\pi}{2}$) as follows
\begin{equation}
\begin{aligned}
{} & \tau^2 = \frac{4r^2}{\left(1+r^2\right)^2}\left(\sin^2 w + \cos^2 I \cos^2 w\right) \\
{} &        = \frac{4a^2}{\left(1+r^2\right)^2}\left[\left(1-e^2\right)\sin^2 u +\frac{k^2}{1-e^2}\left(\cos u-e\right)^2\right] .
\end{aligned}
  \label{eqn:vZ10-nn393-1}
\end{equation}
Note that in Eqs. \eqref{eqn:Z93} and \eqref{eqn:vZ10-nn393-1}
we use $a$ instead of $\alpha$ due to the same formal reason as we stated
on p. \pageref{pg:vZ10-lengthsarenormalized-2}.
\label{pg:vZ10-lengthsarenormalized-3}

Here, \citeauthor{vonzeipel1910} assumes that $k$ is a small quantity of
the first-order. Then he introduces a finite, non-zero, non-dimensional
variable $\rho$ as follows:
\begin{equation}
  1-e^2 = \rho k ,
  \label{eqn:vZ10-nn393-2}
\end{equation}
which is equivalent to
\begin{equation}
  e = \sqrt {1-\rho k} .
  \label{eqn:vZ10-nn393-3}
\end{equation}

Now \citeauthor{vonzeipel1910} states that
$\tau^2$ at $g=\pm\frac{\pi}{2}$ in Eq. \eqref{eqn:vZ10-nn393-1} is
a small quantity of the first-order.
This is true because $k^2$ is a small quantity of the second-order
from his assumption here, and
$1-e^2$ is a small quantity of the first-order due to Eq. \eqref{eqn:vZ10-nn393-2}.
Then, using Eq. \eqref{eqn:vZ10-nn393-3} \citeauthor{vonzeipel1910} expands
$\tau^2$ at $g=\pm\frac{\pi}{2}$ into the series of $k$ as follows (p. Z393):
\begin{equation}
  \tau^2 = 4\alpha^2 \Phi^2 \left[ \rho \sin^2 u + \frac{\left(\cos u -1\right)^2}{\rho} \right] k + \cdots ,
  \label{eqn:vZ10-nn393-5}
\end{equation}
where $\Phi$ is a function of $u$ defined as
\begin{equation}
  \Phi = \frac{1}{1+\alpha^2(1-\cos u)^2} .
  \label{eqn:vZ10-nn393-4}
\end{equation}
Note that in Eqs. \eqref{eqn:vZ10-nn393-5} and \eqref{eqn:vZ10-nn393-4}
we use $\alpha$ instead of $a$ for formal consistency,
as $\Phi$ seems non-dimensional in \citeauthor{vonzeipel1910}'s discussion
in Section Z20.

Adopting Eqs. \eqref{eqn:vZ10-nn393-5} and
              \eqref{eqn:vZ10-nn393-4} to Eq. \eqref{eqn:Z93},
\citeauthor{vonzeipel1910} shows an expanded form of $R$
in the series of $k$ at $g = \pm\frac{\pi}{2}$ as follows:
\begin{equation}
  R = R_0 + \left(P\rho + Q\frac{1}{\rho}\right) k + \cdots ,
  \label{eqn:vZ10-nn393-6}
\end{equation}
where
\begin{equation}
  R_0 = \frac{1}{2\pi}\int_0^{2\pi} \Phi^\frac{1}{2} \left(1-\cos u\right) du,
  \label{eqn:vZ10-nn393-7R}
\end{equation}
\begin{equation}
\begin{aligned}
  P   &= \frac{1}{4\pi}   \int_0^{2\pi} \Phi^\frac{3}{2}    \cos u  du \\
      &\quad
       + \frac{3\alpha^2}{8\pi}\int_0^{2\pi} \Phi^\frac{5}{2} \sin^2 u \left(1-\cos u\right) du ,
\end{aligned}
  \label{eqn:vZ10-nn393-7P}
\end{equation}
\begin{equation}
  Q   = \frac{3\alpha^2}{8\pi}\int_0^{2\pi} \Phi^\frac{5}{2} \left(1-\cos u\right)^3 du .
  \label{eqn:vZ10-nn393-7Q}
\end{equation}

At this point, \citeauthor{vonzeipel1910} claims that
$P$ defined in Eq. \eqref{eqn:vZ10-nn393-7P} and
$Q$ defined in Eq. \eqref{eqn:vZ10-nn393-7Q} are both positive.
Let us cite what \citeauthor{vonzeipel1910} wrote:
\begin{quote}
``$Q$ is a positive quantity.
In the expression of $P$, the second term is obviously positive.
The first term is also positive because the quantity $\Phi$ decreases with $\cos u$.
We therefore have $P>0$, $Q>0$.'' (p. Z394)
\end{quote}
\label{pg:CallAppenPQpositive}

The fact that both $P$ and $Q$ are positive turns out to be important later.
But we think that his above statement
``We therefore have $P>0$, $Q>0$'' needs more confirmation and clarification.
For this purpose,
we have made a little more exposition and put it in Appendix \ref{appen:PQpositive}.

Now,
\citeauthor{vonzeipel1910} moves on to a consideration of a solution of the following equation
\begin{equation}
  \left. \DP{R}{e} \right|_{g=\pm\frac{\pi}{2}} = 0 .
  \label{eqn:vZ10-nn394-dpRe}
\end{equation}
Using the relationship \eqref{eqn:vZ10-nn393-2} between $e$ and $\rho$
together with the expression of $R$ in Eq. \eqref{eqn:vZ10-nn393-6},
he deforms the equation \eqref{eqn:vZ10-nn394-dpRe} as (p. Z394):
\begin{equation}
  -2 \left(P-\frac{Q}{\rho^2}\right) + A_1 k + A_2 k ^2 + \cdots = 0 ,
  \label{eqn:vZ10-nn394-1}
\end{equation}
where $A_1$, $A_2$, $\cdots$ are polynomials of $\rho$ or $\rho^{-1}$
whose actual form he did not show.
Right after stating Eq. \eqref{eqn:vZ10-nn394-1},
\citeauthor{vonzeipel1910} tries to obtain the solution of Eq. \eqref{eqn:vZ10-nn394-dpRe}.
He writes:
\begin{quote}
``It [Eq. \eqref{eqn:vZ10-nn394-1}] has a positive root developed in powers of $k$, and being reduced to $\sqrt{Q:P}$, when $k=0$.
\par
\hspace*{1em}
As a result, the equations
$$
  \DP{R}{e} = 0, \quad
  \DP{R}{g} = 0
$$
are satisfied by putting
\begin{equation}
\begin{aligned}
  g &= \pm \frac{\pi}{2}, \\
  e &= e_{0.2} \\
    &= 1-\frac{1}{2}\sqrt{\frac{Q}{P}}\cdot k + E_2 k^2 + E_3 k^3 + \cdots .
\end{aligned}
  \tag{Z94-\arabic{equation}}
  \stepcounter{equation}
  \label{eqn:Z94}
\end{equation}
\par
\hspace*{1em}
The semi-major axis $a$ being fixed arbitrarily, the series giving $e_{0.2}$ converges if $k$ is small enough.'' (p. Z394)
\end{quote}

Here we see that the solution of Eq. \eqref{eqn:vZ10-nn394-1},
$\rho = \sqrt{Q:P}$ $\bigl( = \sqrt{Q/P} \bigr)$,
can exist as a real number
owing to the fact that both $P$ and $Q$ are positive, as previously shown.
Note that the second equation in \eqref{eqn:Z94} is a Taylor-series expansion
of the eccentricity $e$ by $k$ using the relationship \eqref{eqn:vZ10-nn393-3}.
See its similarity to, and difference from, Eq. \eqref{eqn:Z81} in the $\alpha \ll 1$ approximation.
Note also that
$E_2$, $E_3$, $\cdots$ in Eq. \eqref{eqn:Z94} must be polynomials of $P$ and $Q$,
although \citeauthor{vonzeipel1910} does not mention anything about them.

Now, \citeauthor{vonzeipel1910} moves on to a calculation of
the second derivatives of $R$ at the points
$(e,g) = (e_{0.2}, \pm\frac{\pi}{2})$, and
tries to show that these points are local minima of $R$ when $k$ is small.
From the definition of $\rho$ in Eq. \eqref{eqn:vZ10-nn393-2},
at these points (where $\rho = \sqrt{Q/P}$) he obtains a relation (p. Z394):
\begin{equation}
\begin{aligned}
  \DP[2]{R}{e}
  &= -\frac{2}{k} \DP{R}{\rho} + \frac{4e^2}{k^2}\DP[2]{R}{\rho} \\
  &=  \frac{4}{k} \frac{2Q}{\rho^3} + \cdots \\
  &=  \frac{8P}{k} \sqrt{\frac{P}{Q}} + \cdots > 0 .
\end{aligned}
  \label{eqn:vZ10-nn394-2}
\end{equation}
Note that in the second line of Eq. \eqref{eqn:vZ10-nn394-2},
the first     term  is  at the order of $O\left(k^{-1}\right)$, and
the remaining terms (specifically writing, $-\frac{6Q}{\rho^2} - 2P + \cdots$)
                    are at the order of $O\left(k^{ 0}\right)$.

\citeauthor{vonzeipel1910} also claims that (p. Z394)
\begin{equation}
  \frac{\partial^2 R}{\partial e \partial g} = 0 ,
  \label{eqn:vZ10-nn394-4}
\end{equation}
which is true from the general definition of $R$ in Eq. \eqref{eqn:Z72}
when $g = \pm\frac{\pi}{2}$.

Then from Eqs. \eqref{eqn:Z48} and \eqref{eqn:Z93},
he calculates the following quantity at $g=\pm\frac{\pi}{2}$ (pp. Z393--Z394):
\begin{equation}
\begin{aligned}
\DP[2]{R}{g}
=      & \frac{1}{2\pi}
  \int_0^{2\pi} \frac{r}{a}\frac{1}{\sqrt{1+r^2}} \frac{8\sin^2 I}{\left(1+r^2\right)^2} \\
\times &
  \left(
    \DP{F}{\left(\tau^2\right)}  r^2
  -2\DP{F}{\left(\tau^2\right)}r^2 \sin^2 w \right. \\
& \left.
  + \DP[2]{F}{\left(\tau^2\right)} \frac{8\sin^2 I}{\left(1+r^2\right)^2}
        r^2 \sin^2 w
  \cdot r^2 \cos^2 w \right) du .
\end{aligned}
  \label{eqn:vZ10-nn394-5}
\end{equation}
Note that the dimension of $\tau^2$ is that of $r^{-2}$ from
Eq. \eqref{eqn:vZ10-nn393-1}.
This guarantees the consistency of the dimensions of the left- and right-hand sides of Eq. \eqref{eqn:vZ10-nn394-5}

Next,
\citeauthor{vonzeipel1910} expands $\DP[2]{R}{g}$ into a power series of $k$
by applying the following relationship
between eccentric anomaly and true anomaly to Eq. \eqref{eqn:vZ10-nn394-5}:
\begin{alignat}{1}
  r^2 \sin^2 w &= a^2 \left(1-e^2\right) \sin^2 u ,
  \label{eqn:vZ10-nn395-1-sin} \\
  r^2 \cos^2 w &= a^2 \left(\cos u-e\right)^2 .
  \label{eqn:vZ10-nn395-1-cos}
\end{alignat}
Note that in Eqs. \eqref{eqn:vZ10-nn394-5}, \eqref{eqn:vZ10-nn395-1-sin} and \eqref{eqn:vZ10-nn395-1-cos},
we use $a$ instead of $\alpha$
due to the same formal reason as mentioned before
(see p. \pageref{pg:vZ10-lengthsarenormalized-2} and
     p. \pageref{pg:vZ10-lengthsarenormalized-3} of this monograph).

Among the expanded terms,
\citeauthor{vonzeipel1910} picks only those that are independent of $k$.
He eventually shows an inequality
\begin{equation}
  \DP[2]{R}{g} = 2Q + \cdots > 0 .
  \label{eqn:vZ10-nn395-2}
\end{equation}

The signs of the second derivatives of $R$ presented as
Eqs. \eqref{eqn:vZ10-nn394-2},
     \eqref{eqn:vZ10-nn394-4}, and
     \eqref{eqn:vZ10-nn395-2} tell us that the points
$(e,g) = (e_{0.2}, \pm \frac{\pi}{2})$ are local minima of $R$
for arbitrary values of $\alpha$ as long as $k$ is small.
Here \citeauthor{vonzeipel1910} makes a statement on the equivalence of
the expansion of $e_{0.2}$ in Eq. \eqref{eqn:Z94} to that in Eq. \eqref{eqn:Z81} when $\alpha$ is small,
although no rigorous demonstration is given:
\begin{quote}
``For small values of $\alpha$, the quantity $e_{0.2}$ of the formula (Z94) can obviously be expanded in powers of $\alpha^2$.
The development thus obtained coincides with that given by the equation (Z81).
Indeed, we have seen in no. Z16, that such formula (Z81) gives all the minima
of the function $R$ when $\alpha$ and $k$ are small.'' (p. Z395)
\end{quote}

As we learned in \citeauthor{vonzeipel1910}'s Chapters II and III,
the existence of a local minimum on the equi-potential surface indicates
a possibility to construct the Lindstedt series around it.
\citeauthor{vonzeipel1910} mentions the characteristics of
the perturbed body's orbit around the local minimum
at $(e,g) = (e_{0.2}, \pm \frac{\pi}{2})$.
We literary cite his words:
\begin{quote}
``We can apply the results of this issue by calculating, according to no. Z10,
the Lindstedt series, which exists when the elements are in the vicinity of the points (Z94).
The corresponding orbits belong to a certain class of comets in a stable motion.
In these orbits,
the semimajor axis is arbitrary;
the eccentricity is close to unity;
the distance from the perihelion to the node is close to $\pm\frac{\pi}{2}$;
the inclination is considerable;
Finally, the parameter $a\left(1-e^2\right)$ is small, so that the two nodes are located inside the orbit of Jupiter.'' (p. Z395)
\end{quote}

Note that in the above, ``the parameter $a\left(1-e^2\right)$'' denotes
the semilatus rectum $\ell$ of the perturbed body.
\citeauthor{vonzeipel1910}'s above statement depicts the dynamical characteristics of 
``a certain class of comets in a stable motion''
trapped around one of the stationary points,
$(e,g) = (e_{0.2}, \pm \frac{\pi}{2})$.

The next section (Z21) is devoted to
describing the dynamical characteristics of an example object
that has such an orbit described above:
The comet 1P/Halley whose $a \sim 18$ au and $\ell \sim 1.15$ au
(these values are due to \citeauthor{vonzeipel1910}).
However, we do not go into this subject right now.
Instead, we will return to it later again
(Section \ref{sssec:Halley} on p. \pageref{sssec:Halley} of this monograph)
after we have reviewed the case when $\alpha > 1$ (Sections Z22--Z23)
and the case when the orbits of the perturbed and perturbing bodies intersect
(Sections Z24--Z25).
This is mainly because 1P/Halley satisfies the condition $\alpha > 1$
with respect to the most massive perturbing planet, Jupiter.
In this regard,
we wonder why \citeauthor{vonzeipel1910} brought up the subject
on the motion of 1P/Halley in a section that discusses
the inner CR3BP where $\alpha < 1$.
\label{pg:Z21}

\subsection{Secular disturbing function: Outer case\label{ssec:ocr3bp}}
One of the biggest differences between
\citeauthor{vonzeipel1910}'s work and those of
\citeauthor{lidov1961} or \citeauthor{kozai1962b} is that,
\citeauthor{vonzeipel1910} dealt with the so-called outer version of
the doubly averaged CR3BP in his \citeyear{vonzeipel1910} publication.
We would dare to say that
detailed studies of the outer CR3BP did not seriously begin until the 1990s
when the actual ``outer'' objects were recognized in the solar system,
such as the transneptunian objects (TNOs) perturbed by Neptune
\citep[e.g.][]{jewitt1993,jewitt1999,thomas1996,lykawka2012}.
Therefore
the relevant publications are much fewer than those on the inner CR3BP.
In his Sections Z22 and Z23, \citeauthor{vonzeipel1910} focuses on
describing the basic theoretical framework of the doubly averaged outer CR3BP
with particular emphasis on the search for local extremums of
the secular disturbing function.
Calculations for locating local extremums of
the doubly averaged disturbing function for the outer CR3BP is substantially more
complicated than those for the inner CR3BP, as we will see in what follows.
\label{pg:outerproblemhasnotbeenstudied}

\subsubsection{Expansion of $R$ by $\alpha'$\label{sssec:oCR3BP-expansion-R}}
Similar to the inner problem that we described
in Section \ref{ssec:icr3bp} (p. \pageref{ssec:icr3bp} of this monograph),
\citeauthor{vonzeipel1910}'s treatment of the doubly averaged outer CR3BP
begins with an expansion of the disturbing function with respect
to the ratio of semimajor axis of the perturbed and perturbing bodies.
Now the variable for the expansion is $\alpha' = \frac{a'}{a}$,
which is smaller than 1.

As for the expansion using $\alpha'$,
\citeauthor{vonzeipel1910} again follows what \citet{tisserand1889} did
with the Hansen coefficients. The resulting expansion form is as follows:
\begin{equation}
\begin{aligned}
  R &= \sum_{m=1}^\infty
       \Bigl( A_{0.0}^{(2m)} X_0^{-(2m+1),0} \\
    &\quad +
        2\sum_{i=1}^m A_{i.i}^{(2m)} X_0^{-(2m+1),2i} \cos 2ig
       \Bigr) {\alpha '}^{2m+1} ,
\end{aligned}
  \tag{Z96-\arabic{equation}}
  \stepcounter{equation}
  \label{eqn:Z96}
\end{equation}
where
\begin{equation}
\begin{aligned}
{} & X_0^{-(2m+1),2i} \\
{} & = \frac{(2m-1)!}{(2m-2i-1)!(2i)!} \left( \frac{e^2}{4} \right)^i
      \left(1-e^2\right)^{-2m+\frac{1}{2}} \\
{} & \quad \times
      F\left(i-m+1,i-m+\frac{1}{2},2i+1,e^2 \right) ,
\end{aligned}
  \tag{Z95-\arabic{equation}}
  \stepcounter{equation}
  \label{eqn:Z95}
\end{equation}
are the Hansen coefficients.
$k_{i,j}^{(2m)}$ and $A_{i,j}^{(2m)}$ are already defined in Eqs. \eqref{eqn:Z73} and \eqref{eqn:Z74}.

Then \citeauthor{vonzeipel1910} rewrites the expansion of $R$ in Eq. \eqref{eqn:Z96} as follows:
\begin{equation}
  R = R'_3 {\alpha '}^3 + R'_5 {\alpha '}^5 + \cdots ,
  \tag{Z97-\arabic{equation}}
  \stepcounter{equation}
  \label{eqn:Z97}
\end{equation}
where
\begin{equation}
\begin{aligned}
  R'_3 &= A_{0.0}^{(2)} X_0^{-3,0} \\
       &= \frac{1}{4}\left( 1-\frac{3}{2} \sin^2 I\right) \left( 1-e^2 \right)^{-2+\frac{1}{2}} , \\
  R'_5 &= A_{0.0}^{(4)} X_0^{-5,0} + 2 A_{1.1}^{(4)} X_0^{-5,2} \cos 2g \\
       &=
      \frac{9}{64} \left[
         \left( 1 - 5\sin^2 I +\frac{35}{8} \sin^4 I \right)
         \left( 1 + \frac{3}{2}e^2 \right) \right. \\
       & \left.
        +\frac{15}{4} \sin^2 I \left( 1-\frac{7}{6}\sin^2 I \right)e^2 \cos 2g
       \right] \left(1-e^2\right)^{-4+\frac{1}{2}} .
\end{aligned}
  \tag{Z98-\arabic{equation}}
  \stepcounter{equation}
  \label{eqn:Z98}
\end{equation}

In Eqs. \eqref{eqn:Z97} and \eqref{eqn:Z98},
we should notice that $R$'s leading term $(R'_3)$ does not depend on $g$.
The $g$-dependence of $R$ first shows up in the next order term $(R'_5)$.
This is a stark difference from the inner problem, and
it makes the secular perturbation in the outer CR3BP
subtler than in the inner problem \citep[e.g.][]{ito2016}.
Note also that in the expression of $R$ in Eq. \eqref{eqn:Z97},
the lowest-order term is of the order of $\Oaldcub$, not $\Oaldsqr$.
Compare Eq. \eqref{eqn:Z97} with the corresponding expansions
in the inner problem (Eq. \eqref{eqn:Z75}) where
the lowest-order term is at the order of $\Oalsqr$ except for a constant.
This is another factor that makes the secular perturbation in the outer CR3BP
weaker than in the inner problem.
\label{pg:outerissubtle}

In \citeauthor{vonzeipel1910}'s original paper,
he uses $a'$ instead of $\alpha' = \frac{a'}{a}$
throughout the discussion on the outer problem.
He poses an unnumbered equation on p. Z397 as
\begin{equation}
  a' = \frac{1}{a} .
  \label{eqn:adash-def-zeipel}
\end{equation}
The numerator $(1)$ in Eq. \eqref{eqn:adash-def-zeipel} denotes the value of the semimajor axis of the perturber (Jupiter).
Hence, his original form of Eq. \eqref{eqn:Z97} is as follows:
\begin{equation}
  R = R'_3 {a'}^3 + R'_5 {a'}^5 + \cdots .
  \label{eqn:Z97-original}
\end{equation}
However,
we think that the use of $a'$ defined by Eq. \eqref{eqn:adash-def-zeipel} is
very confusing for the readers of this monograph,
because we have basically used $a'$ as the semimajor axis of the perturbing body.
For avoiding the confusion, and
maintaining compatibility with the descriptions in the previous sections,
in what follows we use the variable $\alpha'$
instead of \citeauthor{vonzeipel1910}'s $a'$.
We also need to be aware that in Eq. \eqref{eqn:Z97-original}
(therefore in Eq. \eqref{eqn:Z97})
\citeauthor{vonzeipel1910} uses the notation $R$ for
the doubly averaged disturbing function without adding a superscript dash.
Only its expanded components on the right-hand side $\left(R'_3, R'_5, \cdots\right)$
have a superscript dash.

Similar to the discussion on the inner problem (see p. \pageref{eqn:Z97}),
\citeauthor{vonzeipel1910} next expands $R$ of Eq. \eqref{eqn:Z97}
into a two-variable Taylor series using the variables
$x^2 = e^2 \cos^2 g$ and $y^2 = e^2 \sin^2 g$ as follows:
\begin{equation}
  R = R'_{0.0} + R'_{2.0} x^2 + R'_{0.2} y^2 + \cdots .
  \tag{Z99-\arabic{equation}}
  \stepcounter{equation}
  \label{eqn:Z99}
\end{equation}
The coefficients $R'_{2.0}$ and $R'_{0.2}$ themselves are expanded into a power series of $\alpha'$ as
\begin{equation}
\begin{aligned}
  R'_{2.0} &= -\frac{3}{16}\left(1-5k^2\right){\alpha '}^3 \\
           +& \frac{45}{256}\left(1-14k^2+21k^4\right){\alpha '}^5 + P'_7{\alpha '}^7 + \cdots , \\
  R'_{0.2} &= -\frac{3}{16}\left(1-5k^2\right){\alpha '}^3 \\
           +& \frac{45}{256}\left(2-22k^2+28k^4\right){\alpha '}^5 + Q'_7{\alpha '}^7 + \cdots ,
\end{aligned}
  \tag{Z100-\arabic{equation}}
  \stepcounter{equation}
  \label{eqn:Z100}
\end{equation}
where $P'_{2i+1}$ and $Q'_{2i+1}$ are certain polynomials in $k^2$,
although \citeauthor{vonzeipel1910} did not give their actual forms.

\citeauthor{vonzeipel1910} did not show the specific form of the first term
in Eq. \eqref{eqn:Z99}, $R'_{0.0}$.
But we can calculate it by carrying out the two-variable Taylor expansion.
The result is:
\begin{equation}
  R'_{0.0} = -\frac{1  -3k^2}{8} {\alpha '}^3
             +\frac{27-270k^2 +315k^4}{512} {\alpha '}^5
             +\cdots .
\end{equation}
However, similar to the inner case
(Eq. \eqref{eqn:R00-inner-maple} on p. \pageref{eqn:R00-inner-maple}),
constant terms such as $R'_{0.0}$ do not matter in the following discussions.
This is because \citeauthor{vonzeipel1910}'s major concern was to search for local extremums of $R$
by calculating its partial derivatives.
Constant terms such as $R'_{0.0}$ 
in the disturbing function
will be all gone by the differentiation.

\subsubsection{Search of local extremums\label{sssec:oCR3BP-extremum}}
In the rest of Section Z22,
\citeauthor{vonzeipel1910} tries to locate local extremums of $R$ in Eq. \eqref{eqn:Z99}.
The logical structure of this part is the same as
in Section Z16 (see p. \pageref{sssec:iCR3BP-extremum} of this monograph).
He picks several possible regions where $R$ can take local extremums:
At the origin, on the outer boundary, and along the $y$- and $x$-axis
on the $(x,y) = (e \cos g, e \sin g)$ plane.
Then he moves on to checking out
the type of the local extremums that he discovered in these regions.
Following \citeauthor{vonzeipel1910},
in this subsection we assume $\alpha' \ll 1$,

Similar to the discussion on the inner problem,
\citeauthor{vonzeipel1910} begins with obtaining the solution of the pair of equations
\begin{alignat}{1}
  R'_{2.0} &= 0 ,
  \label{eqn:vZ10-Rd20=0} \\
  R'_{0.2} &= 0 .
  \label{eqn:vZ10-Rd02=0}
\end{alignat}

In the inner problem where we have Eqs. \eqref{eqn:Z78} and \eqref{eqn:vZ10-R02=0},
we learned that only $R_{0.2}$ can be zero, but $R_{2.0}$ cannot.
On the other hand in the outer case, we will see that
both $R'_{2.0}$ and $R'_{0.2}$ can be zero.
This fact makes the outer problem more complicated than the inner problem.

\citeauthor{vonzeipel1910} denotes
the root of Eq. \eqref{eqn:vZ10-Rd20=0} as ${k'}^2_{2.0}$, and
the root of Eq. \eqref{eqn:vZ10-Rd02=0} as ${k'}^2_{0.2}$.
Then, using Eq. \eqref{eqn:Z100} he expands ${k'}^2_{2.0}$ and ${k'}^2_{0.2}$ up to $\Oaldsqr$ as
\begin{equation}
\begin{aligned}
  {k'}^2_{2.0} &= \frac{1}{5} + \frac{ 9}{50}{\alpha '}^2 + \cdots, \\
  {k'}^2_{0.2} &= \frac{1}{5} + \frac{12}{50}{\alpha '}^2 + \cdots .
\end{aligned}
  \tag{Z101-\arabic{equation}}
  \stepcounter{equation}
  \label{eqn:Z101}
\end{equation}
It is obvious that both ${k'}^2_{2.0}$ and ${k'}^2_{0.2}$ would approach
$\frac{1}{5}$ in the limit of $\alpha' \to 0$.
Also, we see that
\begin{equation}
  {k'}^2_{2.0} < {k'}^2_{0.2},
  \label{eqn:vZ10-kd20llkd02}
\end{equation}
although the difference is not large when $\alpha'$ is small.

\paragraph{At the origin\label{par:oCR3BP-origin}}
The equations \eqref{eqn:Z99}, \eqref{eqn:Z100}, and \eqref{eqn:Z101} tell us
the following fact as long as ${\alpha '}^2 \ll 1$:
\begin{alignat}{2}
  R'_{2.0} \lessgtr 0 & \iff & k^2 \lessgtr {k'}^2_{2.0} , 
  \label{eqn:vZ10-Rd20-kd20} \\
  R'_{0.2} \lessgtr 0 & \iff & k^2 \lessgtr {k'}^2_{0.2} .
  \label{eqn:vZ10-Rd02-kd02}
\end{alignat}
Combining the facts of
\eqref{eqn:vZ10-Rd20-kd20}, \eqref{eqn:vZ10-Rd02-kd02}, and \eqref{eqn:vZ10-kd20llkd02},
\citeauthor{vonzeipel1910} claims that
the doubly averaged disturbing function $R$ in Eq. \eqref{eqn:Z99}
has a local extremum around the origin $(x,y)= (0,0)$.
He also mentions that the origin $(0,0)$ has the following characteristics
as long as $\alpha' \ll 1$ (p. Z399):
\begin{itemize}
  \item $(0,0)$ is a local minimum when ${k'}^2_{0.2} < k^2 < 1$,
  \item $(0,0)$ is a saddle point  when ${k'}^2_{2.0} < k^2 < {k'}^2_{0.2}$,
  \item $(0,0)$ is a local maximum when $ 0 \leq     k^2 < {k'}^2_{2.0}$.
\end{itemize}
We can accept these conclusions without any doubt because of the expansion form shown in Eq. \eqref{eqn:Z99}.

\paragraph{At the outer boundary\label{pg:oCR3BP-kd}}
\citeauthor{vonzeipel1910} then moves on to a consideration of $R$
on its outer boundary, $e=k'$.
In the inner problem, the boundary circle is a kind of
a local maximum of $R$ (see p. \pageref{pg:iCR3BP-kd}).
Similarly,
by calculating the partial derivative $\frac{\partial R}{\partial (e^2)}$
and substituting $e^2 = {k'}^2= 1-k^2$ into it,
on the boundary circle he obtains for the outer case (p. Z399)
\begin{eqnarray}
\left. \DP{R}{\left(e^2\right)} \right|_{e^2={k'}^2}
  &=& \left. \DP{R'_3}{\left(e^2\right)} \right|_{e^2={k'}^2} {\alpha '}^3+ \cdots \nonumber \\
  &=& \frac{3}{4} \left(k^2\right)^{-\frac{5}{2}} {\alpha '}^3 + \cdots > 0 .
  \label{eqn:vZ10-nn399-1}
\end{eqnarray}

As long as $\alpha'$ is so small that we can ignore
the higher-order terms of $\alpha'$,
the right-hand side of Eq. \eqref{eqn:vZ10-nn399-1} is always positive.
Hence $\frac{\partial R}{\partial (e^2)} > 0$ is true on the outer boundary.
This indicates that,
just inside the outer boundary,
$R$ continues to increase along the radial direction
on the $(e \cos g, e \sin g)$ plane until it reaches $e = k'$.
In this sense, the outer boundary $x^2 + y^2 = {k'}^2$ can be regarded
as a kind of local maximum of $R$, similar to the inner case.
\label{pg:R_outerboundary-o}

\paragraph{Along the $y$-axis and $x$-axis\label{pg:oCR3BP-xy=on-xyaxis}}
Next, \citeauthor{vonzeipel1910} moves on to search for
$R$'s local extremums that may be located somewhere
between the origin $(x,y)=(0,0)$ and the outer boundary
$\bigl( x^2+y^2={k'}^2 \bigr)$.
Similar to the discussion on the inner case (see p. \pageref{pg:iCR3BP-xy=0e02}),
it is necessary to find solutions that satisfy both of the following equations
\begin{equation}
  \DP{R}{\left(e^2\right)} = 0, \quad
  \DP{R}{g} = 0 ,
  \tag{Z102-\arabic{equation}}
  \stepcounter{equation}
  \label{eqn:Z102}
\end{equation}
which is exactly the same as Eq. \eqref{eqn:Z80}.

It is obvious that the operation $\DP{R}{g}$ eliminates the
entire $R'_3$ and the first term of $R'_5$ in Eq. \eqref{eqn:Z98},
and just leaves the second term of $R'_5$.
The second term of $R'_5$ in Eq. \eqref{eqn:Z98} yields a factor of $\sin 2g$
through the operation $\DP{R}{g}$, so the roots of the equation
$\DP{R}{g} = 0$ in Eq. \eqref{eqn:Z102} take place when $\sin 2g = 0$.
This means $\cos 2g = +1$ or $-1$.

From the function form of $R$ in Eqs. \eqref{eqn:Z96} and \eqref{eqn:Z98},
$\DP{R}{\left(e^2\right)}$ becomes as follows when $\cos 2g = +1$:
\begin{align}
& \DP{R}{\left(e^2\right)}
 = {\alpha '}^3 \frac{3}{16}\left(1-e^2\right)^{-\frac{7}{2}} \left(-1+5k^2+e^2\right) \nonumber \\
& \quad
  -{\alpha '}^5 \frac{45\left(1-e^2\right)^{-\frac{13}{2}}}{2048}
   \left( -8 + 112 k^2 - 168 k^4 + 17e^2  \right. \nonumber \\
& \quad \quad \left. 
         -98 e^2 k^2 - 63 e^2 k^4 - 10e^4 -14 e^4 k^2 + e^6 \right) .
  \label{eqn:vZ10-nn399-2-sup2}
\end{align}
When $\cos 2g = -1$, $\DP{R}{\left(e^2\right)}$ becomes as:
\begin{align}
& \DP{R}{\left(e^2\right)}
 = {\alpha '}^3 \frac{3}{16}\left(1-e^2\right)^{-\frac{7}{2}} \left(-1+5k^2+e^2\right) \nonumber \\
& \quad
  +{\alpha '}^5 \frac{45\left(1-e^2\right)^{-\frac{13}{2}}}{2048}
   \left( 16 - 176 k^2 + 224 k^4 - 13e^2  \right. \nonumber \\
& \quad \quad \left. 
       - 62 e^2 k^2 + 315 e^2 k^4 - 22 e^4 + 238 e^4 k^2 + 19e^6 \right) .
  \label{eqn:vZ10-nn399-2-sup1}
\end{align}

Note that \citeauthor{vonzeipel1910}'s original equation
(an unnumbered one on p. Z399) that corresponds to
Eqs. \eqref{eqn:vZ10-nn399-2-sup2} and \eqref{eqn:vZ10-nn399-2-sup1} is
truncated at the order of ${\alpha '}^3$.
Only the first terms in the right-hand side of Eqs.
\eqref{eqn:vZ10-nn399-2-sup2} and \eqref{eqn:vZ10-nn399-2-sup1} are presented
as:
\begin{align}
\DP{R}{\left(e^2\right)}
 & = {a'}^3 \DP{R'_3}{\left(e^2\right)} + \cdots \nonumber \\
 & = {a'}^3 \frac{3}{16}\left(1-e^2\right)^{-\frac{7}{2}} \left(-1+5k^2+e^2\right) + \cdots .
  \label{eqn:vZ10-nn399-2-original}
\end{align}

Note also that
we have ourselves calculated and added the terms of $\Oaldpen$ to
this equation (we do not show their specific forms here),
and then derived Eqs. \eqref{eqn:vZ10-nn399-2-sup2} and \eqref{eqn:vZ10-nn399-2-sup1}.
It is because the higher-order terms of $\alpha'$
are necessary for obtaining the expressions of $e'_{2.0}$ and $e'_{0.2}$
in what follows.

Next, \citeauthor{vonzeipel1910} expresses the solution of
the equation $\DP{R}{\left(e^2\right)} = 0$.
When $\cos 2g = +1$, it becomes:
\begin{align}
  e^2 &= {e'}^2_{2.0} \nonumber \\
      &= 1-5k^2 +\frac{3}{100}\frac{7-5k^2}{k^2}{\alpha '}^2 + \cdots .
  \tag{Z103-\arabic{equation}}
  \stepcounter{equation}
  \label{eqn:Z103}
\end{align}
When $\cos 2g = -1$, it becomes as follows:
\begin{align}
  e^2 &= {e'}^2_{0.2} \nonumber \\
      &= 1-5k^2 +\frac{3}{200}\frac{41-125k^2}{k^2}{\alpha '}^2 + \cdots .
  \tag{Z104-\arabic{equation}}
  \stepcounter{equation}
  \label{eqn:Z104}
\end{align}

Similar to the consideration in the inner case (see p. \pageref{eqn:Z81}),
\citeauthor{vonzeipel1910} then proposes the equation
\begin{equation}
  {e'}^2_{2.0} = 0 ,
  \label{eqn:vZ10-ed20=0}
\end{equation}
for the case of $\cos 2g = +1$ (equivalent to $g=0$ or $\pi$).
Although he does not show the function form of $k^2$
that satisfies Eq. \eqref{eqn:vZ10-ed20=0}, we can calculate the form as
\begin{equation}
  k^2 = \frac{1}{5} + \frac{9}{50}{\alpha '}^2 + \cdots
      \equiv \overline{k'_{2.0}}^2 .
  \label{eqn:vZ10-ed20=0-sol}
\end{equation}

Equation \eqref{eqn:vZ10-ed20=0-sol} means that,
when the value of $k^2$ approaches $\overline{k'_{2.0}}^2$,
both the two solutions
$(e,g) = (e'_{2.0}, 0)$ and $(e'_{2.0}, \pi)$ of the equation
\begin{equation}
  \left. \DP{R}{\left(e^2\right)} \right|_{\cos 2g = +1} = 0 ,
  \label{eqn:vZ10-dpRdpe2+1=0}
\end{equation}
would approach the origin at $e=0$.
In other words, the two solutions $(x,y)=(\pm e'_{2.0}, 0)$
on the $(e \cos g, e \sin g)$ plane would approach $(x,y)=(0,0)$.
We have already seen the behavior of $R$ at the origin $(0,0)$
on p. \pageref{par:oCR3BP-origin}.

Note that \citeauthor{vonzeipel1910} uses the notation
$\overline{k^2}$ (p. Z400) for expressing the solution of
Eq. \eqref{eqn:vZ10-ed20=0}.
However, we prefer to use $\overline{k'_{2.0}}^2$ in this monograph.
This is for distinguishing this quantity from a similar one that shows up
in the following discussion on $e'_{0.2}$
(see the descriptions around Eqs. \eqref{eqn:vZ10-ed02=0-sol} and \eqref{eqn:vZ10-dpRdpe2-1=0}).
It is also for maintaining a consistency with the discussion on the inner problem (see p. \pageref{eqn:vZ10-nn382-3}).

At this point,
\citeauthor{vonzeipel1910} mentions the equivalence of
${k'}^2_{2.0}$            in Eq. \eqref{eqn:Z101} and
$\overline{k'_{2.0}}^2$ in Eq. \eqref{eqn:vZ10-ed20=0-sol}.
Let us cite what he says.
Note that in the quoted part below,
we use his original notation $\overline{k^2}$,
instead of our own $\overline{k'_{2.0}}^2$:
\begin{quote}
``The quantity ${e'}^2_{2.0}$ is not always positive.
Indeed, it vanishes for a certain value of $k^2$ adjacent to $\frac{1}{5}$.
Now let $\overline{k^2}$ [be] a root of the equation ${e'}^2_{2.0}=0$.
I say that $\overline{k^2}$ inevitably coincides with the quantity ${k'}^2_{2.0}$,
which cancels the coefficient $R'_{2.0}$.
Indeed, when $k^2$ passes the value $\overline{k^2}$, the two solutions
$$
  x = \pm e'_{2.0}, \quad
  y = 0
$$
of equations
$$
 \DP{R}{x} = \DP{R}{y} = 0
$$
coincide with the solution already studied
$$
  x = y = 0 .
$$

It is therefore necessary that $\overline{k^2}$ cancels the coefficient $R'_{2.0
}$ and that, as a result, $\overline{k^2}$ coincides with the quantity ${k'}^2_{2.0
}$, given by the first [equation] of formulas (Z101).'' (p. Z400)
\end{quote}

\citeauthor{vonzeipel1910}'s statement above and
Eq. \eqref{eqn:vZ10-ed20=0-sol} thus tell us that $\overline{k^2}$
(or $\overline{k'_{2.0}}^2$ in our notation)
is a key parameter for confirming
or denying the existence of local extremums of $R$ on the $x$-axis
of the $(e \cos g, e \sin g)$ plane.

Next he moves on to another, similar equation
\begin{equation}
  {e'}^2_{0.2} = 0 ,
  \label{eqn:vZ10-ed02=0}
\end{equation}
for the case of $\cos 2g = -1$ (equivalent to $g=\pm\frac{\pi}{2}$).
This equation corresponds to Eq. \eqref{eqn:vZ10-e02=0}
in the discussion on the inner problem (see p. \pageref{eqn:vZ10-e02=0} of this monograph).
Although \citeauthor{vonzeipel1910} does not show the function form of $k^2$
that satisfies Eq. \eqref{eqn:vZ10-ed02=0}, we can again calculate the form as
\begin{equation}
  k^2 = \frac{1}{5} + \frac{12}{50}{\alpha '}^2 + \cdots 
      \equiv \overline{k'_{0.2}}^2 .
  \label{eqn:vZ10-ed02=0-sol}
\end{equation}

Equation \eqref{eqn:vZ10-ed02=0-sol} means that,
when $k^2$ approaches $\overline{k'_{0.2}}^2$,
both the solutions
$(e,g) = ({e'}^2_{2.0}, \pm\frac{\pi}{2})$ of the equation
\begin{equation}
  \left. \DP{R}{\left(e^2\right)} \right|_{\cos 2g = -1} = 0 ,
  \label{eqn:vZ10-dpRdpe2-1=0}
\end{equation}
approach the origin $e=0$.
In other words, the two solutions $(x,y)=(0, \pm e'_{0.2})$
on the $(e \cos g, e \sin g)$ plane would approach $(x,y)=(0,0)$.
In addition, we can deduce the equivalence of
${k'}^2_{0.2}$            in Eq. \eqref{eqn:Z101} and
$\overline{k'_{0.2}}^2$ in Eq. \eqref{eqn:vZ10-ed02=0-sol}
due to the same reason as in the case of $\cos 2g = +1$.
Somehow, however, \citeauthor{vonzeipel1910}'s statement about it is very short:
\begin{quote}
``In the same manner it is demonstrated that the quantity ${k'}^2_{0.2}$,
given by the second of the formulas (Z101), is the only root of the equation
$$
  {e'}^2_{0.2} = 0 .
$$

\hspace*{1em}
It remains for us to examine whether the function $R$ is [a] maximum, minimum or minimaxima at the points (Z103) and (Z104).'' (p. Z400)
\end{quote}

\citeauthor{vonzeipel1910}'s above statement and
Eq. \eqref{eqn:vZ10-ed02=0-sol} thus tell us that
$\overline{k'_{0.2}}^2$ is a key value for confirming
(or denying) the existence of local extremums of $R$ on the $y$-axis
of the $(x,y)=(e \cos g, e \sin g)$ plane.

As a summary, let us itemize the conclusions that \citeauthor{vonzeipel1910}
obtained as follows:
\begin{itemize}
\item When $k^2 < {k'}^2_{2.0}$,
  ${e'}^2_{2.0}$ becomes positive, and
  $R$ has two local extremums at $(x,y)=(\pm e'_{2.0}, 0)$.
\item When $k^2 < {k'}^2_{0.2}$,
  ${e'}^2_{0.2}$ becomes positive, and
  $R$ has two local extremums at $(x,y)=(0, \pm e'_{0.2})$.
\item If $\alpha'$ is small, we have
  ${k'}^2_{2.0} < {k'}^2_{0.2}$ from Eq. \eqref{eqn:Z101}.
  And when $k^2 > {k'}^2_{0.2}$,
  neither of ${e'}^2_{2.0}$ or ${e'}^2_{0.2}$ can remain positive, 
  and $R$ has no local extremum along the $x$- or $y$-axis.
\end{itemize}

\begin{figure*}[tbhp]\centering
\ifepsfigure
 \includegraphics[width=\dualmedfigwidth\textwidth]{fig_Z8-10.eps}%fig17
\else
 \includegraphics[width=\dualmedfigwidth\textwidth]{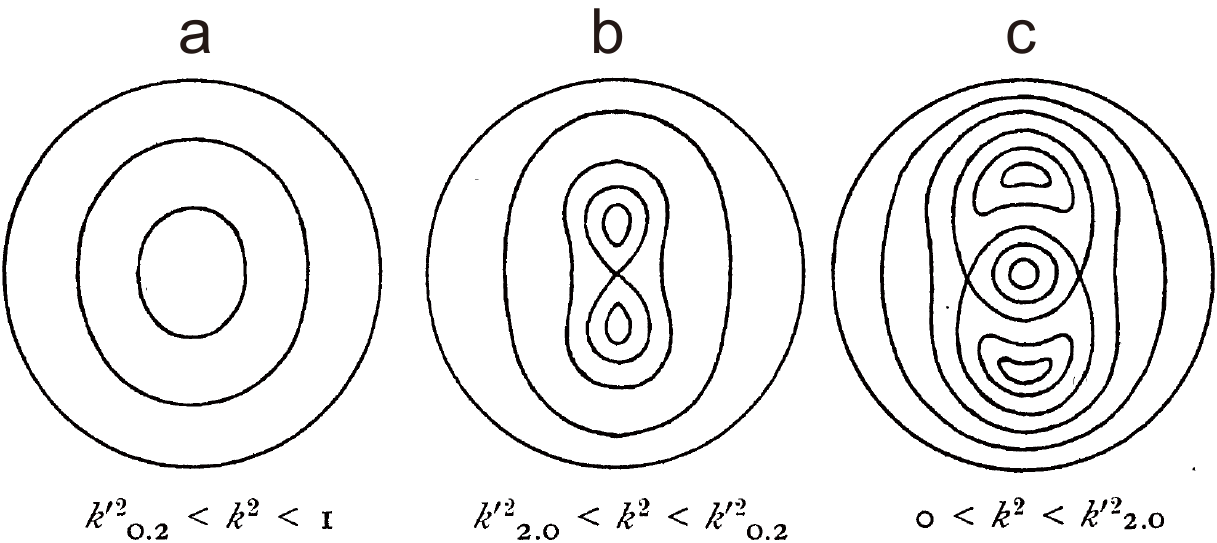}%fig17
\fi
  \caption{%
  Transcription of \citeauthor{vonzeipel1910}'s \mysymfigS Z8, Z9, Z10 on his pp. Z401--Z402
(Credit: John Wiley and Sons. Reproduced with permission).
  They represent schematic trajectories of equi-$R$ contours of
  the doubly averaged outer CR3BP on the $(e \cos g, e \sin g)$ plane.
  \mtxtsf{a}: \citeauthor{vonzeipel1910}'s \mysymfigO Z8
   for the case when ${k'}^2_{0.2} < k^2 < 1$.
  \mtxtsf{b}: His                          \mysymfigO Z9
   for the case when ${k'}^2_{2.0} < k^2 < {k'}^2_{0.2}$.
  \mtxtsf{c}: His                          \mysymfigO Z10
   for the case when $0          < k^2 < {k'}^2_{2.0}$.
  }
  \label{fig:vZ10-f8-10}
\end{figure*}

\paragraph{Type of the local extremums\label{par:oCR3BP-type-ext}}
Similar to the discussion on the inner case (p. \pageref{par:iCR3BP-type-ext}),
\citeauthor{vonzeipel1910} next investigates the characteristics of the points
$(x,y)=(\pm e'_{2.0}, 0)$ and $(0, \pm e'_{0.2})$ as local extremums of $R$
(pp. Z400--Z401).
For this purpose
he focuses on inspecting the sign of
the three second derivatives $\DP[2]{R}{g}$, $\DPsd{R}{e}{g}$, $\DP[2]{R}{e}$
when $\cos 2g = \pm 1$.

As for $\DP[2]{R}{g}$,
from Eq. \eqref{eqn:Z98} we get straightaway its general form as
\begin{equation}
  \DP[2]{R}{g} =
 -{\alpha '}^5
  \frac{135}{64} \frac{e^2 \sin^2 I \left( 1-\frac{7}{6}\sin^2 I\right)}
                      {\left( 1-e^2 \right)^{\frac{7}{2}}} \cos 2g .
  \label{eqn:vZ10-nn400-1}
\end{equation}
From Eqs. \eqref{eqn:Z103} and \eqref{eqn:Z104}, we know that
$k^2 \sim \frac{1-e^2}{5}$
both at $e^2 = {e'}^2_{2.0}$ and
              ${e'}^2_{0.2}$ if $\alpha' \ll 1$. This means
\begin{equation}
\begin{aligned}
  \sin^2 I &= 1 - \cos^2 I \\
           &= 1 - \frac{k^2}{1-e^2} \sim \frac{4}{5},
\end{aligned}
\label{eqn:vZ10-nn400-original1}
\end{equation}
and therefore
\begin{equation}
  1 - \frac{7}{6}\sin^2 I \sim \frac{1}{15} > 0 .
  \label{eqn:vZ10-nn400-original2}
\end{equation}
From Eqs. \eqref{eqn:vZ10-nn400-1} and \eqref{eqn:vZ10-nn400-original2},
we can conclude that
\begin{itemize}
\item When $\cos 2g = +1$, $\DP[2]{R}{g} < 0$ at $(x,y)=(\pm e'_{2.0}, 0)$.
\item When $\cos 2g = -1$, $\DP[2]{R}{g} > 0$ at $(x,y)=(0, \pm e'_{0.2})$.
\end{itemize}

$\DPsd{R}{e}{g}$ has the simplest form. From Eq. \eqref{eqn:Z98} we get
\begin{equation}
  \DPsd{R}{e}{g} = 0 ,
  \label{eqn:vZ10-DP2Rog2}
\end{equation}
when $\cos 2g = \pm 1$. This is because all the terms of $\DPsd{R}{e}{g}$
contain $\sin 2g$ as a factor which is zero when $\cos 2g = \pm 1$.

\label{pg:oCR3BP-dp2Rde2}
$\DP[2]{R}{e}$ is more complicated.
In addition, similar to the case of the inner problem
(p. \pageref{pg:iCR3BP-dp2Rde2}),
\citeauthor{vonzeipel1910}'s original description seems very terse.
Let us add a complementary description as to
how we should deal with $\DP[2]{R}{e}$.

We again adopt the identity \eqref{eqn:vZ10-nn383-1} for $\DP[2]{R}{e}$ of the outer problem.
The resulting $\DP[2]{R}{e}$ in its general form turns out to be rather complicated,
and we have considered it in Appendix \ref{appen:DP2R}.
Then we do the following:
\begin{itemize}
\item Substitute $e^2 = {e'}^2_{2.0}$ of \eqref{eqn:Z101} into $\left. \DP[2]{R}{e} \right|_{\cos 2g = +1}$
\item Substitute $e^2 = {e'}^2_{0.2}$ of \eqref{eqn:Z101} into $\left. \DP[2]{R}{e} \right|_{\cos 2g = -1}$
\end{itemize}
As a result, we get
\begin{equation}
  \DP[2]{R}{e}
  = -\frac{3\sqrt{5}\left(5k^2 -1 \right)}{2500k^{\frac{7}{2}}} {\alpha '}^3
  + \Oaldpen ,
  \label{eqn:vZ10-DP2Roe2-kk-ad2}
\end{equation}
at both the points $(x,y)=(\pm e'_{2.0}, 0)$ and $(x,y)=(0, \pm e'_{0.2})$.
If we ignore the terms of $\Oaldpen$ in the right-hand side,
it is clear that $\DP[2]{R}{e}$ in Eq. \eqref{eqn:vZ10-DP2Roe2-kk-ad2}
monotonically decreases with the increase of $k^2$.
And yet, it remains positive while $k^2 < \frac{1}{5}$.
We should recall that $k^2 < \frac{1}{5}$ is necessary for
${e'}^2_{2.0}$ and ${e'}^2_{0.2}$ to remain positive
(see Eqs. \eqref{eqn:Z103} and \eqref{eqn:Z104}).
Therefore we can conclude that
\begin{equation}
  \DP[2]{R}{e} > 0 ,
  \label{eqn:vZ10-DP2Roe2-kk-gt0}
\end{equation}
at all the four points $(x,y)=(\pm e'_{2.0}, 0)$ and $(0, \pm e'_{0.2})$,
if $\alpha' \ll 1$.
In Appendix \ref{appen:DP2R}
we show the actual expression of the terms
at $\Oaldpen$ in the right-hand side of
Eq. \eqref{eqn:vZ10-DP2Roe2-kk-ad2} which \citeauthor{vonzeipel1910} omitted.
It turned out that the conclusion of Eq. \eqref{eqn:vZ10-DP2Roe2-kk-gt0}
remains true even if we include the $\Oaldpen$ terms.

Collecting all the results presented in Section Z22,
\citeauthor{vonzeipel1910} states the following proposition
about the location of local extremums of $R$ in the outer case
when $\alpha'$ is small:
\begin{quote}
\textit{%
``If ${k'}^2_{0.2}<k^2<1$, the function $R$ does not possess maxima or minimaxima in the domain $e<k'$, and takes a single minimum at $e=0$.}

--- \textit{If ${k'}^2_{2.0}<k^2<{k'}^2_{0.2}$, the function $R$ has no maximum in the domain $e<k'$ and possesses only one minimaximum in this area, situated at the origin $e=0$, and only two minima at the points $g=\pm\frac{\pi}{2}$, $e=e'_{0.2}$.}

--- \textit{Finally, if $0<k^2<{k'}^2_{2.0}$, the function $R$ possesses a single maximum in the domain $e<k'$ situated at the origin $e=0$, only two minimum at the points $g=\pm\frac{\pi}{2}$, $e=e'_{0.2}$ and only two minimaxima at the points $g=0$ or $\pi$, $e=e'_{2.0}$.}'' (p. Z402)
\end{quote}

After these considerations, 
\citeauthor{vonzeipel1910} shows schematic plots of equi-$R$ curves
on the $(e \cos g, e \sin g)$ plane as his \mysymfigS Z8, Z9, and Z10
for illustrating the circumstances.
We transcribed them here as our \mysymfigO \ref{fig:vZ10-f8-10}.
Through \citeauthor{vonzeipel1910}'s figures, it is obvious that
the topology of the equi-$R$ contours qualitatively changes across $k^2 = {k'}^2_{2.0}$ and $k^2 = {k'}^2_{0.2}$.
Since the difference between ${k'}^2_{2.0}$ and ${k'}^2_{0.2}$ is
generally slight (see Eq. \eqref{eqn:Z101}),
the parameter space that realizes the second case
$\bigl( {k'}^2_{2.0} < k^2 < {k'}^2_{0.2} \bigr)$
is narrow (\mysymfigO \ref{fig:vZ10-f8-10}\mtxtsf{b}).
We will present our own numerical demonstrations
in Section \ref{sssec:oCR3BP-num-confirm} to reproduce his schematic plots.

\subsubsection{When $\alpha'$ is not so small\label{sssec:oCR3BP-nosmall-alpha}}
In Section Z23,
\citeauthor{vonzeipel1910} deals with the doubly averaged outer CR3BP when $\alpha'$
(or $a'$ in his notation) is not negligibly small.
In this regard, Section Z23 comprises the counterpart of Z17 that is on the inner problem.
\citeauthor{vonzeipel1910} takes care of this task again through a numerical method.
The particular focus of this section lies in showing that
the doubly averaged disturbing function $R$ presented as Eq. \eqref{eqn:Z99}
takes a local minimum at the origin $(x,y)=(0,0)$ under some condition,
even when $\alpha'$ is arbitrary.
His method to derive numerical values of $R$ seems to follow
what was presented in Section Z17.
Probably due to this reason, his presentation of the mathematical procedures
used in Section Z23 is brief and concise.
Let us show its summary in what follows.

Similar to Eqs. \eqref{eqn:vZ10-def-bik}, \eqref{eqn:vZ10-def-cik}, and \eqref{eqn:vZ10-def-eik} presented in his Section Z17,
\citeauthor{vonzeipel1910} first introduces new coefficients
${b'}^{i.j}$, ${c'}^{i.j}$, ${e'}^{i.j}$ through the following definitions:
\begin{alignat}{1}
& \left[ 1+{\alpha '}^2-2\alpha'\left(\mu\cos M + \nu\cos N\right)\right]^{-\frac{1}{2}} \nonumber \\
& \quad = {b'}^{0.0} +2{b'}^{1.0}\cos M +2{b'}^{0.1}\cos N \nonumber \\
& \quad\quad + 4{b'}^{1.1}\cos M \cos N + \cdots ,
  \label{eqn:vZ10-def-bik-dash} \\
& \left[ 1+{\alpha '}^2-2\alpha'\left(\mu\cos M + \nu\cos N\right)\right]^{-\frac{3}{2}} \nonumber \\
& \quad = {c'}^{0.0} +2{c'}^{1.0}\cos M +2{c'}^{0.1}\cos N \nonumber \\
& \quad\quad + 4{c'}^{1.1}\cos M \cos N + \cdots ,
  \label{eqn:vZ10-def-cik-dash} \\
& \left[ 1+{\alpha '}^2-2\alpha'\left(\mu\cos M + \nu\cos N\right)\right]^{-\frac{5}{2}} \nonumber \\
& \quad = {e'}^{0.0} +2{e'}^{1.0}\cos M +2{e'}^{0.1}\cos N \nonumber \\
& \quad\quad + 4{e'}^{1.1}\cos M \cos N + \cdots .
  \label{eqn:vZ10-def-eik-dash}
\end{alignat}

The coefficients
${b'}^{i.j}$, ${c'}^{i.j}$, ${e'}^{i.j}$ defined for the outer problem in
Eqs. \eqref{eqn:vZ10-def-bik-dash}, \eqref{eqn:vZ10-def-cik-dash} and \eqref{eqn:vZ10-def-eik-dash}, and
$b^{i.j}$,  $c^{i.j}$,  $e^{i.j}$ defined for the inner problem in
Eqs. \eqref{eqn:vZ10-def-bik}, \eqref{eqn:vZ10-def-cik} and \eqref{eqn:vZ10-def-eik},
are connected through the following relationship:
\begin{equation}
\begin{aligned}
  \alpha^\frac{1}{2} b^{i.j} &= {\alpha'}^\frac{1}{2} {b'}^{i.j}, \\
  \alpha^\frac{3}{2} c^{i.j} &= {\alpha'}^\frac{3}{2} {c'}^{i.j}, \\
  \alpha^\frac{5}{2} e^{i.j} &= {\alpha'}^\frac{5}{2} {e'}^{i.j} .
\end{aligned}
  \tag{Z105-\arabic{equation}}
  \stepcounter{equation}
  \label{eqn:Z105}
\end{equation}

Then, \citeauthor{vonzeipel1910} expresses $R'_{2.0}$ and $R'_{0.2}$
in Eq. \eqref{eqn:Z99}
using ${b'}^{i.j}$, ${c'}^{i.j}$, ${e'}^{i.j}$ and $\alpha'$ as follows:
\begin{equation}
\begin{aligned}
  R'_{2.0} &= -\alpha' {c'}^{1.1}+\frac{1}{2}{\alpha '}^2 \left({c'}^{1.0}+{c'}^{0.1}\right), \\
  R'_{0.2} &=  \alpha'{c'}^{1.1}-\frac{1}{4}{\alpha '}^2 \left({c'}^{1.0}+{c'}^{0.1}\right) \\
           &\quad
            +\frac{3}{4}{\alpha '}^2k\left({c'}^{1.0}-{c'}^{0.1}\right) \\
           &\quad\quad
            -\frac{3}{4}{\alpha '}^3 \left(1-k^2\right)
                                  \left({e'}^{0.0}-{e'}^{1.1}\right) .
\end{aligned}
  \tag{Z106-\arabic{equation}}
  \stepcounter{equation}
  \label{eqn:Z106}
\end{equation}

\begin{figure}[tbhp]\centering
\ifepsfigure
 \includegraphics[width=\singlefigwidth\textwidth]{Rd202.eps} %fig18
\else
 \includegraphics[width=\singlefigwidth\textwidth]{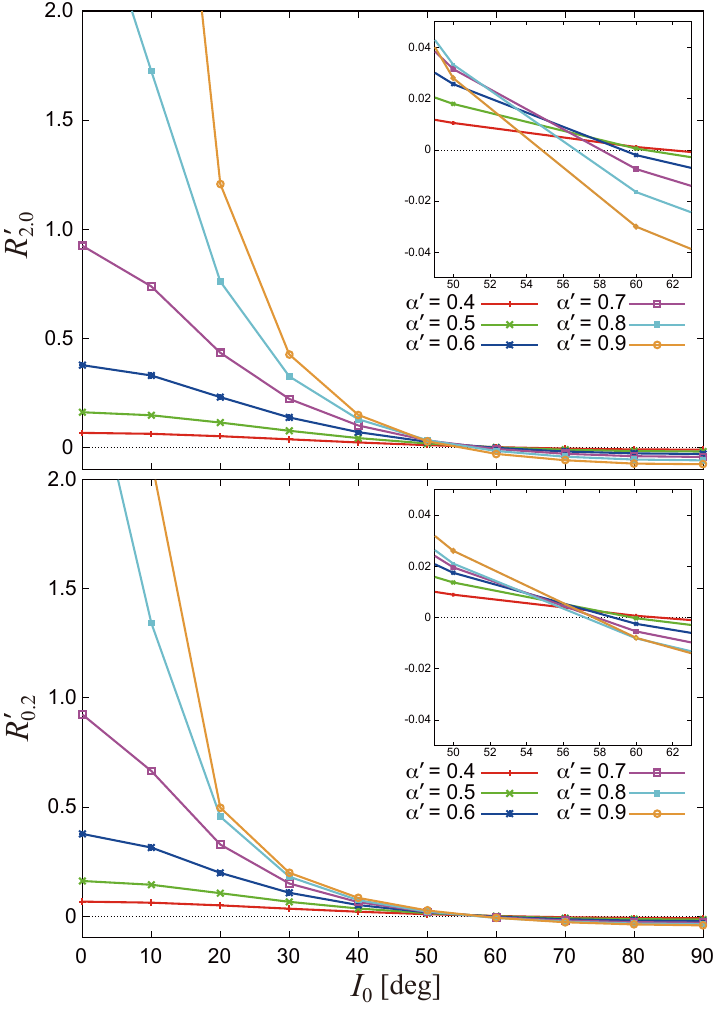} %fig18
\fi
  \caption{%
  Dependence of the values of
  $R'_{2.0}$ (upper panel) and
  $R'_{0.2}$ (lower panel) on $I_0 = \cos^{-1} k$ and on $\alpha'$.
  For magnifying the near-zero part of $R'_{2.0}$ and $R'_{0.2}$,
  we placed inlets at the top right of both the panels.
  All the plots are based on the tabulated numerical values
  on an unnumbered table in \citet[][pp. Z403--Z404]{vonzeipel1910}.
  }
  \label{fig:Rd020-table}
\end{figure}

Using the formulas \eqref{eqn:Z105}, \eqref{eqn:Z106} and others,
\citeauthor{vonzeipel1910} calculates the actual numerical values of
$R'_{2.0}$ and $R'_{0.2}$.
He tabulated them on an unnumbered table (pp. Z403--Z404).
The table shows the values of $R'_{2.0}$ and $R'_{0.2}$,
using $I_0 (= \cos^{-1} k)$ and $\alpha'$ as parameters.
Similar to the unnumbered table for the inner case (pp. Z389--Z390),
the range of $I_0$ is from 0 to $90^\circ$ with an interval of $10^\circ$, and
the range of $\alpha'$ is from 0.4 and 0.9 with an interval of 0.1.
Instead of transcribing the table in this monograph,
we make visual plots of their values in our \mysymfigO \ref{fig:Rd020-table}.

\clearpage

The largest difference between the outer case and the inner case is that
both the components ($R'_{2.0}$ and $R'_{0.2}$) can be zero in the outer case
(\mysymfigO \ref{fig:Rd020-table}).
Meanwhile in the inner case,
one of the components $(R_{2.0})$ always stays positive, and
it is likely that $R_{2.0}$ never subceeds zero (\mysymfigO \ref{fig:R020-table}).
Let us cite \citeauthor{vonzeipel1910}'s description of his own result:
\begin{quote}
``We see from the table that
the equation $R'_{2.0}=0$ always has a root $I_0 = I'_{2.0} = \arccos k'_{2.0}$ if $\alpha' \leq 0.9$, and
the equation $R'_{0.2}=0$ also   has a root $I_0 = I'_{0.2} = \arccos k'_{0.2}$ if $\alpha' \leq 0.9$.'' (p. Z403)
\end{quote}
This statement is followed by his observation:
\begin{quote}
``It also appears that
$$
\begin{array}{ccc}
  R'_{2.0} > 0 & {\rm if} & {k'}^2_{2.0} < k^2 < 1            \\
  R'_{2.0} < 0 & {\rm if} & 0          < k^2 < {k'}^2_{2.0}   \\
  R'_{0.2} > 0 & {\rm if} & {k'}^2_{0.2} < k^2 < 1            \\
  R'_{0.2} < 0 & {\rm if} & 0          < k^2 < {k'}^2_{0.2} \mbox{.''} \\
\end{array}
$$
(p. Z403)
\end{quote}
\label{pg:def-Id202}

By interpolating the numerical values of $R'_{2.0}$ and $R'_{0.2}$,
\citeauthor{vonzeipel1910} calculated the numerical values of
$I'_{2.0}$ and $I'_{0.2}$ which satisfy the equations \eqref{eqn:vZ10-Rd20=0} and \eqref{eqn:vZ10-Rd02=0} at each $\alpha'$.
He tabulated the result in an unnumbered table at the bottom of p. Z403.
We transcribed the unnumbered table as our Table \ref{tbl:vZ10-table-5}
(we will also visualize his results later).

\begin{table}[t]\centering
\footnotesize
 \setlength{\tabcolsep}{5pt} % given by Ito
  \caption[]{%
Transcription of an unnumbered table in \citet[][p. Z403]{vonzeipel1910}.
The unit of $I'_{2.0}$ and $I'_{0.2}$ is the degree.
See also \mysymfigO \ref{fig:Id202comp} for a visualization of the tabulated values.
}
  \label{tbl:vZ10-table-5}
\begin{tabular}{cccccccc} \hline
$\alpha'$  & $0.0$   & $0.4$   & $0.5$   & $0.6$   & $0.7$   & $0.8$   & $0.9$ \\
\hline
$I'_{2.0}$ & $63.43$ & $61.43$ & $60.34$ & $59.06$ & $57.56$ & $55.95$ & $54.1$ \\
$I'_{0.2}$ & $63.43$ & $60.89$ & $59.67$ & $58.38$ & $57.25$ & $56.6 $ & $57.2$ \\
\hline
\end{tabular}
\end{table}

From the results tabulated in our Table \ref{tbl:vZ10-table-5},
\citeauthor{vonzeipel1910} makes an observation as follows:
\begin{quote}
``We see that the quantity $I'_{2.0} - I'_{0.2}$ is positive when $0 < \alpha' < 0.74\cdots$ and negative if $\alpha' > 0.74\cdots$.'' (p. Z404)
\end{quote}

Then, through the above observation, he subsequently states the following theorem:
\begin{quote}
``The functions ${k'}^2_{2.0}$ and ${k'}^2_{0.2}$ being thus defined, we can state, as demonstrated, the following theorem:

\hspace*{1em}
\textit{At the origin $x=y=0$,
the function $R$ becomes minimum
if ${k'}^2 > {k'}^2_{2.0}$ and if ${k'}^2 > {k'}^2_{0.2}$;
[$R$] becomes a saddle point
if ${k'}^2$ is between ${k'}^2_{2.0}$ and ${k'}^2_{0.2}$;
[and $R$] becomes maximum if $k^2 < {k'}^2_{2.0}$ and $k^2 < {k'}^2_{0.2}$.}''
(p. Z404)
\end{quote}

As a natural consequence of the above theorem,
\citeauthor{vonzeipel1910} immediately states another theorem as follows:
\begin{quote}
``\textit{The necessary and sufficient condition for the eccentricity of
the orbit of an infinitely small mass located outside 
the disturbing planet to be always small, if it is small at a given
moment, is that the inclination of the orbit is not located between
the two angles $I'_{2.0}$ and $I'_{0.2}$.}
(In this statement we have neglected the disturbing mass and the square of the eccentricity.)'' (p. Z404)
\end{quote}

As for the last sentence in parentheses
``In this statement we have neglected the disturbing mass and the square of the eccentricity,''
we interpret this that
\citeauthor{vonzeipel1910} assumes that the mass of the perturbing body (Jupiter)
is much smaller than the mass of the central object (Sun) in his setting.
Also, we are aware that the terms higher than
the square of the eccentricity $\left(e^2\right)$ have been neglected in this section,
because \citeauthor{vonzeipel1910} just uses the lowest-order components
($R'_{2.0} x^2$ and $R'_{0.2} y^2$)
in the two-variable Taylor expansion \eqref{eqn:Z99}.

\citeauthor{vonzeipel1910} concludes Section Z23 by making the
following statement.
This statement seems to be a summary of what he discovered
through the discussion in this section:
\begin{quote}
``Asteroids exterior at the disturbing planet thus belong to two categories.
In the first category,
the inclinations are smaller than ${I'}^2_{2.0}$ and ${I'}^2_{0.2}$;
in the second, they are larger than these quantities.

\hspace*{1em}
We have demonstrated at no.~Z22 that the function $R$ possesses minimum values at two symmetric points
$$
  x = 0, \quad y=\pm e'_{0.2}
$$
if $\alpha'$ is small and if $k^2 < {k'}^2_{0.2}$.
It seems very likely that these minima exist for appropriate values of ${k'}^2$, as long as $\alpha' < 0.74 \cdots$ (value of $\alpha'$ which makes $k'_{0.2}=k'_{2.0}$).
If instead $\alpha' > 0.74 \cdots$, it is probable that the two symmetric minima are on the axis of $x$, at least for the appropriate values of ${k'}^2 < {k'}^2_{2.0}$.
But for rigorously demonstrating the existence of these minima and for determining the position, it would be necessary to resort to numerical calculations.''
(pp. Z404--Z405)
\end{quote}
\label{pg:alphad074}

It is not surprising for us to know that \citeauthor{vonzeipel1910}
did not actually carry out the ``numerical calculations''
for demonstrating the existence of the local minima of $R$
along the $x$-axis when $\alpha'$ is large.
Such a numerical analysis must have been formidable in his time.
Since studies on the doubly averaged outer CR3BP are less often encountered
than the inner one even in modern days,
we have not found relevant literature that confirms \citeauthor{vonzeipel1910}'s theorem (or conjecture) stated above.
Therefore in the next part of this monograph (Section \ref{sssec:oCR3BP-num-confirm}),
we carry out a numerical confirmation of the theorem and conjecture
that \citeauthor{vonzeipel1910} stated for the outer case
when $\alpha'$ is not negligibly small.

\subsubsection{Numerical confirmation\label{sssec:oCR3BP-num-confirm}}
Here we carry out several numerical experiments
in order to confirm how accurate (or inaccurate)
\citeauthor{vonzeipel1910}'s theory is on the doubly averaged outer CR3BP
that has been presented in this monograph.
Naturally, the contents in what follows are not included in
\citeauthor{vonzeipel1910}'s original work.

\begin{figure*}[thbp]\centering
\ifepsfigure
 \includegraphics[width=\dualfigwidth\textwidth]{eqR_vzoex_a0.eps} %fig19
\else
 \includegraphics[width=\dualfigwidth\textwidth]{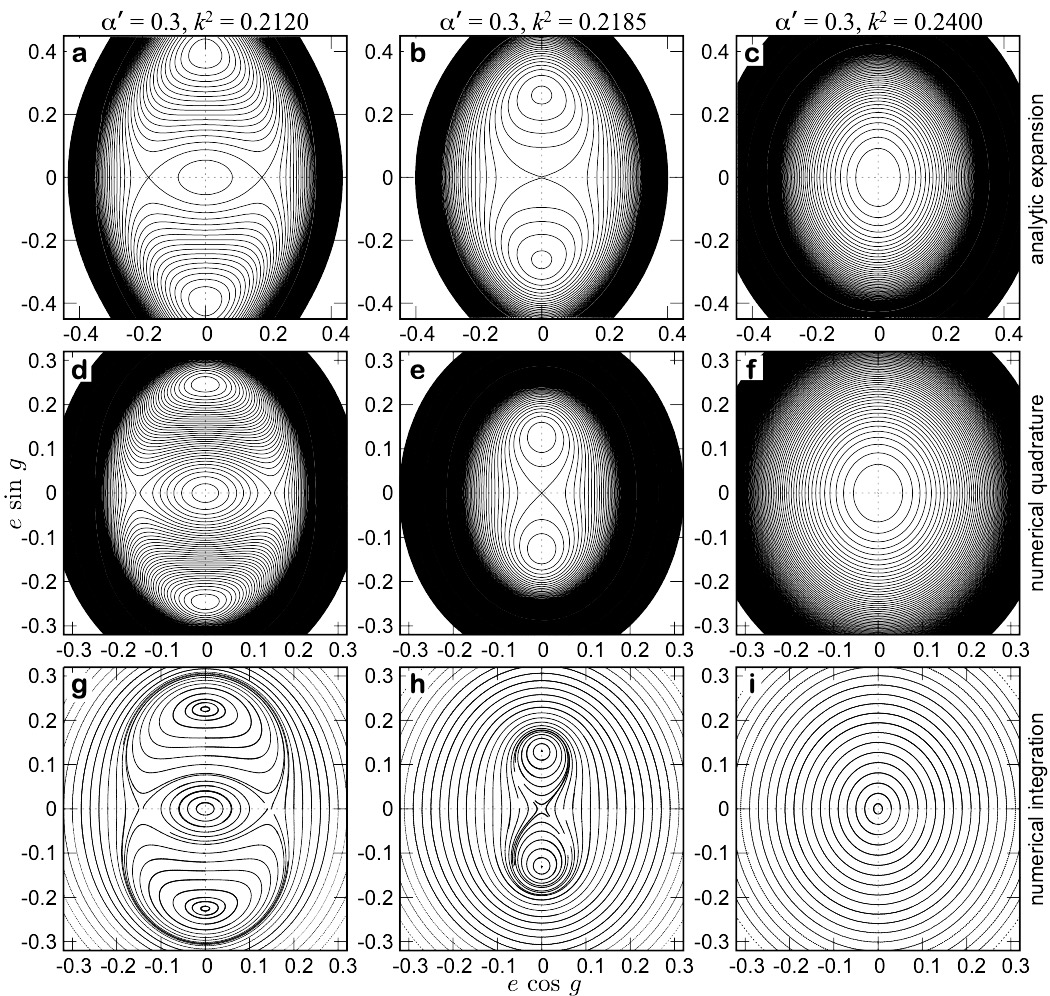} %fig19
\fi
  \caption{%
Equi-potential contours of the outer CR3BP plotted
on the $(x,y)=(e \cos g, e \sin g)$ plane.
The top row (\mtxtsf{a}, \mtxtsf{b}, \mtxtsf{c}):
 Results obtained from \citeauthor{vonzeipel1910}'s analytic expansion
of $R$ presented as Eqs. \eqref{eqn:Z97} and \eqref{eqn:Z98}.
The middle row (\mtxtsf{d}, \mtxtsf{e}, \mtxtsf{f}):
 Results obtained from the numerical quadrature defined by Eq. \eqref{eqn:K09}.
The bottom row (\mtxtsf{g}, \mtxtsf{h}, \mtxtsf{i}):
 Results obtained from the direct numerical integration of the equations of motion.
The columns are categorized by the parameter sets
$(\alpha', k^2)$ that are used:
$(\alpha', k^2) = (0.3, 0.2120$) for the panels in the left   column
(\mtxtsf{a}, \mtxtsf{d}, \mtxtsf{g}),
                 $(0.3, 0.2185$) for the panels in the middle column
(\mtxtsf{b}, \mtxtsf{e}, \mtxtsf{h}), and
                 $(0.3, 0.2400$) for the panels in the right  column
(\mtxtsf{c}, \mtxtsf{f}, \mtxtsf{i}).
Note that the $x$- and $y$-ranges are
both from $-0.45$ to $+0.45$ only in the top row
(\mtxtsf{a}, \mtxtsf{b}, \mtxtsf{c}),
while they are from $-0.32$ to $+0.32$ in the middle and bottom rows
(\mtxtsf{d}, \mtxtsf{e}, \mtxtsf{f}, \mtxtsf{g}, \mtxtsf{h}, \mtxtsf{i}).
}
  \label{fig:eqR-vzoex-a0}
\end{figure*}

\paragraph{Equi-potential contours (when $\alpha' \ll 1$)} %
First, let us reproduce the equi-potential contours
given in \citeauthor{vonzeipel1910}'s schematic \mysymfigS Z8, Z9, and Z10
(transcribed as our \mysymfigO \ref{fig:vZ10-f8-10}).
We use the doubly averaged disturbing function $R$
in Eq. \eqref{eqn:Z97} that \citeauthor{vonzeipel1910} derived.
As we learned,
the leading-order term of $R$ ($R'_3$ in Eq. \eqref{eqn:Z98}) does not depend on the variable $g$.
$R$'s dependence on $g$ shows up in $R'_5$ and the higher-order terms.
Therefore, to see the dependence of $R$ on $g$,
we need to choose $\alpha'$ which is small but not too small.
If we adopt too small a value of $\alpha'$,
the higher-order terms would become extremely tiny, and
the dependence of $R$ on $g$ would vanish.
For this reason we chose $\alpha'=0.3$ as an example here.
This yields ${k'}^2_{2.0} \sim 0.2162$ and ${k'}^2_{0.2} \sim 0.2216$ from Eq. \eqref{eqn:Z101}.

We reproduced the equi-potential contours of $R$
using Eqs. \eqref{eqn:Z97} and \eqref{eqn:Z98}, and
presented the result as the top three panels
(\mtxtsf{a}, \mtxtsf{b}, \mtxtsf{c}) of \mysymfigO \ref{fig:eqR-vzoex-a0}
labeled as ``analytic expansion'' at the right edge.
The three typical patterns of equi-potential curves that we saw in
\citeauthor{vonzeipel1910}'s original drawing
(\mysymfigO \ref{fig:vZ10-f8-10}) are well reproduced.
When $k^2 < {k'}^2_{2.0}$ (\mysymfigO \ref{fig:eqR-vzoex-a0}\mtxtsf{a}),
we see a local maximum of $R$ at the origin $(x,y)=(0,0)$ together with
a pair of local minima  along the $y$-axis, as well as
a pair of saddle points along the $x$-axis.
When ${k'}^2_{2.0} <k^2<  {k'}^2_{0.2}$ (\mysymfigO \ref{fig:eqR-vzoex-a0}\mtxtsf{b}),
we find the origin being a saddle point of $R$, and
a pair of local minima along the $y$-axis shows up.
When ${k'}^2_{0.2} < k^2$ (\mysymfigO \ref{fig:eqR-vzoex-a0}\mtxtsf{c}),
the origin becomes a local minimum of $R$ which is its only local extremum.
These results are consistent with \citeauthor{vonzeipel1910}'s statement
given with 
his \mysymfigS Z8--Z10 (our \mysymfigO \ref{fig:vZ10-f8-10}).

We would like readers to note that each of the panels
\mtxtsf{a}, \mtxtsf{b}, \mtxtsf{c} in \mysymfigO \ref{fig:eqR-vzoex-a0}
has an individual and different
contour interval. They represent a fixed energy interval of
the doubly averaged disturbing potential in each of the systems.
Hence, regions with dense contours imply that the potential gradient is
steep there. Then, the results in these panels indicate that the
potential gradient around the origin $(0,0)$ is much less steep than
in other regions where the contour density is much higher.
Note also that we did not draw all the contours in the outermost part
of the plots (near the panel edge)
mainly because the contour intervals would become too narrow
and it would make the plots too busy.

Second, we draw equi-$R$ curves using the same parameter set
$\left(\alpha', k^2\right)$ but through the numerical quadrature defined by
Eq. \eqref{eqn:K09}. The result is presented as
the middle three panels (\mtxtsf{d}, \mtxtsf{e}, \mtxtsf{f})
of \mysymfigO \ref{fig:eqR-vzoex-a0}
labeled as ``numerical quadrature'' at the right edge.
We clearly see that each of them has similar topological patterns
as those in the top three panels (\mtxtsf{a}, \mtxtsf{b}, \mtxtsf{c}).
There seems to be a slight difference in the locations of local extremums on the $x$- and $y$-axis:
The local extremums in panels \mtxtsf{d}, \mtxtsf{e}, \mtxtsf{f}
in \mysymfigO \ref{fig:eqR-vzoex-a0}
are located closer to the origin compared with the panels
\mtxtsf{a}, \mtxtsf{b}, \mtxtsf{c}.
Readers should note that we adopted the smaller plot ranges
(from $-0.32$ to $+0.32$ for both $x$ and $y$)
in the middle three panels \mtxtsf{d}, \mtxtsf{e}, \mtxtsf{f}
than those in the top three panels
\mtxtsf{a}, \mtxtsf{b}, \mtxtsf{c} (from $-0.45$ to $+0.45$).
This is to show the detailed structure of the equi-$R$ curves
near the origin in the middle three panels.
The slight difference between the result of the numerical quadrature and
that of the analytic expansion is probably caused by the fact that
the expression of
the analytic expansions in Eqs. \eqref{eqn:Z97} and \eqref{eqn:Z98} does not
include higher-order terms such as $R'_7, R'_9, \cdots$ .
We have confirmed that the locations of the local extremums
obtained from the analytic expansion become more similar to those
obtained from the numerical quadrature
when we use expansions of even higher-orders \citep{ito2016}.

We also carried out a set of direct numerical integration of the
equations of motion for the outer CR3BP using the same three parameter sets of $\left(\alpha', k^2\right)$.
The numerical integration scheme is the same as what we used
when drawing \mysymfigO \ref{fig:xy-inner}.
We placed a perturber on a circular orbit with the semimajor axis $a' = 5.2042$ au.
The integration result is presented as
the bottom three panels (\mtxtsf{g}, \mtxtsf{h}, \mtxtsf{i})
of \mysymfigO \ref{fig:eqR-vzoex-a0}
labeled as ``numerical integration'' at the right edge.
  The nominal stepsize of the integration is 20 days, and
  the total integration time is 1200 million years with
  a data output interval of 5000 years.
Note that here we set the ratio of the perturber's mass and the central mass
as $9.5479194 \times 10^{-5}$ which is close to
$\frac{1}{10}$ of the mass ratio between Jupiter and the Sun.
This is to avoid the inclusion of short periodic oscillations in the plots, and
to clearly show the detailed structure of the equi-$R$ contours.
The influence of the mass ratio is limited to the timescale of the orbital evolution of the perturbed body,
and it does not affect the orbit's secular topology
(see the discussion in Section \ref{ssec:CR3BP-averaging} on p. \pageref{pg:limitedmasseffect} of this monograph).

In the panels \mtxtsf{g}, \mtxtsf{h}, \mtxtsf{i} of
\mysymfigO \ref{fig:eqR-vzoex-a0} that show the result of the
direct numerical integration, the contour interval does not
represent fixed intervals of disturbing potential.
We placed the perturbed bodies initially along the $x$-axis on the $(e \cos g, e \sin g)$ plane.
More specifically,
we placed the perturbed bodies along the $x$-axis with an interval of
initial eccentricity ($e_0$) as $\Delta e_0 = 0.02$.
The minimum of $e_0$ in the panel \mtxtsf{g}                 is $e_{0,\mathrm{min}} = 0.02$, and
the minimum of $e_0$ in the panels \mtxtsf{h} and \mtxtsf{i} is $e_{0,\mathrm{min}} = 0.01$.
We also placed several more perturbed bodies
along the $y$-axis around the local minima with the same interval $\Delta e_0$.
As for the initial argument of pericenter $g$,
we basically selected $g=0$ and $\frac{\pi}{2}$,
as well as $g=\pi$ and $\frac{3\pi}{2}$ when necessary.
As we see,
the trajectories obtained from the numerical integration (\mtxtsf{g}, \mtxtsf{h}, \mtxtsf{i}),
those obtained from the numerical quadrature (\mtxtsf{d}, \mtxtsf{e}, \mtxtsf{f}), and 
those obtained from the analytic expansion of the doubly averaged disturbing function (\mtxtsf{a}, \mtxtsf{b}, \mtxtsf{c})
agree well with each other.
Note that some of the trajectories on the bottom two panels
(\mtxtsf{g} and \mtxtsf{h}) are not completely closed.
These trajectories are very close to the separatrix
where it principally takes an infinite amount of time for the perturbed body
to reach the saddle points.
It seems that the integration period that we chose for the numerical
orbit propagation to draw the panels \mtxtsf{g} and \mtxtsf{h} was not long enough
for the trajectories to be closed.

\paragraph{Values of ${k'}^2_{2.0}, {k'}^2_{0.2}, I'_{2.0}, I'_{0.2}$
           (when $\alpha'$ is not small)} %
Now, we move on to the case when $\alpha'$ is not so small.
First, let us calculate the critical inclinations
$I'_{2.0} = \cos^{-1} k'_{2.0}$ and
$I'_{0.2} = \cos^{-1} k'_{0.2}$, as well as their dependence on $\alpha'$.
They were discussed on p. \pageref{pg:def-Id202} of this monograph,
and \citeauthor{vonzeipel1910}'s calculation results are transcribed in our Table \ref{tbl:vZ10-table-5}.
He wrote that the sign of
$I'_{2.0} - I'_{0.2}$ changes from positive to negative at $\alpha' \sim 0.74$
(see p. \pageref{pg:alphad074} of this monograph).
As for the inner CR3BP,
we compared \citeauthor{vonzeipel1910}'s $I_{0.2}$ with
\citeauthor{kozai1962b}'s limiting inclination $i_0$
(see our \mysymfigO \ref{fig:I02-table} and the accompanying discussion on p. \pageref{fig:I02-table}).
As for the outer CR3BP, we calculate the critical inclination values
that are equivalent to $I'_{2.0}$ and $I'_{0.2}$
through numerical quadrature as follows.

For calculating $I'_{2.0}$ and $I'_{0.2}$,
we need to obtain the values of $R$ along the $x$- and $y$-axis
on the $(e \cos g, e \sin g)$ plane.
Considering the symmetry of $R$,
we carry out a series of numerical quadrature defined in Eq. \eqref{eqn:K09} only on
the positive $x$-axis $(g=0)$ and
the positive $y$-axis $(g=\frac{\pi}{2})$
using several parameter sets $\left(\alpha', k^2\right)$.
Then we observe the behavior of $R$, and estimate the sign of
$\DP{R}{x}$ and $\DP{R}{y}$ around the origin $(0,0)$.
As we saw in Section \ref{sssec:oCR3BP-nosmall-alpha},
there can be four patterns in the appearance of local extremums.

\begin{mylist}{1.0em}
\item {\textit{Pattern 1.\/}\label{pg:le-pattern1}} \\
The origin $(0,0)$ makes a local maximum, and
there shows up a pair of local minima  on the $y$-axis,
    as well as a pair of saddle points on the $x$-axis.
This pattern is seen in the left column panels
(\mtxtsf{a}, \mtxtsf{d}, \mtxtsf{g}) of \mysymfigO \ref{fig:eqR-vzoex-a0}.
In this pattern,
$R$ decreases both along the $x$- and $y$-directions from the origin $(0,0)$.
Therefore we have
\begin{equation}
  \left. \DP{R}{x} \right|_{x=y=0} < 0, \quad
  \left. \DP{R}{y} \right|_{x=y=0} < 0 .
  \label{eqn:DPRcond-0}
\end{equation}
\item {\textit{Pattern 2.\/}\label{pg:le-pattern2}} \\
The origin $(0,0)$ makes a saddle point, and
there shows up a pair of local minima only on the $y$-axis.
This pattern is seen in the middle column panels
(\mtxtsf{b}, \mtxtsf{e}, \mtxtsf{h}) of \mysymfigO \ref{fig:eqR-vzoex-a0}.
In this pattern,
$R$ decreases along the $y$-axis, and it increases along the $x$-axis from the origin.
So we have
\begin{equation}
  \left. \DP{R}{x} \right|_{x=y=0} > 0, \quad
  \left. \DP{R}{y} \right|_{x=y=0} < 0 .
  \label{eqn:DPRcond-1}
\end{equation}
\item {\textit{Pattern 3.\/}\label{pg:le-pattern3}} \\
The origin $(0,0)$ makes a local minimum, and
no other local extremum exists.
This pattern is seen in the right column panels
(\mtxtsf{c}, \mtxtsf{f}, \mtxtsf{i}) of \mysymfigO \ref{fig:eqR-vzoex-a0}.
In this pattern,
$R$ monotonically increases from the origin along any directions.
Therefore we have
\begin{equation}
  \left. \DP{R}{x} \right|_{x=y=0} > 0, \quad
  \left. \DP{R}{y} \right|_{x=y=0} > 0 .
  \label{eqn:DPRcond-2}
\end{equation}
\item {\textit{Pattern 4.\/}\label{pg:le-pattern4}} \\
There shows up a pair of local minima on the $x$-axis, and
the origin $(0,0)$ makes a saddle point.
This pattern is what \citeauthor{vonzeipel1910} predicted in his Section Z23
(see p. \pageref{pg:alphad074}),
but no figure has been given yet.
In this pattern,
$R$ would decrease from the origin along the $x$-axis, and
it would increase along the $y$-axis.
Therefore we would have
\begin{equation}
  \left. \DP{R}{x} \right|_{x=y=0} < 0, \quad
  \left. \DP{R}{y} \right|_{x=y=0} > 0 .
  \label{eqn:DPRcond-3}
\end{equation}
\end{mylist}

Note that as long as the approximation is up to $O\left(e^2\right)$ as
\citeauthor{vonzeipel1910} assumed (see p. \pageref{eqn:vZ10-nn384-04}),
the $R$ values along the $x$- and the $y$-axis are equivalent to
$R'_{2.0}$ and $R'_{0.2}$ themselves in Eqs. \eqref{eqn:Z99} and \eqref{eqn:Z100},
respectively. In other words, we have in this approximation
\begin{alignat}{1}
   \left. \DP{R}{x}        \right|_{x=y=0}
&= \left. \DP{R'_{2.0}}{x} \right|_{x=y=0} ,
\label{eqn:dpRx=dpR202x} \\
   \left. \DP{R}{y}        \right|_{x=y=0}
&= \left. \DP{R'_{0.2}}{y} \right|_{x=y=0} .
\label{eqn:dpRy=dpR202y}
\end{alignat}

\begin{figure*}[htbp]\centering
\ifepsfigure
 \includegraphics[width=\dualfigwidth\textwidth]{RxRy.eps}%fig20
\else
 \includegraphics[width=\dualfigwidth\textwidth]{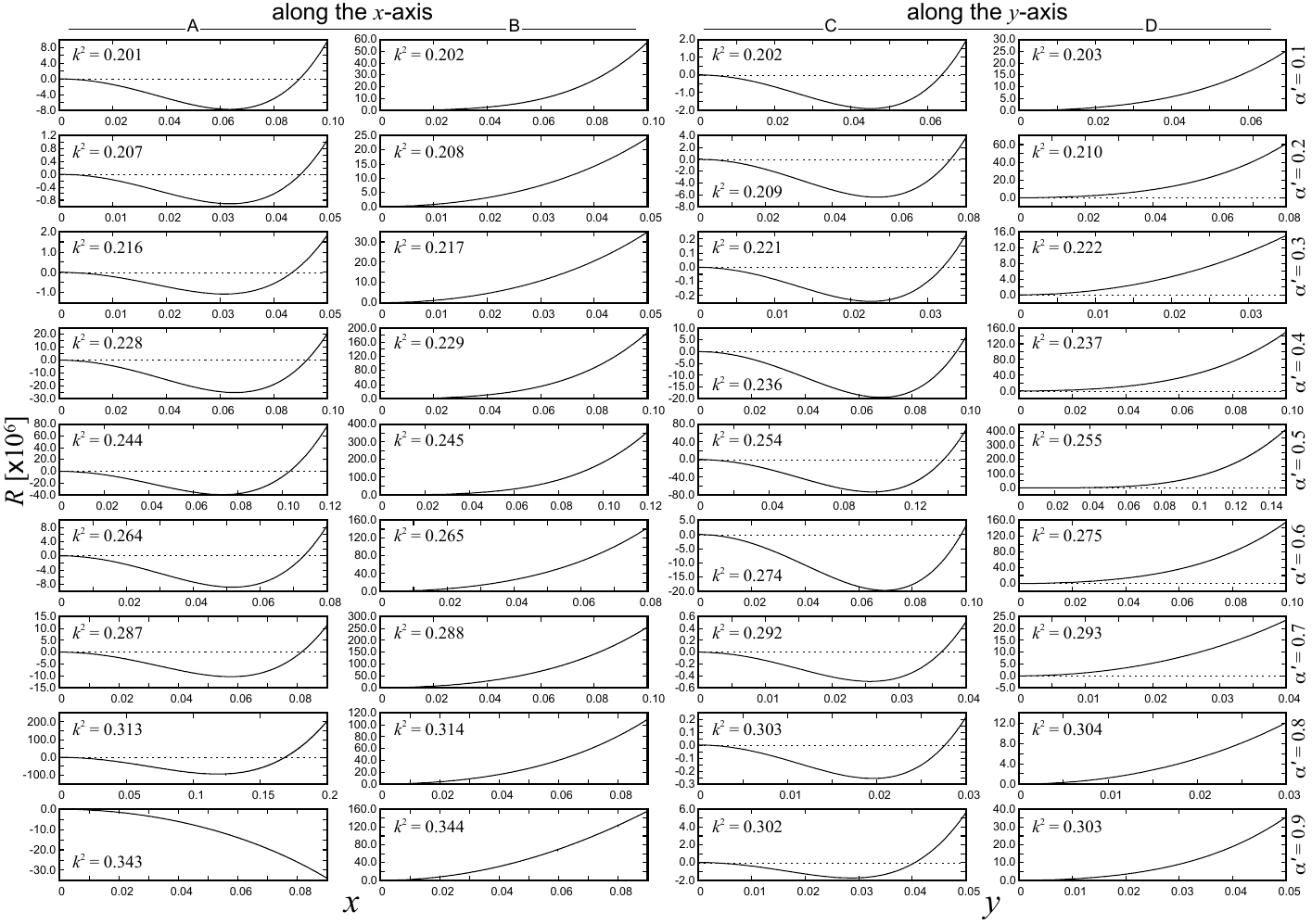}%fig20
\fi
  \caption{%
  Values of the doubly averaged outer disturbing function $R$
  obtained through the numerical quadrature \protect\eqref{eqn:K09}
  along the positive $x$-axis $(g=0)$ and
  along the positive $y$-axis $(g=\frac{\pi}{2})$
  with several $(\alpha', k^2)$ sets.
  The rows are, from the top, for $\alpha' = 0.1, 0.2, 0.3, \ldots , 0.9$.
  The left  two columns (\mtxtsf{A} and \mtxtsf{B}) are for $R$ along the $x$-axis, and
  the right two columns (\mtxtsf{C} and \mtxtsf{D}) are for $R$ along the $y$-axis.
  Note that the $R$ values are magnified by $10^6$ in all the panels:
  Their original magnitude is as small as $O(10^{-6})$ to $O(10^{-4})$
  in our calculation.
  See \mysymfigO \protect{\ref{fig:Rx-a09}} for comparison.
  }
  \label{fig:RxRy}
\end{figure*}

The purpose of our series of numerical quadrature is to investigate
which set of $\left(\alpha', k^2\right)$ realizes which of the above patterns
1, 2, 3, 4 (or the conditions
\eqref{eqn:DPRcond-0},
\eqref{eqn:DPRcond-1},
\eqref{eqn:DPRcond-2},
\eqref{eqn:DPRcond-3}).
One of \citeauthor{vonzeipel1910}'s theoretical predictions
that we learned in Section \ref{sssec:oCR3BP-nosmall-alpha} tells us the following:
\begin{itemize}
\item The condition \eqref{eqn:DPRcond-0} is satisfied
  while $k^2$ is smaller than both ${k'}^2_{2.0}$ and ${k'}^2_{0.2}$.
(\textit{Pattern 1\/})
\item When ${k'}^2_{2.0} < {k'}^2_{0.2}$,
if we fix $\alpha'$ and gradually increase the value of $k^2$,
we would reach a point where the status of the system changes
from the condition \eqref{eqn:DPRcond-0}
  to the condition \eqref{eqn:DPRcond-1}.
The value of $k^2$ at this point would be equal to ${k'}^2_{2.0}$, and
the limiting inclination $I'_{2.0}$ is calculated as $I'_{2.0} = \cos^{-1} k'_{2.0}$.
(\textit{Pattern 2\/})
\item If we further increase $k^2$ from this point,
we would reach another point where the status of the system changes
from the condition \eqref{eqn:DPRcond-1}
  to the condition \eqref{eqn:DPRcond-2}.
The value of $k^2$ at this point would be equal to ${k'}^2_{0.2}$,
and the limiting inclination $I'_{0.2}$ is calculated as $I'_{0.2} = \cos^{-1} k'_{0.2}$.
(\textit{Pattern 3\/})
\item
If ${k'}^2_{2.0} > {k'}^2_{0.2}$ is realized when $\alpha'$ is large
(as \citeauthor{vonzeipel1910} predicted),
the condition \eqref{eqn:DPRcond-3} can take place
in a certain range of $k^2$.
(\textit{Pattern 4\/})
\end{itemize}

Bearing the above circumstances in mind,
we calculated the values of $R$ along the $x$- and $y$-axis
through the series of numerical quadrature.
As for the value of $\alpha'$,
we varied it from $\alpha'=0$ to $0.95$ in steps of $0.05$.
As for the value of $k^2$,
we started from $k^2 = 0.200$ and increased it in steps of $0.001$.
We plotted the resulting values of $R$ along the $x$- and $y$-axis
with several $\alpha'$ and $k^2$ in \mysymfigO \ref{fig:RxRy}.

In \mysymfigO \ref{fig:RxRy},
the panels in the left  two columns depict $R$ along the $x$-axis just before (the column \mtxtsf{A}) and just after (the column \mtxtsf{B})
$\left. \DP{R}{x} \right|_{x=y=0}$ changes its sign.
The panels in the right two columns depict $R$ along the $y$-axis just before (the column \mtxtsf{C}) and just after (the column \mtxtsf{D})
$\left. \DP{R}{y} \right|_{x=y=0}$ changes its sign.
For example, when $\alpha'=0.7$ (see the seventh row from the top),
$\left. \DP{R}{x} \right|_{x=y=0}$ seems negative while $k^2 \leq 0.287$ (the panel in the column \mtxtsf{A})
but it seems to turn positive while $k^2 \geq 0.288$
(the panel in the column \mtxtsf{B}).
Thus, we conclude that the ${k'}^2_{2.0}$ value lies somewhere between 0.287 and 0.288 when $\alpha'=0.7$.
Similarly,
$\left. \DP{R}{y} \right|_{x=y=0}$ seems negative while $k^2 \leq 0.292$ (the panel in the column \mtxtsf{C})
but it seems to turn positive while $k^2 \geq 0.293$
(the panel in the column \mtxtsf{D}).
Thus, we conclude that the ${k'}^2_{0.2}$ value lies somewhere between 0.292 and 0.293 when at $\alpha'=0.7$.

\begin{figure}[t]\centering
\ifepsfigure
 \includegraphics[width=\singlefigwidth\textwidth]{Rx_a09.eps} %fig21
\else
 \includegraphics[width=\singlefigwidth\textwidth]{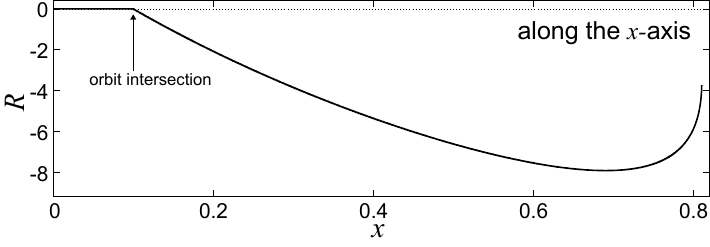} %fig21
\fi
  \caption{%
  An extended plot of the bottom panel in the left-most column
  (\mtxtsf{A}) of \mysymfigO \protect{\ref{fig:RxRy}}.
  This shows the changing values of $R$ along the $x$-axis when $(\alpha', k^2) = (0.9, 0.343$).
  The location of orbit intersection of the perturbing body and the perturbed body
  at $x \sim 0.09$ is indicated by an arrow.
  Note that, unlike in \mysymfigO \protect{\ref{fig:RxRy}},
  the vertical scale is not magnified in this panel.
  This means that the magnitude of $R$
                       is $O(10^{-5})$ when $x \lesssim 0.09$,
              while it is $O(1)$       when $x \gtrsim  0.09$.
  }
  \label{fig:Rx-a09}
\end{figure}

Looking at the panels in the columns \mtxtsf{A} and \mtxtsf{C} of
\mysymfigO \ref{fig:RxRy}, we can visually locate the local minima of $R$ with our eyes.
However, we do not see any local extremums in the bottom panel of the column \mtxtsf{A} when $(\alpha', k^2) =(0.9, 0.343)$.
$\left. \DP{R}{x} \right|_{x=y=0}$ seems to remain negative throughout this panel.
From the analogy of other panels in this column,
we may want to imagine the existence of a local minimum somewhere along the $x$-axis beyond the range of this plot.
However, we cannot actually identify the location of a local minimum in this case.
This is because, in the large $\alpha'$ range such as $\alpha'=0.9$,
the line of orbit intersection takes place relatively close to the origin, $(0,0)$.
To better illustrate the circumstance, we extended the $x$-range of
this panel and redrew it as \mysymfigO \ref{fig:Rx-a09}.
In this figure, we see that
the line of orbit intersection happens just before $x=0.1$, and
$R$ shows a steep and rapid change beyond this point.
Readers should pay attention to the huge magnitude difference of $R$
between inside and outside this border.
See Section \ref{ssec:orbitintersection} of this monograph
(p. \pageref{ssec:orbitintersection})
for more detail about the local extremums that occur beyond the line of orbit intersection.

\begin{figure}[t]\centering
\ifepsfigure
 \includegraphics[width=\singlefigwidth\textwidth]{Id202comp.eps} %fig22
\else
 \includegraphics[width=\singlefigwidth\textwidth]{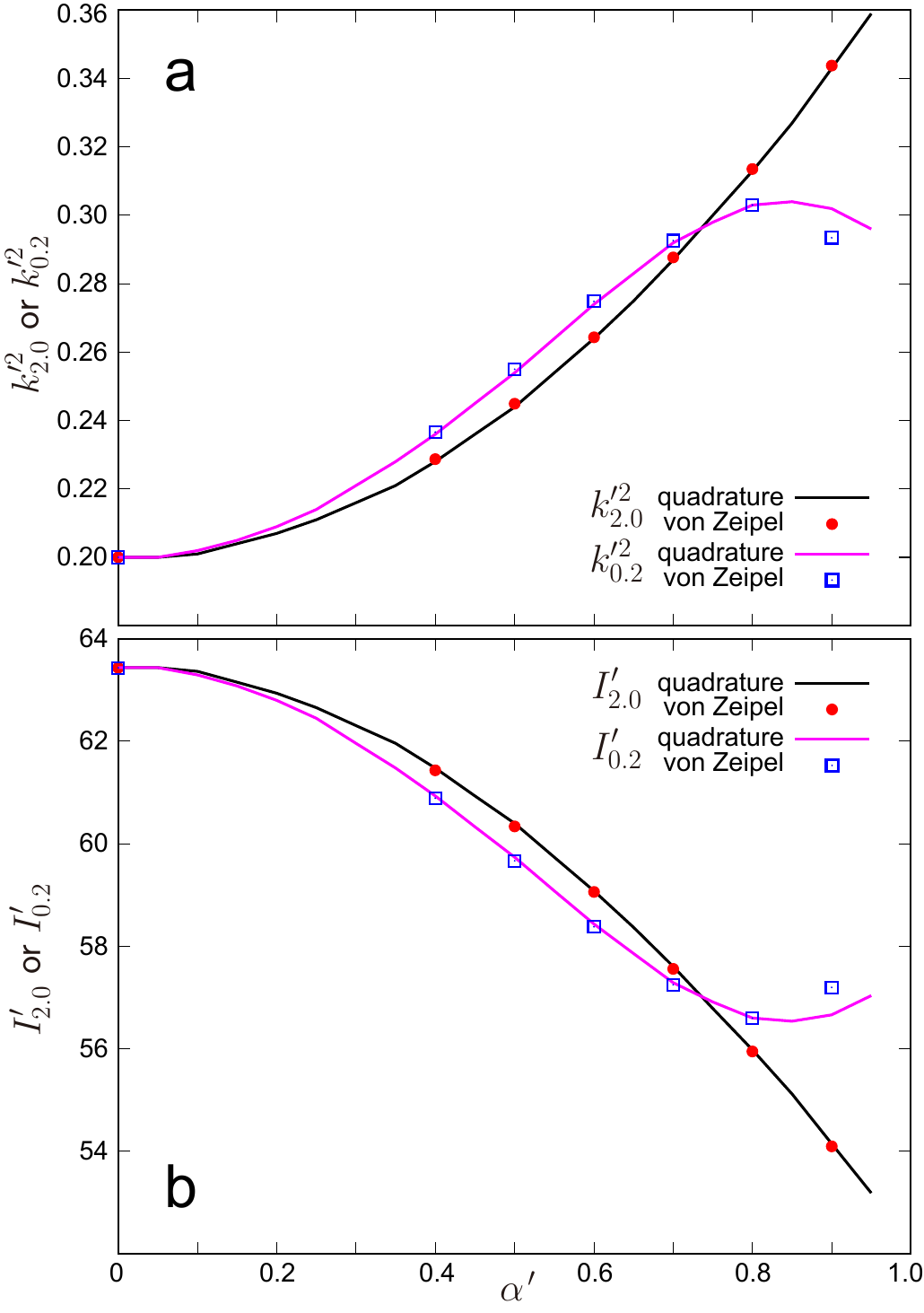} %fig22
\fi
  \caption{%
\mtxtsf{a}: Actual values of
${k'}^2_{2.0}$ (the black solid line) and
${k'}^2_{0.2}$ (the magenta solid line) and their dependence on
$\alpha'$ estimated from our numerical quadrature.
We also converted \citeauthor{vonzeipel1910}'s calculation result
on $I'_{2.0}$ and $I'_{0.2}$ (see Table \ref{tbl:vZ10-table-5})
into ${k'}^2_{2.0} = \cos^2 I'_{2.0}$ (the filled red circle)
and  ${k'}^2_{0.2} = \cos^2 I'_{0.2}$ (the open blue square with a central dot),
and plotted them in this panel.
\mtxtsf{b}: The corresponding values of
$I'_{2.0}$ (the black solid line) and
$I'_{0.2}$ (the magenta solid line) converted from
${k'}^2_{2.0}$ and ${k'}^2_{0.2}$ plotted in the panel \mtxtsf{a}.
\citeauthor{vonzeipel1910}'s numerical estimate summarized in Table \ref{tbl:vZ10-table-5} is also plotted:
$I'_{2.0}$ (the filled red circle) and
$I'_{0.2}$ (the open blue square with a central dot).
}
  \label{fig:Id202comp}
\end{figure}

Using the numerical quadrature result that is partly presented in \mysymfigO \ref{fig:RxRy},
we calculated ${k'}^2_{2.0}$ and ${k'}^2_{0.2}$,
   as well as $I'_{2.0} (= \cos^{-1} k'_{2.0})$ and
              $I'_{0.2} (= \cos^{-1} k'_{0.2})$,
all as functions of $\alpha'$.
Our calculation result is summarized in \mysymfigO \ref{fig:Id202comp}
together with the values that \citeauthor{vonzeipel1910} presented
for comparison.
\mysymfigO \ref{fig:Id202comp}\mtxtsf{a} shows the dependence of ${k'}^2_{2.0}$ and ${k'}^2_{0.2}$ on $\alpha'$.
We plotted \citeauthor{vonzeipel1910}'s ${k'}^2_{2.0}$ and ${k'}^2_{0.2}$
in the range of $0.4 \leq \alpha' \leq 0.9$.
We obtained them by converting his $I'_{2.0}$ and $I'_{0.2}$
tabulated in Table \ref{tbl:vZ10-table-5}
into ${k'}^2_{2.0}$ and ${k'}^2_{0.2}$.
\mysymfigO \ref{fig:Id202comp}\mtxtsf{b} shows the dependence of $I'_{2.0}$ and $I'_{0.2}$ on $\alpha'$.
\citeauthor{vonzeipel1910}'s $I'_{2.0}$ and $I'_{0.2}$ are extracted
from Table \ref{tbl:vZ10-table-5} and plotted in this panel.
These panels clearly indicate that the agreement between
\citeauthor{vonzeipel1910}'s result and our numerical result is excellent,
with just a slight difference when $\alpha' = 0.90$.
This figure exemplifies the correctness of \citeauthor{vonzeipel1910}'s
theory and his calculations on the doubly averaged outer CR3BP,
even when $\alpha'$ is not small.
To our knowledge, this type of quantitative comparison has never been done with the doubly averaged outer CR3BP over this extensive range of $\alpha'$.

\begin{figure*}[htbp]\centering
\ifepsfigure
 \includegraphics[width=\dualfigwidth\textwidth]{eqR_vzoex_aL.eps} %fig23
\else
 \includegraphics[width=\dualfigwidth\textwidth]{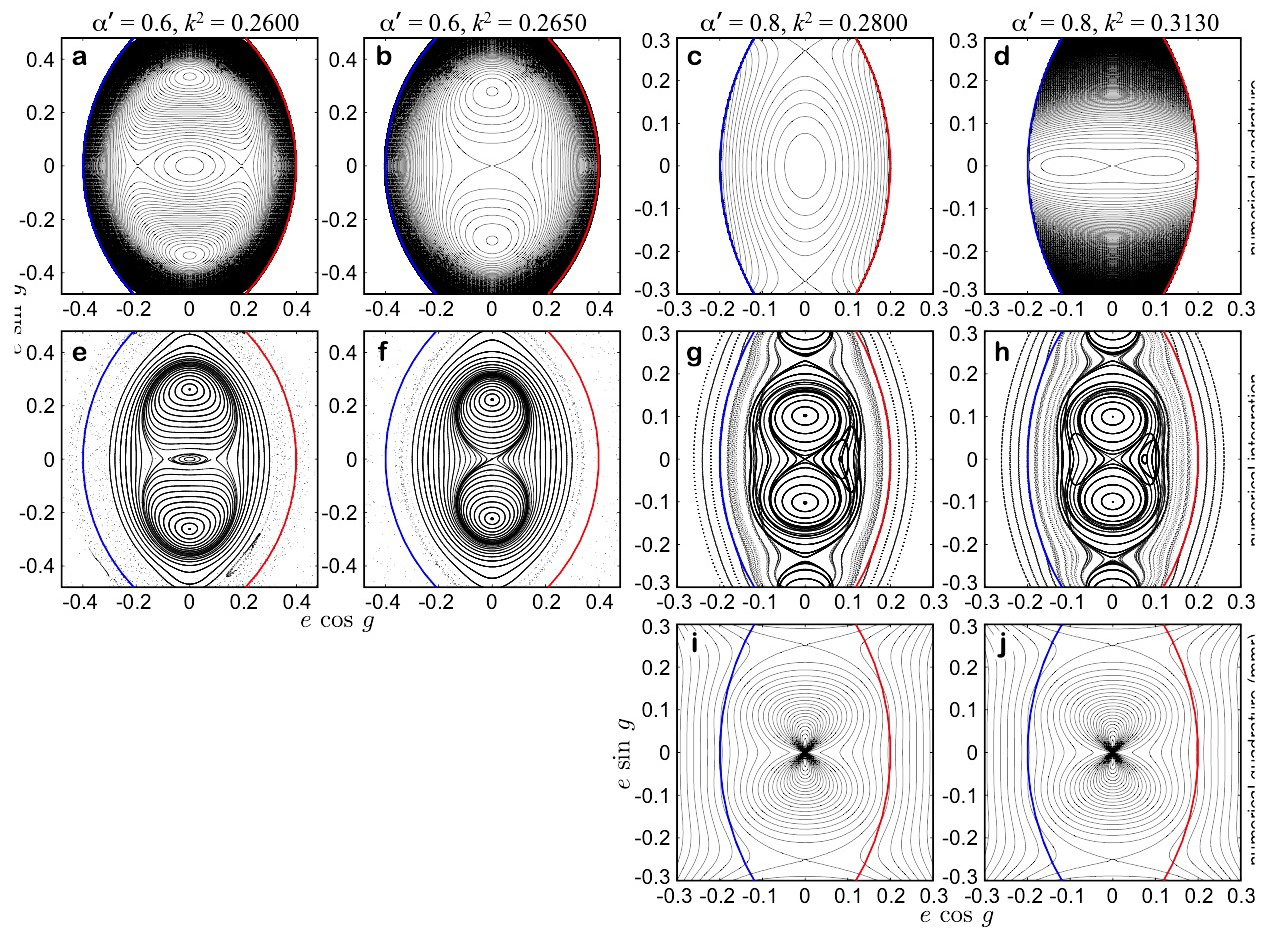} %fig23
\fi
  \caption{%
Equi-potential contours of the doubly averaged outer CR3BP
on the $(x,y)=(e \cos g, e \sin g)$ plane when $\alpha'$ is not small.
The four panels in the top row
(\mtxtsf{a}, \mtxtsf{b}, \mtxtsf{c}, \mtxtsf{d}) are the results
obtained from the numerical quadrature defined by Eq. \eqref{eqn:K09}.
The four panels in the middle row (\mtxtsf{e}, \mtxtsf{f}, \mtxtsf{g}, \mtxtsf{h}) are the results
obtained from the direct numerical integration of the equations of motion.
The two panels in the bottom row (\mtxtsf{i}, \mtxtsf{j}) are the results
obtained from the numerical quadrature that assumes
the 5:7 mean motion resonance between the perturbed and perturbing bodies with a fixed critical argument of
$5\lambda' - 7\lambda +2g = \pi$.
The columns are categorized by the parameter set $(\alpha', k^2)$.
We chose
$(\alpha', k^2) = (0.6, 0.2600$) for the leftmost column (the panels \mtxtsf{a}, \mtxtsf{e}),
$(\alpha', k^2) = (0.6, 0.2650$) for the second left column (the panels \mtxtsf{b}, \mtxtsf{f}),
$(\alpha', k^2) = (0.8, 0.2800$) for the second right column (the panels \mtxtsf{c}, \mtxtsf{g}, \mtxtsf{i}), and
$(\alpha', k^2) = (0.8, 0.3130$) for the rightmost column (the panels \mtxtsf{d}, \mtxtsf{h}, \mtxtsf{j}).
Note that the $x$- and $y$-ranges are from $-0.48$ to $+0.48$ only
in the $2 \times 2$ panels in the left two columns
(\mtxtsf{a}, \mtxtsf{b}, \mtxtsf{e}, \mtxtsf{f}),
while they are from $-0.30$ to $+0.30$
in the $2 \times 3$ panels in the right two columns
(\mtxtsf{c}, \mtxtsf{d}, \mtxtsf{g}, \mtxtsf{h}, \mtxtsf{i}, \mtxtsf{j}).
The partial red circles in each of the panels correspond to the negative sign
in the first term of the left-hand side of Eq. \eqref{eqn:Z45},
representing the location where the orbits of the perturbed and perturbing bodies
intersect each other at the ascending node of the perturbed body.
The partial blue circles correspond to the positive sign
in the first term of the left-hand side of Eq. \eqref{eqn:Z45},
representing the location where the orbits of the perturbed and perturbing bodies
intersect each other at the descending node of the perturbed body.
}
  \label{fig:eqR-vzoex-aL}
\end{figure*}

\paragraph{Equi-potential contours (when $\alpha'$ is not small)} %
From the results presented in
\mysymfigS \ref{fig:RxRy}, \ref{fig:Rx-a09}, and \ref{fig:Id202comp},
we have confirmed that \citeauthor{vonzeipel1910}'s theory
seems valid and accurate in the doubly averaged outer CR3BP,
even when $\alpha'$ is not small.
Let us further move on and make another confirmation in this case.
In what follows, our discussion goes along with a new figure
(\mysymfigO \ref{fig:eqR-vzoex-aL}).
This figure is a counterpart of \mysymfigO \ref{fig:eqR-vzoex-a0}
that we had made for the small $\alpha'$ systems.
The new figure contains ten panels,
showing various trajectories of perturbed bodies in the outer CR3BP
on the $(x,y)=(e \cos g, e \sin g)$ plane.
We employed two methods to draw the panels:
Numerical quadrature of two kinds, and
direct numerical integration of the equations of motion.

We first set $\alpha' = 0.6$, and chose relatively small $k^2$ values
($k^2 = 0.2600$ and $k^2 = 0.2650$).
The $2 \times 2$ panels in the left two columns
(\mtxtsf{a}, \mtxtsf{b}, \mtxtsf{e}, \mtxtsf{f})
of \mysymfigO \ref{fig:eqR-vzoex-aL} show the
equi-potential contours in this parameter set.
The panels \mtxtsf{a} and \mtxtsf{b} are based on the result obtained from numerical quadrature, and
the panels \mtxtsf{e} and \mtxtsf{f} are based on the result obtained from direct numerical integration of the equations of motion.
The methods of numerical quadrature and numerical integration are the same
as what we used in \mysymfigO \ref{fig:eqR-vzoex-a0}.
We placed a perturber on a circular orbit with the semimajor axis $a' = 5.2042$ au.
In the numerical integration result shown in the panels \mtxtsf{e} and \mtxtsf{f},
  the nominal stepsize is 20 days, and
  the total integration time is 200 million years with
  a data output interval of 1000 years.
The ratio of the perturber's mass and the central mass is
$4.7739597 \times 10^{-5}$ which is close to
$\frac{1}{20}$ of the mass ratio of Jupiter and the Sun.

The left $2 \times 2$ panels
(\mtxtsf{a}, \mtxtsf{b}, \mtxtsf{e}, \mtxtsf{f})
of \mysymfigO \ref{fig:eqR-vzoex-aL}
reproduce well the topological characteristics of the equi-$R$ curves
that \citeauthor{vonzeipel1910} predicted.
From the preceding calculation result presented in \mysymfigO \ref{fig:RxRy},
when $\alpha' = 0.6$ we found
${k'}^2_{2.0}$ lies somewhere in $\mathopen{]}0.264, 0.265\mathclose{[}$, and
${k'}^2_{0.2}$ lies somewhere in $\mathopen{]}0.274, 0.275\mathclose{[}$.
Therefore,
the plots with $k^2 = 0.260$ (panels \mtxtsf{a} and \mtxtsf{e})
correspond to the topological pattern of $R$ when $k^2 < {k'}^2_{2.0}$
(Pattern 1 on p. \pageref{pg:le-pattern1}).
In this case
the origin $(0,0)$ makes a local maximum, and we see
a pair of local minima  along the $y$-axis together with
a pair of saddle points along the $x$-axis.
The topological pattern looks the same as that observed in the $\alpha' \ll 1$ case
(\mysymfigS \ref{fig:eqR-vzoex-a0}\mtxtsf{a},
         \ref{fig:eqR-vzoex-a0}\mtxtsf{d},
         \ref{fig:eqR-vzoex-a0}\mtxtsf{g}),
although the locations of
the local extremums along the $x$- and $y$-axis seem slightly different.
On the other hand,
the plots with $k^2 = 0.265$ (panels \mtxtsf{b} and \mtxtsf{f})
correspond to the topological pattern of $R$ when ${k'}^2_{2.0} < k^2 < {k'}^2_{0.2}$
(Pattern 2 on p. \pageref{pg:le-pattern2}).
In this case,
we see a saddle point at the origin $(0,0)$ as well as
a pair of local minima along the $y$-axis.
Again, the topological pattern seems the same as that observed in the $\alpha' \ll 1$ case
(\mysymfigS \ref{fig:eqR-vzoex-a0}\mtxtsf{b},
         \ref{fig:eqR-vzoex-a0}\mtxtsf{e},
         \ref{fig:eqR-vzoex-a0}\mtxtsf{h}).

So far, the predictions that \citeauthor{vonzeipel1910} made about the
dynamical behavior of the perturbed body in the doubly averaged outer CR3BP seem correct.
However if we go on and further increase $\alpha'$,
the situation changes drastically:
We find a limitation, not only of \citeauthor{vonzeipel1910}'s predictions,
but of the double averaging approximation itself---the influence of mean motion resonance emerges.

\clearpage

As an example of our further investigation,
we chose two more parameter sets of 
$(\alpha', k^2) = (0.8, 0.2800)$  and
$(\alpha', k^2) = (0.8, 0.3130)$,
and again carried out numerical quadrature.
The calculation results are presented in the upper two panels in the right
two columns of \mysymfigO \ref{fig:eqR-vzoex-aL}
(the panels \mtxtsf{c} and \mtxtsf{d}).
From the preceding calculation result presented in \mysymfigO \ref{fig:RxRy},
when $\alpha' = 0.8$ we found
${k'}^2_{2.0}$ lies somewhere in $\mathopen{]}0.313, 0.314\mathclose{[}$, and
${k'}^2_{0.2}$ lies somewhere in $\mathopen{]}0.303, 0.304\mathclose{[}$.
Note that we have ${k'}^2_{2.0} > {k'}^2_{0.2}$ at this large value of $\alpha'$
(see \mysymfigO \ref{fig:Id202comp}) which is a clear difference from the systems with small $\alpha'$
that satisfy the opposite condition, Eq. \eqref{eqn:vZ10-kd20llkd02}.
This collectively means that
the value $k^2 = 0.280$ employed in the panel \mtxtsf{c} is in the range of $k^2 < {k'}^2_{0.2}$, while
the value $k^2 = 0.313$ employed in the panel \mtxtsf{d} is (although marginally) in the range of ${k'}^2_{0.2} < k^2 < {k'}^2_{2.0}$.

From the analogy of the result in the small $\alpha'$ systems
where ${k'}^2_{2.0} < {k'}^2_{0.2}$ is satisfied,
we may want to anticipate that the topological pattern of the equi-$R$ curves
in the systems with ${k'}^2_{2.0} > {k'}^2_{0.2}$ would be obtained by
a $90^\circ$ rotation of the case of ${k'}^2_{2.0} < {k'}^2_{0.2}$.
This anticipation is mostly correct, and
the circumstance is realized in our numerical quadrature:
\mysymfigO \ref{fig:eqR-vzoex-aL}\mtxtsf{c} is close to a $90^\circ$ topological rotation of \mysymfigO \ref{fig:eqR-vzoex-aL}\mtxtsf{a}, and
\mysymfigO \ref{fig:eqR-vzoex-aL}\mtxtsf{d} is close to a $90^\circ$ topological rotation of \mysymfigO \ref{fig:eqR-vzoex-aL}\mtxtsf{b}.
The truncation of equi-$R$ curves by the red and blue lines
in the panels \mtxtsf{c} and \mtxtsf{d} (the lines representing orbit intersection)
produces an apparent difference between them.
For example, in the panel \mtxtsf{c}
we do not see any local minima along the $x$-axis
that are supposed to correspond to the local minima
observed in the panel \mtxtsf{a} along its $y$-axis.
Therefore,
at this point we may want to believe that the following conjecture
that \citeauthor{vonzeipel1910} stated at the end of his Section Z23
(p. \pageref{pg:alphad074} of this monograph) is true:
\begin{quote}
``If instead $\alpha' > 0.74 \cdots$, it is probable that the two symmetric minima are on the axis of $x$, at least for the appropriate values of ${k'}^2 < {k'}^2_{2.0}$.''
 (p. Z405)
\end{quote}

However,
the motion of the perturbed body in the actual (i.e. not averaged) CR3BP
sometimes exhibits very different behavior from the averaged system.
The panels \mtxtsf{g} and \mtxtsf{h} of \mysymfigO \ref{fig:eqR-vzoex-aL} are
typical examples of this sort. These panels are drawn through the results
obtained from our direct numerical integration of the equations of motion.
The parameters $(\alpha', k^2)$ used in the panels
\mtxtsf{g} and \mtxtsf{h} are identical to those used in the panels
\mtxtsf{c} and \mtxtsf{d}, respectively.
Here is the detail of the numerical integration
used to draw the panels \mtxtsf{g} and \mtxtsf{h}:
  the nominal stepsize is 4 days, and
  the total integration time is 30 million years with
  a data output interval of 500 years.
The ratio of the perturber's mass and the central mass is
$9.5479194 \times 10^{-6}$ which is close to
$\frac{1}{100}$ of the mass ratio of Jupiter and the Sun.
The reason for the small mass ratio is to avoid the occurrence of
orbital instability of the perturbed body due to close encounters with
the perturber near the orbit intersection lines.
This is necessary because now $\alpha'$ is as large as 0.8.
We also chose a small value for the stepsize of numerical integration
(4 days) for the same reason.

As is very clear in both the panels \mtxtsf{g} and \mtxtsf{h},
stationary points (local minima) show up along the $y$-axis.
We do not see them in the panels \mtxtsf{c} or \mtxtsf{d} that are
obtained from the result of numerical quadrature of the doubly averaged system.
The overall trajectory shapes are very different
between the panels \mtxtsf{c} and \mtxtsf{g}, and also
between the panels \mtxtsf{d} and \mtxtsf{h}.
In short, the doubly averaged approximation does not seem to work out
in the system sampled here.

As far as we have investigated, the trajectory shape seen in
\mysymfigS \ref{fig:eqR-vzoex-aL}\mtxtsf{g} and 
        \ref{fig:eqR-vzoex-aL}\mtxtsf{h} are
related to the existence of the 5:7 mean motion resonance
between the perturbing and perturbed bodies.
The perturbed body's secular motion in R3BP
with mean motion resonance has been intensively studied since
the mid-twentieth century \citep[see an extensive review by][]{moons1996}.
Their relation to the
\citeauthor{lidov1961}--\citeauthor{kozai1962b} {\mainword},
as well as several practical methods
to draw equi-$R$ contours for resonant systems, has been further developed
\citep[e.g.][]{kozai1985,yoshikawa1989,yoshikawa1990,yoshikawa1991,morbidelli2002,gomes2005b,gallardo2006a,gomes2011}.
For visualizing the influence of the mean motion resonance in this case,
we implemented a simple method devised by \citet{kozai1985} which is
based on a more general theory on the secular motion of perturbed body
in resonant dynamical systems \citep{giacaglia1968,giacaglia1969}.
The method of \citet{kozai1985} enables us to reduce the degrees of freedom
of a resonant system to unity and draw its equi-$R$ trajectory.
This method assumes that the critical argument $\sigma$ of mean motion resonance stays at a stable equilibrium,
and always takes a single fixed value without an amplitude.
In general, the critical argument $\sigma$ in CR3BP is defined as
\begin{equation}
  \sigma = j_1 \lambda' + j_2 \lambda + j_3 g,
  \label{eqn:def-criticalangle}
\end{equation}
where $\lambda'$ is the mean longitude of the perturbing body, and
$\lambda$ is that of the perturbed body.
We also have a relationship $j_1 + j_2 + j_3 = 0$,
one of the so-called d'Alembert rules \citep[e.g.][]{hamilton1994,murray1999}.
It is needless to say that $\sigma$ is not a constant in general resonant systems:
Usually $\sigma$ exhibits libration with a certain amplitude and period.
\citeauthor{kozai1985}'s \citeyearpar{kozai1985} method does not consider
the libration amplitude of $\sigma$
(i.e. it assumes the libration amplitude of $\sigma$ to be zero) for simplicity.
This makes the calculation of doubly averaged disturbing function easier.

By inspecting the time variation of the perturbed body's orbital elements in the systems presented in
\mysymfigS \ref{fig:eqR-vzoex-aL}\mtxtsf{g} and
        \ref{fig:eqR-vzoex-aL}\mtxtsf{h},
we confirmed that an argument $5\lambda' - 7\lambda +2g$ stays around $\pi$,
although it oscillates with a certain amplitude.
Therefore we presume that
\begin{equation}
  \sigma = 5\lambda' - 7\lambda +2g = \pi,
  \label{eqn:def-criticalangle-5to7}
\end{equation}
holds true in this case, and
we carried out the numerical quadrature \eqref{eqn:K09}
under the constraint of Eq. \eqref{eqn:def-criticalangle-5to7},
following \citeauthor{kozai1985}'s \citeyearpar{kozai1985} method.
We show the resulting equi-potential trajectories in
\mysymfigS \ref{fig:eqR-vzoex-aL}\mtxtsf{i} and
        \ref{fig:eqR-vzoex-aL}\mtxtsf{j}.
Although the approximation is crude,
topological patterns of equi-potential curves seen in
\mysymfigS \ref{fig:eqR-vzoex-aL}\mtxtsf{i} and
        \ref{fig:eqR-vzoex-aL}\mtxtsf{j}
seem similar to those in
\mysymfigS \ref{fig:eqR-vzoex-aL}\mtxtsf{g} and
        \ref{fig:eqR-vzoex-aL}\mtxtsf{h}:
The origin $(0,0)$ is a saddle point, and
another pair of saddle points shows up at $y \sim \pm 0.25$.
Hence we believe this model explains the situation.
A more sophisticated method for secular dynamics
when mean motion resonance is at work,
including the influence of $\sigma$'s time variation,
will better reproduce the behavior of perturbed bodies exhibited in
\mysymfigO \ref{fig:eqR-vzoex-aL}\mtxtsf{g} and \ref{fig:eqR-vzoex-aL}\mtxtsf{h}.
For recent, more advanced studies on
the perturbed body's secular motion in R3BP with mean motion resonances, consult
\citet{gallardo2012}, \citet{brasil2014}, or
\citet{saillenfest2016,saillenfest2017a,saillenfest2017b}.

Incidentally,
note that the influence of mean motion resonance exists
not only in the outer problem but also in the inner problem,
although we do not give any examples.
\citet{kozai1985} showed several examples along this line.
However, we suspect the influence of mean motion resonance is
more prominent in the outer problem than in the inner problem.
This is because of the subtlety of the secular perturbation that causes the 
\citeauthor{lidov1961}--\citeauthor{kozai1962b} {\mainword}
in the outer problem
compared with that in the inner problem.
We had mentioned this point on p. \pageref{pg:outerissubtle} of this monograph.

As we have seen in this subsection,
the conjectures that \citeauthor{vonzeipel1910} made
for the doubly averaged outer CR3BP are largely justified through our numerical confirmation,
except when major mean motion resonance is at work.
The influence of mean motion resonance such as seen in
\mysymfigS \ref{fig:eqR-vzoex-aL}\mtxtsf{i} and
        \ref{fig:eqR-vzoex-aL}\mtxtsf{j} is relatively more apparent
near the origin $(0,0)$ where the gradient of $R$ is small.
On the other hand the influence seems weaker in regions further away from the origin,
and the trajectories of the perturbed bodies largely follow
what the doubly averaged disturbing function forecasts,
particularly near and beyond the lines of orbit intersection (i.e. the red and blue curves in \mysymfigS \ref{fig:eqR-vzoex-aL}).

\subsection{Cases of orbit intersection\label{ssec:orbitintersection}}
We would say that, up to the previous subsection (Section \ref{ssec:ocr3bp})
we have already written most of the things that we intended to write
in this monograph.
From here until the end of Section \ref{sec:vonzeipel},
we mention some of the additional features in
\citeauthor{vonzeipel1910}'s work that are worth noting.
Compared with what we have already reviewed,
the later part (Sections Z24--Z29) of \citeauthor{vonzeipel1910}'s paper
seems under development and largely remaining speculative,
relying on rough approximations and assumptions.
\citeauthor{vonzeipel1910}'s descriptions get lame and very terse,
such as using a new function without giving its definition.
Therefore, in what follows we do not present much detail of \citeauthor{vonzeipel1910}'s mathematical expositions;
we just introduce his major conclusions.

\citeauthor{vonzeipel1910}'s Section Z24 (pp. Z405--Z413) is devoted to
a series of mathematical demonstration in order to show that
the doubly averaged disturbing function $R$ for CR3BP can have local minima
in the third case that he mentioned in his Section Z16
(p. Z378, also p. \pageref{pg:orbit-threecases} of this monograph)---when
the two orbits intersect and they ``behave like rings of a chain.''
This includes a comprehensive study of $R$ in the domain $B$ or $B'$ in \mysymfigO \ref{fig:vZ10-f1-5},
assuming that $k'$ $\bigl( = \sqrt{1-k^2} \bigr)$ is small
and also $\alpha' - 1$ (or its absolute value) is small.

\citeauthor{vonzeipel1910}'s major goals in Section Z24 seem to be as follows:
First, expressing $R$ in the domain $B$ or $B'$ of \mysymfigO \ref{fig:vZ10-f1-5}
using some holomorphic functions,
and second, seeking all the solutions of the equations $\DP{R}{e} = \DP{R}{g} = 0$ located in the domain $B$ or $B'$.
This section begins with an expression of $R$ in Eq. \eqref{eqn:Z93}
using a single integral in the complex domain.
\citeauthor{vonzeipel1910} gives the expression
\begin{equation}
   R = \frac{\alpha'}{2\pi\sqrt{-1}}\int_{|z=1|}\frac{r}{\sqrt{1+r^2}}F(\tau)\frac{dz}{z} ,
  \tag{Z110-\arabic{equation}}
  \stepcounter{equation}
  \label{eqn:Z110}
\end{equation}
where $\sqrt{-1}$ denotes the imaginary unit, and
$z$ is a complex variable defined as
\begin{equation}
\!\!\!\!\!\!\!
  z = \exp \sqrt{-1} u ,
  \tag{Z109-\arabic{equation}}
  \stepcounter{equation}
  \label{eqn:Z109}
\end{equation}
with eccentric anomaly $u$.
The variable $\tau$ and the function $F(\tau)$ seen in Eq. \eqref{eqn:Z110}
have already been defined and used in Eqs. \eqref{eqn:Z48} and \eqref{eqn:Z54}.
\citeauthor{vonzeipel1910} assumes a condition on $\alpha'$
(or $a'$ in his notation) as
\begin{equation}
  \alpha' = 1 + s k' .
  \tag{Z107-\arabic{equation}}
  \stepcounter{equation}
  \label{eqn:Z107}
\end{equation}
with a new parameter $s$ in the range of
\begin{equation}
  -1 < s < +1 .
  \label{eqn:Z107-append-s}
\end{equation}
He also places a condition on eccentricity $e$ as
\begin{equation}
  e = \varepsilon k' ,
  \tag{Z108-\arabic{equation}}
  \stepcounter{equation}
  \label{eqn:Z108}
\end{equation}
with a new parameter $\varepsilon$ in the range of
\begin{equation}
  0 \leq \varepsilon \leq 1 .
  \label{eqn:Z108-append-epsilon}
\end{equation}

Using the foregoing,
\citeauthor{vonzeipel1910} shows that $R$ in Eq. \eqref{eqn:Z110} has the form
\begin{equation}
  R = R_1 \log {k'}^2 + R_2 + R_3 + R_4 + R_5 ,
  \label{eqn:vZ10-126-pre}
\end{equation}
or
\begin{equation}
  R = R_1 \log {k'}^2 + R' ,
  \tag{Z130-\arabic{equation}}
  \stepcounter{equation}
  \label{eqn:vZ10-130}
\end{equation}
where $R_2$, $R_3$, $R_4$, $R_5$ and $R'$ are holomorphic functions of
orbital elements that are expressed by complex integrals.
Their actual forms are shown as Eq. (Z126) on p. Z409
(which we do not reproduce here).
After going through complicated calculations for obtaining
actual but approximate function forms of $R_1$, $R_2$, $R_3$, $R_4$, $R_5$,
\citeauthor{vonzeipel1910} eventually reaches the conclusion that
$R$ in Eq. \eqref{eqn:vZ10-130} has a minimum at $g=0$ when $k'$ is small
(i.e. when $\alpha' \sim 1$ due to Eq. \eqref{eqn:Z107}).
We cite what he says about it:
\begin{quote}
``The research of this issue were aimed at finding all the solutions of equations
\begin{equation}
  \DP{R}{e} = \DP{R}{g} = 0
  \tag{Z132-\arabic{equation}}
  \stepcounter{equation}
\end{equation}
that exist in the domain $B$, assuming that $k'$ is small.
\par
\hspace*{1em}
[$\cdots$]
\par
\hspace*{1em}
Returning now to equations (Z132).
For small values of $k'$ they possess a single solution at the interior of domain $0 \leq e \leq k'$, $-\frac{\pi}{2} \leq g \leq +\frac{\pi}{2}$.
This solution, which can be written
\begin{equation}
  e = e_1 = k'\left( \frac{1}{\sqrt{2}} + {\cal B}\left(k',k'\log k'\right) \right);
  \quad
  g=0
  \tag{Z134-\arabic{equation}}
  \stepcounter{equation}
\end{equation}
where ${\cal B}(0,0)=0$, corresponds to a minimum value of the function $R$,
since $A(0)<0$, as we saw on page Z371.
\par
\hspace*{1em}
Consider the equation
\begin{equation}
  s = \frac{e_1}{k'}.
  \nonumber
\end{equation}

For small values of $k'$ it has a single root
\begin{equation}
  s = s_1 = \frac{1}{\sqrt{2}} + {\cal B}_1 \left(k', k'\log k'\right)
  \nonumber
\end{equation}
where ${\cal B}(0,0)=0$.'' (pp. Z411--Z412)
\end{quote}

Note that
\citeauthor{vonzeipel1910} did not give any definitions of the functions
${\cal B}$ and ${\cal B}_1$ in the above except their boundary conditions
(${\cal B} (0,0)=0$ and ${\cal B}_1 (0,0)=0$).
Note also that the quantity $A(0)$ in above once showed up in Section \ref{ssec:R-general}.
See Eqs. \eqref{eqn:Z54}, \eqref{eqn:Z55}, \eqref{eqn:Z56} in p. \pageref{pg:def-A0}.

\citeauthor{vonzeipel1910} continues and states a theorem as follows:
\begin{quote}
''We can now state the following theorem as shown:
\par
\hspace*{1em}
\textit{%
For sufficiently small values of $k'$, the function $R$ possesses neither maxima nor minimaxima in the domain $B$.
If $|\alpha'-1| > s_1 k'$, $R$ has no minima in this domain.
If instead $|\alpha'-1| < s_1 k'$, the function $R$ possesses in the domain $B$ only one minimum situated at the point (Z134).}'' (p. Z412)
\end{quote}

Right after stating the above theorem,
\citeauthor{vonzeipel1910} admits that he did not succeed in showing the
existence of these minima when $\alpha'-1$ or $k'$ gets large:
\begin{quote}
``If the parameters $\alpha'$ and $k'$ vary, the points
\begin{equation}
  e=e_1, \quad g=0 \:\: \mathrm{or} \:\: \pi,
  \tag{Z135-\arabic{equation}}
  \stepcounter{equation}
\end{equation}
where the function $R$ is minima in the domain $B$ or $B'$, move.
But it did not succeed in pursuing them analytically if $\alpha'-1$ and $k'$ cease to be small.
My goal was only to direct the attention on the existence of these minima [that are] quite remarkable.
Obviously it does not offer serious difficulty [in] finding their positions by calculating numerical values given for the parameters $\alpha'$ and $k'$.''
(p. Z413)
\end{quote}

As we have learned,
the existence of local minima in $R$ means the existence of the Lindstedt series in their vicinity.
In solar system dynamics,
the corresponding periodic orbits would belong to
a certain class of small bodies in a stable motion.
\citeauthor{vonzeipel1910} specifically depicts the characteristics of
the orbits in this state as follows:
\begin{quote}
``We can apply the result of this issue by calculating, according to no. Z7, Z8 and Z9, the series of Lindstedt, which exist when the elements are in the vicinity of the points (Z135).
The corresponding orbits belong to a certain class of comets in stable motion.
In these orbits, the semimajor axis, eccentricity and inclination are more or less almost constant;
the perihelion distance to the node is still close to 0 or $\pi$;
the line of nodes has a retrograde motion;
Finally, the orbits of the comet and of the disturbing planet are situated relative to the other like the rings of a chain.'' (p. Z413)
\end{quote}

\begin{figure*}[htbp]\centering
\ifepsfigure
 \includegraphics[width=\dualfigwidth\textwidth]{ocross_ex.eps} %fig24
\else
 \includegraphics[width=\dualfigwidth\textwidth]{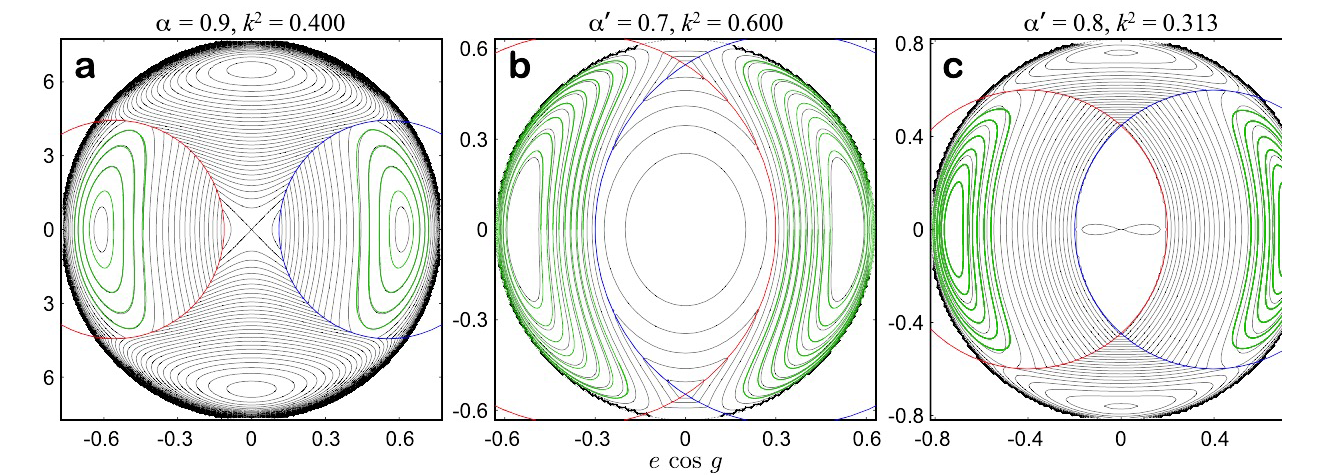} %fig24
\fi
  \caption{%
  Equi-potential contours and trajectories of fictitious perturbed bodies
  that realize the circumstance of ``rings of a chain.''
  \mtxtsf{a}: An inner case with $(\alpha,  k^2) = (0.9, 0.400)$.
  \mtxtsf{b}: An outer case with $(\alpha', k^2) = (0.7, 0.600)$.
  \mtxtsf{c}: An outer case with $(\alpha', k^2) = (0.8, 0.313)$.
  The solid black lines (equi-potential contours) are drawn
  based on the numerical quadrature defined as Eq. \eqref{eqn:K09}.
  The green dots are the actual trajectories of the perturbed bodies
  obtained through direct numerical integration of the equations of motion,
  starting from several initial locations beyond the lines of orbit intersection.
The conditions of the numerical integration of the equations of motion are as follows:
For the panels \mtxtsf{a} and \mtxtsf{b},
  the nominal stepsize is 2 days, and
  the total integration time is 10 million years with
  a data output interval of 500 years.
  The ratio of the perturber's mass and the central mass is
  $3.0404326 \times 10^{-6}$ which is close to
  the mass ratio of the Earth$+$Moon and the Sun.
For the panel \mtxtsf{c},
  the nominal stepsize is 4 days, and
  the total integration time is 30 million years with
  a data output interval of 500 years.
  The ratio of the perturber's mass and the central mass is
  $9.5479194 \times 10^{-6}$ which is close to
  $\frac{1}{100}$ of the mass ratio of Jupiter and the Sun.
  The meanings of the red and the blue partial circles (the lines of orbit intersection),
  as well as that of the black dashed circles at the outer boundary
  (although it is hard to see in the panels), are the same as in
  \mysymfigO \protect{\ref{fig:xy-inner}}.
  }
  \label{fig:ocross-example}
\end{figure*}

Let us present a numerical demonstration of some typical orbits that 
\citeauthor{vonzeipel1910} depicts as ``rings of a chain.''
Using the numerical quadrature defined by Eq. \eqref{eqn:K09},
we produced equi-$R$ contours of three CR3BP systems with
different parameter sets:
$(\alpha,  k^2) = (0.9, 0.400)$,
$(\alpha', k^2) = (0.7, 0.600)$, and
$(\alpha', k^2) = (0.8, 0.313)$.
We show the results in \mysymfigO \ref{fig:ocross-example}.
Here we intentionally chose large $\alpha$ or $\alpha'$
in order to realize the circumstance that \citeauthor{vonzeipel1910} considered.
In addition to the numerical quadrature,
we carried out direct numerical integration of
the equations of motion of typical objects that stay in the area
corresponding to the domain $B$ or $B'$ in \mysymfigO \ref{fig:vZ10-f1-5},
and plotted the result.
In all the three panels of \mysymfigO \ref{fig:ocross-example},
the existence of periodic orbits along the direction of $g=0$ and $g=\pi$ in the domains $B$ and $B'$ is obvious.
These orbits literally possess the characteristics that \citeauthor{vonzeipel1910} mentions on p. Z413:
The eccentricity $e$ and inclination $I$ are more or less constant, and
the argument of pericenter $g$ stays around 0 or $\pi$ in the domains $B$ and $B'$.
The longitude of ascending node $h$ has a retrograde motion
(although we do not show their time series here).
Most importantly, these orbits in $B$ or $B'$ realize
the geometric state that \citeauthor{vonzeipel1910} described as
``the orbits of the comet and of the disturbing planet are situated relative to the other like the rings of a chain,''
beyond the lines of orbit intersection (the red and blue partial circles drawn in \mysymfigO \ref{fig:ocross-example}).

Note that in the panels \mtxtsf{a} and \mtxtsf{c} in \mysymfigO \ref{fig:ocross-example},
we find trajectories around a pair of stationary points along the axis of $g = \pm \frac{\pi}{2}$.
However, these trajectories and the stationary points are located in the areas of $A$ or $A'$
in \mysymfigO \ref{fig:vZ10-f1-5}, not $B$ or $B'$.
There, the orbits of the perturbed and perturbing bodies do not compose ``rings of a chain.''
Therefore, they are out of \citeauthor{vonzeipel1910}'s consideration.

\begin{figure}[t!]\centering
\ifepsfigure
 \includegraphics[width=\singlefigwidth\textwidth]{Rmap_deVico.eps} %fig25
\else
 \includegraphics[width=\singlefigwidth\textwidth]{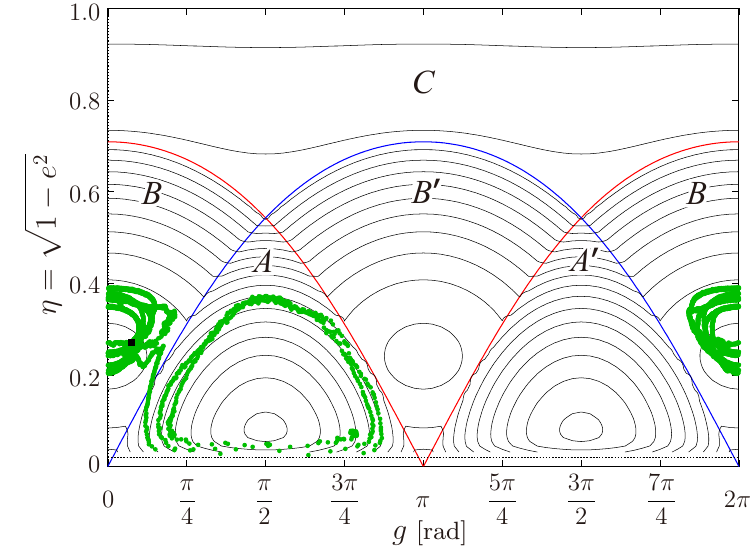} %fig25
\fi
  \caption{%
  The equi-potential contours and numerical trajectories of
  the comet 122P/de Vico under the CR3BP framework
  plotted on the $(g, \eta)$ plane.
  The comet has the parameters of $(\alpha', k^2) \sim (0.295, 0.000437)$.
  The green dots are the trajectories of this comet
  obtained through direct numerical integration of the equations of motion
  where we assume Jupiter on a circular orbit with the current
  semimajor axis and the current mass to be the perturbing body.
As for the numerical integration,
  the nominal stepsize is 4 days, and
  the total integration time is 5 million years with
  a data output interval of 500 years.
  The initial starting point of the numerical integration is shown as
  a filled black square at $(g, \eta) = (0.0746\pi, 0.271)$.
  The red and the blue curves are topologically equivalent to
  the red and the blue partial circles
  presented on the $(e \cos g, e \sin g)$ plane,
  such as in \mysymfigO \protect{\ref{fig:ocross-example}}.
  The black dashed line near the bottom of the plot implies
  the upper boundary of the perturbed body's eccentricity,
  equivalent to the black dashed circles
  presented on the $(e \cos g, e \sin g)$ plane,
  such as in \mysymfigO \protect{\ref{fig:ocross-example}}.
  The red and the blue curves divide the plane into five domains
  which correspond to the domains $A$, $A'$, $B$, $B'$, and $C$
  designated in \mysymfigO \protect{\ref{fig:vZ10-f1-5}}.
  }
  \label{fig:Rmap-deVico}
\end{figure}

In the current solar system
we find some objects whose orbits are approximately
located in the domain $B$ or $B'$ in \mysymfigO \ref{fig:vZ10-f1-5}.
One of the examples is the comet 122P/de Vico (1846 D1)
mentioned in \citet[][Table 2 on their p. 319]{bailey1992}.
This comet has $\alpha' \sim 0.295$, and 
its $k^2$ value is very small $\left(k^2 \sim 0.000437\right)$
due to its large eccentricity $(e \sim 0.963)$.
Since its inclination is also large $(i \sim 85^\circ.38)$, and
since its argument of perihelion is relatively close to zero
$(g \sim 0.0746 \pi)$,
this comet's orbit is expected to have a state of
``a ring of a chain'' with respect to its main perturber, Jupiter.
We once tried to draw its equi-$R$ contours
on the usual $(e \cos g, e \sin g)$ plane
in the framework of CR3BP having Jupiter as the perturber.
However, it turned out that the polar coordinates $(e \cos g, e \sin g)$ is not
quite suitable for drawing trajectories of this comet
because its eccentricity is too large.
The trajectory that this comet makes is located too close to the outer
boundary of the $(e \cos g, e \sin g)$ plane,
and the equi-potential contours become very busy.
Therefore, instead,
we used the rectangular coordinates $(g, \eta)$ where $\eta = \sqrt{1-e^2}$,
which is similar to \citeauthor{kozai1962b}'s $(2g, \eta^2)$ plane.
Using these coordinates,
in \mysymfigO \ref{fig:Rmap-deVico} we plotted equi-$R$ contours for this comet
obtained through the numerical quadrature defined in Eq. \eqref{eqn:K09}.
Here, we assume that Jupiter with the present mass on a circular orbit
with the present semimajor axis is the only perturber.

We also carried out direct numerical integration of the equations of motion of this comet in the framework of CR3BP,
and plotted its trajectory superposed on the equi-$R$ contours
in \mysymfigO \ref{fig:Rmap-deVico}.
This figure evidently tells us that,
for a while this comet stays around the stationary point along the $g = 0$ axis,
which corresponds to the domain $B$ in \mysymfigO \ref{fig:vZ10-f1-5}.
These are the very characteristics that \citeauthor{vonzeipel1910} described.
However as we see in \mysymfigO \ref{fig:Rmap-deVico},
the cometary trajectory soon (within about 2.3 million years) moves away toward
another stationary point centered at $g=\frac{\pi}{2}$,
which corresponds to the area $A$ in \mysymfigO \ref{fig:vZ10-f1-5}.
There, the cometary orbit no longer composes a chain ring with Jupiter's orbit.

Readers must be aware that the motion of the comet 122P/de Vico
described in \mysymfigO \ref{fig:Rmap-deVico} is a demonstration
within the framework of CR3BP.
The actual motion of the comet in the actual solar system with many perturbers is much more
complicated than what is shown in \mysymfigO \ref{fig:Rmap-deVico},
as \citet{bailey1992} showed by their numerical integration.

As far as we are aware, currently
the modern standard of the secular dynamical theory
that can deal with orbit intersection of this kind is the work of \citet{gronchi1999a,gronchi1999b}.
Some numerical demonstrations had existed before their work,
but just for several special cases \citep[e.g.][]{quinn1990b,bailey1992}.
\citet{gronchi1999a} showed that it is possible to define a generalized
averaged equation of motion even for a planet-crossing asteroid with singularities.
They actually obtained generalized solutions that are unique and piecewise smooth
with the help of an approximation proposed by \citet{wetherill1967} and
the method of Kantorovich of extraction of singularities \citep[e.g.][]{demidovich1966}.
Following this achievement,
\citet{gronchi1999b} proved that 
there is a stable region of orbits of the perturbed body
for any values of $a$ and $k^2$ of an asteroid (or comet) and
for any number of perturbing bodies with planar and circular orbits.
While \citeauthor{vonzeipel1910} tried to obtain an expression of
the disturbing function in each of the separated domains $A$, $B$, $C$, $\cdots$ in \mysymfigO \ref{fig:vZ10-f1-5},
\citeauthor{gronchi1999b}'s approach is to obtain
an expression of the disturbing function that is common to all the regions.
Although it is out of the scope of this monograph,
it would be interesting to quantitatively compare their theory with
\citeauthor{vonzeipel1910}'s
to examine their difference and similarity in detail.

\subsubsection{Motion of 1P/Halley\label{sssec:Halley}}
In the process of portraying their theoretical framework,
\citet[][their \mysymfigO 6 on p. 934]{gronchi1999b} took one of the most
famous comets---1P/Halley---as an example, and
discussed its motion under the perturbation from major planets.
According to one of their equi-potential diagrams
(the bottom right panel in their \mysymfigO 6),
the motion of a 1P/Halley-like object has a stable equilibrium point
around $g = \frac{\pi}{2}$ with a large eccentricity,
although its equi-potential surface is irregularly shaped
due to the lines of orbit intersection with major planets.
Here, we want readers to recall that \citeauthor{vonzeipel1910} also
carried out a quantitative demonstration on the secular motion of 1P/Halley
earlier in his Section Z21 (see p. \pageref{pg:Z21} of this monograph).
Let us cite \citeauthor{vonzeipel1910}'s words about it:
\begin{quote}
``It might be interesting to apply the results of the previous issue
  to a case in the nature,'' (p. Z395).
\end{quote}
Note that ``the results of the previous issue'' indicates
his theoretical development on the doubly averaged inner CR3BP
(where $\alpha < 1$, as we have seen in his Sections Z16--Z20).

\citeauthor{vonzeipel1910} begins his Section Z21 with a calculation of
1P/Halley's eccentricity at one of its stationary points, $e_{0.2}$,
using its orbital elements known at that time.
He used the CR3BP framework including Jupiter as the perturbing planet
on a circular orbit.
He estimates 1P/Halley's semimajor axis as
$a = 17.9676$ au according to \citet{galle1894}, saying:
\begin{quote}
``The semi-major axis of the orbit of this comet is
\begin{center}
  17.9676.
\end{center}
This is the average distance [obtained from] all the apparitions of the comet from 1378 until 1835.
(See J. G. Galle, List of elements of the previously calculated comets.)
Choosing the semimajor axis of the orbit of Jupiter as unit length, we thus obtain
$$
  \alpha = 17.9676:5.202800 = 3.45345 . % \mbox{''}
$$
'' (p. Z395)
\end{quote}

For calculating $k$ and $k'$ of 1P/Halley,
\citeauthor{vonzeipel1910} used its osculating orbital elements
known at that time published in
\citet[][pp. 50--51]{galle1894} and \citet{rosenberger1835a,rosenberger1835b}.
They are tabulated on an unnumbered table on p. Z395 as
\begin{equation}
\begin{aligned}
  g & = 110^\circ 37'59'' , \\
  h & =  55^\circ  9'47'' , \\
  I & = 162^\circ 14'43'' , \\
  e & = 0.96738879 ,
\end{aligned}
  \label{eqn:Halley-gWIe}
\end{equation}
with a notice of
``equinox through 1835 November 15.94542 t. m. of Paris.''
\citeauthor{vonzeipel1910} converts these orbital elements into
those on the solar system invariable plane, saying:
\begin{quote}
``By comparing the elements at the invariable plane of the solar system and by taking the value of the acute angle for the inclination, we obtained
$$
  I =  18^\circ 47' 44'', \quad
  g = 114^\circ 27' 26'' .
$$

We see that $g$ is not very distant from $90^\circ$.

[From] the values indicated for $e$ and $I$, we give, for $k$ and $k'$, the numbers as follows
$$
\begin{aligned}
  k  = \sqrt{1-e^2} \cos I         &= 0.239789, \\
  k' = \sqrt{1-k^2} \;\;\;\;\;\;\; &= 0.970825,
\end{aligned}
$$
'' (p. Z396)
\end{quote}

\citeauthor{vonzeipel1910} then moves on to the calculation of
1P/Halley's $e_{0.2}$.
For this purpose
he did not use his own approximation formula presented as Eq. \eqref{eqn:Z94}.
Instead, he resorted to a numerical quadrature of $R$
along the axis of $g = \frac{\pi}{2}$.
We literally cite \citeauthor{vonzeipel1910}'s words:
\begin{quote}
``Since $k$ is not very small, we have not employed the development (Z94) to calculate $e_{0.2}$.
We have preferred to determine this quantity by calculating, by means of mechanical quadrature, a certain number of values of the function $R$ on the axis, where $g=\frac{\pi}{2}$.
The value of $e$, which renders $R=$ minimum, is the sought value of $e_{0.2}$.
\par
\hspace*{1em}
In these calculations, we used the formulas (Z48), (Z51) and (Z53).
The field of integration was divided by $10^\circ$ to $10^\circ$.
To calculate the complete elliptical of the first kind, the tables of Legendre were employed.'' (p. Z396)
\end{quote}

The resulting values of $e$ and $R$ are tabulated on an unnumbered table
on p. Z396. From the tabulated values,
\citeauthor{vonzeipel1910} draws the following conclusion:
\begin{quote}
``These numbers are found by means of interpolation calculations that
$$
  e_{0.2} = 0.9537 .
$$
\par
\hspace*{1em}
It is therefore a value very close to the actual eccentricity of the orbit of Halley's comet.'' (p. Z396)
\end{quote}

\begin{figure}[t]\centering
\ifepsfigure
 \includegraphics[width=\singlefigwidth\textwidth]{Halley_eR.eps} %fig26
\else
 \includegraphics[width=\singlefigwidth\textwidth]{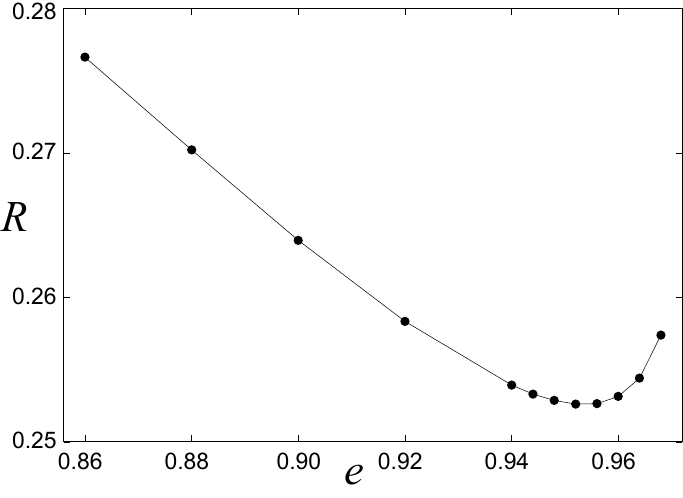} %fig26
\fi
  \caption{%
  A visualization of \citeauthor{vonzeipel1910}'s unnumbered table on p. Z396
  that shows the dependence of comet 1P/Halley's $R$ on its eccentricity $e$
  along the axis of $g = \frac{\pi}{2}$.
  We simply connected the filled black circles
  (\citeauthor{vonzeipel1910}'s tabulated values) by straight line segments.
  }
  \label{fig:Halley-eR}
\end{figure}

\begin{figure}[t]\centering
\ifepsfigure
 \includegraphics[width=\singlefigwidth\textwidth]{Rmap_halley.eps}%fig27
\else
 \includegraphics[width=\singlefigwidth\textwidth]{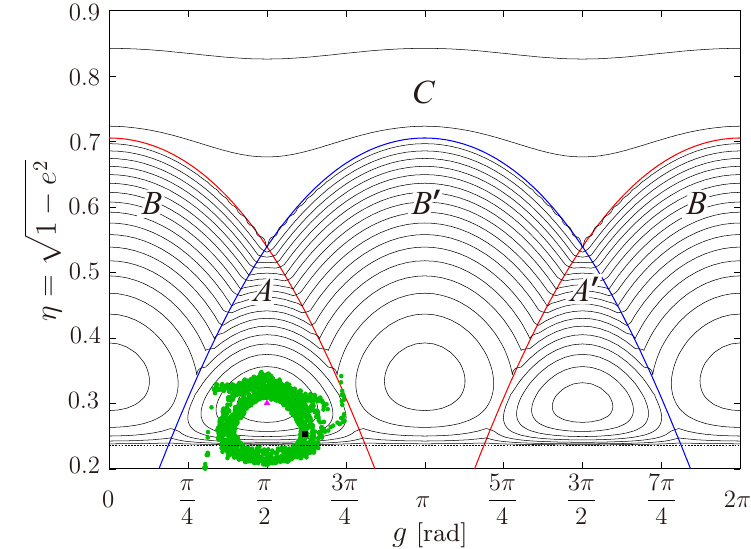}%fig27
\fi
  \caption{%
  The equi-potential contours and numerical trajectories of
  the comet 1P/Halley under the CR3BP framework
  plotted on the $(g, \eta)$ plane.
  The comet has the parameters of $(\alpha', k^2) \sim (0.292, 0.0558)$.
  The green dots are the trajectories of this comet
  obtained through direct numerical integration of the equations of motion
  where we assume Jupiter on a circular orbit with the current
  semimajor axis and the current mass to be the perturbing body.
As for the numerical integration,
  the nominal stepsize is 1 day, and
  the total integration time is 1 million years with
  a data output interval of 500 years.
  The initial starting point of the numerical integration is shown as
  a filled black square at $(g, \eta) = (0.621\pi, 0.254)$.
  We also plotted 1P/Halley's $e_{0.2}$ value that
  \citeauthor{vonzeipel1910} calculated as the magenta filled triangle at
  $(g, \eta) = (\frac{\pi}{2}, 0.301)$ based on his estimate,
  $e_{0.2} = 0.9537$.
  The meanings of the red and the blue curves,
  that of the black dashed line near the bottom, and
  the designation of the domains $A$, $A'$, $B$, $B'$, and $C$,
  are all common to those of \mysymfigO \protect\ref{fig:Rmap-deVico}.
  Note that the range of the vertical axis $(\eta)$ is narrower here
  than in \mysymfigO \protect\ref{fig:Rmap-deVico}.
  This is because 1P/Halley's eccentricity is smaller than de 122P/Vico.
  }
  \label{fig:Rmap-halley}
\end{figure}

Instead of transcribing \citeauthor{vonzeipel1910}'s table,
we have made a visual plot of the numbers tabulated in this table
on an $(e,R)$ diagram (\mysymfigO \ref{fig:Halley-eR}).
We see a local minimum of $R$ in this figure, and
the eccentricity value at that place seems to match
what \citeauthor{vonzeipel1910} estimated as $e_{0.2}$ above (0.9537).
Although he did not explicitly state this,
we guess that \citeauthor{vonzeipel1910} wanted to claim that 1P/Halley's argument of perihelion is
currently located in the libration area centered at
$(e,g) = (e_{0.2}, \frac{\pi}{2})$.
As a more straightforward demonstration,
we carried out a set of numerical quadrature
and direct numerical integration of the equations of motion of this comet,
and drew the result on the $(g, \eta)$ plane.
We placed Jupiter on a circular orbit with the current semimajor axis
and the current mass as the perturbing body.
The result is shown in \mysymfigO \ref{fig:Rmap-halley}.
Under the CR3BP framework, we see that 1P/Halley's $g$ librates at
$g = \frac{\pi}{2}$ as \citeauthor{vonzeipel1910} predicted.
In addition, the eccentricity value at the libration center that
our calculation indicates
(the small magenta triangle in \mysymfigO \ref{fig:Rmap-halley})
seems almost the same as what he calculated
($e = 0.9537$, or $\eta \sim 0.301$). Here again,
we believe that the accuracy of \citeauthor{vonzeipel1910}'s theory is confirmed.

Readers should be aware that, unlike 122P/de Vico,
1P/Halley is not in the state of ``rings of a chain'' with the perturber
(Jupiter) in this dynamical setting.
As the pericenter of this comet is inside  the perturber's orbit, and
as the apocenter  of this comet is outside the perturber's orbit,
it apparently seems that the orbit of this comet is located
across the line of orbit intersection.
However as seen in \mysymfigO \ref{fig:Rmap-halley},
the motion of this comet is confined in the domain $A$ defined
in \mysymfigO \ref{fig:vZ10-f1-5}
with its argument of perihelion $g$ librating around $\frac{\pi}{2}$.
Therefore its orbit does not make a ``chain'' with the perturber's orbit.
For a perturbed body's orbit to compose a ring of a chain with perturber's orbit,
the perturbed body's motion must be confined in the domain $B$ or $B'$
with its argument of perihelion $g$ librating around $0$ or $\pi$.

Note also that our numerical result shown
in \mysymfigO \ref{fig:Rmap-halley} is just a numerical demonstration
for confirming the validity and accuracy of
\citeauthor{vonzeipel1910}'s theoretical estimate on 1P/Halley's motion.
In the actual solar system,
close encounters with major planets frequently change this comet's orbital elements,
and its trajectory cannot remain regular.
Even in the CR3BP approximation that we used to draw \mysymfigO \ref{fig:Rmap-halley}
we see an irregular oscillation in its trajectory, and
this comet could not stay in the libration area
centered at $g = \frac{\pi}{2}$ for more than one million years.
Hence, it is easy to imagine that the actual motion of 1P/Halley
in the actual solar system is much more complicated and irregular.
This has been demonstrated by many previous studies
\citep[e.g.][]{chirikov1989,lohinger1995,bailey1996,munoz-gutierrez2015,boekholt2016}.
It is worth noting that \citet[][his \mysymfigO 4 in p. 234]{kozai1979}
analyzed 1P/Halley's orbital motion through an equi-potential diagram,
assuming four giant planets (Jupiter, Saturn, Neptune, and Uranus)
as perturbers on circular orbits.

\subsection{Some other extensions\label{ssec:Z10-extention}}
The last section in \citeauthor{vonzeipel1910}'s Chapter IV (Section Z25),
as well as his last chapter (Chapter V, Sections Z26--Z28), are both short.
They are devoted to describing some other extensions of his theory.

 \subsubsection{Secular motion near the $e = k'$ circle\label{sssec:motionneare=kd}}
Section Z25 is about a treatment of the
doubly averaged motion of a perturbed body near the outer boundary of the
$(e \cos g, e \sin g)$ plane; in other words, near the circle $e = k'$.
\citeauthor{vonzeipel1910} had mentioned this issue once in Section Z15 in his Chapter III in a general form
(see p. \pageref{pg:motionneare=kd-general} of this monograph).
His discussion in Section Z25 eventually goes toward a statement at the end of this section 
(pp. Z414--415. We cite it later on p. \pageref{pg:zeipelZ25conclusion}) that,
there exists a stable motion of perturbed body with $e \sim k'$
on nearly a planar orbit.
Below, let us quickly follow \citeauthor{vonzeipel1910}'s argument.

As we have seen, 
perturbed body's inclination $I$ becomes small
when its eccentricity $e$ approaches its largest value $(k')$
in the doubly averaged CR3BP.
This is due to the conservation of the quantity
$k^2 = \left( 1-e^2 \right) \cos^2 I$.
In order to deal with the motion of perturbed bodies with large $e$ and small $I$,
\citeauthor{vonzeipel1910} brings up a different set of variables, $(\xi', \eta')$.
Actually,
he had already defined them in Section Z3 (his Chapter I, pp. Z349--Z350. See also p. \pageref{pg:def-xidashetadash} of this monograph).
See Eqs. \eqref{eqn:vZ10-xyzvars-dash} and \eqref{eqn:vZ10-xi+eta-dash}
in this monograph for the definitions of $\xi', \eta'$
as well as those of $x'_1, x'_2, x'_3, y'_1, y'_2, y'_3$.
He first mentions the fact that
the motion of the perturbed body near the outer boundary $e=k'$
on the $(e \cos g, e \sin g)$ plane is translated into the motion
around the origin $(0,0)$ on the $(\xi', \eta')$ plane.
At the same time, \citeauthor{vonzeipel1910}
reminds readers of the fact that
the outer boundary $e=k'$ on the $(e \cos g, e \sin g)$ plane
behaves like a local maximum of the doubly averaged disturbing function $R$,
particularly
when $\alpha$ is small
(see p. \pageref{pg:R_outerboundary-i} of this monograph) or
when $\alpha'$ is small
(see p. \pageref{pg:R_outerboundary-o} of this monograph).
He came to the conclusion that on the $(\xi', \eta')$ plane,
the origin $(0,0)$ becomes a local maximum.
Let us cite his original description about this:
\begin{quote}
``By studying the function $R$ for small values of $\alpha$ in no. Z16 and
  for small values of $\alpha'$ at no. Z22,
  we have already remarked that, in both cases, the function $R$ increases if,
  $g$ remaining constant, the eccentricity $e$ increase toward the value $k'$.
  The formula (Z42) shows that the inclination $I$ decreases toward zero
  when the eccentricity $e$ increases toward $k'$.
  For values of $e$ close to $k'$,
  it is therefore advantageous to introduce the variables of no. Z3.
  Among the variables $\xi'$, $\eta'$ of this no. Z3 and Keplerian variables,
  we obviously have the following relations
$$
\begin{aligned}
  {\xi '}^2 + {\eta '}^2 & = 2\sqrt{a}\sqrt{1-e^2} - 2\sqrt{a}\sqrt{1-e^2}\cos I \\
                   & = 2\sqrt{a}\left( \sqrt{1-e^2} - k \right)
\end{aligned}
$$
$$
  \eta' : \xi' = \tan g .
$$
\par
\hspace*{1em}
Therefore if, $g$ remaining constant, the eccentricity $e$ increases towards $k'$, the corresponding point $\xi'$, $\eta'$ approaches the origin on a straight line.
\par
\hspace*{1em}
It is therefore necessary to conclude that the function $R$, regarded as a function of variables $\xi'$ and $\eta'$ of no. Z3, has a maximum value at point $\xi'=\eta'=0$, if $\alpha$ is small enough or if $\alpha'$ is small enough.''
(p. Z413)
\end{quote}
Note that $\eta' : \xi'$ means $\frac{\eta'}{\xi'}$ in the above.

\citeauthor{vonzeipel1910} then reminds readers of the condition
for $R$ to be holomorphic (regular) over the entire region of $e \leq k'$.
This condition was already presented around \mysymfigS Z1--Z5 on his p. Z369
(see \mysymfigO \ref{fig:vZ10-f1-5} of this monograph,
  Eq. \eqref{eqn:vZ10-a-crossing}, and the discussions presented
  on pp. \pageref{pg:entirelyholomorphic1}--\pageref{pg:entirelyholomorphic2}),
but \citeauthor{vonzeipel1910} repeats it again here.
Let us cite his words:
\begin{quote}
``We already know that $R$ is holomorphic in the neighborhood of $\xi'=\eta'=0$ as soon as
\begin{equation}
\begin{aligned}
{}  & 0 < \alpha  < \frac{1}{1+k'} \\
\mbox{ or else } & \\
{}  & 0 < \alpha' < 1-k' ,
\end{aligned}
  \tag{Z136-\arabic{equation}}
  \stepcounter{equation}
  \label{eqn:Z136}
\end{equation}
(see Chapter III).'' (p. Z413)
\end{quote}
The first  inequality of Eq. \eqref{eqn:Z136} corresponds to the circumstance
that is depicted in \mysymfigO \ref{fig:vZ10-f1-5}\mtxtsf{a} for the inner problem.
The second inequality corresponds to  the circumstance depicted in
\mysymfigO \ref{fig:vZ10-f1-5}\mtxtsf{e} for the outer problem.

Now, \citeauthor{vonzeipel1910} assumes that $R$ can be Taylor-expanded
using ${\xi '}^2$ and ${\eta '}^2$ around the origin $(\xi',\eta') = (0,0)$,
similar to the case when he expanded $R$
using $x^2 = e^2 \cos^2 g$ and $y^2 =e^2 \sin^2 g$
(see Eq. \eqref{eqn:Z77} or Eq. \eqref{eqn:Z99}).
Let us cite his words again:
\begin{quote}
``Then, given the properties of symmetry of the disturbing function exhibited at no. Z4, it is necessary that $R$ is of the form
$$
  R = R''_{0.0} + R''_{2.0} {\xi '}^2 + R''_{0.2} {\eta '}^2 + \cdots .
$$

From what has been said earlier, the inequalities
\begin{equation}
  R''_{2.0} < 0, \quad
  R''_{0.2} < 0
  \tag{Z137-\arabic{equation}}
  \stepcounter{equation}
  \label{eqn:Z137}
\end{equation}
exist as soon as, $k'$ having been arbitrarily fixed but $<1$, we choose $\alpha$ or $\alpha'$ small enough.''
(p. Z413)
\end{quote}

So far, \citeauthor{vonzeipel1910}'s statement has been about the case when
$\alpha$ or $\alpha'$ ($a$ or $a'$ in his notation) is small.
Then, would the two inequalities in Eq. \eqref{eqn:Z137} both hold true
even when $\alpha$ (or $\alpha'$) is not small?
He questions this issue right after the previously quoted part:
\begin{quote}
``One may wonder if these inequalities [in Eq. \eqref{eqn:Z137}] always take place at the same time as one or other of the inequalities [in Eq.] \eqref{eqn:Z136}.'' (p. Z414)
\end{quote}

To answer to this question, \citeauthor{vonzeipel1910} first considers
the second inequality of Eq. \eqref{eqn:Z137}, $R''_{0.2} < 0$.
Although we do not reproduce his specific calculations,
in the end he concludes that $R''_{0.2} < 0$ is always satisfied
if either of the inequalities in Eq. \eqref{eqn:Z136} is satisfied,
i.e. in the cases of
\mysymfigO \ref{fig:vZ10-f1-5}\mtxtsf{a} or
\mysymfigO \ref{fig:vZ10-f1-5}\mtxtsf{e}.
Note that \citeauthor{vonzeipel1910} develops this consideration (on p. Z414)
without showing the actual function form of $R ({\xi '}^2, {\eta '}^2)$.

As for the first inequality of Eq. \eqref{eqn:Z137}, $R''_{2.0} < 0$,
\citeauthor{vonzeipel1910} admits the difficulty of its proof
when $\alpha$ or $\alpha'$ is not small.
Instead, he considers this inequality from its dynamical aspect.
Here is how he describes his thought:
\begin{quote}
``It seems to be more difficult to rigorously discuss the sign of the coefficient $R''_{2.0}$ when $\alpha$ or $\alpha'$ is no longer a relatively small quantity.
But, see what would happen if the coefficient $R''_{2.0}$ could be positive, given one or other of the inequalities (Z136).
Then the function $R$ has a minimaxima value at the point $\xi' = \eta' = 0$.
As a result, an orbit infinitely less inclined to the orbit of the disturbing planet and [yet] can never meet the orbit of this planet, could, by the effect of secular perturbations, obtain a finite
inclination.
It seems very unlikely.
I therefore thought that it would probably be useless to examine the sign of the coefficient $R''_{2.0}$ by means of numerical calculations [that would be] inevitably quite long.'' (p. Z414)
\end{quote}

From the above considerations, at the end of Section Z25
\citeauthor{vonzeipel1910} makes a prediction on the existence of
a class of cometary orbits. His words are as follows:
\begin{quote}
``It must be concluded from these studies and those of no. Z12 that there exist orbits of comet[s] in stable motion, whose inclinations are small, and for which the distance from the perihelion to the node possesses a positive mean motion (of the value $+\sigma$ from no. Z7).
  In these orbits the eccentricity is approximately constant $(=k')$ as well as the semi-major axis.
  Finally, the nodes rotate in the opposite direction to the motion of the comet.
  Among these orbits, two classes are distinguished.
  In one [of them], the two nodes are located at the inside of the orbit of the disturbing planet, [and] in the other, the nodes are located, on the contrary, outside the orbit [of the disturbing planet].'' (pp. Z414--Z415)
\end{quote}
\label{pg:zeipelZ25conclusion}

\begin{figure*}[htbp]\centering
\ifepsfigure
 \includegraphics[width=\dualmedfigwidth\textwidth]{emax_ex.eps}%fig28
\else
 \includegraphics[width=\dualmedfigwidth\textwidth]{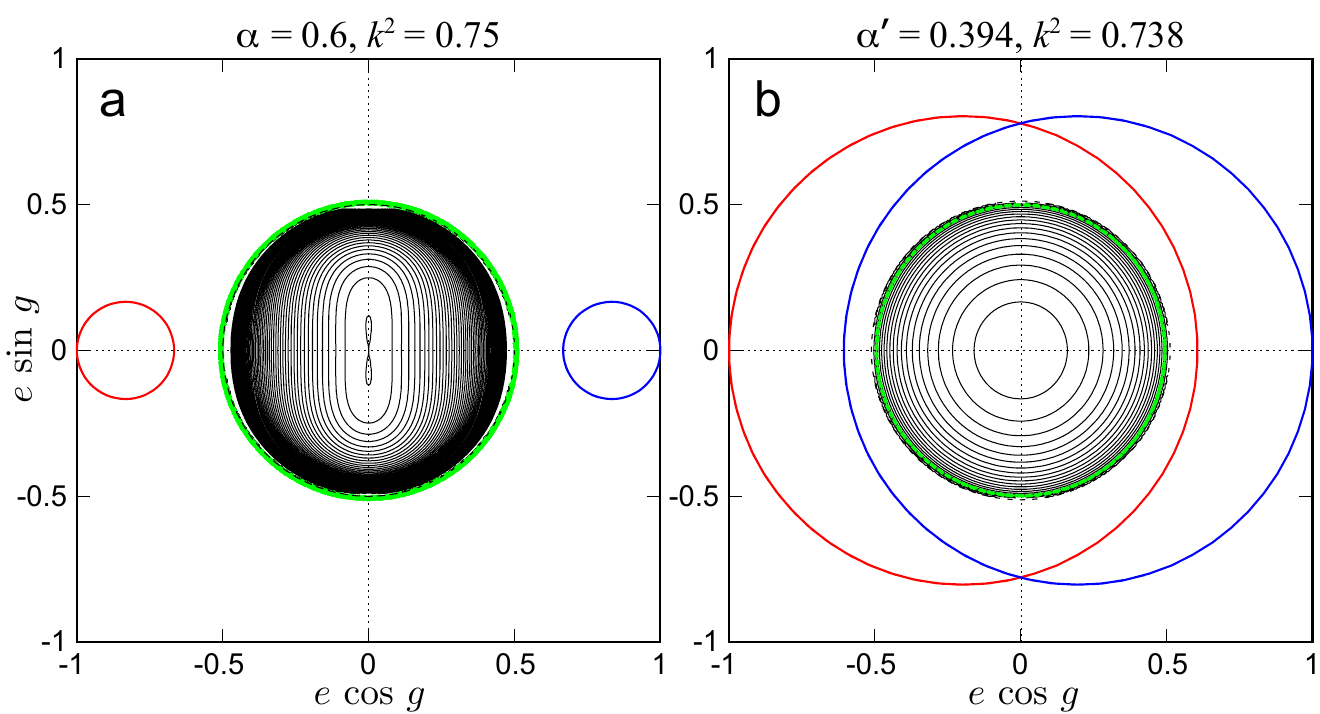}%fig28
\fi
  \caption{%
The equi-potential contours and numerical trajectories of perturbed bodies
in two CR3BP systems whose eccentricities are close to their maximum, $k'$.
\mtxtsf{a}: A fictitious body in an inner CR3BP system
whose $\alpha  = 0.6$   and $k^2 = 0.75$.
This means $k' = \sqrt{1-k^2} = 0.5$.
The ratio of the perturber's mass and the central mass is
$9.5479194 \times 10^{-5}$ which is close to
$\frac{1}{10}$ of the mass ratio of Jupiter and the Sun.
This system models a main belt asteroid orbiting inside Jupiter's orbit.
In this system, the perturbed body's initial inclination is zero
because $e = k'$ is satisfied at the beginning.
Thus its inclination remains zero (i.e. the planar orbit)
throughout the calculation period because both the perturbed body's orbit
and the perturbing body's orbit stay in the same plane.
\mtxtsf{b}: A fictitious body in an outer CR3BP system
whose $\alpha' = 0.394$ and $k^2 = 0.733$.
This means $k' = \sqrt{1-k^2} \sim 0.516$.
The ratio of the perturber's mass and the central mass is the same as
in the panel \mtxtsf{a}.
This system models a TNO orbiting outside Neptune's orbit.
In fact, the perturbed body in the system \mtxtsf{b} is a proxy of
an actual TNO, 2000 PH${}_{30}$
whose orbital elements are
$a = 77.07718637$ au,
$e = 0.50203455$, and
$I = 8^{\circ}.048740$ 
according to
\mtxtsf{astorb.dat}
 provided by Lowell Observatory as of September 28, 2018
(see \supinfo{3} for its specific website).
The value $\alpha' = 0.394$ adopted here is close to the semimajor axis ratio between Neptune and this object.
In both the panels,
the green dots near the outer boundary $(e = k')$ indicate
the trajectories of perturbed bodies obtained through
direct numerical integration of the equations of motion.
Parameters for the numerical integration are common to
both \mtxtsf{a} and \mtxtsf{b}:
  the nominal stepsize is 4 days,
  the total integration time is 8 million years with
  a data output interval of 500 years.
The meanings of the red and the blue circles,
as well as the black dashed circles near the outer boundaries,
are common to those in \mysymfigO \protect{\ref{fig:Rmap-deVico}} and
in \mysymfigO \protect{\ref{fig:Rmap-halley}}.
As usual, equi-potential contours of the perturbed body
obtained through the numerical quadrature defined by Eq. \eqref{eqn:K09}
are drawn in the black solid lines in both panels.
}
  \label{fig:emax-example}
\end{figure*}

For confirming \citeauthor{vonzeipel1910}'s statement and prediction
described above,
particularly when $\alpha$ or $\alpha'$ is not small,
we carried out direct numerical integration of the equations of motion
of two hypothetical perturbed bodies that are in the above category.
One of them composes an inner CR3BP $(\alpha  = 0.6)$, and
the other   composes an outer CR3BP $(\alpha' = 0.394)$.
The numerical results are presented in \mysymfigO \ref{fig:emax-example}
together with the equi-potential contours for these objects obtained
through the numerical quadrature defined by Eq. \eqref{eqn:K09}.
Note that in \mysymfigO \ref{fig:emax-example},
we chose the conventional coordinates $(x,y)=(e \cos g, e \sin g)$ in this monograph, not $(\xi', \eta')$.
This is for emphasizing how close the actual trajectories of the perturbed bodies
(colored in green) are from their maximum eccentricity boundaries, $e=k'$.
If we use the coordinates $(\xi', \eta')$, all the numerical trajectories
would concentrate on the vicinity of the origin $(0,0)$ of the plane,
and would be very hard to distinguish.

By looking at \mysymfigO \ref{fig:emax-example},
we can say that our numerical calculation reproduces well the orbital
characteristics of the perturbed bodies that \citeauthor{vonzeipel1910} described.
In both the panels \mtxtsf{a} (the inner case) and \mtxtsf{b} (the outer case),
the perturbed body's motion is stable for a long time with
a large, nearly constant eccentricity $(e \sim k')$ and
a small, nearly constant inclination.
The argument of pericenter $g$ circulates.
Therefore we can conclude that
\citeauthor{vonzeipel1910}'s prediction above
(the possible existence of a stable motion of a perturbed body
 with a large, nearly constant eccentricity
 on nearly a planar orbit with the circulating argument of pericenter)
is correct, even when $\alpha$ or $\alpha'$ is as large as
that shown in \mysymfigO \ref{fig:emax-example}.

In the actual solar system, there is a category of small objects that
possesses the characteristics that \citeauthor{vonzeipel1910} mentioned above.
For example, as pointed out in the caption of \mysymfigO \ref{fig:emax-example},
the perturbed body used in the system \mtxtsf{b} is
a proxy of an actual TNO (transneptunian object), 2000 PH${}_{30}$.
In the system \mtxtsf{b},
we placed a Neptune-like planet on a circular orbit as the perturbing body
so that they make an outer CR3BP.
Therefore the value $\alpha' = 0.394$ adopted here is close to the semimajor axis ratio between Neptune and 2000 PH${}_{30}$.
This object has the parameter
$k^2                \sim 0.733$, therefore
$k' = \sqrt{1- k^2} \sim 0.516$.
Hence the ratio between $e$ and $k'$ of the perturbed body is quite large:
$\frac{e}{k'} > 0.972$. This indicates that this object's eccentricity is very close
to the theoretical maximum in the CR3BP framework.

TNOs that possess both a large semimajor axis and
a large eccentricity are collectively called the scattered TNOs
\citep[e.g.][]{gladman2008,gomes2008,lykawka2007a,lykawka2007b,lykawka2008,lan2019}.
Since many of the scattered TNOs have a much larger inclination than 2000 PH${}_{30}$, 
this object may perhaps be an exception among this group.
However, it is still interesting to know that
there is at least one transneptunian object that fulfills \citeauthor{vonzeipel1910}'s prediction, and possibly many more.
If we dare define this as 
\citeauthor{vonzeipel1910}'s ``unintentional prediction'' of a class of TNOs,
we could say that it was achieved
more than 30 years before 
Kenneth Edgeworth's prediction of a small body population beyond Neptune's orbit \citep{edgeworth1943,edgeworth1949}.
The number of objects in this category will surely increase
along with the progress of ongoing and future observational surveys
\citep[e.g.][]{lsst2009,yoshida2011a,bannister2016},
which will further illuminate the foresight of \citeauthor{vonzeipel1910} on this subject.
See p. \pageref{pg:zeipelandtnos} for a further discussion on this subject.

\subsubsection{Two or more perturbers\label{sssec:pluralperturbers}}
\citeauthor{vonzeipel1910}'s final chapter is very short, entitled
``\textit{Chapitre V. G\'en\'eralisations. Applications aux orbites instables,\/}'' and includes four sections (Z26--Z29) over just four pages.
This chapter is about a generalization of 
\citeauthor{vonzeipel1910}'s theory in order to incorporate $N$
perturbing planets $(N \geq 2)$, assuming that their eccentricities and
inclinations are so small that we can ignore them.
Although it may be practically a useful extension for actual problems
in solar system dynamics, we think \citeauthor{vonzeipel1910}'s
descriptions in this chapter are too short to accomplish
his original intention. It is also hard for us to confirm the validity or
accuracy of his discussion just from the description he put in this chapter.
Therefore,
what follows is our own interpretation of what he intended to claim.

Let us begin by telling readers an incidental but interesting fact.
According to the ADS records as of September 15, 2019,
the only citation of \citet{vonzeipel1910} in the modern literature was made by \citet{bailey1996}.
And, the reason why they cited \citeauthor{vonzeipel1910}'s work is due to the subject in his Chapter V.
The description in \citet{bailey1996} reads:
\begin{quote}
``It is well known that if $N_P$ planets are assumed to be on coplanar
  circular orbits, the problem of the motion of a small body under the
  influence of the secular part of the disturbing function is integrable
  (Zeipel 1910; Moiseev 1945; Kozai 1962; Vashkov'yak 1981a,b), and may be
  reduced to a description of the integrable curve in the $(e,\omega)$ plane.''
  \citep[][the left column on p. 1095, the beginning of Section 5]{bailey1996}
\end{quote}

From what we have seen, we learned that
the addition to CR3BP of a perturbing planet orbiting on a planar and circular orbit $(e'=I'= 0)$ would not increase the degrees of freedom
after averaging, and the system would remain integrable.
This principle would not be different even when $N$ perturbing planets
are included. The only issue would be how to specifically calculate
the secular perturbation from the multiple perturbers.
At the beginning of his Section Z26, 
\citeauthor{vonzeipel1910} states the issue as follows:
\begin{quote}
``In what preceded, we have assumed that there was one disturbing planet, and that the eccentricity of its orbit was zero.
But it is possible to generalize [this study] by assuming that the number of disturbing planets is any $N$ and that their masses as well as the eccentricities and inclinations of their orbits are small.'' (p. Z415)
\end{quote}

After some preparation such as extending the variables and secular
equations of motion for the $N$-perturber system,
\citeauthor{vonzeipel1910} claims that the doubly averaged disturbing function $R$ that contains perturbations
from multiple perturbing planets is composed by
a sum of multiple disturbing functions.
Let us cite his original description:
\begin{quote}
``$\ldots$
By neglecting these first modules, which is equivalent to neglecting the eccentricities and the inclinations, the function $R$ has the form
\begin{equation}
  R = R^{(0)} = \sum_{i=1}^N \frac{\beta_i}{a_i} R_i .
  \tag{Z140-\arabic{equation}}
  \stepcounter{equation}
  \label{eqn:Z140}
\end{equation}

The $\beta_i$ are the masses of the planets, the mass of Jupiter being chosen as the unit;
the $a_i$ are the zero-order secular parts of the semimajor axes of the orbits of the planets;
finally $R_i$ are obtained by the formula (Z44) by putting only
$$
  \frac{a}{a_i} \mbox{ instead of } a .
$$
'' (p. Z416)
\end{quote}

As a result of his consideration and approximation,
it turns out that we can again use Eq. \eqref{eqn:Z12}
as the equations of motion of the multiple perturber system.
The only difference from the single perturber system is that we would have to
replace $R$ in Eq. \eqref{eqn:Z12} for $R^{(0)}$ in Eq. \eqref{eqn:Z140}.
Consequently, it would be possible to
construct the Lindstedt series if $R^{(0)}$ possesses the local extremum.

In Section Z27
\citeauthor{vonzeipel1910} studies the characteristics of $R^{(0)}$
in Eq. \eqref{eqn:Z140},
and tries to locate its local extremums if any.
He first mentions the fact that, when there are $N$ perturbers,
the phase space $(x, y) = (e \cos g, e \sin g)$ is divided into pieces by $N$ pairs of circles.
The $N$ pairs of circles are expressed by the following equations:
\begin{equation}
    \left( x \pm \frac{a_i}{2a}\right)^2 + y^2
  = \left( 1 -   \frac{a_i}{2a}\right)^2
  \quad
   i = 1, 2, \cdots N .
  \tag{Z141-\arabic{equation}}
  \stepcounter{equation}
  \label{eqn:Z141}
\end{equation}
This is topologically different from the single perturber system
where the phase space is divided just by a pair of circles expressed as
Eq. \eqref{eqn:Z45}.

Although \citeauthor{vonzeipel1910} did not show any figures
that demonstrate the phase space division by $N$ perturbers represented by Eq. \eqref{eqn:Z141},
\citeauthor{gronchi1999b}'s \citeyearpar{gronchi1999b} \mysymfigO 1 on their p. 929 is
an exact example of how the multiple circles \eqref{eqn:Z141} divide the phase space.
\citeauthor{vonzeipel1910} insists that each of the divided domains has
a corresponding disturbing function.
Literary citing his words:
\begin{quote}
``By means of these circles, the domain (Z46) will be divided into a certain number of parts $(D_j)$.
In any of these domains $(D_j)$ [there] corresponds to a function $R_j$, holomorphic in this domain and on its boundaries (except the circle $e=k'$), and giving the value of $R^{(0)}$ in this domain.''
(p. Z416)
\end{quote}

\citeauthor{vonzeipel1910} continues his discussion
in the subsequent section (Z28).
He particularly focuses on the search for, and evaluation of,
local extremums of $R^{(0)}$ in each of the divided domains.
He begins the study in this section with a description of the
behavior of $R^{(0)}$ at the coordinate origin, $(\xi,\eta)=(0,0)$.
Citing his words:
\begin{quote}
``Now consider the maximum and minimum values of the function $R^{(0)}$.
\par
\hspace*{1em}
At the origin $\xi = \eta = 0$, the function $R^{(0)}$ can be minima, maxima or minimaxima according to the values of parameters $a$ and $k$.
It is obvious that $R^{(0)}$ is always minimum at the origin if $k^2$ is close enough to unity.''
(p. Z417)
\end{quote}

In the analogy of what \citeauthor{vonzeipel1910} has demonstrated
over the past chapters, we find the above statement correct.
Note that the variables $(\xi,\eta)$ were defined
earlier in his Chapter I
(p. Z349; see Eqs. \eqref{eqn:vZ10-xyzvars} and \eqref{eqn:vZ10-xi+eta}).
If we use the ordinary Delaunay elements,
they are expressed as follows:
\begin{alignat}{1}
   \xi &= \:\:\: \sqrt{2(L-G)} \cos g ,
  \label{eqn:vZ10-xi-redef} \\
  \eta &=       -\sqrt{2(L-G)} \sin g .
  \label{eqn:vZ10-eta-redef}
\end{alignat}
From the definitions in 
Eqs. \eqref{eqn:vZ10-xi-redef} and \eqref{eqn:vZ10-eta-redef},
we anticipate that $R^{(0)}$'s behavior
on the $(\xi,\eta)$ plane is topologically the same as that
on the $(x,y)=(e\cos g, e\sin g)$ plane.

In the second and the third paragraphs of Section Z28,
\citeauthor{vonzeipel1910} gives a description of a system where
all the disturbing planets are located outside the perturbed body.
This kind of system can be regarded as an extension of the inner CR3BP.
We cite his description:
\begin{quote}
``Consider first the orbits where $a$ is smaller than the semimajor axis of the orbit of Jupiter.
Given the dominant mass of this planet, it is clear, looking at the tables giving $R_{2.0}$, $R_{0.2}$, $R'_{2.0}$ and $R'_{0.2}$, that the function $R^{(0)}$ is minima at the origin $\xi=\eta=0$ if
$$
  k^2_{0.2} < k^2 < 1
$$
and minimaxima if
$$
  0 < k^2 < k^2_{0.2} .
$$
\par
\hspace{1em}
The proposition on the page Z390 and the remark on [the] revolution of planetary system of the page Z392 are still valid for the parts of the solar system at the interior of the orbit of Jupiter.
Only, by the effect of other planets, the limit $I_{0.2}$ (of the page Z389) will be a little high.''
(p. Z417)
\end{quote}
Later in this monograph
(p. \pageref{pg:remark-on-planetformation})
we will briefly state   what we think   he meant by
``the remark on [the] revolution of planetary system of the page Z392''.

In \citeauthor{vonzeipel1910}'s above statement,
the last sentence on the change of the limiting inclination $I_{0.2}$
``Only, by the effect of other planets, the limit $I_{0.2}$ (of the page Z389) will be a little high''
draws our particular attention. Let us interpret it as follows.
Consider a typical inner CR3BP system, Sun--asteroid--Jupiter.
Then, the addition of yet another disturbing planet orbiting exterior to
Jupiter's orbit (such as Saturn) would change the shape of
the total disturbing potential that the perturbed asteroid feels.
It means that the condition for the equilibrium points
to occur could be less easily achieved.
Let us choose an asteroid with $\alpha = 0.65$ as an example of perturbed body.
According to our \mysymfigO \ref{fig:I02-table} (p. \pageref{fig:I02-table}),
its limiting inclination $I_{0.2}$ is roughly $27^{\circ}.6$
(equivalent to $k^2_{0.2} \sim 0.785$).
Now, consider yet another disturbing planet orbiting exterior to Jupiter's orbit.
The $\alpha$ value of the perturbed asteroid with respect to this new, exterior perturber (referred to as $\alpha^{\rm e}$)
would then become smaller than 0.65 (i.e. $\alpha^{\rm e} < \alpha$).
This means that the limiting inclination with respect to the new
exterior perturber (referred to as $I^{\rm e}_{0.2}$)
would be larger than $I_{0.2}$.
If we place the two perturbers (Jupiter and the exterior planet)
at the same time in this system, the total perturbation from
the two perturbers would be a combination of the two CR3BPs
that have a limiting inclination of 
$I_{0.2} (\alpha)$ and $I^{\rm e}_{0.2} (\alpha^{\rm e})$, respectively.
And, since $I^{\rm e}_{0.2} > I_{0.2}$,
the resulting limiting inclination of the perturbed asteroid
in the two perturber system would be larger than $I_{0.2}$.
This is our interpretation of \citeauthor{vonzeipel1910}'s above sentence.

In the following paragraph of Section Z28,
\citeauthor{vonzeipel1910} gives a description of a system where
all the disturbing planets are located inside the orbit of the perturbed body.
This kind of system can be regarded as an extension of the outer CR3BP.
At the same time, he also considers a system where
the orbit of the perturbed body is located just outside the orbit of
one of the disturbing planets
(so that the influence of other disturbing planets gets relatively smaller).
His remarks about this are as follows:
\begin{quote}
``Consider then the values of $a$, which are larger than the semimajor axis of the orbit of Jupiter.
If $a$ is between certain limits, especially if $a$ is larger than the semimajor axis of the orbit of the last planet (Neptune) or slightly larger than the semimajor axis of the orbit of Uranus, Saturn or Jupiter, then $R^{(0)}$ can be maximum for small values of $k^2$.
Conversely, if $a$ is a bit smaller than the semimajor axis of the orbit of Saturn, $R^{(0)} $ is minimaximum for small values of $k^2$.''
(p. Z417)
\end{quote}

We already learned that,
the doubly averaged disturbing function for the outer CR3BP
takes a local maximum at the origin $(x,y)=(0,0)$ when $k^2$ is small enough
(see \mysymfigO \ref{fig:vZ10-f8-10} or \mysymfigO \ref{fig:eqR-vzoex-a0} of this monograph).
\citeauthor{vonzeipel1910}'s first conclusion in the above paragraph
(``$\cdots$ if $a$ is larger than the semimajor axis of the orbit of the last planet (Neptune) $\cdots$ then $R^{(0)}$ can be maximum for small values of $k^2$'')
seems consistent with this fact.
On the other hand, the last sentence of the above paragraph
(``Conversely, if $a$ is a bit smaller than the semimajor axis of the orbit of Saturn, $R^{(0)} $ is minimaximum for small values of $k^2$'')
seems similar to the discussion that we had at the inner CR3BP
where $R$ takes a saddle point at the origin $(0,0)$
when $k^2$ is small enough.

As a consequence of the above considerations,
\citeauthor{vonzeipel1910} reaches a specific conjecture, and writes:
\begin{quote}
``We can conclude from this that, in the solar system, there may be orbits which are always not very eccentric and have inclinations close to $90^\circ$.
These orbits may exist somewhat outside the orbit[s] of Jupiter, Saturn or Uranus.
But above all, they might be outside the last planet of the system.''
(p. Z417)
\end{quote}
\label{pg:justoutsidethelastplanet}

As we saw in Section \ref{ssec:ocr3bp} of this monograph
about the doubly averaged outer CR3BP,
this kind of orbit---not very eccentric, with a large inclination
(if not nearly $90^\circ$), and located somewhat outside the outermost
perturbing body---can stably exist if $k^2$ is small enough.
An orbit of this kind would be confined near the origin $(0,0)$
of the $(e \cos g, e \sin g)$ plane.
For example,
the CR3BP systems employed in \mysymfigS \ref{fig:eqR-vzoex-aL}\mtxtsf{a} and
                                         \ref{fig:eqR-vzoex-aL}\mtxtsf{e}
with $(\alpha', k^2) = (0.6, 0.2600)$ predict the existence of
stable orbits of perturbed bodies whose inclination is nearly $60^\circ$
around the origin where $e=0$.

If we turn our attention to the actual solar system,
we recognize a certain number of objects in this category---small $e$,
large $I$, located somewhat outside the orbit of perturbing planets,
and whose $k^2$ is small.
We chose several small objects of this kind and listed them in Table \ref{tbl:ex-Z28}.
As seen in this table, 
we may want to classify the candidate objects in two categories.
The first category comprises objects with very small $k^2$
and large inclination, but with moderate eccentricity.
The ten objects in the upper part of Table \ref{tbl:ex-Z28}
denoted as ``small $k^2$'' correspond to these.
See the caption of Table \ref{tbl:ex-Z28} for details of
our empirical determination criteria.
The second category contains objects with a very small eccentricity,
but with a mildly high inclination,
therefore their $k^2$ is not so small.
The six objects in the lower part of Table \ref{tbl:ex-Z28}
denoted as ``small $e$ and large $I$'' correspond to these.
Existence of these two orbital characteristics may imply something about how
these objects were formed and how their orbits have been dynamically evolved.
But here, let us just emphasize that a substantial number of small bodies
in the actual solar system satisfy the orbital conditions
that \citeauthor{vonzeipel1910} has depicted, at least partially.
We believe this corroborates the correctness of his theory.
These objects are either Centaurs or TNOs by definition,
and their discoveries were made many decades later than
\citeauthor{vonzeipel1910}'s era---%
Centaurs were not discovered until 1977 when (2060) Chiron was found, and
TNOs     were not discovered until 1992 when 1992 QB${}_{1}$ was found,
except for Pluto which was discovered in 1930.
As the number of objects in these categories increases
along with the progress of observational surveys,
the validity of \citeauthor{vonzeipel1910}'s theory
will be more rigorously verified and substantiated.

\newpage
\renewcommand{\baselinestretch}{0.9}
\begin{table}[hbtp]\centering
\setlength{\tabcolsep}{8pt}
\caption{%
Osculating orbital elements $a$, $e$, $I$ and the parameter $k^2$ of
several small objects in the actual solar system
that can match \citeauthor{vonzeipel1910}'s description in his Section Z28:
Not very eccentric, with an inclination close to $90^\circ$,
and located somewhat outside the orbit of perturbing planet.
The data is taken from the file
\mtxtsf{astorb.dat}
provided by Lowell Observatory as of September 28, 2018.
We chose and put the objects into two categories.
The objects in the first category are those with very small $k^2$
with a large inclination and moderate eccentricity.
They are denoted as ``small $k^2$'' on the left-end of the table.
Our empirical determination criteria for this category is $k^2 < 0.2$.
The objects in the second category have a very small eccentricity
but with a mildly high inclination;
therefore their $k^2$ is not quite small
(denoted as ``small $e$ and large $I$'').
Our           determination criteria for this category is
$e \lesssim 0.08$ and $I$ (or $\pi - I$) ranges from $40^\circ$ to $50^\circ$.
The orbit type (Centaur or TNO) is based on the categorization by
the JPL Small-Body Database Browser:
Centaurs are defined as objects with orbits between Jupiter and Neptune $(5.5 \: \mathrm{au} < a < 30.1 \: \mathrm{au})$, and
TNOs are defined as objects with orbits outside Neptune $(a > 30.1 \: \mathrm{au})$.
The ratio of the semimajor axes between each small object and its major perturbing planet
(in the framework of the outer CR3BP) is denoted as $\alpha'$.
This is either
$\alpha'_\mathrm{J} = \frac{a_\mathrm{J}}{a}$,
$\alpha'_\mathrm{S} = \frac{a_\mathrm{S}}{a}$,
$\alpha'_\mathrm{U} = \frac{a_\mathrm{U}}{a}$, or
$\alpha'_\mathrm{N} = \frac{a_\mathrm{N}}{a}$,
where
$a_\mathrm{J}$,
$a_\mathrm{S}$,
$a_\mathrm{U}$,
$a_\mathrm{N}$
are the semimajor axes of Jupiter, Saturn, Uranus and Neptune, respectively.
Note that, for each small object, we just put the value of $\alpha'$ for the closest perturbing planet.
For example, we only showed the value of $\alpha' = \alpha'_\mathrm{S} = 0.82219$ for 2008 YB${}_{3}$,
not $\alpha'_\mathrm{J}$, $\alpha'_\mathrm{U}$, or $\alpha'_\mathrm{N}$.
This is because Saturn's orbit is closer to this object than any of the other three planetary orbits.
Near the right-end of the table,
``arc (d)'' denotes the number of days spanned by the observational data arc for the object, and
$n_\mathrm{obs}$ indicates the number of observations used to determine the orbital elements.
We included these two columns to gauge the accuracy (or inaccuracy) of the orbit determination for each object.
Here is a note on the symbol $(\ast 1)$ for the orbit type of 2015 KG${}_{157}$:
This object's orbital elements have been determined from just twelve sets of observations
$(n_\mathrm{obs} = 12)$ spanning just two days $(\mathrm{arc} = 2)$.
Consequently, they contain a non-negligible amount of uncertainties.
In fact, on
the JPL Small-Body Database Browser
its semimajor axis is denoted as
$a = 5.928588171390333 \pm 2.3908$ au (as of November 14, 2018)
which is slightly different from the value listed on
\mtxtsf{astorb.dat},
and its orbital type is categorized as Centaur.
Similar uncertainties lie in the orbital elements of
2006 HU${}_{122}$ and 2016 FM${}_{59}$ whose observational arcs are short.
Note also that
some of the TNOs in the table may be in the category of resonant TNOs
if their orbits are accurately determined.
For example,
2014 XZ${}_{40}$'s orbital location is similar to that of Plutinos
(those in the 2:3 mean motion resonance with Neptune), and
that of 2006 HU${}_{122}$ is rather close to the 4:7 mean motion resonance
with Neptune.
We would need a long-term direct numerical integration for
accurately characterizing their dynamical categorization.
For a better understanding of what kind of orbits these objects actually have,
in \supinfo{5}
we put figures for the orbital diagrams of all the objects listed in this table.
}
\label{tbl:ex-Z28}
\begin{scriptsize}
\begin{tabular}{clcrrrrrrr}
\Hline
                           &
\multicolumn{1}{c}{object} &
\multicolumn{1}{c}{type}   &
\multicolumn{1}{c}{$a$ (au)}  &
\multicolumn{1}{c}{$e$}       &
\multicolumn{1}{c}{$I$ (deg)} &
\multicolumn{1}{c}{$k^2$}     &
\multicolumn{1}{c}{$\alpha'$} &
\multicolumn{1}{c}{arc (d)} &
\multicolumn{1}{c}{$n_\mathrm{obs}$} \\
\Hline
\parbox[t]{2mm}{\multirow{14}{*}{\rotatebox[origin=c]{90}{small $k^2$}}}
& \multicolumn{9}{c}{outside Jupiter} \\
\cline{2-10}
& 2015 KG${}_{157}$ &$(\ast 1)$& 5.39893316 & 0.42080926 &  72.904103 & 0.07112 & 0.96363 $(\alpha'_\mathrm{J})$ &     2 &  12 \\
& 2007 VW${}_{266}$ &         &  5.44387890 & 0.39031355 & 108.310551 & 0.08366 & 0.95568 $(\alpha'_\mathrm{J})$ &    38 &  59 \\
& 2010 CR${}_{140}$ & Centaur &  5.62365608 & 0.40883143 &  74.667719 & 0.05823 & 0.92513 $(\alpha'_\mathrm{J})$ &  1418 &  43 \\
& 2012 YO${}_{6}$   & Centaur &  6.31511335 & 0.47807149 & 106.886748 & 0.06509 & 0.82383 $(\alpha'_\mathrm{J})$ &   142 &  47 \\
& 2014 JJ${}_{57}$  & Centaur &  6.99082311 & 0.29313977 &  95.922691 & 0.00973 & 0.74420 $(\alpha'_\mathrm{J})$ &  2815 &  30 \\
\cline{2-10}
& \multicolumn{9}{c}{outside Saturn} \\
\cline{2-10}
& 2008 YB${}_{3}$   & Centaur & 11.62121366 & 0.44058710 & 105.059221 & 0.05440 & 0.82219 $(\alpha'_\mathrm{S})$ &  3236 & 502 \\
\cline{2-10}
& \multicolumn{9}{c}{outside Uranus} \\
\cline{2-10}
& 2011 MM${}_{4}$   & Centaur & 21.14652789 & 0.47287733 & 100.490295 & 0.02574 & 0.90882 $(\alpha'_\mathrm{U})$ &  2552 & 144 \\
& 2007 BP${}_{102}$ & Centaur & 23.96197862 & 0.25905974 &  64.680786 & 0.17062 & 0.80204 $(\alpha'_\mathrm{U})$ &  3396 &  51 \\
\cline{2-10}
& \multicolumn{9}{c}{outside Neptune} \\
\cline{2-10}
& 2011 KT${}_{19}$  & TNO     & 35.58572066 & 0.33046594 & 110.102428 & 0.10523 & 0.84614 $(\alpha'_\mathrm{N})$ &  1779 & 132 \\
& 2008 KV${}_{42}$  & TNO     & 41.58639205 & 0.49226472 & 103.398712 & 0.04068 & 0.72404 $(\alpha'_\mathrm{N})$ &  3610 &  55 \\
\Hline
\parbox[t]{2mm}{\multirow{8}{*}{\rotatebox[origin=c]{90}{small $e$ and large $I$}}}
& \multicolumn{9}{c}{outside Jupiter} \\
\cline{2-10}
& 2016 LN${}_{8}$   & Centaur &  5.76477817 & 0.04542763 &  43.360457 & 0.52751 & 0.90248 $(\alpha'_\mathrm{J})$ &  6290 &  94 \\
\cline{2-10}
& \multicolumn{9}{c}{outside Neptune} \\
\cline{2-10}
& 2014 XZ${}_{40}$  & TNO     & 39.65156553 & 0.05013667 &  44.564654 & 0.50632 & 0.75937 $(\alpha'_\mathrm{N})$ &    77 &  10 \\
& 2013 SA${}_{87}$  & TNO     & 41.67277625 & 0.07141871 &  40.868092 & 0.56895 & 0.72254 $(\alpha'_\mathrm{N})$ &  1400 &  53 \\
& 2004 DF${}_{77}$  & TNO     & 43.55225696 & 0.02231651 &  43.670818 & 0.52293 & 0.69136 $(\alpha'_\mathrm{N})$ &    17 &   4 \\
& 2006 HU${}_{122}$ & TNO     & 44.84803376 & 0.08128279 &  46.907861 & 0.46364 & 0.67139 $(\alpha'_\mathrm{N})$ &     1 &   3 \\
& 2016 FM${}_{59}$  & TNO     & 45.12950220 & 0.05981495 & 131.522580 & 0.43788 & 0.66720 $(\alpha'_\mathrm{N})$ &     3 &   9 \\
\Hline
\end{tabular}
\end{scriptsize}
\end{table}
\clearpage
\renewcommand{\baselinestretch}{1.0}

The next section (Z29) that closes
\citeauthor{vonzeipel1910}'s work \citeyearpar{vonzeipel1910} is composed of
a very brief consideration on the secular motion of a comet, 2P/Encke.
This section seems to be an exposition of his theory
on the doubly averaged restricted systems with more than one perturbing body
stated in this chapter.
However the description in this section is too concise, and
it does not seem to contain much quantitative material
that we should deal with in this monograph.
Therefore we do not go into the contents of this section.

\section{Summary and Discussion\label{sec:discussion}}
Since we are in the final part of this monograph,
we bring up several issues to conclude our study.
In Sections \ref{ssec:summary-kozai}, \ref{ssec:summary-lidov}, and \ref{ssec:summary-zeipel},
we try to organize and compare the achievements presented in the works of
\citeauthor{kozai1962b},
\citeauthor{lidov1961}, and
\citeauthor{vonzeipel1910}.
After that, 
in Section \ref{ssec:earlier-zeipel} we introduce
\citeauthor{vonzeipel1910}'s earlier publications that can be regarded as
a preparatory step
toward his 1910 paper.
Finally in Sections \ref{ssec:historicEX} and \ref{ssec:ZLKcycle},
we express our opinion as to how the secular dynamical effect
discussed in this monograph should be referred to,
in comparison with several historical examples.

\begin{table}[t]\centering
\setlength{\tabcolsep}{4pt}
\caption{A comparison of the achievements shown in 
\citeauthor{kozai1962b}'s,
\citeauthor{lidov1961}'s, and
\citeauthor{vonzeipel1910}'s work.
Some symbols have superscript numbers that have the following meanings:
1: There are some descriptions, but they are very terse and do not seem complete.
2: Mentioned in \citet{lidov1963a,lidov1963b} just for a limited case when $i=\frac{\pi}{2}$.
3: Described in \citet{lidov1963a,lidov1963b} for a limited case.
4: Orbital variation of the satellites of Uranus is discussed in \citet{lidov1963a,lidov1963b}.
5: Described in \citet{lidov1963a,lidov1963b} briefly.
6: Drawn by hand, but topologically correct.
7: $I_{0}$ and $I_{0.2}$ of several asteroids such as (2) Pallas or (531) Zerlina are calculated and discussed.
8: Motions of 1P/Halley and 2P/Encke are quantitatively discussed.
9: Detailed mathematical exposition is presented, but without tables or figures.
10: Mathematical exposition is presented without tables or figures.
}
\label{tbl:comparison}
\begin{tabular}{rccc}
\Hline
  & \citeauthor{kozai1962b} & \citeauthor{lidov1961} & \citeauthor{vonzeipel1910} \\
\Hline
\multicolumn{4}{c}{\textit{General features and treatment of CR3BP\/}} \\
\hline
  Hamiltonian formalism        & $\bigcirc$ & ---         & $\bigcirc$ \\
  Gauss's form of equations    & ---        & $\bigcirc$  & ---        \\
  single averaging             & ---        & $\bigcirc$  & ---        \\
  double averaging             & $\bigcirc$ & $\bigcirc$  & $\bigcirc$ \\
  conservation of $\left(1-e^2\right) \cos^2 i$
                               & $\bigcirc$ & $\bigcirc$  & $\bigcirc$ \\
  conservation of total energy & $\bigcirc$ & $\bigcirc$  & $\bigcirc$ \\
  numerical quadrature         & $\bigcirc$ & ---         & $\bigcirc$ \\
  direct numerical integration & ---        & $\bigcirc$  & ---        \\
\Hline
\multicolumn{4}{c}{\textit{Doubly averaged inner CR3BP\/}} \\
\hline
  libration of $g$ around $\pm\frac{\pi}{2}$ & $\bigcirc$  & $\bigcirc$  & $\bigcirc$  \\
  conservation of $c_2$-like variable      & --- & $\bigcirc$  & --- \\
  solutions in special cases               & $\bigtriangleup^1$ & $\bigcirc$  & --- \\
  time-dependent analytic solution         & $\bigcirc$  & $\bigtriangleup^2$ & --- \\
  equi-potential contours                  & $\bigcirc$  & $\bigtriangleup^3$ & $\bigcirc^6$ \\
  treatment when $\alpha$ is not small     & $\bigcirc$  & --- & $\bigcirc$ \\
  mention of actual objects                & $\bigcirc$ & $\bigtriangleup^4$ & $\bigcirc^7$  \\
  oblateness of central mass               & --- & $\bigtriangleup^5$ & --- \\
\Hline
\multicolumn{4}{c}{\textit{Doubly averaged outer CR3BP\/}} \\
\hline
  libration of $g$ around $\pm\frac{\pi}{2}$ & --- & --- & $\bigcirc$   \\
  libration of $g$ around $0$ or $\pi$       & --- & --- & $\bigcirc$   \\
              equi-potential contours        & --- & --- & $\bigcirc^6$ \\
     treatment when $\alpha'$ is not small   & --- & --- & $\bigcirc$   \\
                  mention of actual objects  & --- & --- & $\bigcirc^8$ \\
\Hline
\multicolumn{4}{c}{\textit{Other features and treatment\/}} \\
\hline
\multicolumn{1}{r}{Treatment of orbit intersection} & --- & --- & $\bigcirc^9$  \\
\multicolumn{1}{r}{Multiple perturbing bodies}      & --- & --- & $\bigtriangleup^{10}$  \\
\Hline
\end{tabular}
\end{table}

For comparing the works of the three authors,
we have made a list of similarities and differences
between the works of \citeauthor{kozai1962b}, \citeauthor{lidov1961}, and
\citeauthor{vonzeipel1910} in Table \ref{tbl:comparison}.
In this table,
the open circles $(\bigcirc)$ denote ``achieved and clearly presented'' by the work.
The long hyphens (---) denote ``not achieved or mentioned.''
The open triangles $(\bigtriangleup)$ denote
``partially achieved or just mentioned, or mentioned in relevant publications.''
We admit that our evaluation for these three categories
($\bigcirc$, ---, $\bigtriangleup$) are subjective, and sometimes ambiguous.
For example, we graded ``numerical integration of orbits'' in
\citeauthor{lidov1961}'s work as $\bigcirc$,
but it can be arguable whether or not the numerical integration
presented in \citeauthor{lidov1961}'s work is worth a discussion
in the context of modern celestial mechanics.

\subsection{Achievements of \citeauthor{kozai1962b}'s work\label{ssec:summary-kozai}}
\citeauthor{kozai1962b}'s work has been recognized as a classic in this line of study.
As we learned,
he began with a general Hamiltonian formalism of the inner CR3BP, and
derived the secular disturbing function
through the double averaging procedure.
Through the analysis of the doubly averaged disturbing function,
\citeauthor{kozai1962b} showed the possibility of libration of the perturbed body's argument of pericenter $g$ around $\pm \frac{\pi}{2}$ under a certain condition.
He clarified the condition to be
$\Theta = \left( 1-e^2 \right) \cos^2 i < \frac{3}{5}$
at the quadrupole level approximation.
As we have seen, \citeauthor{kozai1962b}'s $\Theta$ is equivalent to
\citeauthor{lidov1961}'s $c_1$ and
\citeauthor{vonzeipel1910}'s $k^2$.
\citeauthor{kozai1962b} did not mention any parameters equivalent to
\citeauthor{lidov1961}'s $c_2$ whose sign explicitly tells us
whether or not the perturbed body's argument of pericenter librates
at the quadrupole level approximation.

One important accomplishment in \citeauthor{kozai1962b}'s work is that
he employed diagrams with equi-potential contours of
the doubly averaged disturbing function as a fundamental subject of discussion.
\citeauthor{kozai1962b}'s aim of exploiting the equi-potential diagrams was
to show the global structure of possible trajectories that the perturbed body draws,
and to visually confirm the existence of equilibrium points.
Using the equi-potential diagrams,
\citeauthor{kozai1962b} found that the argument of perihelion $g$ of
(1373) Cincinnati librates around $g=\frac{\pi}{2}$.
It seems that \citeauthor{kozai1962b} reached this conclusion
just from the topological pattern of the equi-potential contours and
the initial location of this asteroid (\mysymfigO K7 on p. K597),
not following the time variation of its orbital elements.
Nowadays
we see this kind of equi-potential diagram ubiquitously in the literature,
not only in CR3BP studies, but in other dynamical studies
where a system's degrees of freedom can be somehow reduced to unity.
In this regard, \citeauthor{kozai1962b} was a pioneer.

For calculating the equi-potential contours,
\citeauthor{kozai1962b} carried out the numerical quadrature
defined by Eq. \eqref{eqn:K09}.
\citeauthor{kozai1962b} called the quadrature
``numerical harmonic analysis'' (p. K593, the left column,
four lines below Eq. (K17). Also in the caption of his \mysymfigO K1).
It is not hard to imagine that carrying out numerical quadrature
in the early 1960s was a formidable task when the use of high-speed digital
computers was extremely limited. However, \citeauthor{kozai1962b} completed
the numerical work using digital computers available in the United States
while he was a long-term visiting scholar at Smithsonian Astrophysical
Observatory, Cambridge, Massachusetts (Kozai 2017, personal communication).
Although
we have not confirmed if \citeauthor{kozai1962b}'s work is
the very first one that quantitatively exploits
equi-potential diagrams in celestial mechanics,
we are convinced that \citeauthor{kozai1962b} was one of the first
to recognize the importance and usefulness of the equi-potential contours
in this line of study.

Incidentally, note that \citet[][his Section 3.2.4 on p. 39]{shevchenko2017}
calls the equi-potential diagrams ``the \citeauthor{lidov1961}--\citeauthor{kozai1962b} diagram.''
We do not agree with his opinion mainly because there is no such
equi-potential contour plot in \citeauthor{lidov1961}'s original paper
\citeyearpar{lidov1961}.
There is an equi-potential diagram in his later papers \citep{lidov1963a,lidov1963b}
but it was just used for a special case.
Moreover, as we have seen in this monograph,
\citeauthor{vonzeipel1910} had already drawn this kind of equi-potential diagrams much earlier
(see \mysymfigS \ref{fig:vZ10-f1-5} and \ref{fig:vZ10-f6-7} in this monograph)
even though they look somewhat schematic.

We should also recall that
\citeauthor{kozai1962b} derived explicit expressions of
high-order analytic expansions of the doubly averaged disturbing function
for the inner CR3BP up to $\Oaloct$,
which is reproduced as Eq. \eqref{eqn:K23}.
This is not achieved in \citeauthor{lidov1961}'s work or
in \citeauthor{vonzeipel1910}'s work.
Using \citeauthor{kozai1962b}'s analytic expansion,
we can draw equi-potential curves of perturbed bodies accurately
without going through numerical quadrature, even when $\alpha$ is not small.

Yet another unique achievement seen only in \citeauthor{kozai1962b}'s work is
that he derived a time-dependent analytic solution for orbital elements
as function of time by employing Weierstrass's elliptic functions $\wp$
(see p. \pageref{ssec:Kozai-quadrupole} of this monograph).
The equi-potential trajectories calculated from
the doubly averaged disturbing function show us the dynamical structure of perturbed bodies at a glance,
but they do not give us information about their time variation,
such as $e(t)$ or $g(t)$.
Having the time-dependent analytic solution that \citeauthor{kozai1962b} constructed,
we can calculate approximate time variation of perturbed body's orbital elements,
even though the approximation is limited to the quadrupole level.

\subsubsection{Later developments\label{sssec:later-kozai}}
\citeauthor{kozai1962b}'s work has been checked, confirmed, and developed
by numerous authors including \citeauthor{kozai1962b} himself
since its publication in \citeyear{kozai1962b}.
We can pick the following developments as for \citeauthor{kozai1962b}'s
four remarks on future prospects that
we mentioned in Section \ref{ssec:kozai-remarks}.

As for the issue (i) on the effect of perturber's non-zero eccentricity $e'$,
its influence has turned out to be quite significant. It is now well known as
the eccentric
    \citeauthor{lidov1961}--\citeauthor{kozai1962b}
(or \citeauthor{kozai1962b}--\citeauthor{lidov1961}) {\mainword}
\citep[e.g.][]{naoz2016,sidorenko2018}.
It can substantially change the long-term orbital motion of
the perturbed body, sometimes invoking sudden orbital flips
as we demonstrated in \mysymfigO \ref{fig:eLK-example}
(p. \pageref{fig:eLK-example}).

As for the issue (ii) on the incorporation of the indirect perturbation from other planets,
\citeauthor{kozai1962b} himself succeeded in incorporating the effect of
other planets by assuming that they have circular orbits \citep{kozai1979}.
This was possible because the inclusion of perturbers on planar and circular orbits
would not increase the total degrees of freedom of the system after
the averaging procedure.
The same subject was studied by \citet{vashkovyak1976} and \citet{vashkovyak1981c}.
\citeauthor{kozai1979}'s method was followed, and used by many later studies
\citep[e.g.][]{nakai1985,ito1999a,ito2001a,ito2001b}.

As for the issue (iii) on the inclusion of planetary oblateness on the motion of satellites,
\citet{lidov1963a,lidov1963b} studied it shortly later than \citeauthor{kozai1979}.
\citeauthor{kozai1962b} extended his theory by himself by including the
oblateness of central body such as $J_2$, $J_3$, and $J_4$
\citep{kozai1969b}.
Much later, \citet{kinoshita1991c} worked on the point that
\citeauthor{kozai1962b} raised:
Possible decrease and disappearance of the threshold value of
perturbed body's $\left(1-e^2\right) \cos^2 i$
due to planetary oblateness.
\citeauthor{kinoshita1991c} quantitatively formulated the influence of
Uranian $J_2$ on the secular motion of its fictitious satellites.
They found that when the Uranian satellites are located inside the region
where the perturbation caused by planetary oblateness is dominant,
their argument of pericenter does not librate,
and stationary points do not occur.

As for the issue (iv) on the motion of satellites for which the
orbital period of the Sun may not be regarded as short,
we presume that the subject is in the realm of the
singly averaged three-body system
such as \citeauthor{moiseev1945a} or \citeauthor{lidov1961} dealt with.
Although the single averaging procedure is not very popular now,
mostly because they can be replaced for direct numerical integration
in many cases,
there is still a certain flow of studies along this line in modern
celestial mechanics \citep[e.g.][]{domingos2013,elshaboury2013,nie2019}.
There are even studies of singly averaged hyperbolic \citep{sorokovich1982}
and parabolic \citep{memedov1989} restricted three-body problems.

Efforts to obtain the time-dependent analytic solutions of perturbed body's
orbital elements in the doubly averaged CR3BP using elliptic functions,
as \citeauthor{kozai1962b} demonstrated,
has been continued and extended.
However Weierstrass's elliptic functions $\wp$
that \citeauthor{kozai1962b} employed are not used anymore in this respect.
A major extension of this line of work was achieved by a pair of Japanese
celestial mechanists, Hiroshi Kinoshita and Hiroshi Nakai,
both of who worked close to \citeauthor{kozai1962b}.
\citet{kinoshita1991c} succeeded in deriving an analytic solution
for this problem using Jacobi's elliptic function that can be used
when the perturbed body's $g$ circulates.
\citet{kinoshita1999} later gave another type of analytic solution
expressed by Jacobi's elliptic function
that can be used when the perturbed body's $g$ librates around $\frac{\pi}{2}$.
Finally
\citet{kinoshita2007a} reached a general analytic solution that
can be used either when the perturbed body's $g$ librates or when it circulates.
We should note that \citet{vashkovyak1999} had also reached a general solution
of this kind independently, through a slightly different way,
for the motion of distant satellites of Uranus.

So far, these analytic solutions are derived from the doubly averaged disturbing function
of the inner CR3BP at the quadrupole level approximation,
and their practical usefulness
does not seem to be as significant as direct numerical integration.
However, the way of deriving the time-dependent analytic solutions in these works
seems to be inherited from dynamical studies of comets with very large perihelion distances under the galactic tide.
More specifically speaking,
the dynamical evolution of small bodies that compose the Oort Cloud
\citep{oort1950,oort1951} under the galactic tide has been often discussed and
calculated using analytically obtained approximate solutions
\citep[e.g.][]{heisler1986,fouchard2005,higuchi2007}.
These works can be regarded as an extension of \citeauthor{kozai1962b}'s work.
Note also that, in a small body system that is under the perturbation from the galactic tide,
the quantity $\sqrt{ 1-e^2 } \cos I_\mathrm{g}$ becomes constant after averaging over small body's mean anomaly
if we only take the vertical component of the galactic tide
(where $I_\mathrm{g}$ denotes orbital inclination of the small body with respect to the galactic plane).
This is due to the rotational invariance of the averaged galactic tidal potential around the normal of the galactic plane \citep[e.g.][]{saillenfest2019a}.
This point is similar to the doubly averaged CR3BP that we have considered in this monograph.
\citet[][their p. 1222]{breiter2005} presents a concise review of the historic development of this line of studies.

At the end of this subsection,
we would like readers to pay attention to a critical study
that highlights incorrect applications of Jacobi's elimination of the nodes
in several past works \citep{naoz2013a}.
\citeauthor{naoz2013a} pointed to a description in \citet[][p. K592]{kozai1962b}
as an example of the false arguments.
Citing their paragraphs:
\begin{quote}
``Since the total angular momentum is conserved, the ascending nodes
relative to the invariable plane follow a simple relation,
$h_1(t) = h_2(t) - \pi$.
If one inserts this relation into the Hamiltonian, which
only depends on $h_1 - h_2$, the resulting `simplified' Hamiltonian
is independent of $h_1$ and $h_2$. One might be tempted to conclude
that the conjugate momenta $H_1$ and $H_2$ are constants of the motion.
However, that conclusion is false. This incorrect argument has been
made by a number of authors. $\cdots$
\par
[In a footnote]
For example, Kozai (1962, p. 592) incorrectly argues that
`as the Hamiltonian $F$ depends on $h$ and $h'$ as a combination
$h-h'$, the variables $h$ and $h'$ can be eliminated from $F$
by the relation (5). Therefore, $H$ and $H$ are constant'.''
\citep[][their Appendix C on p. 2171]{naoz2013a}
\end{quote}
Consult \citeauthor{naoz2013a}'s Appendix C for the details of their reasoning.
Fortunately, the final conclusion described in \citet{kozai1962b} is correct
because \citeauthor{kozai1962b} used the test particle approximation
(i.e. restricted three-body problem) in the most part of his discussion.

\subsection{Achievements of \citeauthor{lidov1961}'s work\label{ssec:summary-lidov}}
Compared with \citeauthor{kozai1962b}'s work whose objective was
to establish a secular dynamical theory of asteroids,
\citeauthor{lidov1961}'s work had the motivation to construct a
secular dynamical theory of Earth-orbiting artificial satellites.
In his works published between 1961 and 1963,
\citeauthor{lidov1961} did not start from the Hamiltonian formalism.
His works started from the classical Gauss's form of equations
where the time derivatives of orbital elements are expressed
by three components of perturbing forces.
Note that \citeauthor{lidov1961} used the Hamiltonian formalism
in his later works \citep[e.g.][]{lidov1974,lidov1976}.

Similar to the line of studies previously presented by \citeauthor{moiseev1945a},
\citeauthor{lidov1961} first went through the single averaging procedure of
disturbing forces. He then moved on to the double averaging procedure,
and finally reached the doubly averaged disturbing function.
\citeauthor{lidov1961} just dealt with the inner CR3BP,
and his approximation was at the quadrupole level.
Although it does not have any influence on his conclusion,
\citeauthor{lidov1961}'s formulation included the effect of the perturber's
eccentricity $e_k$ as a constant factor multiplied with perturbing forces
(see Eq. \eqref{eqn:L55} on p. \pageref{eqn:L55} of this monograph).

Unlike \citeauthor{kozai1962b},
\citeauthor{lidov1961} did not present time-dependent analytic solutions
of orbital elements in specific forms,
except for a very special case when the perturbed body's orbit is
vertically inclined, $i = \frac{\pi}{2}$
(see Section \ref{ssec:lidov1963} of this monograph).
Instead, he showed a general guideline as to how we can obtain
time-dependent solutions for orbital elements by quadrature
(see Eqs. \eqref{eqn:L60} and \eqref{eqn:L61}, and the discussion there).
An interesting point is that \citeauthor{lidov1961} described
a practical computation method of a satellite's orbital evolution
based on his theory.
He even tried to confirm the accuracy of his analytic theory by
carrying out direct numerical integration of equations of motion.
Although this part of \citeauthor{lidov1961}'s work
(his Sections 8, 9, 10) in his \citeyear{lidov1961} paper
may be already obsolete from a modern viewpoint,
it typically exemplifies the practical importance of his work
in the field of artificial Earth satellites at that time.

An achievement seen only in \citeauthor{lidov1961}'s work,
and not in \citeauthor{kozai1962b}'s or in \citeauthor{vonzeipel1910}'s work,
is that \citeauthor{lidov1961} introduced a parameter
$c_2$ in Eq. \eqref{eqn:L59},
not only $c_1 = \left(1-e^2 \right) \cos^2 i$ in Eq. \eqref{eqn:L58}.
Using the combination of $c_2$ and $c_1$,
\citeauthor{lidov1961} showed that it is possible to predict
whether or not the argument of pericenter of a particular perturbed body
librates around $\pm\frac{\pi}{2}$ at the quadrupole level approximation.
\citeauthor{kozai1962b}'s parameter $\Theta$ is equivalent to
\citeauthor{lidov1961}'s $c_1$, and
\citeauthor{kozai1962b} also
showed a condition for an asteroid's argument of pericenter to librate,
$\Theta \leq \frac{3}{5}$
(Eq. \eqref{eqn:Theta-max} of this monograph).
However as we saw,
having $c_1$ (or $\Theta$) that is smaller than $\frac{3}{5}$ is
not a sufficient condition for
the libration of the perturbed body's argument of pericenter to occur.
Whether or not the argument of pericenter librates depends on
how large the (averaged) total potential energy is; in other words,
which equi-potential contour the perturbed body moves along.
For this purpose \citeauthor{lidov1961} devised a parameter $c_2$,
a combination of the total potential energy and the vertical component
of the perturbed body's angular momentum,
although he did not leave any specific derivations as to
how he devised the expression of $c_2$ shown in Eq. \eqref{eqn:L59}.
As a result, \citeauthor{lidov1961} succeeded in 
explicitly showing that the condition $c_2 < 0$ is necessary
for the perturbed body's argument of pericenter to librate around $\pm\frac{\pi}{2}$
(see p. \pageref{pg:lidov-specialsolutions} of this monograph).
Actually this is not only the necessary condition but also the sufficient condition,
recalling the fact that $c_1 < \frac{3}{5}$ is automatically fulfilled when $c_2 < 0$.
We showed a visualization of the theoretically possible range of $(c_1, c_2)$ on the \citeauthor{lidov1961} diagram (\mysymfigO \ref{fig:lidovdiagram}).
Although the function form of $c_2$ in Eq. \eqref{eqn:L59} is valid
only at the quadrupole level approximation,
it is principally possible to find a dependence of $c_2$
(or other equivalent parameters) on $\alpha$
at a higher-order approximation, such as $\Oalqua$ or $\Oalhex$.
It is also the case for $c_1$, as \citeauthor{kozai1962b} has already done
on his $\Theta$ (see Eq. \eqref{eqn:K25} on p. \pageref{eqn:K25} of this monograph).

Let us also mention that in \citeauthor{lidov1961}'s work,
solutions that take place on the borders of the \citeauthor{lidov1961} diagram
(\mysymfigO \ref{fig:lidovdiagram}) are elaborately illustrated and
quantitatively explained (see Section \ref{ssec:Lidovdiagram}).
\citeauthor{lidov1961}'s detailed and precise characterization
of the solutions in the special and extreme cases,
together with his visualization realized in the \citeauthor{lidov1961} diagram,
provides us with a comprehensive understanding of the dynamical structure
of CR3BP at the quadrupole level approximation.
We believe this part significantly enhances the value of
\citeauthor{lidov1961}'s work in comparison with other similar studies.

\subsubsection{Later developments in Russia\label{sssec:later-lidov}}
The Soviet Union (currently Russian) academic community has developed a rich flow of three-body problem studies.
\citeauthor{lidov1961}'s work on this subject in the 1960s
has been later extended to a great deal in this community.
Let us cite a few examples from the early days:
\citet{orlov1965a,orlov1965b} extended \citeauthor{lidov1961}'s framework,
and obtained approximate solutions of doubly averaged CR3BP 
using the disturbing function with higher-order terms
through a canonical perturbation method.
\citet{orlov1972} extended his own study and calculated the short-term
periodic perturbation from the Sun on the motion of planetary satellites.
\citet{gordeeva1968} obtained an analytic solution of a variable
$\varepsilon = 1-e^2$ in the doubly averaged CR3BP at the quadrupole level approximation
using a complete elliptic integral and the Jacobi elliptic function.
Let us cite an example from recent days:
\citet{prokhorenko2001} gave a geometric interpretation,
as well as a topological illustration,
of the class of orbits that appear on the \citeauthor{lidov1961} diagram utilizing cylindrical and spherical coordinates.
\citeauthor{prokhorenko2001} has pursued this line of study in detail and
published many papers \citep[e.g.][]{prokhorenko2002a,prokhorenko2002b,prokhorenko2010,prokhorenko2014}.

At this point,
we feel obliged to mention the series of works achieved by Mikhail A. \citeauthor{vashkovyak1999} who has been intensively working on this problem for a long time.
Subjects of \citeauthor{vashkovyak1999}'s publications that stem from and extend \citeauthor{lidov1961}'s work range over a large variety.
Let us cite several examples:
Extension of \citeauthor{orlov1965a}'s work using so-called the numerical--analytic method \citep{vashkovyak2005a,vashkovyak-teslenko2009,vashkovyak2010},
analysis and classification of families of orbits on the $(e, \omega)$ and $(\alpha, c_1)$ planes including the cases of $\alpha > 1$ \citep{vashkovyak1981a,vashkovyak1981b},
obtaining special solutions in the singly averaged CR3BP \citep{vashkovyak2005b,vashkovyak-teslenko2005,vashkovyak-teslenko2008},
addition of more than one perturber on planar and circular orbits \citep{vashkovyak1976,vashkovyak1981c},
construction of analytic solutions of doubly averaged system using elliptic integrals and elliptic functions \citep{vashkovyak1999},
a study of the orbital evolution of certain types of artificial Earth satellites for searching suitable orbits of space VLBI \citep{vashkovyak1990},
and more.
In addition,
the preprint \citep{vashkovyak2008} that we cited earlier
(p. \pageref{pg:vashkovyak2008})
provides a unique and extensive review of the past and modern studies of
the averaged CR3BP.
Readers of this monograph are strongly encouraged to read \citet{vashkovyak2008}
for gaining a comprehensive understanding of the historical development of the subject.

It seems that
\citeauthor{lidov1961}'s work attracted the world's attention from the early days,
and it has been developed not only by the Soviet (Russia) community
but also by the western academic community.
Several examples:
\citet{kovalevsky1964,kovalevsky1966} achieved a similar accomplishment to
\citeauthor{orlov1965a} as early as \citet{orlov1965a,orlov1965b}.
\citet{lorell1965} categorized regions in the \citeauthor{lidov1961} diagram and
tabulated the type of motions of satellites in a great detail.
\citet{felsentreger1968},
while categorizing the lunar satellite orbits,
devised a version of the \citeauthor{lidov1961} diagram
when the perturbing body exerts a gravitational effect due to its equatorial bulge $(J_2)$.
Although it is not an early work,
\citeauthor{antognini2015} exactly reproduced the \citeauthor{lidov1961} diagram
using the variables $\Theta$ (equivalent to $c_1$) and
$C_{\rm KL}$ (equivalent to $\frac{5}{2}c_2$)
with additional information on the period values of the $g$-oscillation
\citep[][his \mysymfigO 1 on p. 3614]{antognini2015}.

Let us also mention that \citeauthor{lidov1961} himself kept working on
the subject and left many publications.
For example,
he dealt with the doubly averaged CR3BP when $\alpha$ is large \citep{lidov1974}.
Also, his work on the doubly averaged general (non-restricted) 3BP is
now famous for the term ``happy coincidence''
(see p. \pageref{pg:happycoincidence-intro} of this monograph).
Let us cite the corresponding paragraph:
\begin{quote}
``Let us note, that the integrability of the non-restricted problem under consideration is, in a way, a happy coincidence.
If one would like take into account the following terms in the expansion by the parameter $|r_{01}|/|r_2|$,
then $\overline{\cal H}$ will be dependent on $g_2$, and the problem will no more be integrable. A similar situation takes place already in the restricted three body problem.''
\citep[][the second line from the bottom on p. 475]{lidov1976}.
\end{quote}

In \citeauthor{lidov1976}'s three-body system with point masses $(m_0, m_1, m_2)$,
$r_{01}$ denotes the radial vector of the mass $m_1$ relative to the mass $m_0$,
$r_{2}$  denotes the radial vector of the mass $m_2$ relative to the barycenter of $m_0$ and $m_1$,
$\overline{\cal H}$ is doubly averaged disturbing Hamiltonian (their Eq. (13) on p. 474), and
$g_2$ is argument of pericenter of the mass $m_2$.

Talking about the term ``happy coincidence,''
readers of this monograph might be interested in a paradoxical remark that \citet{lithwick2011} made about it:
\begin{quote}
``The fact that the exterior body's argument of periapse does not appear to
  quadrupole order has been called `a happy coincidence'
  because it makes the system integrable
  (Lidov {\&} Ziglin 1976; Laskar {\&} Bou\'e 2010).
  However, it is perhaps more of an unhappy coincidence
  in view of the fact that it has misled some researchers into neglecting
  the role of the planet's eccentricity.''
  \citep[][their footnote 2 on p. 2]{lithwick2011}
\end{quote}
\label{pg:happycoincidence}

\subsubsection{$c_2$-like parameter in later work\label{sssec:c2-like}}
We presume that the discovery and use of the parameter $c_2$ is
\citeauthor{lidov1961}'s unique achievement
which is not seen in the works by \citeauthor{kozai1962b},
\citeauthor{moiseev1945a}, or \citeauthor{vonzeipel1910}.
The combination of $c_1$ and $c_2$ is still employed in modern
solar system dynamics. We see examples in studies of the identification of
object groups possibly with a common dynamical origin,
such as among the near-Earth asteroids or meteoroids
\citep[e.g.][]{ryabova2006,ohtsuka2006,ohtsuka2007,babadzhanov2015,babadzhanov2017}.

In association with this subject,
we would like to note that some later authors independently
``discovered'' $c_2$-like parameters as a flag of libration of
perturbed body's argument of pericenter.
For example, \citet[][their pp. 68--69]{kinoshita2007a} showed that
the separatrix between the libration and circulation regions
on the equi-potential diagram is expressed by the quantity
\begin{equation}
  C_{\rm se} = 2 (3 h  -1) ,
  \label{eqn:KN2007-8}
\end{equation}
where $h$ is equivalent to \citeauthor{lidov1961}'s $c_1$.
\citeauthor{kinoshita2007a} claimed that the libration of perturbed body's
argument of pericenter $(\omega)$ happens when both the conditions
\begin{equation}
  h < 0.6, \quad
  C < C_{\rm se},
  \label{eqn:KN2007-p68-b}
\end{equation}
are satisfied (their condition (b) on p. 69).
$C$ in Eq. \eqref{eqn:KN2007-p68-b} is equivalent to
$W^\ast_{\Oalsqr}$ in Eq. \eqref{eqn:R2-final},
the factor $ \frac{1}{16} \frac{{\cal G}}{a'} \left( \frac{a}{a'} \right)^2$
being stripped off. It is easy to show that their condition 
$C < C_{\rm se}$ in Eq. \eqref{eqn:KN2007-p68-b} is equivalent to
\citeauthor{lidov1961}'s condition, $c_2 < 0$.
The specific form of $C$ in \citet[][their Eq. (6) on p. 68]{kinoshita2007a} is as follows:
\begin{equation}
  C = \left(2 + 3 e^2\right) \left(3 \cos^2 i -1\right)
    + 15 e^2 \sin^2 i \cos 2 \omega ,
  \label{eqn:KN2007-6}
\end{equation}
while \citeauthor{lidov1961}'s $c_2$ can be rewritten from Eq. \eqref{eqn:L59} as
\begin{equation}
  30 c_2 = 12 e^2 - 15 e^2 \sin^2 i + 15 e^2 \sin^2 i \cos 2 \omega .
  \label{eqn:lidov-c2-30}
\end{equation}
Subtracting Eq. \eqref{eqn:lidov-c2-30} from Eq. \eqref{eqn:KN2007-6}, and
using the definition of $c_1$ in Eq. \eqref{eqn:L58}, we get
\begin{equation}
\begin{aligned}
  C - 30 c_2 &= 4 - 6 \left( \cos^2 i \left(e^2 -1\right) + 1\right) \\
             &= 2 \left( 3 c_1 - 1 \right) .
  \label{eqn:C-30c2-1}
\end{aligned}
\end{equation}
Since  $2 \left( 3 c_1 - 1 \right)$ in Eq. \eqref{eqn:C-30c2-1} is
equivalent to $C_{\rm se}$ defined in Eq. \eqref{eqn:KN2007-8},
we can rewrite Eq. \eqref{eqn:C-30c2-1} as
\begin{equation}
  C - C_{\rm se} = 30 c_2 .
  \label{eqn:C-30c2-3}
\end{equation}
From Eq. \eqref{eqn:C-30c2-3} we see that
the condition $C \lessgtr C_{\rm se}$ is equivalent to $c_2 \lessgtr 0$.
Incidentally, note that in \citet[][p. Z390]{vonzeipel1910} we found a quantity
equivalent to \citeauthor{kinoshita2007a}'s $C$ in Eq. \eqref{eqn:KN2007-6}.
\citeauthor{vonzeipel1910} used the symbol $h$ to indicate this quantity.
Care must be taken with the confusing fact that
\citeauthor{vonzeipel1910}'s  $h$ is different from
\citeauthor{kinoshita2007a}'s $h$:
\citeauthor{vonzeipel1910}'s  $h$ is equal to \citeauthor{kinoshita2007a}'s $C$, and
\citeauthor{kinoshita2007a}'s $h$ is the same as \citeauthor{vonzeipel1910}'s $k^2$ (and \citeauthor{lidov1961}'s $c_1$).

Let us mention another example of the rediscovery of $c_2$.
It seems that
\citet[][their \mysymfigO 16 on p. 506]{nagasawa2008} independently
devised an equivalent plot to the \citeauthor{lidov1961} diagram.
They employed a pair of variables $C$
(equivalent to $C$ in Eq. \eqref{eqn:KN2007-6}) and
$h$ (equivalent to \citeauthor{lidov1961}'s $c_1$)
as parameters, and drew the possible motion area of the perturbed body
on the $(\sqrt{h},C)$ plane in the doubly averaged CR3BP.
They explicitly showed the form of $C$ as a function of $h$
on the boundaries of the diagram:
$C = 10 - 6h$ on the upper boundary of the diagram,
$C = 6h -  2$ on the boundary between circulation and libration (which is equivalent to Eq. \eqref{eqn:KN2007-8}), and
$C = -20 -24h +12\sqrt{15h}$ on the lower boundary.
See the caption of their \mysymfigO 16 for more detail.
\citeauthor{nagasawa2008}
also obtained an explicit form of the attainable maximum eccentricity
as a function of $C$ and $h$
(see their Eq. (13) on p. 506.
 A similar relationship is presented in
\citet[][his Eqs. (29) and (31) on p. 3613]{antognini2015}).

\subsubsection{Citations of \citeauthor{lidov1961}'s work\label{sssec:lidov-citation}}
\begin{figure}[t]\centering
\ifepsfigure
 \includegraphics[width=\singlefigwidth\textwidth]{citations.eps} %fig29
\else
 \includegraphics[width=\singlefigwidth\textwidth]{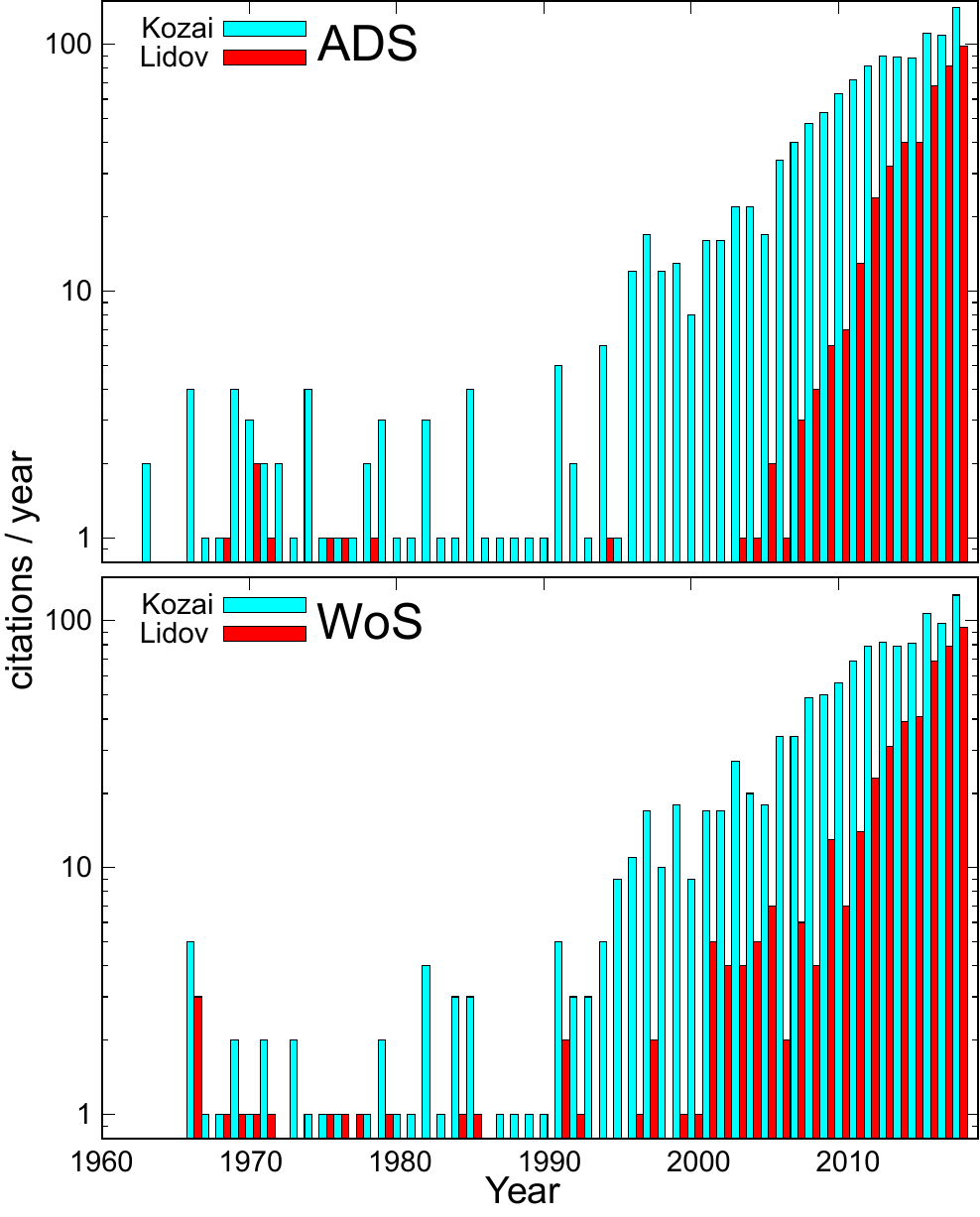} %fig29
\fi
  \caption{%
  Time series of citation frequencies of
  \citeauthor{kozai1962b}'s publications (cyan bars) and
  \citeauthor{lidov1961}'s  publications (red  bars)
  between 1961 and 2018
  based on ADS (upper panel) and WoS (lower panel) as of May 1, 2019.
  As for the citation data for ``\citeauthor{kozai1962b},''
  we bunched the citation frequencies of two publications:
  The main paper \citep{kozai1962b} and 
  a meeting abstract by the same author with the same title
  published in the same issue of the same journal
  (abstracts of papers presented at the 111th Meeting of
   the American Astronomical Society at Yale University,
   New Haven, Connecticut, August 26--29, 1962,
   on p. 579 of \textit{The Astronomical Journal,\/} \textit{67}, 1962).
  Similarly, as for the citation data for ``\citeauthor{lidov1961},''
  we bunched the citation frequencies of closely relevant publications:
  The original publication \citep{lidov1961},
  its English translations \citep{lidov1962,lidov1963},
  and the papers with the same subject \citep{lidov1963a,lidov1963b}.
}
  \label{fig:citations-LK}
\end{figure}

As we have seen, 
\citeauthor{lidov1961}'s achievements
on doubly averaged CR3BP at the quadrupole level approximation is
practically equivalent to \citeauthor{kozai1962b}'s work.
Nevertheless, it seems that
\citeauthor{lidov1961}'s work had not been cited as frequently as
\citeauthor{kozai1962b}'s.
We guess this is mainly because of the disparity of popularity of
the journals that published their papers.
The first publication (as far as we found on ADS) that cites
both \citet{lidov1962} and \citet{kozai1962b} at the same time is \citet{lowrey1971}.
This is a paper on the orbital evolution of a meteorite named Lost City,
where \citeauthor{kozai1962b}'s work is cited as
``\textit{Kozai\/} [1962] found that the argument of perihelion of the minor
  planet (1373) Cincinnati librates about $90^\circ$ $\ldots$,''
while \citeauthor{lidov1962}'s work is introduced as
``\textit{Lidov\/} [1962] has obtained two simple formulas that are
  surprisingly successful in predicting the over-all variations,''
together with the definition of $c_1$ and $c_2$ \citep[][p. 4086]{lowrey1971}.
Nevertheless for the next 33 years since \citet{lowrey1971},
it seems that not only the practical equivalence of these two works
but \citeauthor{lidov1961}'s work itself has been largely forgotten,
inconspicuous, and perhaps even ostracized,
judging from the frequency of citations seen on ADS.
However, after the publication of a paper about the secular dynamics of
irregular satellites of the giant planets \citep{cuk2004}
which cites \citet{lidov1962} and \citet{kozai1962b} at the same time
for the first time in the twenty-first century,
\citeauthor{lidov1961}'s work began rapidly gaining attention.
Nowadays, more and more people have come to know the equivalence of
\citeauthor{lidov1961}'s work and \citeauthor{kozai1962b}'s work.
Consequently, the citation frequency of \citeauthor{lidov1961}'s work
has been soaring. Inspired
by a figure in \citet[][his \mysymfigO 1 of Preface on p. vi]{shevchenko2017} and
by a figure in \citet[][his \mysymfigO 2 on p. 366]{takahashi2015e-3},
we have produced a pair of plots concerning the time series of the citation frequency of
\citeauthor{lidov1961}'s and
\citeauthor{kozai1962b}'s publications
between 1961 and 2018 using the citation database stored
in ADS and in WoS (\mysymfigO \ref{fig:citations-LK}).
Note that the vertical axis of the plots in the figure has a logarithmic scale.
Therefore, from these plots, we clearly see that the citation frequencies of
\citeauthor{kozai1962b}'s work, and also
\citeauthor{lidov1961}'s work, have exponentially increased in the recent era.

In \mysymfigO \ref{fig:citations-LK},
we also find that \citeauthor{lidov1961}'s work was cited
a non-negligible number of times in the twentieth century.
We can particularly see this fact on the citation database stored on WoS
(the lower panel of \mysymfigO \ref{fig:citations-LK}).
Consider the fact that citation databases such as ADS or WoS are not always complete for old literature published in the 1960s or earlier.
Also, recall that the total number of publications at that time was
much smaller than the present.
Then,
we should say that \citeauthor{lidov1961}'s work has been rather well recognized in the academic community since then.
As seen in his obituary that we mentioned on p. \pageref{pg:lidovobituary},
Michail L'vovich Lidov's social status was (and still is)
very high in the former Soviet Union and in the Russian Federation.
Therefore, we may find many more publications that cites his work
if we were to survey literature published in the Soviet-related domestic
academic community, and in particular, those written in Russian.

According to the rapid increase of attention to \citeauthor{lidov1961}'s work,
more and more people began using the prefix ``\citeauthor{lidov1961}--\citeauthor{kozai1962b},''
not just ``Kozai,'' for designating this subject.
As far as we have done searches on ADS and WoS,
\citet[][in their abstract]{michtchenko2006} seems to be the first publication
that used the prefix ``\citeauthor{lidov1961}--\citeauthor{kozai1962b}'' (as ``the \citeauthor{lidov1961}--\citeauthor{kozai1962b} resonance'').
On the other hand, the prefix ``Kozai--Lidov'' showed up
first in \citet[][in their abstract, as ``Kozai--Lidov resonance'']{sidlichovsky2005}.
Recently, we even find publications that use the prefix
``\citeauthor{lidov1961}--\citeauthor{kozai1962b}'' but cite only \citet{lidov1961}, not \citet{kozai1962b},
such as \citet{ulivieri2013}.

\subsubsection{Interrelation to \citeauthor{kozai1962b}'s work\label{sssec:i11n2kozai}}
Here, we would like to mention some interrelation between
\citeauthor{lidov1961}'s work and \citeauthor{kozai1962b}'s work.
These two works have been so far presumed to have been
independently carried out. However, this may not be entirely true.

\citeauthor{lidov1961}'s work (including its English translations)
on this subject, as well as \citeauthor{kozai1962b}'s work,
were mainly published between 1961 and 1964.
Let us itemize some major relevant events that happened
during this period along a timeline:
\begin{itemize}
\item Sometime in 1961,
\citet{lidov1961} was published in \textit{Iskusstvennyye Sputniki Zemli\/}.
Note that the publication of this paper could be earlier or later than
the international conference in Moscow that we mention below.
We could not identify the exact publication date.
\item On November 20--25, 1961,
there was a conference on
general and applied problems of theoretical astronomy in Moscow.
\citet{grebenikov1962} made a detailed report of this conference.
\citeauthor{lidov1961} participated in it and gave a presentation
on the subject that we have discussed in this monograph.
\citeauthor{kozai1962b} was also invited to this conference
as a part of the US delegation (at that time \citeauthor{kozai1962b} was
working at the Smithsonian Astrophysics Observatory, Massachusetts, USA), and
gave a talk about the motion of close artificial satellites.
\citeauthor{kozai1962b} and \citeauthor{lidov1961} met each other
at this conference, and had a short conversation
(Kozai 2017, personal communication).
It was their first and their last direct encounter.
\item On May 28--30, 1962,
an IUTAM (International Union of Theoretical and Applied Mechanics) symposium on the dynamics of satellites was held in Paris.
This time \citeauthor{lidov1961} did not attend the conference,
but his presentation was delivered by a delegate
\citep[][his footnote 4 on p. 7]{shevchenko2017}.
Meanwhile, \citeauthor{kozai1962b} participated in the symposium and gave a presentation
on the gravitational potential of the Earth derived from satellite motions.
Note also that
L. I. Sedov who worked on the spacecraft Luna-3's peculiar motion
(see p. \pageref{pg:sedovswork} of this monograph) was present at this symposium.
In \citet{obi1979e} we find a group photo of the symposium participants
including both \citeauthor{kozai1962b} and Sedov.
\item On August 26--29, 1962, \citeauthor{kozai1962b}'s work
on this subject was (presumably) first unveiled
as a paper presented at the 111th Meeting of the American Astronomical Society.
See the caption of \mysymfigO \ref{fig:citations-LK} for more details.
\item On August 29, 1962, \citet{kozai1962b} was received by \textit{The Astronomical Journal.\/}
\item In October 1962,
the first English translation of \citet{lidov1961} was published in \textit{Planetary and Space Science\/} \citep{lidov1962}.
\item In November 1962, \citet{kozai1962b} was published in \textit{The Astronomical Journal.\/}
\item In August 1963, the second English translation of \citet{lidov1961} was published in \textit{AIAA Journal Russian Supplement\/} \citep{lidov1963}.
\item
Sometime in 1963,
two proceedings volumes were published.
One of them \citep{subbotin1963} was the proceedings volume of the 1961 Moscow conference that contains \citet{lidov1963a} and \citet{kozai1963b},
both of which were written in Russian.
The other one \citep{roy1963}    was the proceedings volume of the 1962 Paris symposium   that contains \citet{lidov1963b} and \citet{kozai1963a},
both of which were written in English.
\item Sometime in 1964,
the Moscow 1961 conference proceedings were translated into English and published \citep{lidov1964,kozai1964}\footnote{%
According to the foreword of the proceedings volume,
the translation from Russian into English in this volume was achieved by a machine translation.
This is quite a surprising fact if we consider the publication year \citeyearpar{lidov1964}.
Let us literally cite the entire foreword:
\begin{quote}
``This document is a machine translation of Russian text which has been
processed by the AN/GSQ--16 (XW--2) Machine Translator, owned and operated
by the United States Air Force. The machine output has been fully post-edited.
Ambiguity of meaning, words missing from the machine's dictionary, and
words out of the context of meaning have been corrected.
The sentence word order has been rearranged for readability due to the
fact that Russian sentence structure does not follow the English
subject--verb--predicate sentence structure.
The fact of translation does not guarantee editorial accuracy,
nor does it indicate USAF approval or disapproval of the material translated.''
\end{quote}
}.
\end{itemize}

It seems that, at some later point \citeauthor{lidov1961} recognized
\citeauthor{kozai1962b}'s work when he extended his study on this subject.
For example, \citet{lidov1974} cites \citet{kozai1962b}.
Also, \citeauthor{lidov1961} uses the Hamiltonian formalism
in his later publication such as \citet{lidov1974,lidov1976},
similar to \citeauthor{kozai1962b}.
Meanwhile \citeauthor{kozai1962b} had recognized
\citeauthor{lidov1961}'s work earlier, and
cited \citeauthor{lidov1961}'s presentation given at the 1962 Paris symposium
in \citet[][p. K591]{kozai1962b} as
``A lunar problem with high inclination was similarly studied by Lidov (1962),''
with the corresponding reference of
``Lidov, M. L. 1962, Paper presented at the International Symposium on Dynamics of Satellites, Paris'' (p. K598).
In addition, the AIAA version of the English translation of
\citeauthor{lidov1961}'s work seems to be
officially ``reviewed'' by \citeauthor{kozai1962b}.
A footnote in \citet{lidov1963} says:
\begin{quote}
``Translated from \textit{Iskusstvennye Sputniki Zemli\/} (\textit{Artificial Earth Satellites\/})
(Academy of Science Press, Moscow, 1961), No.~8, pp. 5--45.
Translated by Jean Findlay, Green Bank, West Va.
Reviewed by Yoshide Kozai, Smithsonian Astrophysical Observatory, Cambridge, Mass.'' (p. 1985)
\end{quote}
Therefore,
it is possible that the five extra paragraphs found at the leading part of
\citet{lidov1963} that we mentioned in Section \ref{ssec:publicationsbylidov}
(p. \pageref{ssec:publicationsbylidov} of this monograph),
which the original version \citep{lidov1961} and the other translation
\citep{lidov1962} do not contain,
may have been added by the reviewer, \citeauthor{kozai1962b}.
Note that \citeauthor{kozai1962b}'s first name is wrongly typed in the above:
``Yoshide'' should be ``Yoshihide''.

Bearing these circumstances in mind,
we collectively judge that \citeauthor{lidov1961}'s work and
\citeauthor{kozai1962b}'s work thus interacted with each other,
unlike the common view that they were independently carried out
without any interaction.
Some people may have been already aware of the personal interrelation
between \citeauthor{lidov1961} and \citeauthor{kozai1962b} before we did.
For example, \citet{tremaine2014} presents a description
of the history of the \citeauthor{lidov1961}--\citeauthor{kozai1962b} {\mainword} study:
\begin{quote}
``The nonlinear trajectories of the linear instabilities we have described
are known as Kozai, Kozai--Lidov, or Lidov--Kozai oscillations.
Although Laplace had all of the tools needed to investigate this phenomenon,
it was only discovered in the early 1960s by Lidov in the Soviet Union
and brought to the West by Kozai.''
(at the beginning of their Section IV, in the left column of p. 776)
\end{quote}

We are not sure if \citeauthor{tremaine2014} knew
that \citeauthor{lidov1961} and \citeauthor{kozai1962b} met
each other at the conference in Moscow in 1961.
But the above description literally claims that
\citeauthor{lidov1961} recognized the dynamical phenomenon first, and then
\citeauthor{kozai1962b} spread it over the western community after that.
However recall that, in this monograph, we are not going to judge
which should be called the first and which should be the second
between \citeauthor{lidov1961} and \citeauthor{kozai1962b}.
We do not think this kind of discussion has a significant meaning anymore,
as we now know that \citeauthor{vonzeipel1910} had recognized
this dynamical mechanism much earlier than
\citeauthor{lidov1961} or \citeauthor{kozai1962b} did.
We are not yet familiar with what Laplace did along this line of work
as \citeauthor{tremaine2014} pointed out, but
we will continue to investigate this historical theme.

\subsubsection{Choice of terms\label{sssec:mainword}}
At the end of this subsection let us deviate from
\citeauthor{lidov1961}'s work, and think more generally about the terms
that people use for describing the secular dynamical phenomena
that we deal with.
In this monograph,
we have basically used the term ``the \citeauthor{lidov1961}--\citeauthor{kozai1962b} \textit{{\mainword},\/}''
because we want to emphasize the oscillating nature of the phenomenon.
However, as readers are well aware,
different people make different choice of words on this same phenomenon
other than \textit{{\mainword}.\/}
So it may be interesting to consider how people have collectively
called this secular dynamics in the past literature.

To this end,
we carried out a simple search on the abstracts of
all bibliographic sources registered on ADS
(i.e. refereed and non-refereed publications in astronomy and physics)
using a series of keywords,
``\textit{Kozai ---\/}'' and ``\textit{Lidov ---\/}''.
Here, ``---'' is either of the following words:
\textit{mechanism,\/}
\textit{resonance,\/}
\textit{cycle,\/}
\textit{oscillation,\/}
\textit{effect,\/}
\textit{dynamics,\/}
\textit{perturbation,\/} or
\textit{libration.\/}
Note that this is not an \mtxtsf{AND} (logical conjunction) search:
We just used fixed, consecutive phrases such as
``\textit{Kozai mechanism\/}'',
not ``\textit{Kozai\/}'' \mtxtsf{AND} ``\textit{mechanism\/}''.
Thus we are supposed to count the number of the ADS abstracts that contain
the phrase ``$\ast$\textit{Kozai mechanism\/}$\ast$''
where $\ast$ denotes the wildcard match.
Naturally,
the search result would include ``\textit{Lidov--Kozai mechanism\/}''.
Similarly, a phrase search ``\textit{Lidov mechanism\/}'' on ADS is
equivalent to the search of ``$\ast$\textit{Lidov mechanism\/}$\ast$''
which would also find ``\textit{Kozai--Lidov mechanism\/}''.
We applied the same procedure to other words
(\textit{resonance,\/} \textit{cycle,\/} \textit{oscillation,\/} $\cdots$).
We are aware that this kind of simple search overlooks relevant expressions such as
``\textit{Lidov--Kozai secular dynamics\/}'' or
``\textit{Kozai--Lidov resonant effect\/}''.
However, although it is a crude statistical method,
through this search
we can find a general tendency
as to which term the authors of past literature have preferred for describing this secular dynamics.

\begin{table}[htbp]\centering
\caption[]{%
Result of our abstract search through ADS as of May 1, 2019.
Let us explain how the table should be read
by taking the line for the word \textit{mechanism\/} as an example.
On this line,
the first number (157) means that
         we found 157 abstracts containing the expression
``$\ast$\textit{Kozai mechanism\/}$\ast$''.
This includes the abstracts that contain expressions such as ``\textit{Lidov--Kozai mechanism\/}''.
The second number (66) means that
         we found  66 abstracts containing the expression
``$\ast$\textit{Lidov mechanism\/}$\ast$''.
This includes the abstracts that contain expressions such as ``\textit{Kozai--Lidov mechanism\/}''.
The total number of the abstracts that contain the expression
``$\ast$\textit{Kozai mechanism\/}$\ast$'' or
``$\ast$\textit{Lidov mechanism\/}$\ast$'' is thus
$157 + 66 = 223$,
as indicated in the right-hand column.
Note that our search targets were all bibliographic sources in astronomy and physics registered on ADS.
These include non-refereed publications such as conference abstracts, proceedings, and the \mtxtsf{arXiv} e-prints.
Note also that we carried out the same search using several other words as well
(\textit{circulation,\/}
 \textit{state,\/}
 \textit{phenomenon,\/}
 \textit{behavior,\/}
 \textit{motion\/}), and ended up with a null result.
}
\label{tbl:choiceofword}
\begin{tabular}[hbtp]{lrrr}
\hline
word in ``---'' &
  \multicolumn{1}{r}{\textit{Kozai ---\/}} &
  \multicolumn{1}{r}{\textit{Lidov ---\/}} &
  \multicolumn{1}{r}{total} \\
\hline
\textit{mechanism\/}    & 157 & 66 & 223 \\
\textit{resonance\/}    & 157 & 10 & 167 \\
\textit{cycle\/}        & 103 & 18 & 121 \\
\textit{oscillation\/}  &  76 & 43 & 119 \\
\textit{effect\/}       &  39 & 13 &  52 \\
\textit{perturbation\/} &   7 &  6 &  13 \\
\textit{dynamics\/}     &   9 &  3 &  12 \\
\textit{libration\/}    &   7 &  0 &   7 \\
\hline
\end{tabular}
\end{table}

We summarized our search result in Table \ref{tbl:choiceofword}.
Note that the resulting statistics would be different if we do
the same kind of search through the full text of the literature.
But as the full text search of all the literature is practically impossible,
we just have to presume that the abstracts overall represent
the common preference in the academic world.

Among the words tabulated in the leftmost column of
Table \ref{tbl:choiceofword}, particularly the most popular ones
(\textit{mechanism,\/}
 \textit{resonance,\/}
 \textit{cycle,\/}
 \textit{oscillation,\/} and
 \textit{effect\/}),
we prefer to use \textit{{\mainword}\/} in this monograph.
\textit{Mechanism,\/} which is seen most frequently for expressing the subject,
generally emphasizes (in our opinion) an internalized dynamical structure
rather than the apparent, manifested characteristics it realizes. 
We think that the accomplishment of
\citeauthor{lidov1961} and \citeauthor{kozai1962b} should be acknowledged
in the first place by their quantitative and accurate recognition of
the apparent, overt phenomenon itself (\textit{oscillation\/}),
followed by the detailed inspection of what causes it (\textit{mechanism\/}).
\textit{Resonance\/} is also common according to the statistics shown in
Table \ref{tbl:choiceofword},
but we think this term is not entirely appropriate for denoting this subject.
It is because a resonance in general is supposed to work between a frequency
that one object (or phenomenon) possesses and another frequency that another object (or phenomenon) possesses.
For example, mean motion resonance between Jupiter and an asteroid,
spin-orbit resonance between planetary rotation and its revolution,
secular resonance between a forced orbital frequency and an intrinsic (proper) orbital frequency,
and so forth.
But in the considered problem (CR3BP) that practically has only two objects,
the perturber's motion does not have any frequencies after averaging:
All of its orbital elements become constant or vanish.
If we purposely search any resonance-related argument in the system,
it would be the relation $g = \pm\frac{\pi}{2}$
seen in the libration of perturbed body's argument of pericenter.
However, this involves just one variable of just an object.
Therefore we would like to avoid using the word \textit{resonance\/} here.
\textit{Effect\/} sounds okay, but we think this word's meaning can be too broad
for the purpose.
\textit{Cycle\/} sounds as fine as \textit{{\mainword},\/} and
we could bring this word up as our best preference instead.
The only concern of ours is that
we are not quite sure if we can safely use the term
\textit{cycle\/} for a non-cyclic phenomenon such as what emerges
when the eccentricity of the perturbing body is in effect
(see \mysymfigO \ref{fig:eLK-example} of this monograph).
In the end,
the word \textit{{\mainword}\/} sounds relatively better for us and
more appropriate than others.
This word describes an externally manifested characteristic of the phenomenon.
Also, we can probably use \textit{{\mainword}\/} for non-cyclic phenomena,
such as seen in \mysymfigO \ref{fig:eLK-example},
more safely than \textit{cycle.\/}

However, there is obviously a large uncertainty and degrees of freedom
in the choice of terms here.
Although we will continue preferring the term \textit{{\mainword}\/} in this monograph,
readers can choose other appropriate words at their discretion
as long as the selected word does not introduce unnecessary mix-up or misunderstanding.

\subsection{Achievements of \citeauthor{vonzeipel1910}'s work\label{ssec:summary-zeipel}}
Published more than fifty years before the works of \citeauthor{lidov1961} and \citeauthor{kozai1962b},
we can regard \citeauthor{vonzeipel1910}'s work as a pioneering manifestation
of the application of the Lindstedt series to the theoretical framework of
the doubly averaged CR3BP.
Standing on the fundamentals of celestial mechanics that had been established by his time,
\citeauthor{vonzeipel1910} clearly manifested, in the form of the Lindstedt series,
the existence of stationary points and periodic trajectories
that show up in perturbed body's motion in the doubly averaged CR3BP.

After making a general and detailed mathematical preparation for
how he deals with the Lindstedt series,
\citeauthor{vonzeipel1910} moves on to considering
the actual doubly averaged disturbing function for CR3BP.
One of the interesting aspects of his work is that
he focuses on the topology of the disturbing potential's surface.
He first searches local extremums of
the doubly averaged disturbing function $R$
by calculating the first derivatives of $R$ such as  $\DP{R}{\left(e^2\right)}$.
Then he calculates the second derivatives such as $\DP[2]{R}{\left(e^2\right)}$ to
confirm whether the discovered local extremum is a local minimum, a local maximum, or a saddle point.
As a result, \citeauthor{vonzeipel1910} found that
the doubly averaged disturbing function for the inner CR3BP $(\alpha < 1)$
possesses a pair of local minima along the axis of $g = \pm\frac{\pi}{2}$
when $k^2 = \left(1-e^2\right) \cos^2 I < \frac{3}{5}$
in the limit of $\alpha \ll 1$
(Section \ref{ssec:icr3bp} of this monograph).
Evidently, this result is equivalent to those found by
\citeauthor{lidov1961} and \citeauthor{kozai1962b} years later.
Although handwritten,
\citeauthor{vonzeipel1910} drew a pair of equi-potential diagrams for two possible modes:
One       is for the case when perturbed body's argument of pericenter $g$ circulates from 0 to $2\pi$, and
the other is for the case when $g$ librates around $g = \pm\frac{\pi}{2}$
(\mysymfigO \ref{fig:vZ10-f6-7} of this monograph).
In the limit of $\alpha \ll 1$,
he gave an explicit function form of
perturbed body's eccentricity at the equilibrium points
($e_{0.2}$ in Eq. \eqref{eqn:Z81}).
In \citeauthor{vonzeipel1910}'s work,
there is no mention on parameters relevant to \citeauthor{lidov1961}'s $c_2$.

\citeauthor{vonzeipel1910} extended his theory to a more general
inner CR3BP where $\alpha$ is not very small.
Based on a set of formulas presented in 
\citeauthor{tisserand1889}'s work,
\citeauthor{vonzeipel1910} obtained numerical values of $R$
as a function of $\alpha$ and $I_0 \equiv \cos^{-1} k$
(\mysymfigO \ref{fig:R020-table} of this monograph).
This enabled him to estimate numerical values of
the smallest inclination $I_{0.2}$ and its dependence on $\alpha$
for the local minima of $R$ to show up
along the axis of $g = \pm\frac{\pi}{2}$.
His result turned out almost identical to what
\citeauthor{kozai1962b} later obtained
(\mysymfigO \ref{fig:I02-table} of this monograph).

What makes \citeauthor{vonzeipel1910}'s work unique and different
from \citeauthor{kozai1962b}'s or \citeauthor{lidov1961}'s is the fact
that he dealt with not only the inner CR3BP
but also the outer CR3BP where $\alpha > 1$
(Section \ref{ssec:ocr3bp} of this monograph).
Similar to the way he studied the inner problem,
\citeauthor{vonzeipel1910} paid attention to
the topology of the disturbing potential's surface, and
searched local extremums of the doubly averaged disturbing function $R$
for the outer CR3BP by calculating its first and second derivatives.
As a result, he found that
in the limit of very small $\alpha'$ $\left( =\alpha^{-1} \right)$,
$R$ possesses various local extremums when $k^2 < \frac{1}{5}$:
The origin $(e \cos g, e \sin g)=(0,0)$ can be a local maximum, a local minimum, or a saddle point.
Also, a pair of local minima along the axis of $g = \pm\frac{\pi}{2}$ can show up.
In addition, a pair of saddle points along the axis of $g = 0$ or $g = \pi$ can take place.
\citeauthor{vonzeipel1910} also left a triplet of handwritten equi-potential diagrams for three possible modes
(\mysymfigO \ref{fig:vZ10-f8-10} of this monograph).
In the limit of $\alpha' \ll 1$,
\citeauthor{vonzeipel1910} gave explicit function forms of perturbed body's eccentricity at the equilibrium points
($e'_{2.0}$ in Eq. \eqref{eqn:Z103} and $e'_{0.2}$ in Eq. \eqref{eqn:Z104})
in the same way as he did for the inner problem.

Similar to his treatment of the inner problem,
\citeauthor{vonzeipel1910} then extended his theory to a more general outer
CR3BP where $\alpha'$ is not very small.
As a result, he obtained numerical values of $R$
as a function of $\alpha'$ and $I_0$
(\mysymfigO \ref{fig:Rd020-table} of this monograph).
This enabled him to estimate the numerical values of
the smallest inclinations ($I'_{2.0}$ and $I'_{0.2}$) and
their dependence on $\alpha'$ for the local extremums of $R$ to show up.
His results agree well with our numerical confirmation
(\mysymfigO \ref{fig:Id202comp} of this monograph),
although influence of mean motion resonance was not taken into account
in his work.

As we mentioned on p. \pageref{pg:outerproblemhasnotbeenstudied} of this monograph,
studies of the outer CR3BP lagged behind those of the inner problem until the 1990s,
when people came to seriously realize the existence of TNOs and extrasolar planets.
After the discoveries of these objects,
we find many more serious studies on the outer CR3BP.
But in \citeauthor{vonzeipel1910}'s era,
the only ``outer'' objects that satisfied the condition $\alpha > 1$
were just some comets,
while many more ``inner'' objects (main belt asteroids) were recognized.
In this regard, we are impressed by the fact that \citeauthor{vonzeipel1910}
dealt with the outer problem in an equivalent manner and depth to the inner problem.

\citeauthor{vonzeipel1910} applied his theory to the actual asteroids
in the solar system.
Among the 665 asteroids that were recognized at that time,
he chose six of them as candidates
that may have a large secular oscillation of eccentricity because
they have large inclinations
(p. \pageref{pg:vonzeipel-actualasteroids} of this monograph).
Although none of the six asteroids turns out to be a $g$-librator
(\mysymfigO \ref{fig:aasteroids-ecosg}),
we found, from a modern viewpoint, that two of the six asteroids
that \citeauthor{vonzeipel1910} picked
((2) Pallas and (531) Zerlina)
belong to the Pallas family.
The other four asteroids also have similar orbital elements.
Therefore in hindsight, we may want to say that
\citeauthor{vonzeipel1910}
unintentionally approached the theoretical recognition of
asteroid families earlier than Kiyotsugu Hirayama \citep{hirayama1918,hirayama1922}
who is regarded the first to advocate the concept of asteroid families
based on the theory of proper orbital elements.
However, we are sure that \citeauthor{vonzeipel1910}'s
interest was just confined to finding unusual asteroids having a large secular
oscillation of eccentricity and inclination, and that finding or recognizing
family-like groups of small bodies was not his priority%
\footnote{%
According to Kiyotsugu Hirayama,
there had been several attempts to detect asteroidal groups prior to his,
such as
\citet{kirkwood1877,kirkwood1890,kirkwood1891},
\citet{tisserand1891b}, or
\citet{mascart1899,mascart1902}.
\citeauthor{hirayama1922} judged that they did not meet with success, writing:
\begin{quote}
``Anyhow these attempts were not successful,
if they were not a complete failure.
This was mainly due to the reason that
the actual orbits were taken for comparison,
whereas these orbits are varied remarkably by the action of the planets.
The variation is in fact very slow,
but the effect steadily accumulates during a long interval of time.''
\citep[][the first paragraph on his p. 56]{hirayama1922}
\end{quote}

From a modern viewpoint, we interpret that
the preceding attempts failed because they did not use
the proper elements when comparing asteroidal orbits.
In this regard, readers may want to add \citeauthor{vonzeipel1910} to
the list of failures,
as he too did not bring asteroidal proper elements in his calculation.
However, we would like to emphasize that
what \citeauthor{vonzeipel1910} used was not ``the actual orbits'' 
(as Hirayama wrote in the above) of asteroids either:
In \citeauthor{vonzeipel1910}'s Section Z18 where he dealt with six asteroids
as candidates of $g$-librator
(see p. \pageref{pg:Z1910-actualasteroids} of this monograph),
he calculated their $I_0$ and $I_{0.2}$ through his own secular dynamical theory.
Thus, although we admit that he did not reach the discovery of
asteroid families at the depth that \citeauthor{hirayama1922} achieved,
we claim that
\citeauthor{vonzeipel1910}'s calculation on this matter should be
better known and recognized in the community
as a part of the establishment process of asteroid family studies.
Readers can find a number of good reviews \citep[e.g.][]{yoshida1997e,yoshida2011,knezevic2016,yoshida2019e}
as for historical background of the concept of asteroid family
with a focus on \citeauthor{hirayama1922}'s work.%
}% End of \footnote
.

\citeauthor{vonzeipel1910} also discussed the motion of two actual comets in the solar system:
1P/Halley    (Section Z21. See our Section \ref{sssec:Halley} of this monograph),
and 2P/Encke (Section Z29).
The discussion on the motion of 1P/Halley is particularly quantitative and detailed.
\citeauthor{vonzeipel1910} made an accurate calculation of
the location of the equilibrium points of the doubly averaged disturbing function for 1P/Halley
(\mysymfigS \ref{fig:Halley-eR} and \ref{fig:Rmap-halley} of this monograph),
although his treatment was limited to the framework of a doubly averaged CR3BP
where Jupiter on a circular orbit serves as the only perturber.

\citeauthor{vonzeipel1910} further extended his theory for investigating more subjects,
and came to further quantitative conclusions.
One of them is the calculation of the doubly averaged disturbing function of CR3BP
when the orbits of the perturbed and perturbing bodies act like rings in a chain.
As we mentioned in Section \ref{ssec:orbitintersection} of this monograph,
it seems that this subject had not been seriously investigated until
\citet{gronchi1999a,gronchi1999b} worked on it.
Another extension is a treatment of the motion of the perturbed body using a different set of canonical
variables when its eccentricity is close to the maximum, $k'$
(Section \ref{sssec:motionneare=kd} of this monograph).
Here, \citeauthor{vonzeipel1910} stated that a perturbed body's orbit
with a large eccentricity and small inclination
outside the perturber's orbit can stably exist.
Again in a modern viewpoint,
we might want to regard this statement as his unintentional prediction
of the existence of the scattered TNOs (with a small inclination).
Yet another example of his theoretical extensions are
the outline of the method to deal with systems with multiple perturbing bodies
(Section \ref{sssec:pluralperturbers} of this monograph).
To date, this approximation has been refined and made more sophisticated,
and is extensively used in modern solar system dynamics
as we mentioned in Section \ref{sssec:pluralperturbers}.

\subsection{Earlier studies by \citeauthor{vonzeipel1910}\label{ssec:earlier-zeipel}}

Earlier in this monograph (p. \pageref{pg:fromprototypetocomplete}),
we wrote that
\citeauthor{vonzeipel1910}'s work in \citeyear{vonzeipel1910},
particularly the subject in its Chapter II,
can be regarded as a prototype study
that was later developed into a more sophisticated
canonical perturbation theory.
From a broader historical perspective,
we understand that \citet{vonzeipel1910} occupies an important midpoint
in the major flow of celestial mechanics that showed remarkable development
from the late nineteenth century to the early twentieth century.
The flow was established by the giants of that era
\citep[examples of major publications:][]{%
tisserand1889,tisserand1891,tisserand1894,tisserand1896,%
picard1891,picard1893,picard1896,%
poincare1892,poincare1893,poincare1899,poincare1905,poincare1907,poincare1909,poincare1910,%
birkhoff1913,birkhoff1915,birkhoff1922,birkhoff1926}.
And, \citeauthor{vonzeipel1910} himself undoubtedly belongs to the giants
due to his own outstanding achievements.

Although we wrote about \citeauthor{vonzeipel1910}'s essential contribution to the historical flow of celestial mechanics,
we are sure that very few people have been aware of his work published in \citeyear{vonzeipel1910}.
And, we are surer that even fewer people know that
there are two more studies by \citeauthor{vonzeipel1910}
published earlier than \citeyear{vonzeipel1910}:
\citet{vonzeipel1898,vonzeipel1901}.
We regard these publications as being forerunners of later, more complete products
\citep[e.g.][]{vonzeipel1905,vonzeipel1910,vonzeipel1915,vonzeipel1916a,vonzeipel1916b,vonzeipel1917a,vonzeipel1917b}.
In the present subsection,
we make a brief summary of \citeauthor{vonzeipel1901}'s achievement in the two early papers.
He may have published more work along these lines prior to \citeyear{vonzeipel1898},
but so far we have not found any others than these two.

\subsubsection{\citet{vonzeipel1898}\label{sssec:Zeipel1898}}
\citeauthor{vonzeipel1910} dealt with the full three-body problem
in both his \citeyear{vonzeipel1898} and \citeyear{vonzeipel1901} papers,
and regarded the restricted problem as an extreme case.
The first paper \citep{vonzeipel1898} is entitled
``Sur la forme g\'e\'nerale des \'el\'ements elliptiques dans le probl\`eme des trois corps,''
written in French and published in
\textit{Bihang till Kongl Svenska Vetenskaps--Akademiens Handlingar.\/}
This paper begins with a series of equations that defines the system
that he considers, without any descriptions on
the purpose or background of his intentions.
Literally citing the first part of the paper, we see that
\citeauthor{vonzeipel1898} deals with the general three-body system
described in the Jacobi coordinates:
\begin{quote}
``It is assumed that the masses $m_1$ and $m_2$ are small with respect to the mass $m_0$ and that $m_1$ and $m_2$ are of the same order of magnitude.
The motion of $m_1$ is referred to a system of coordinates whose origin is situated in $m_0$ and whose axes retain the same directions.
The coordinates of $m_1$ in this system are $x_1$, $y_1$, $z_1$.
The center of gravity of the masses $m_0$ and $m_1$ is the origin of another system of coordinates, the axes of which are parallel to the axes of the first.
The coordinates of $m_2$ in the latter system are $x_2$, $y_2$, $z_2$.
If the masses $m_1$, $m_2$, $m_3$ attract according to Newton's law, the equations of motion are [$\ldots$],'' (p. 3)
\end{quote}
After this, a description of the standard canonical equations of motion
using the three-body Hamiltonian follows.

\citeauthor{vonzeipel1898}'s intention in his \citeyear{vonzeipel1898} paper is to show that,
as long as the orbits in a three-body system are roughly circular
(i.e. both the bodies have small eccentricity),
it is possible to construct approximate solutions for their orbital variation
in the form of trigonometric series without generating terms that secularly increase.
Although \citeauthor{vonzeipel1898} does not use the word
Lindstedt at all in this paper, it is obvious that
this series is nothing but the Lindstedt series.
Also, from his assumption that the eccentricities of the bodies are small,
we presume that he tried to construct
the Lindstedt series around the origin $(0,0)$
on the $(e \cos g, e\sin g)$ plane.

In this paper
\citeauthor{vonzeipel1898} utilizes the result given in \citet{tisserand1889}
for obtaining the function form of the averaged Hamiltonian
(referred to as $[F_1]$)
when the eccentricities of both the bodies are small (Eq. (5) on his p. 12).
Going through similar mathematical procedures that
we have seen in Section \ref{ssec:Z10eom-secular} of this monograph,
\citeauthor{vonzeipel1898} searches conditions for the series to
exist---conditions that the orbital elements are expressed in
trigonometric series that do not contain secularly increasing terms.

\label{pg:Zeipelscondition}
\citeauthor{vonzeipel1898} eventually found that the series he seeks exists
if the pair of solutions of a quadratic equation become unequal and negative.
The quadratic (actually, a quartic) equation that he mentions appears on his p. 23 as follows:
\begin{align}
& \alpha^{(1)4} +
\left\{
   a_{11}^{(1)} c_{11}^{(1)}
+ 2a_{12}^{(1)} c_{21}^{(1)}
+  a_{22}^{(1)} c_{22}^{(1)}
\right\} \alpha^{(1)2} \nonumber \\
& \quad\quad\quad +
\left|
\begin{array}{cc}
a_{11}^{(1)} & a_{12}^{(1)} \\
a_{21}^{(1)} & a_{22}^{(1)} \\
\end{array}
\right|
\left|
\begin{array}{cc}
c_{11}^{(1)} & c_{12}^{(1)} \\
c_{21}^{(1)} & c_{22}^{(1)} \\
\end{array}
\right|
= 0 .
\tag{$\alpha$}
\end{align}
An equivalent equation shows up again on his p. 47 as:
\begin{align}
x^4 & + \left\{ a_{11}c_{11} + 2a_{12}c_{12} + a_{22}c_{22} \right\} x^2
\nonumber \\
    &
\quad \quad
+
\left|
\begin{array}{cc}
a_{11} & a_{12} \\
a_{21} & a_{22} \\
\end{array}
\right|
\left|
\begin{array}{cc}
c_{11} & c_{12} \\
c_{21} & c_{22} \\
\end{array}
\right|
= 0 ,
\tag{$\alpha$}
\end{align}
with the coefficients defined as:
\begin{equation}
  a_{ik} = \frac{\partial^2 [F_1]}{\partial \xi_i \partial \xi_k},
\:\:
  c_{ik} = \frac{\partial^2 [F_1]}{\partial \eta_i \partial \eta_k} ,
\:\:
  (i, k = 1, 2)
\end{equation}
together with other variables defined as:
\begin{align}
  \xi_k     &= \:\:\: \sqrt{2\left(\Lambda_k - H_k\right)} \cos g_k, \\
  \eta_k    &=       -\sqrt{2\left(\Lambda_k - H_k\right)} \sin g_k, \\
  \Lambda_k &= \beta_k L_k, \:\:
   H_k       = \beta_k G_k,
\end{align}
where $L_k, G_k, g_k$ $(k=1,2)$ are Delaunay elements with
\begin{alignat}{2}
 \beta_1 &= \frac{\mu_0}{\mu_1}, & \tpspcE
 \beta_2 &= \frac{\mu_1}{\mu_2} \frac{m_2}{m_1}, \\
 \mu_1   &= m_0 + m_1,           & \tpspcE
 \mu_2   &= m_0 + m_1 + m_2 ,
\end{alignat}
and $\mu_0 = m_0$.
\citeauthor{vonzeipel1898} only showed approximate forms of
the coefficients $a_{11}, c_{11}, a_{22}, \cdots$ on his p. 50.
Exhibiting specific function forms of these coefficients is a major subject
of his next paper \citep{vonzeipel1901}.
Here he just yields his final result---a pair of solutions of Eq. $(\alpha)$ as:
\begin{alignat}{1}
  x_1^2 &= -A^2 \left\{ \left(
  5 \cos^2 J_0 - 1 + 2\frac{m}{m'} \sqrt{\alpha} \cos J_0 \right)^2 \right. \nonumber \\
& \quad\quad\quad\quad\quad\quad\quad\quad\quad\quad
  \left. -25 \sin^4 J_0 \biggr\}\right. + \cdots ,
  \label{eqn:vZ1898-x1} \\
  x_2^2 &= -A^2 \left\{ \left( 5 \cos^2 J_0 - 1 \right) \frac{m}{m'} \sqrt{\alpha} + 2 \cos J_0 \right\}^2 + \cdots ,
  \label{eqn:vZ1898-x2}
\end{alignat}
where $J_0$ is the initial mutual inclination of the orbits of the two bodies,
$A$ is a positive coefficient that is a function of $\alpha^2$ where
$\alpha$ is the ratio of semimajor axis, $\frac{a}{a'}$ or $\frac{a_1}{a_2}$.
We presume that $m$ and $m'$ are the masses of the inner and outer bodies,
respectively,
although \citeauthor{vonzeipel1898} did not leave any specific definitions of them in this paper.

Using
$x_1$ in Eq. \eqref{eqn:vZ1898-x1} and
$x_2$ in Eq. \eqref{eqn:vZ1898-x2},
the condition for the Lindstedt series to exist that
\citeauthor{vonzeipel1898} mentioned
(``the pair of solutions of a quadratic equation become unequal and negative'') is expressed as:
\begin{equation}
 x_1^2 < 0, \quad
 x_2^2 < 0, \quad
 x_1^2 \neq x_2^2 .
\end{equation}
Then, he makes a conclusion when $\alpha$ is very small as:
\begin{quote}
``We can conclude that
$$
\begin{aligned}
  x_1^2 < 0 &        \quad\mbox{\rm if $\displaystyle{\cos^2 J_0 > \frac{6}{10}}$} \\
  x_2^2 < 0 &        \quad\mbox{\rm for all the values of $J_0$}  \\
  x_1^2 \neq x_2^2 & \quad\mbox{\rm when $\displaystyle{\cos^2 J_0 \neq \frac{2}{3}}$}
\end{aligned}
$$

We have
$$
  \cos^2 J_0 = \frac{6}{10}
$$
for
$$
  |J_0| = 39^\circ 14' \cdots
$$
When
$$
  |J_0| > 39^\circ 14' \cdots
$$
and [when] $\alpha$ is a small quantity,
the series of the previous section are in default.'' (p. 50)
\end{quote}

\citeauthor{vonzeipel1898} thus claims that the initial mutual inclination
$J_0$ cannot exceed $\cos^{-1} \sqrt{\frac{6}{10}} \sim 39^\circ 14'$
for the Lindstedt series to exist.
Note that for \citeauthor{vonzeipel1898} at that time,
the non-existence of the Lindstedt series literally meant the dynamical instability of
the entire three-body system.

At the end of his \citeyear{vonzeipel1898} paper,
\citeauthor{vonzeipel1898} leaves a paragraph that closes the paper as follows:
\begin{quote}
``The same thing has been shown before in other ways.
Tisserand has generalized the method of Delaunay for the Moon by making a series of transformations, each of which eliminates a part of the disturbing function.
In these methods of demonstration, Mr. Poincar\'e has always used the partial differential equation of Hamilton--Jacobi.'' (p. 51)
\end{quote}

We are not completely sure what \citeauthor{vonzeipel1898} meant by 
``the same thing'' (``la m{\^e}me chose'') here,
but we presume that he describes the historic use of canonical transformation
in perturbation theories ever since Delaunay.
This is particularly true when we mention the elimination of
fast-oscillating variables from the disturbing function by averaging.

The critical value of the initial mutual inclination that \citeauthor{vonzeipel1898} found
in this paper, $J_0 = \cos^{-1} \sqrt{\frac{6}{10}}$, is
evidently equivalent to what \citeauthor{kozai1962b} and
\citeauthor{lidov1961} found in the doubly averaged inner CR3BP.
We have not confirmed whether or not \citeauthor{vonzeipel1898}'s third condition,
$\cos^2 J_0 = \frac{2}{3}$ (that would realize $x_1^2 = x_2^2$),
really causes a problem in this study.
But we can at least say that \citeauthor{vonzeipel1898} was already aware
of the critical value of $\cos^{-1} \sqrt{\frac{6}{10}} \sim 39^\circ 14'$
at the end of the nineteenth century---more than sixty years
before the works of \citeauthor{lidov1961} and \citeauthor{kozai1962b} came out.

Although \citeauthor{vonzeipel1898} dealt with the non-restricted
three-body problem with point masses $m_0$, $m_1$, $m_2$
in his \citeyear{vonzeipel1898} paper,
it seems that he did not give deep consideration
on the mass ratio such as $\frac{m_1}{m_2}$.
His solutions
$x_1^2$ in Eq. \eqref{eqn:vZ1898-x1} and
$x_2^2$ in Eq. \eqref{eqn:vZ1898-x2}
contain terms with the mass ratio $\frac{m}{m'}$ as a coefficient.
But these terms are ignored when he calculates
$J_0$ by taking the limit of $\alpha = \frac{a}{a'} \to 0$.

\subsubsection{\citet{vonzeipel1901}\label{sssec:Zeipel1901}}
The next paper \citep{vonzeipel1901} is entitled
``Recherches sur l'existence des s\'eries de M. Lindstedt,''
written in French, and again published in
\textit{Bihang till Kongl Svenska Vetenskaps--Akademiens Handlingar.\/}
It seems that now \citeauthor{vonzeipel1901} pays particular
attention to the influence of the ratio of the semimajor axes
of the two bodies $\left(\alpha = \frac{a}{a'}\right)$
on the critical inclination values, when $\alpha$ is not so small.
The example used in this study is the Sun--Jupiter--Saturn system
where $\alpha \sim 0.5431$.

For better facilitating an understanding of
what \citeauthor{vonzeipel1901} did in his \citeyear{vonzeipel1901} publication,
let us take a look at a short introductory article, \citet{callandreau1902},
which briefly summarizes what is done in \citet{vonzeipel1901}.
Let us cite the article in its entirety, since it is very short.
Note that \citet{callandreau1902} was originally written in French, and
the following English translation is ours:
\begin{quote}
``Mr. v. Zeipel thus states the conclusions of his work:
If the inclination of the orbit of an asteroid exceeds a certain limit (about $30^\circ$, slightly variable with the ratio of the axes), the series of Mr. Lindstedt does not exist.
We cannot avoid the secular terms, and the orbit is likely unstable $\ldots$ .
Perhaps this is the cause of this surprising fact that, among about 500 small planets, there is only one, Pallas, whose inclination exceeds $30^\circ$.
It was first necessary to make the study of the series possible, no matter how great the mutual inclination of the orbits [is];
Mr. v. Zeipel took advantage of the work of Jacobi and Tisserand.''
\end{quote}

As is seen in its title,
\citeauthor{vonzeipel1901} frequently uses the term ``Lindstedt series''
in his \citeyear{vonzeipel1901} paper.
This paper begins with the conclusion of his previous work \citeyearpar{vonzeipel1898} as follows.
Note that the partial use of the italic characters in the following is
by the original author (\citeauthor{vonzeipel1901}):
\begin{quote}
``By means of Mr. Lindstedt's series, one can, as we know, give the coordinates of the three bodies a form, where time never leaves the signs [of] sin and cos.

\hspace*{1em}
In an earlier M\'emoire (\textit{On the general form of elliptical elements in the problem of three bodies. Bihang Till K. Svenska Vetenskapsakademiens handlingar.\/} Band 24. Afd I. N:o 8)
I gave a new exhibition of the series in question by a little generalizing the known theory of Mr. Poincar\'e for the periodic and asymptotic solutions in the problems of the dynamics.

\hspace*{1em}
By eliminating the nodes, as did Jacobi, it was possible to study the series,
no matter how great the mutual inclination of the orbits is.
So I found (loc. cit.) as necessary and sufficient conditions for the existence of the series of Lindstedt:
\begin{description}
\item[] \textit{1:0 that the orbits are roughly circular.\/}
\item[] \textit{2:0 that the roots of the equation of the fourth degree\/}
\end{description}
\begin{equation}
\begin{aligned}
  \sigma^4
&- \left( a_{1,1}c_{1,1} + 2a_{1,2}c_{1,2} + a_{2,2}c_{2,2} \right) \sigma^2 \\
& \quad \quad
+
\left|
\begin{array}{cc}
a_{1,1} & a_{1,2} \\
a_{2,1} & a_{2,2} \\
\end{array}
\right|
\left|
\begin{array}{cc}
c_{1,1} & c_{1,2} \\
c_{2,1} & c_{2,2} \\
\end{array}
\right|
= 0
\end{aligned}
  \tag{Zb1-\arabic{equation}}
  \stepcounter{equation}
  \label{eqn:Zb1}
\end{equation}
\textit{where\/}
\begin{equation}
\begin{aligned}
  a_{ik} &= \left. \frac{\partial^2 \left[\frac{fm_2}{\Delta}\right]}{\partial \xi_i \partial \xi_k} \right|_{\xi_r = \eta_r = 0}, \\
  c_{ik} &= \left. \frac{\partial^2 \left[\frac{fm_2}{\Delta}\right]}{\partial \eta_i \partial \eta_k} \right|_{\xi_r = \eta_r = 0}, \\
& (i, k = 1, 2)
\end{aligned}
  \tag{Zb2-\arabic{equation}}
  \stepcounter{equation}
  \label{eqn:Zb2}
\end{equation}
\textit{are real and unequal.\/}'' (p. 3)%
\end{quote}

Note that in the above,
we renamed the equation numbers in the original publication
from (1) to \eqref{eqn:Zb1}, and from (2) to \eqref{eqn:Zb2},
for avoiding potential confusions.
Note also that the small subscript $r$ placed together with the partial derivatives in the right-hand side of
Eq. \eqref{eqn:Zb2} as $\xi_r = \eta_r = 0$ means,
although \citet{vonzeipel1901}
did not explicitly explain, that we must substitute
$\xi_r = \eta_r = 0$ (where $r = 1$ and $2$) after 
completing the partial differentiation
\citep[see p. 22 and p. 47 of][]{vonzeipel1898}.

Readers might find that Eq. \eqref{eqn:Zb1} is close to Eq. $(\alpha)$
in \citet{vonzeipel1898}. But there are differences:
\begin{itemize}
  \item The variable $x$ in Eq. $(\alpha)$ has been replaced for the variable $\sigma$ in Eq. \eqref{eqn:Zb1}.
  \item The subscripts of the coefficients in Eq. \eqref{eqn:Zb1} have
   a comma such as $a_{1,1}$ instead of $a_{11}$ in Eq. $(\alpha)$.
  \item The sign of the second term in the left-hand side is positive in Eq. $(\alpha)$, while it is negative in Eq. \eqref{eqn:Zb1}.
\end{itemize}

Among the above-mentioned differences, the third one is the most significant.
It changes the condition for the existence of the Lindstedt series
that expresses the periodic solutions
which \citeauthor{vonzeipel1901} seeks.
More specifically saying,
\begin{itemize}
  \item \citet[][p. 23]{vonzeipel1898} depicts the condition as,
right after Eq. $(\alpha)$,
``The two values of $\alpha^{(1)2}$ are unequal and $<0$''.
This means that all solutions $\alpha^{(1)}$ (or $x$ on p. 47) would need to be imaginary.
  \item \citet[][p. 3]{vonzeipel1901}  depicts the same condition as
\textit{``the roots of the equation of the fourth degree (Eq. \eqref{eqn:Zb1}) are real and unequal.''\/}
Here, \textit{the roots\/} designate $\sigma$ in Eq. \eqref{eqn:Zb1}.
\end{itemize}

Although we can show that there is no practical contradiction between the above two%
\footnote{%
Although the above mentioned two conditions are not identical in general,
we can practically see that both of them are satisfied in this case.
Let us symbolically denote Eq. $(\alpha)$ and Eq. \eqref{eqn:Zb1}
respectively as follows:
\begin{alignat}{1}
  x^4      + 2p x^2      + q &= 0 , 
  \label{eqn:vZ1898-alpha-rev} \\
  \sigma^4 - 2p \sigma^2 + q &= 0 .
  \label{eqn:vZ1901-eqZb1-rev}
\end{alignat}
Then, their solutions would respectively have the following form:
\begin{alignat}{1}
  x^2      &= - p \pm \sqrt{p^2 - q},
  \label{eqn:x2-solution} \\
  \sigma^2 &= + p \pm \sqrt{p^2 - q}.
  \label{eqn:sigma2-solution}
\end{alignat}
Here, let us recall a statement in \citet[][p. 23]{vonzeipel1898} which claims
that $q > 0$ and that the discriminant $p^2 - q > 0$.
This means that $\sqrt{p^2 - q}$ is always real and smaller than $\left| p \right|$.
Therefore, both of the above two conditions
($x^2 < 0$      as for Eq. $(\alpha)$, and
 $\sigma^2 > 0$ as for Eq. \eqref{eqn:Zb1}) are satisfied
by means of Eqs. \eqref{eqn:x2-solution} and \eqref{eqn:sigma2-solution},
as long as \citeauthor{vonzeipel1898}'s statement is reliable.
},
we have no idea why \citeauthor{vonzeipel1898} made this change between the two publications.

He then moves on to stating the objective of the entire paper.
\citeauthor{vonzeipel1898} wrote as follows:
\begin{quote}
``In the case where the ratio $\alpha$ of the two major axes is very small, the discussion of the equation \eqref{eqn:Zb1} was very simple.
We found in this case (loc. cit. p. 50), as the upper limit of the inclination, the value $39^{\circ}.14 \ldots$

\hspace*{1em}
In this M\'emoire I want to give the analytic form of the coefficients $a_{i,k}$ and $c_{i,k}$ in detail.
I also want to apply the formulas in a special case by choosing, for $\alpha$, the ratio of the major axes of the orbits of Jupiter and Saturn.
The calculation shows that the second condition is no longer fulfilled when the inclination $J_0$ exceeds a certain limit, depending on the two ratios $\frac{m_1}{m_2}$ and $\frac{a_1}{a_2}$.

\hspace*{1em}
We will see, that the limit in question is quite low.'' (p. 4)
\end{quote}

When he expands the direct part of the disturbing function $\left( \frac{1}{\Delta} \right)$ in this publication,
\citeauthor{vonzeipel1901} again assumes that the eccentricity of both the
two bodies are small. We find his assumption in the following description
in his {\S}1:
\begin{quote}
``To find the form of the coefficients $a_{i,j}$ and $c_{i,k}$ as functions of $m_1, m_2$ and $J_0$,
we want to develop the function $\frac{1}{\Delta}$ in the increasing powers of eccentricities $e_1$ and $e_2$.
In this development it suffices to write only of the terms of the second degree, as in formulas \eqref{eqn:Zb2} one must place $\xi_r=\eta_r=0$, i.e. $e_1 = e_2 =0$ after differentiation.'' (p. 5)
\end{quote}

As \citeauthor{vonzeipel1901} states above,
in his \citeyear{vonzeipel1901} paper he first aims at expressing the coefficients
$a_{i,k}$ and $c_{i,k}$ in Eq. \eqref{eqn:Zb1} using orbital elements.
After that, he calculates their numerical values by substituting the
actual orbital elements of objects such as Jupiter or Saturn.
\citeauthor{vonzeipel1901}'s eventual purpose in this paper is
to know the dependence of the critical value of the initial mutual
inclination $J_0$ on the semimajor axis ratio, $\alpha$.
In \citeauthor{vonzeipel1901}'s theory, as developed at this point,
construction of the Lindstedt series would be impossible
when $J_0$ exceeds the critical value.

The remaining part of his \citeyear{vonzeipel1901} paper is full of detailed mathematical expositions
where \citeauthor{vonzeipel1901} expands the disturbing function in
a way that is already familiar to us.
In the procedure,
he employs the same technique that was later
used in his \citeyear{vonzeipel1910} paper---use of the
Laplace-like coefficients $b^{i,j}$, $c^{i,j}$, $e^{i,j}$.
We already browsed through how he achieved it
in Section \ref{sssec:iCR3BP-nosmall-alpha} of this monograph.

\begin{table}[t]\centering
  \caption[]{%
A reproduction of an unnumbered table in \citet[][p. 22]{vonzeipel1901}.
Note that this table is presented as an equation numbered (32)
in \citeauthor{vonzeipel1901}'s original paper, not as a table.
Note also that 
\citeauthor{vonzeipel1901}'s definition yields the relations
$a_{1,2} = a_{2,1}$ and $c_{1,2} = c_{2,1}$.
This is based on the symmetry of second derivatives (see Eq. \eqref{eqn:Zb2}).
}
\begin{tabular}[b]{c|rrr} \hline
          & $J_0=0$    & $J_0=30^\circ$ & $J_0=45^\circ$ \\
\hline
$a_{1,1}$ & $+32''.96$ & $+21''.70$     & $+15''.64$ \\
$c_{1,1}$ & $+32''.96$ & $+11''.27$     & $+ 1''.78$ \\
$a_{2,2}$ & $+43''.80$ & $+19''.83$     & $+ 8''.39$ \\
$c_{2,2}$ & $+43''.80$ & $+17''.16$     & $+ 6''.94$ \\
$a_{1,2}$ & $ -7''.58$ & $- 3''.32$     & $- 1''.29$ \\
$c_{1,2}$ & $ -7''.58$ & $+ 1''.91$     & $+ 3''.64$ \\
\hline
\end{tabular}
\label{tbl:Zb32}
\end{table}

After going through a large amount of algebra,
\citeauthor{vonzeipel1901} reached a set of numerical values of the
coefficients on his p. 22. We reproduced these as our Table \ref{tbl:Zb32}.
Using the coefficients,
\citeauthor{vonzeipel1901} finally attains the solution $\sigma$ of
the quadratic equation \eqref{eqn:Zb1}
for three values of initial mutual inclination ($J_0 = 0, 30^\circ, 45^\circ$)
under the actual mean motions $(n_1, n_2)$ and masses $(m_1, m_2)$
of Jupiter and Saturn.
Let us rewrite his results that are summarized
as an equation (on his p. 22) in a slightly different format:
\begin{equation}
\begin{aligned}
{} & J_0=0        : &\tpspcD \sigma &= \pm 47''.68,         &\tpspcD \pm & 29''.07 \\
{} & J_0=30^\circ : &\tpspcD \sigma &= \pm 18''.26,         &\tpspcD \pm & 15''.43 \\
{} & J_0=45^\circ : &\tpspcD \sigma &= \pm 1''.19\sqrt{-1}, &\tpspcD \pm &  8''.84 \\
\end{aligned}
  \tag{Zb33-\arabic{equation}}
  \stepcounter{equation}
  \label{eqn:Zb33}
\end{equation}
where $\sqrt{-1}$ denotes the imaginary unit.

From Eq. \eqref{eqn:Zb33} we know that
the existence of the Lindstedt series is assured
when $J_0 = 0$ and $30^\circ$ because all four $\sigma$ are real and unequal.
However, this is not the case
when $J_0 = 45^\circ$         because two of the four $\sigma$ are imaginary
$(\sigma = \pm 1''.19\sqrt{-1})$.
\citeauthor{vonzeipel1901} depicts this outcome as follows.
Note that the partial use of italic characters in the following is
by the original author (\citeauthor{vonzeipel1901}):
\begin{quote}
``By thus increasing the mutual inclination of the orbits of Jupiter and Saturn, the series of Mr. Lindstedt still exist, while $J_0$ reaches the value $30^\circ$.
\textit{But $J_0$ exceeding a limit slightly less than $45^\circ$, the series in question cease to exist.
Then it is no longer possible to avoid the secular terms, and the orbits are probably unstable.}'' (p. 23)
\end{quote}

\citeauthor{vonzeipel1901} then moves on to a special case when
one of the planetary masses is zero---the restricted three-body problem.
Citing his description about it:
\begin{quote}
``If the outer planet is Jupiter, the inner planet has the mean diurnal motion $n_1=742'',2$
(because of the value used for $\alpha$ and the diurnal motion $n_1=299'',1$ of Jupiter).
If, moreover, the mass $m_1$ of this planet is $=0$, so that it is an asteroid, the roots of equation \eqref{eqn:Zb1} are
$$
\sigma
\begin{cases}
 \pm m_2 n_1 \sqrt{ a_{1,1}^{(2)} c_{1,1}^{(2)}} \\
 \pm m_2 n_1 \sqrt{ a_{2,2}^{(2)} c_{2,2}^{(2)}} \\
 \quad\quad
 = \pm m_2 n_1 \frac{\alpha^2}{2} \left( c^{1,0} - c^{0,1} \right)
\end{cases}
$$

The two first roots become imaginary when $J_0$ exceeds the limit $31^\circ.1 \ \ldots$,
because, then, $c^{(2)}_{1.1}$ going through zero becomes negative.'' (p. 23)
\end{quote}

Here,
$m_2$ is the mass of the perturbing body, and
$n_1$ is the mean motion of the perturbed body.
$a_{1,1}^{(2)}$, $a_{2,2}^{(2)}$, $c_{1,1}^{(2)}$, $c_{2,2}^{(2)}$ are equivalent to
$a_{1,1}$, $a_{2,2}$, $c_{1,1}$, $c_{2,2}$ respectively,
when $m_1 = 0$ by \citeauthor{vonzeipel1901}'s definitions
(Eq. (13) on his p. 14, although we do not reproduce them here).
$c^{1,0}$ and $c^{0,1}$ are respectively equivalent to the numerical coefficients
$c^{1.0}$ and $c^{0.1}$ that we studied before
(Section \ref{sssec:iCR3BP-nosmall-alpha} of this monograph).

\citeauthor{vonzeipel1901}'s above conclusion on the critical value of $J_0$
in the restricted case $(J_0 \sim 31^\circ.1)$ must be consistent
with what he obtained in his \citeyear{vonzeipel1910} paper.
The ratio of mean motions between the inner and the outer bodies
in the above example is $\frac{742.2}{299.1}$, which yields the ratio
of the semimajor axes $\alpha = \frac{a}{a'} = 0.545587$.
Remember that we have made a plot of the quantity equivalent to $J_0$
(which we have designated as $I_{0.2}$ or $i_0$ in this monograph)
as \mysymfigO \ref{fig:I02-table}.
There, we saw that $I_{0.2}$ takes a value just above $30^\circ$
when $\alpha \sim 0.55$. This is fairly well consistent with
\citeauthor{vonzeipel1901}'s statement, $J_0 \sim 31^\circ.1$.
Recalling that the method that he employed for this purpose was common
between his \citeyear{vonzeipel1901} and \citeyear{vonzeipel1910} papers,
the agreement is no surprise.
The surprise is the fact that the dependence of
the limiting inclination value ($J_0$ or $I_{0.2}$) on
the ratio of the semimajor axes $\alpha$ was already discussed, and
its precise numerical estimate was accomplished
in 1901, the first year of the twentieth century.

The largest difference between the conclusions of \citeauthor{vonzeipel1901}'s
\citeyear{vonzeipel1901} paper and
\citeyear{vonzeipel1910} paper is that,
by the \citeyear{vonzeipel1910} work
he had made substantial progress in his theory, and
showed that a perturbed body's periodic orbit can stably exist
around a disturbing potential's local minima or maxima
even when the initial mutual inclination is larger than the critical value.
In his \citeyear{vonzeipel1910} paper,
he proved that we can construct the Lindstedt series
providing the doubly averaged disturbing function possesses local minima or maxima.
This is what we demonstrated in Section \ref{sec:vonzeipel} of this monograph.
Let us remark that, after his \citeyear{vonzeipel1901} paper,
\citeauthor{vonzeipel1901} completed his PhD thesis entitled
``Recherches sur les solutions p\'eriodiques de la troisi\`em sorte dans le probl\`eme des trois corps,'' at Uppsala University \citep{vonzeipel1904}.
Then he moved to the Paris Observatory and stayed there for about two years
(from June 1904 through September 1906)
to study celestial mechanics more deeply under the supervision of Henri Poincar\'e and Paul Painlev\'e before his \citeyear{vonzeipel1910} paper came out
\citep{mcgehee1986,barrow-green1996}.

At the end of his \citeyear{vonzeipel1901} paper,
\citeauthor{vonzeipel1901} made the following concluding statement.
Note again that the partial use of the italic characters is due to him:
\begin{quote}
``We have previously found (see my M\'emoire \textit{On the general form\/} etc. p. 50) that for a planet very close to the Sun, the corresponding limit was $39^\circ 2 \ldots$ .

\hspace*{1em}
We can therefore state the following theorem:

\hspace*{1em}
\textit{If the inclination of the orbit of an asteroid exceeds a certain limit (about $30^\circ$, slightly variable with $\alpha$), the series of Mr. Lindstedt (the absolute orbit of Gyld\'en) does not exist.
Secular terms cannot be avoided, and the orbit is likely unstable.\/}

\hspace*{1em}
\textit{Although the series of Mr. Lindstedt are only semi-convergent, it is possible to see in this theorem, the cause of this surprising fact, that among about 500 small planets there is only one (Pallas) whose inclination exceeds $30^\circ$.\/}'' (p. 23)
\end{quote}

The above conjecture on the inclination distribution of
 \textit{small planets\/}
(``\textit{petite plan\`etes\/}'' in the original expression. Most of them were the main belt asteroids at his time) is
probably associated with a description
in his \citeyear{vonzeipel1910} paper (Section Z19 in Chapter IV.
See p. \pageref{pg:Z1910-actualasteroids} of this monograph).
Nowadays, more and more asteroids are being recognized with inclination larger than $30^\circ$,
although they do not constitute the majority of the population
\citep[e.g.][]{ivezic2001,warner2009b,novakovic2011}.
Also, it is now widely believed that the orbital distribution of asteroids is sculptured not only by the secular three-body dynamics
but also by complicated processes including the radial migration of major planets
\citep[e.g.][]{morbidelli2010,walsh2011,lykawka2013,lykawka2017,lykawka2019,roig2015}.
Therefore from a modern viewpoint,
\citeauthor{vonzeipel1910}'s above statement may not be totally accurate.

However, we should also note that \citeauthor{vonzeipel1910}'s
statement is true about some part of the actual solar system structure.
One of the typical examples is the irregular satellites of the jovian giant planets \citep[e.g.][]{jewitt2005,sheppard2006,bottke2010}.
In particular,
\citet{nesvorny2003b} carried out numerical simulations and an analytic estimate
for explaining the orbital inclination distribution of
the irregular satellites of the jovian giant planets
that are rather concentrated near ecliptic.
As a result, \citeauthor{nesvorny2003b} found that
the satellite orbits that are highly inclined
with respect to ecliptic can become unstable due to ``Kozai resonance.''
As we learned,
this radially stretches the orbits of the satellites until they
get out of planetary Hill spheres,
collide with other massive satellites, or
hit the mother planet.
We consider this as a typical manifestation of \citeauthor{vonzeipel1910}'s conjecture.

Another example is
the near-Earth asteroids with a small perihelion distance.
These objects are sometimes called near-Sun asteroids \citep[e.g.][]{emelyanenko2017} or near-Sun comets \citep[e.g.][]{jones2018}.
Using the latest, self-consistent stationary dynamical model of the
near-Earth asteroid distribution,
\citet{granvik2016,granvik2018} showed that the actually observed near-Earth asteroids
with a small perihelion distance (such as $q = a(1-e) < 0.2$ au) are much fewer
than that predicted by conventional dynamical models.
This result can be interpreted
as a consequence of catastrophic disruption of the asteroids
by the thermal effect when they approach the Sun very closely.
Also, many of these asteroids are known to be in the ``\citeauthor{lidov1961}--\citeauthor{kozai1962b} state''
that enhances their eccentricity and inclination
\citep[e.g.][]{ohtsuka2006,ohtsuka2007,ohtsuka2008,ohtsuka2009,urakawa2014},
which is possibly causing their thermal metamorphism and disintegration
in the end \citep[e.g.][]{german2010,delbo2014,ito2018a}.
It is also worth a mention that the distribution of the argument of perihelion
of the Apollo near-Earth asteroids
(Earth-crossing objects with a semimajor axis $a>1.0$ au and perihelion distance $q<1.017$ au)
shows a non-uniform distribution,
which can be attributed to the ``Kozai effect'' arising from
Jovian perturbation \citep{jeongahn2014}.
An orbital alignment pattern 
that was recently recognized among the main belt comet population
\citep{kim2018} may have relevance to this.
Therefore,
we would say that \citeauthor{vonzeipel1910}'s conjecture that 
the secular three-body dynamical mechanism he found has reduced
the number of ``small planets'' with a large inclination is partially true.

 \subsection{Consideration on historical examples\label{ssec:historicEX}}
Going through \citeauthor{vonzeipel1910}'s series of works,
we became strongly impressed by how accurate his theories and calculations are
on the motion of perturbed bodies in the doubly averaged CR3BP
whether it is the inner problem or the outer one.
We should recall the fact that electronic computers did not
exist at all in his era. Therefore
his ``numerical'' calculations must have been done manually.
Nevertheless
his numerical results show a beautiful agreement with much later studies
that exploited high-speed digital computers,
as we have confirmed
(see \mysymfigS \ref{fig:I02-table}, \ref{fig:Id202comp}, \ref{fig:Halley-eR},
 and \ref{fig:Rmap-halley}).

In addition,
we should recall that during \citeauthor{vonzeipel1910}'s era,
observational evidence on the small solar system bodies was still very limited.
In Berliner Astronomisches Jahrbuch f\"ur 1911 that
\citeauthor{vonzeipel1910} cited,
only about 650 main belt asteroids,
a handful of Jupiter Trojans, and
just one near-Earth asteroid ((433) Eros)
were listed.
The number of comets that were recognized at that time was about 400
(this number comes from our own estimate
 using the list of comets and dates of their first observations
 registered in
 the JPL Small-Body Database Search Engine).
The asteroid (1373) Cincinnati
that \citeauthor{kozai1962b} brought up as an example of the $g$-librators was not discovered until 1935.
There was no artificial satellite like Luna-3 whose dynamical
behavior became a motivation of \citeauthor{lidov1961}'s work.
Discoveries of Centaurs did not come until the 1970s.
It is needless to mention TNOs or extrasolar planets that were not discovered until the 1990s.

Recognizing the limitations of scientific knowledge and tools at his time,
we are impressed by how strongly motivated \citeauthor{vonzeipel1910} was
on advancing celestial mechanics
toward the foundation of theories that can deal with the orbital motion
of the small solar system bodies in a comprehensive way.
We surmise that his eventual motivation was oriented
toward an understanding of the origin and evolution of the entire solar system.
In vindication, let us pick some of his occasional statements
on the dynamical evolution of the solar system bodies in his publications.
One example is \citeauthor{vonzeipel1901}'s \citeyearpar[][p. 23]{vonzeipel1901}
statement on the fact that most asteroids (recognized at his time) have
a smaller inclination than $30^\circ$ together with a conjecture on this.
We mentioned it in Section \ref{sssec:Zeipel1901} of this monograph.
As another example, let us take up a paragraph
placed at the end of Section Z19 of \citet{vonzeipel1910}
that describes possible dynamical processes that 
he thought took place in the early solar system.
This paragraph appears after he presented the
calculation results of $I_0$ and $I_{0.2}$ for several actual asteroids
(p. \pageref{pg:vonzeipel-actualasteroids} of this monograph).
Citing the entire paragraph:
\begin{quote}
``We shall here make an important remark from the point of view of the evolution of the planetary system.
Let us suppose that there is resistance in space against the motion.
It is well known that it will have the effect of making the eccentricities (and the major axes) smaller and smaller, finally reducing them to zero.
But then, by virtue of what preceded, the combined effect of the resistance and of secular perturbations of an outer planet will also inevitably diminish the inclinations, and lower them more or less below the limit $I_{0.2}$.
Then, if the resistance is greater against the retrograde motion, the particles that turn [around the Sun] in the reverse direction will be drawn earlier towards the Sun than the others.
There will thus come a time when the originally chaotic nebula will have, at the interior of a disturbing and dominant mass (Jupiter), the present aspect of our solar system.'' (p. Z392)
\end{quote}

\label{pg:remark-on-planetformation}
We can presume that, through the above paragraph,
\citeauthor{vonzeipel1910} intended to give an explanation on the
current status of the solar system where most of the bodies are
on prograde and only slightly or moderately inclined orbits.
At present, it is known that the solar system formation was a
combination of many physical processes that interacted with each other.
The couple of secular orbital change and nebula gas dynamics that
\citeauthor{vonzeipel1910} mentioned in the above must have been at work too,
at least partially, on forming some part of the current solar system status.
Moreover, it is becoming better known that
the secular orbital oscillation that \citeauthor{vonzeipel1910}
discussed efficiently works on shaping some sort of extrasolar planetary systems
in combination with tidal interactions with central star
and scatterings between planets
\citep[e.g.][]{wu2003,fabrycky2007,narita2009,hebrard2011}.
Thus, one may want to say that \citeauthor{vonzeipel1910}'s statements
on the solar system evolution based on his own theory hold true to a certain extent even today.

At this point, we would like to express the following opinion.
Considering the fact that \citeauthor{vonzeipel1910} far preceded \citeauthor{lidov1961} and \citeauthor{kozai1962b},
we believe it is perfectly reasonable to call the
theoretical framework and its outcome that we have discussed,
``the \citeauthor{vonzeipel1910}--\citeauthor{lidov1961}--\citeauthor{kozai1962b} {\mainword},''
instead of using the term
\citeauthor{lidov1961}--\citeauthor{kozai1962b}
or
\citeauthor{kozai1962b}--\citeauthor{lidov1961}.
To reinforce our opinion, let us pick up some historical examples for reference:
  Rodrigues' formula,
  Laplace--Runge--Lenz vector,
  Yarkovsky effect,
and
  Kuiper belt.

\paragraph{Rodrigues' formula} %
Rodrigues' formula is a famous formula that generates the Legendre polynomial $P_n (x)$.
It is one of the typical examples whose name changed
as its earlier pioneer got recognized later.  % With a complex number $z$, 
The formula is expressed as follows \citep[e.g.][p. 334]{stegun1965}:
\begin{equation}
  P_n (x) = \frac{1}{2^n n!}\DD[n]{}{x} \left( x^2 - 1 \right)^n .
  \label{eqn:rodriguesformula}
\end{equation}
According to \citet[][pp. 105--106]{askey2005},
this formula first appeared in \citet{rodrigues1816} as Olinde Rodrigues' doctoral dissertation.
However,
\citet{ivory1824} and \citet{jacobi1827} made equivalent discoveries of this formula later.
And, somehow the formula was called the Ivory--Jacobi formula for decades.
Much later, Eduard Heine paid attention to Rodrigues' work as the pioneer, and
began calling the formula Rodrigues' formula \citep{heine1878}.
See also \citet[][p. 135]{simon2015} for a summary of the history of this formula.

\paragraph{Laplace--Runge--Lenz vector} %
A vectorial quantity called the Laplace--Runge--Lenz vector is a constant vector 
appearing in the motion of a body under the potential of $\frac{1}{r}$,
where $r$ is the distance from the force center
\citep[e.g.][]{goldstein1980,vilasi2001}.
The vector is often denoted as $\bm{A}$ or $\bm{e}$, and
it shows up in the Keplerian problem too.
Using the standard notation in celestial mechanics
\citep[e.g.][p. 132]{boccaletti1996}, it is expressed as follows:
\begin{equation}
  \bm{e} = \frac{\bm{v} \times \bm{h}}{\mu} - \frac{\bm{r}}{r} ,
  \label{eqn:LRLvector}
\end{equation}
where
$\bm{v}$ is the velocity vector of the body on a Keplerian orbit,
$\bm{r}$ is its position vector,
$\bm{h}$ is its angular momentum per unit mass, and
$\mu$ is a factor including the mass and the gravitational constant.
Although nowadays it seems that the majority of the literature calls this quantity
the Laplace--Runge--Lenz vector,
this is another example of quantities whose name has changed as
the study of history goes further back.
Readers can consult \citet{goldstein1975,goldstein1976} and \citet{alemi2009}
for a detailed and complicated historical background as to
who derived this vector first, and how it has been called.
After some length of period in the mid-twentieth century
during which this quantity was referred to as the Runge--Lenz vector
\citep[e.g.][]{redmond1964,dahl1968,collas1970,heintz1974},
\citet{goldstein1975} found that Pierre-Simon de Laplace
correctly identified this vector in a much earlier era than Runge and Lenz.
\citet{goldstein1976} discovered an even earlier history where
this vector was also identified by William Rowan Hamilton,
as well as by Jakob Hermannn and Johann I. Bernoulli.
\citeauthor{goldstein1976}'s \citeyearpar{goldstein1976} recommendation was
to call this vector the Hermann--Bernoulli--Laplace vector.
\citet{subramanian1991} went even further and called it
the Hermann--Bernoulli--Laplace--Hamilton--Runge--Lentz vector.
Note that this quantity is also known as the eccentricity vector
in solar system dynamics \citep[e.g.][]{murray1999}.
This name came from the original publication of Hamilton in 1845,
according to \citet{goldstein1976}.

\paragraph{Yarkovsky effect} %
The Yarkovsky effect is a momentum transfer mechanism that works on rotating bodies.
It is caused by the anisotropic absorption and emission of thermal energy.
It works effectively on small bodies in the solar system,
causing their radial orbit migration over a long timescale
\citep[e.g.][]{bottke2001,bottke2002c,bottke2006,vokrouhlicky2015}.
The story of the discovery and re-discovery of this effect is now rather popular,
but let us briefly summarize it.
Ivan Osipovich Yarkovsky, a Polish civil engineer working in Russia,
published a private ``pamphlet'' about his idea
on this effect at the end of the 19th century.
However, the effect was practically forgotten
until Ernst Julius \"Opik (who read the pamphlet around 1909) recalled it just from his memory.
\citet[][Section \textit{9. The Yarkovsky Effect\/} on p. 194]{opik1951} made a quantitative estimate
on how the effect works in a course of studies of the collision probability of
small particles with planets, and named it the Yarkovsky effect.
Nearly at the same time, Vladimir Vyacheslavovich Radzievskii gave a study
on the same effect in his publication \citep{radzievskii1952}.
In the 1990s, detailed and quantitative studies followed along with the
actual detection of this effect in asteroidal motion
\citep[e.g.][]{farinella1998,farinella1999,vokrouhlicky1998a,vokrouhlicky1998b,chesley2003}.
In a word, we can practically say that most of the quantification of
this effect was accomplished by the people in a much later era than Yarkovsky.
However, the effect has been just called as the Yarkovsky effect;
not the \"Opik effect or the Radzievskii effect, or the \"Opik--Radzievskii effect,
nor anything else.
Note also that there is another aspect of the Yarkovsky effect that influences
the spin rate of small bodies through their irregular shape, which is now
referred to as the YORP effect \citep[e.g.][]{rubincam2000,vokrouhlicky2003}.
``YORP'' is an abbreviation of four people
(Yarkovsky, O'Keefe, Radzievskii, Paddack), and again,
Yarkovsky's name comes first.
See \citet{beekman2005,beekman2006} for an elaborate review of
the life of Ivan Osipovich Yarkovsky,
as well as how Yarkovsky came up with the idea of this effect.

As for Rodrigues' formula and the Laplace--Runge--Lentz vector,
the earlier pioneer(s) had been forgotten or were not recognized
for a long time despite their achievement.
The formula and the vector were once named and called after the
``re''-discoverers, until someone drew attention to the achievements of
the original discoverers and an appropriate renaming was made.
As for the Yarkovsky effect,
the original achievement was just about a rough, conceptual idea.
Yarkovsky's original aim was to reinforce his hypothesis about the existence of ether,
which is, in hindsight, plainly wrong in the context of modern science.
Therefore, we should say that \citeauthor{opik1951} was so humble that
he mentioned Yarkovsky's original work and even named the effect
after Yarkovsky.
In this regard, we may say that the Poynting--Robertson effect,
working on small dust grains through radiation pressure from the Sun
\citep{poynting1904,robertson1937}, has had a similar history.
Although \citet{poynting1904} presented a quantitative description of this
drag effect, the original description of \citeauthor{poynting1904} was
rather incomplete \citep[e.g.][]{klacka2014}.
\citeauthor{opik1951} even wrote as follows:
\begin{quote}
``\textit{8. The Poynting--Robertson effect.\/} $[\cdots]$ %
A secular decrease of the major axis and of the eccentricity of the
orbit of the particle results. The effect is connected with
Poynting's name (1903), although his reasoning was based on a
misconcept of absolute motion, and his predicted effect amounted to
only one-third of the correct value. Improvement in the theory was
introduced by Larmor. The problem has been more recently revived by
H. N. Russell and thoroughly investigated by Robertson (21). For the
sake of brevity we will refer to it below as the Robertson effect. ''
\citep[][the first paragraph on his p. 191--192]{opik1951}
\end{quote}

\citeauthor{poynting1904}'s theory was later substantially rewritten and
sophisticated not only by \citet{robertson1937} but also by many other authors
\citep[e.g.][]{burns1979,klacka2014,burns2014}
until it acquired the modern, correct formulation.
And yet, the drag effect keeps the name of the historical pioneer
(\citeauthor{poynting1904}) first.

\label{pg:Kuiperbelthistory}
\paragraph{Kuiper belt} %
The previous examples that we browsed through
(Rodrigues' formula, the Laplace--Runge--Lenz vector, and the Yarkovsky effect)
all have their pioneer's name first.
We think they are excellent examples
that illustrate how these kinds of discoveries should be named.
On the other hand,
we might want to add another, arguable example: the Kuiper belt.
The name came from a famous Dutch--American astronomer, Gerard Peter Kuiper.
The point is that, Kuiper himself did not particularly claim the existence of
the small body population in the Neptune--Pluto region
($30 \mbox{ au} < r < 50 \mbox{ au}$) of the present solar system
\citep{kuiper1951a,kuiper1951b}.
And yet, the small body population has been often called the ``Kuiper belt''
\citep[see][for critical reviews]{green1999,green2004a,green2004b}.
Today it is generally recognized
that many pioneers, other than Kuiper, had anticipated the existence of
the small body population in this region
\citep[e.g.][]{edgeworth1943,edgeworth1949,cameron1962,whipple1964a,whipple1964b}.
One of the earliest predictions was already made right after the discovery of Pluto,
which was stated as follows:
\begin{quote}
``Is it not likely that in Pluto there has come to light
the \textit{first\/} of a \textit{series\/} of ultra-Neptunian bodies,
the remaining members of which still await discovery but which are
destined eventually to be detected?'' \citep[][his p. 124]{leonard1930}
\end{quote}
Note that the partial use of the italic characters is due to the original author (\citeauthor{leonard1930}).
Note also that \citeauthor{leonard1930}
considered Pluto and these bodies to be
as massive as the four terrestrial planets, as was the case in that era.

As the perception that there are more pioneers than Kuiper spread,
the small body population is more commonly and collectively referred to as
transneptunian objects (TNOs) without including Kuiper's name.
See \citet{davies2008} for a detailed historic background on this issue.
Also, recall that
\citeauthor{vonzeipel1910} had predicted in the CR3BP framework
the possible existence of the stable orbits of small bodies
outside the perturbing planet near the boundary of $e=k'$
with a small inclination
(see Section \ref{sssec:motionneare=kd} of this monograph,
 p. \pageref{pg:zeipelZ25conclusion},
 in particular the citation from \citeauthor{vonzeipel1910}'s pp. Z414--Z415).
He also made a prediction for the possible orbits of small bodies
with a small eccentricity and a large inclination,
located somewhat outside the outermost perturbing planet
when two or more perturbing planets are at work
(see Section \ref{sssec:pluralperturbers} of this monograph,
 p. \pageref{pg:justoutsidethelastplanet},
 in particular the citation from \citeauthor{vonzeipel1910}'s p. Z417).
If we regard Neptune as the outermost perturbing planet,
these small bodies would be nothing but transneptunian objects.
\label{pg:zeipelandtnos}

\subsection{\citeauthor{vonzeipel1910}--\citeauthor{lidov1961}--\citeauthor{kozai1962b} \mainword\label{ssec:ZLKcycle}}
We hope that
now readers understand the reason why we advocate using the term
``\citeauthor{vonzeipel1910}--\citeauthor{lidov1961}--\citeauthor{kozai1962b}''
for what we have discussed.
The earliest pioneer's name, \citeauthor{vonzeipel1910}, 
should be included and put first.
Also,
the correctness and completeness of \citeauthor{vonzeipel1910}'s series of works
have the same merit as
the later works by \citeauthor{lidov1961} and \citeauthor{kozai1962b},
and even better in some aspects.
This is what we have rendered throughout this monograph.
It is certainly possible that someone in the future discovers
even earlier literature than \citeauthor{vonzeipel1898}'s work
along this line of study.
But until then, let us use the prefix
``\citeauthor{vonzeipel1910}--\citeauthor{lidov1961}--\citeauthor{kozai1962b}''
for commemorating this pioneer
and his revelations on this subject.
We believe it is proper, fair, and more justifiable than using
the currently common terms such as
\citeauthor{lidov1961}--\citeauthor{kozai1962b} or
\citeauthor{kozai1962b}--\citeauthor{lidov1961}.

Currently, a front line of the hierarchical three-body problem
(including R3BP) is the motion of a perturbed body
when the perturbing body has a non-zero eccentricity, $e' > 0$
\citep[e.g.][]{naoz2016}.
In general, the inclusion of perturber's non-zero eccentricity
in the (restricted) three-body problem increases the system's degrees of freedom.
This is because it makes the disturbing potential non-axisymmetric,
and the vertical component of perturbed body's angular momentum
$\bigl( \propto \sqrt{1-e^2}\cos i \bigr)$
would not remain constant anymore even in the doubly averaged system%
\footnote{%
As described in \citet[][Appendix A on their p. 7]{lithwick2011},
we can regard that perturbing body's argument of pericenter $g'$
takes the form of $g' = \pi - h$ in the disturbing function
of the (restricted) three-body problem when $e' > 0$.
$h$ is the longitude of ascending node of the perturbed body.
Therefore $h$ remains in the disturbing function
even after the combination $h - h'$ is eliminated
through Jacobi's elimination of the nodes.
Hence $h$'s conjugate momentum,
$H = \sqrt{\mu a \left( 1-e^2 \right)} \cos i$,
could not remain constant even after the double averaging operation.%
} % End of \footnote
unless we truncate the disturbing function at the quadrupole level
so that ``the happy coincidence'' takes place.
It turned out that the perturbed body's dynamical behavior
can be significantly different in this case
as we already demonstrated in Section \ref{ssec:CR3BP-examples} of this monograph.
This phenomenon is now called the \textit{eccentric\/} \citeauthor{lidov1961}--\citeauthor{kozai1962b} {\mainword}, and
it can give us answers to many questions in solar system dynamics
that the classic framework (assuming $e'=0$) could not,
such as the origin of objects on retrograde orbits
\citep[e.g.][]{ford2000,katz2011,naoz2011,naoz2013a,delafuentemarcos2014}.
Inclusion of the general relativity into the framework of
the eccentric \citeauthor{lidov1961}--\citeauthor{kozai1962b} {\mainword} is also going on
\citep[e.g.][]{naoz2013b,will2017,sekhar2017}.
The eccentric \citeauthor{lidov1961}--\citeauthor{kozai1962b} {\mainword} in the outer problem is
also formulated, and applied to solar system dynamics
\citep[e.g.][]{naoz2017,zanardi2017,vinson2018,deElia2019}.
Here again, we insist that this line of orbital phenomenon should be called,
consistent with our discussion,
the eccentric \citeauthor{vonzeipel1910}--\citeauthor{lidov1961}--\citeauthor{kozai1962b} {\mainword}.
Note that the use of the term \textit{cycle\/} is not valid any longer here,
because the motion of the perturbed body is not exactly cyclic with
a rigid period when $e'>0$.

Let us say one thing in passing at the end of this monograph.
Readers might want to remember the fact that
a short but important publication by \citeauthor{vonzeipel1910}
on the singularities in $N$-body dynamics \citep{vonzeipel1908} had been
practically forgotten for a long time,
until Richard McGehee made a clarification of \citeauthor{vonzeipel1910}'s
fundamental contribution to singularity theory \citep{mcgehee1986,diacu1996}.
We hope that this monograph becomes a trigger for the community to reinstate
\citeauthor{vonzeipel1910}'s yet another epoch-making accomplishment
on an issue that ubiquitously shows up in celestial mechanics and
dynamical astronomy.

\paragraph{Acknowledgements.} %
The creation of this monograph began in December 2012
when the authors came across a publication \citep{bailey1996} which explicitly
mentions \citeauthor{vonzeipel1910}'s contribution on this subject.
We are deeply indebted to Junko Oguri, an excellent librarian
at National Astronomical Observatory of Japan (NAOJ),
for searching, finding, and interpreting old literature
that are necessary for our work.
Without her great expertise and capability,
we could never have accomplished the publication of this monograph.
We are also grateful to the NAOJ library itself
that made many historical
publications necessary for this monograph easily available to the authors.
We would like to thank Tadashi Mukai, an editor of this monograph series
(MEEP), not only for providing us with a unique opportunity to publish our work
but also for giving us detailed and constructive comments on our manuscript.
We have also benefited from stimulating enlightenment through discussions with many other people. Let us name just a handful of them:
Ruslan Salyamov supplied us a lot of information for interpreting literature published in the Russian language.
Simon Guez and Marc Fouchard
suggested to us some hints for interpreting antique French phrases that sometimes show up in the publications of \citeauthor{vonzeipel1910}.
Arika Higuchi informed us of the nature of the \citeauthor{vonzeipel1910}--\citeauthor{lidov1961}--\citeauthor{kozai1962b} {\mainword} caused by the galactic tidal potential.
Her aid to interpret literature written in German was also crucial for us to understand the historic background of \citeauthor{vonzeipel1910}'s work, particularly concerning \citeauthor{lindstedt1882a}'s publications.
Toshio Fukushima helped us decode historic ephemeris tables written in non-English languages.
Kiyotaka Tanikawa gave us instructions on several issues that need particular care when we deal with the restricted three-body problem compared with the full (non-restricted) three-body problem.
Hiroshi Kinoshita told us how the term \textit{resonance\/} is not appropriate for expressing the secular orbital phenomenon that this monograph deals with.
He also notified us of the inclusion of typographic errors in some equations
of \citet[][see p. \pageref{pg:typosinKozai1962b} of this monograph]{kozai1962b}.
Patryk Sofia Lykawka provided us with information on the orbital characteristics of the scattered TNOs in the solar system and their potential relationship to a class of orbits that \citeauthor{vonzeipel1910} mentioned.
Akihiko Fujii discussed with us how we should deal with the expansion of the satellite-type disturbing function in high-orders.
Yuko Kimura,
Ibuki Kawamoto, 
Junko Baba,
Makoto Yoshikawa,
Sachiko Honma,
Hiroko Kikkawa, and
Michiru Goto
provided us with the resources and environment
that were necessary for preparing this monograph.
Fumi Yoshida lectured us about several observational aspects of the small solar system bodies that are in the \citeauthor{vonzeipel1910}--\citeauthor{lidov1961}--\citeauthor{kozai1962b} {\mainword}. She also gave us a personal contact point for reaching Yoshihide Kozai.
And,
Yoshihide Kozai personally provided us with historic information about his own experience in this line of study, as well as
about his encounter with M.~L.~Lidov at the Moscow symposium in November, 1961.
We sincerely pray his soul may rest in peace.
T.I. personally thanks
Renu Malhotra,
Ayako Abe-Ouchi,
Yukiko Higuchi,
Yu Arakawa,
and
Hideo Ito
for their spiritual support while he worked on this monograph.
T.I. also appreciates the perpetual encouragement from Takiko Ito over her lifetime.
Constructive editing by Yolande McLean improved the presentation of the manuscript,
and
a very thorough review by an anonymous proofreader has remarkably refined its English expressions.
In addition, intensive shaping of our {\LaTeX} manuscript by Michio Tagawa significantly improved the appearance of the monograph.
Numerical quadrature operations,
orbit propagation by numerical integration, and
algebraic manipulations using Maple{\texttrademark}
carried out for this monograph were largely performed at
Center for Computational Astrophysics (CfCA), NAOJ.
This study has made use of
NASA's Astrophysics Data System (ADS) Bibliographic Services and
Web of Science{\texttrademark} (an online citation indexing service maintained by Clarivate Analytics).
This study is also supported by
the JSPS Kakenhi Grant
(JP25400458 [2013--2017], JP16K05546 [2016--2019], JP18K03730 [2018--2021])
and
the JSPS bilateral open partnership joint research project [2014--2015].
Finally, we wish to greatly thank
David J. Asher for providing us with detailed, critical, and fundamental reviews.
These suggested directions that immensely better the presentation and quality of
 the monograph.

\addcontentsline{toc}{section}{References}
\label{pg:beginbibliography}

\vspace{1.0\baselineskip}
\noindent
{\footnotesize
\mtxtsf{Competing interests}: 
The authors declare no competing interests.

}

\vspace{0.5\baselineskip}
\noindent
{\footnotesize
\mtxtsf{Data availability}: 
The authors declare that the data supporting this study's findings are available within the article and its Supplementary Information.

}

\vspace{0.5\baselineskip}
\noindent
{\footnotesize
\mtxtsf{Notes on References}: 
As we already explained in p. \pageref{pg:reasonforanotherbib},
some of the embedded hyperlinks to the URLs presented in the following section are truncated and will not properly function,
even though the URLs themselves are correct.
This is due to one of the technical limitations in this monograph's {\LaTeX} 
typesetting process by the publisher.
See \supinfo{2} for a complete bibliography with fully functional hyperlinks.

}

{\footnotesize
\ifterrapub

\else
 
\fi
\label{pg:endbibliography}

} % End of \footnotesize

\clearpage
\appendix
\section{\protect{\citeauthor{vonzeipel1910}}'s $P, Q$ and Their Signs\label{appen:PQpositive}}
\par
In p. Z394 (Section \ref{sssec:iCR3BP-nosmall-alpha} of this monograph, p. \pageref{pg:CallAppenPQpositive}),
\citeauthor{vonzeipel1910} gives a statement on the functions $P$ and $Q$,
``We therefore have $P>0$, $Q>0$.''
It is easy to understand the reason why $Q>0$.
By its definition in Eq. \eqref{eqn:vZ10-nn393-4},
the function $\Phi$ is always positive.
Then we see the integrand on the right-hand side of Eq. \eqref{eqn:vZ10-nn393-7Q},
$\Phi^\frac{5}{2} \left( 1-\cos u \right)^3$, is always positive.
Hence we know $Q$ is always positive.

\begin{figure}[htbp]\centering
\ifepsfigure
 \includegraphics[width=\singlefigwidth\textwidth]{Phi-p393flat.eps}%fig30
\else
 \includegraphics[width=\singlefigwidth\textwidth]{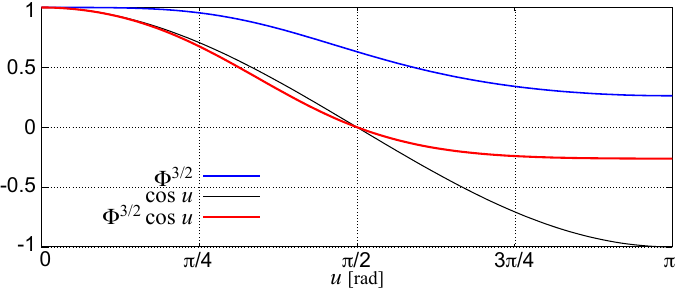}%fig30
\fi
  \caption{%
  An example plot of 
  $\Phi^\frac{3}{2}$, $\cos u$, and $\Phi^\frac{3}{2} \cos u$
  that shows their dependence on $u$.
  $\alpha = 0.6$ in this plot.
  }
  \label{fig:Phi-p393}
\end{figure}

As for $P$, we see that
the second term of $P$ on the right-hand side of
Eq. \eqref{eqn:vZ10-nn393-7P} is always positive
because the integrand of this term,
$\Phi^\frac{5}{2} \sin^2 u \left( 1-\cos u \right)$, is positive.
As for the first term,
it is enough to consider only the range $u = [0, \pi]$
due to the symmetry of the integrand, $\Phi^\frac{3}{2} \cos u$.
In addition, we can easily see the following:
\begin{itemize}
\item $\cos u$ changes its sign at $u = \frac{\pi}{2}$, but
  its behavior is point-symmetric around $(u, \cos u) = (\frac{\pi}{2}, 0)$.
\item $\Phi^\frac{3}{2}$ monotonically decreases as $u$
  from $\Phi=1$ (at $u=0$). This is because
  $\DP{\Phi^\frac{3}{2}}{u}$ is always negative unless $\alpha = 0$.
  More specifically writing:
  \begin{equation}
    \DP{\Phi^\frac{3}{2}}{u}
    = -3 \Phi^\frac{5}{2} \alpha^2 \sin u \left( 1-\cos u\right) .
    \label{eqn:dp-Phi32-u}
  \end{equation}
\item Nevertheless, $\Phi^\frac{3}{2}$ remains positive in $u = [0, \pi]$.
\end{itemize}
The above facts indicate that
the first half of the integral in question
$\left( \int_0^{\frac{\pi}{2}}   \Phi^\frac{3}{2} \cos u du \right)$
is positive, and the second half of the integral in question
$\left( \int_\frac{\pi}{2}^{\pi} \Phi^\frac{3}{2} \cos u du \right)$
is negative.
And, the former's absolute value is larger than the latter's.
Therefore their sum
\begin{equation}
  \int_0^\pi \Phi^\frac{3}{2} \cos u  du
=
  \int_0^{\frac{\pi}{2}}   \Phi^\frac{3}{2} \cos u du +
  \int_\frac{\pi}{2}^{\pi} \Phi^\frac{3}{2} \cos u du ,
  \label{eqn:dp-Phi-sum2ints}
\end{equation}
becomes positive, irrespective of $\alpha$'s value.
This proves that the first term of $P$
in Eq. \eqref{eqn:vZ10-nn393-7P} is always positive.
For illustrating the circumstance,
we made a plot of
$\Phi^\frac{3}{2}$,
$\cos u$, and
$\Phi^\frac{3}{2} \cos u$ in \mysymfigO \ref{fig:Phi-p393}
using $\alpha = 0.6$ as an example.

\section{{\protect{\citeauthor{vonzeipel1910}}}'s $\frac{\partial^2 R}{\partial e^2}$ in the Outer Problem\label{appen:DP2R}}
\par
Here we show the actual expressions of the second derivative of
the doubly averaged disturbing function $R$ for the outer CR3BP by eccentricity $e$
up to $\Oaldpen$ when $\cos 2g = \pm 1$.
This is omitted in our Eq. \eqref{eqn:vZ10-DP2Roe2-kk-ad2} on p. \pageref{eqn:vZ10-DP2Roe2-kk-ad2}.
Our calculation is based on the expressions of $R'_3$ and $R'_5$ in Eq. \eqref{eqn:Z98}.
When $\cos 2g=+1$, the derivative is
\begin{equation}
\begin{aligned}
{} & \DP[2]{R}{e}
= \left. \DP[2]{}{e} \left( R'_3 {\alpha '}^3 + R'_5 {\alpha '}^5 \right) \right|_{\cos 2g=+1} \\
{} &
= \frac{1}{1024 \left(1 - e^2\right)^\frac{15}{2}}
  \left. \Bigl\{
  \left[
                             -1536 e^{10} \right. \right. \\
{} & + \left(-11520 k^2 + 5760\right) e^8 
     + \left( 32640 k^2 - 7680\right) e^6 \\
{} & \quad
     + \left(-28800 k^2 + 3840\right) e^4 \\
{} & \quad \quad
  \left.
                          + 5760 k^2  e^2
     + 1920 k^2 -384
\right] {\alpha '}^3 \\
{} & + \left[
                             -270 e^8
              + \left(5040 k^2 + 3285\right) e^6  \right. \\
{} & \quad
+ \left(28350 k^4  + 47250 k^2 - 5400\right) e^4 \\
{} & \quad \quad
+ \left(99225 k^4  - 47250 k^2 + 2025\right) e^2 \\
{} &  \quad \quad \quad \left. \left.
+ 7560 k^4  - 5040 k^2 + 360
\right] {\alpha '}^5
\right\} .
\end{aligned}
  \label{eqn:appen:DP2Re-cos+1}
\end{equation}

When $\cos 2g=-1$, it is
\begin{equation}
\begin{aligned}
{} & \DP[2]{R}{e}
= \left. \DP[2]{}{e} \left( R'_3 {\alpha '}^3 + R'_5 {\alpha '}^5 \right) \right|_{\cos 2g=-1} \\
{} &
= \frac{1}{1024 \left(1 - e^2\right)^\frac{15}{2}}
  \left. \Bigl\{
  \left[
                             -1536 e^{10} \right. \right. \\
{} & + \left(-11520 k^2 + 5760\right) e^8 
     + \left( 32640 k^2 - 7680\right) e^6 \\
{} & \quad
     + \left(-28800 k^2 + 3840\right) e^4 \\
{} & \quad \quad
  \left.
                          + 5760 k^2  e^2
     + 1920 k^2 -384
\right] {\alpha '}^3 \\
{} & + \left[
                                5130 e^8
                + (85680 k^2 - 1935) e^6 \right. \\
{} & \quad
 + (141750 k^4  + 25650 k^2 - 10800) e^4 \\
{} & \quad \quad
 + (163485 k^4  - 103410 k^2 + 6885) e^2 \\
{} & \quad \quad \quad
  \left. \left.
 + 10080 k^4  - 7920 k^2 + 720
\right] {\alpha '}^5
\right\} .
\end{aligned}
  \label{eqn:appen:DP2Re-cos-1}
\end{equation}

As we see, the terms at $\Oaldcub$ are common between 
Eq. \eqref{eqn:appen:DP2Re-cos+1} and
Eq. \eqref{eqn:appen:DP2Re-cos-1}.
Now substituting $e = e'_{2.0}$ of Eq. \eqref{eqn:Z103}
into Eq. \eqref{eqn:appen:DP2Re-cos+1},
$\DP[2]{R}{e}$ becomes as follows at $\left(\pm e'_{2.0}, 0\right)$ up to $\Oaldpen$:
\begin{equation}
\begin{aligned}
  \left. \DP[2]{R}{e} \right|_{(\pm e'_{2.0},0)}
& \!\!\!
=-\frac{3\sqrt{5}\left(5k^2 -1 \right)}{2500k^{\frac{7}{2}}}
   {\alpha '}^3 \\
& \!\!\!\,\, % \quad
  +\frac{9\sqrt{5}\left(425k^4 - 210k^2 + 49 \right)}{5000000k^{\frac{11}{2}}}   {\alpha '}^5 .
\end{aligned}
\label{eqn:vZ10-DP2Roe2-kk-pi1}
\end{equation}

Similarly, substituting $e = e'_{0.2}$ of Eq. \eqref{eqn:Z104}
into Eq. \eqref{eqn:appen:DP2Re-cos-1},
$\DP[2]{R}{e}$ becomes as follows at $\left(0, \pm e'_{0.2}\right)$:
\begin{equation}
\begin{aligned}
  \left. \DP[2]{R}{e} \right|_{(0, \pm e'_{0.2})}
& \!\!\!
= -\frac{3\sqrt{5}\left(5k^2 -1 \right)}{2500k^{\frac{7}{2}}}
    {\alpha '}^3 \\
& \!\!\!\,\, % \quad
   +\frac{9\sqrt{5}\left(1375k^4 - 520k^2 + 81 \right)}{5000000k^{\frac{11}{2}}} {\alpha '}^5 .
\end{aligned}
\label{eqn:vZ10-DP2Roe2-kk-pi2}
\end{equation}

\begin{figure}[htbp]\centering
\ifepsfigure
 \includegraphics[width=\singlefigwidth\textwidth]{appenB_add.eps}%fig31
\else
 \includegraphics[width=\singlefigwidth\textwidth]{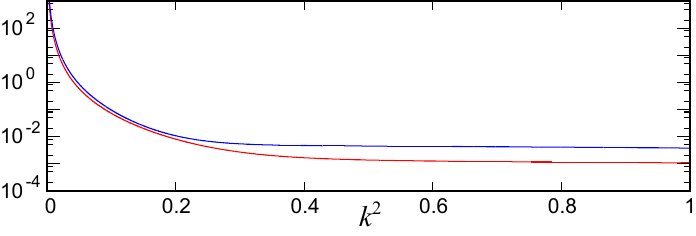}%fig31
\fi
  \caption{%
Red:  the second term in Eq. \protect\eqref{eqn:vZ10-DP2Roe2-kk-pi1}.
Blue: the second term in Eq. \protect\eqref{eqn:vZ10-DP2Roe2-kk-pi2}.
$\alpha' = 1$ is assumed for both the terms.
See \supinfo{6} for more expository figures.
}
  \label{fig:AppenB}
\end{figure}

The first terms on the right-hand side of
Eqs. \eqref{eqn:vZ10-DP2Roe2-kk-pi1} and
     \eqref{eqn:vZ10-DP2Roe2-kk-pi2} remain positive while $k^2 < \frac{1}{5}$.
It is also easy to confirm that the second terms of both the equations are positive
in the entire region of $0 \leq k^2 \leq 1$
(see \mysymfigO \ref{fig:AppenB} for a simple illustration).
Therefore, the conclusion we stated (hence \citeauthor{vonzeipel1910} stated)
as Eq. \eqref{eqn:vZ10-DP2Roe2-kk-gt0},
i.e. $\DP[2]{R}{e} > 0$,
stands true
even when the terms of $\Oaldpen$ are taken into account.

\label{finalpage}

\clearpage
%\section*{Supplementary Information}
%\href{https://www.terrapub.co.jp/onlinemonographs/meep/pdf/07/0701_2.pdf}%
%{Supplementary Information is available from the publisher's webpage.}

\section*{Appendix for \textsf{arXiv}}

In the spring of 2022, the publisher of this monograph (Terrapub) went bankrupt and the official website disappeared ({\large \url{http://www.terrapub.co.jp/}}).
Since there is no printed version of this monograph series, there is currently no way to access any of the papers published in this monograph.
Naturally, the DOI for our paper is also not linked anywhere
({\large \textsf{10.5047/meep.2019.00701.0001}}).
We have no idea if this publisher's website will be restored or when the DOI information will be corrected.

As a temporary solution to this problem, we have placed the PDF files of the main text and its supplementary material of this paper at the following sites so that interested readers can access our paper.

\begin{itemize}
\item ResearchGate
  \begin{itemize}
  \item \href{https://www.researchgate.net/publication/337164032_The_Lidov-Kozai_Oscillation_and_Hugo_von_Zeipel}{main text and supplementary information}
  \end{itemize}
\item The author's homepage
  \begin{itemize}
  \item \href{https://www.cfca.nao.ac.jp/~tito/ftp/psdoc/meep2019/}{a whole archive of the publisher's webpage}
    \begin{itemize}
    \item \href{https://www.cfca.nao.ac.jp/~tito/ftp/psdoc/0701_1.pdf}{PDF main text}
    \item \href{https://www.cfca.nao.ac.jp/~tito/ftp/psdoc/0701_2.pdf}{PDF supplementar information}
    \end{itemize}
  \end{itemize}
\item Figshare
    \begin{itemize}
    \item \href{https://doi.org/10.6084/m9.figshare.19620609}{pointers to the above material}
    \end{itemize}
\end{itemize}

\end{document}

\noindent
$\hrulefill$

(Delete before publication)
\begin{itemize}
\item Textwidth $\to$ \the\textwidth
\item Columnwidth $\to$ \the\columnwidth
\item Columnsep $\to$ \the\columnsep
\end{itemize}

\end{document}